\pdfoutput=1
\documentclass[letterpaper,hyperref]{cernyrep}
\usepackage[T1]{fontenc}
\usepackage{lmodern}
\usepackage{graphicx}
\usepackage{rotating}
\usepackage{amsmath}
\usepackage{amssymb}
\usepackage[dvipsnames]{xcolor}
\usepackage{booktabs}
\usepackage{arydshln}
\usepackage{lscape}
\usepackage{floatpag}
\floatpagestyle{plain}
\usepackage{axodraw}
\usepackage{lineno}
\usepackage{graphpap}
\usepackage{cite}
\usepackage{import}
\usepackage{enumitem}

\usepackage{caption}
\usepackage{xspace}

\usepackage{commath}

\usepackage[caption=false]{subfig}

\usepackage{siunitx}
\usepackage{multirow}

\newcommand{\eV}{\ensuremath{\text{e\kern-0.15ex{}V}}\xspace}
\newcommand{\MeV}{\ensuremath{\text{M\eV}}\xspace}
\newcommand{\GeV}{\ensuremath{\text{G\eV}}\xspace}
\newcommand{\TeV}{\ensuremath{\text{T\eV}}\xspace}

\makeatletter
\renewcommand{\thechapter}{\@Roman\c@chapter}
\makeatother

\newcommand{\NLO}[1]{N${}^{#1}$LO}

\newcommand{\NLOH}[1]{N${}^{#1}$LO${}_{\rm HEFT}$}
\newcommand{\NLOQ}[1]{N${}^{#1}$LO${}_{\rm QCD}$}
\newcommand{\NLOE}[1]{N${}^{#1}$LO${}_{\rm EW}$}
\newcommand{\NLOHone}{NLO${}_{\rm HEFT}$}
\newcommand{\NLOQone}{NLO${}_{\rm QCD}$}
\newcommand{\NLOEone}{NLO${}_{\rm EW}$}
\newcommand{\NLOQE}[2]{N${}^{(#1,#2)}$LO${}_{{\rm QCD}\otimes{\rm EW}}$}
\newcommand{\LOQ}{LO${}_{\rm QCD}$}
\newcommand{\NLOQonetb}{NLO${}_{\rm QCD}$}
\newcommand{\NLOQtb}[1]{N${}^{#1}$LO${}_{\rm QCD}$}
\newcommand{\NLOQzzero}[1]{N${}^{#1}$LO${}_{\rm QCD}^{(z\to0)}$}

\newcommand{\NLOQVBF}[1]{N${}^{#1}$LO${}_{\rm QCD}^{(\rm VBF)}$}

\newcommand{\NLOEoneVBF}{NLO${}_{\rm EW}^{(\rm VBF)}$}

\newcommand{\NLOQVBFstar}[1]{N${}^{#1}$LO${}_{\rm QCD}^{(\rm VBF^{*})}$}

\newcommand{\NLOggHVtb}[1]{N${}^{#1}$LO${}_{gg\to HZ}^{(t,b)}$}

\newcommand{\tb}{\bar{t}}
\newcommand{\bb}{\bar{b}}
\newcommand{\qb}{\bar{q}}

\newcommand{\wdecay}{}
\newcommand{\wodecays}{(w/o decays)}
\newcommand{\wdecays}{}
\newcommand{\wleptdecays}{}

\newcommand{\WishListMGaMC}{\textsc{Madgraph5}\_a\textsc{MC@NLO}\xspace}

\begin{document}

{\centering{\LARGE{\bf{Les Houches 2017: Physics at TeV Colliders\\ Standard Model Working Group Report \par }}}}

\pagenumbering{roman}

\vspace{0.7cm}
\leftline{\bf Conveners}


\noindent \emph{Higgs physics: SM issues} \\
         J.~Bendavid (CMS), \,
         F.~Caola (Theory),\,
         R.~Harlander (Theory), \,
         K.~Tackmann (ATLAS) \\
\vspace{0.1cm}

\noindent \emph{SM: Loops and Multilegs}  \\	
         G.~Heinrich (Theory), \,
         J.~Huston (ATLAS), \,
         S.~Kallweit (Theory), \,
         J.~Thaler (Jets contact),\,
         K.~Theofilatos (CMS)\\
\vspace{0.1cm}

\noindent \emph{Tools and Monte Carlos} \\	
         V.~Ciulli (CMS), \,
         E.~Re (Theory), \,
         S.~Prestel (Theory) \\
\vspace{1.0cm}

{\leftline{\bf{Abstract}}}

\vspace{0.5cm}
This Report summarizes the proceedings of the 2017 Les Houches workshop 
on Physics at TeV Colliders. Session 1 dealt with 
(I) new developments relevant for high precision Standard Model calculations, 
(II) theoretical uncertainties and dataset dependence of parton distribution functions, 
(III) new developments in jet substructure techniques,
(IV) issues in the theoretical description of the production of Standard 
Model Higgs bosons and how to relate experimental measurements, 
(V) phenomenological studies essential for comparing LHC data 
from Run II with theoretical predictions and projections for future 
measurements, 
and (VI) new developments in Monte Carlo event generators.

\vspace{0.5cm}

{\leftline{\bf{Acknowledgements}}}

\vspace{0.5cm}
We would like to thank the organizers (N.~Berger, F.~Boudjema,
C.~Delaunay, M.~Delmastro, S.~Gascon, P.~Gras, J.~P.~Guillet,
B.~Herrmann, S.~Kraml, G.~Moreau, E.~Re, P.~Slavich and D.~Zerwas) and
the Les Houches staff for the stimulating environment always present
at Les Houches. We thank the Labex ENIGMASS, CNRS, LAPTh, LAPP, the
Universit\'e Savoie Mont Blanc, and the Minist\`ere des Affaires
\'etrang\`eres for support.



\newpage
{\centering{\bf{Authors\par}}}
\begin{flushleft}
  J.~R.~Andersen$^{1}$,
  J.~Bellm$^{2}$,
  J.~Bendavid$^{3}$,
  N.~Berger$^{4}$,
  D.~Bhatia$^{5}$,
  B.~Biedermann$^{6}$,
  S.~Br{\"a}uer$^{7}$,
  D.~Britzger$^{8}$,
  A.~G.~Buckley$^{9}$,
  R.~Camacho$^{3}$
% ...
\end{flushleft}
\begin{itemize}
\item[$^{1}$] Institute for Particle Physics Phenomenology, Durham University, Durham, DH1 3LE, UK
\item[$^{2}$] Department of Astronomy and Theoretical Physics, Lund University, Sweden
\item[$^{3}$] Experimental Physics Department, CERN, CH-1211 Geneva 23, Switzerland
\item[$^{4}$] LAPP, CNRS/IN2P3 \& Universit\'e Savoie Mont Blanc, 74940 Annecy, France
\item[$^{5}$] Tata Institute of Fundamental Research, Mumbai, India
\item[$^{6}$] Institut f{\"u}r Theoretische Physik und Astrophysik, Universit{\"a}t W{\"u}rzburg, 97074 W{\"u}rzburg, Germany
\item[$^{7}$] Institute for Theoretical Physics, Georg-August-University G{\"o}ttingen, 37077 G{\"o}ttingen, Germany
\item[$^{8}$] Physikalisches Institut, Universit{\"a}t Heidelberg, Germany
\item[$^{9}$] School of Physics and Astronomy (SUPA), University of Glasgow, Glasgow, G12 8QQ, Scotland, UK
\item[$^{10}$] Instituto de F\'{\i}sica Te\'orica UAM/CSIC \& Universidad Aut\'onoma de Madrid, 28049\linebreak Madrid, Spain
\item[$^{11}$] Physik Institut, Universit{\"a}t Z{\"u}rich, CH-8057 Z{\"u}rich, Switzerland
\item[$^{12}$] Dipartimento di Fisica e Astronomia, Universit\`a di Firenze and INFN Sezione di Firenze, 50019 Sesto Fiorentino, Firenze, Italy
\item[$^{13}$] Center for Theoretical Physics, Massachusetts Institute of Technology, Cambridge, MA 02139, USA
\item[$^{14}$] Instituto de F\'{\i}sica Corpuscular, Universitat de Val\`{e}ncia -- Consejo Superior de Investigaciones Cient\'{\i}ficas, Parc Cient\'{\i}fic, 46980 Paterna, Valencia, Spain
\item[$^{15}$] Tif Lab, Dipartimento di Fisica, Universit\`{a} di Milano and INFN, Sezione di Milano, 20133 Milan, Italy
\item[$^{16}$] II. Institute for Theoretical Physics, Hamburg University, 22761 Hamburg, Germany
\item[$^{17}$] Institute for Theoretical Physics, Karlsruhe Institute of Technology (KIT), 76131 Karls\-ruhe, Germany
\item[$^{18}$] IRFU, CEA, Universit\'{e} Paris-Saclay, Gif-sur-Yvette, France
\item[$^{19}$] Department of Physics and Astronomy, University College London, London, WC1E 6BT, UK
\item[$^{20}$] The University of Oxford, Oxford, UK
\item[$^{21}$] Institute for Theoretical Particle Physics and Cosmology, RWTH Aachen University, 52056 Aachen, Germany
\item[$^{22}$] Max Planck Institute for Physics, 80805 M{\"u}nchen, Germany
\item[$^{23}$] University of Wisconsin, Madison, WI 53706, USA
\item[$^{24}$] Institute for Theoretical Physics, ETH Z{\"u}rich, CH-8093 Z{\"u}rich, Switzerland
\item[$^{25}$] Particle Physics, Faculty of Physics, University of Vienna, 1090 Vienna, Austria
\item[$^{26}$] SLAC National Accelerator Laboratory, Menlo Park, CA 94025, USA
\item[$^{27}$] Theoretical Physics Department, CERN, CH-1211 Geneva 23, Switzerland
\item[$^{28}$] Department of Physics and Astronomy, Michigan State University, East Lansing, MI 48824, USA
\item[$^{29}$] University of the Witwatersrand, Johannesburg, South Africa
\item[$^{30}$] Dipartimento di Fisica, Universit\`{a} di Torino e INFN, Sezione di Torino, 10125 Turin, Italy
\item[$^{31}$] Physics Department, Reed College, Portland, OR 97202, USA
\item[$^{32}$] Department of Physics, University of Arizona, Tucson, AZ 85721, USA
\item[$^{33}$] Physikalisches Institut, Albert-Ludwigs-Universit{\"a}t Freiburg, 79104 Freiburg, Germany
\item[$^{34}$] Dipartimento di Fisica, Universit\`{a} di Genova and INFN, Sezione di Genova, 16146 Genoa, Italy
\item[$^{35}$] Berkeley Center for Theoretical Physics, University of California, Berkeley, CA 94720, USA
\item[$^{36}$] Theoretical Physics Group, Lawrence Berkeley National Laboratory, Berkeley, CA 94720, USA
\item[$^{37}$] Institute of Experimental Particle Physics, Karlsruhe Institute of Technology (KIT), 76131 Karlsruhe, Germany
\item[$^{38}$] Fermi National Accelerator Laboratory, Batavia, IL, 60510-0500, USA
\item[$^{39}$] Physics Division, Lawrence Berkeley National Laboratory, Berkeley, CA 94720, USA
\item[$^{40}$] IPhT, CEA Saclay, CNRS, Universit\'e Paris-Saclay, 91191 Gif-sur-Yvette, France
\item[$^{41}$] Nikhef, Theory Group, Science Park 105, 1098 XG, Amsterdam, The Netherlands
\item[$^{42}$] Institute for Theoretical Physics Amsterdam and Delta Institute for Theoretical Physics, University of Amsterdam, Science Park 904, 1098 XH Amsterdam, The Netherlands
\item[$^{43}$] Institute for Particle Physics and Astrophysics, ETH Z{\"u}rich, CH-8093 Z{\"u}rich, Switzerland
\item[$^{44}$] Centro de F{\'i}sica Te{\'o}rica de Part{\'i}culas (CFTP), Instituto Superior T{\'e}cnico IST, Universidade de Lisboa, 1049-001 Lisboa, Portugal
\item[$^{45}$] Physics Department, Florida State University, Tallahassee, FL 32306, USA
\item[$^{46}$] LAPTh, CNRS, Universit\'e Savoie Mont Blanc, 74940 Annecy, France
\item[$^{47}$] Department of Physics and Astronomy, VU University, 1081 HV Amsterdam
\item[$^{48}$] Institute for Theoretical Particle Physics, Karlsruhe Institute of Technology (KIT), 76131 Karlsruhe, Germany
\item[$^{49}$] Rudolf Peierls Centre for Theoretical Physics, University of Oxford, OX1 3NP Oxford, UK
\item[$^{50}$] Indian Institute of Technology, Kanpure, India
\item[$^{51}$] Laboratoire de Physique Th\'eorique et Hautes \'Energies (LPTHE), UMR 7589, Sorbonne Universit\'e et CNRS, 75252 Paris Cedex 05, France
\item[$^{52}$] The Henryk Niewodnicza{\'{n}}ski Institute of Nuclear Physics in Cracow, Polish Academy of Sciences
\item[$^{53}$] School of Physics and Astronomy, Monash University, Clayton, VIC 3800, Australia
\item[$^{54}$] Higgs Centre for Theoretical Physics, University of Edinburgh, Edinburgh EH9 3FD, UK
\item[$^{55}$] The University of Sussex, Falmer, Brighton, UK
\item[$^{56}$] Deutsches Elektronen-Synchrotron (DESY), 22607 Hamburg, Germany
\item[$^{57}$] Institut f{\"u}r Physik, Universit{\"a}t Mainz, 55099 Mainz, Germany
\item[$^{58}$] Institute of High Energy Physics, Chinese Academy of Sciences, Beijing, China
\item[$^{59}$] Nuclear Science Division, Lawrence Berkeley National Laboratory, Berkeley, CA 94720, USA
\end{itemize}


\newpage
{\parskip=1.0ex \tableofcontents}


\newpage
\pagenumbering{arabic}
\setcounter{footnote}{0}

As the experimental precision continues to improve during the 13 TeV running at the LHC, the requirements on the corresponding theoretical predictions have increased as well. The predictions include those defined at fixed-order, those resumming large logarithms due to kinematic thresholds and boundaries, and those involving parton showering, and subsequent hadronization. The latter allows for direct comparison at the hadron level to data. All
levels of theoretical predictions are needed for a full exploration of LHC physics. We 
continue in these proceedings to discuss advances in theoretical predictions, while also examining their connections, their limitations, and their prospects for improvement. 

Calculations for $2\rightarrow 2$ processes at NNLO start to become the new standard, with $2\rightarrow3$ calculations on the horizon. One of the complications of such calculations is the dissemination of the results. In contrast to NLO, most of these programs are too difficult/lengthy 
for non-author users to be able to run independently. Two contributions discuss flexible storage possibilities for the results of calculations, either as ROOT ntuples or in grid form, both of which techniques have been very successful at NLO. 
Part of the impressive progress in NNLO calculations is due to the development of more efficient methods to treat infrared divergent real radiation at NNLO. 
Among these methods is a novel scheme, called  ``nested soft-collinear subtraction scheme'', which is presented in these proceedings.
A different approach aims to avoid the occurence of (dimensionally regulated) poles which need to be isolated and cancelled between real and virtual parts by combining all the contributions at integrand level and then performing the integrations completely numerically in four dimensions. 
Such four-dimensional frameworks are promising as they avoid some tedious technicalities related to calculations in $D$ dimensions.
However these methods face hurdles of different type, related to the purely numerical approach. Progess how to overcome these is reported here.
Automation of electroweak NLO corrections has seen lots of progress in recent years,
and a tuned comparison to validate the different available tools for obtaining EW one-loop amplitudes
and complete fixed-order results at NLO EW accuracy are presented for off-shell $ZZ$ and $WW$
production, at the level of amplitudes as well as integrated and differential cross sections.

A key ingredient for any theoretical prediction at hadron colliders are parton distribution functions
(PDFs). In the recent past, improved fit methodologies in combination with the use of new LHC data in
the fits have led to a reduction of the nominal PDFs uncertainties. 
In many phenomenologically
relevant cases, they are now at the level of few percent. Nevertheless, so far PDFs errors only
reflect uncertainties coming from the fitted data, and not from the theoretical predictions used
in the fit. Given the reduction of the former, the latter is no longer negligible. To address
this issue, we perform preliminary studies of theory uncertainty in PDFs fits. More precisely, we
study the convergence of PDFs fits at different orders in perturbation theory, and use perturbative
stability as a handle on theoretical uncertainties. 

Jet substructure techniques are widely applied at the LHC for testing QCD in extreme regions of phase space and for searching beyond the Standard Model for new physical phenomena.
In these proceedings, we perform two studies that could extend the LHC physics program in both of these areas.
To improve our understanding of QCD using jet substructure, we study the feasibility of combining precision jet substructure measurements and calculations to extract the strong coupling constant, $\alpha_s$.  We find that 10\% precision may already be feasible using existing technology in a region of phase space complementary to other methods for measuring $\alpha_s$.
To improve searches for beyond the Standard Model physics using jet substructure, we study the trade-off between performance and robustness for tagging boosted $W$ bosons.  We identify two-prong tagging strategies that could enhance the signal significance and robustness to non-perturbative effects compared with the default methods used at ATLAS and CMS.

One of the pillars of the LHC program is the detailed study of the Higgs
boson. Understanding if the discovered particle has the properties as
predicted by the Standard Model, or if deviations from the Standard
Model predictions point to beyond-the-Standard-Model effects in the
Higgs sector, requires increased precision both in theoretical
predictions as well as experimental measurements. One contribution
studies the soft gluon resummation effect in the Higgs boson plus
two-jet production in the weak boson fusion process at the LHC. 
The
results are also compared with the prediction of the Monte Carlo event
generator {\tt Pythia8}. 
We also include updated results for fully differential Higgs production
in weak boson fusion at NNLO, with a particular emphasis on the structure
of jet dynamics in the experimental fiducial region.
Gluon fusion production with two jets is an important background to measurements
of weak boson fusion. One of the contributions compares perturbative predictions for 
gluon fusion production with two jets and studies the suppression of this background.
Another contribution studies the potential to extract
signals from beyond the Standard Model by comparing the strongly related
processes of $WH$ and $ZH$ production.
In several loop induced processes, in particular $H$, $HH$ and $ZZ$ production in gluon fusion, a substantial hardening of the $p_T$ spectrum of the Higgs boson or the boson pair has been observed when matching the fixed order NLO result to a parton shower. 
An investigation of the reasons for this observation in the case of Higgs boson pair production, matched to {\tt Powheg} combined with both {\tt Pythia6} and {\tt Pythia8}, {\tt MG5\_aMC@NLO+Pythia8} and {\tt Sherpa} with both {\tt CS}-shower and {\tt Dire}-shower, is presented here.
Furthermore, we develop a scheme for the parametrization of theory uncertainties
for weak boson fusion and associated production with weak bosons in the
context of the Simplified Template Cross Section (STXS) framework and discuss in which form
to report theory uncertainties on STXS measurements to allow for their coherent treatment
in the interpretation of the measurements.

Relevant progress on theoretical calculations has been achieved
recently on many other important LHC physics processes. The results of several phenomenological studies are presented.
A study for top quark pair production incorporating electroweak and QCD higher-order corrections
into the multi-jet merging framework of {\tt Sherpa} is presented. This is in particular targeted
to allow for a reliable modelling of the top transverse momentum distribution, which is crucial
for a multitude of new physics searches.
Monte Carlo predictions at particle level are compared against a recent ATLAS measurement in the
lepton+jet channel for the reconstructed top transverse momentum spectrum. Excellent agreement
between Monte Carlo predictions and data is found when the electroweak corrections are included.
Top quark pair production in the di-lepton channel is investigated comparing four different theoretical descriptions.
The full NLO corrections to $pp\rightarrow W^+W^-b\bar b\rightarrow
(e^+ \nu_e)\,(\mu^- \bar{\nu}_{\mu})\,b\bar b$ production are compared
to calculations in the narrow width approximation, where the production
of a top quark pair is calculated at NLO and combined with 
different descriptions of the top quark decay: LO, NLO and via a parton shower.
The study works out differences in the shape of the $m_{lb}$ and $m_{Wb}$ distributions in view of top quark mass determinations, 
in particular showing that corrections beyond the leading order in the decay play a significant role.
A further study addresses the final state $\ell'^+  \nu_{\ell'}  \ell^+ \ell^- {\rm j} {\rm j}$,
which is of particular interest as it proceeds via diagrams featuring WZ scattering.
The different contributions to this final state 
are first evaluated showing that the QCD contributions are overwhelming over the EW ones.
Then, a comparison of theoretical predictions at LO accuracy and at LO matched with various
parton showers is performed.
While at LO all predictions are in good agreement, the inclusion of parton shower effects can
introduce large differences especially for observables defined beyond LO
(\emph{e.g.}\ third-jet observables).
%
In another contribution, we include a short pedagogical discussion of the treatment of the underlying event in measurements at high luminosity at the LHC, 
a treatment that is unknown even to most members of the experimental collaborations. 
We also examine how well fixed-order predictions, which often have 
the highest available accuracy, can describe the jet R-dependence for the inclusive jet, Z+jet and Higgs+jet final states. 
This tests how well fixed-order predictions, containing an increasing number of parton emissions as the perturbative order increses, 
can describe jet shapes, as compared to parton shower predictions. 
On the other hand,  the non-perturbative corrections
needed by the fixed-order predictions for best comparison to data, can only be obtained from 
the parton shower Monte Carlo predictions for the relevant processes. This continues the
study from Les Houches 2015, where good agreement was shown between fixed order and 
parton shower Monte Carlo predictions (for non-Sudakov observables) for Higgs+jets 
production with one value of R (0.4).

As parton shower Monte Carlo programs are crucial for physics at the LHC, we continue
the direct comparisons of the different parton shower algorithms started in Les Houches 2015, in
order to gain a better understanding of any differences that may result in intrinsic uncertainties
for the resultant predictions. 
This concerns \emph{e.g.}\ parton shower variations due to renormalization
scale variations and their interplay with hadronization, or the
correlation between PDF choices and the modeling of underlying event
via multiparton interactions.  Both of these points are
addressed in a study that shows that correlations exist and that a
sensible description of data sensitive to non-perturbative effects
requires a retuning of the event generator when varying
renormalization scales. This retuning does not necessarily mean a loss
of predictivity in perturbatively dominated phase-space
regions.
Another study investigates how scale variations in the parton
shower can be defined when considering known higher-order
corrections. Keeping these higher-order corrections intact when
performing scale variations requires the introduction of compensating
terms. The use of similar compensation schemes leads to closer
agreement of the variation bands of different parton-shower
predictions, and interpolates between previous aggressive or
conservative uncertainty estimates.
The influence of Monte Carlo modelling on the extraction and
calculation of nonperturbative correction factors is also considered
in another study, where it is addressed if the choice of PDF sets and
Monte Carlo models used for the extraction of nonperturbative
correction factors may bias other measurements in which these
correction factors are subsequently used.
A short study that compares different Monte Carlo generators for a
process involving $B$-hadrons in the final state is also included,
with the aim of assessing to which extent NLO-based tools are
successful in simulating kinematic configurations sensitive to the
gluon splitting to pair of (massive) $b$-quarks.


\newpage

\chapter{NLO automation and (N)NLO techniques}
\label{cha:nnlo}
\section{Update on the precision Standard Model wish list~\protect\footnote{
  F.~Caola,
  G.~Heinrich,
  J.~Huston,
  S.~Kallweit,
  K.~Theofilatos}{}}
\label{sec:SM_wishlist}

Identifying key observables and processes that require improved theoretical input has been
a key part of the Les Houches programme. In this contribution we briefly summarise progress since the previous
report in 2015 and explore the possibilities for further advancements.
For the first time, we also provide an estimate of the experimental uncertainties for key processes.

\subsection{Introduction}

The period since the Les Houches 2015 report~\cite{Badger:2016bpw} is
marked by significant progress in the automation of electroweak
corrections and the production of NNLO results in an almost industrial way.
The latter is mainly due to the development of methods which allow to treat the 
doubly unresolved real radiation parts occurring at NNLO in a largely automated manner,
as well as due to the availability of two-loop integrals with an increasing number of kinematic scales.

On the parton shower side,  
NLO QCD matched results and matrix element improved multi-jet merging techniques have become a standard
level of theoretical precision. The automation of full SM corrections including NLO
electroweak predictions has also seen major improvements.

Another challenge is to make the NNLO $2\to2$ predictions or complex NLO predictions publicly
available to experimental analyses, and there has been major progress to achieve this goal.
\textsc{Root Ntuples} have been a useful tool for complicated final states at NLO and
allow for very flexible re-weighting and analysis. The cost for this is
the large disk space required to store the event information.  A feasibility
study using \textsc{Root NTuples} to store the much larger NNLO events in $pp\to
2$ jets~\cite{Currie:2017eqf} is described in Sec.~\ref{cha:nnlo}.\ref{sec:SM_NNLOntuples}.  
An extension of ApplGrid~\cite{Carli:2010rw} and FastNLO~\cite{Kluge:2006xs} offers a simpler, but less flexible method to
distribute higher order predictions. The latter option is likely to be used heavily in
precision PDF fits, and new developments in the APPLfast project are described in Sec.~\ref{cha:nnlo}.\ref{sec:SM_fastNNLO}.

\subsection{Developments in theoretical methods}

Precision predictions require a long chain of various tools and methods, all
of which demand highly technical computations. 

Computational methods for the amplitude level ingredients have seen substantial
progress in the last few years.  Scattering amplitudes at $L$ loops are
generally decomposed into a basis of integrals together with rational coefficients,
\begin{equation}
  A^{(L)}_{2\to n} = \sum (\text{coefficients})_i (\text{integrals})_i,
\end{equation}
one must then remove infrared singularities to obtain a finite cross-section,
\begin{equation}
  d\sigma_{2\to n} \text{\NLO{k}} = {\rm IR}_k(A^{k}_{2\to n}, A^{k-1}_{2\to n+1},\cdots, A^{0}_{2\to n+k}).
\end{equation}
where the function ${\rm IR}_k$ represents an infrared subtraction technique; a recently developed one is presented in Sec.~\ref{cha:nnlo}.\ref{sec:SM_nestedSCsubtraction}. 
Ultraviolet renormalisation must also be performed but in a (semi-)analytic approach presents no technical difficulties.
There are also fully numerical approaches, aiming to calculate higher order corrections without the separation into individually divergent components, 
such that 4-dimensional methods can be applied. Sections~\ref{cha:nnlo}.\ref{sec:SM_looptreeduality} and~\ref{cha:nnlo}.\ref{sec:SM_loopnumerical} are dedicated to a description of such methods.

\subsubsection{Loop integrals}

Most of the new analytic results for two-loop integrals and beyond have been calculated employing the differential equations technique~\cite{Kotikov:1990kg,Gehrmann:1999as}, 
which got a significant boost through Henn's canonical form~\cite{Henn:2013pwa}.
Since the last Les Houches report, several tools to find a canonical basis automatically have been developed: {\tt epsilon}~\cite{Prausa:2017ltv}, {\sc Fuchsia}~\cite{Gituliar:2017vzm} and {\sc Canonica}~\cite{Meyer:2017joq}.
Important new developments concerning the differential equations technique to calculate multi-loop integrals can be found in  Refs.~\cite{Zeng:2017ipr,Harley:2017qut,Lee:2017qql,Bosma:2017hrk}.
For a review on the method of differential equations we refer to Ref.~\cite{Henn:2014qga}.

Major progress since the last report has been made in the calculation of two-loop
master integrals with massive  propagators, for example the planar ones entering 
Higgs+jet~\cite{Bonciani:2016qxi,Primo:2016ebd}, $gg\to \gamma\gamma$ via massive top quark loops~\cite{Becchetti:2017abb},  
or HH~\cite{Davies:2018ood}.

Classes of integrals with one additional mass scale appearing in the
propagators also have been calculated in the context of massive Bhabha
scattering~\cite{Henn:2013woa}, electron-muon scattering 
(with $m_e=0,m_{\mu}\not =0$)~\cite{Mastrolia:2017pfy},  the mixed QCD-EW
corrections to the Drell-Yan process~\cite{Bonciani:2016ypc,vonManteuffel:2017myy} and 
three-loop corrections to the heavy flavour Wilson coefficients in DIS
with two different masses~\cite{Ablinger:2017err,Ablinger:2017xml}.

The integrals entering top quark pair production at NNLO~\cite{Czakon:2013goa} 
have been calculated numerically~\cite{Baernreuther:2013caa}, while analytic results are only partially available~\cite{Bonciani:2013ywa,Abelof:2015lna}. 

A major complication related to the (2-loop) integrals with massive propagators is related to the fact that 
the basis for an analytic representation of such integrals may go beyond the function class of generalized polylogarithms (GPLs), 
i.e.  integrals of elliptic type occur. The latter have been subject of intense studies recently, see e.g. Refs.~\cite{Bogner:2016qbf,Bonciani:2016qxi,Primo:2016ebd,Adams:2017tga,Bogner:2017vim,Remiddi:2017har,Ablinger:2017bjx,Hidding:2017jkk,Lee:2017qql,Chen:2017soz,Chen:2018dpt,Broedel:2017kkb,Broedel:2017siw,Adams:2018yfj}.

For integrals which do not leave the class of GPLs, improvements in the understanding of the basis of multiple polylogarithms through symbol calculus 
and Hopf algebras (see e.g. \cite{Duhr:2014woa,Abreu:2017enx}) has led to a high degree of
automation for these integral computations. This is a necessary step
in order to apply such techniques to phenomenologically relevant cases, most
notably e.g. of $pp\to H$ at \NLO3~\cite{Dulat:2017prg,Mistlberger:2018etf}, 
 four-loop contributions to the cusp anomalous dimension or 
N$^3$LO splitting functions~\cite{vonManteuffel:2015gxa,Davies:2016jie,vonManteuffel:2016xki,Lee:2016ixa,Ruijl:2017eht,Lee:2017mip,Moch:2017uml}.
At the multi-loop front, remarkable recent achievements are also the calculation of the five-loop QCD beta-function~\cite{Baikov:2016tgj,Herzog:2017ohr,Luthe:2017ttg,Chetyrkin:2017bjc} and Higgs decays to hadrons and the R-ratio at N$^{4}$LO~\cite{Herzog:2017dtz}.

There have also been developments in the direct evaluation of Feynman integrals with fewer scales, but higher loops.
The {\sc HyperInt}~\cite{Panzer:2014caa} and {\sc mpl}~\cite{Bogner:2015nda} packages
have focused mainly on zero and one scale integrals with a high number of loops,
but the algorithms employed have potential applications to a wider class of integrals.
Another newly developed tool is {\sc Dream}~\cite{Lee:2017ftw}, a program for the computation of multiloop integrals within the {\sc dra} (Dimensional Recurrence \& Analyticity) method.

In order to facilitate the search for analytic results for multi-loop integrals in the literaure, a new database  
Loopedia~\cite{Bogner:2017xhp} has been created. At \url{https://loopedia.mpp.mpg.de} results for integrals can be searched for by topology. 
The webpage also allows to upload results for newly calculated integrals and liteature information.

Direct numerical evaluation remains a powerful technique.
Two contributions in these proceedings explain methods where loop and phase space integrations can be combined to cancel all poles at integrand level, such that the amplitudes can be evaluated in 4 dimensions, see Secs.~\ref{cha:nnlo}.\ref{sec:SM_loopnumerical} and \ref{cha:nnlo}.\ref{sec:SM_looptreeduality}.

Calculating multi-loop integrals numerically is also a promising strategy for integrals with a rather large number of kinematic scales.
The sector decomposition algorithm~\cite{Binoth:2000ps} has seen a number of optimisations, implemented into the publicly available 
updates of the codes {\sc (py)SecDec}~\cite{Borowka:2017idc,Borowka:2015mxa} and {\sc Fiesta}~\cite{Smirnov:2015mct}.
For example, computations of analytically unknown 2-loop integrals entering $pp\to HH$~\cite{Borowka:2016ehy,Borowka:2016ypz} 
and $pp\to H+$\,jet~\cite{Jones:2018hbb} at NLO including the full top quark mass dependence have been completed using a fully numerical approach.

\subsubsection{Loop amplitudes and integrands}

Following the analytic calculation of two-loop integrals with two massive legs, a complete set of helicity amplitudes has been obtained by two independent groups for $pp\to VV'$ \cite{Caola:2014iua,Caola:2015ila,vonManteuffel:2015msa,Gehrmann:2015ora}.  The results are publicly 
available from
\url{http://vvamp.hepforge.org/}. Both approaches relied heavily on
efficient implementations of integration-by-parts (IBP) reduction identities~\cite{Smirnov:2008iw,Smirnov:2013dia,Smirnov:2014hma,Lee:2012cn,Studerus:2009ye,vonManteuffel:2012np}.
There are also new ideas how to improve multivariate functional reconstruction in the context of amplitude reduction~\cite{vonManteuffel:2014ixa,Peraro:2016wsq,Boehm:2017wjc}. 
Recent developments of tools in the IBP context are e.g. {\sc Kira}~\cite{Maierhoefer:2017hyi}, a new program for IBP reduction identities, and {\sc Forcer}, 
a {\tt FORM} program for the reduction of four-loop massless propagator diagrams~\cite{Ruijl:2017cxj}.

\subsubsection{Generalised unitarity and integrand reduction}

Extending the current multi-loop methods to higher multiplicity still represents
a serious challenge. The increased complexity in the kinematics, and large amount
of gauge redundancy in the traditional Feynman diagram approach, at one-loop has been solved numerically through on-shell
and recursive off-shell methods. This breakthrough has led to the development
of the now commonly used automated one-loop codes~\cite{Berger:2008sj,Bevilacqua:2011xh,Cascioli:2011va,Badger:2012pg,Cullen:2014yla,Alwall:2014hca,Actis:2016mpe}.

The $D$-dimensional generalised unitarity cuts algorithm~\cite{Bern:1994zx,Bern:1994cg,Britto:2004nc,Giele:2008ve,Forde:2007mi} has
been extended to multi-loop integrands using integrand reduction~\cite{Ossola:2006us,Ellis:2008ir}\footnote{We do not attempt a complete review of integrand reduction here. Further information can be found in the review article \cite{Ellis:2011cr} and references therein.} 
and elements of computational algebraic geometry~\cite{Mastrolia:2011pr,Badger:2012dp,Zhang:2012ce,Kleiss:2012yv,Feng:2012bm,Mastrolia:2012an,Mastrolia:2013kca,Badger:2013gxa,Mastrolia:2016dhn}.
In contrast to the one-loop case, the basis of integrals obtained through this
method is not currently known analytically and is much larger than the set of
basis functions defined by standard integration-by-parts identities.
The maximal unitarity method~\cite{Kosower:2011ty}, which incorporates IBP identities, has been applied to a
variety of two-loop examples in four dimensions~\cite{Larsen:2012sx,CaronHuot:2012ab,Johansson:2012zv,Johansson:2013sda,Johansson:2015ava}.
Efficient algorithms to generate unitarity compatible IBP identities are a key ingredient
in both approaches and have been the focus of on-going investigations~\cite{Gluza:2010ws,Schabinger:2011dz,Ita:2015tya,Larsen:2015ped,Hirschi:2016mdz,Mastrolia:2016dhn}.
Very recently, automated tools for IBP reductions based on algebraic geometry have been developed, see e.g. {\sc Cristal} and {\sc Azurite}~\cite{Georgoudis:2016wff,Larsen:2017aqb,Georgoudis:2017iza}.

Using cutting edge numerical unitarity methods, a result for the full 2-loop 4-gluon amplitude has been achieved~\cite{Abreu:2017xsl}.
Results for 2-loop 5-gluon amplitudes based on numerical unitarity are also advancing rapidly~\cite{Badger:2015lda,Abreu:2017hqn,Badger:2017jhb}.

\subsubsection{NNLO infrared subtraction methods for differential cross-sections}

The construction of fully differential NNLO cross-sections for $2\to2$
processes has been a major theoretical challenge over the last years.
This programme has been a remarkable success with many different approaches now
applied to LHC processes. We give a brief characterisation of the main methods below, 
as well as some of their LHC applications.

\begin{itemize}
\item Antenna subtraction~\cite{GehrmannDeRidder:2005cm,Currie:2013vh}:\\
  Analytically integrated counter-terms, applicable to hadronic initial and final states. Almost
  completely local, requires averaging over azimuthal angles. Applied
  to $e^+e^-\to 3j$~\cite{Ridder:2014wza,Gehrmann:2017xfb},  $pp\to j+X$~\cite{Currie:2016bfm}, $pp\to 2j$~\cite{Currie:2017eqf}, 
  $pp\to Z+j$~\cite{Ridder:2015dxa,Ridder:2016nkl}, $pp\to W+j$~\cite{Gehrmann-DeRidder:2017mvr}, $pp\to H+j$~\cite{Chen:2016zka}, dijets in DIS~\cite{Currie:2017tpe} and Higgs production in VBF~\cite{Cruz-Martinez:2018rod}.

\item Sector Improved Residue Subtraction~\cite{Czakon:2010td,Czakon:2011ve,Boughezal:2011jf}:\\
  Fully local counter-terms, based on a sector decomposition~\cite{Binoth:2000ps} approach
  for IR divergent real radiation~\cite{Heinrich:2002rc,Anastasiou:2003gr,Binoth:2004jv} and an extension of the FKS approach at NLO \cite{Frixione:1995ms,Frederix:2009yq}.
  Numerically integrated counter-terms, for hadronic initial and final states.
  Recently formulated in a four-dimensional setting~\cite{Czakon:2014oma}. Applied to 
  top-quark processes~\cite{Czakon:2013goa,Czakon:2014xsa,Czakon:2015owf,Czakon:2016ckf,Brucherseifer:2013iv,Brucherseifer:2014ama}
 and to $pp\to H+j$~\cite{Boughezal:2015dra,Caola:2015wna}.

\item $q_T$ \cite{Catani:2007vq}:\\
  Phase-space slicing approach for colourless final states, applied to many
  $pp\to VV'$ processes. An extension for $t\tb$ final states has been
  proposed \cite{Bonciani:2015sha}. A full list of processes is available in
  {\sc Matrix}~\cite{Grazzini:2017mhc}, see also phenomenological studies~\cite{Catani:2009sm,Catani:2011qz,Ferrera:2013yga,Grazzini:2013bna,Ferrera:2014lca,Gehrmann:2014fva,Cascioli:2014yka,Grazzini:2015nwa,Grazzini:2016swo,Grazzini:2015hta,Grazzini:2016ctr,Grazzini:2017ckn,Catani:2018krb,Grazzini:2018bsd}.
\\
Also applied to obtain NNLO differential results for $VH$~\cite{Ferrera:2011bk,Ferrera:2014lca} 
and $HH$~\cite{deFlorian:2016uhr}, 
as well as for $HHW$~\cite{Li:2016nrr} 
and $HHZ$~\cite{Li:2017lbf}.

\item $N$-jettiness \cite{Boughezal:2015eha,Boughezal:2015dva,Gaunt:2015pea}:\\
  Extension of the $q_T$ method to final states including a jet,
  matching to soft-collinear effective theory (SCET)
  below the $N$-jettiness cut-off parameter. Applied to $2\to 2$
   processes containing vector bosons or a boson plus one jet in the final
  state~\cite{Boughezal:2015dva,Boughezal:2015aha,Boughezal:2015ded,Boughezal:2016yfp,Boughezal:2016dtm,Campbell:2016jau,Campbell:2016lzl,Campbell:2017aul};
  for colourless final states see also MCFM version\,8~\cite{Boughezal:2016wmq}.
  Similar techniques also applied to top decay~\cite{Gao:2012ja} and $t$-channel single top production~\cite{Berger:2016oht}.

\item ColorFull \cite{DelDuca:2015zqa}:\\
  Fully local counter-terms extending the Catani-Seymour dipole method \cite{Catani:1996vz}.
  Analytically integrated for infrared poles, numerical integration for finite parts.
  Currently developed for hadronic final states such as $H\to b\bb$
  \cite{DelDuca:2015zqa} and $e^+e^-\to$ 3 jets~\cite{DelDuca:2016csb,DelDuca:2016ily,Tulipant:2017ybb}.

\item Nested Soft-Collinear Subtraction~\cite{Caola:2017dug}:\\
Fully local subtraction terms, (partially) numerical cancellation of
IR poles, allows matrix elements to be evaluated in four dimensions.
Described in detail in Sec.~\ref{cha:nnlo}.\ref{sec:SM_nestedSCsubtraction}.
Recently applied to $pp\to WH$ with $H\to b\bar{b}$~\cite{Caola:2017xuq}.

\item Analytic local sector subtraction~\cite{Magnea:2018jsj}:\\
Local subtraction, aiming at  the minimal counterterm structure arising from a sector partition of the radiation
phase space. Analytic integration of the counterterms. Proof of
principle example from $e^+e^-\to 2$\,jets.

\item Structure function approach/projection to
  Born~\cite{Cacciari:2015jma}:\\
Range of applicability limited (it requires the knowledge of inclusive NNLO corrections).
Applied to VBF Higgs production~\cite{Cacciari:2015jma}, and $t$-channel single top production~\cite{Berger:2016oht}.

\end{itemize}


\subsection{The precision wish list}
\label{sec:SM_wishlist:precision_wish_list}
We break the list of precision observables into
four sections: Higgs,  jets, vector bosons and top quarks.

Corrections are defined with respect to the leading order, and we organise
the perturbative expansion into QCD corrections, electroweak (EW) corrections and
mixed QCD$\otimes$EW,
\begin{equation}
  d\sigma_X = d\sigma_X^{\rm LO} \left(1 +
      \sum_{k=1} \alpha_s^k d\sigma_X^{\delta \text{\NLOQ{k}}}
    + \sum_{k=1} \alpha^k d\sigma_X^{\delta \text{\NLOE{k}}}
    + \sum_{k,l=1} \alpha_s^k \alpha^l d\sigma_X^{\delta \text{\NLOQE{k}{l}}}
    \right).
  \label{eq:SM_wishlist:dsigmapertexp}
\end{equation}
We explicitly separate the mixed QCD and EW corrections to distinguish between additive predictions
QCD+EW and mixed predictions QCD$\otimes$EW. The definition above only applies in the case where the leading order
process contains a unique power in each coupling constant. For example, in the case of $q\qb\to q\qb Z$ two
leading order processes exist: via gluon exchange of $\mathcal{O}(\alpha_s^2\alpha)$, via electroweak boson 
exchange of $\mathcal{O}(\alpha^3)$ and the interference $\mathcal{O}(\alpha_s\alpha^2)$. 
In these cases it is customary to classify the Born process with highest power in 
$\alpha_s$ (and typically the largest cross section) as the leading order, and 
label the others as subleading Born processes. The above classification is then 
understood with respect to the leading Born process, unless otherwise stated.

In the following we attempt to give a current snapshot of the
available  calculations of higher (fixed) order corrections in both QCD and EW theory.
The main aim is to summarise computations that appeared in the 2015
wish list and that have now been completed, as well as to identify
processes with a large mismatch between the (expected)
experimental precision and the theoretical uncertainties~\footnote{Unfortunately, time has allowed a discussion of 
experimental uncertainties only for the Higgs sector and the inclusive $W, Z$ and $t\tb$ processes. This will be rectified in future updates. Extrapolating to a data
sample of 3000~fb$^{-1}$ can be problematic. Assuming a center-of-mass energy of 14 TeV for most of the running leads to a decrease in statistical errors
by a factor of 10. We make the assumption that the systematic errors stay the same; this may be optimistic given the environment 
in the high luminosity LHC, so take this with a grain of salt. In almost all cases, the systematic errors will dominate over
the statistical ones for this large data sample. We assume a luminosity uncertainty of 2\%, as current.}.
We are aware that there are obvious difficulties in compiling such lists, which make it difficult to address every possible relevant computation. 
Specific approximations and/or extensions beyond fixed order are often necessary when comparing theory to data.

Following the 2015 wishlist we clarify that it is desirable to have a prediction that combines all the known corrections.
For example \NLOQ2+\NLOEone~refers to a single code that produced differential predictions including
$\mathcal{O}(\alpha_s^2)$ and $\mathcal{O}(\alpha)$ corrections. In most cases this is a non-trivial task and when considered
in combination with decays can lead to a large number of different sub-processes.

%
\paragraph*{Electroweak corrections}

Complete higher order corrections in the SM can quickly become technically complicated in comparison
to the better known corrections in QCD. 
As a basic rule of thumb $\alpha_s^2 \sim \alpha$, so corrections at \NLOQ2~and \NLOEone~
are desirable together. Moreover, for energy scales that are large compared to
the $W$-boson mass, EW corrections are enhanced by large logarithms (often called
Sudakov logarithms). There has been progress towards a complete automation of
\NLOEone~corrections within one-loop programs such as
\textsc{OpenLoops}, \textsc{GoSam}, \textsc{Recola}, \textsc{MadLoop} and \textsc{NLOX},
which has led to the completion of many items from the 2013 list.
A detailed tuned comparison of these amplitude generators
for the production of the $4\ell$ and $2\ell2\nu$ final states (off-shell $ZZ$ and $WW$ production)
is presented at the level of amplitudes, and in combination with the Monte Carlo
integration frameworks \textsc{BBMC}, \textsc{MoCaNLO}, \textsc{Munich}, \textsc{Sherpa} and
\WishListMGaMC at the level of integrated and differential cross sections ---
is presented in Sec.~\ref{cha:nnlo}.\ref{sec:SM_ew_comparison}.

%
\paragraph*{Heavy top effective Higgs interactions and finite mass effects}
  
Many calculations of SM processes involving Higgs bosons use the effective
gluon-Higgs couplings that arise in the $m_t\to\infty$ limit, also called ``Higgs Effective Field Theory'' (HEFT).
At high energy hadron colliders, gluon fusion is the most dominant production process for Higgs bosons. 
However, at high momentum transfers, where the top quark loops are resolved, the
approximation will break down. 

For the data collected during Run II, and even more so at the HL/HE LHC,  it is certainly true that they
probe regions where the HEFT approximation becomes invalid and finite mass effects are important.
Calculating the complete top mass dependence of such loop-induced processes at NLO is difficult
since it involves two-loop integrals with several mass scales. 
While the analytic calculations of such integrals 
have seen much progress in the last two years, as reported here, the phenomenological results available so far for this class of processes 
mostly rely on either numerical methods or approximations.
We list processes in the wishlist as \NLOH{k}$\otimes$\NLOQ{l} when re-weighting including the full top mass dependence up to
order $l$ has been performed. 

\paragraph*{Resummation}
                
We do not attempt a complete classification of all possible resummation
procedures that have been considered or applied to the processes in the list.
In many cases precision measurements will require additional treatment beyond
fixed order, and since resummed predictions always match onto fixed order
outside the divergent region it would be desirable for most predictions to be available
this way. Since this is not feasible, some specific cases are highlighted in
addition to the fixed order.

There are several important kinematic regions where perturbative predictions
are expected to break down. Totally inclusive cross-sections often have large
contributions from soft-gluon emission in which higher order logarithms can be
computed analytically. The $q_T$ and $N$-jettiness subtraction methods naturally
match on to resummations of soft/collinear gluons, in the latter case through
soft-collinear effective theory. A study using the $q_T$ method has been
applied in the case of $pp\to ZZ$ and $pp\to W^+ W^-$ \cite{Grazzini:2015wpa} where further details can be found. 
$0$-jettiness resummations within SCET have also been
considered for Higgs boson production~\cite{Alioli:2015toa}, recently also extending to next-to-leading-logarithmic power corrections~\cite{Moult:2016fqy,Moult:2017jsg,Beneke:2017ztn,Boughezal:2018mvf}.

Observables with additional restrictions on jet transverse momenta can also
introduce large logarithms and jet veto resummations have been studied extensively in the case of 
$pp\to H$ and $pp\to H+j$~\cite{Boughezal:2013oha,Boughezal:2014qsa,Banfi:2015pju}.
More in general, the logarithmic structure of Higgs production in gluon fusion has been recently investigated
in details, see e.g.~\cite{Monni:2016ktx,Ebert:2016gcn,Caola:2016upw,Ebert:2017uel,Bizon:2017rah,Melnikov:2016emg,Liu:2017vkm}.

With increasing precision of both experimental data and fixed order
calculations other regions may also begin to play a role. A method for the
resummation of logarithms from small jet radii has been developed
e.g. in Refs.~\cite{Dasgupta:2014yra,Banfi:2015pju,Dasgupta:2016bnd,Kolodrubetz:2016dzb,Liu:2017pbb,Liu:2018ktv}.
A clear understanding of these effects is important as the most popular jet radius
for physics analyses at the LHC is 0.4, a size for which resummation may start to 
become noticeable. These logarithms are implicitly resummed in parton shower Monte Carlos. 
In Sec.~\ref{cha:pheno}.\ref{sec:SM_Higgs_jet_R}, we continue our comparisons of fixed order and matrix element plus parton shower
predictions for Higgs + jet production, started in Les Houches 2015, paying close attention to the
dependence upon the jet radius. 

These represent only a tiny fraction of the currently available tools and predictions
with resummed logarithms. For a review the interested reader may refer to~\cite{Luisoni:2015xha}
and references therein.

%
\paragraph*{Parton showering}

As in the case of resummation - we refrain from listing all processes in the wishlist to be desired with
matching to a parton shower (PS). \NLOQone+PS predictions are available in a largely automated way within \WishListMGaMC,
\textsc{Sherpa}, \textsc{Powheg} and \textsc{Herwig7}. There have been many recent efforts in matching \NLOQ2 corrections to parton showers
for single boson production processes~\cite{Hamilton:2013fea,Alioli:2013hqa,Hoeche:2014aia,Astill:2016hpa} and there are good prospects for extending these techniques to $2\to2$ processes.

%
\paragraph*{Decay sub-processes}

The description of decay sub-processes is incomplete though we do list a few
notable cases. Ideally all on-shell (factorised) decays would be available up
to the order of the core process. In some cases this is potentially an
insufficient approximation and full off-shell decays including background
interference would be desirable, but are often prohibitive. The $t\tb$ final state
is an obvious example where the off-shell decay to $WWb\bb$ at NNLO is beyond the scope
of current theoretical methods.

Decays in the context of electroweak corrections are usually much more
complicated. Full off-shell effects at NLO are expected to be small, but higher order
corrections within factorisable contributions to the decay can be important, see e.g. Sec.~\ref{cha:pheno}.\ref{sec:MC_WWbb}.

The case of vector boson pair production is particularly important given the
completion of the \NLOQ2 computation, and corrections are known at NLO  within the double pole
approximation \cite{Billoni:2013aba} and beyond \cite{Biedermann:2016yvs,Biedermann:2016guo,Biedermann:2016yds,Biedermann:2016lvg,Biedermann:2017bss}.

\subsection{Higgs boson associated processes}
An overview of the status of Higgs boson associated processes is given in Table~\ref{tab:SM_wishlist:wlH}.

\begin{table}
  \renewcommand{\arraystretch}{1.5}
\setlength{\tabcolsep}{5pt}
  \begin{center}
  \begin{tabular}{lll}
    \hline
    \multicolumn{1}{c}{process} & \multicolumn{1}{c}{known} &
    \multicolumn{1}{c}{desired} \\
    \hline
    $pp\to H$ &
    \begin{tabular}{l}
      \NLOH3 (incl.) \\
      \NLOQE11 \\
      \NLOH2$\otimes$\NLOQone
    \end{tabular} &
    \begin{tabular}{l}
      \NLOH3 (partial results available) \\
      \NLOQtb2
    \end{tabular} \\
    \hline
    $pp\to H+j$ &
    \begin{tabular}{l}
      \NLOH2 \\
      \NLOQone
    \end{tabular} &
    \begin{tabular}{l}
      \NLOH2$\otimes$\NLOQonetb+\NLOEone
    \end{tabular} \\
    \hline
    $pp\to H+2j$ &
    \begin{tabular}{l}
      \NLOHone$\otimes$\LOQ \\
      \NLOQVBFstar3 (incl.) \\
      \NLOQVBFstar2 \\
      \NLOEoneVBF
    \end{tabular} &
    \begin{tabular}{l}
      \NLOH2$\otimes$\NLOQone+\NLOEone\\
      \NLOQ2+\NLOEoneVBF 
    \end{tabular} \\
    \hline
    $pp\to H+3j$ &
    \begin{tabular}{l}
      \NLOHone \\
      \NLOQVBF1
    \end{tabular} &
    \begin{tabular}{l}
      \NLOQone+\NLOEone \\
    \end{tabular} \\
    \hline
    $pp\to H+V$ &
    \begin{tabular}{l}
      \NLOQ2+\NLOEone \\
    \end{tabular} &
    \begin{tabular}{cl}
      \NLOggHVtb{} \\
    \end{tabular} \\
    \hline
    $pp\to HH$ &
    \begin{tabular}{l}
      \NLOH2$\otimes$\NLOQone \\
    \end{tabular} &
    \begin{tabular}{cl}
      \NLOEone \\
    \end{tabular} \\
    \hline
    $pp\to H+t\tb$ &
    \begin{tabular}{l}
      \NLOQone+\NLOEone\\
    \end{tabular} &
    \begin{tabular}{l}
      \\
    \end{tabular} \\
    \hline
    $pp\to H+t/\tb$ &
    \begin{tabular}{l}
      \NLOQone\\
    \end{tabular} &
    \begin{tabular}{l}
      \NLOQone+\NLOEone
    \end{tabular} \\
    \hline
  \end{tabular}
  \caption{Precision wish list: Higgs boson final states. \NLOQVBFstar{x} means a
   calculation using the structure function approximation.}
  \label{tab:SM_wishlist:wlH}
  \end{center}
\renewcommand{\arraystretch}{1.0}
\end{table}

\begin{itemize}[leftmargin=2cm]
  \item[$H$:] 
Among the remarkable recent developments
are first steps towards differential results for Higgs production in gluon fusion at \NLO3~\cite{Dulat:2017brz}
as well as results going beyond the threshold approximation~\cite{Dulat:2017prg,Mistlberger:2018etf,Dulat:2018rbf}.
  Another important achievement is the recent calculation of the mixed
  QCD-EW corrections to this process at order $\alpha\alpha_s^2$ in
  the soft gluon approximation~\cite{Bonetti:2017ovy,Bonetti:2018ukf}. The latter corrections were found to increase the NLO QCD corrections by 5.4\%.

The NNLO HEFT result has been supplemented by an expansion in $1/m_t^n$, and matched to a calculation of the high energy limit~\cite{Harlander:2009my,Pak:2011hs}.

A comprehensive phenomenological study has been presented in~\cite{Anastasiou:2016cez}, 
and an updated version of the program {\tt iHixs}~\cite{Dulat:2018rbf} is available which combines the effective theory with:
    \begin{itemize}
      \item complete mass dependence at NLO including top, bottom and charm loops,
      \item ${m_H}/{m_t}$ corrections at NNLO,
      \item electro-weak corrections at NLO,
      \item re-scaling of the \NLOH3 with the \LOQ~top loop.
    \end{itemize}
Including the mixed order $\alpha\alpha_s^2$ corrections calculated in~\cite{Bonetti:2017ovy,Bonetti:2018ukf} should lead to a further improvement, 
such that the dominant uncertainties currently are expected to be PDF uncertainties and finite mass effects, estimated  to be below 2\%.
Both of these effects can (probably) be further reduced. 

The NNLO+PS computations~\cite{Hamilton:2013fea,Hoche:2014dla} have been extended to include finite top
and bottom mass corrections at NLO~\cite{Hamilton:2015nsa}.
    %


The experimental uncertainty on the total Higgs boson cross section is currently of the order of 10-15\%~\cite{deFlorian:2016spz}, based on a
data sample of 36~fb$^{-1}$,
and is expected to reduce to the order of 3\% or less with a data sample of 3000~fb$^{-1}$~\cite{Campbell:2286381}. 
To achieve the desired theoretical uncertainty, it may be necessary to calculate the finite mass effects to NNLO,
as well as fully differential \NLOH3 corrections.

  \item[$H+j$:] Known through to \NLOQ2 in the infinite top mass limit~\cite{Chen:2014gva,Chen:2016zka,Boughezal:2015dra,Boughezal:2015aha,Caola:2015wna}.
The residual scale uncertainty in the  NNLO HEFT is of the order of 5\%.
    Very recently, this process has been calculated at NLO with full top quark mass dependence~\cite{Jones:2018hbb}, based on numerical methods~\cite{Borowka:2015mxa,Borowka:2017idc}, settling a longstanding question about the impact of the top quark mass dependence. 

The top-bottom interference effects also have been calculated~\cite{Melnikov:2016qoc,Lindert:2017pky}, as well as the mass effects in the large transverse momentum expansion~\cite{Lindert:2018iug,Neumann:2018bsx}.
Using high-energy resummation techniques at leading logarithmic accuracy, the Higgs boson transverse momentum spectrum with finite quark mass effects 
beyond the leading (fixed) order has been calculated in Ref.~\cite{Caola:2016upw}.
Parton shower predictions including finite mass effects in various approximations are also available~\cite{Frederix:2016cnl,Neumann:2016dny,Hamilton:2015nsa,Buschmann:2014sia}

The full NLO calculation~\cite{Jones:2018hbb} revealed that the K-factor (NLO/LO) is fairly constant over the Higgs
boson transverse momentum range above the top quark threshold when using the scale $H_T/2$.
The full result is roughly 9\% larger than in the HEFT approximation and 6\% larger
than  in FT$_{\rm{approx}}$, an approximation where the real radiation contains the full mass dependence, while the virtual part is calculated in the rescaled HEFT approximation. The Higgs $p_T$ distribution in the full theory is significantly different from the one in the HEFT beyond $p_T^H\sim 500$\,GeV.

The current experimental uncertainty on the Higgs + $\ge$ 1 jet differential cross section is of the order of 25-30\%, dominated by 
the statistical error, for example the fit statistical errors for the case of the $H\rightarrow \gamma \gamma$ analysis~\cite{Aaboud:2018xdt}. 
With a sample of
3000 fb$^{-1}$, the statistical error will nominally decrease by about a factor of 10, resulting in a statistical error of 
the order of 2.5\%.  If the remaining systematic errors 
(dominated for the diphoton analysis by the spurious signal systematic error) remain the same, 
the resultant systematic error would be of the order of 9\%, leading to a 
total error of approximately 9.5\%. This is similar enough to the current theoretical uncertainty that it may motivate
improvements on the $H+j$ cross section calculation. 
The improvements could entail a combination of the NNLO HEFT results with the full NLO results, 
similar to the reweighting procedure that has been done one perturbative order lower.

\item[$H+\geq 2j$:] QCD corrections are an essential background to Higgs
 production in vector boson fusion (VBF). VBF production of a Higgs boson has recently been computed
 differentially to \NLOQ2 accuracy~\cite{Cacciari:2015jma,Cruz-Martinez:2018rod} in the ``DIS'' approximation. 
For the total cross section, results at \NLOQ3 accuracy are also available~\cite{Dreyer:2016oyx}.

In the gluon fusion channel, a detailed phenomenological study of Higgs boson production in association with up to 3 jets can be 
found in~\cite{Greiner:2016awe}. 
An assessment of the mass dependence of the various jet multiplicities has been performed in~\cite{Greiner:2015jha}.

In the VBF channel, full NLO QCD corrections are available~\cite{Campanario:2013fsa,Campanario:2018ppz}.

EW corrections to VBF stable Higgs boson production have been calculated in Ref.~\cite{Ciccolini:2007jr} and are available in {\sc Hawk}~\cite{Denner:2014cla}.
Complete NLO QCD+EW corrections to W$^{+}$W$^{+}$ scattering have been calculated recently~\cite{Biedermann:2017bss}.

The current experimental error on the $H+\geq 2j$ cross section is on the order of 35\%~\cite{Aaboud:2018xdt}, again dominated by statistical errors, 
and again for the diphoton final state, by the fit statistical error. With the same assumptions as above, for 3000 fb$^{-1}$, the statistical 
error will reduce to the
order of 3.5\%. If the systematic errors remain the same, at approximately 12\%, (in this case the largest systematic error is from the jet energy scale
uncertainty and the jet energy resolution uncertainty), a total uncertainty of approximately 12.5\% would result, less than the
current theoretical uncertainty. 
To achieve a theoretical uncertainty less than this
value would require the calculation of $H+\geq 2j$ to NNLO in the HEFT. 


  \item[$VH$:] Associated production of a Higgs boson with a vector boson is important to pin down the EW couplings of the Higgs, 
   and also to access the $H\to b\bar{b}$ coupling.
   First predictions at \NLOQ2 have been available for some time~\cite{Brein:2003wg,Brein:2011vx} 
   and are implemented in the program {\tt vh@nnlo}~\cite{Brein:2012ne}, 
   where version 2 also can be used for calculations in the 2HDM and MSSM~\cite{Harlander:2013mla,Harlander:2018yio}.
  NLO EW corrections have been calculated in~\cite{Ciccolini:2003jy,Denner:2011id,Obul:2018psx,Granata:2017iod}, 
where Ref.~\cite{Granata:2017iod} contains combined QCD+EW predictions including parton shower effects.
Soft gluon resummation effects have been calculated and found to be small compared to the NNLO fixed order result~\cite{Dawson:2012gs}.
  Differential predictions at \NLOQ2 for $WH$~\cite{Ferrera:2011bk} and $ZH$~\cite{Ferrera:2014lca} have recently been extended to include 
Higgs boson decays to bottom quarks~\cite{Ferrera:2017zex}.
The gluon initiated processes is particularly sensitive to new physics effects such as new particles in the loop or resonant additional Higgs bosons.
It has been calculated at NLO in Ref.~\cite{Altenkamp:2012sx}, where the K-factor is obtained in the limit $m_t\to \infty$  and
$m_b = 0$, and then used to rescale the full (one-loop)  LO cross section.
Threshold resummation for $gg\to ZH$ has been calculated in Ref.~\cite{Harlander:2014wda}.
Top quark mass effects at NLO, in the framework of an $1/m_t$ expansion, have been considered in Ref.~\cite{Hasselhuhn:2016rqt}.

Fully-differential NNLO QCD results for associated $VH$ production based on N-jettiness subtraction have been calculated in
Ref.~\cite{Campbell:2016jau} and implemented in MCFM including  decays. 
Fully differential NNLO corrections to $pp\to WH$ with $H\to b\bar{b}$ based on nested soft-collinear subtraction~\cite{Caola:2017dug} have been presented in~\cite{Caola:2017xuq}.
A parton shower matched prediction using the MiNLO procedure in {\tt POWHEG} has also been completed \cite{Astill:2016hpa}.
An implementation of dimension-six SMEFT operators related to $VH$ production, which can be used for
NLO QCD+PS accurate Monte Carlo event generation within the MG5\_aMC@NLO framework, is described in~\cite{Degrande:2016dqg}.
An implementation of the
Higgs Pseudo-Observables framework for electroweak Higgs
production in the {\sc HiggsPO} UFO model for Monte Carlo event generation at NLO in QCD is available in~\cite{Greljo:2017spw}.

The total inclusive cross-section has been considered in the threshold limit at \linebreak \NLOQ3, extracted from the inclusive Higgs cross-section \cite{Kumar:2014uwa}.
  \item[$HH$:] The \NLOQ2
    corrections were first computed in the infinite top mass limit~\cite{deFlorian:2013jea} and have since been improved with
    threshold resummation to NLO+NNLL~\cite{Shao:2013bz} and NNLO+NNLL~\cite{deFlorian:2015moa}. 
Differential NNLO results in the $m_t\to\infty$ limit are also available~\cite{deFlorian:2016uhr}.
Power corrections to the NLO and NNLO cross sections in the $m_t\to \infty$ limit have been computed~\cite{Grigo:2013rya,Grigo:2015dia}. 
A complete computation at NLO including all finite top quark mass effects has been
    achieved using numerical methods~\cite{Borowka:2016ehy,Borowka:2016ypz}. 
This calculation also has been matched to parton shower Monte Carlo programs~\cite{Heinrich:2017kxx,Jones:2017giv} 
and is publicly available within the {\tt POWHEX-BOX-V2} framework.
On the analytical side, the planar two-loop integrals entering $gg\to HH$ have been recently computed in the
high-energy limit~\cite{Davies:2018ood}.
An interesting approach to reconstruct the top-quark mass dependence of the two-loop
virtual amplitudes for HH production in gluon fusion (and possibly also other loop-induced processes) is presented in Ref.~\cite{Grober:2017uho}, 
where with Pad{\'e} approximants based on the large-$m_t$ expansion of the amplitude in combination with analytic results near the top
threshold leads to a result which comes very close to the full result.
Very recently, top quark mass effects have been incorporated in the \NLOQ2 calculation, combining
one-loop double-real corrections with full top mass dependence with suitably reweighted
real-virtual and double-virtual contributions evaluated in the large-$m_t$ approximation~\cite{Grazzini:2018bsd}.
The residual uncertainty from missing $m_t$ effects is estimated to be below 3\% at 14\,TeV and below 5\% at 100\,TeV.

  \item[$t\bar{t}H$:] \NLOEone~corrections have been considered within the MadGraph5\_aMC@NLO\linebreak framework~\cite{Frixione:2014qaa,Frixione:2015zaa}.  
Moreover, NLO QCD corrections have been calculated for the process including the top quark decays~\cite{Denner:2015yca}. 
Very recently, the calculation of combined NLO QCD and EW corrections to $t\bar{t}H$ production, including top quark decays and full off-shell effects, has been achieved~\cite{Denner:2016wet}. 
NLO+NNLL resummation for this process has been calculated in Ref.~\cite{Kulesza:2017ukk}.
For results in the  in the Standard Model Effective Field Theory at NLO in QCD see Ref.~\cite{Maltoni:2016yxb}.
\\
NLO QCD corrections to $tH$ associated production are known~\cite{Campbell:2013yla,Demartin:2015uha}.
\end{itemize}

\subsection{Jet final states}
An overview of the status of jet final state is given in Table~\ref{tab:SM_wishlist:wljets}.

\begin{table}
  \renewcommand{\arraystretch}{1.5}
\setlength{\tabcolsep}{5pt}
  \centering
  \begin{tabular}{lll}
    \hline
    \multicolumn{1}{c}{process} & \multicolumn{1}{c}{known} & \multicolumn{1}{c}{desired} \\
    \hline
    $pp\to 2$\,jets &
    \begin{tabular}{l}
      \NLOQ2 \\
      \NLOQone+\NLOEone
    \end{tabular} &
    \begin{tabular}{cl}
      \\
    \end{tabular} \\
    \hline
    $pp\to 3$\,jets &
    \begin{tabular}{l}
      \NLOQone
    \end{tabular} &
    \begin{tabular}{l}
      \NLOQ2 
    \end{tabular} \\
    \hline
  \end{tabular}
  \caption{Precision wish list: jet final states.}
  \label{tab:SM_wishlist:wljets}
  \renewcommand{\arraystretch}{1.0}
\end{table}

\begin{itemize}
\item[j+X:] Differential \NLOQ2 corrections have been calculated in Ref.~\cite{Currie:2016bfm}, scale choices have been studied in~\cite{Currie:2017ctp}.
\item[2 jets:] The \NLOQ2 corrections have been calculated in Ref.~\cite{Currie:2017eqf}. Complete NLO QCD+EW corrections are also available~\cite{Frederix:2016ost}. 
A remarkable result is also the calculation of 2-loop 4-gluon scattering based on numerical unitarity~\cite{Abreu:2017xsl}.
\item[3 jets:] A rapidly increasing number of results on 5-point two-loop amplitudes can be found in~\cite{Badger:2013gxa,Gehrmann:2015bfy,Papadopoulos:2015jft,Dunbar:2016aux,Badger:2017jhb,Abreu:2017hqn,Chicherin:2017dob,Boels:2018nrr}.  
\end{itemize}

\subsection{Vector boson associated processes}

The numerous decay channels for vector bosons and the possible inclusion
of full off-shell corrections versus factorised decays in the narrow width approximation
make vector boson processes complicated to classify.
A full range of decays in the narrow width approximation would be a desirable minimum precision.
In the meanwhile, for leptonic decays this goal is met for essentially all processes in the list.
In terms of QCD corrections, full off-shell decays don't mean a significant complication of
the respective QCD calculations and are available almost everywhere.
This is no longer true for EW corrections, where leptonic decays increase the complexity of
the calculation, and are thus not availalbe for many high-multiplicity processes (involving more than
four final-state particles) yet.
Hadronic decays are even harder to classify because they are formally part of subleading Born
contributions to processes involving jets and possibly further leptonically decaing vector bosons.
Including higher-order corrections in a consisistent way here will usually require full
SM corrections to the complete tower of Born processes,
as briefly discussed in Sec.~\ref{sec:SM_wishlist:precision_wish_list}.
An overview of the status of vector boson associated processes is given in Table~\ref{tab:SM_wishlist:wlV},
where leptonic decays are understood if not stated otherwise.
Also $\gamma$ induced processes become increasingly important in cases where EW
corrections are highly relevant. While often included only at their leading order, first
computations involving also full EW corrections to $\gamma$-induced channels were recently achieved.

\begin{table}
  \renewcommand{\arraystretch}{1.5}
\setlength{\tabcolsep}{5pt}
  \centering
  \begin{tabular}{lll}
    \hline
    \multicolumn{1}{c}{process} & \multicolumn{1}{c}{known} & \multicolumn{1}{c}{desired} \\
    \hline
    $pp\to V$ &
    \begin{tabular}{l}
      \NLOQzzero3 (incl.) \\
      \NLOQ2 \\
      \NLOEone 
    \end{tabular} &
    \begin{tabular}{l}
      \NLOQ3+\NLOE2+\NLOQE11 \wdecay{}
    \end{tabular} \\
    \hline
    $pp\to VV'$ &
    \begin{tabular}{l}
      \NLOQ2 \wleptdecays{} \\
      \NLOEone{ }\wleptdecays{} \\
      \NLOQone{ }($gg$ channel) \wleptdecays{} \\
    \end{tabular} &
    \begin{tabular}{l}
      \NLOQ2+\NLOEone{ }\wleptdecays{} \\
      \NLOQone{ }($gg$ channel, w/ massive loops) \\
    \end{tabular} \\
    \hline
    $pp\to V+j$ &
    \begin{tabular}{l}
      \NLOQ2+\NLOEone{ }\wleptdecays{} \\
    \end{tabular} &
    \begin{tabular}{l}
      hadronic decays
    \end{tabular} \\
    \hline
    $pp\to V+2j$ &
    \begin{tabular}{l}
      \NLOQone+\NLOEone{ }\wleptdecays{} \\
      \NLOEone{ }\wleptdecays{}
    \end{tabular} &
    \begin{tabular}{l}
      \NLOQ2 \wdecays{} \\
    \end{tabular}\\
    \hline
    $pp\to V+b\bar{b}$ &
    \begin{tabular}{l}
      \NLOQone{ }\wleptdecays{} \\
    \end{tabular} &
    \begin{tabular}{l}
      \NLOQ2 +\NLOEone{ }\wdecays{} \\
    \end{tabular} \\
    \hline
    $pp\to VV'+1j$ &
    \begin{tabular}{l}
      \NLOQone{ }\wdecays{} \\
      \NLOEone{ }\wodecays{}
    \end{tabular} &
    \begin{tabular}{l}
      \NLOQone+\NLOEone{ }\wdecays{} \\
    \end{tabular} \\
    \hline
    $pp\to VV'+2j$ &
    \begin{tabular}{l}
      \NLOQone \wleptdecays{} \\
    \end{tabular} &
    \begin{tabular}{l}
      \NLOQone+\NLOEone{ }\wdecays{} \\
    \end{tabular} \\
    \hline
    $pp\to W^+W^++2j$ &
    \begin{tabular}{l}
      \NLOQone+\NLOEone{ }\wleptdecays{} \\
    \end{tabular} &
    \begin{tabular}{l}
      \\
    \end{tabular} \\
    \hline
   $pp\to VV'V''$ &
    \begin{tabular}{l}
      \NLOQone \\
      \NLOEone{ }\wodecays{}
    \end{tabular} &
    \begin{tabular}{l}
      \NLOQone+\NLOEone \wdecays{} \\
    \end{tabular} \\
    \hline
    $pp\to \gamma\gamma$ &
    \begin{tabular}{l}
      \NLOQ2+\NLOEone
    \end{tabular} &
    \begin{tabular}{l}
      \\ 
    \end{tabular} \\
    \hline
    $pp\to \gamma+j$ &
    \begin{tabular}{l}
      \NLOQ2+\NLOEone
    \end{tabular} &
    \begin{tabular}{l}
      \\
    \end{tabular} \\
    \hline
    $pp\to \gamma\gamma+j$ &
    \begin{tabular}{l}
      \NLOQone
    \end{tabular} &
    \begin{tabular}{cl}
      \NLOQ2+\NLOEone
    \end{tabular} \\
    \hline
  \end{tabular}
  \caption{Precision wish list: vector boson final
    states. $V=W,Z$ and $V',V''=W,Z,\gamma$.
    Full leptonic decays are understood if not stated otherwise.}
  \label{tab:SM_wishlist:wlV}
  \renewcommand{\arraystretch}{1.0}
\end{table}

\begin{itemize}[leftmargin=2cm]
  \item[$V$:] Inclusive cross-sections and rapidity distributions in the
    threshold limit have been extracted from the $pp\to V$ results
    \cite{Ahmed:2014cla,Ahmed:2014uya}. Parton shower matched \NLOQ2
    computations using both the MiNLO method \cite{Karlberg:2014qua}, SCET
    resummation \cite{Alioli:2015toa} and via the UN${}^2$LOPS technique \cite{Hoeche:2014aia}. Completing the inclusive \NLOQ3 computation
    beyond the threshold limit is an important step for phenomenological studies.
    The dominant factorisable corrections at $\mathcal{O}(\alpha_s \alpha)$ (\NLOQE11) are also now available \cite{Dittmaier:2015rxo}.
   
    The inclusive production cross section for $W$ and $Z$ bosons has been measured at the LHC using the leptonic decays of the vector bosons. 
    The precision in those measurements already reached the barrier of the luminosity uncertainty $\sim 2\%$, which is not easy to further improve. 
    For example, the most precise measurement of the $W$ and $Z$ bosons integrated fiducial cross sections 
    is for the $\sqrt{s} = 7$~\TeV sample having 
    $\Delta\sigma_W/\sigma_W = 1.87\%$ and $\Delta\sigma_Z/\sigma_Z = 1.82\%$ uncertainty, with the luminosity uncertainty ($\sim1.8\%$) 
    accounting for most of it~\cite{Aaboud:2016btc}.

    While the inclusive integrated cross sections have been already measured and compared fairly well with the present theoretical predictions, this is not the case for 
    differential distributions.
    A key observable, both for precision studies as well as for new physics searches, is the transverse momentum of the vector bosons, as well as the $\phi^\ast$ variable which is 
    also very much related with the momentum of the vector boson, without being affected by the leptons' energy scale uncertainties. 
    For neutral Drell--Yan, those have been measured at $8$~\TeV with precision that is $<1\%$ for $0<p_\mathrm{T}<20$~\GeV~\cite{Aad:2015auj, Sirunyan:2017igm}. 
    At the same time, the high energy tail of the measured transverse momentum distribution is dominated by statistical uncertainties due to the sample size of the data.
    Refined measurements are expected both for the low and the high $p_\mathrm{T}$ part of the transverse momentum distribution in Drell--Yan.
    Special runs with very low pileup have been taken from the LHC, with the experiments targeting to measure with $<1\%$ accuracy in very fine grained bins 
    the low $p_\mathrm{T}<20$~\GeV part of the distribution, seeking to understand the origin of the data over theory discrepancies in this part of the spectrum.
    For high $p_\mathrm{T}$ part of the $p_\mathrm{T}$ distribution, 
    more experimental accuracy is also to be expected with higher luminosity data at $\sqrt{s}=13$~\TeV. 
    Data with high $p_\mathrm{T}$ vector bosons could be used to study the strong coupling constant at \NLOQ2 accuracy. For that, a complete \NLOQ3 
    calculation is needed, given that at tree level the Drell-Yan production is not sensitive to the QCD coupling. 




  %
  \item[$V/\gamma+j$:] Both $Z+j$~\cite{Ridder:2015dxa,Boughezal:2015ded,Boughezal:2016isb,Boughezal:2016yfp,Gehrmann-DeRidder:2017mvr} and
    $W+j$~\cite{Boughezal:2015dva,Boughezal:2016dtm,Boughezal:2016yfp,Gehrmann-DeRidder:2017mvr} have been completed through
    \NLOQ2 including leptonic decays, via antenna subtraction and $N$-jettiness slicing.
    Also $\gamma+j$ was calculated more recently through \NLOQ2 in the $N$-jettiness slicing
    approach~\cite{Campbell:2016lzl}.
    All processes of this class, and in particular their ratios were investigated in great
    detail in Ref.~\cite{Lindert:2017olm}, combining \NLOQ2 predictions with full NLO EW and
    leading NNLO EW effects in the Sudakov approximation, including also approximations for leading
    NLO QCD$\otimes$EW effects. Particular attention was devoted here to error estimates and
    correlations between the processes.
  \item[$V+\geq2j$:] While fixed order \NLOQone~computations of $V+\geq2$ jet final states have been known for many years
    recent progress has been made for \NLOEone~corrections~\cite{Denner:2014ina} including merging and showering \cite{Kallweit:2014xda,Kallweit:2015dum}.
  \item[$VV'$:] Complete \NLOQ2 are now available for all vector-boson pair production
    processes, namely
    $WW$~\cite{Gehrmann:2014fva,Grazzini:2016ctr},
    $ZZ$~\cite{Cascioli:2014yka,Grazzini:2015hta},
    $WZ$~\cite{Grazzini:2016swo,Grazzini:2017ckn},
    $Z\gamma$~\cite{Grazzini:2013bna,Grazzini:2015nwa},
    $W\gamma$~\cite{Grazzini:2015nwa}, using the $q_T$ subtraction method.
    All these results have recently become publicly available in the \textsc{Matrix} Monte Carlo
    framework~\cite{Grazzini:2017mhc}: Leptonic decays of the bosons are included
    throughout, consistently accounting for all resonant and non-resonant diagrams,
    off-shell effects and spin correlations.
    More recently, \NLOQ2 results were achieved for $Z\gamma$~\cite{Campbell:2017aul} and
    $ZZ$~\cite{Heinrich:2017bvg} within the $N$-jettiness method.
    Leading \NLOQ3 corrections, namely the \NLOQone corrections to the loop-induced $gg$ channels,
    became available for the neutral final states $ZZ$~\cite{Caola:2015psa} and
    $WW$~\cite{Caola:2015rqy} involving full off-shell leptonic dacays,
    based on the two-loop amplitudes of
    Refs.~\cite{Caola:2015ila,vonManteuffel:2015msa}.
    More recently, also their interference effects with off-shell Higgs contributinos were
    investigated~\cite{Caola:2016trd,Campbell:2016ivq}.
        There has been good progress in calculating NLO EW corrections, and in the meanwhile
    all vector-boson pair production processes have been completed including full leptonic
    decays~\cite{Denner:2014bna,Denner:2015fca,Biedermann:2016yvs,Biedermann:2016guo, Biedermann:2016lvg, Biedermann:2017oae}.
    In Ref.~\cite{Kallweit:2017khh}, the combined NLO QCD+EW corrections
    were presented for all $2\ell2\nu$ final states, taking into account the complete NLO corrections
    to the photon-induced channels for the first time.
    In these calculations, the recently developed automated approaches have been employed,
    which are validated against each other in a tuned comparison for the off-shell production
    of $WW$ and $ZZ$ pairs in Sec.~\ref{cha:nnlo}.\ref{sec:SM_ew_comparison}.
  \item[$VV'+j$:] \NLOQone~corrections have been known for
    many years. More recently, \NLOEone~corrections became available for the some
    on-shell processes, with subsequent leptonic decays treated in narrow-width
    approximation~\cite{Li:2015ura,Yong:2016njr}. Full \NLOEone~corrections
    including decays are clearly in reach of the automated tools described in
    Sec.~\ref{cha:nnlo}.\ref{sec:SM_ew_comparison}.
  \item[$VV'+2j$:] \NLOQone~corrections have been known for
    several years, both in the QCD production and the EW VBS production modes.
    More recently, a first calculation of the full NLO SM corrections,
    i.e.\ \NLOQone, \NLOEone{} and mixed NLO corrections to all
    production channels, became available for $W^+W^++2j$ production including full
    leptonic decays~\cite{Biedermann:2016yds,Biedermann:2017bss}: In particular the
    NLO EW corrections to the EW VBS production mode turn out to be remarkably
    large and negative if typical VBF kinematics are considered.
  \item[$VV'V''$:] \NLOQone~corrections have been known for
    many years. More recently, \NLOEone~corrections became available for the
    on-shell processes involving
    three~\cite{Nhung:2013jta,Yong-Bai:2015xna,Yong-Bai:2016sal,Hong:2016aek,Dittmaier:2017bnh}
    and two~\cite{Chong:2014rea,Yong:2017nag} massive vector bosons,
    in some cases with their subsequent leptonic decays treated in narrow-width approximation.
    Given the comparable complexity to $W^+W^++2j$ production, where full NLO SM corrections
    could already be achieved, \NLOEone~corrections to all these processes including full
    leptonic decays may be considered in reach. 
    The processes involving two and three photons in the final state were
    completed at \NLOQone{} and \NLOEone~accuracy~\cite{Greiner:2017mft}, taking into
    account full leptonic decays in case of $V\gamma\gamma$ production.
    
  \item[$\gamma\gamma,\gamma\gamma+j$:] This process remains an important
  ingredient in Higgs measurements at Run II. Originally computed at \NLOQ2
  with $q_T$ subtraction \cite{Cieri:2015rqa}, it has recently been re-computed
  \cite{Campbell:2016yrh} using the $N$-jettiness subtraction implemented
  within MCFM. The $q_T$ resummation at NNLL requested on the 2013 wish list
  are also now available \cite{Cieri:2015rqa}. Given the recent excitement in
  di-photon production a detailed understanding of these processes at high $q_T$ will
  be important in the coming years. Prospects for \NLOQ3 corrections remain
  closely connected with differential Higgs and Drell-Yan production at \NLOQ3.
  At high transverse momentum it may also be interesting to have \NLOQ2
  predictions for $\gamma\gamma+j$. Given that this is of equivalent complexity
  to $3j$ production we add this process to the wish list.


\end{itemize}

\subsection{Top quark associated processes}
An overview of the status of top quark associated processes is given in Table~\ref{tab:SM_wishlist:wlTJ}
\begin{table}
  \renewcommand{\arraystretch}{1.5}
\setlength{\tabcolsep}{5pt}
  \centering
  \begin{tabular}{lll}
    \hline
    \multicolumn{1}{c}{process} & \multicolumn{1}{c}{known} &
    \multicolumn{1}{c}{desired} \\
    \hline
    $pp\to t\tb$ &
    \begin{tabular}{l}
      \NLOQ2+\NLOEone \\
      \NLOQone{ }(w/ decays, off-shell effects)\\
      \NLOEone{ }(w/ decays, off-shell effects)\\
    \end{tabular} &
    \begin{tabular}{l}
      \NLOQ2{ }(w/ decays)
    \end{tabular} \\
    \hline
    $pp\to t\tb+j$ &
    \begin{tabular}{l}
      \NLOQone{ }(w/ decays) \\
      \NLOEone
    \end{tabular} &
    \begin{tabular}{l}
      \NLOQ2+\NLOEone{ }(w/ decays)
    \end{tabular} \\
    \hline
    $pp\to t\tb+2j$ &
    \begin{tabular}{l}
      \NLOQone{ }(w/ decays)
    \end{tabular} &
    \begin{tabular}{l}
      \NLOQone+\NLOEone{ }(w/ decays)
    \end{tabular} \\
    \hline
    $pp\to t\tb+Z$ &
    \begin{tabular}{l}
      \NLOQone+\NLOEone{ }(w/ decays)
    \end{tabular} &
    \begin{tabular}{cl}
      \\
    \end{tabular} \\
    \hline
    $pp\to t\tb+W$ &
    \begin{tabular}{l}
      \NLOQone \\
      \NLOEone
    \end{tabular} &
    \begin{tabular}{l}
      \NLOQone+\NLOEone{ }(w/ decays)
    \end{tabular} \\
    \hline
    $pp\to t/\tb$ &
    \begin{tabular}{l}
      \NLOQ2{*}(w/ decays)
    \end{tabular} &
    \begin{tabular}{l}
      \NLOQ2+\NLOEone{ }(w/ decays)
    \end{tabular} \\
    \hline
  \end{tabular}
  \caption{Precision wish list: top quark  final states. \NLOQ2$^{*}$ means a
   calculation using the structure function approximation.}
  \label{tab:SM_wishlist:wlTJ}
  \renewcommand{\arraystretch}{1.0}
\end{table}

\begin{itemize}[leftmargin=2cm]
  \item[$t\tb$:] 
Fully differential predictions for $t\tb$ production at \NLOQ2 are available~\cite{Czakon:2015owf,Czakon:2016ckf} are available, as well as {\tt fastNLO} tables~\cite{Czakon:2017dip} and a study on scale choices~\cite{Czakon:2016dgf}.
The \NLOQ2 corrections have recently been combined with NLO electroweak corrections~\cite{Czakon:2017wor}.
Polarized double-virtual amplitudes for heavy-quark pair production are also available~\cite{Chen:2017jvi}.
Complete \NLOEone\ corrections have been calculated in Ref.~\cite{Denner:2016jyo} for both the on-shell case and with complete off-shell effects.
Electroweak corrections to multi-jet merged on-shell top quark pair production have been presented in Ref.~\cite{Gutschow:2018tuk}.
 
Radiative corrections to top quark decays have been calculated in
Refs.~\cite{Bernreuther:2004jv,Melnikov:2009dn,Campbell:2012uf}, and have been extended up to NNLO QCD~\cite{Gao:2012ja,Brucherseifer:2013iv}.
Resummation has been accomplished up to NNLL, together with other improvements going beyond fixed
order~\cite{Beneke:2011mq,Cacciari:2011hy,Ferroglia:2013awa,Broggio:2014yca,Kidonakis:2015dla,Pecjak:2016nee}. 

NLO QCD corrections to $W^+W^- b\bar{b}$ production with full off-shell effects have been performed in
Refs.~\cite{Denner:2010jp,Denner:2012yc,Bevilacqua:2010qb,Heinrich:2013qaa}  including leptonic
decays of the $W$ bosons, and in Ref.~\cite{Denner:2017kzu} in the lepton plus jets channel.
In Refs.~\cite{Frederix:2013gra,Cascioli:2013wga}, NLO calculations in the 4-flavour scheme, i.e. with massive $b$-quarks, have been performed.

In the narrow-width-approximation (NWA) the NLO QCD calculation has been matched 
to a parton shower in Ref.~\cite{Campbell:2014kua}
within an extension of the {\tt PowHeg}~\cite{Frixione:2007vw,Alioli:2010xd} framework, called {\tt ttb\_NLO\_dec}
in the {\tt POWHEG-BOX-V2}.
Within the {\tt Sherpa} framework, NLO QCD predictions for top quark
pair production with up to three jets matched to a parton shower are
also available~\cite{Hoeche:2014qda,Hoche:2016elu}.
A new NLO multi-jet merging algorithm relevant to top quark pair
production is also available in {\tt Herwig\,7.1}~\cite{Bellm:2017idv}.

Based on an NLO calculation of $W^+W^- b\bar{b}$ production combined
with the {\tt Powheg} framework, first results of the $W^+W^-
b\bar{b}$ calculation in the 5-flavour scheme matched to a parton
shower have been presented in Ref.~\cite{Garzelli:2014dka}. 
However, it has been noticed later that the matching of NLO matrix elements
involving resonances of coloured particles to parton showers poses
problems which can lead to artefacts in the top quark lineshape~\cite{Jezo:2015aia}.
As a consequence, an improvement of the resonance treatment has been
implemented in {\tt POWHEG-BOX-RES}, called ``resonance aware matching'', 
and combined with NLO matrix elements from OpenLoops~\cite{Cascioli:2011va}, to arrive at 
the most complete description so far~\cite{Jezo:2016ujg}, based on the
framework developed in Ref.~\cite{Jezo:2015aia} and the 4FNS calculation of
Ref.~\cite{Cascioli:2013wga}. An alternative algorithm to treat
radiation from heavy quarks in the {\tt Powheg} NLO+PS framework has been
presented in Ref.~\cite{Buonocore:2017lry}.

NLO and off-shell effects in top quark mass determinations, comparing fixed order results with full off-shell effects, results based on the NWA including NLO corrections to both production and decay, and results based on NLO $t\tb$ matched to a parton shower, have been studied in Ref.~\cite{Heinrich:2017bqp}.
Various aspects of the definition and extraction of the top quark mass have also been studied recently in Refs.~\cite{Beneke:2016cbu,Butenschoen:2016lpz,Kawabata:2016aya,Hoang:2017btd,Hoang:2017kmk,Bevilacqua:2017ipv,Corcella:2017rpt,Ravasio:2018lzi}.

In terms of experimental precision, the inclusive $t\tb$ production cross section has been measured by ATLAS and CMS Collaborations 
at $\sqrt{s} = 7, 8$ and $13$~\TeV.
\begin{table}[t]
\begin{center}
\begin{tabular}{cccc}
$\sqrt{s}$ & ATLAS & CMS & NNLO+NNLL \\\hline 
7 TeV  & 3.9\% & 3.6\% & 4.4\%     \\
8 TeV  & 3.6\% & 3.7\% & 4.1\%   \\
13 TeV & 4.4\% & 5.3\% & 5.5\%   \\
\end{tabular}
\caption{
Experimental uncertainty $\Delta\sigma_{t\tb}/\sigma_{t\tb}$ on the inclusive $t\tb$ production cross section measurements, in the electron-muon channel at the 
LHC~\cite{Aad:2014kva, Aaboud2016136, Khachatryan:2016mqs, Khachatryan:2016kzg} compared to the precision of the NNLO+NNLL calculation~\cite{Czakon:2013goa,Czakon:2011xx}.}
\label{tab:SM_wishlist:expTTbar}
\end{center}
\end{table}
The measurements' uncertainty is a bit smaller than the corresponding theoretical calculations (Table~\ref{tab:SM_wishlist:expTTbar}). 
Significant part of the theory uncertainty  stems from PDFs and $\alpha_\mathrm{s}$. For example, in the $13$~\TeV calculation
$\sim 4.2$\% comes from PDFs and $\alpha_\mathrm{s}$, while the scale uncertainty is about $3.5$\%.
In terms of the total production cross section, the measurements agree with the theoretical predictions within the quoted uncertainties. 
However, a long standing problem related to the discrepancy observed in the transverse momentum distribution of the top-quarks 
(see e.g.,~\cite{Khachatryan:2015oqa, Aad:2015mbv}), still misses from a complete resolution though higher
order effects~\cite{Czakon:2016ckf} seem to alleviate at least partially the effect. 
Understanding the origin of this discrepancy is important for the LHC physics programme
since it affects directly (or indirectly) many physics analyses for 
which the $t\tb$ is a dominant source of background.

  \item[$t\tb\,j$:] NLO QCD corrections have been calculated in~\cite{Dittmaier:2007wz,Melnikov:2010iu,Melnikov:2011qx} for on-shell top quarks, 
results matched to a parton shower are also available~\cite{Kardos:2011qa,Alioli:2011as}.
In Ref.~\cite{Bevilacqua:2015qha,Bevilacqua:2016jfk}  full off-shell effects have been included. 
Recent studies about the extraction of the top quark mass based on $t\tb$\,jet production have been performed in~\cite{Fuster:2017rev,Bevilacqua:2017ipv}.

  \item[$t\tb V$:]  Will help to improve the constraints on anomalous EW couplings in the top quark sector during Run II.
  \NLOQone~corrections to $t\tb Z$ including decays have been considered in Ref.~\cite{Rontsch:2014cca,Rontsch:2015una}.
  NLO QCD corrections to $t\tb\gamma\gamma$ production matched to a parton shower, 
  with focus on observables which are sensitive to the polarisation of the top quarks, have been calculated in~\cite{vanDeurzen:2015cga}.
    Both \NLOEone\ and QCD corrections have been
    computed within the  \WishListMGaMC framework~\cite{Frixione:2015zaa}.
   Recent developments include the calculation of the NNLL corrections to the associated production of a top pair and a W boson~\cite{Broggio:2016zgg} 
   resp. a Z boson~\cite{Broggio:2017kzi}, and an investigation of the relative sizes of various types of QCD and EW corrections for $t\tb W$ and $t\tb t\tb$ production~\cite{Frederix:2017wme}.

\item[$t$/$\tb$:] Fully differential \NLOQ2~corrections have been completed for the dominant $t$-channel production process, first for stable tops~\cite{Brucherseifer:2014ama}, and more recently including top-decays to 
NNLO accuracy, in the NWA~\cite{Berger:2016oht}. Both the computations~\cite{Brucherseifer:2014ama,Berger:2016oht}
were performed in the structure function approximation. 
NLO QCD corrections to $t$-channel electroweak $W+bj$ production, with finite top-width effects taken into account are available within MG5\_aMC@NLO~\cite{Papanastasiou:2013dta}, see also~\cite{Frixione:2005vw} for earlier work within MC@NLO.
NLO QCD corrections to single top production in the in the $t, s$ and $tW$ channels are also available in {\sc Sherpa}~\cite{Bothmann:2017jfv} and in {\tt POWHEG}~\cite{Alioli:2009je,Re:2010bp}.
Differential distributions for $t$-channel single top production and decay at NNLO in QCD have been presented recently in~\cite{Berger:2017zof}.
A determination of the top-quark mass from hadro-production of single top-quarks has been performed in~\cite{Alekhin:2016jjz}, 
and in~\cite{Martini:2017ydu} using the Matrix Element Method at NLO QCD. 
\end{itemize}





\section{NTuples for NNLO processes~\protect\footnote{
      D.~Ma\^{\i}tre}{}}
\label{sec:SM_NNLOntuples}


NTuples files have proven very useful for the dissemination of NLO results for high multiplicity processes \cite{Bern:2013zja}. This contribution investigates the possibility of extending the use of event files for NNLO processes. As an example we investigate the dijet production at NNLO \cite{Currie:2017eqf}.

\subsection{Introduction}
A first investigation of event files for NNLO was performed in the previous Les Houches workshop \cite{Badger:2016bpw} where the process $e^+e^-\rightarrow 3j$ was considered at NNLO using {\tt EERAD3} \cite{GehrmannDeRidder:2007jk}. The size of the event files were found to be in an acceptable region. 

In this contribution we turn to the inclusive hadronic production of two jets \cite{Currie:2017eqf}. Hadronic processes introduce a range of new complication compared to the original study:
\begin{itemize}
\item the scale variation is more complex,
\item we need to include information about the initial state flavours and momentum fractions.
\item the infrared structure of the process is much more complex and therefore the number of subtractions is much enhanced.
\end{itemize} 
The goal of this study is to ascertain whether using event files for hadronic processes is tractable from the point of view of the storage space needed to gather enough statistics. 

\subsection{Storage format}
For this experiment we used a ROOT \cite{Brun:1997pa} file backend for storage. In the previous study the layout of the NLO nTuples described in \cite{Bern:2013zja}. For this study we have decided to modify the file layout from a quite specialised layout to a more general one. For a given phase-space point the form of a weight is given by:
\begin{equation}\label{eq:SM_nnlontuples:weight}
\omega=\alpha_S^n \;{\rm pdf}(x_1, id_1,\mu)\;{\rm pdf}(x_2, id_2,\mu)\times\left(c_0+c_1 \log(\mu^2)+c_2 \log^2(\mu^2)+\dots\right)\;,
\end{equation}
where we have set the factorisation and renormalisation scale to be the same. Keeping track of both dependence separately is possible but not considered in this study. This omission is not relevant for the purpose of the study: the contribution that requires the largest storage is the double real radiation contribution, which has no logarithmic dependence on the scale. 

So for each phase-space point we save the final state kinematic and flavour information, and an array of entries
\[n,x_1,x_2,id_1,id_2,j,c_j\]
corresponding to every coefficient of a logarithm in the weight in Eq.~\eqref{eq:SM_nnlontuples:weight}.
 
\subsection{Event file sizes}

The program {\tt NNLOjet} \cite{Currie:2017eqf} we used to estimate the file sizes separates the calculation is several parts: 
\begin{itemize}
\item the born cross section (LO), 
\item the NLO virtual (V) and real (R) parts
\item the two-loop contribution (VV)
\item the squared one-loop three jet contribution (RV) 
\item the double real radiation contribution (RR), separated in two parts (RRa, RRb).
\end{itemize}
Each part is integrated separately and has different characteristics that impact on the storage size. 
Table~\ref{tab:SM_nnlontuples:size} shows the size of the storage required for each part of the calculation per event. These numbers were obtained by creating a small event file for each part. We use this estimate to extrapolate the size of the storage capacity needed for a realistic scenario. 
\begin{table}
\begin{center}
\begin{tabular}{ c c c c c}
 part & $\alpha_s$ order & size [kB/event] & est. need [$10^9$ events] & est. size [TB] \\\hline 
LO     & 2  &                 0.21  &        10.0 &        1.9  \\
V      & 3  &                 3.86  &         5.0 &       18.0  \\
R      & 3  &                 8.95  &         5.0 &       41.7  \\
VV     & 4  &                 5.51  &        10.0 &       51.3  \\
RV     & 4  &               154.09  &         2.0 &      287.0  \\
RRa    & 4  &               463.91  &         5.0 &     2160.2  \\
RRb    & 4  &               124.09  &         2.5 &      288.9  \\
\end{tabular}
\end{center}
\caption{Size of the event files produced using NNLOjet for the dijet production process at NNLO.}\label{tab:SM_nnlontuples:size}
\end{table}

Table~\ref{tab:SM_nnlontuples:size} shows the size writing out the weights and the momentum configurations in the order {\tt NNLOjet} produces them. We can reduce the size of the event file by collecting all weights that share the same final state phase-space configuration and collect weights corresponding to the same value of $n,x_1,x_2,id_1,id_2,j,c_j$, the factor in size gained and the resulting estimated storage capacities needed are listed in Table~\ref{tab:SM_nnlontuples:compact}. From this table we can see that the order of magnitude of the storage space needed for the full process is of the order of 100 TB. This is somewhat higher than a comfortable size to work with, but not completely unmanageable.       
\begin{table}
\begin{center}
\begin{tabular}{ c c c c c c}
 part & $\alpha_s$ order & size [kB/event] & est. need [$10^9$ events] & est. size [TB] & gain factor\\\hline 

LO     & 2  &                 0.13  &        10.0 &        1.2 & 1.6 \\
V      & 3  &                 0.86  &         5.0 &        4.0 & 4.5 \\
R      & 3  &                 1.79  &         5.0 &        8.3 & 5.0 \\
VV     & 4  &                 1.32  &        10.0 &       12.3 & 4.2 \\
RV     & 4  &                12.52  &         2.0 &       23.3 & 12.3 \\
RRa    & 4  &                12.84  &         5.0 &       59.8 & 36.1 \\
RRb    & 4  &                 8.13  &         2.5 &       18.9 & 15.3 \\
\end{tabular}
\end{center}
\caption{Size of the event files produced using {\tt NNLOjet} for the dijet production process at NNLO, reordering the phase-space points and collecting weights with the same final state. The factor in storage space gained with respect to the na\"ive strategy is given in the last column.}\label{tab:SM_nnlontuples:compact}
\end{table}

There are optimisations that could be used to further reduce the storage requirements that we have not considered yet and are left for further studies. 

\subsection{Conclusions}
In this contribution we investigated the size of the storage needed to produce event files for the dijet production at NNLO accuracy. Our current implementation forecasts storage requirements in the region of $100\,{\rm  TB}$. This requirement is quite high but not completely intractable. The code we used for the analysis is optimised for CPU consumption and not minimising the number of weights produced, there are therefore further improvement that can be considered to reduce the size requirement without affecting the accuracy, these will be considered in further studies. It should be noted that the information collected in the event files is exactly what is needed to produce fastNLO/APPLgrid/MCgrid tables \cite{Carli:2010rw,Britzger:2012bs,Bothmann:2015dba} for specific histograms. NNLO event files can therefore be seen as a useful intermediate format. 

\subsection*{Acknowledgements}
We would like to thank Nigel Glover and Juan Cruz-Martinez for providing us with the {\tt NNLOjet} program and guidance for its usage.











\definecolor{maroon}{rgb}{0.8, 0.3, 0.4}
\definecolor{linkblue}{rgb}{0.0, 0.2, 0.7}

\section{Progress in grid techniques for the fast reproduction of QCD calculations at NNLO~\protect\footnote{
      D.~Britzger,
      C.~Gwenlan,
      A.~Huss,
      K.~Rabbertz,
      M.~R.~Sutton}{}}
\label{sec:SM_fastNNLO}


Techniques for generating interpolation grids for perturbative 
QCD calculations offer a fast and flexible way to reproduce those calculations with any choice of parton 
distribution function set or any value of the strong coupling constant.
Such grids are suitable for the iterative fitting of parton distributions and standard model parameters, 
or detailed studies of scale dependence.  
Recent developments are briefly discussed for the APPLfast project 
which implements a common interface for the APPLgrid and fastNLO fast interpolation grid libraries with the NNLOJET 
calculation for general
cross section calculations at next-to-next-to-leading order.   
Progress towards a more generic common lower level interface for all interaction with the fastNLO 
and APPLgrid backends suitable for use with other QCD calculations is also briefly discussed. 

\subsection{Introduction}

The LHC performed extremely well both during Run 1 -- with 7 and 8 TeV collisions -- and, 
since 2015, during Run 2 with collisions at 13 TeV centre of mass energy. This has allowed measurements of the $pp$ 
interaction cross sections over an unprecedentedly large kinematic region of many orders of 
magnitude in both the hard process scale, and proton momentum fraction, $x$. 
These increasingly precise measurements highlight the need
for a more precise understanding of parton distributions functions (PDF) within the proton. 
Several important channels -- such as searches for new, massive particles and  Higgs
production -- suffer from large theoretical uncertainties dominated by
PDFs, which also play a r\^{o}le in limiting the precision of various important
parameters such as the $W$-boson mass, or strong coupling, $\alpha_s(M_Z^2)$. 
As LHC Run 2 continues, the expected increase in
experimental precision in the new phase space region afforded by the 13~TeV collisions, 
renders it evermore important to have access to precise QCD calculations at high orders,
and a correspondingly precise knowledge of the proton PDF. For a recent review, see elsewhere~\cite{Gao:2017yyd}. 

Higher order calculations at hadronic colliders generally require the  
numerical integration over the kinematics of the final state particles necessary for 
the cancellation of the infrared and collinear singularities. To achieve the required statistical precision, these 
calculations  typically take many 
thousands of CPU hours. For an iterative fit for the proton PDF these calculations would 
need to be performed many hundreds, or thousands of times, once for each of the
points in the PDF minimisation space, thus precluding the use of these higher order QCD calculations directly in 
PDF fits.

Since the turn of the current century, fast grid techniques have been developed which allow the storage of the 
weights of these higher order calculations on an interpolation grid. In this way, the convolution of the weights 
with the PDF can be performed {\em a posteriori}, such that the time consuming QCD calculation need
be performed only once. These techniques were first used for jet production at next-to-leading order 
(NLO) in $ep$ collisions at HERA~\cite{Adloff:2000tq,Chekanov:2005nn}, and were then extended to jet production 
at NLO at the LHC with the APPLgrid~\cite{Carli:2005ji,Carli:2010rw} and the fastNLO~\cite{Kluge:2006xs,Britzger:2012bs} projects.
The complete range of two-to-two processes for NLO QCD processes at the LHC became available in 
2013, and in principle any NLO process can now be used with such grids.

The next-to-next-to-leading order (NNLO) corrections for inclusive electroweak boson production 
have been known for some 
time~\cite{Hamberg:1990np,Anastasiou:2003yy,Catani:2010en,Catani:2009sm,Gavin:2012sy,Gavin:2010az,Anastasiou:2003ds,Ridder:2016nkl}. 
The recent, much anticipated completion of the 
inclusive jet~\cite{Currie:2016bfm} and dijet~\cite{Currie:2017eqf} cross sections, together with the 
NNLO cross section for Z+jets production~\cite{Ridder:2015dxa} implemented within the unified 
framework of the NNLOJET calculation provides a valuable opportunity to be able to make use 
of these new QCD calculations in a PDF fit.

To this end, a project to implement a combined interface between the NNLOJET calculation and 
the fast grid technology in both the fastNLO and APPLgrid projects is underway. 
Known as APPLfast -- a portmanteau of the APPLgrid and fastNLO project names -- the intention 
is to provide an implementation of an interface that will allow for the production of full 
NNLO grids for any of the processes implemented within the NNLOJET program.

\subsection{The APPLfast project}

The APPLfast project was initiated at the QCD@LHC workshop at the end of 2015. The first 
promising results regarding the closure of the grids for the different contributions to 
the cross section have been shown previously. Here a very brief update on the project and the current status is provided 
together with some potential directions for future development. 

The APPLfast code is written in C++ with Fortran callable sections. It is structured as a lightweight 
library which can be used to bridge between the NNLOJET code and the specific code for 
booking, and filling the grids. In line with many other unix based applications, including 
the APPLgrid and fastNLO packages themselves, the code is automatically configured and 
compiled with autotools. 

Two of the major design principles for the interface to the NNLOJET code are that it 
should unify as much as possible the code that is required to interface with either of 
the underlying APPLgrid or fastNLO technologies, and that the interface itself 
should be as unobtrusive as possible in the NNLOJET code. In particular this last objective 
means that the interface should 
provide no additional performance overhead in terms of execution time when not filling 
a grid, and when filling a grid, should keep any additional overhead as low as possible.
This is achieved by the use 
of a minimal set of hook functions that can be called from within the NNLOJET code itself
and which can be left within the code with no impact on performance if the grid filling 
functionality is not required. This is in line with the implementation of the interfaces 
with other QCD calculations~\cite{Bertone:2014zva} and it is hoped that this may eventually be 
generalisable to simplify the production of an interface with any future calculations.

Fast interpolating grids themselves work by taking the PDF evaluated at each 
$x_1$, $x_2$, $\mu_F^2$ phase space 
point, and generating interpolating coefficients which can be used to generate this PDF 
value using a smaller number of PDF evaluations at specific grid nodes. By then storing 
each weight from the hard process on each of the relevant grid nodes, but additionally weighted 
by these interpolating coefficients, summing the product of the PDF evaluated at the grid nodes 
with the stored weights, will result in the generation of the full cross section. 
This is a quadrature method the precision of which is determined predominantly by the
quality of the interpolation.  Typically a grid {\em warm up} stage is required in order 
to more precisely determine more optimal limits for the phase space in $x$ and $\mu_F^2$ of the grids.  
During this stage, the NNLOJET code runs in a custom mode which generates a uniform weight 
for each phase space point rather than the full weights from the matrix elements.

In the endeavour towards the implementation of a common interface, in line with the first 
principle,  opportunities 
have been taken for the reuse of existing tools or implementation of common tools. 
One of the most significant aspects has been in the configuration of the grids themselves, 
and the mapping of the internal processes of the NNLOJET code to the different parton 
luminosity contributions within the grids in an efficient way, discussed briefly in the following section.

\subsubsection{Implementation of parton luminosity contributions}

In order to map between the internal processes of the NNLOJET code and the smaller space of parton-parton 
luminosity processes stored in 
the grids, both APPLgrid and fastNLO make use of the parton luminosity class
from APPLgrid. In this class the different parton-parton contributions are each assigned to a specific
{\em subprocess}. Ignoring top, in principle there can be 121 different subprocceses. However, the presence of 
different terms in a calculation which might have the same parton-parton input, means that the 
class must be able to map between the different internal processes in the NNLO code to these individual 
{\em subprocesses}. In addition, the presence of terms in the matrix elements which contain for example
different CKM matrix terms, means that for {\em a posteriori} setting of the CKM matrix, these different 
contributions should be mapped to different {\em subprocesses} even if their initial parton-parton
contributions are the same.

The ability to map different internal NNLO calculation sub-contributions to the same, or different 
{\em subprocesses} is implemented in the class. Although separation into the different CMK contributions 
is not yet fully implemented, it need only be performed at the point of determining the mapping of 
processes in the NNLOJET initialisation, and should be considered to be straightforward.

\begin{figure}[t]
\includegraphics[width=\textwidth]{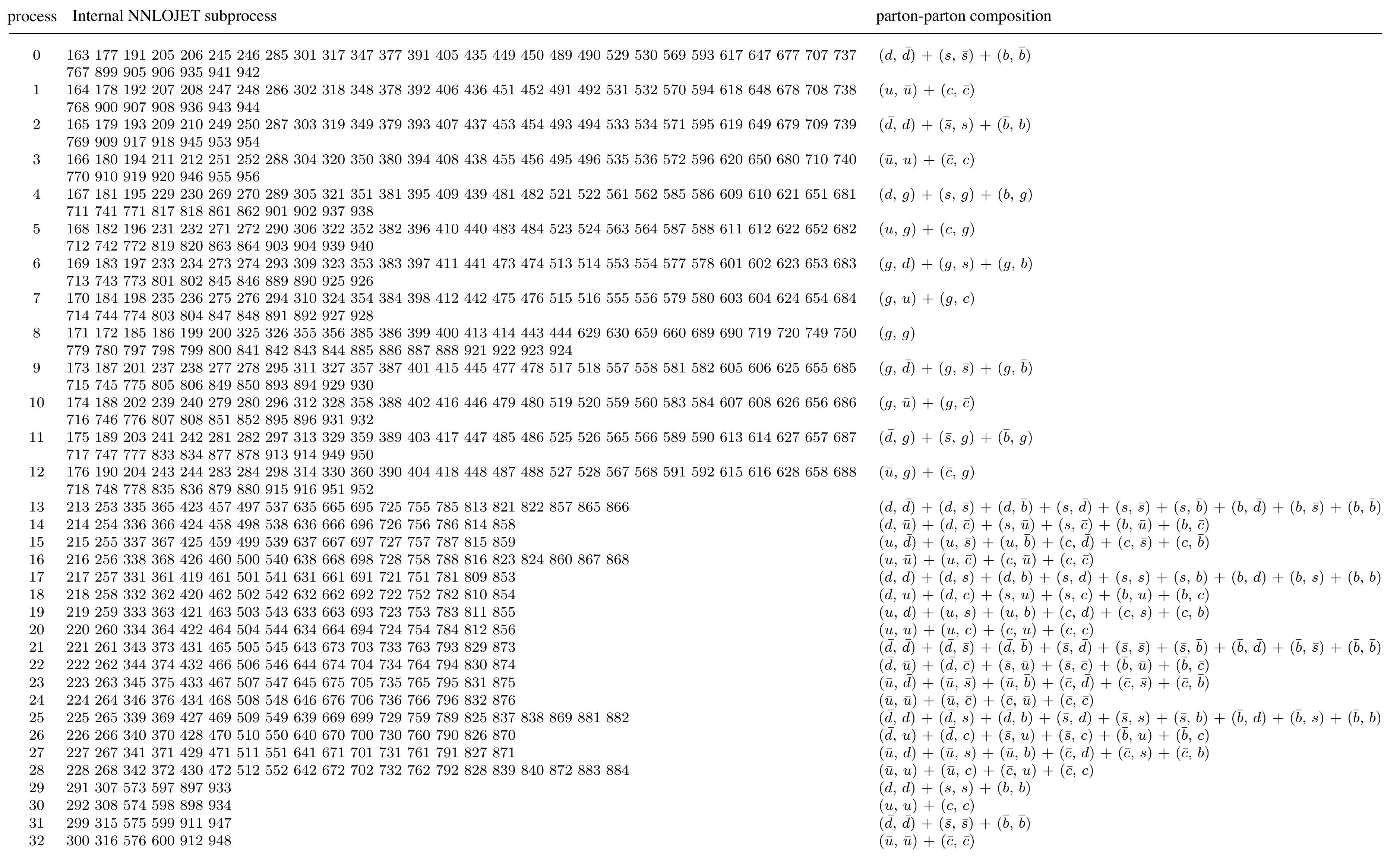}
\caption{The NNLO contrbutions to the Z+jets calculation illustrating the 33 separate combinations of parton-parton 
input terms and their mapping to 768 separate internal NNLOJET contributions.}
\label{fig:SM_fastNNLO:one}
\end{figure}

By way of example, Fig.~\ref{fig:SM_fastNNLO:one} shows the set of parton-parton luminosities from the NNLO contribution to the $Z$+jets 
calculation, showing how the 768 internal NNLOJET processes are mapped to the separate 33 parton-parton 
luminosities stored in the grid.

\subsubsection{Contributions at NNLO}

The NNLOJET code itself generates the various contributions to the cross section individually,  
generating each of the the leading order (LO), the NLO real (R), NLO virtual (V), and the NNLO 
double-real (RR), double-virtual (VV),  and real-virtual (RV) contributions using a distinct 
run of the executable. The grids generated after each of these stages must then be combined to 
render the full cross section.

To produce a stable cross section at higher orders it is necessary to run a calculation generating 
a very large number of weights. 
This is particularly 
true for the double-real contribution since the large number of partons in the final state and the correspondingly
more complicated infrared structure, typically requires 
hundreds of thousands of CPU hours. Because of this, it is necessary that many hundreds or thousands of 
separate jobs are required for each sub-contribution. The resulting grids for each cross section from 
each job is typically ${\cal O}(10 - 100)$ MBytes in size, depending on the number of bins in the observable and details 
of the interpolation used. Fortunately, the grids obtained by summing the output from 
the many thousands of separate jobs is typically of a similar size to the largest single grid.

In order to reproduce the cross section from the calculation with adequate precision, it is necessary that
the grids themselves be combined using an analogous procedure to that used in the NNLOJET calculation itself.
In NNLOJET a sophisticated combination procedure is used~\cite{tamorganthesis,Ridder:2016rzm} which allows 
individual cross section bins to 
be weighted independently from their neighbours. It was necessary to implement this functionality 
in the combination of the output grids before any detailed comparisons of the fast grid convolution and 
the NNLO cross sections could be made.

\begin{figure}[t]
  \centering
  \includegraphics[width=0.48\textwidth]{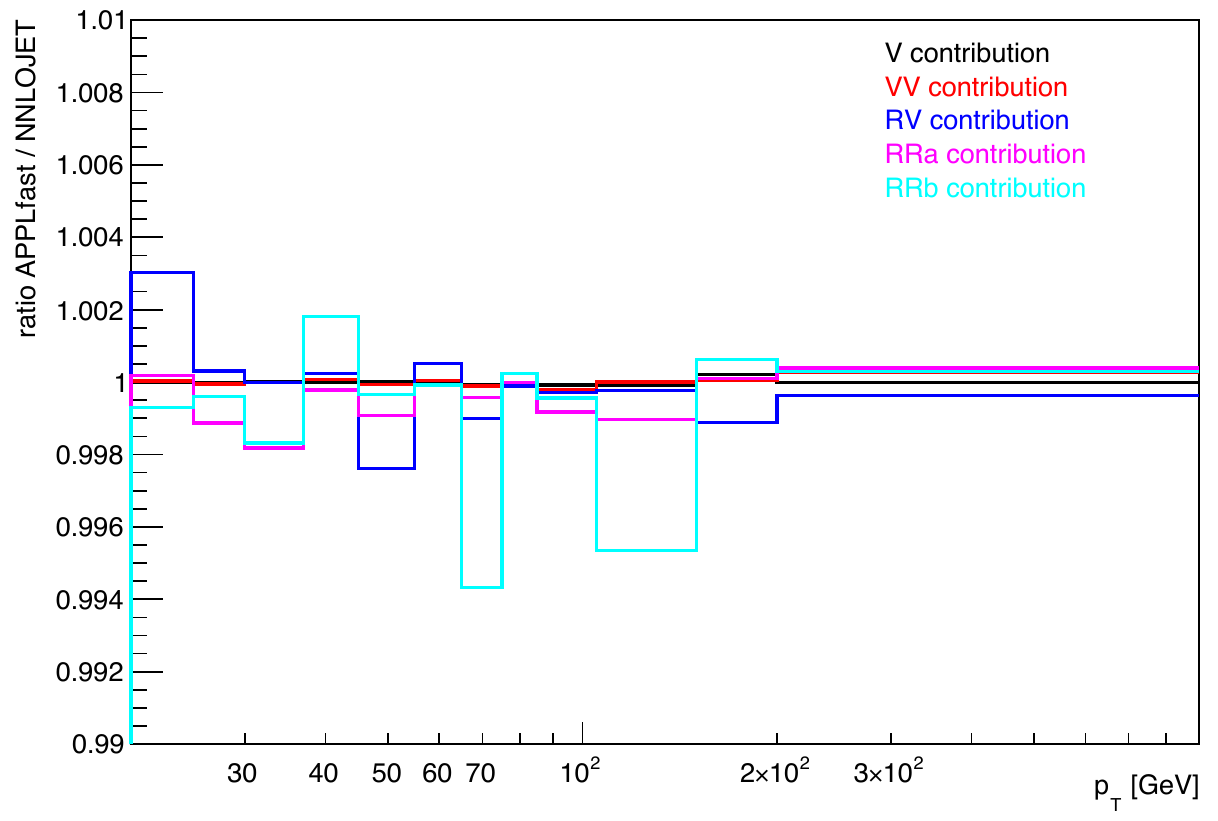}
  \caption{The ratio of the fast convolution from APPLfast with respect to the original NNLOJet calculation 
    for various contributions to the Z+jets cross section at NLO and NNLO.}
  \label{fig:SM_fastNNLO:two}
\end{figure}

Figure~\ref{fig:SM_fastNNLO:two} shows the ratio of the fast convolution from APPLfast with respect to the original 
NNLOJET calculation for the NNLO contributions from the $Z$+jets cross section, together with 
the NLO virtual contribution shown for reference.
In all cases the agreement is much better than 6\textperthousand.
Here the ratio is taken after combination of the grids following the NNLOJET combination procedure.
It is worth mentioning the apparent worse performance of the ratio for the real contributions -- 
particularly the double-real part. Although this is a quadrature method -- each individual weight should
be equally well stored on the grid -- the precision of the convolution is affected by how well optimised 
the grid is after the initial warm up stage. For the real-virtual and double-real contributions shown here, 
a longer warmup phase would be beneficial.

\begin{figure}[t]
  \includegraphics[width=0.48\textwidth]{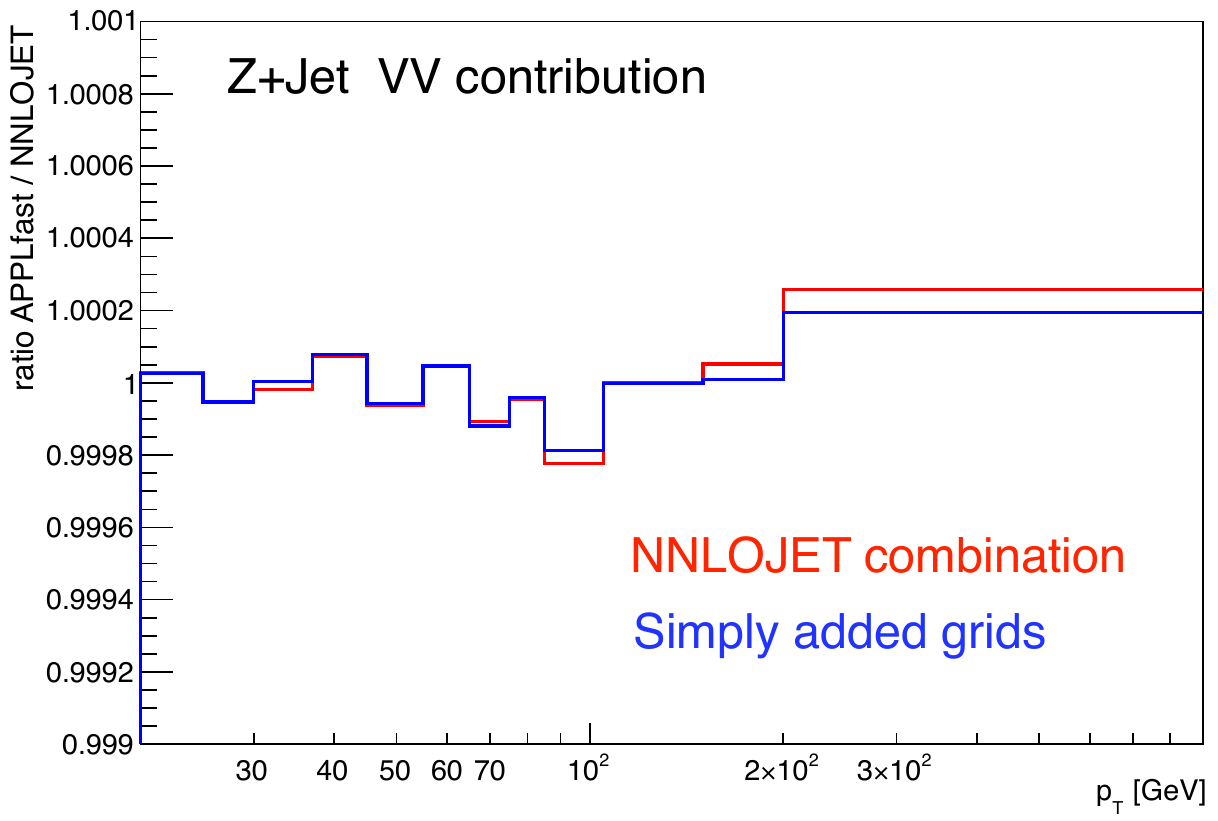}
  \hfill
  \includegraphics[width=0.48\textwidth]{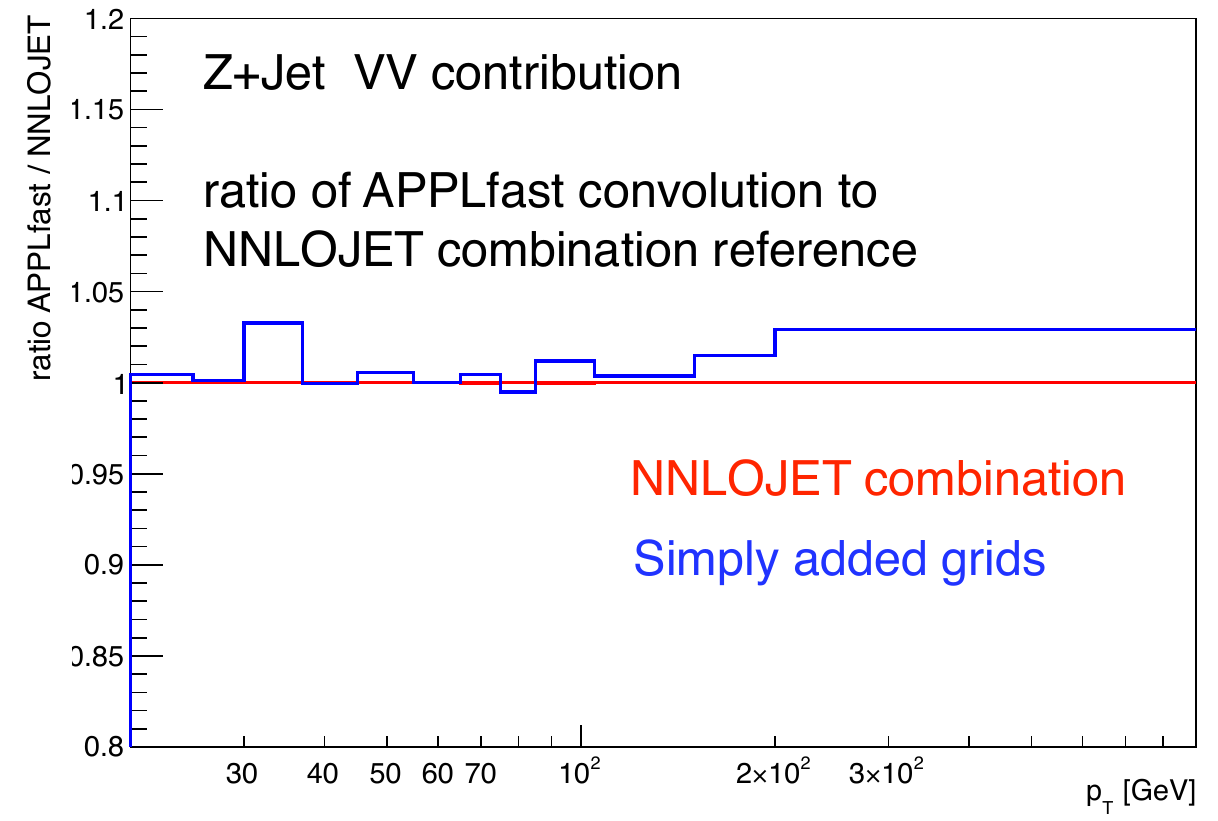}
  \caption{Ratios of the combined APPLfast convolution to the NNLOJET calcualtion for the NNLO double-virtual contribution. Different combination 
    schemes have been used for illustration: on the left the grid convolution with either particular scheme 
    is shown in ratio with the NNLOJET cross section, combined with the same scheme, while on the left, the 
    different grid combinations are shown always with respect to the NNLOJET combination using the NNLOJET 
    prescription. }
  \label{fig:SM_fastNNLO:three}
\end{figure}

Figure~\ref{fig:SM_fastNNLO:three} shows the combination in more detail specifically for the $Z$+jets double-virtual 
contribution. Note the reduced 
1\textperthousand\
maximum range of the vertical-axis. 
In the first panel, the ratio of the fast convolution with respect to the NNLOJET calculation is shown 
for two different types of combination; the first, a basic simple addition, where each grid and each bin is added 
with a weight of one; and the second, with the procedure from NNLOJET. In the case of the simply added grids the 
NNLOJET cross section has also been added with weight one. In both cases, the cross section is reproduced to 
within 0.4\textperthousand. In the second panel, the fast convolution with both of these addition schemes is 
compared with the NNLOJET cross section combined with the full NNLOJET prescription. Clearly in this case, 
the simply added grid is seen to agree with the NNLOJET cross section only to within approximately 4\%. 
This illustrates the correct function of the implementation the NNLOJET combination prescription when applied to the grids.

\begin{figure}[t]
  \centering
  \includegraphics[width=0.48\textwidth,height=0.35\textwidth]{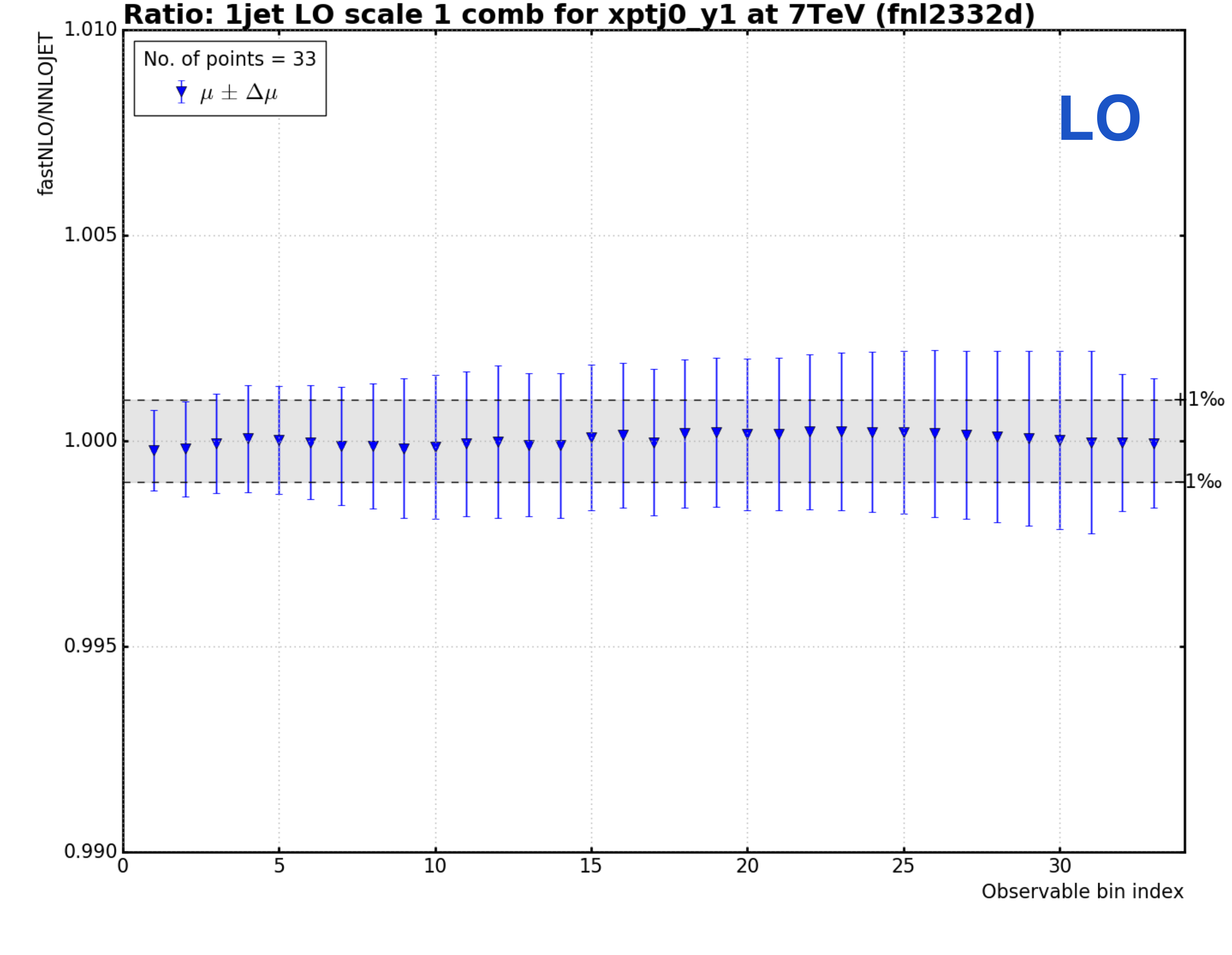}
  \hfill
  \includegraphics[width=0.48\textwidth,height=0.35\textwidth]{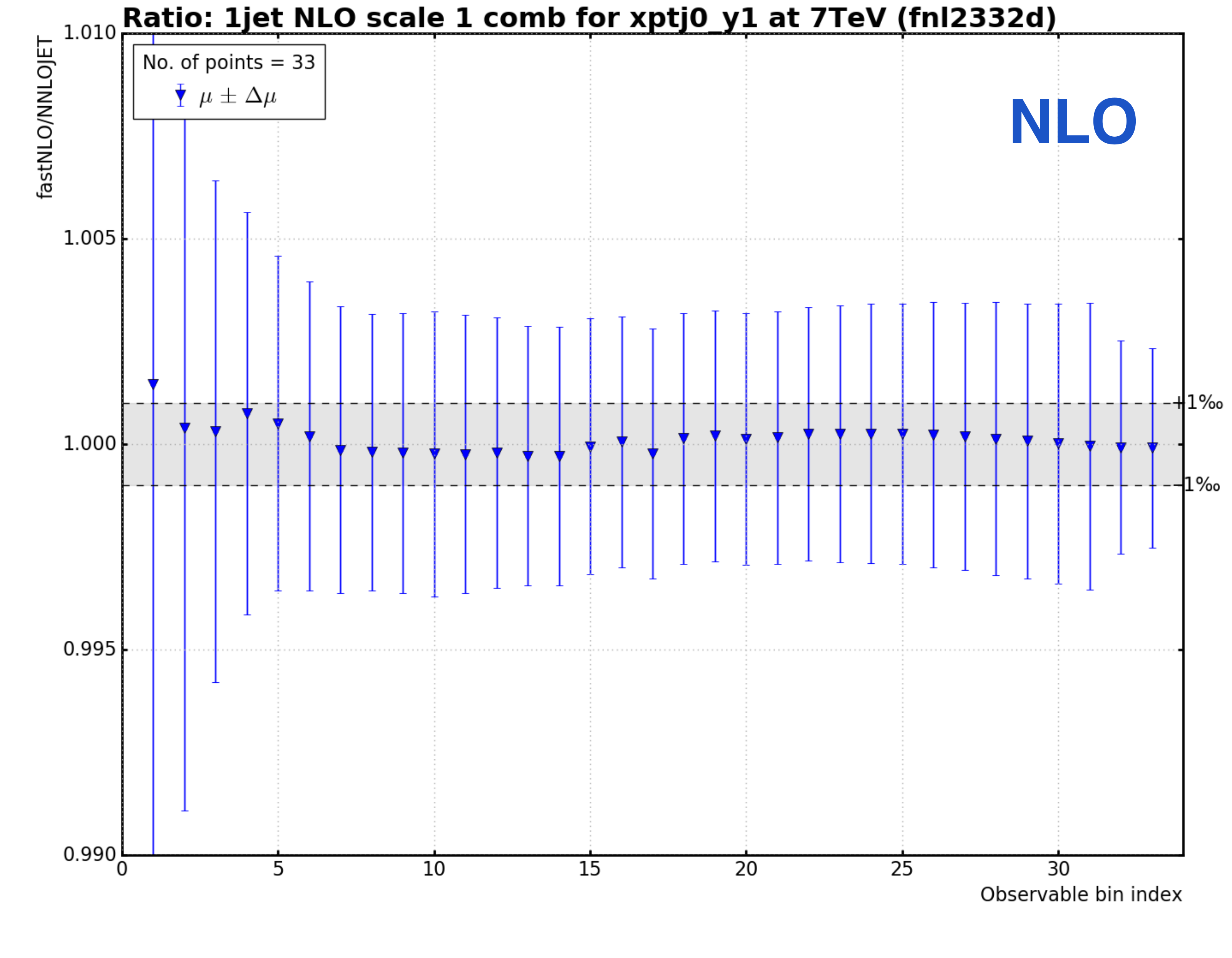}\\

  \includegraphics[width=0.48\textwidth,height=0.35\textwidth]{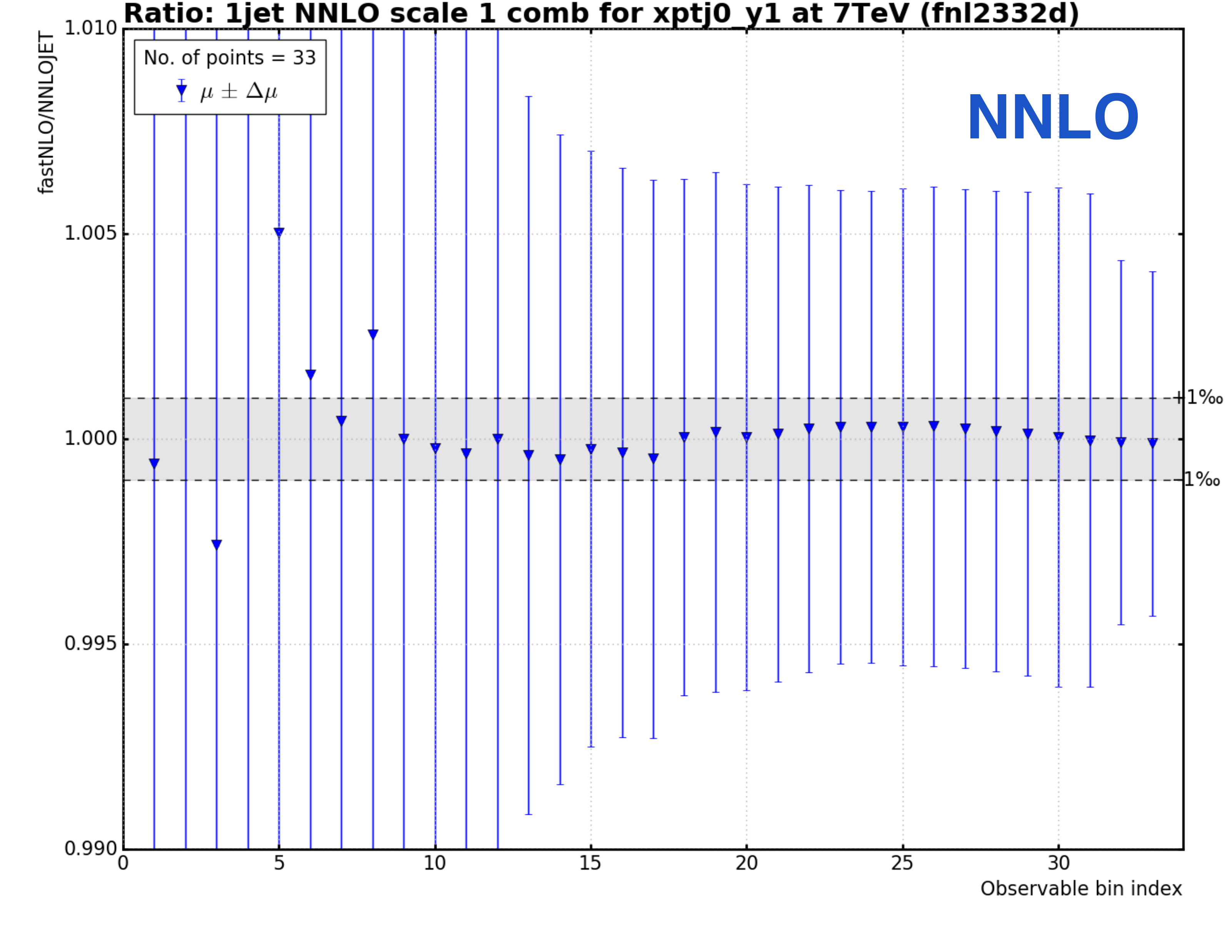}
  \caption{The ratio of the fast convolution from APPLfast with respect to the original NNLOJet calculation 
    for LO, NLO and NNLO contributions from the inclusive jet cross section. The statistical error on the calculation 
    is shown, with the 1\textperthousand\ region shown as the shaded band.}
  \label{fig:SM_fastNNLO:four}
\end{figure}

Figure~\ref{fig:SM_fastNNLO:four} shows the ratio of the the full convolution procedure for the combined 
cross section for each of the LO, NLO, and NNLO components for the inclusive jet 
cross section at the LHC, evaluated with the leading jet $p_T$ scale. Note the 1\% maximum vertical-axis 
range. Shown are the statistical uncertainties on the combined cross section and the inner shaded 
band represents the one per mille limit. The statistical uncertainties are larger on the NNLO 
contribution due to the NNLO real contributions. The fast convolution reproduces the NNLOJET cross section 
reliably at each of the orders, well within the statistical uncertainty on the calculation, 
but again, reasonably large excursions of up to 5\textperthousand\ are observed in the NNLO contribution 
which have since been seen to improve with higher numbers of weights generated during the phase space warmup.

What can be concluded from the results so far is that all the major conceptual 
hurdles have been overcome and the primary development is reasonably complete. Currently a 
campaign of high statistical precision grid production is underway. This campaign is a significant undertaking  
given the number of different physics processes and the different phase space and binning schemes 
used by the different experimental collaborations.

\subsection{Towards a standard grid interface}

A combined interface capable of filling grids in either the fastNLO or APPLgrid formats, 
while useful, is only the beginning. A more beneficial approach would be to implement in each grid 
technology a common, lower level API for generic grid interactions between formats such that the 
bridging code or any client PDF fitting code could be fully agnostic about which format was being used.

In this way, instead of a single package with the bridging code including two distinct interfaces sharing 
common elements, the bridge code itself would consist of only a truly single interface interacting with 
the common API. Small differences would be handled within the implementation of the API for the particular 
grid technology being used.
This should then render any change of the underlying grid technology essentially immaterial.

It is necessary to ensure some degree of backwards compatibility to ensure that 
existing grids which represent many thousands of hours of processing are still usable. Any such 
proposed new interface should then necessarily be seen as an evolving development implementation in the first instance, rather than 
the standard into which it will hopefully evolve.

In principle, stateless classes are to be preferred for the implementation of specific functionalities.
Consequently, the interface or the filling of the grid should take all the parameters that are needed so 
that it can present the filling operation as an atomic action.  

For a grid to be used for the fast convolution within a PDF fit, an interface for initiating the fast convolution 
to be called by the fitting code is also required,  
and it would be useful to define a separate, common interface for this purpose. This aspect of the interface will not be 
discussed here, except to say that is should fit into the general scheme. 
In this way, a number of application specific interfaces in a common namespace might be envisaged, for instance a common 
namespace,  {\tt lhgrid}, containing the interface to fill the grid in a specific API, {\tt lhgrid::fillgrid\_api}.
Any eventual interface for the fast convolution could for instance be {\tt lhgrid::convolute\_api}. The full grid 
implementation would then inherit both these interfaces, or handle interaction between helper classes each 
of which could implement only one of the interfaces.  

\begin{figure}[htp]
{\small \bf
{\color{maroon}
\begin{verbatim}
  /// production run  filling interface

  /// fill the grid with a single weight for a single parton luminosity
\end{verbatim}
}
\vspace{-4mm}
\begin{verbatim}
  virtual int fill( const double& x1, const double& x2, const double& Q2, 
                    const int&    order, 
                    const int&    process_index, 
                    const double& observable, 
                    double weight ) = 0;
\end{verbatim}
{\color{maroon}
\begin{verbatim}
  /// fill the grid with a single weight for a vector of weights - one 
  /// entry for each parton luminosity
\end{verbatim}
}
\vspace{-4mm}
\begin{verbatim}
  virtual int fill( const double& x1, const double& x2, const double& Q2, 
                    const int&    order, 
                    const double& observable,
                    const std::vector<double>& weights ) = 0;
\end{verbatim}
{\color{maroon}
\begin{verbatim}
  /// filling interface for the phase space warmup

  /// fill the grid with a single weight for a single parton luminosity
\end{verbatim}
}
\vspace{-4mm}
\begin{verbatim}
  virtual int fill_phasespace( const double& x1, const double& x2, 
                               const double& Q2, 
                               const int&    order, 
                               const int&    process_index, 
                               const double& observable, 
                               double weight ) = 0;
\end{verbatim}
{\color{maroon}
\begin{verbatim}
  /// fill the grid with a single weight for a vector of weights - one 
  /// entry for each parton luminosity
\end{verbatim}
}
\vspace{-4mm}
\begin{verbatim}
  virtual int fill_phasespace( const double& x1, const double& x2, 
                               const double& Q2, 
                               const int&    order, 
                               const double& observable,
                               const std::vector<double>& weights ) = 0;
\end{verbatim}
}
\caption{Proposed interface functions for filling of a standardised grid, including a warm up stage}
\label{fig:SM_fastNNLO:codeone}
\end{figure}

Figure~\ref{fig:SM_fastNNLO:codeone} illustrates an early implementation of the filling interface with filling methods 
to fill a single weight for a given phase space point for a given parton luminosity, of for filling 
the complete set of weights for all parton luminosities. A method to set a caching flag is also provided
although not shown here.

This design has been chosen to facilitate grid filling with the different weight generation schemes 
implemented by the many different calculations currently interfaced but also be flexible enough 
to accommodate new calculations. The different schemes considered are:
\vspace{2mm}  
\begin{itemize}
\item[i.] some calculations 
generate the weights for all parton luminosities at the same time for each phase space point
\item[ii.] some 
generate  only one weight, 
for a single phase space point for a single parton luminosity
\end{itemize}
\vspace{2mm}
The interface must be able to handle each case. Both of these cases may be operated in two modes, where 
the code {\em a)} generates these weights in sequences for a number of different contributions but all with the 
same, or related phase points, or {\em b)} where each phase space point is completely distinct. For instance, 
many general purpose Monte Carlo generators~\cite{Gleisberg:2008ta}, behave as in {\em iib)}, whereas many 
NLO calculations~\cite{Nagy:1998bb,Nagy:2001fj,Campbell:1999ah,Campbell:2000bg,Campbell:2010ff}
behave as in {\em ia)}.

The opportunity for caching is then required since the grid filling operation itself is quite costly --  
for each phase space point, the interpolating coefficients in each of $x_1$, $x_2$ and $\mu_F^2$ must be calculated.
For case {\em ia)}, no caching is required, but for {\em ib)} and particularly {\em iib)} it becomes very 
important to avoid the recalculation of the interpolating coefficients each time a weight is passed to the grid 
filling.      

Unfortunately, the use of a cache to some extent breaks the desired stateless operation of the interface, but
the filling process would  still be presented as an atomic operation.  There are methods for the filling of the 
grid during the phase space warmup stage, however these may in the end not be necessary, since the grid itself 
will know whether it is in a phase space warmup stage or not, and so can decide to fill with the weights or not. 
In principle the actual weights are not required during the phase space warmup -- only the values of the $x$, $\mu_F^2$ etc.
Similarly, the parton luminosity process index may not be required if the phase space is common to each parton
luminosity.

\begin{figure}
{\small \bf 
{\color{maroon}
\begin{verbatim}
  /// fill the reference histogram
\end{verbatim}
}
\vspace{-4mm}
\begin{verbatim}
  virtual int fill_reference( const double& observable, 
                              const double& weight ) = 0;
\end{verbatim}
{\color{maroon}
\begin{verbatim}
  /// fill the reference histogram for a particular parton luminosity
\end{verbatim}
}
\vspace{-4mm}
\begin{verbatim}
  virtual int fill_reference( const int&  process_index, 
                              const double& observable, 
                              const double& weight ) = 0;
\end{verbatim}
}
\caption{Proposed interface methods for filling of reference histograms.}
\label{fig:SM_fastNNLO:codetwo}
\end{figure}

To facilitate the closure testing of a grid, methods to fill a reference histogram are also provided, shown in 
Fig.~\ref{fig:SM_fastNNLO:codetwo}, 
such that the full weights, convoluted with the PDF when the grid was being filled can also be stored. 

By storing in the general grid state information the identity of the PDF used in the calculation for the grid construction, 
it will be possible for the grid itself to perform the closure test on the quality of the implementation directly
without input from the user.

\subsection{Outlook} 

The prospect for creation of fast interpolation  grids using the NNLOJET code is good -- essentially all major hurdles 
have been overcome and high statistical precision grids are being produced. 

An outline proposal for a common grid filling scheme that should simplify the process of creating an interface  
for any new calculations with the grid filling back ends has been proposed. Tests implementing this interface 
in development code have been made and the interface could be  used in the next major release 
of both APPLgrid and fastNLO. 
Operational issues may lead to a necessary period of evolution of the interface until it is fully stable. 
  
\subsection*{Acknowledgements}
The authors would like to thank the Institute of Particle Physics Phenomenology (IPPP) at Durham for the award 
of an IPPP Associateship to support this work, and gratefully acknowledges support from the state of Baden-W\"{u}rttemberg 
through bwHPC and the German Research Foundation (DFG) through grant no INST 39/963-1 FUGG. 




\section{The nested soft-collinear subtraction scheme~\protect\footnote{
  R.~R{\"o}ntsch}{}}
\label{sec:SM_nestedSCsubtraction}

\subsection{Introduction}

We consider NNLO subtraction schemes based on
the framework of residue-improved sector decomposition,
first developed in Refs.~\cite{Czakon:2010td,Czakon:2011ve}.
This framework allows the highly successful NLO subtraction procedure
of Frixione, Kunszt and Signer~\cite{Frixione:1995ms,Frixione:1997np}
to be extended to NNLO by using sector decomposition to separate overlapping
divergences.
At present, it is the only approach to NNLO subtractions which is
fully local in multi-particle phase space,
and subtraction schemes based on this framework,
such as  {\tt STRIPPER}~\cite{Czakon:2010td,Czakon:2011ve},
have been used in a variety of non-trivial NNLO calculations
~\cite{Czakon:2013goa,Czakon:2014xsa,Czakon:2015owf,Czakon:2016ckf,Brucherseifer:2014ama,Boughezal:2015dra,Caola:2015wna}.

There are, however, several aspects of the residue-improved sector decomposition
framework which can be improved upon.
First, the division of the phase space into sectors is useful to separate individual singular configurations,
but it somewhat obscures the simplicity of the final result, and 
one expects simplifications upon recombining the sectors.
Second, in the original method of Refs.~\cite{Czakon:2010td,Czakon:2011ve}
an additional (artificial) sector was introduced to separate overlapping soft-collinear
divergences which appear in Feynman diagrams but not in full amplitudes,
and this further complicates the construction.
As a result, the cancellation of the infrared poles is not
transparent in this implementation.
Initially this meant that the matrix elements had to be computed in $d$-dimensions,
which adds to the computational difficulty,
although a later development~\cite{Czakon:2014oma} allowed the computation of the matrix elements  in four dimensions.

These issues can be addressed simultaneously
by focusing on full matrix elements,
and using the color coherence property of QCD amplitudes,
which implies the decoupling of soft and collinear radiation.
The soft and collinear limits may then be subtracted independently of one another,
removing the need for the sector with correlated soft-collinear limits.
Once this sector is discarded,
there are clear simplifications on recombining sectors, which allow  the
integration over the unresolved phase space to be performed analytically for most of the singular limits,
with one exception that requires numerical integration.
This leads to an explicit (although partially numerical) cancellation of the IR poles for different kinematic structures,
a compact expression for the finite integrated subtraction terms, and a straightforward method to remove singular regions,
and  allows all matrix elements to be evaluated in four dimensions.
This approach to residue-improved sector decomposition,
called the \textit{nested soft-collinear subtraction scheme}~\cite{Caola:2017dug}, 
will be explained briefly in the following subsection.
We refer the interested reader to Ref.~\cite{Caola:2017dug} for a more detailed discussion.

\subsection{Nested soft-collinear subtractions at NNLO}
In the discussion of the subtraction scheme, we will focus on the
hadroproduction of a color singlet final state $V$. This process allows us
to confront most of the complexities of infrared singularities at a hadron collider,
while also avoiding unnecessary complications.
Our treatment is independent of the matrix element, so that the
procedure may readily be adapted for any color singlet production process
(e.g. Drell-Yan, diboson, associated $VH$ production, etc.).
We will focus on double-real emissions $q\bar{q} \to V+g g$,
as this partonic configuration contains the most intricate singularity structures.
We have performed similar calculations for all other partonic
channels relevant for Drell-Yan production, and these will be presented in a future publication.
We will, however, make some brief comments on these calculations at the end of this subsection, and
show some preliminary results.

The differential cross section for the process $q_1 \bar{q}_2 \to V +g_4 g_5$ is
\begin{equation}
  2 s \cdot {\rm d} \sigma^{\rm RR} =\frac{1}{2!} \int [{\rm d} g_4] [{\rm d} g_5]  F_{\rm LM}(1,2,4,5) ,
  \label{eq:SM_nestedSCsubtraction:rr1}
\end{equation}
where  $s$ is the partonic center-of-mass energy, the phase space integration measure of the emitted gluon $i$ is
\begin{equation}
  [{\rm d}g_i] = \frac{{\rm d}^{d-1} p_i}{(2\pi)^{d-1} 2 E_i} \theta(E_{\rm max}-E_i),
  \label{eq:SM_nestedSCsubtraction:dg}
\end{equation}
with $E_{\rm max}$ an arbitrary energy parameter defined in the partonic center-of-mass frame, and  
\begin{equation}
F_{\rm LM} (1,2,4,5) = {\rm d}{\rm Lips}_V\; |{\cal M}(1,2,4,5,V)|^2\; {\cal F}_{\rm kin}(1,2,4,5,V).
\end{equation}
In the above, ${\rm d}{\rm Lips}_V$ is the Lorentz-invariant phase space for the colorless particles,
including the momentum-conserving delta-function;
${\cal M}(1,2,4,5,V)$ is the matrix element for the process $q_1\bar{q}_2 \to V + g_4 g_5$,
and ${\cal F}_{\rm kin}$ defines an infrared-safe observable.
This process has singularities which appear when either $g_4$ or $g_5$ becomes soft,
or when either becomes collinear to either initial state parton,
or when $g_4$ and $g_5$ become collinear to each other.
A combination of these configurations may also occur, so one needs to consider different approaches to each singular limit.
Thus, the key to extracting the poles is to separate these singular regions.

As mentioned above, the soft and collinear regions can be treated independently, as a consequence of color coherence
(provided that one considers physical, i.e. gauge invariant and onshell, QCD amplitudes).
We will therefore first regularize the soft singularities, followed by the collinear singularities.
To do this, it is convenient to introduce the energy ordering $E_4 > E_5$. This removes the $1/2!$ factor in Eq.~\eqref{eq:SM_nestedSCsubtraction:rr1},
which can then be rewritten as
\begin{equation}
  2 s \cdot {\rm d} \sigma^{\rm RR} =\int [{\rm d} g_4] [{\rm d} g_5]  F_{\rm LM}(1,2,4,5) \theta(E_4>E_5) \equiv \left\langle F_{\rm LM}(1,2,4,5) \right\rangle,
  \label{eq:SM_nestedSCsubtraction:rr2}
\end{equation}
where we have introduced the averaging sign $\langle \ldots \rangle$ to indicate integration over the final state phase space.
We also introduce the soft and double-soft operators
\begin{equation}
S_i A = \lim_{E_i \to 0} A,\;\;\;\; {S{\hspace{-5.0pt}}S} A = \lim_{E_4,E_5\to 0} A, {\rm~ at~ fixed~ }E_5/E_4.
\end{equation}
We only need to consider the limits corresponding to ${S{\hspace{-5.0pt}}S}$ and $S_5$; the limit $S_4$ does not occur due to the energy ordering.
We can then write
\begin{equation}
  \begin{split}
  \big\langle F_{\rm LM}(1,2,4,5) \big\rangle 
& = 
\big\langle {S{\hspace{-5.0pt}}S}    F_{\rm LM}(1,2,4,5) \big\rangle  +  \big\langle S_5 ( I -  {S{\hspace{-5.0pt}}S}) F_{\rm LM}(1,2,4,5) \big\rangle 
\\
&+
\big\langle (I - S_5) ( I -  {S{\hspace{-5.0pt}}S}) F_{\rm LM}(1,2,4,5) \big\rangle.
\label{eq:SM_nestedSCsubtraction:softreg}
  \end{split}
\end{equation}
The first term on the left-hand side in Eq.~\eqref{eq:SM_nestedSCsubtraction:softreg} corresponds to the double-soft limit, in which both gluons decouple completely.
The second term captures the limit where $g_5$ is soft but singularities from $S_4$ are removed.
In both terms, we can integrate over the phase space of the unresolved gluons to obtain explicit IR poles in $1/\epsilon$.
In doing so, we note that the soft limits acting on $F_{\rm LM}$ remove the corresponding momentum from the momentum-conserving delta-function inside $F_{\rm LM}$.
However, the parameter $E_{\rm max}$ in the gluonic phase space, cf.~Eq.~\eqref{eq:SM_nestedSCsubtraction:dg}, prevents the energy integral from becoming unbounded from above.
The final term of Eq.~\eqref{eq:SM_nestedSCsubtraction:softreg} has all soft singularities removed.
All three terms, however, contain (potentially overlapping) collinear singularities, which must be disentangled and then subtracted.

We now introduce collinear and double-collinear operators
\begin{equation}
  C_{ij} A = \lim_{\rho_{ij} \to 0} A, \; \; \; \; {C{\hspace{-6.0pt}C}}_i A = \lim_{\rho_{4i},\rho_{5i}\to 0} A, {\rm~with~non~vanishing~} \rho_{4i}/\rho_{5i},\rho_{45}/\rho_{4i},\rho_{45}/\rho_{5i},
\end{equation}
where  $\rho_{ij} = 1 - n_{i} \cdot n_{j} $ and  $n_{i}$ is a unit vector that describes the direction of the momentum of the $i$-th particle in $(d-1)$-dimensional space.
The task of separating the collinear singularities then proceeds in two stages.
First, we partition the phase space by writing
\begin{equation}
  w^{14,15} + w^{24,25} + w^{14,25} + w^{15,24}=1,
\end{equation}
where the phase space partitions $w$ are functions of the angles between the partons.
They have the property that they vanish in various collinear limits,
so that only certain limits need to be considered in each partition of the phase space.
The triple-collinear partitions $w^{i4,i5}$ (for $i=1,2$) only have singularities corresponding to operators $C_{4i}$, $C_{5i}$ and $C_{45}$, while the double-collinear partitions $w^{14,25}$ and $w^{15,24}$ only have singularities corresponding to, respectively, $C_{41}$ and $C_{52}$, and  $C_{51}$ and $C_{42}$.

The two double-collinear partitions are now free of overlapping singularities, but the triple-collinear partitions still have overlapping singularities that can occur,
for example, in partition $w^{14,15}$ when $\vec{p}_1||\vec{p}_4||\vec{p}_5$. To disentangle these, we make use of a sector decomposition based on angular ordering.
For partition  $w^{i4,i5}$ (where $i=1,2$), we write
\begin{equation}
\begin{split}
1 =& 
\theta\left(\eta_{5i} < \frac{\eta_{4i}}{2}\right) + 
\theta\left(\frac{\eta_{4i}}{2} < \eta_{5i} < \eta_{4i}\right) 
 + 
\theta\left(\eta_{4i} < \frac{\eta_{5i}}{2}\right) + 
\theta\left(\frac{\eta_{5i}}{2} < \eta_{4i} < \eta_{5i}\right) \\
 \equiv & \theta^{(a)} + \theta^{(b)} + \theta^{(c)} + \theta^{(d)}.
\end{split}
\end{equation}
A parametrization of the angular phase space which allows both this decomposition
and the factorization of singularities in hard amplitudes is given in Refs.~\cite{Czakon:2010td,Czakon:2011ve}.
It is clear that the decomposition results in each sector containing only one collinear singularity:
in partition $w^{14,15}$, for example, sectors $a$ and $c$ have limits $C_{51}$ and $C_{41}$, respectively,
while sectors $b$ and $d$ only have the $C_{45}$ limit.
Note that sectors $b$ and $d$ are not the same, as the energies of $g_4$ and $g_5$ are ordered.

Thus we have divided the phase space in such a way as to completely separate all the overlapping singularities.
Note that since soft and collinear radiation is treated separately,
the use of sectors is only required to separate overlapping \textit{collinear} singularities,
and there is no need for a sector in which
the energies and emission angles of the radiated gluons vanish in a correlated manner.
This should allow for additional flexibility is constructing the phase space for radiation,
which we intend to explore in future work.

Using the sector decomposition described above, we can write the soft-regulated term as
\begin{equation}
\big\langle (I - S_5) ( I -   {S{\hspace{-5.0pt}}S} ) F_{\rm LM}(1,2,4,5) \big\rangle =  \big\langle F_{\rm LM}^{s_rc_s}(1,2,4,5) \big\rangle + \big\langle F_{\rm LM}^{s_rc_t}(1,2,4,5) \big\rangle  
+ \big\langle F_{\rm LM}^{s_rc_r}(1,2,4,5) \big\rangle,
\label{eq:SM_nestedSCsubtraction:collreg}
\end{equation}
where the soft-regulated, single-collinear term $\langle F_{\rm LM}^{s_rc_s} \rangle$ reads
\begin{equation}
\begin{split}
& \langle F_{\rm LM}^{s_rc_s} \rangle =
\sum_{(ij)\in dc}
\left\langle
\big[I- {S{\hspace{-5.0pt}}S} \big]\big[I-S_5\big]
\bigg[ C_{4i} [{\rm d}g_4]  + C_{5j} [{\rm d}g_5]  \bigg] w^{i4,j5} F_{\rm LM}(1,2,4,5)
\right\rangle\\
&\quad\quad
+\sum_{i\in tc} 
\bigg\langle
\big[I-{S{\hspace{-5.0pt}}S}  \big]\big[I-S_5\big]
\bigg[
\theta^{(a)} C_{5i} + \theta^{(b)} C_{45} + \theta^{(c)} C_{4i} + \theta^{(d)} C_{45}
\bigg]\\
&\quad\quad\quad\quad\quad
\times[{\rm d}g_4] [{\rm d}g_5] w^{i4,i5} F_{\rm LM}(1,2,4,5)
\bigg\rangle,
\label{eq:SM_nestedSCsubtraction:srcs}
\end{split} 
\end{equation}
the soft-regulated, triple-collinear terms $ \langle F_{\rm LM}^{s_rc_t} \rangle $ reads
\begin{equation}
\begin{split}
& \langle F_{\rm LM}^{s_rc_t} \rangle =
-\sum_{(ij)\in dc}\bigg\langle
\big[I-{S{\hspace{-5.0pt}}S}  \big]\big[I-S_5\big]
C_{4i}C_{5j} [{\rm d}g_4]  [{\rm d}g_5]  w^{i4,j5} F_{\rm LM}(1,2,4,5)
\bigg\rangle\\
&\quad\quad
+\sum_{i\in tc} 
\bigg\langle
\big[I- {S{\hspace{-5.0pt}}S} \big]\big[I-S_5\big]
\bigg[
\theta^{(a)}  {C{\hspace{-6.0pt}C}}_i\big[I-C_{5i}\big] + \theta^{(b)} {C{\hspace{-6.0pt}C}}_i\big[I-C_{45}\big]  \\
&\quad\quad\quad\quad~~
 + \theta^{(c)}  {C{\hspace{-6.0pt}C}}_i\big[I-C_{4i}\big]+ \theta^{(d)} {C{\hspace{-6.0pt}C}}_i\big[I-C_{45}\big]
\bigg][{\rm d}g_4]  [{\rm d}g_5]  w^{i4,i5} F_{\rm LM}(1,2,4,5)
\bigg\rangle,
\label{eq:SM_nestedSCsubtraction:srct}
\end{split} 
\end{equation}
and the fully regulated term $\langle F_{\rm LM}^{s_rc_r} \rangle $ reads
\begin{equation}
\begin{split}
& \langle F_{\rm LM}^{s_rc_r} \rangle  = 
\sum_{(ij)\in dc}\bigg\langle
\big[I- {S{\hspace{-5.0pt}}S} \big]\big[I-S_5\big]
\bigg[(I- C_{5j})(I-C_{4i})\bigg]\\
&\quad\quad\quad\quad\quad\quad
\times [{\rm d}g_4] [{\rm d}g_5]w^{i4,j5} F_{\rm LM} (1,2,4,5)
\bigg\rangle\\
&\quad\quad
+\sum_{i\in tc} 
\bigg\langle
\big[I-{S{\hspace{-5.0pt}}S} \big]\big[I-S_5\big]
\bigg[
\theta^{(a)} \big[I- {C{\hspace{-6.0pt}C}}_i\big]\big[I-C_{5i}\big] + 
\theta^{(b)} \big[I- {C{\hspace{-5.pt}C}}_i\big]\big[I-C_{45}\big]  \\
&\quad\quad\quad\quad~~
 + \theta^{(c)} \big[I-{C{\hspace{-6.0pt}C}}_i\big]\big[I-C_{4i}\big]+ 
\theta^{(d)} \big[I-{C{\hspace{-6.0pt}C}}_i\big]\big[I-C_{45}\big]
\bigg]\\
&\quad\quad\quad\quad~~
\times [{\rm d}g_4] [{\rm d}g_5] w^{i4,i5} F_{\rm LM}(1,2,4,5)
\bigg\rangle .
\end{split}
\label{eq:SM_nestedSCsubtraction:srcr}
\end{equation}
This last term is manifestly finite, as all singularities are removed through the nested subtractions.
As a result, it can be evaluated in four space-time dimensions and integrated numerically.
It is the only term which contains the fully resolved double-real matrix element.
The two terms describing the collinear limits, Eqs.~\eqref{eq:SM_nestedSCsubtraction:srcs} and \eqref{eq:SM_nestedSCsubtraction:srct},
can be treated along the same lines as the soft and double-soft limits,
i.e. the first two terms of Eq.~\eqref{eq:SM_nestedSCsubtraction:softreg}.
In each of these terms, the radiated gluons either decouple completely (in the case of a soft limit),
or partially (in the case of a collinear limit).
If a gluon decouples completely, we integrate over its angles and energies.
If a gluon decouples partially, we integrate over its angles and rewrite its energy integral as a convolution with
a splitting function (which is related but not identical to one of the standard Altarelli-Parisi splitting functions).
As mentioned before,  this integration over the unresolved gluons can be performed analytically in most cases.
The exception is the triple-collinear limit in the triple-collinear partitions.
For this configuration, the integration could be performed analytically with some effort, 
but is straightforward to do numerically, which we did.
\footnote{The double-soft limit was initially treated in the same way~\cite{Caola:2017dug}, but we have since computed this limit analytically
  for color singlet production.}
In some cases (e.g. when we consider the $S_5$ limit), the resulting expression has NLO-like kinematics, including leftover singular regions;
in these cases, we repeat the subtraction procedure as at NLO.
The result is expressions which have explicit IR poles in $1/\epsilon$ multiplying manifestly finite $F_{\rm LM}$ structures of lower particle multiplicity convoluted with splitting functions.
These expressions completely describe the singular structure of the double-real emission.

Having removed  the soft-collinear correlated sector, it is now quite easy
to see similarities between intermediate expressions for collinear singularities that appear in different partitions or sectors.
By combining these expressions in a judicious manner, the results for the integrated subtraction terms
can be simplified substantially.
For example, restrictions on the phase space related to the parameter $E_{\rm max}$ in $F_{\rm LM}^{s_r c_s}$ can be entirely
removed by combining intermediate results from the two double-collinear partitions with those from sectors $a$ and $c$ of the
two triple-collinear partitions.
As a result of this, the cancellation of the IR poles between double-real, real-virtual, and double-virtual corrections
can be verified analytically to order $1/\epsilon^2$, and numerically at order $1/\epsilon$,
and compact expressions are obtained for the finite part of the integrated subtraction terms.

The description presented in this subsection has focused on the partonic channel $q\bar{q}\to V+g g$,
but we stress that similar calculations have been performed for all partonic channels contributing to Drell-Yan
production. In all cases, the cancellation of the IR poles is demonstrated, and compact formulae are
derived for the finite integrated subtraction terms. Indeed, the singularity structure of the
remaining partonic channels is simpler than that considered above. The full results will be presented in a forthcoming
publication; here we make some brief comments about each channel:
\begin{itemize}
\item In the  $qg \to V+q'g $ channel, there are no divergences corresponding to the final state quark becoming soft.
  As a result, there is no need for energy ordering,
  and hence we parametrize the phase space in a slightly different way in order to take advantage of this.
  Moreover, there are no divergences when the final state quark is collinear to the initial state quark,
  and as a result, the partitioning of the phase space is done differently too.
\item In the $gg \to V+q\bar{q}$ channel, there are no soft singularities.
  The collinear singular structure is also simple, with only single collinear limits and no overlapping singularities.
  As a result, phase space partitioning is not needed, and the subtraction procedure is trivial.
\item The four-quark channel $q\bar{q}' \to V+q_1\bar{q}_2$ (including all crossings) may be divided into a singlet and non-singlet contribution.
  The latter contributes only to $q\bar{q}$ and $\bar{q}q$ channels, while the former contributes to these as well as the $qq$ and $\bar{q}\bar{q}$ channels.
  If the outgoing quarks are identical, there is an interference between the singlet and non-singlet contributions which has a triple-collinear singularity.
  These features make the bookkeeping for this channel quite complicated, but the actual singularities are relatively simple -- for example, there are no single-soft divergences.
  \end{itemize}

\subsection{Numerical results for proof-of-concept calculation}
\begin{figure}[t]
\centering
\includegraphics[width=0.48\textwidth]{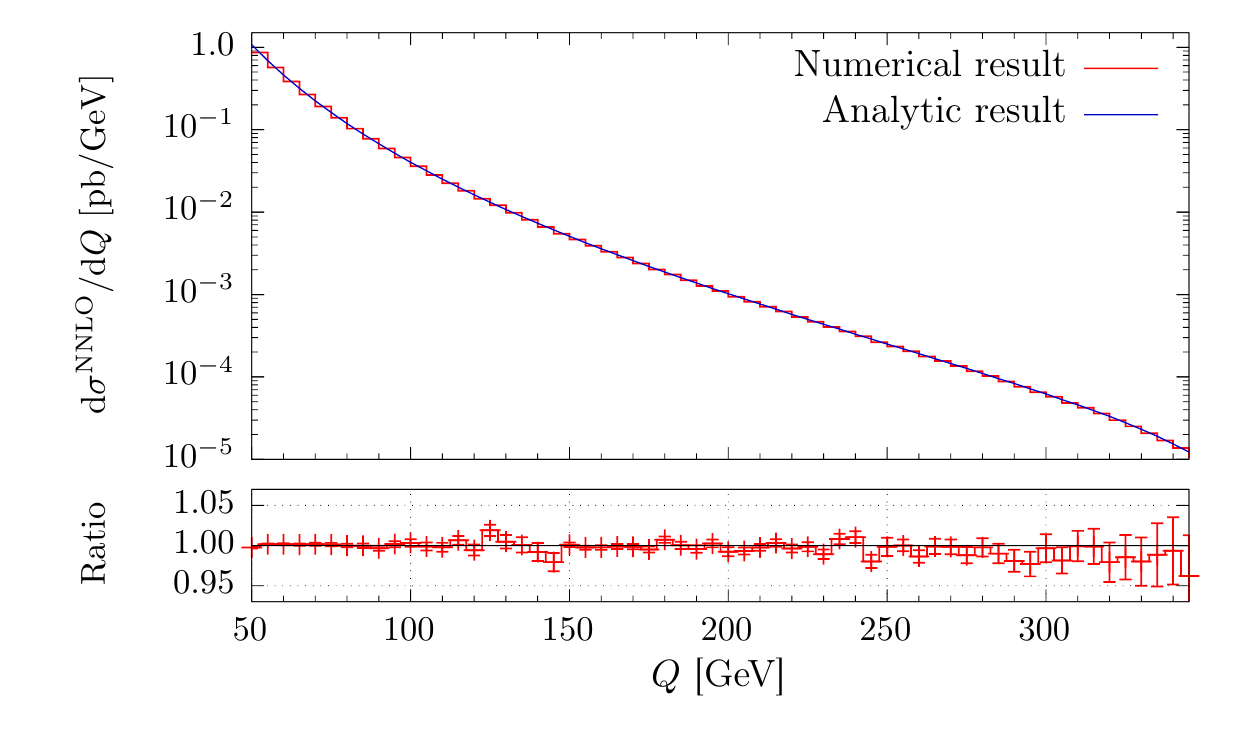}\hfill
\includegraphics[width=0.48\textwidth]{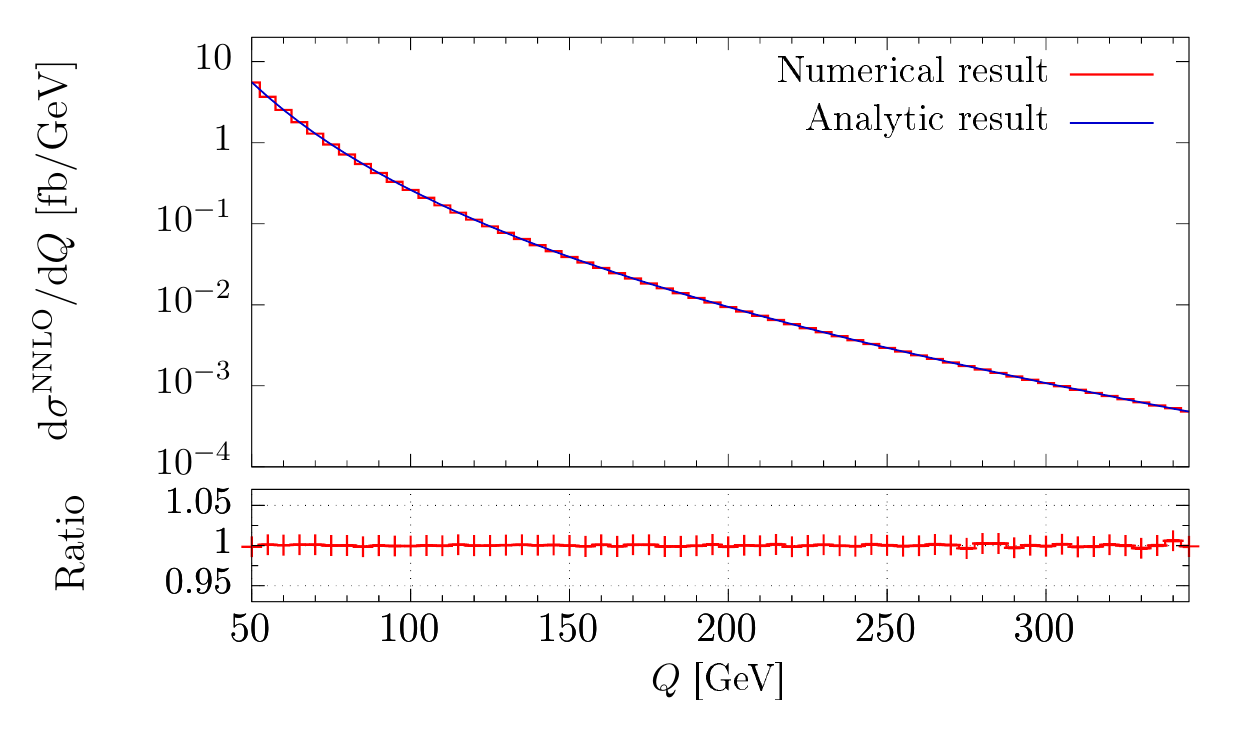}
\caption{Comparison of the NNLO QCD contribution ${\rm d}\sigma^{\rm NNLO}/{\rm d}Q$ 
computed using the nested soft-collinear subtraction scheme with 
the analytic results in Ref.~\cite{Hamberg:1990np}, for the $q\bar{q} \to V+gg$ partonic channel (left), and
for the $4q{\rm AC}$ contribution (right).
The ratios between the results are shown in the lower panes.}
\label{fig:SM_nestedSCsubtraction:vNcomp}
\end{figure}

In this section, we present some results for the process $pp \to \gamma^* \to e^- e^+ + X$ at the 14 TeV LHC,
comparing these with the analytic results of Ref.~\cite{Hamberg:1990np}.
We consider lepton pairs with an invariant mass $50~{\rm GeV} < Q < 350~{\rm GeV}$, and use NNPDF3.0 parton distribution functions~\cite{Ball:2014uwa}
with the renormalization and factorization scale $\mu =100$ GeV.
We show representative results for the $q\bar{q} \to V +gg$ partonic channel,
as well as a subset of the four-quark contributions corresponding to the interference of $s$-channel and $t$-channel amplitudes labeled `A' and `C' in Ref.~\cite{Hamberg:1990np},
which we call the $4qAC$ contribution.
The NNLO \textit{contributions} ${\rm d} \sigma^{\rm NNLO}$ for these channels are
\begin{align}
  {\rm d}\sigma^{\rm NNLO}_{q\bar{q}}&=14 471(4)~{\rm fb} &   {\rm d}\sigma^{\rm NNLO}_{4qAC}&= -96.867(17)~{\rm fb} \\
  {\rm d}\sigma^{\rm NNLO,analytic}_{q\bar{q}}&=14 470~{\rm fb} &   {\rm d}\sigma^{\rm NNLO,analytic}_{4qAC}&= -96.866~{\rm fb}.
\end{align}
The NNLO contributions from the two channels differ by two orders of magnitude;
nevertheless, the agreement between our results and the analytic results is better than one per mille.
This indicates that our subtraction scheme is able to give extremely precise predictions even for numerically tiny contributions.
This may appear unnecessary at first glance, but we also observe large cancellations between different partonic channels,
in which case this degree of precision becomes important.
The differential distributions in $Q$ for these two NNLO contributions are shown in Fig.~\ref{fig:SM_nestedSCsubtraction:vNcomp}.
In both cases, the agreement is at the level of a few per mille to a few percent across several orders of magnitude.
This level of agreement on the NNLO contribution indicates that this
subtraction scheme is well suited to the task of providing high precision results for
physical cross sections and kinematic distributions.

\subsection*{Acknowledgments}
I thank the organizers of the Les Houches Workshop series and the SM Working Group conveners in particular for a stimulating workshop. I am grateful for support from the German Federal Ministry for Education and Research (BMBF) under grant 05H15VKCCA.






\newcommand\bp{p\hspace{-.42em}/\hspace{-.07em}}
\newcommand\bq{q\hspace{-.42em}/\hspace{-.07em}}
\newcommand\bl{\ell\hspace{-.42em}/\hspace{-.07em}}

\section{Loop-tree duality and the four-dimensional unsubtraction~\protect\footnote{
      G.~Chachamis,
      F.~Driencourt-Mangin,
      G.~Rodrigo,
      G.~Sborlini}{}}
\label{sec:SM_looptreeduality}


In the context of four-dimensional methods for higher-order computations, we describe the main features of the four-dimensional unsubtraction (FDU) and the loop-tree duality (LTD) theorem. They grant a very powerful framework that allows to compute Feynman integrals and physical observables in four dimensions through a purely numerical implementation. Moreover, due to some remarkable mathematical properties of the intermediate expressions, the formalism is well suited for performing asymptotic expansions, circumventing many technical issues that may arise in the traditional approach. Here we will focus on the one-loop case, although we provide some insights of possible NNLO developments.

\subsection{Introduction}
The Loop-Tree Duality (LTD) formalism~\cite{Catani:2008xa,Bierenbaum:2010cy,Bierenbaum:2012th,Bierenbaum:2013nja,Buchta:2014dfa,Buchta:2014fva,Buchta:2015jea,Sborlini:2015uia,Buchta:2015wna,Buchta:2016wfg,Hernandez-Pinto:2015ysa,Sborlini:2016gbr,Sborlini:2016hat,Driencourt-Mangin:2017gop,Chachamis:2016olm,Chachamis:2017yzl,Ramirez-Uribe:2017gbf,Jurado:2017xut} turns $N$-leg loop integrals and amplitudes into a sum of connected tree-level-like diagrams with a remaining integration measure very similar to the  $(N+1)$--body phase-space~\cite{Catani:2008xa}. Loop and tree-level radiative corrections of the same order in the perturbative expansion then, may in principle be computed numerically under a common integral sign  with the use of a suitable integrator (usually a Monte Carlo routine)~\cite{Hernandez-Pinto:2015ysa,Sborlini:2016gbr}. The LTD approach fits into a wider effort in phenomenology aiming at fully automated next-to-leading order (NLO) computations. Many steps toward that direction have been taken in the recent years~\cite{Soper:1998ye,Soper:1999xk,Soper:2001hu,Kramer:2002cd,Ferroglia:2002mz,Nagy:2003qn,Nagy:2006xy,Moretti:2008jj,Gong:2008ww,Kilian:2009wy,Becker:2010ng,Binoth:2010nha,Becker:2012aqa,DeRoeck:2009id,AlcarazMaestre:2012vp,Becker:2012nk,Seth:2016hmv,Gnendiger:2017pys,Bevilacqua:2011xh,Cascioli:2011va,Cullen:2014yla,Frixione:2008ym,Gleisberg:2008ta,Alwall:2014hca}. Substantial progress has also been made at higher orders~\cite{Passarino:2001wv,Anastasiou:2007qb,Becker:2012bi}.

In dimensional regularization, a one-loop scalar diagram can be represented by 
\begin{equation}
L^{(1)}(p_1, p_2,\dots , p_N) =  -i\int \frac{d^d\ell}{(2 \pi)^d} \prod\limits_{i=1}^NG_F(q_i)~,
\label{eq:SM_looptreeduality:a}
\end{equation}
where $\ell = (\ell_0, {\boldsymbol{\ell}})$ is the loop momentum, $G_F(q_i) = 1/(q_i^2-m_i^2+i0)$ are Feynman propagators and $q_i$ are the momenta of the internal lines which depend on $\ell$. By applying the LTD, one essentially integrates over the energy component $\ell_0$ using Cauchy's residue theorem. The loop diagram then turns into a sum of integrals over the three-momentum $\boldsymbol{\ell}$, each of which is called a \emph{dual contribution}. They emerge from the original integral after cutting one of the internal lines:
\begin{equation}
L^{(1)}(p_1,p_2,\dots ,p_N)= \raisebox{-11ex}{\includegraphics[scale=1]{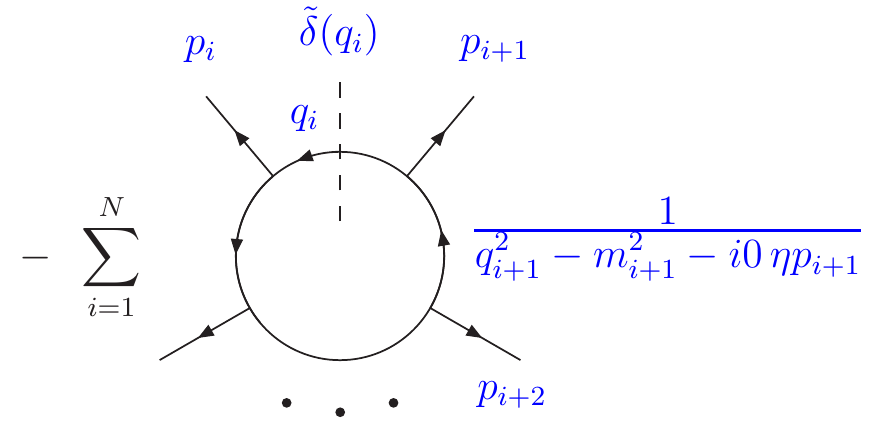}}\;,
\end{equation}
where $\tilde{\delta}(q_i)=2\pi i\delta_+(q_i^2-m_i^2)$. The $\tilde{\delta}(q_i)$ sets the internal lines on-shell by selecting the pole of the Feynman propagators with positive energy and negative imaginary part. Integrating the dual contributions over $\boldsymbol{\ell}$ requires most of the times a contour deformation due to the presence of the so-called {\it ellipsoid} and {\it hyperboloid} singularities~\cite{Buchta:2015wna}. These singularities are, in general, present at the integrand level.

On the other hand, the application of LTD implies modifying the propagators associated with the lines that are not being set on-shell. We define the dual propagators as
\begin{equation}
G_D(q_i;q_j) = \frac{1}{q_j^2 -m_j^2 - \imath 0 \, \eta \cdot k_{ji}} \, ,
\end{equation}
in such a way that the original scalar integrals reads
\begin{equation}
L^{(1)}(p_1,p_2,\dots ,p_N)= - \sum_{i=1}^N \, \int_\ell \, \tilde{\delta}(q_i) \, \prod_{j \neq i} G_D(q_i;q_j)\, ,
\end{equation}
with $i,j \in \{1,2,\ldots N\}$, $k_{ji} = q_j - q_i$ and $\eta$ an arbitrary future-like or light-like vector, $\eta^2 \ge 0$, with positive definite energy $\eta_0 > 0$. This modified prescription is very important, since it is responsible for the cancellation of non-physical singularities manifesting in the different dual contributions. Moreover, as explained in Ref. \cite{Catani:2008xa}, the dual prescription encodes the information contained in the different multiple cuts that appear when using the Feynman-Tree theorem (FTT)~\cite{Feynman:1963ax}, thus allowing to show the equivalence between both theorems. It is worth to notice that we can chose $\eta^\mu = (1,{\bf 0})$ to simplify the computations; this is equivalent to applying Cauchy's residue theorem in the energy component of the loop momentum. 

Assuming that there are only single powers of the Feynman propagators, the dual representation is straightforwardly valid for loop scattering amplitudes. Due to the fact that the single-cuts do not affect numerators, then the dual representation of scattering amplitudes is obtained by adding all possible \emph{dual} single-cuts of the original loop amplitudes, and replacing the uncut Feynman propagators by dual propagators. In the case that higher powers of the propagators are present, we have to use the extended version of the LTD theorem~\cite{Bierenbaum:2012th}, which consists in applying Cauchy's residue theorem with the well-known formula for poles of order $n$, i.e.
\begin{eqnarray}
{\rm Res}({\cal A},q_{i,0}^{(+)}) &=& \frac{1}{(n-1)!} \, \left. \frac{\partial^{n-1}}{\partial^{n-1} \, 
q_{i,0}}\left({\cal A}(q_{i,0})\, (q_{i,0}-q_{i,0}^{(+)})^{n}\right)\right|_{q_{i,0}=q_{i,0}^{(+)}},
\label{eq:SM_looptreeduality:RESMULTIPLE}
\end{eqnarray}
where $q_{i,0}^{(+)} = \sqrt{\textbf{q}_i^2+m_i^2-\imath 0}$ is the positive energy solution of the corresponding on-shell dispersion relation. In that case, the explicit form of the scattering amplitude is relevant because the numerator is affected by the derivative. Specifically, this observation leads to non-trivial consequences when looking into the local form of the renormalization factors \cite{Sborlini:2016hat}.

It is important to mention that the LTD theorem is valid beyond the one-loop level. For a generic $L$-loop integral, the theorem leads to a dual decomposition involving $L$ cuts, in such a way that the remaining structure is topologically equivalent to real radiation processes with $L$ additional final-state particles.
 
\subsection{Computation of Feynman integrals} 
The numerical implementation of the LTD had been initially used to compute a huge amount of integrals with up to six external legs and the results were compared against reference values obtained with {\tt LoopTools 2.10}~\cite{Hahn:1998yk} and {\tt SecDec 3.0}~\cite{Borowka:2015mxa}. Subsequently, results for one-loop diagrams with up to seven (heptagons) and eight (octagons) external legs were presented~\cite{Buchta:2015wna,Chachamis:2017yzl}. Here we present the result for a non-trivial decagon (one-loop, ten external legs) and we comment on how the numerical implementation of the method can be optimised. 

The LTD method is implemented in a C\texttt{++} code~\cite{Buchta:2015wna} and relies on the {\tt Cuba} library~\cite{Hahn:2004fe} for the numerical integration routines. At runtime, the code initially reads in and properly assigns internal masses and external momenta. Then it proceeds with an analysis of the ellipsoid and hyperboloid singularity structure to set up the details of the contour deformation, and finally performs the numerical integration using either the routine {\tt Cuhre} or {\tt Vegas} as are implemented in the {\tt Cuba} library. We let our one-loop decagon take both a scalar and a tensor (rank-two) numerator. The external momenta configuration (the same for both the scalar and tensor case) is shown below:\\
\begin{equation}
\begin{aligned}
&\hspace{-1.5cm}p_1 = (-2.50000,  \hspace{1.cm}0,   \hspace{2.cm}        0,   \hspace{1.7cm}       -2.50000)     \\[-2pt] 
&\hspace{-1.5cm}p_2 = (-2.50000,  \hspace{1.cm}0,   \hspace{2.cm}        0,   \hspace{2.cm}         2.50000)            \\[-2pt] 
&\hspace{-1.5cm}p_3 = (-0.95848,  \hspace{.7cm}-0.38291, \hspace{.95cm}  0.86652,  \hspace{.65cm}  -0.14559) \\[-2pt] 
&\hspace{-1.5cm}p_4 = (-0.26804,  \hspace{1.cm} 0.18418,  \hspace{.65cm}-0.17115,  \hspace{.65cm} -0.09288) \\[-2pt] 
&\hspace{-1.5cm}p_5 = (-0.90712,  \hspace{.7cm} -0.17547, \hspace{.65cm} -0.15156, \hspace{.95cm} 0.87699)  \\[-2pt] 
&\hspace{-1.5cm}p_6 = (-0.79290, \hspace{1.cm} 0.75301, \hspace{.95cm} 0.21387,   \hspace{.95cm} 0.12617)  \\[-2pt] 
&\hspace{-1.5cm}p_7 = (-0.09296,  \hspace{.7cm} -0.02540,  \hspace{.65cm} -0.04121,  \hspace{.95cm} 0.07935) \\[-2pt]
&\hspace{-1.5cm}p_8 = (-0.72985,  \hspace{.7cm} -0.64952,  \hspace{.65cm} -0.24701,  \hspace{.95cm} 0.22314) \\[-2pt] 
&\hspace{-1.5cm}p_9 = (-0.58078,  \hspace{1.cm} 0.08323,  \hspace{.65cm} -0.36994,  \hspace{.65cm} -0.43990) \\[-2pt]
&\hspace{-1.5cm}p_{10} = - p_1 - p_2 - p_3 - p_4 - p_5 - p_6 - p_7 - p_8 - p_9\,.
\end{aligned}\,
\end{equation}
We work with internal propagators that have all different masses:\\
\begin{equation}
\begin{aligned}
&\hspace{-1.2cm} m_1 = 5.2020\,, m_2 = 4.2031\,, m_3 = 3.2042\,, m_4 = 7.2053\,, m_5 = 3.2064 \\
&\hspace{-1.2cm} m_6 = 1.2075\,, m_7 = 6.2086\,, m_8 = 8.2097\,, m_9 = 3.2008\,, m_{10} = 3.2019 .
\end{aligned}\,
\end{equation}
In Table~\ref{tab:SM_looptreeduality:LTDdecagon}, we summarise the results for the scalar and tensor decagon. The running time in order to obtain the results on a typical Desktop machine (Intel i7 @ 3.4 GHz processor, 4-cores 8-threads), is around 30 seconds.

\begin{table}[htb]
\begin{center}
\begin{tabular}{|c|c|c|} \hline
Propagator   & Real Part  &  Imaginary Part \\
\hline
       1       & $~~~~~~2.530(4)\times 10^{-14}$ & $~~+ i~8.514(1)\times 10^{-14}$ \\
\hline
      $\ell . p_3 \times \ell . p_5$      & $~~~~~8.08(4)\times 10^{-15}$ & $~~+ i~6.144(5)\times 10^{-13}$ \\
\hline
\end{tabular}
\caption{Scalar and tensor decagon with all internal masses different.
\label{tab:SM_looptreeduality:LTDdecagon}}
\end{center}
\end{table}
\vspace{-.3cm}

Recently, a new implementation code of the LTD was developed in {\tt MATHEMATICA} which takes advantage of the system's internal numerical integration routines. The implementation in {\tt MATHEMATICA} seems to present favorable features in various cases over the C\texttt{++} code. However, a thorough comparison is needed to select the strong points of both codes into a unified implementation. This will not only make the computation of one-loop diagrams much faster but it will also allow for rapid progress in attacking non-trivial two-loop diagrams.  

\subsection{Application to cross-section computations}
Besides the possibility of computing Feynman integrals in an alternative way, LTD provides a robust framework to tackle cross-section computations in four space-time dimensions. As we mentioned in the Introduction, the dual contributions closely resembles PS integrals. This is a crucial point because the calculation of IR-safe observables involves adding both virtual and real corrections, with the last ones defined in a PS containing additional real particles. In consequence, if the on-shell internal lines could be interpreted as part of the real radiation, then the structure of the dual contributions might be directly related with the real-emission one. 

The four-dimensional unsubtraction (FDU) framework is a fully local regularization \linebreak method that can be used to compute physical observables directly in four dimensions. Since physical IR-safe quantities are associated with finite results, all the singularities appearing in intermediate steps of the calculation must be canceled in the final expression. So, the core idea of FDU is to achieve this cancellation directly at integrand level, and obtain a smooth integrable function by mapping the singular regions to the same points. This approach can be summarized in three steps:
\begin{enumerate}
\item Compute the dual contributions through the application of LTD to the virtual amplitudes. Locate the regions of the dual integration domain which are responsible of originating the physical IR in the loop amplitudes and demonstrate that these regions are compact.
\item Relate the real-radiation kinematics with the dual ones, forcing the internal lines that are on-shell to play the role of the real emission. Through a proper momentum mapping, describe the real and the dual contributions using the same integration variables: the combined expressions are smooth in any possible IR limit.
\item Rewrite the renormalization factors and the UV counter-terms in a local form, adjusting the sub-leading terms to reproduce the selected renormalization scheme. Adding these terms to the results obtained in the second step leads to an integrable behavior in the high-energy region.
\end{enumerate} 
In the following, we will center the discussion in the implementation of NLO corrections. In particular, we emphasize the importance of the momentum mapping and the local form of renormalization factors (i.e. their proper integrand-level expressions). The computation of the local UV subtraction counter-terms is obtained through an expansion around the UV propagator, as explained in Refs. \cite{Sborlini:2016gbr,Sborlini:2016hat}. Finally, we conclude the section with a recompilation of benchmark results and a brief discussion about the NNLO extension.
 
\subsubsection{Real-virtual momentum mapping}
Let's consider a pedagogical example: a generic decay process involving $m$ particles in the final state. From the point of view of perturbation theory, the Born-level contribution is defined in a $m$-particle PS. On the other hand, the NLO corrections are obtained from the sum of the one-loop amplitudes integrated in a $m$-particle PS and the real-emission tree-level amplitudes in a \mbox{$(m\!+\!1$)}-particle PS. We denote the momenta associated to the Born kinematics as $\{p_i^{\mu}\}_{i=1,\ldots,m}$, whilst we use $\{p_i'^{\mu}\}_{i=1,\ldots,m+1}$ for those momenta involved in the real radiation. Since the dual decomposition of the one-loop amplitudes includes an additional integration variable (namely the internal line that is set on-shell), both the real and the dual-virtual contributions can be written using \mbox{$(m\!+\!1$)} physical momenta. In order to relate both sets of momenta in such a way that the singularities are mapped to the same integration points, we follow the strategy applied in the dipole subtraction formalism~\cite{Catani:1996jh,Catani:1996vz}. 

In the first place, we introduce a proper partition of the real-emission PS to isolate the IR singularities. So, we define
\begin{equation}
{\cal R}_i = \{{y'}_{i r} < {\rm min}\, {y'}_{jk} \} \, , \quad \quad
\sum_{i=1}^{m} {\cal R}_i =1 \, ,
\label{eq:SM_looptreeduality:regionpartition} 
\end{equation}
where ${y'}_{ij}=2\, p_i' \cdot p_j'/Q^2$, $r$ is the radiated parton from parton $i$, and $Q$ is the typical hard scale of the scattering process. According to this definition, the only
allowed collinear/soft configurations inside ${\cal R}_i$ are $i\parallel r$ or $p_r'^{\mu}\to 0$. So, different collinear singularities appear in non-overlapping regions of the real-emission PS.

When studying the virtual contribution through the LTD, we find $m$ dual contributions, each one associated with a single cut of an internal line. It is worth appreciating that the different dual amplitudes include different IR singularities, which should be matched to those present in the real part. However, in the PS splitting defined in Eq.~\eqref{eq:SM_looptreeduality:regionpartition}, the IR singularities are isolated in disjoint partitions; thus we need to identify a connection among \emph{cuts} and \emph{regions}. And, to infer the solution, we rely on the diagrammatic identification in the soft/collinear limit.
\begin{figure}[t]
\begin{center}
\includegraphics[width=0.6\textwidth]{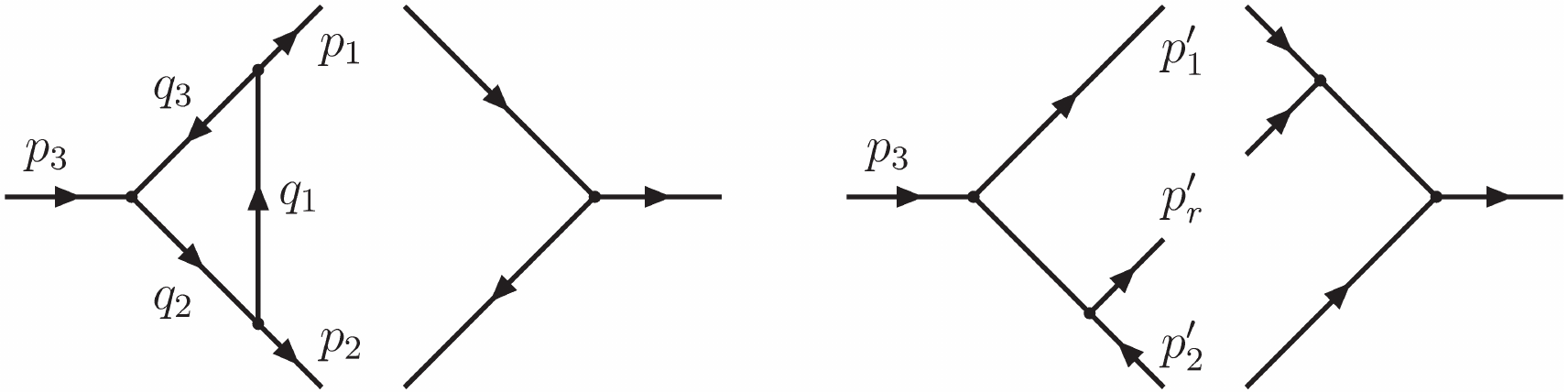}
\vspace*{3ex}

\caption{\label{fig:SM_looptreeduality:factorizacion}
Topological correspondence among one-loop (left) and real-emission amplitudes (right). In this illustrative example, we consider a $1 \to 2$ decay process at NLO. When $q_1$ is set on-shell, the virtual diagram factorizes in the limit $q_1 \parallel p_1$ exactly in the same way that the real contribution does in the limit $p^{'}_r \parallel p^{'}_1$.}
\end{center}
\end{figure}
As an explicit example, let's consider the $1 \to 2$ decay process shown in Fig.~\ref{fig:SM_looptreeduality:factorizacion}. First, we interpret the on-shell internal state in the dual amplitude as the extra-radiated particle in the real contribution, namely $q_i \leftrightarrow p_r'$. Then, we settle in one of the partitions, for instance ${\cal R}_i$. Because the only collinear singularity allowed is originated by $i \parallel r$, we distinguish particle $i$ and call it
the {\it emitter}. After that, we single out all the squared amplitude-level diagrams in the real contribution that become singular when $i \parallel r$ and cut the line $i$. Following Fig.~\ref{fig:SM_looptreeduality:factorizacion}, these have to be topologically compared with the dual-Born interference diagrams whose internal momenta $q_i$ are on-shell. Due to the factorization properties \cite{Buchta:2014dfa,Catani:2011st} and the topological analysis performed, we can guarantee that the dual contribution $i$ and the real-emission in the region ${\cal R}_i$ give rise to the same IR singular structure.

Finally, we propose an explicit connection among the dual and the real-emission momenta. Let us take the
$(m\!+\!1)$-particle real-emission kinematics, with $i$ as the emitter and $r$ as the radiated particle, and we introduce a reference momentum, associated to the spectator $j$. Then, the momentum mapping with $q_i=\ell+p_1+\ldots+p_i$ on-shell is given by 
\begin{eqnarray}
&& p_i'^\mu
= p_i^\mu - q_i^\mu + \alpha_i \, p_j^\mu\,, \quad
\quad p_j'^\mu = (1-\alpha_i) \, p_j^\mu\,, \quad \quad
p_k'^\mu = p_k^\mu \ \ k \ne i,j \ \, ,
\nonumber \\*
&& p_r'^\mu
= q_i^\mu~, \qquad \qquad \quad \quad \quad \ \,
\alpha_i = \frac{(q_i-p_i)^2}{2 p_j\cdot(q_i-p_i)}\,,
\label{eq:SM_looptreeduality:mappingmassless}
\end{eqnarray}
where all the partons are considered massless, i.e. $p_i^2=0$ and $p_i'^2=p_j'^2=p_r'^2=0$. This construction fulfills momenta conservation, since the original Born-level kinematics also fulfills this physical constraint. Of course, this mapping can be generalized to multi-leg processes involving arbitrary masses \cite{Sborlini:2016hat}.

\subsubsection{Local expressions for renormalization counter-terms}
Let's start with the well-known expression for the wave-function renormalization constant in the Feynman gauge with on-shell renormalization conditions, i.e.
\begin{equation}
\Delta Z_2 = \frac{g_S^2}{16\pi^2} \, C_F \left(-\frac{1}{\epsilon_{{\rm UV}}}-\frac{2}{\epsilon_{{\rm IR}}}+ 3\,  \ln{\frac{M^2}{\mu^2}}-4\right)~,
\label{eq:SM_looptreeduality:DeltaZ2expressionINTEGRADA}
\end{equation}
where we keep track of the IR and UV origin of the $\epsilon$-poles within DREG. This distinction is relevant because $\Delta Z_2$ includes contributions originated in the calculation of self-energies, which partially cancels the IR singularities present in the squared terms of the real-emission contributions. Thus, we need to provide a proper \emph{unintegrated} expression to locally cancel these singularities, i.e. 
\begin{eqnarray}
\Delta Z_2(p_1) &=& -g_S^2 \, C_F \, \int_{\ell} G_F(q_1) \, G_F(q_3) \, \left((d-2)\frac{q_1 \cdot p_2}{p_1 \cdot p_2} \right. \nonumber
\\ &+& \left.4 M^2 \left(1- \frac{q_1 \cdot p_2}{p_1 \cdot p_2}\right)G_F(q_3)\right)~,
\label{eq:SM_looptreeduality:DeltaZ2expression}
\end{eqnarray}
which includes higher-order powers of the propagators \cite{Sborlini:2016hat} and reproduces the result shown in Eq.~\eqref{eq:SM_looptreeduality:DeltaZ2expressionINTEGRADA} after integration\footnote{More details about this calculation can be found in Refs. \cite{Sborlini:2016gbr,Sborlini:2016hat}.}. As we mentioned, the IR pole in $\Delta Z_2$ will cancel when combined with the real contributions, but there are still UV divergences that need to be removed. To achieve the full regularization, we perform an expansion around the UV propagator, i.e. \begin{equation}
G_F(q_{{\rm UV}}) = \frac{1}{q_{{\rm UV}}^2-\mu_{{\rm UV}}^2+\imath 0} \, ,
\end{equation}
with $\mu_{\rm UV}$ the renormalization scale. Thus, the UV counter-term for the wave-function renormalization constant is given by 
\begin{eqnarray}
\Delta Z_2^{\rm UV}(p_1) &=& -(d-2)\, g_S^2 \, C_F \, \, \int_{\ell} (G_F(q_{\rm UV}))^2 \, \left(1+\frac{q_{\rm UV} \cdot p_2}{p_1 \cdot p_2}\right) \nonumber
\\ &\times& \left(1- G_F(q_{\rm UV})(2\, q_{\rm UV} \cdot p_1 + \mu^2_{\rm UV})\right) \, ,\\ 
&\equiv& - \widetilde{S}_{\epsilon} \, \frac{g_S^2}{16 \, \pi^2} \, C_F \,  \left( \frac{\mu_{\rm UV}^2}{\mu^2}\right)^{-\epsilon} \, \frac{1-\epsilon^2}{\epsilon} ~,
\label{eq:SM_looptreeduality:ParteUV}
\end{eqnarray}
whose integrated form exactly reproduces the UV pole present in Eq.~\eqref{eq:SM_looptreeduality:DeltaZ2expressionINTEGRADA}. The finite remainders depend on the sub-leading terms proportional to $\mu_{\rm UV}^2$, which can be adjusted to work in an specific renormalization scheme.

On the other hand, the vertex renormalization factors need also to be expressed in an \emph{unintegrated} form. In the Feynman gauge, the generic vertex UV counter-term reads 
\begin{equation}
{\Gamma}^{(1)}_{A,{\rm UV}} = g_S^2 \, C_F\, 
\int_\ell \left( G_F(q_{\rm UV})\right)^3 \, 
\left[\gamma^{\nu} \, \bq_{{\rm UV}} \, {\Gamma}^{(0)}_A \, \bq_{\rm UV} \, \gamma_{\nu} - 
d_{A, {\rm UV}} \, \mu_{\rm UV}^2 \, {\Gamma}^{(0)}_{A}\right]~,
\label{eq:SM_looptreeduality:UVGenericvertex}
\end{equation}
where ${\Gamma}^{(0)}_{A}$ represents the tree-level vertex. In the numerator, the term proportional to $\mu_{\rm UV}^2$ is sub-leading in the UV-limit. In this way, the coefficient $d_{A,{\rm UV}}$ can be adjusted in order to implement the desired renormalization scheme. For instance, we can tweak $d_{A,{\rm UV}}$ in order to reproduce the $\overline{\rm MS}$ scheme, in which the counter-term only cancels the $\epsilon_{\rm UV}$ pole in DREG. 
 
It is important to emphasize that this construction of local UV-subtraction counter-terms is completely general and that the sub-leading terms can be always adjusted to reproduce the desired scheme-dependent contributions. Moreover, the tweaking of sub-leading terms in the UV-limit is \emph{universal}, i.e. it only depends on the nature of the particles involved in the interaction (vertex or wave-function renormalization) but not on the specific process under consideration \cite{Sborlini:2016gbr,Sborlini:2016hat,Seth:2016hmv}.

\subsubsection{Selected examples}
\begin{figure}[t]
\begin{center}
\includegraphics[width=0.48\textwidth]{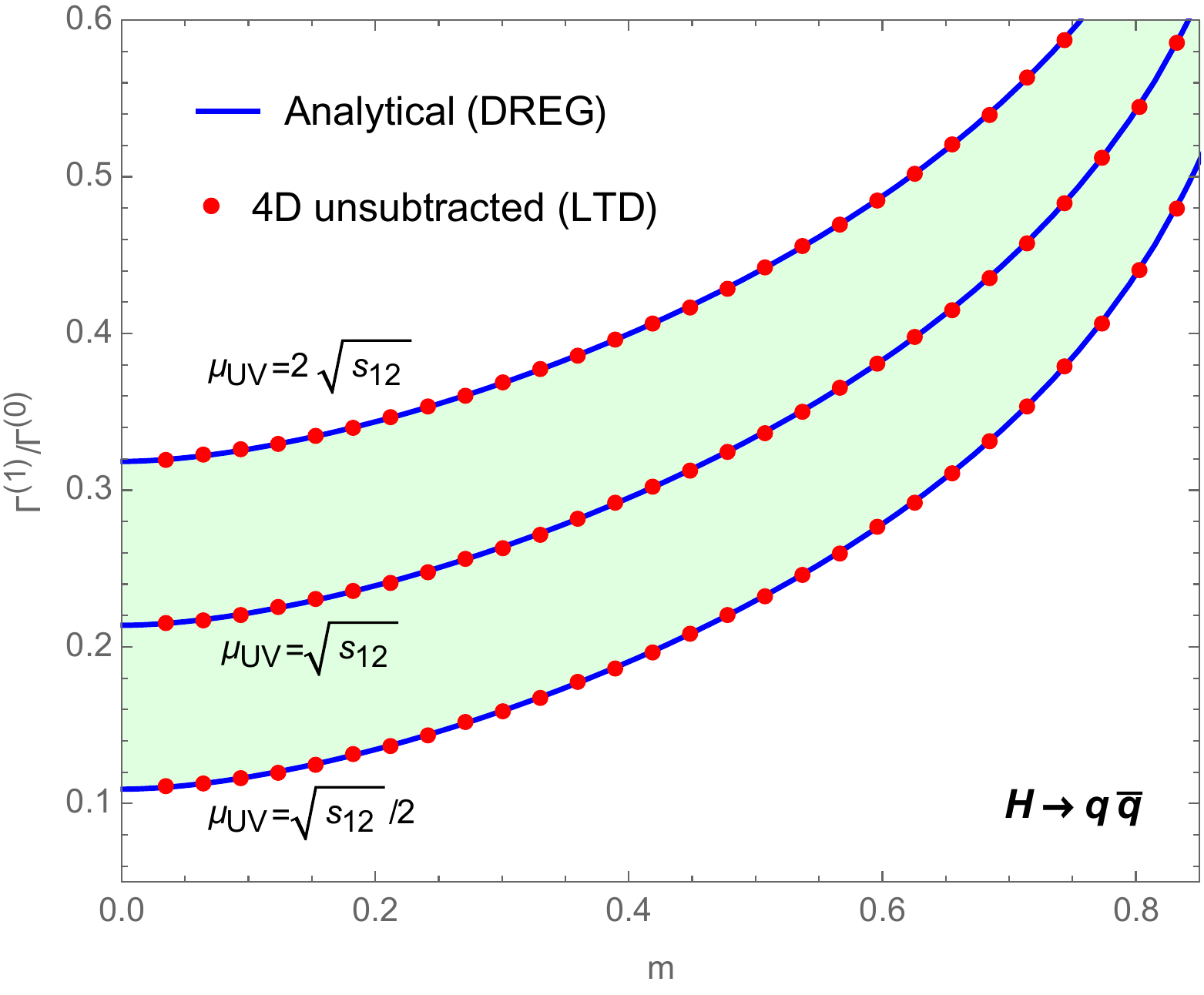}\hfill
\includegraphics[width=0.48\textwidth]{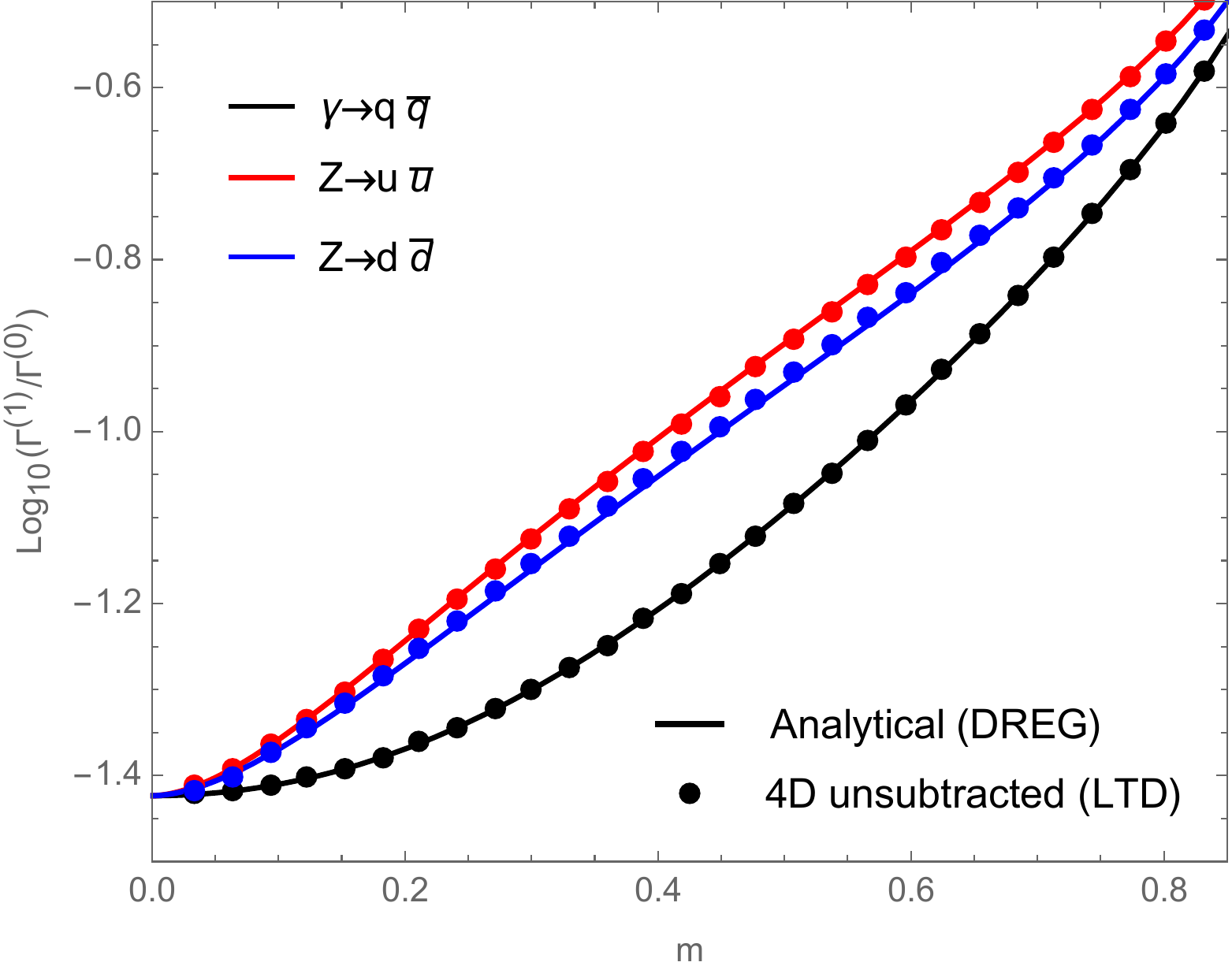}
\caption{\label{fig:SM_looptreeduality:NLOEXAMPLES}
NLO QCD corrections to the decay rates $Z,\gamma^* \to q \bar q$ (right) and $H \to q \bar q$ (left), as a function of the normalized quark mass $m=2 M/\sqrt{s_{12}}$. The solid lines represent the results obtained within the DREG approach, and the colored dots were computed numerically through the application of the FDU technique. We find a complete agreement between these approaches and a smooth massless-limit transition. Moreover, the scale dependence for the Higgs decay is exactly reproduced, thanks to the introduction of local UV counter-terms.}
\end{center}
\end{figure}

In order to have a proof of concept, we apply the FDU framework to compute the decay processes $H \to q \bar q$ and ${Z,\gamma^*} \to q \bar q$ at NLO in QCD with massive and massless quarks \cite{Sborlini:2016gbr,Sborlini:2016hat}. The mentioned framework allows to obtain a combined real-virtual integrand with a non-singular behavior, i.e. numerically integrable in four space-time dimensions. The results are compared with the known expressions computed within the DREG framework. In Fig.~\ref{fig:SM_looptreeduality:NLOEXAMPLES}, the solid lines denote the analytical results computed in DREG as a function of $m=2 M/\sqrt{s_{12}}$, with $M$ the quark mass and $s_{12}$ the virtuality of the decaying particle. The colored dots are the values obtained through the FDU implementation. 

In the first place, we emphasize that the FDU implementation is \emph{purely numerical}. The agreement between both approaches is excellent; moreover, we can appreciate that the massless transition is smooth in both cases. Of course, the massless limit of the analytical DREG result is straightforward and leads to another analytical expression. The expressions for $m>0$ involve some logarithmic-enhanced terms in the real and virtual contributions, separately. Within DREG, these logarithms transform into $\epsilon$-poles when considering the limit $m \to 0$. However, the FDU implementation manages to overcome the $m \to 0$ limit in a surprisingly smooth way, as a consequence of the local regularization of the integrand and the smoothness of the real-virtual mapping in the massless limit. Finally, it is worth noticing that the renormalization scale dependence is successfully reproduced with the four-dimensional framework (left panel of Fig.~\ref{fig:SM_looptreeduality:NLOEXAMPLES}).

\subsubsection{Towards a fully local NNLO framework}
Previously, we have explicitly mentioned the details behind the implementation of FDU at NLO. However, it could be extended to deal with NNLO computations as well. Let's remember that the total NNLO cross-section is generically written as
\begin{equation}
\sigma^{\rm NNLO} = \int_{m} d\sigma_{{\rm V}{\rm V}}^{(2)} + \int_{m+1} d\sigma_{{\rm V}{\rm R}}^{(2)} + \int_{m+2} d\sigma_{{\rm R}{\rm R}}^{(2)}~,
\label{eq:SM_looptreeduality:refo}
\end{equation}
where the double virtual cross-section $d\sigma_{{\rm V}{\rm V}}^{(2)}$ contains the interference of the two-loop with the Born scattering amplitudes and the square of the one-loop scattering amplitude with $m$ final-state particles; the real-virtual cross-section $d\sigma_{{\rm V}{\rm R}}^{(2)}$ includes the contributions from the interference of one-loop and tree-level scattering amplitudes with one extra external particle; and the double real cross-section $d\sigma_{{\rm R}{\rm R}}^{(2)}$ are tree-level contributions integrated in a $m+2$ PS. The LTD representation of the two-loop scattering amplitude is obtained by setting two internal lines on-shell~\cite{Bierenbaum:2010cy}, whilst the squared one-loop introduces two independent loop three-momenta. In both contributions, the PS integration involves $m$ external particles. On the other hand, the dual representation of the real-virtual contribution $d\sigma_{{\rm V}{\rm R}}^{(2)}$ includes an additional particle from the real radiation and one independent loop-three momentum from the application of LTD. The double-real terms, contained within $d\sigma_{{\rm R}{\rm R}}$, directly involve two additional real-particles (that, however, are constrained by momentum conservation). In consequence, since the dual representation of the individual terms appearing in Eq.~\eqref{eq:SM_looptreeduality:refo} always include two additional three-vectors to be integrated, we can find a proper momentum mapping to displace the different IR singularities to the same integration points. In this way, also a local cancellation will take place.

\subsection{Asymptotic expansions of dual contributions}
As stated before, the LTD theorem reduces the $d$-dimensional integration domain with a \linebreak Minkowski metric into a $(d-1)$-dimensional one with an Euclidian metric. This is another remarkable property of the FDU formalism because it allows to perform, \textit{at integrand level}, naive and straightforward asymptotic expansions in any ratio of scales present in the process, without having to deal with most difficulties the traditional approach may encounter. This is a very interesting property because for processes involving many scales, and given a specific kinematical configuration, it is possible to perform the appropriate expansion, and consider and integrate only relevant terms, whose expression is much less complicated than the full integral. This is possible because the sum of all integrated terms of the expansion is the same as the expansion of the full integral, i.e. there is a perfect commutativity between integrating and expanding.

As an example, let's consider the two-point scalar process without applying LTD, and perform a naive expansion
\begin{equation}
\int_\ell\frac{1}{(\ell^2-M^2+\imath0)((\ell+p)^2-M^2+\imath0)}=\int_{\ell}\frac{1}{(\ell-M^2+\imath0)^2}\left(1+\frac{2\ell\cdot p+p^2}{\ell-M^2+\imath0}+...\right)\,,
\end{equation}
for $0<p^2\ll M^2$ and $p_0>0$. The issue here is that this expansion is not valid when $\ell^2\approx M^2$ because of the presence of $p^2$ in the second Feynman propagator, therefore in the traditional approach, an additional step is required.

On the other hand, applying LTD and considering the first cut, the expansion
\begin{equation}
-\int_\ell\frac{\tilde{\delta}(\ell)}{2\ell\cdot p+p^2-\imath0}=-\int_\ell\frac{\tilde{\delta}(\ell)}{2\ell\cdot p}\sum\limits_{n=0}^{\infty}\left(\frac{-p^2}{2\ell\cdot p}\right)^2
\end{equation}
becomes valid for any value of $\ell$, because $0<2\ell\cdot p=\mathcal{O}(M)$, thanks to the presence of $\tilde{\delta}(\ell)$.

In Ref. \cite{Driencourt-Mangin:2017gop}, the Higgs boson production through gluon fusion and the Higgs decay to two photons processes have been computed using LTD. Asymptotic expressions have been obtained for the large and small mass limit of the particle inside the loop, reproducing well-known results in the literature, while showing the efficiency of this method.

\subsection{Conclusions}
The loop-tree duality is a very useful technique to decompose loop into tree-level amplitudes, with the additional property that the \emph{dual} contributions are evaluated in an Euclidean space. In this article we discuss feasible applications that point towards the direction of skipping DREG and achieve a fully numerical implementation of higher-order computations in four space-time dimensions.

In the first place, we use this formalism to compute Feynman integrals in a more efficient way. By decomposing the loop integration into a sum of dual contributions we manage to simplify the treatment of intermediate expressions, since we avoid the presence of Gram determinants. We have tested this approach with one-loop multi-leg tensorial integrals, and compared the results with those provided by the traditional algorithms.

On the other hand, LTD allows to interpret the on-shell internal particles in each cut as a real external particle being radiated from the Born-level process. So, the formalism is well suited for performing an integrand-level combination of the real and virtual contributions. Moreover, through the introduction of a physically-motivated momentum mapping, the IR singular regions present in each term can be routed to the same integration points. This leads to a completely local cancellation of IR singularities, thus avoiding the introduction of IR counter-terms. Also, a local regularization of UV divergences is possible by performing an expansion around the UV propagator and calculating the dual expressions. The combination of all these ingredients leads to a four-dimensional framework, the FDU, which was successfully tested with some simple processes ($H \to q \bar q$ and $A^\mu \to q \bar q$ at NLO).

A spin-off of the local regularization properties of LTD is the possibility of simplifying asymptotic expansions. Since the dual contributions are defined in an Euclidean space, and all the IR/UV singularities have been locally regularized (thus transforming the integrands into actual \emph{integrable} functions), the commutation among integration and series expansions is fulfilled. In this way, the formalism is expected to be more efficient than the traditional expansion-by-regions approach.

Hence, in the context of four-dimensional methods, the LTD/FDU approach provides many advantages for the implementation of physical computations at higher orders. Besides that, it could also shed light into deeper mathematical structures hidden behind the presence of IR/UV singularities.

\subsection*{Acknowledgements}
This work is supported by the Spanish Government and ERDF funds from European Commission (Grants No. FPA2014-53631-C2-1-P and SEV-2014-0398), by Generalitat Valenciana (Grant No. PROMETEO/2017/057), and by Consejo Superior de Investigaciones Cient\'{\i}ficas (Grant No. PIE-201750E021). GC acknowledges support from the Spanish Government grants FPA2015-65480-P, FPA2016-78022-P and the Spanish MINECO Centro de Excelencia Severo Ochoa Programme (SEV-2016-0597). FDM acknowledges support from Generalitat Valenciana (GRISOLIA/2015/035) and GS from Fondazione Cariplo under the Grant No. 2015-0761.

\let\bp\undefined
\let\bq\undefined
\let\bl\undefined





\section{Loop amplitudes: The numerical approach~\protect\footnote{
        S.~Seth,
        S.~Weinzierl}{}}
\label{sec:SM_loopnumerical}


\subsection{Introduction}

Numerical methods enjoy a great flexibility. 
In most precision calculations they are used at least partially.
A first example is the phase space integration through Monte Carlo integration.
This provides the flexibility for arbitrary infrared-safe observables.
A second example is the numerical computation of tree amplitudes through Berends-Giele recurrence relations.
The approach based on recurrence relations is very efficient: The required CPU time scales polynomially
like $n^3$ with the number of external particles of the amplitude.
The numerical approach is therefore the preferred method for processes with a large number of particles,
starting already at moderate values of $n$.

Let us now look at higher-order corrections.
Here we face the occurrence of divergences in intermediate stages of the calculation.
It is common practice in NLO computations to use the subtraction method for the real part.
This allows to treat the real part numerically.
What about the virtual part?
In the virtual part one has to supplement the subtraction method with an algorithm for
contour deformation in loop momentum 
space \cite{Soper:1998ye,Soper:1999xk,Soper:2001hu,Kramer:2002cd,Nagy:2003qn,Gong:2008ww,Assadsolimani:2009cz,Assadsolimani:2010ka,Becker:2010ng,Becker:2011vg,Becker:2012aq,Becker:2012nk,Becker:2012bi,Goetz:2014lla,Seth:2016hmv}.
The contour deformation avoids regions, where individual loop propagators become singular, but
Feynman's $i\delta$-prescriptions allows a deformation into the complex plane.
Related approaches are discussed in \cite{Catani:2008xa,Bierenbaum:2012th,Buchta:2014dfa,Hernandez-Pinto:2015ysa,Buchta:2015wna,Sborlini:2016gbr,Pittau:2012zd,Pittau:2013qla,Donati:2013voa,Page:2015zca,Gnendiger:2017pys}.
In a condensed notation we have
\begin{eqnarray}
 \int\limits_{n+1} d\sigma^{\mathrm{R}} + \int\limits_n d\sigma^{\mathrm{V}} 
 & = & 
 \underbrace{\int\limits_{n+1} \left( d\sigma^{\mathrm{R}} - d\sigma^{\mathrm{A}}_{\mathrm{R}} \right)}_{\mathrm{convergent}} 
 + \underbrace{\int\limits_n  \left( {\bf I} + {\bf L} \right) \otimes d\sigma^B}_{\mathrm{finite}} 
 + \underbrace{\int\limits_{n+\mathrm{loop}} \left( d\sigma^{\mathrm{V}} - d\sigma^{\mathrm{A}}_{\mathrm{V}} \right).}_{\mathrm{convergent}}  
\end{eqnarray}
The subtracted real part and the subtracted virtual part are integrable in four space-time dimensions.
These integrations can be performed by Monte Carlo techniques.
For a process with $n$ final state particles at Born level, the integration for the subtracted real part is over
the phase space of $(n+1)$ final state particles, the integration for the subtracted virtual part is over 
the phase space of $n$ final state particles and a four dimensional loop momentum space.
The combination $({\bf I} + {\bf L})$ contains the subtraction terms added back.
At NLO, all subtraction terms are rather simple and the integration over the unresolved phase space (for the real subtraction terms)
and the integration over loop momentum space (for the virtual subtraction terms) can be performed analytically within
dimensional regularisation.
In the combination $({\bf I} + {\bf L})$ all explicit poles in the dimensional regularisation parameter $\varepsilon$ cancel.

\subsection{Cancellations at the integrand level}

We are interested in extending the numerical approach towards NNLO.
We will need the integrals over the subtraction terms.
While at NLO the analytic integration of the subtraction terms is rather simple, this is no longer
the case at NNLO.
Since the sum of all subtraction terms is finite, we may ask if a numerical approach for the sum of these integrals
is feasible.
We therefore ask if a cancellation at the integrand level is possible.
Since the individual subtraction terms live on different spaces, a few technical difficulties have to be addressed.
It is therefore best, to study this issue at NLO first \cite{Seth:2016hmv}.
In this report we therefore focus on the sum of subtraction terms from the real and the virtual part at NLO.
We are interested in the integral
\begin{eqnarray}
 \int\limits_n  \left( {\bf I} + {\bf L} \right)
 & = & 
 \int\limits_{n} 
 \left[ 
        \int\limits_{1} d\sigma^{\mathrm{A}}_{\mathrm{R}} 
      + \int\limits_{\mathrm{loop}} d\sigma^{\mathrm{A}}_{\mathrm{V}} 
      + d\sigma_{\mathrm{CT}}^{\mathrm{V}} 
      + d\sigma^{\mathrm{C}} 
 \right],
\end{eqnarray}
where $d\sigma^{\mathrm{A}}_{\mathrm{R}}$ is related to the real subtraction terms, 
$d\sigma^{\mathrm{A}}_{\mathrm{V}}$ to the virtual subtraction terms, 
$d\sigma_{\mathrm{CT}}^{\mathrm{V}}$ to the ultraviolet counterterm from renormalisation and 
$d\sigma^{\mathrm{C}}$ to the collinear counterterm from factorisation for initial-state partons.
The first problem we face is that the unresolved phase space is $(D-1)$-dimensional, while
the loop momentum space is $D$-dimensional.
This problem can be overcome with the help of the 
loop-tree duality \cite{Catani:2008xa,Bierenbaum:2012th,Buchta:2014dfa,Hernandez-Pinto:2015ysa,Buchta:2015wna,Sborlini:2016gbr}:
A cyclic-ordered one-loop amplitude
\begin{eqnarray}
 A_n 
 & = & 
 \int \frac{d^Dk}{(2\pi)^D} \frac{P(k) }{\prod\limits_{j=1}^n \left( k_j^2 - m_j^2 + i \delta \right)}
\end{eqnarray}
can be written with Cauchy's theorem as a sum of $n$ integrations over $(D-1)$-dimensional forward hyperboloids 
\begin{eqnarray}
 A_n 
 & = &
 - i \sum\limits_{i=1}^n
 \int \frac{d^{D-1}k}{(2\pi)^{D-1} \; 2 k_i^0} 
 \frac{P(k) }{\prod\limits_{\stackrel{j=1}{j \neq i}}^n \left[ k_j^2 - m_j^2 - i \delta \left(k_j^0-k_i^0\right)\right]}
 \left.
 \vphantom{
 \frac{P(k) }{\prod \left[ k_j^2 - m_j^2 - i \delta \left(k_j^0-k_i^0\right)\right]}
 }
 \right|_{k_i^0=\sqrt{\vec{k}_i^2+m_i^2}}.
\end{eqnarray}
There exists an analogous formula for backward hyperboloids.
The modified $i\delta$-prescription should be noted.
This reduces the integration over $D$-dimensional loop momentum space to an integration in an $(D-1)$ dimensional space.
In the next step we relate points in this space with points
in real unresolved phase space:
Given a set $\{p_1,p_2,...,p_n\}$ of external momenta and an on-shell loop momentum $k$ there is 
an invertible map
\begin{eqnarray}
\label{eq:SM_loopnumerical:def_map}
 \{p_1,p_2,...,p_n\} \times \{k \}
 & \rightarrow &
 \{p_1',p_2',...,p_n',p_{n+1}'\}.
\end{eqnarray}
Note that taking the inverse map and projecting on the $\{p_1,p_2,...,p_n\}$-subspace 
\begin{eqnarray}
 \{p_1',p_2',...,p_n',p_{n+1}'\}
 & \rightarrow &
 \{p_1,p_2,...,p_n\}
\end{eqnarray}
is the standard Catani-Seymour projection.
The map in Eq.~\eqref{eq:SM_loopnumerical:def_map} singles out an emitter $i$ and a spectator $k$.
It therefore meshes well with dipole subtraction. 
For each dipole we have a map which maps all singular regions of this dipole onto each other.

Equipped with this map we may therefore map the singularities from the real subtraction terms onto the singularities
of the virtual subtraction terms.
Do they cancel locally? Not yet, there is a problem with the collinear singularities.
Let us first focus on final state collinear singularities.
The problem is easily understood in physical terms:
Let us first consider the collinear singularities in the real subtraction terms.
Since it is a real process, both partons have transverse polarisations.
Aside from the collinear singularities in the splittings $g \rightarrow g g$ and $q \rightarrow q g$
there is also a collinear divergence in the splitting $g \rightarrow q \bar{q}$.
Now let us look at the collinear singularities in the virtual subtraction terms.
Here one finds that in the collinear limit one parton has a longitudinal polarisation.
Furthermore there is no divergence in the splitting $g \rightarrow q \bar{q}$.
Thus one cannot expect that the singularities cancel in the sum.
The solutions of this problem comes from the field renormalisation constants.
For massless particles, the field renormalisation constants are $1$.
However, they are only $1$ due to a cancellation between (divergent) contributions from the ultraviolet and
infrared region. We may write
\begin{eqnarray}
 Z_2 
 & = & 
 1 
 \;\; = \;\;
 1 
 + \frac{\alpha_s}{4 \pi} C_F 
    \left( 
          \frac{1}{\varepsilon_{\mathrm{IR}}} - \frac{1}{\varepsilon_{\mathrm{UV}}}
    \right),
\nonumber \\
 Z_3 
 & = & 
 1 
 \;\; = \;\;
 1 
 + \frac{\alpha_s}{4 \pi}
     \left( 2 C_A - \beta_0 \right) \left( \frac{1}{\varepsilon_{\mathrm{IR}}} - \frac{1}{\varepsilon_{\mathrm{UV}}} \right).
\end{eqnarray}
The infrared part of the field renormalisation constants will provide the missing terms
which cancel against the collinear singularities from the real and virtual part.

Let us now try to implement the contributions from the field renormalisation constants into our numerical approach.
The field renormalisation constants derived from self-energies and we are tempted to consider the integral representation
for the self-energies.
Here the next problem appears:
For self-energy corrections on external lines, we have an internal (non-loop-like) propagator, which is on-shell.
\begin{figure}
\begin{center}
\includegraphics[width=0.55\textwidth]{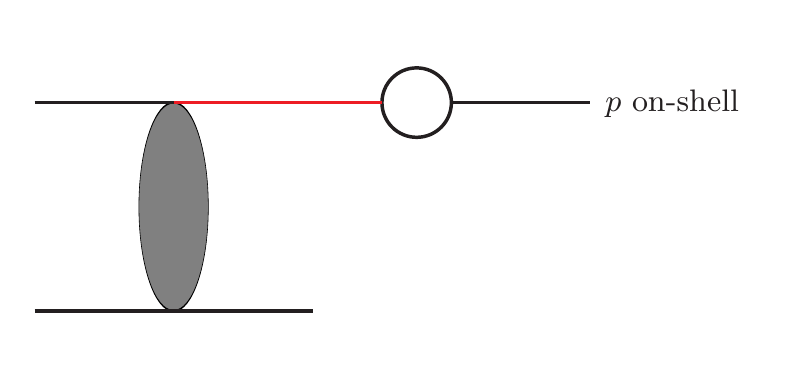}
\caption{
Self-energy corrections on external lines lead to an internal on-shell propagator (shown in red).
}
\label{fig:SM_loopnumerical:fig1}
\end{center}
\end{figure}
This is shown in Fig.~\ref{fig:SM_loopnumerical:fig1}.
This problem can be solved with the help of a dispersion relation.
\begin{figure}
\begin{center}
\includegraphics[width=0.55\textwidth]{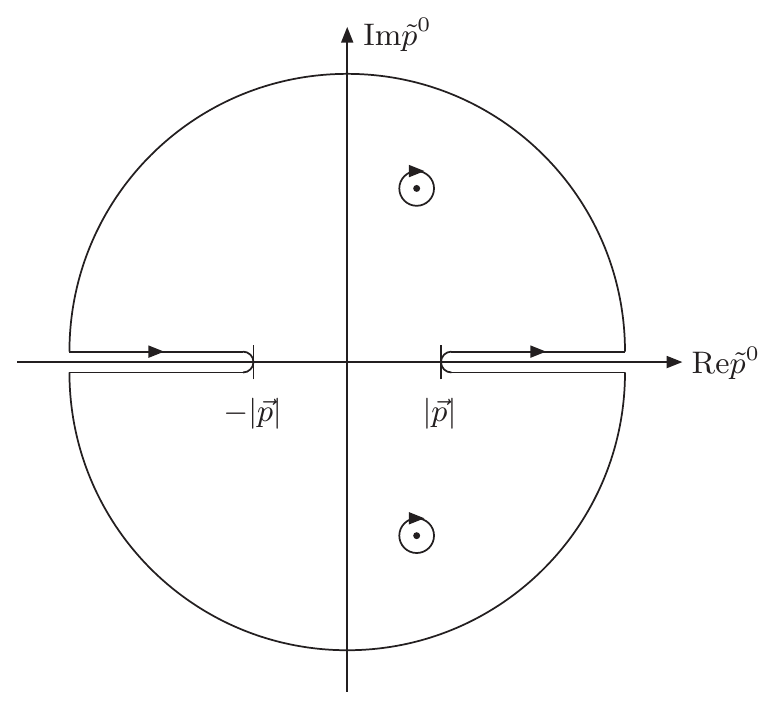}
\caption{
The contour for the dispersion integral for the self-energies.
}
\label{fig:SM_loopnumerical:fig2}
\end{center}
\end{figure}
We re-write the contribution from the self-energy corrections as a dispersion integral with a contour
as shown in Fig.~\ref{fig:SM_loopnumerical:fig2}.

For processes with only final-state infrared singularities (i.e electron-positron annihilation) 
we now achieve a cancellation of all singularities at the integrand level.
It is instructive to visualise the cancellations in loop momentum space.
\begin{figure}
\begin{center}
\includegraphics[width=0.48\textwidth]{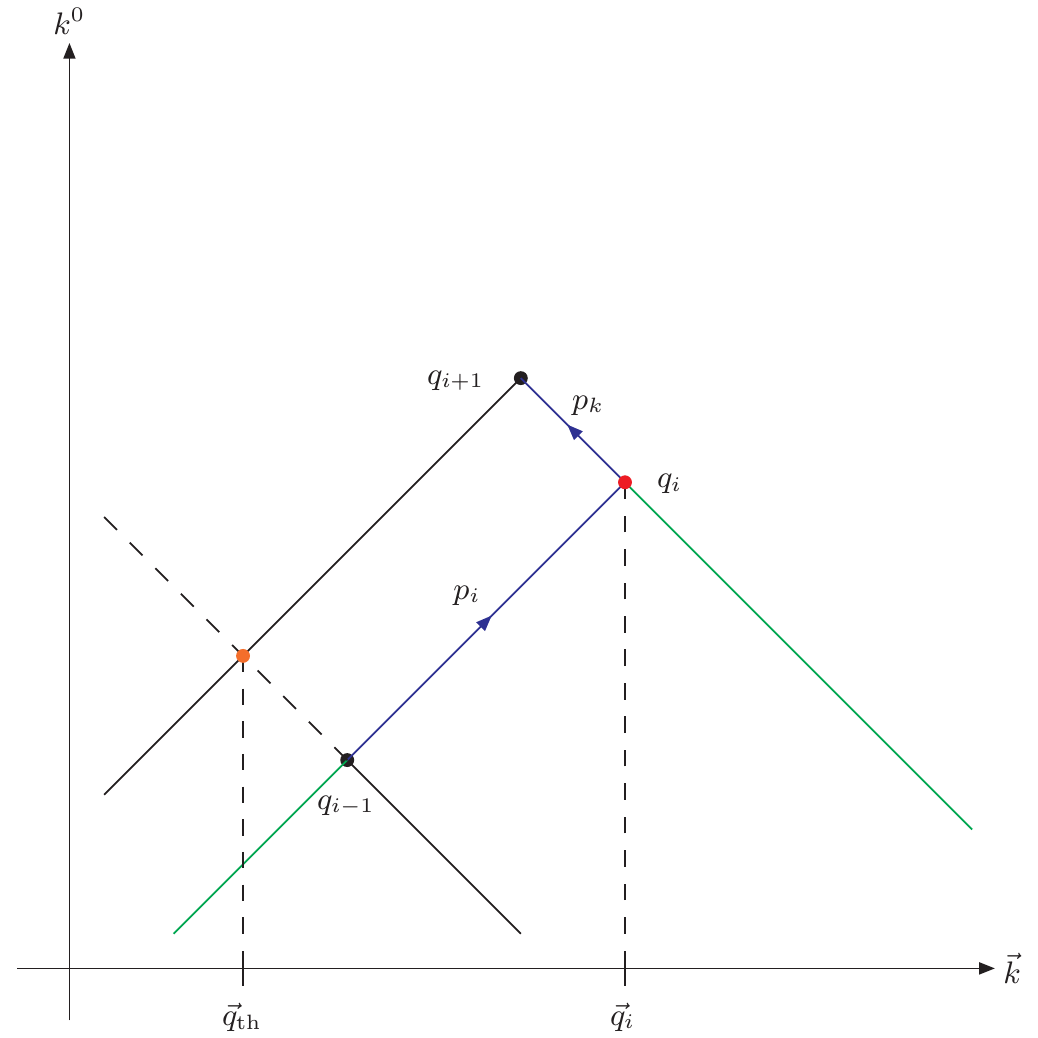}\hfill
\includegraphics[width=0.48\textwidth]{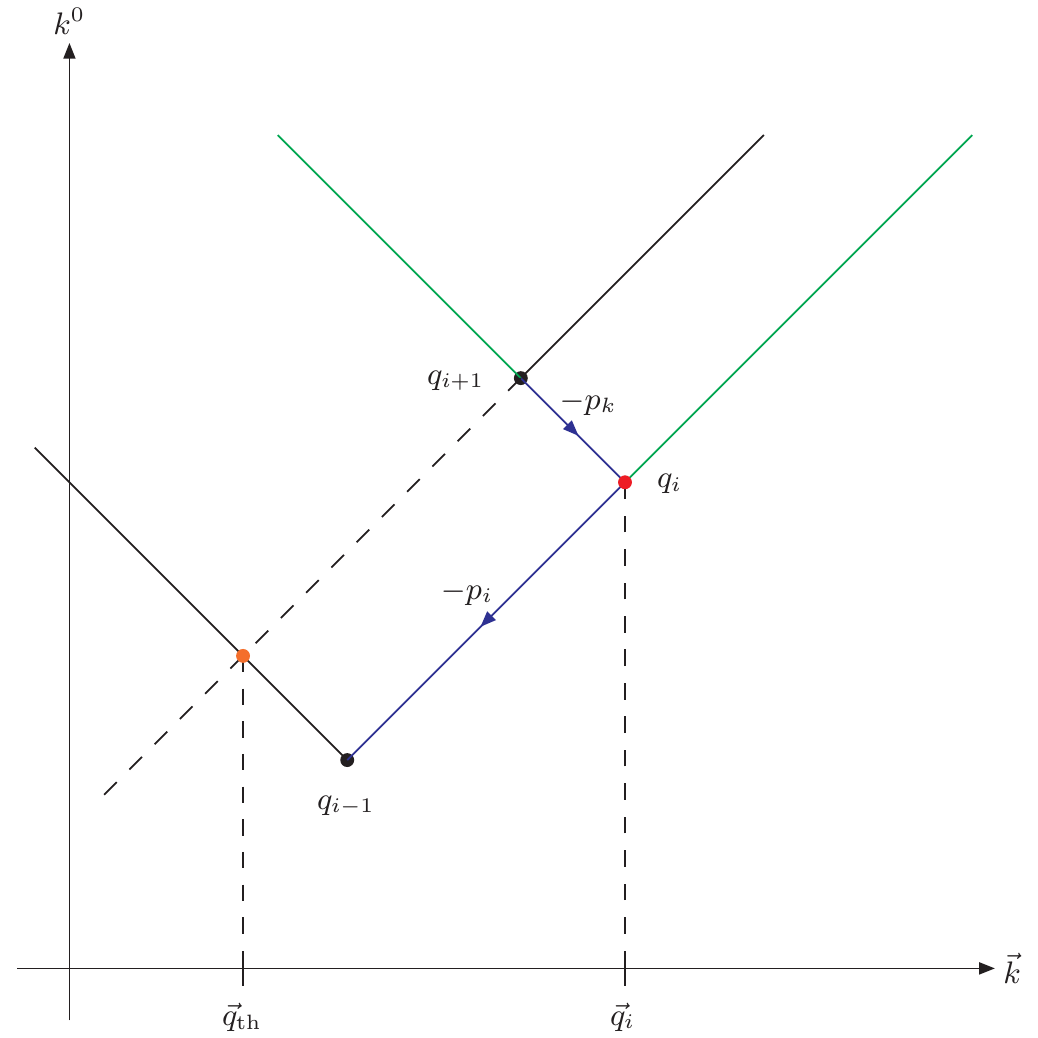}
\caption{
The integration regions for a final-final antenna.
The left picture corresponds to the dipole with emitter $i$ and spectator $k$,
the right picture corresponds to the dipole with emitter $k$ and spectator $i$.
The soft region is indicated by a red dot, the collinear regions by blue line segments.
There is a cancellation of singularities within the virtual dual contributions in the regions where two propagators are
on-shell and have the same sign in the energy component. These regions are indicated in green.
There is a threshold singularity (indicated by an orange dot) at $\vec{q}_{\mathrm{th}}$. 
The threshold singularity is avoided by contour deformation.
}
\label{fig:SM_loopnumerical:fig3}
\end{center}
\end{figure}
This is shown in Fig.~\ref{fig:SM_loopnumerical:fig3}.

Now let us turn to processes with initial-state partons (i.e. hadron colliders like the LHC).
For initial-state partons the local cancellation of collinear singularities is more involved.
Here we face the problem that the regions of the collinear singularities from the real part
and of the collinear singularities from the virtual part do not match in our unified space.
The solution comes through the counterterm from factorisation.
This counterterm has the form
\begin{eqnarray}
 d\sigma^{\mathrm{C}}
 & = &
 \frac{\alpha_s}{4\pi} 
 \;
 \int\limits_0^1 dx_a 
 \;
  \frac{2}{\varepsilon} \left( \frac{\mu_F^2}{\mu^2} \right)^{-\varepsilon}
  P^{a'a}\left(x_a\right) d\sigma^{\mathrm{B}}\left(..., x_a p_a', ... \right),
\end{eqnarray}
where $P^{a'a}$ denotes a splitting function.
For example, for the splitting $g \rightarrow gg$ the splitting function $P^{gg}$ is given by
\begin{eqnarray}
 P^{gg}
 & = &
 2 C_A \left[ \left. \frac{1}{1-x}\right|_+ + \frac{1-x}{x} - 1 + x \left(1-x\right) \right]
 + \frac{\beta_0}{2} \delta\left(1-x\right).
\end{eqnarray}
This splitting function consists of an $x$-dependent part related to the real corrections and
an end-point contribution proportional to $\delta(1-x)$ related to the virtual corrections.
We may now again use integral representations for these two contributions.
\begin{figure}
\begin{center}
\includegraphics[width=0.48\textwidth]{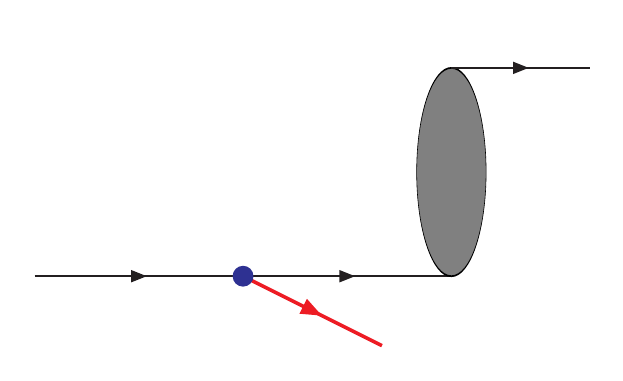}\hfill
\includegraphics[width=0.48\textwidth]{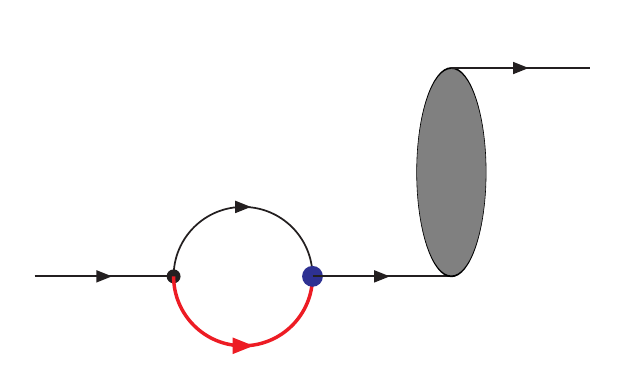}
\caption{
The integral representation of the collinear counterterm for initial-state partons can be split into two
contributions: The $x$-dependent part (left) matches on the real contribution, the end-point contribution (right)
matches on the virtual contribution.
}
\label{fig:SM_loopnumerical:fig5}
\end{center}
\end{figure}
This is sketched in Fig.~\ref{fig:SM_loopnumerical:fig5}.
Taking these into account, one achieves a local cancellation of all singularities.

Let us summarise how the singularities cancel locally.
\begin{figure}
\begin{center}
\includegraphics[width=0.48\textwidth]{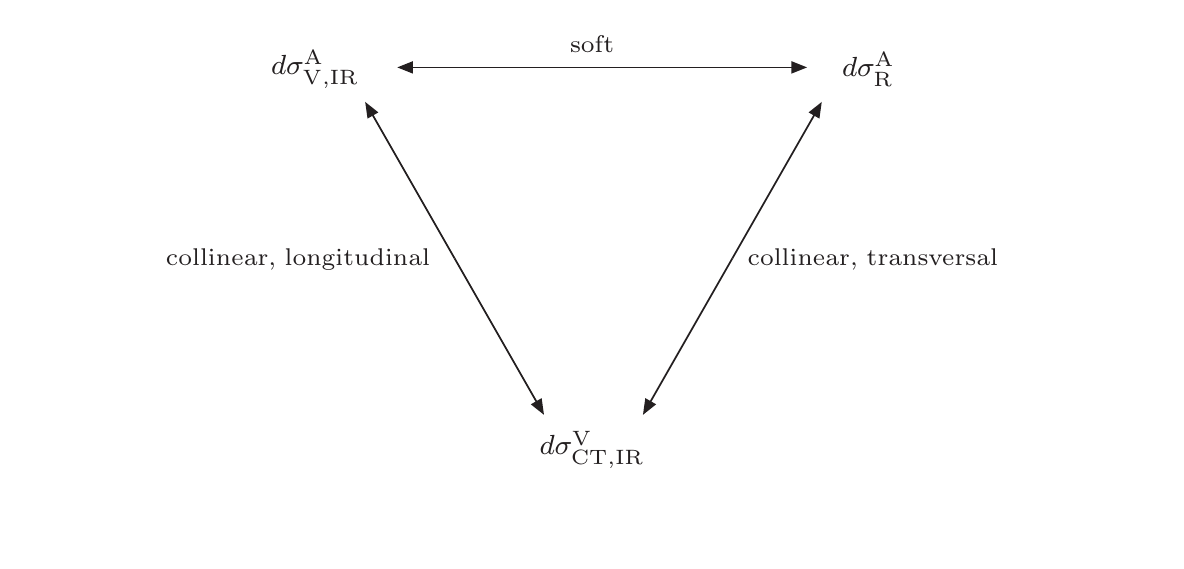}\hfill
\includegraphics[width=0.48\textwidth]{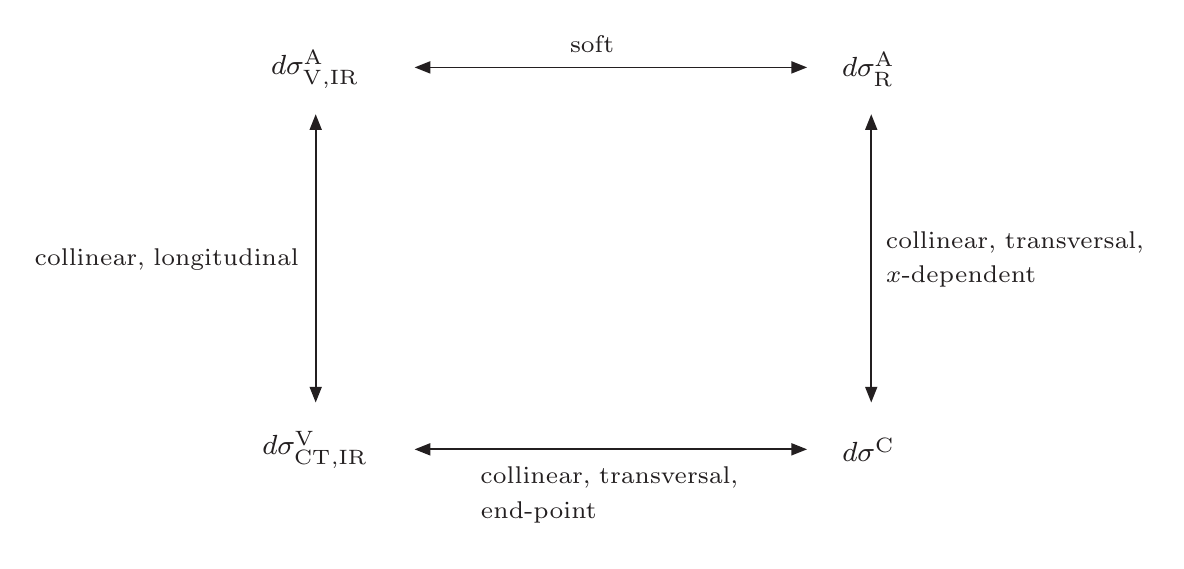}
\caption{
The local cancellation of infrared singularities for final-state singularities (left) and initial-state singularities (right).
}
\label{fig:SM_loopnumerical:fig8}
\end{center}
\end{figure}
Soft singularities cancel between the virtual subtraction terms $d\sigma^{\mathrm{A}}_{\mathrm{V},\mathrm{IR}}$ 
and the real subtraction terms $d\sigma^{\mathrm{A}}_{\mathrm{R}}$.
The longitudinal parts of the collinear singularities present in
virtual subtraction terms $d\sigma^{\mathrm{A}}_{\mathrm{V},\mathrm{IR}}$ cancel against similar parts
in the term $d\sigma^{\mathrm{V}}_{\mathrm{CT},\mathrm{IR}}$ from field renormalisation.
For the transverse parts of the collinear singularities we have to distinguish final-state collinear singularities and
initial-state collinear singularities.
In the former case these singularities cancel between $d\sigma^{\mathrm{A}}_{\mathrm{R}}$
and $d\sigma^{\mathrm{V}}_{\mathrm{CT},\mathrm{IR}}$.
In the latter case the $x$-dependent transverse part of the collinear singularities
cancels between $d\sigma^{\mathrm{A}}_{\mathrm{R}}$ and the counterterm $d\sigma^{\mathrm{C}}$ from factorisation, while
the end-point contribution cancels between $d\sigma^{\mathrm{V}}_{\mathrm{CT},\mathrm{IR}}$
and $d\sigma^{\mathrm{C}}$.
This is shown in Fig.~\ref{fig:SM_loopnumerical:fig8}.
To complete the discussion let us turn to the ultraviolet divergences.
These are unproblematic, the singularities cancel locally between $d\sigma^{\mathrm{A}}_{\mathrm{V},\mathrm{UV}}$
and $d\sigma^{\mathrm{V}}_{\mathrm{CT},\mathrm{UV}}$.

\subsection{Conclusions}

We have discussed the local cancellation of ultraviolet and infrared (i.e. soft and collinear) singularities
at the integrand level at NLO for electron-positron colliders and hadron colliders.
Although we have focused in the discussion on massless particles, the extension towards massive final-state particles
is straightforward and does not pose any further conceptional problem.
We may therefore perform the integrals over the subtraction terms numerically.
At NLO there is no real advantage, as all integrated subtraction terms are known analytically.
The application is the extension towards NNLO, where analytic integration of local subtraction terms
is very challenging.
We have seen that the cancellation at the integrand level involves several subtleties, which are best studied at NLO.

\section{NLO EW automation and technical comparisons~\protect\footnote{
      S.~Kallweit (section coordinator); 
      B.~Biedermann,
      S.~Br\"auer,
      M.~Chiesa,
      A.~Denner,
      N.~Greiner,  
      V.~Hirschi,
      J.-N.~Lang,
      J.~M.~Lindert,
      P.~Maierh\"ofer,
      M.~Pellen,
      S.~Pozzorini,
      S.~Quackenbush,
      C.~Reuschle,
      M.~Sch\"onherr,
      S.~Schumann,
      H.-S.~Shao,
      S.~Uccirati}{}}
\label{sec:SM_ew_comparison}



\subsection{Introduction}
Adequate predictions for scattering processes at particle colliders
such as the Large Hadron Collider~(LHC) are known to require the inclusion of at least next-to-leading order~(NLO)
perturbative corrections of the strong interactions.
With the increasing experimental precision, and the extension of the kinematic reach
thanks to larger collision energies,
not only beyond-NLO predictions in QCD are required, but
also the inclusion of perturbative corrections
from electroweak~(EW) interactions becomes inevitable.
In particular, in the high-energy and more precisely
the high transverse-momentum tails of kinematic distributions which are currently probed, EW corrections become sizeable.

The automated calculation of NLO QCD corrections started to be established
almost a decade ago, and nowadays several independent and well-tested tools exist
\cite{Hahn:1998yk,Arnold:2008rz,Berger:2008sj,Becker:2010ng,Badger:2010nx,Hirschi:2011pa,Bevilacqua:2011xh,Cullen:2011ac,Cascioli:2011va,Actis:2012qn},
  based on different approaches and implementations.
In the recent years, also the automation of NLO EW corrections was
pushed forward, and in the meanwhile a few tools are able to produce
predictions at full NLO Standard Model~(SM) accuracy
\cite{Actis:2012qn,Kallweit:2014xda,Frixione:2014qaa,Chiesa:2015mya}.

In this contribution we review the status of the existing tools for the
automated calculation of EW one-loop amplitudes and their implementations
into integration frameworks.
We also give an overview of phenomenological studies at NLO EW precision
carried out with these tools in the recent years.
After the more qualitative comparison of some selected processes in the 
proceedings of the previous Les Houches 2015 workshop~\cite{Badger:2016bpw},
we perform a detailed technical comparison for two $2\to4$ processes with
complicated resonance structures, namely the off-shell production of leptonically decaying
$\rm{ZZ}$ and $\rm{WW}$ pairs: we start with a point-wise comparison of the EW 1-loop matrix elements
between the different amplitude generators, and we study integrated cross sections and
distributions at NLO EW accuracy.
%

\subsection{Automated calculation of electroweak NLO corrections}
\label{sec:SM_ew_comparison:integratorcodes}
Similar to NLO QCD calculations, the automation of EW corrections at NLO accuracy
requires two main ingredients: a framework that takes care of the bookkeeping
of partonic subprocesses contributing at the coupling
order(s) under consideration, provides a subtraction procedure to treat IR
QED singularities, and performs a stable phase-space integration to provide
predictions for any observables of interest; and a one-loop provider~(OLP)
that generates one-loop transition amplitudes including appropriate EW
renormalization procedures, and moreover guarantees their stable numerical evaluation
in the complete phase space.

Since both tasks are to a wide extent complementary, and the combination of any
integration framework with any OLP is in principle possible, we discuss them separately
in the following. In the first part of this section, we give an overview
of the existing Monte Carlo integration frameworks that have been applied for NLO EW calculations.
In the second part, we briefly introduce the available OLPs, and discuss their recent
applications in phenomenological studies.

\subsubsection{Integration frameworks for automated electroweak NLO corrections}

\paragraph*{\textsc{BBMC}}
\textsc{BBMC}\footnote{\textsc{BBMC} has been written by B.~Biedermann, and is so far only
a working title as the program is not yet ready for publication.}
is a multi-channel Monte Carlo integrator that has been
designed for NLO corrections to processes with identified massless leptons
and jets in the final state. The multi-channel kernel is a further development
of the implementation in Ref.~\cite{Denner:1999gp,Dittmaier:2002ap}. The selection of
the partonic channels for a given hadronic process and the NLO dipole
subtraction~\cite{Catani:1996vz,Dittmaier:1999mb} is done in a fully automated way.
The implementation supports typical NLO EW features like handling of collinear-unsafe
photon radiation off charged leptons~\cite{Dittmaier:2008md},
fragmentation functions and photon-induced contributions.
The program is interfaced with \textsc{Recola}~\cite{Actis:2016mpe}
as matrix-element generator for both one-loop and tree-level amplitudes.
\textsc{BBMC} has been used for all kinds of vector-boson pair-production
processes with leptonic decays at NLO EW
accuracy~\cite{Biedermann:2016yvs,Biedermann:2016lvg,Biedermann:2016guo,Biedermann:2017oae}
and for the calculation of the NLO EW and QCD corrections to vector-boson
scattering in the same-sign WW channel~\cite{Biedermann:2016yds,Biedermann:2017bss}.
The code is intended to be publicly available in the future.

\paragraph*{\textsc{MoCaNLO}}
\textsc{MoCaNLO}\footnote{\textsc{MoCaNLO} stands for `` MOnte CArlo event generator for NLO computation``. It was originally written by Robert Feger and further developed by M.~Pellen. In preparation.} is a flexible Monte Carlo program that can compute arbitrary
processes in the SM with NLO QCD and EW accuracy.
The fast integration is ensured by using similar phase-space mappings to
those of Refs.~\cite{Berends:1994pv,Denner:1999gp,Dittmaier:2002ap}.
The infrared~(IR) divergences appearing in virtual and real corrections are handled with the help of the
Catani--Seymour dipole formalism~\cite{Catani:1996vz,Dittmaier:1999mb}.
These tools have been successfully used for the computation of NLO corrections for high-multiplicity
processes~\cite{Denner:2015yca,Denner:2016jyo,Denner:2016wet,Biedermann:2016yds,Biedermann:2017bss,Denner:2017kzu}.

\paragraph*{\textsc{Munich}}
\textsc{Munich}\footnote{\textsc{Munich} is the abbreviation of 
              ``MUlti-chaNnel Integrator at swiss (CH) precision''---
              an automated parton level NLO generator by S.~Kallweit. 
              In preparation.} 
is a fully general  and very fast parton-level Monte Carlo integrator, written in \textsc{C++},
which was originally developed in the context of multi-leg NLO QCD
calculations~\cite{Denner:2010jp,Denner:2012yc,Denner:2012dz,Cascioli:2013wga}.
It automatically performs the bookkeeping of partonic subprocesses,
phase-space integration based on multi-channel techniques,
and the treatment of IR singularities at NLO by means of
dipole subtraction for massless~\cite{Catani:1996vz} and massive~\cite{Catani:2002hc}
partons.
Subsequently, this framework was extended to deal with full SM corrections
at NLO accuracy, by imposing a generalized bookkeeping for subprocesses at arbitrary
coupling orders. The subtraction of IR QED singularities is performed in 
the dipole-subtraction approach as detailed in Refs.~\cite{Kallweit:2014xda,Kallweit:2017khh}.
These developments allow not only NLO EW corrections to be performed, but full SM NLO
corrections including all sub-leading coupling orders and photon-induced contributions for 
arbitrary SM processes.
\textsc{Munich} relies on external amplitudes throughout:
So far it has been applied in combination with \textsc{OpenLoops}, where the latter
provides not only the one-loop amplitudes of QCD, EW, and mixed types,
but also the tree amplitudes with all relevant colour and helicity correlations,
as required in the IR subtraction. Corresponding applications
of \textsc{Munich+OpenLoops}~\cite{Kallweit:2014xda,Kallweit:2015dum,Lindert:2017olm,Kallweit:2017khh}
are detailed in the next section.
\textsc{Munich} also constitutes the fundament of the public NNLO QCD framework
\textsc{Matrix}\footnote{\textsc{Matrix} is available for
  download from: \url{http://matrix.hepforge.org}.}~\cite{Grazzini:2017mhc}, which was applied to perform
the first calculations for almost all di-boson processes at NNLO QCD
accuracy~\cite{Grazzini:2013bna,Grazzini:2015nwa,Cascioli:2014yka,Grazzini:2015hta,Gehrmann:2014fva,Grazzini:2016ctr,Grazzini:2015wpa,Grazzini:2016swo,Grazzini:2017ckn,deFlorian:2016uhr}.
These features will facilitate combined NNLO QCD+NLO EW predictions
based on the \textsc{Munich/Matrix+OpenLoops} framework in the near future.

\paragraph*{\textsc{Sherpa}}
\textsc{Sherpa}\footnote{The \textsc{Sherpa} program is available from \url{http://sherpa.hepforge.org}.}~\cite{Gleisberg:2003xi,Gleisberg:2008ta} is a multi-purpose Monte Carlo event generator
aiming for the full simulation of high-energy collider scattering events. The \textsc{Sherpa}
framework contains modules and algorithms for the various different phases of event evolution.
This includes methods for the generation and integration of hard-scattering matrix elements,
QCD parton showers~\cite{Schumann:2007mg,Hoche:2015sya}, and phenomenological models for the
parton-to-hadron transition and the underlying event. 

For the hard process generation it contains two tree-level matrix element generators
for the full SM and New Physics scenarios~\cite{Hoche:2014kca}, 
called \textsc{Amegic}~\cite{Krauss:2001iv} and \textsc{Comix}~\cite{Gleisberg:2008fv}. 
Both matrix element generators are equipped with an automated subtraction of NLO QCD divergences 
in the Catani--Seymour dipole-factorization scheme~\cite{Catani:1996vz,Catani:2002hc,Gleisberg:2007md}.
This allows for the automated evaluation of NLO QCD corrections to arbitrary processes. 
Recently, \textsc{Amegic} was updated to be able to handle processes with NLO EW divergences as 
well~\cite{Kallweit:2014xda,Schonherr:2017qcj}, which is used in these comparisons. To facilitate full NLO
computations, interfaces to various one-loop matrix element codes are provided. 
In the following, \textsc{Sherpa} is used in combination with \textsc{GoSam}~\cite{Chiesa:2017gqx,Greiner:2017mft}, 
\textsc{OpenLoops}~\cite{Kallweit:2014xda,Kallweit:2015dum,Kallweit:2017khh,Lindert:2017olm}
and \textsc{Recola}~\cite{Biedermann:2017yoi}. In these combinations of tools, \textsc{Sherpa} provides 
the tree-level matrix elements for both the Born and real emission contributions, the IR subtraction, 
the process management and phase-space integration of all contributions to all processes considered.
To analyse the generated parton-level events the \textsc{Rivet}~\cite{Buckley:2010ar} package is used. 
On-the-fly scale variations are available through an extension of the algorithm of Ref.~\cite{Bothmann:2016nao}.

\paragraph*{\textsc{MadGraph5\_\-aMC@NLO}}
\textsc{MadGraph5\_\-aMC@NLO}\footnote{The latest \textsc{MadGraph5\_\-aMC@NLO} version can be downloaded from \url{https://launchpad.net/mg5amcnlo}.}~\cite{Alwall:2014hca} is a single framework automating the
computation of both LO (tree-level and loop-induced~\cite{Hirschi:2015iia}) and NLO accuracy
(differential) cross sections, including their matching to parton shower programs via the
MC@NLO method~\cite{Frixione:2002ik}. It provides all the necessary elements for the SM and
beyond SM~(BSM) phenomenology studies. In particular, virtual contributions are numerically
calculated by the \textsc{MadLoop} module~\cite{Hirschi:2011pa}, which will be described later,
while the real emission contributions are regulated using the Frixione-Kunszt-Signer~(FKS)
subtraction method~\cite{Frixione:1995ms,Frixione:1997np,Frederix:2009yq}.
Inclusive samples accurate at the NLO QCD level across many jet multiplicities can be obtained
using the \textsc{FxFx} merging method~\cite{Frederix:2012ps}.
External OLPs can also be interfaced to \textsc{MadGraph5\_\-aMC@NLO}, using the
Binoth Les Houches Accord~(BLHA) standard~\cite{Binoth:2010xt,Alioli:2013nda}
(as for example done with \textsc{GoSam}~\cite{Cullen:2014yla} in Ref.~\cite{vanDeurzen:2015cga}).

More recently, the \textsc{MadGraph5\_\-aMC@NLO} framework has been extended in order to support
the computation of NLO EW corrections, including the case of mixed QCD and EW coupling order
perturbative expansions. This allowed the computation of the EW corrections to the
process $t\bar{t}+\PH/\PZ/\PW^{\pm}$~\cite{Frixione:2014qaa, Frixione:2015zaa},
and of the \emph{complete} set of mixed-order NLO corrections to dijet~\cite{Frederix:2016ost},
$t\bar{t}\PW^{\pm}$ and $t\bar{t}t\bar{t}$~\cite{Frederix:2017wme}.
The framework was also instrumental in combining NNLO QCD and NLO EW corrections to
$t\bar{t}$~\cite{Czakon:2017wor,Czakon:2017lgo} production at both the LHC collision energies
and 100 TeV. The corresponding new NLO EW capable version of \textsc{MadGraph5\_\-aMC@NLO}
is planned for an imminent public release, together with a forthcoming reference
paper~\cite{MG5aMC_EW}.
Two NLO UFO models~\cite{Degrande:2011ua} with all the necessary rational
$R_2$~\cite{Ossola:2008xq,Draggiotis:2009yb,Garzelli:2009is,Garzelli:2010qm,Shao:2011tg}
and ultraviolet~(UV) renormalization counterterms will be included, allowing for the computation of the
complete NLO corrections to arbitrary scattering and decay processes in the SM.
The two models correspond to two different EW renormalization schemes:
$\alpha(M_Z)$~\cite{Dittmaier:2001ay} and $G_\mu$~\cite{Denner:1991kt,Dittmaier:2001ay}.
For both setups, the use of renormalization conditions from either the on-shell or the
complex-mass scheme~\cite{Denner:1999gp,Denner:2005fg} has been validated.
The introduction of the complex-mass scheme allows to include off-shell effects of
unstable particles while retaining both gauge invariance and a well-defined perturbative expansion.
A new syntax is introduced in \textsc{MadGraph5\_\-aMC@NLO} for specifying the type of
NLO corrections (QCD and/or EW) that must be accounted for: 
\begin{verbatim}
  MG5_aMC> set complex_mass_scheme True
  MG5_aMC> import model <NLOmodel_with_qcd_qed>
  MG5_aMC> generate <process> QCD=n QED=m [QCD QED]
  MG5_aMC> output; launch
\end{verbatim}
The above syntax implies that complex-mass scheme renormalization conditions are considered
and that the following LO and NLO contributions must be included:
\begin{eqnarray}
\sigma^{\rm LO}&=&\sum_{i\le n, j\le m,i+j=k_0}{\alpha_s^i\alpha^j\sigma^{\rm LO}_{(i,j)}},\nonumber\\
\delta \sigma^{\rm NLO}&=&\sum_{i\le n+1, j\le m+1,i+j=k_0+1}{\alpha_s^i\alpha^j\delta\sigma^{\rm NLO}_{(i,j)}}.
\end{eqnarray}
One can choose the different integer values $n$ and $m$ and coupling order names in
squared brackets to select specific terms in the perturbative series.

Also, numerous efforts in last few years have pushed the \textsc{MadGraph5\_\-aMC@NLO}
framework so as to be able to perform various BSM studies at the NLO QCD accuracy~\cite{Artoisenet:2013puc,Maltoni:2013sma,Degrande:2014tta,Demartin:2014fia,Degrande:2014sta,Durieux:2014xla,Degrande:2015vaa,Backovic:2015soa,Demartin:2015uha,Degrande:2015vpa,Degrande:2015xnm,Mattelaer:2015haa,Neubert:2015fka,Franzosi:2015osa,Das:2016pbk,Fuks:2016ftf,Degrande:2016hyf,Arina:2016cqj,Bylund:2016phk,Maltoni:2016yxb,Zhang:2016omx,Degrande:2016dqg,Degrande:2016aje}
with the help of \textsc{FeynRules}~\cite{Christensen:2008py,Alloul:2013bka}
and \textsc{NLOCT}~\cite{Degrande:2014vpa}, which highlights the flexibility of the
framework\footnote{The available NLO QCD-ready UFO models for BSM studies are listed and can be downloaded at \url{http://feynrules.irmp.ucl.ac.be/wiki/NLOModels}.}.
Indeed, these studies necessitate the support of many novel aspects absent in SM physics,
like non-renormalizable operators~\cite{Degrande:2016dqg},
Majorana fermions~\cite{Degrande:2015vaa} and spin-2 particles~\cite{Das:2016pbk}.

\subsubsection{Generators for electroweak one-loop amplitudes and applications}
\label{sec:SM_ew_comparison:amplitudecodes}

\paragraph*{\textsc{Recola --- BBMC/MoCaNLO/Sherpa+Recola}}
\textsc{Recola}~\cite{Actis:2012qn,Actis:2016mpe} is a general one-loop
amplitude provider which is publicly available.\footnote{\textsc{Recola}
  can be obtained at \url{http://recola.hepforge.org}.}
The library combines the process generation and computation of matrix
elements in a fully
automated and recursive way, pushing limitations on the numbers of external
particles to a yet unmatched level, while remaining a very flexible tool.  In
particular, \textsc{Recola} provides the functionality to compute tree and one-loop
matrix elements, squared matrix elements, optionally summed over spin and
colour, and spin- and/or colour-correlated matrix elements in the
't Hooft--Feynman gauge for arbitrary initial and final states. Standard
renormalization schemes for the electroweak and strong gauge coupling are supported.
Unstable particles are treated in the
complex-mass scheme~\cite{Denner:1999gp,Denner:2005fg,Denner:2006ic}.
The recent upgrade, dubbed \textsc{Recola2}~\cite{Denner:2017vms,Denner:2017wsf},
extends the original version by new model files, as well as the possibility to
perform computations in the Background-Field method, and a \textsc{Python}
interface.
A key aspect of \textsc{Recola}/\textsc{Recola2}, compared to other automated NLO tools, is
that no process source files are generated, and, thus, no intermediate
compilation is required. Therefore, it naturally fits into the framework of
modern event generators requiring flexible amplitude providers which can be used
as black boxes.

Internally, \textsc{Recola} is based on the so-called Berends-Giele
recursion~\cite{Berends:1987me} (BGR) allowing to construct tree-level
amplitudes without referring to Feynman diagrams in the generation
or in the computation phase. A.~van
Hameren showed~\cite{vanHameren:2009vq} that, based on the
decomposition of one-loop amplitudes $\mathcal{M}_1$ in terms of tensor
coefficients $c$ and tensor integrals $T$ as
\begin{align}
  \mathcal{M}_1 = \sum_k c_{k,\mu_1\ldots} T_k^{\mu_1\ldots},
  \label{eq:SM_ew_comparison:loopamp}
\end{align}
a recursion for tensor coefficients can be derived, similar to the BGR.
\textsc{Recola} and \textsc{Recola2} implement such an algorithm for computing tensor
coefficients numerically for arbitrary processes in the SM and beyond,
respectively, while the tensor integrals are obtained by means of the
\textsc{Collier} library~\cite{Denner:2014gla,Denner:2016kdg}. Finally, the complete
1-loop renormalized amplitude $\mathcal{M}_{\rm ren}$ is constructed using the purely
4-dimensional part of Eq.~\eqref{eq:SM_ew_comparison:loopamp} in addition to counterterm and rational
parts,
\begin{align}
  \mathcal{M}_{\rm ren} = \mathcal{M}_1 + \mathcal{M}_\mathrm{CT} + \mathcal{M}_{\mathrm{R}_2},
  \label{eq:SM_ew_comparison:renloopamp}
\end{align}
with $\mathcal{M}_\mathrm{CT}$ and $\mathcal{M}_{\mathrm{R}_2}$ being computed
on equal footing with the tree-level amplitude, but using special Feynman rules.

\textsc{Recola} has been used 
with several Monte Carlo programs for the calculation of NLO QCD and EW corrections.
In combination with the parton-level integration frameworks \textsc{BBMC} and \textsc{MoCaNLO},
it has demonstrated to be particularly efficient for high-multiplicity processes up to $2\to7$ scattering. 


A first class of processes which have been calculated with the help of \textsc{Recola}
is the production of massive di-bosons at NLO EW: WW~\cite{Biedermann:2016guo},
ZZ~\cite{Biedermann:2016lvg,Biedermann:2016yvs}, and WZ~\cite{Biedermann:2017oae}.
All these computations consider the off-shell production as well as all interferences and non-resonant contributions.

A second class of processes evaluated with \textsc{Recola} concerns the off-shell production of top quarks.
Thus, the EW corrections~\cite{Denner:2016jyo} to the off-shell production of top--antitop pairs have been computed.
Recently, QCD corrections to the same process but in the lepton+jets channel~\cite{Denner:2017kzu} have also been obtained.
In addition, the associated production of a Higgs boson with off-shell top-antitop pairs
has been calculated at both NLO QCD and EW accuracy~\cite{Denner:2015yca,Denner:2016wet}.
This constituted the first $2\to7$ process obtained at NLO.
For the EW corrections, for the first time, 9-point functions appear in the loop amplitude.

A further class covered is vector-boson scattering (VBS).
First, NLO EW corrections to the same-sign WW VBS processes have been computed~\cite{Biedermann:2016yds}.
These turn out to be particularly large as an intrinsic feature of VBS processes at the LHC.
Then, the full NLO QCD and EW corrections to the VBS process and its irreducible background have been obtained~\cite{Biedermann:2017bss}.
As the signature possesses three LO contributions, this amounts to compute four separate NLO contributions.

In order to fully exploit the high level of automation of \textsc{Recola}, it has been interfaced
to the multi-purpose Monte Carlo generator \textsc{Sherpa}~\cite{Gleisberg:2003xi,Gleisberg:2008ta}.
Based on an automated subtraction of both QCD and EW divergences~\cite{Gleisberg:2007md,Schonherr:2017qcj},
\textsc{Sherpa+Recola}~\cite{Biedermann:2017yoi} can compute any process at NLO QCD and EW accuracy in the SM.
Some examples of the capabilities have been demonstrated by studying vector-boson production
in association with jets\footnote{The first application of \textsc{Recola} was the computation of
NLO EW corrections to lepton pair production in association with two hard jets~\cite{Actis:2012qn,Denner:2014ina}.}, 
on-shell ${\rm t}\bar{\rm t}{\rm H}$ production, and off-shell ZZ production for both QCD and EW corrections.
This framework has been used \emph{e.g.}\ for the computation of NLO QCD corrections to on-shell
top-quark pair production~\cite{BuarqueFranzosi:2017qlm}.
Note that \textsc{Recola} has also been recently interfaced to \textsc{Whizard}~\cite{Kilian:2007gr,Kilian:2018onl}.

\paragraph*{\textsc{OpenLoops --- Munich/Sherpa+OpenLoops}}
\textsc{OpenLoops}\footnote{The \textsc{OpenLoops} one-loop generator 
              is publicly available at \url{http://openloops.hepforge.org}.}
provides scattering amplitudes at NLO QCD+EW 
based on the open-loops algorithm~\cite{Cascioli:2011va} -- a fast hybrid tree-loop  recursion for the
numerical evaluation of tree and one-loop scattering amplitudes.
At NLO QCD more than one hundred processes are publicly provided in the form of an automatically generated
library that supports all interesting LHC processes, and which can be easily extended upon user
request. A public NLO EW  library is under development and will very soon be released as part
of \textsc{OpenLoops2}~\cite{OL2}.

The recently achieved automation of EW
corrections~\cite{Kallweit:2014xda,Kallweit:2015dum,Kallweit:2017khh} is based on the
well established QCD implementations and allows for NLO QCD+EW
simulations for a vast range of SM processes, up to high particle
multiplicity, at current and future colliders.
To be precise, the new implementations allow for NLO calculations at
any given order $\alpha_s^n\alpha^m$, with all relevant QCD--EW
interference effects.  Full NLO SM calculations that include all
possible $\mathcal{O}(\alpha_s^{n+k} \alpha^{m-k})$ contributions to a
certain process are also supported.

The extension to NLO EW corrections required the
implementation of all $\mathcal{O}(\alpha)$ EW Feynman rules in the
framework of the numerical open-loops recursion including counterterms
associated with so-called $R_2$ rational parts~\cite{Garzelli:2009is}
and with the on-shell renormalization of UV
singularities~\cite{Denner:1991kt}.  Additionally, for the treatment
of heavy unstable particles the complex-mass
scheme~\cite{Denner:2005fg} has been implemented in a fully general way.
Combined with the \textsc{Collier} tensor-reduction
library~\cite{Denner:2014gla},Denner:2016kdg, which implements the Denner--Dittmaier
reduction techniques~\cite{Denner:2002ii,Denner:2005nn} and the scalar
integrals of Ref.~\cite{Denner:2010tr}, or with \textsc{CutTools}~\cite{Ossola:2007ax},
which implements the OPP method~\cite{Ossola:2006us}, together with the \textsc{OneLOop}
library~\cite{vanHameren:2010cp}, the employed recursion permits to
achieve very high CPU performance and a high degree of numerical
stability.
A new method for the automated construction of one-loop amplitudes and
their on-the-fly reduction to scalar integrals building on the open-loops algorithm
was introduced in Ref.~\cite{Buccioni:2017yxi}, and will be part of \textsc{OpenLoops2}~\cite{OL2}.
This improved algorithms will further significantly enhance the numerical stability of the
amplitude evaluation in particular relevant for real-virtual contributions in NNLO computations. 

%

The frameworks \textsc{Munich+OpenLoops} and \textsc{Sherpa+OpenLoops}
automate the full chain of operations -- from process definition to
collider observables -- that enter NLO QCD+EW simulations at parton
level.
Employing the described frameworks, in Ref.~\cite{Kallweit:2014xda} the
simulation of $\PW^+$+1,2,3jets production at NLO EW+QCD was presented.
To facilitate the calculation of the process with the highest jet
multiplicity, these processes were factorized into
a production part and a decay part.
All sub-leading Born channels including interference-based and photon-induced ones
were studied, and NLO corrections were calculated at relative $\mathcal{O}(\alpha_{\rm{s}})$
and $\mathcal{O}(\alpha)$ with respect to the leading Born contribution.
The latter also involves interferences with sub-leading Born diagrams beside genuine EW contributions.
A careful implementation of the narrow-width approximation is required in
order to control numerical stability, given the appearance of
pseudo-resonances for two or more associated jets:
Here a pole regularization with a technical width parameter was chosen, which
corresponds to a smooth and numerically negligible deformation with respect to
the gauge-invariant on-shell limit.
Phenomenologically it was found that $V+$multijet final states feature genuinely
different EW effects as compared to the case of $V+$1jet.

Subsequently in Ref.~\cite{Kallweit:2015dum} NLO QCD+EW simulations were presented for
$\PZ+$jets and $\PW^{\pm}+$jets production including off-shell leptonic
decays with 0, 1, and 2 charged leptons in the final states, applying the
complex-mass scheme throughout.
A naive-exclusive-sums approach
was applied to $V+$1jet and $V+$2jets NLO QCD+EW predictions in order to derive
an approximation that combines exact EW virtual corrections with an inclusive treatment
of bremsstrahlung effects.
This so-called \textit{VI approximation} 
made it possible 
to include NLO EW corrections
into the \textsc{MEPS@NLO} multijet merging framework of \textsc{Sherpa}, and
so to recover perturbative convergence in a merged $V+$jet calculation.
Recently this framework has also been applied for the calculation of NLO EW corrections to $t\bar t(+$jet) production 
and a corresponding MEPS@NLO QCD+EW$_{\rm VI}$ multijet merging~\cite{Gutschow:2018tuk}, as presented in Sec.~\ref{cha:pheno}.\ref{sec:SM_ewmerging_ttbar} of the proceedings at hand.

The work on $V+$jets was continued in the context of the dark-matter background study
of Ref.~\cite{Lindert:2017olm}, where the main focus was on providing not only the best
available perturbative prediction, but in particular a reasonable uncertainty estimate.
The region of interest --- as main background in dark-matter searches --- is the high-energy
tail of the transverse-momentum distribution of the vector boson in the invisible decay channel
$Z\to\nu\bar\nu$. Most accurate predictions are achieved here by estimating $\rm{Z}(\nu\bar\nu)+$jet
from related $V+$jet processes.
Thus best possible theoretical predictions and uncertainty estimates
were required for ratios between the different $V$+jet processes, with $V=\rm{Z,W,\gamma}$.
The frameworks \textsc{Munich+OpenLoops} and \textsc{Sherpa+OpenLoops}
delivered all results involving EW corrections. The Sudakov approximation was used to formulate
reasonable error estimates at NLO EW. Moreover, approximate NNLO EW predictions were achieved,
based on an implementation of 2-loop Sudakov logarithms in \textsc{OpenLoops},
which was required to reduce the uncertainties originating from EW effects to
a subdominant level.
Moreover, mixed NNLO QCD$\times$EW effects were estimated, based on the universality assumption
of the leading Sudakov effects that promotes a multiplicative combination of QCD and EW corrections.
Dedicated studies of the behaviour of NLO EW corrections on $V+2$jets under variation of a jet
resolution parameter
were performed to achieve an improved uncertainty estimate beyond just taking the full relative
$\mathcal{O}(\alpha_s\alpha)$ contribution as unknown.

In Ref.~\cite{Kallweit:2017khh}, NLO QCD+EW corrections to $2\ell2\nu$ final states were
investigated, both in different-flavour and same-flavour channels.
The first is dominated by $\rm{WW}$ resonances, whereas the latter involves, depending on
neutrino flavours, only $\rm{ZZ}$ resonances or both $\rm{WW}$ and $\rm{ZZ}$ resonances.
In this last case, interference effects turn out to be completely negligible apart from
the region around the $Z\to2\ell2\nu$ peak.
Throughout the calculation, photon-induced channels are included not only at LO, but also
in the corrections at NLO EW accuracy. Both the overall impact of incoming-photon contributions and
the dependence on the different photon PDFs available by then were investigated
on a differential level.
Moreover, the possibility of reproducing the exact NLO EW results by the above-mentioned
EW VI approximation augmented with QED radiation effects was studied, where the latter were
generated via YFS soft-photon resummation or, alternatively, by
the Catani--Seymour dipole-based DGLAP-type resummation of the
\textsc{Csshower}.
Both approaches describe the high-energy regions similarly well with deviations typically below
$10\%$. The YFS resummation implementation in \textsc{Sherpa} also preserves the existing
resonance structure.

In the framework of \textsc{Powheg+OpenLoops} recently Monte Carlo generators for the production
of $HV(+$jet) at NLO QCD+EW have been presented~\cite{Granata:2017iod}.
They provide a consistent matching to the QCD and QED parton showers in \textsc{Pythia8}.
Here $V=\{W^{\pm},Z\}$ denotes the corresponding leptonic off-shell processes, and the application
of the improved \textsc{MiNLO} method to $HV+$jet production allows for  NLO QCD+EW accuracy
for observables with both zero or one resolved jet. These generators have been used to study the
behaviour of EW corrections for various kinematic distributions, relevant for experimental
analyses of Higgs-strahlung processes. The \textsc{OpenLoops} tree and one-loop amplitudes
are accessible in the \textsc{Powheg-Box-Res} framework~\cite{Jezo:2015aia} via a
process-independent interface developed in Ref.~\cite{Jezo:2016ujg}.
This framework allows for the consistent resonance-aware parton-shower matching of
off-shell processes at NLO including radiation off the decay products of (narrow) resonances,
in particular relevant in the context of matched NLO EW corrections.

\paragraph*{\textsc{MadLoop --- MadGraph5\_\-aMC@NLO+MadLoop}}
\textsc{MadLoop}~\cite{Hirschi:2011pa} is a module of the
\textsc{MadGraph5\_\-aMC@NLO}~\cite{Alwall:2014hca} framework in charge of generating the code
for one-loop matrix element computations\footnote{Procedure for obtaining a standalone code for one-loop computations in \textsc{MadLoop} can be found at http://cp3.irmp.ucl.ac.be/projects/madgraph/wiki/MadLoopStandaloneLibrary.}. It uses an original approach to the generation of
one-loop Feynman diagrams that takes advantage of the existing tree-level diagram generation
algorithm of \textsc{MadGraph5\_\-aMC@NLO} to directly generate \textit{L-cut} diagrams,
corresponding to loop diagrams with one loop propagator cut open, effectively turning it into a
tree-level diagram with two additional final states compared to the starting loop topology.

Significant improvements have been made in \textsc{MadLoop5}~\cite{Alwall:2014hca} compared to
its predecessor \textsc{MadLoop4}~\cite{Hirschi:2011pa}. The generation of the loop numerator
is rendered efficient by using an in-house implementation of the open-loops~\cite{Cascioli:2011va}
technology, and is generically applicable to any model (spin-2~\cite{Das:2016pbk},
supersymmetric models~\cite{Degrande:2014sta,Degrande:2015vaa},
EFT~\cite{Zhang:2017mls,Degrande:2014tta,Zhang:2014rja,Maltoni:2016yxb,Zhang:2016omx,Franzosi:2015osa},
vector-like quarks~\cite{Fuks:2016ftf},
dark matter~\cite{Mattelaer:2015haa,Backovic:2015soa,Neubert:2015fka} and
others\footnote{See \url{http://feynrules.irmp.ucl.ac.be/wiki/NLOModels} for a list of
available UFO models ready for NLO QCD computations.}) thanks to the generation of the
vertex functions using \textsc{Aloha}~\cite{deAquino:2011ub}.
Moreover, \textsc{MadLoop} offers the possibility of dynamically switching between interfaces
to three different integrand-level reduction
programmes~\cite{Ossola:2006us,Mastrolia:2008jb,Ossola:2007ax,Mastrolia:2012bu,Peraro:2014cba,Hirschi:2016mdz,Mastrolia:2010nb}
and four different tensor integral reduction
codes~\cite{Passarino:1978jh,Davydychev:1991va,Denner:2005nn,Denner:2016kdg,Binoth:2008uq,Fleischer:2011bi},
each time providing a reliable estimate of the numerical uncertainty\footnote{The one-loop scalar integrals in \textsc{MadLoop} are
numerically evaluated by the program \textsc{OneLOop}~\cite{vanHameren:2010cp} or directly
by \textsc{Collier}~\cite{Denner:2016kdg} when it is used for the loop reduction.}.
Whenever a numerically unstable evaluation is encountered, \textsc{MadLoop} will re-evaluate
the point on-the-fly in quadruple precision, not only within the loop reduction
(in the case of \textsc{Ninja}~\cite{Peraro:2014cba,Hirschi:2016mdz} and
\textsc{CutTools}~\cite{Ossola:2007ax}) but also for the computation of the numerator tensor
coefficients.
Another key feature of \textsc{MadLoop} is the flexibility it offers to the user of selecting only particular contributions of the complete one-loop matrix-element, for instance by limiting the allowed particle content, requiring certain propagators or selecting any arbitrary gauge-invariant subset of diagrams. As long as the input \textsc{UFO} model provides the relevant information, the correct corresponding $R_2$ and UV counterterm contributions of Eq.~\eqref{eq:SM_ew_comparison:renloopamp} will be consistently accounted for.

\textsc{MadLoop} can also provide independent results for multiple terms factorizing specific user-defined sets of couplings. This is useful in the context of BSM NLO QCD computations for studying particular interference contributions, while in the SM it is a necessary feature for computing the \emph{complete} set of NLO corrections of order $\mathcal{O}(\alpha_s^{b-n} \alpha^n)$ with $n \leq b$ (see as for example dijet production~\cite{Frederix:2016ost}, where $b=3$). It is also capable of providing spin- and colour-correlated matrix elements, allowing it to be used in the context of NNLO QCD computations. The internal projection of amplitudes onto colour-ordered ones also renders \textsc{MadLoop} competitive for the computation of loop-induced matrix elements~\cite{Hirschi:2015iia}. Finally, \textsc{MadLoop} is not bound to be used within the \textsc{MadGraph5\_\-aMC@NLO} integration framework, and it can easily be used in standalone mode to generate computer libraries for arbitrary processes, ready to be used within any environment (including facilities for calling these libraries directly from within \textsc{Python}).

\paragraph*{\textsc{GoSam --- Sherpa+GoSam}}
\textsc{GoSam}~\cite{Cullen:2011ac,Cullen:2014yla} is a publicly available tool for the automated generation 
of virtual amplitudes.\footnote{\textsc{GoSam}
  can be obtained at \url{http://gosam.hepforge.org}.} It is based on a Feynman 
diagrammatic approach, where the amplitude is generated algebraically in $D$ dimensions. To generate the Feynman
diagrams it uses \textsc{Qgraf}~\cite{Nogueira:1991ex} to generate all the relevant topologies,
and further employs \textsc{Form}~\cite{Vermaseren:2000nd,Kuipers:2012rf} and
\textsc{Spinney}~\cite{Cullen:2010jv} (a \textsc{Form} library to handle the spinor-helicity
formalism) to apply the appropriate Feynman rules and write an optimized \textsc{Fortran} output.
Due to the algebraic
approach no special treatment or additional Feynman rules for the rational $R_2$ terms are needed. Internally \textsc{GoSam}
uses dimensional reduction~(DRED)~\cite{Siegel:1979wq,Siegel:1980qs,Jack:1994bn,Stockinger:2005gx} 
to construct the amplitude, but it can be 
converted to the 't Hooft-Veltman scheme~\cite{'tHooft:1972fi}.
The output for the loop amplitude
does not require a specific reduction technique but can be used for several reduction methods, either on the integral- or on the
integrand level. At the moment three different reduction techniques are built in and supported. Per default it uses \textsc{Ninja}
\cite{Mastrolia:2012bu,Peraro:2014cba,vanDeurzen:2013saa}, a package that performs the reduction on the integrand
level by applying a Laurent expansion. Alternatively one can choose between an \textsc{OPP}
reduction~\cite{Ossola:2006us,Mastrolia:2008jb,Ossola:2008xq} as implemented in \textsc{Samurai}~\cite{Mastrolia:2010nb},
or the tensor reduction methods contained in the \textsc{Golem95}
library~\cite{Heinrich:2010ax,Binoth:2008uq,Cullen:2011kv,Guillet:2013msa}. The remaining scalar integrals can be evaluated
by either using \textsc{OneLoop}~\cite{vanHameren:2010cp}, or \textsc{QCDLoop}~\cite{Ellis:2007qk,Carrazza:2016gav}. Both
the reduction method and the library for the scalar integrals can be changed at any point on the fly.
 
To calculate physical observables like cross sections and differential distributions at the NLO level \textsc{GoSam} can be 
combined with any Monte Carlo event generator that supports the BLHA interface \cite{Binoth:2010xt,Alioli:2013nda}.
In practice, interfaces have been established with \textsc{Herwig7/\linebreak Herwig++}~\cite{Bellm:2015jjp},
\textsc{MadGraph5\_aMC@NLO}~\cite{vanDeurzen:2015cga}, \textsc{Powheg}~\cite{Luisoni:2013kna,Luisoni:2015mpa}, 
\textsc{Sherpa}~\cite{Gleisberg:2008ta} and \textsc{Whizard}~\cite{Kilian:2007gr} and has been used to calculate a large
variety of different processes at NLO in QCD, EW and BSM. \textsc{GoSam} itself contains model files for various 
versions of the SM (with or without CKM mixing, complex mass scheme, effective Higgs couplings). However,
\textsc{GoSam} is able to interpret model files in the \textsc{UFO} format~\cite{Degrande:2011ua} which allows for an immediate
and straightforward implementation of BSM models.
  
In the context of EW corrections, the combination of \textsc{GoSam} and \textsc{Sherpa} has recently been used
to calculate diphoton processes, in association with $0$, $1$, and $2$ jets~\cite{Chiesa:2017gqx} or in association
with a third vector boson~\cite{Greiner:2017mft}.
  
A recent implementation of incorporating the use of quadruple precision within \textsc{GoSam} in connection with \textsc{Ninja}
allows to automatically switch from double to quadruple precision when numerical instabilities are detected. This switch
allows to maintain numerical stability in extreme regions of phase space when approaching singularities as is the case for
the real--virtual contributions in a two-loop calculation~\cite{Jones:2018hbb}.
  
\textsc{GoSam} is the only automated tool that has been extended to calculate two loop corrections. It has successfully
been applied to calculate the full NLO QCD corrections to double Higgs production, 
taking full top-quark mass dependence into account~\cite{Borowka:2016ehy,Borowka:2016ypz}, and very recently also to calculate
the NLO QCD corrections to $H+$1jet, also including full top-quark mass dependence~\cite{Jones:2018hbb}.

\paragraph*{\textsc{NLOX}}
\textsc{NLOX} is the newest member in the set of matrix element providers for the automated generation of one-loop amplitudes.
A non-public predecessor of the program had been available in the past, 
to calculate one-loop QCD corrections to selected processes~\cite{Reina:2011mb}.
\textsc{NLOX} has seen quite some progress in recent years.
The current version of the program provides fully renormalized QCD and EW one-loop amplitudes in the SM, 
for all possible QCD+EW mixed coupling-power combinations to one-loop accuracy, 
including the full mass dependencies on initial- and final-state masses.
\textsc{NLOX} is based on a Feynman diagrammatic approach, 
utilizing \textsc{Qgraf}~\cite{Nogueira:1991ex}, \textsc{Form}~\cite{Vermaseren:2000nd,Kuipers:2012rf} and \textsc{Python},
to algebraically generate \textsc{C++} code for the virtual contribution to a certain process at a certain order of QCD+EW coupling powers,
in terms of one-loop tensor coefficients.
The tensor coefficients are calculated recursively at runtime through standard tensor reduction methods by the \textsc{C++} library \textsc{Tred},
an integral part of \textsc{NLOX}.
Several reduction techniques are available to \textsc{Tred}, many of which are found in Ref.~\cite{Passarino:1978jh,Denner:2005nn}.
The scalar one-loop integrals are evaluated by either using \textsc{OneLOop}~\cite{vanHameren:2010cp}, 
or \textsc{QCDLoop}~\cite{Ellis:2007qk,Carrazza:2016gav}.
UV and IR singularities are regularized in dimensional regularization. 
UV renormalization in NLOX is carried out by means of counterterm diagrams, 
which provide a flexible way to systematically include mass renormalization for massive propagators as well as Yukawa-type vertices.
The renormalization constants in terms of which the EW counterterms are formulated are derived in the on-shell renormalization scheme, 
based on Ref.~\cite{Denner:1991kt}, or in the complex-mass scheme, based on Ref.~\cite{Denner:2005fg}, 
where the choice in \textsc{NLOX} is to expand self-energies with complex squared momenta around real squared momenta as described
in Ref.~\cite{Denner:2005fg}.
The renormalization constants in terms of which the QCD counterterms are formulated are currently derived in a mixed renormalization scheme:
The field and mass renormalization constants for massive quarks are derived in the on-shell renormalization scheme, 
while the field renormalization constants for massless quarks and gluons are derived in the $\overline{\rm{MS}}$ renormalization scheme,
however, with possible contributions from massive quarks decoupling.
As EW input scheme NLOX provides the $\alpha(0)$ EW input scheme or the $G_\mu$ EW input scheme.
\textsc{NLOX} is not yet publicly available and for more details we refer the reader to Ref.~\cite{NLOX}.
The current version of \textsc{NLOX} can be interfaced to a selection of in-house Monte Carlo integration routines. 
A general interface to the Monte Carlo integrators and event generators discussed in this study is currently under development,
which is why no cross-section level comparison was undertaken with \textsc{NLOX} for this study.
As \textsc{NLOX} is still under development, 
the photon-induced channels 
were not available as of the drafting of this document.

\subsection{Technical comparison for $4\ell$ and $2\ell2\nu$ production at NLO EW accuracy}
\label{sec:SM_ew_comparison:comparison}
In order to validate the automated tools for evaluating NLO EW corrections, described in the
previous section, a technical comparison has been performed, at the level of amplitudes
for a set of phase-space points,
integrated cross sections, and a selection of differential distributions.
In contrast to the study on EW automation in the proceedings of the previous Les
Houches 2015 workshop~\cite{Badger:2016bpw}, where a more qualitative comparison has been performed
--- \emph{i.e.}\ without an exhaustive fine-tuning of all ingredients of the calculations,
like details of \emph{e.g.}\ lepton--photon recombination, jet--photon
clustering descriptions, treatment of fragmentation contributions, etc. ---
the goal of this contribution is to pin down all involved codes as far as possible to
a common set of input parameters and conventions.
In that way, a point-wise agreement only limited
by the numerical precision of the respective matrix-element generators, and a statistical
agreement on the level of integrated cross sections and distributions should be achievable.
To do so, we chose to investigate the off-shell production of $\rm{ZZ}$ and $\rm{WW}$ pairs with
corresponding leptonic decays in the different-flavour channels, \emph{i.e.}\
${\rm p}{\rm p}\to\rm{e}^+\rm{e}^-\mu^+\mu^-$ and ${\rm p}{\rm p}\to\rm{e}^+\nu_{\rm{e}} \mu^-\bar{\nu}_{\mu}$. More precisely, the production of the full 4-lepton final state
is considered and, correspondingly, all double-, single-, and non-resonant contributions as well
as spin correlations and interference effects are fully taken into account without any resonance
approximation applied. These processes are very well suited for such a technical
comparison as they are non-trivial $2\to4$ processes that exhibit involved resonance
structures, so the different implementations of intermediate resonances by means of the
complex-mass scheme are probed. On the other hand, they do not involve jets at their leading
order~(LO), which facilitates the choice of a common definition of observables, as no
treatment of jet--photon configurations is needed, which would in general allow for various
self-consistent implementations. As a side effect, the number of partonic processes is easily
manageable, and the relative $\mathcal{O}(\alpha)$ contains only genuine EW corrections, but no
interference contributions with sub-leading orders as it would be the case if LO jets were involved.

\subsubsection{Setup}
\label{sec:SM_ew_comparison:setup}
In this section the setup of the calculations is detailed. As this study is not
a phenomenological, but a technical one, we refrain from using the most up-to-date input parameters
and parton distribution functions~(PDF). We rather adopt the setup of the detailed comparison
for off-shell $\rm{ZZ}$ production applied in Ref.~\cite{Biedermann:2017yoi}\footnote{The 
  benchmark for the results of Ref.~\cite{Biedermann:2017yoi} has actually been set in
  Ref.~\cite{Biedermann:2016lvg}, where all ingredients of the calculations, in particular the
  virtual matrix elements, have been internally cross-checked by means of at least two fully
  independent implementations. The setup of this original calculation deviates in some subtleties
  from the one of Ref.~\cite{Biedermann:2017yoi}. For convenience, we adopt the setup of the
  latter, as a small part of the comparison presented here has already been carried out therein
  using this setup.}.
For off-shell $\rm{WW}$ production, we stick to the very same setup, and only minimally adapt the
applied phase-space cuts to the new final state.

All predictions are for proton--proton collisions at a centre-of-mass energy of
$\sqrt{s} = 13\,\rm{TeV}$, like in the present LHC run.
Following Ref.~\cite{Biedermann:2017yoi}, we use the NNPDF-2.3QED NLO PDF set~\cite{Ball:2013hta} throughout, \emph{i.e.}\ both for the LO and the NLO EW calculations, and we do not consider incoming photons
in the observables by simply setting the photon PDF to zero (non-vanishing lepton PDFs are
not considered either).
We note that all results presented here are independent of the strong coupling $\alpha_{\rm{s}}$
since no QCD corrections are considered and the leading coupling order is
$\mathcal{O}(\alpha^0_{\rm{s}}\alpha^4)$ for both processes under consideration.
Correspondingly, there is also no dependence on a QCD renormalization scale.
The QCD factorization scale is chosen to be fixed, and we
set $\mu_{\rm{F}}=M_{\rm{Z/W}}$ for $\rm{ZZ}$ and $\rm{WW}$ production, respectively, where $M_{V}$
denotes the pole mass of the respective weak boson (see below).

The electromagnetic coupling $\alpha$ is calculated in the $G_\mu$ scheme \cite{Denner:2000bj},
\begin{equation}
  \alpha =\dfrac{\sqrt{2}}{\pi}\,G_\mu M^2_{\rm{W}}\left(1-\dfrac{M^2_{\rm{W}}}{M^2_{\rm{Z}}}\right)\;,\quad\rm{with}\quad G_\mu=1.16637\times10^{-5}\, GeV^{-2},
\end{equation}
and $M^2_{V}$ corresponds to the real parts of the squared pole masses. We use the
complex-mass scheme~\cite{Denner:1999gp,Denner:2005fg,Denner:2006ic} throughout the calculation, \emph{i.e.}\ 
the weak mixing angle becomes a complex quantity, derived from the ratio
$\mu_{\rm{W}}/\mu_{\rm{Z}}$, where $\mu^2_V=M^2_V-\rm{i}M_V\Gamma_V$. For details on the complex renormalization of the EW
parameters, we refer to Ref.~\cite{Denner:2005fg}. Some subtleties of the EW renormalization
procedure that cause differences of higher order in $\alpha$ in the virtual matrix elements
will be discussed in the next section.

As numerical values of the on-shell gauge-boson masses and widths we use
\begin{equation}
  \begin{array}{lclllcl}
    M^{\rm{OS}}_{\rm{Z}} & = & 91.1876\,\rm{GeV}\;, & \qquad & \Gamma^{\rm{OS}}_{\rm{Z}} & = & 2.4952\,\rm{GeV}\;,\\
    M^{\rm{OS}}_{\rm{W}} & = & 80.385\,\rm{GeV}\;, & \qquad & \Gamma^{\rm{OS}}_{\rm{Z}} & = & 2.085\,\rm{GeV}\;,\\
  \end{array}
\end{equation}
and convert them to pole quantities according to the relations~\cite{Bardin:1988xt}
\begin{equation}
  \begin{array}{lclllcll}
  M_{V}&=&\dfrac{M^{\rm{OS}}_V}{\sqrt{1+\Bigl(\Gamma^{\rm{OS}}_V/M^{\rm{OS}}_V\Bigr)^2}}\;,&\quad&
  \Gamma_{V}&=&\dfrac{\Gamma^{\rm{OS}}_V}{\sqrt{1+\Bigl(\Gamma^{\rm{OS}}_V/M^{\rm{OS}}_V\Bigr)^2}}\;,&\qquad
  V=\rm{W},\rm{Z}\;.
  \end{array}
\end{equation}
For convenience, we also state the numerical values of the resulting gauge-boson pole masses and widths and of $\alpha$ in double-precision accuracy, 
\begin{equation}
  \label{eq:SM_ew_comparison:numerical_masses}
  \begin{array}{rcllrcl}
    M_{\rm{Z}} & = & 91.1534806191828\,\rm{GeV}\;, & \qquad & \Gamma_{\rm{Z}} & = & 2.49426637877282\,\rm{GeV}\;,\\
    M_{\rm{W}} & = & 80.3579736098775\,\rm{GeV}\;, & \qquad & \Gamma_{\rm{W}} & = & 2.08429899827822\,\rm{GeV}\;,\\
  \end{array}
\end{equation}
and
\begin{equation}
  \begin{array}{lcl}
    \alpha & = & 0.00755525416742918\;.
  \end{array}
\end{equation}
Higgs bosons and top quarks do not appear as internal resonances in our calculation\footnote{We
  note that this is no longer true for the top quark if photon-induced processes or QCD corrections
  are considered.}, so we can set their widths equal to zero and use the following values for
their masses,
\begin{equation}
  \begin{array}{rcllrcl}
    M_{\rm{H}} & = & 125\,\rm{GeV}\;, & \qquad & \Gamma_{\rm{H}} & = & 0\,\rm{GeV}\;,\\
    m_{\rm{t}} & = & 173\,\rm{GeV}\;, & \qquad & \Gamma_{\rm{t}} & = & 0\,\rm{GeV}\;.\\
  \end{array}
\end{equation}
All the remaining quarks, $q = \rm{u}, \rm{d}, \rm{c}, \rm{s}, \rm{b}$, and all charged leptons
and neutrinos are considered as light particles with mass equal to zero throughout the calculation.
The Cabibbo-Kobayashi-Maskawa~(CKM) matrix is set to unity.
We note, however, that a block-diagonal
form of Cabibbo type with mixing only between the two light quark generations would not affect our
results.

In contrast to the calculation of Ref.~\cite{Biedermann:2016lvg}, where the deep-ineleastic
scattering~(DIS) factorization scheme~\cite{Diener:2005me,Dittmaier:2009cr} is used,
in our present calculation we apply the $\overline{\rm{MS}}$ factorization scheme.
The numerical impact of this difference turns out to be far below
the permille level and thus phenomenologically irrelevant.
This difference could not be resolved in the comparison of Ref.~\cite{Biedermann:2017yoi}.
However, in order to compare with our present calculations, the results
for $\rm{ZZ}$ production therein have been converted to the $\overline{\rm{MS}}$ factorization scheme.

To be consistent with Ref.~\cite{Biedermann:2017yoi}, we only apply phase-space cuts
on the (dressed) charged leptons.
Namely, we require $4\,(2)$ charged leptons in our calculations of
off-shell $\rm{ZZ}\,(\rm{WW})$ production that fulfil the requirements
\begin{equation}
  p_{\rm{T},\ell_i}>15\,\rm{GeV}\;,\quad |y_{\ell_i}|<2.5\;,
\end{equation}
and are separated in the rapidity--azimuthal-angle plane,
\begin{equation}
  \Delta R(\ell_i,\ell_j)=\sqrt{\left(\Delta y(\ell_i,\ell_j)\right)^2+\left(\Delta\phi(\ell_i,\ell_j)\right)^2}>0.2\;,
\end{equation}
independently of the flavours and charges of the leptons $\ell_i$ and $\ell_j$.
The above cuts are applied on the level of dressed leptons, and the dressing procedure is
defined as follows: if a photon is emitted close to one of the bare charged leptons in
the rapidity--azimuthal-angle plane, such that 
\begin{equation}
  \Delta R(\ell_i,\gamma)=\sqrt{\left(\Delta y(\ell_i,\gamma)\right)^2+\left(\Delta\phi(\ell_i,\gamma)\right)^2}<0.2\;,
\end{equation}
the photon is combined with the respective bare lepton $\ell_i$, and the momentum of the
resulting dressed lepton is defined by simple addition of the involved four-momenta. If
the above condition holds for more than one of the leptons, the photon is combined with the
lepton that is closer in the $y-\phi$ plane.

\subsubsection{Point-wise comparison of virtual EW matrix elements}
\label{sec:SM_ew_comparison:pointwise}

%
\begin{table}[p]
  \footnotesize
  \setlength{\tabcolsep}{4pt}
  \centering
  \begin{tabular}{lrrrrrrr}
    & $p_1$ & $p_2$ &  & $p_3$ & $p_4$ & $p_5$ & $p_6$ \\[2ex]
    a)\hspace*{1em} & $\rm{u}$ & $\bar{\rm{u}}$ & $\to$ & $\rm{e}^+$ & $\rm{e}^-$ & $\mu^+$ & $\mu^-$ \\
    b) & $\rm{u}$ & $\bar{\rm{u}}$ & $\to$ & $\rm{e}^+$ & $\nu_{\rm{e}}$ & $\mu^-$ & $\bar{\nu}_\mu$ \\
    c) & $\gamma$ & $\gamma$ & $\to$ & $\rm{e}^+$ & $\rm{e}^-$ & $\mu^+$ & $\mu^-$ \\
    d) & $\gamma$ & $\gamma$ & $\to$ & $\rm{e}^+$ & $\nu_{\rm{e}}$ & $\mu^-$ & $\bar{\nu}_\mu$ \\
  \end{tabular}
  \vspace*{4ex}
  
\setlength{\tabcolsep}{2pt}
  \begin{tabular}{rrrrr}
    PSP 1 & $E$	& $P_x$	& $P_y$	& $P_z$\\[2ex]
    $p_1$ & $ 5.00000000000000$E$+02	$ & $ 0.00000000000000$E$+00	$ & $ 0.00000000000000$E$+00	$ & $ 5.00000000000000$E$+02 $ \\
    $p_2$ & $ 5.00000000000000$E$+02	$ & $ 0.00000000000000$E$+00	$ & $ 0.00000000000000$E$+00	$ & $ -5.00000000000000$E$+02 $ \\
    $p_3$ & $ 8.85513330545030$E$+01	$ & $ -2.21006902876900$E$+01	$ & $ 4.00803531916853$E$+01	$ & $ -7.58054309569366$E$+01 $ \\
    $p_4$ & $ 3.28329419227098$E$+02	$ & $ -1.03849611883456$E$+02	$ & $ -3.01933755389540$E$+02	$ & $ 7.64949213871659$E$+01 $ \\
    $p_5$ & $ 1.52358109467431$E$+02	$ & $ -1.05880959666592$E$+02	$ & $ -9.77096383269757$E$+01	$ & $ 4.95483852267928$E$+01 $ \\
    $p_6$ & $ 4.30761138250968$E$+02	$ & $ 2.31831261837738$E$+02	$ & $ 3.59563040524830$E$+02	$ & $ -5.02378756570221$E$+01 $ \\
  \end{tabular}
  \vspace*{2ex}
  
  \begin{tabular}{rrrrr}
    PSP 2 & $E$	& $P_x$	& $P_y$	& $P_z$\\[2ex]
    $p_1$ & $ 5.00000000000000$E$+02	$ & $ 0.00000000000000$E$+00	$ & $ 0.00000000000000$E$+00	$ & $ 5.00000000000000$E$+02 $ \\
    $p_2$ & $ 5.00000000000000$E$+02	$ & $ 0.00000000000000$E$+00	$ & $ 0.00000000000000$E$+00	$ & $ -5.00000000000000$E$+02 $ \\
    $p_3$ & $ 1.17747708137171$E$+02	$ & $ -6.07059218463615$E$+01	$ & $ 7.12310458623266$E$+01	$ & $ 7.14524452324688$E$+01 $ \\
    $p_4$ & $ 3.50954173077969$E$+02	$ & $ -3.17881667026108$E$+01	$ & $ 8.39394286931734$E$+01	$ & $ 3.39282354933456$E$+02 $ \\
    $p_5$ & $ 3.49332228573790$E$+02	$ & $ 1.84009303995763$E$+02	$ & $ -5.15277979392370$E$+01	$ & $ -2.92435408257719$E$+02 $ \\
    $p_6$ & $ 1.81965890211070$E$+02	$ & $ -9.15152154467912$E$+01	$ & $ -1.03642676616263$E$+02	$ & $ -1.18299391908206$E$+02 $ \\

  \end{tabular}
  \vspace*{2ex}
  
  \begin{tabular}{rrrrr}
    PSP 3 & $E$	& $P_x$	& $P_y$	& $P_z$\\[2ex]
    $p_1$ & $ 5.00000000000000$E$+02	$ & $  0.00000000000000$E$+00	$ & $  0.00000000000000$E$+00	$ & $  5.00000000000000$E$+02 $ \\
    $p_2$ & $ 5.00000000000000$E$+02	$ & $  0.00000000000000$E$+00	$ & $  0.00000000000000$E$+00	$ & $ -5.00000000000000$E$+02 $ \\
    $p_3$ & $ 2.71200874120175$E$+02	$ & $ -3.73192604194911$E$+01	$ & $  3.27628636758588$E$+01	$ & $ -2.66615419075953$E$+02 $ \\
    $p_4$ & $ 2.28710119031214$E$+02	$ & $  3.01476738097433$E$+01	$ & $ -2.76873495467817$E$+01	$ & $ -2.25017437071458$E$+02 $ \\
    $p_5$ & $ 3.68223563778399$E$+02	$ & $ -3.45262681797353$E$+01	$ & $ -6.56193561878981$E$+00	$ & $  3.66542590606031$E$+02 $ \\
    $p_6$ & $ 1.31865443070211$E$+02	$ & $  4.16978547894831$E$+01	$ & $  1.48642148971272$E$+00	$ & $  1.25090265541380$E$+02 $ \\

  \end{tabular}
  \vspace*{2ex}

  \begin{tabular}{rrrrr}
    PSP 4 & $E$	& $P_x$	& $P_y$	& $P_z$\\[2ex]
    $p_1$ & $ 5.00000000000000$E$+02	$ & $  0.00000000000000$E$+00	$ & $   0.00000000000000$E$+00	$ & $  5.00000000000000$E$+02 $ \\
    $p_2$ & $ 5.00000000000000$E$+02	$ & $  0.00000000000000$E$+00	$ & $   0.00000000000000$E$+00	$ & $ -5.00000000000000$E$+02 $ \\
    $p_3$ & $ 6.57509213776183$E$+01	$ & $ -1.80605085177932$E$+01	$ & $  -3.57198237923879$E$+00	$ & $ -6.31208573766775$E$+01 $ \\
    $p_4$ & $ 4.34316531884259$E$+02	$ & $  9.08209885688625$E$+01	$ & $   8.42255785958083$E$+01	$ & $ -4.16279293039596$E$+02 $ \\
    $p_5$ & $ 4.45341565495701$E$+02	$ & $ -7.55663459270628$E$+01	$ & $ -4.72204111505749$E$+01	$ & $  4.36335960118077$E$+02$ \\
    $p_6$ & $ 5.45909812424219$E$+01	$ & $  2.80586587599355$E$+00	$ & $ -3.34331850659945$E$+01	$ & $  4.30641902981974$E$+01 $ \\

  \end{tabular}
  \vspace*{2ex}
  
  \begin{tabular}{rrrrr}
    PSP 5 & $E$	& $P_x$	& $P_y$	& $P_z$\\[2ex]
    $p_1$ & $ 5.00000000000000$E$+02	$ & $  0.00000000000000$E$+00	$ & $  0.00000000000000$E$+00	$ & $  5.00000000000000$E$+02 $ \\
    $p_2$ & $ 5.00000000000000$E$+02	$ & $  0.00000000000000$E$+00	$ & $  0.00000000000000$E$+00	$ & $ -5.00000000000000$E$+02 $ \\
    $p_3$ & $ 3.17520530505573$E$+02	$ & $  6.02726046666900$E$+01	$ & $  3.06151548398959$E$+02	$ & $ -5.88024645074751$E$+01 $ \\
    $p_4$ & $ 1.82384412506608$E$+02	$ & $  4.00234094798358$E$+01	$ & $  1.75895774032594$E$+02	$ & $  2.68863775258899$E$+01 $ \\
    $p_5$ & $ 2.46378670867285$E$+02	$ & $ -4.83680566904740$E$+01	$ & $ -2.40294803208887$E$+02	$ & $ -2.49276573532170$E$+01 $ \\
    $p_6$ & $ 2.53716386120534$E$+02	$ & $ -5.19279574560520$E$+01	$ & $ -2.41752519222667$E$+02	$ & $  5.68437443348023$E$+01 $ \\

  \end{tabular}
  \vspace*{2ex}

  \begin{tabular}{rrrrr}
    PSP 6 & $E$	& $P_x$	& $P_y$	& $P_z$\\[2ex]
    $p_1$ & $ 5.00000000000000$E$+02	$ & $  0.00000000000000$E$+00	$ & $  0.00000000000000$E$+00	$ & $  5.00000000000000$E$+02 $ \\
    $p_2$ & $ 5.00000000000000$E$+02	$ & $  0.00000000000000$E$+00	$ & $  0.00000000000000$E$+00	$ & $ -5.00000000000000$E$+02 $ \\
    $p_3$ & $ 3.67309882280838$E$+02	$ & $  1.65971559565154$E$+01	$ & $  3.39808610366238$E$+01	$ & $  3.65357886350532$E$+02 $ \\
    $p_4$ & $ 1.32530757794944$E$+02	$ & $ -1.52987445917027$E$+01	$ & $ -3.08889911842803$E$+01	$ & $  1.27969607326260$E$+02 $ \\
    $p_5$ & $ 2.97450839881748$E$+01	$ & $  9.82801281377121$E$+00	$ & $  1.54935079102153$E$+01	$ & $ -2.34122061803670$E$+01 $ \\
    $p_6$ & $ 4.70414275936040$E$+02	$ & $ -1.11264241785838$E$+01	$ & $ -1.85853777625588$E$+01	$ & $ -4.69915287496428$E$+02 $ \\

  \end{tabular}
  
  \vspace*{2ex}
  \caption{Phase-space points used for comparison between different matrix-element generators.}
  \label{tab:SM_ew_comparison:psp}
\end{table}
\begin{table}[t]
  \small
  \centering
  \setlength{\tabcolsep}{4pt}
  \begin{tabular}{lrrrr}
a)\quad PSP 1 & $B/10^{-15}$      & $V_{\rm{finite}}/10^{-16}$	& $V_{1}/10^{-17}$	& $V_{2}/10^{-17}$\\[2ex]
\textsc{MadLoop}   & $ 5.26592465401088 $ & $ \phantom{-}6.60297993618509 $ & $ \phantom{-}2.63915540074976 $ & $ -3.09566543908773 $ \\
\textsc{Recola}    & $ 5.26592465401090 $ & $ \phantom{-}6.60088670209820 $ & $ \phantom{-}2.63915540075328 $ & $ -3.09566543908732 $ \\
\textsc{OpenLoops} & $ 5.26592465401100 $ & $ \phantom{-}6.60088670210145 $ & $ \phantom{-}2.63915540078563 $ & $ -3.09566543905505 $ \\
\textsc{GoSam}     & $ 5.26592465401086 $ & $ \phantom{-}6.60088670209788 $ & $ \phantom{-}2.63915540076095 $ & $ -3.09566543909091 $ \\
\textsc{NLOX}      & $ 5.26592465401084 $ & $ \phantom{-}6.60088670211436 $ & $ \phantom{-}2.63915540076702 $ & $ -3.09566543908783 $ \\
  \end{tabular}
  \vspace*{4ex}
  
  \begin{tabular}{lrrrr}
a)\quad PSP 2 & $B/10^{-12}$      & $V_{\rm{finite}}/10^{-13}$	& $V_{1}/10^{-14}$	& $V_{2}/10^{-14}$\\[2ex]
\textsc{MadLoop}   & $ 2.74057953273116 $ & $ -3.10720743529659 $ & $ \phantom{-}2.47558966660999 $ & $ -1.61109736655361 $ \\
\textsc{Recola}    & $ 2.74057953273120 $ & $ -3.10783717792090 $ & $ \phantom{-}2.47558966661119 $ & $ -1.61109736655360 $ \\
\textsc{OpenLoops} & $ 2.74057953273113 $ & $ -3.10783717792216 $ & $ \phantom{-}2.47558966660688 $ & $ -1.61109736655762 $ \\
\textsc{GoSam}     & $ 2.74057953273109 $ & $ -3.10783717792575 $ & $ \phantom{-}2.47558966661326 $ & $ -1.61109736655355 $ \\
\textsc{NLOX}      & $ 2.74057953273088 $ & $ -3.10783717791578 $ & $ \phantom{-}2.47558966660321 $ & $ -1.61109736655852 $ \\
  \end{tabular}
  \vspace*{4ex}
  
  \begin{tabular}{lrrrr}
a)\quad PSP 3 & $B/10^{-4}$      & $V_{\rm{finite}}/10^{-6}$	& $V_{1}/10^{-7}$	& $V_{2}/10^{-7}$\\[2ex]
\textsc{MadLoop}    & $ 1.21906911746527 $ & $ -4.79121605677418 $ & $ -9.28399419983122 $ & $ -7.16650993758228 $ \\
\textsc{Recola}     & $ 1.21906911746653 $ & $ -4.77231274104044 $ & $ -9.28399419025240 $ & $ -7.16650993468800 $ \\
\textsc{OpenLoops}  & $ 1.21906911746730 $ & $ -4.77231273844357 $ & $ -9.28399415556438 $ & $ -7.16650990014111 $ \\
\textsc{GoSam}      & $ 1.21906911746070 $ & $ -4.77231359778343 $ & $ -9.28399407990066 $ & $ -7.16650993856488 $ \\
\textsc{NLOX}       & $ 1.21906911748497 $ & $ -4.77231281258676 $ & $ -9.28399522232122 $ & $ -7.16651015319136 $ \\
  \end{tabular}
  \vspace*{4ex}
  
  \begin{tabular}{lrrrr}
a)\quad PSP 4 & $B/10^{-6}$      & $V_{\rm{finite}}/10^{-7}$	& $V_{1}/10^{-8}$	& $V_{2}/10^{-8}$\\[2ex]
\textsc{MadLoop}    & $ 4.77962555243898 $ & $ \phantom{-}1.65145000279798 $ & $ -3.61194825362166 $ & $ -2.80978604924244 $ \\
\textsc{Recola}     & $ 4.77962555246723 $ & $ \phantom{-}1.63956377750150 $ & $ -3.61194826344888 $ & $ -2.80978605025647 $ \\
\textsc{OpenLoops}  & $ 4.77962555244817 $ & $ \phantom{-}1.63956377748191 $ & $ -3.61194826326975 $ & $ -2.80978605006768 $ \\
\textsc{GoSam}      & $ 4.77962555243871 $ & $ \phantom{-}1.63956377924796 $ & $ -3.61194825975445 $ & $ -2.80978604980914 $ \\
\textsc{NLOX}       & $ 4.77962555244696 $ & $ \phantom{-}1.63956377842641 $ & $ -3.61194826211540 $ & $ -2.80978605014797 $ \\
  \end{tabular}
  \vspace*{2ex}
  
  \caption{Matrix-element comparison at the phase-space points given in Table~\ref{tab:SM_ew_comparison:psp} for the partonic process $\rm{u}\bar{\rm{u}}\to \rm{e}^+\rm{e}^-\mu^+\mu^-$.}
\label{tab:SM_ew_comparison:uux_emexmx}
\end{table}

\begin{table}[t]
  \small
  \centering
  \setlength{\tabcolsep}{4pt}
  \begin{tabular}{lrrrr}
b)\quad PSP 1 & $B/10^{-13}$      & $V_{\rm{finite}}/10^{-14}$	& $V_{1}/10^{-15}$	& $V_{2}/10^{-16}$\\[2ex]
\textsc{MadLoop}   & $ 1.26112388530353 $ & $ -3.30864987248181 $ & $ \phantom{-}1.21073012235170 $ & $ -4.38084412126839 $ \\
\textsc{Recola}    & $ 1.26112388530350 $ & $ -3.30882625694720 $ & $ \phantom{-}1.21073012235186 $ & $ -4.38084412126840 $ \\
\textsc{OpenLoops} & $ 1.26112388530352 $ & $ -3.30882625694700 $ & $ \phantom{-}1.21073012234675 $ & $ -4.38084412131955 $ \\
\textsc{GoSam}     & $ 1.26112388530352 $ & $ -3.30882625695049 $ & $ \phantom{-}1.21073012235248 $ & $ -4.38084412126866 $ \\
\textsc{NLOX}      & $ 1.26112388530352 $ & $ -3.30882625694586 $ & $ \phantom{-}1.21073012235199 $ & $ -4.38084412126837 $ \\
\end{tabular}
  \vspace*{4ex}
  
  \begin{tabular}{lrrrr}
b)\quad PSP 2 & $B/10^{-12}$      & $V_{\rm{finite}}/10^{-13}$	& $V_{1}/10^{-14}$	& $V_{2}/10^{-14}$\\[2ex]
\textsc{MadLoop}   & $ 9.19168012745872 $ & $ -8.42372887318610 $ & $ \phantom{-}1.52668534752201 $ & $ -3.19297083499976 $ \\
\textsc{Recola}    & $ 9.19168012745880 $ & $ -8.42533896054780 $ & $ \phantom{-}1.52668534751979 $ & $ -3.19297083499981 $ \\
\textsc{OpenLoops} & $ 9.19168012745858 $ & $ -8.42533896054679 $ & $ \phantom{-}1.52668534751231 $ & $ -3.19297083500727 $ \\
\textsc{GoSam}     & $ 9.19168012745844 $ & $ -8.42533896055772 $ & $ \phantom{-}1.52668534752199 $ & $ -3.19297083499963 $ \\
\textsc{NLOX}      & $ 9.19168012745785 $ & $ -8.42533896056472 $ & $ \phantom{-}1.52668534750067 $ & $ -3.19297083500005 $ \\
  \end{tabular}  
 \vspace*{4ex}

  \begin{tabular}{lrrrr}
b)\quad PSP 5 & $B/10^{-6}$      & $V_{\rm{finite}}/10^{-7}$	& $V_{1}/10^{-8}$	& $V_{2}/10^{-9}$\\[2ex]
\textsc{MadLoop}    & $ 2.66444601710804 $ & $ -6.26911810540777 $ & $ \phantom{-}2.59441472204941 $ & $ -9.25565109671323 $ \\
\textsc{Recola}     & $ 2.66444601710734 $ & $ -6.27314341024947 $ & $ \phantom{-}2.59441472291996 $ & $ -9.25565109556038 $ \\
\textsc{OpenLoops}  & $ 2.66444601710704 $ & $ -6.27314341020842 $ & $ \phantom{-}2.59441471256317 $ & $ -9.25565118709011 $ \\
\textsc{GoSam}      & $ 2.66444601710846 $ & $ -6.27314341015885 $ & $ \phantom{-}2.59441472307111 $ & $ -9.25565109542055 $ \\
\textsc{NLOX}       & $ 2.66444601705951 $ & $ -6.27314341916503 $ & $ \phantom{-}2.59441471119468 $ & $ -9.25565111942051 $ \\
  \end{tabular}
  \vspace*{4ex}
  
  \begin{tabular}{lrrrr}
b)\quad PSP 6 & $B/10^{-3}$      & $V_{\rm{finite}}/10^{-5}$	& $V_{1}/10^{-5}$	& $V_{2}/10^{-6}$\\[2ex]
\textsc{MadLoop}    & $ 2.08107330428286 $ & $ -3.94303566134416 $ & $ -1.76896206881657 $ & $ -7.22915309068623 $ \\
\textsc{Recola}     & $ 2.08107330429484 $ & $ -3.96855804139705 $ & $ -1.76896184875300 $ & $ -7.22915318450819 $ \\
\textsc{OpenLoops}  & $ 2.08107330428676 $ & $ -3.96855798524525 $ & $ -1.76896146354940 $ & $ -7.22915318412123 $ \\
\textsc{GoSam}      & $ 2.08107330364288 $ & $ -3.96852693676055 $ & $ -1.76896416151949 $ & $ -7.22915263418978 $ \\
\textsc{NLOX}       & $ 2.08107341495760 $ & $ -3.96853794693293 $ & $ -1.76895854332064 $ & $ -7.22915326559073 $ \\
  \end{tabular}
  \vspace*{2ex}
  
  \caption{Matrix-element comparison at the phase-space points given in Table~\ref{tab:SM_ew_comparison:psp} for the partonic process $\rm{u}\bar{\rm{u}}\to \rm{e}^+\nu_{\rm{e}}\mu^-\bar{\nu}_\mu$.}
  \label{tab:SM_ew_comparison:uux_emxnmnex}
\end{table}

\begin{table}[t]
  \small
  \centering
  \setlength{\tabcolsep}{4pt}
  \begin{tabular}{lrrrr}
c)\quad PSP 1 & $B/10^{-13}$      & $V_{\rm{finite}}/10^{-14}$	& $V_{1}/10^{-15}$	& $V_{2}/10^{-15}$\\[2ex]
\textsc{MadLoop}   & $ 4.63762790127829 $ & $ \phantom{-}6.79330655006349 $ & $ \phantom{-}4.07216839247769 $ & $ -2.23061748556626 $ \\
\textsc{Recola}    & $ 4.63762790127830 $ & $ \phantom{-}6.79163662486900 $ & $ \phantom{-}4.07216839245629 $ & $ -2.23061748556050 $ \\
\textsc{OpenLoops} & $ 4.63762790127838 $ & $ \phantom{-}6.79163662486753 $ & $ \phantom{-}4.07216839246097 $ & $ -2.23061748560388 $ \\
\textsc{GoSam}     & $ 4.63762790127830 $ & $ \phantom{-}6.79163662486761 $ & $ \phantom{-}4.07216839247955 $ & $ -2.23061748556541 $ \\
\end{tabular}
  \vspace*{4ex}
  
  \begin{tabular}{lrrrr}
c)\quad PSP 2 & $B/10^{-10}$      & $V_{\rm{finite}}/10^{-11}$	& $V_{1}/10^{-12}$	& $V_{2}/10^{-11}$\\[2ex]
\textsc{MadLoop}   & $ 2.26737153141645 $ & $ \phantom{-}1.88180804083847 $ & $ \phantom{-}2.38584215397888 $ & $ -1.09056584355518 $ \\
\textsc{Recola}    & $ 2.26737153141650 $ & $ \phantom{-}1.88096074990550 $ & $ \phantom{-}2.38584215406775 $ & $ -1.09056584355509 $ \\
\textsc{OpenLoops} & $ 2.26737153141649 $ & $ \phantom{-}1.88096075053150 $ & $ \phantom{-}2.38584215383146 $ & $ -1.09056584370294 $ \\
\textsc{GoSam}     & $ 2.26737153141644 $ & $ \phantom{-}1.88096075053592 $ & $ \phantom{-}2.38584215397731 $ & $ -1.09056584355520 $ \\
\end{tabular}
\vspace*{4ex}

\begin{tabular}{lrrrr}
c)\quad PSP 3 & $B/10^{-6}$      & $V_{\rm{finite}}/10^{-6}$	& $V_{1}/10^{-9}$	& $V_{2}/10^{-9}$\\[2ex]
\textsc{MadLoop}    & $ 1.37978612284930 $ & $ \phantom{-}1.55018919031339 $ & $ \phantom{-}4.89785291501769 $ & $ -6.63652265273678 $ \\
\textsc{Recola}     & $ 1.37978612284863 $ & $ \phantom{-}1.55013518201790 $ & $ \phantom{-}4.89788114834105 $ & $ -6.63652866815621 $ \\
\textsc{OpenLoops}  & $ 1.37978612284923 $ & $ \phantom{-}1.55013518232261 $ & $ \phantom{-}4.89788114988480 $ & $ -6.63652866830139 $ \\
\textsc{GoSam}      & $ 1.37978612284092 $ & $ \phantom{-}1.55011760547612 $ & $ \phantom{-}4.89816579319493 $ & $ -6.63658173170046 $ \\
  \end{tabular}
  \vspace*{4ex}
  
\begin{tabular}{lrrrr}
c)\quad PSP 4 & $B/10^{-7}$      & $V_{\rm{finite}}/10^{-6}$	& $V_{1}/10^{-10}$	& $V_{2}/10^{-9}$\\[2ex]
\textsc{MadLoop}    & $ 2.19037672578717 $ & $ \phantom{-}1.68165624485106 $ & $ \phantom{-}7.61526199100670 $ & $ -1.05353269570773 $ \\
\textsc{Recola}     & $ 2.19037672578999 $ & $ \phantom{-}1.68164383557005 $ & $ \phantom{-}7.61526198688565 $ & $ -1.05353269571203 $ \\
\textsc{OpenLoops}  & $ 2.19037672578763 $ & $ \phantom{-}1.68164383554897 $ & $ \phantom{-}7.61526198357373 $ & $ -1.05353269576734 $ \\
\textsc{GoSam}      & $ 2.19037672578690 $ & $ \phantom{-}1.68164383554095 $ & $ \phantom{-}7.61526199187791 $ & $ -1.05353269572332 $ \\
\end{tabular}
  \vspace*{2ex}

  \caption{Matrix-element comparison at the phase-space points given in Table~\ref{tab:SM_ew_comparison:psp} for the partonic process $\gamma\gamma\to \rm{e}^+\rm{e}^-\mu^+\mu^-$.}
  \label{tab:SM_ew_comparison:aa_emexmx}
\end{table}

\begin{table}[t]
  \small
  \centering
  \setlength{\tabcolsep}{4pt}
  \begin{tabular}{lrrrr}
d)\quad PSP 1 & $B/10^{-13}$      & $V_{\rm{finite}}/10^{-15}$	& $V_{1}/10^{-15}$	& $V_{2}/10^{-16}$\\[2ex]
\textsc{MadLoop}   & $ 3.67352032122512 $ & $ \phantom{-}6.79828192861256 $ & $ \phantom{-}4.15882308803665 $ & $ -8.83449344853984 $ \\
\textsc{Recola}    & $ 3.67352032122520 $ & $ \phantom{-}6.78860727240780 $ & $ \phantom{-}4.15882308802252 $ & $ -8.83449344853268 $ \\
\textsc{OpenLoops} & $ 3.67352032122515 $ & $ \phantom{-}6.78860727241689 $ & $ \phantom{-}4.15882308802947 $ & $ -8.83449344826758 $ \\
\textsc{GoSam}     & $ 3.67352032122515 $ & $ \phantom{-}6.78860727251103 $ & $ \phantom{-}4.15882308806114 $ & $ -8.83449344857124 $ \\
  \end{tabular}
  \vspace*{4ex}
  
  \begin{tabular}{lrrrr}
d)\quad PSP 2 & $B/10^{-11}$      & $V_{\rm{finite}}/10^{-14}$	& $V_{1}/10^{-12}$	& $V_{2}/10^{-13}$\\[2ex]
\textsc{MadLoop}   & $ 9.78728443151128 $ & $ \phantom{-}2.83565594714426 $ & $ \phantom{-}1.43919663402090 $ & $ -2.35375586979757 $ \\
\textsc{Recola}    & $ 9.78728443151120 $ & $ \phantom{-}2.52764290726850 $ & $ \phantom{-}1.43919663402107 $ & $ -2.35375586979726 $ \\
\textsc{OpenLoops} & $ 9.78728443151125 $ & $ \phantom{-}2.52764290753401 $ & $ \phantom{-}1.43919663402070 $ & $ -2.35375586979729 $ \\
\textsc{GoSam}     & $ 9.78728443151106 $ & $ \phantom{-}2.52764290829874 $ & $ \phantom{-}1.43919663401982 $ & $ -2.35375586979631 $ \\
  \end{tabular}
  \vspace*{4ex}

  \begin{tabular}{lrrrr}
d)\quad PSP 5 & $B/10^{-5}$      & $V_{\rm{finite}}/10^{-5}$	& $V_{1}/10^{-6}$	& $V_{2}/10^{-7}$\\[2ex]
\textsc{MadLoop}    & $ 8.56103162287602 $ & $ -2.08508067115892 $ & $ \phantom{-}1.40092813161870 $ & $ -2.05885284890288 $ \\
\textsc{Recola}     & $ 8.56103162287098 $ & $ -2.08643204830236 $ & $ \phantom{-}1.40092813169621 $ & $ -2.05885284880211 $ \\
\textsc{OpenLoops}  & $ 8.56103162287173 $ & $ -2.08643204831341 $ & $ \phantom{-}1.40092813323908 $ & $ -2.05885283530621 $ \\
\textsc{GoSam}      & $ 8.56103162287777 $ & $ -2.08643204828055 $ & $ \phantom{-}1.40092813175874 $ & $ -2.05885284872070 $ \\
  \end{tabular}
  \vspace*{4ex}
  
  \begin{tabular}{lrrrr}
d)\quad PSP 6 & $B/10^{-1}$      & $V_{\rm{finite}}/10^{-1}$	& $V_{1}/10^{-3}$	& $V_{2}/10^{-4}$\\[2ex]
\textsc{MadLoop}    & $ 2.30984277049480 $ & $ \phantom{-}1.15802573980481 $ & $ \phantom{-}2.61399257786962 $ & $ -5.55496881434693 $ \\
\textsc{Recola}     & $ 2.30984277049406 $ & $ \phantom{-}1.15474814667339 $ & $ \phantom{-}2.61399259319492 $ & $ -5.55496881453983 $ \\
\textsc{OpenLoops}  & $ 2.30984277049366 $ & $ \phantom{-}1.15474813504440 $ & $ \phantom{-}2.61399265092704 $ & $ -5.55496797160493 $ \\
\textsc{GoSam}      & $ 2.30984277049006 $ & $ \phantom{-}1.15420829481100 $ & $ \phantom{-}2.61399257843170 $ & $ -5.55496881427150 $ \\
  \end{tabular}
  \vspace*{2ex}

  \caption{Matrix-element comparison at the phase-space points given in Table~\ref{tab:SM_ew_comparison:psp} for the partonic process $\gamma\gamma\to \rm{e}^+\nu_{\rm{e}}\mu^-\bar{\nu}_\mu$.}
   \label{tab:SM_ew_comparison:aa_emxnmnex}
\end{table}

We first validate the different OLPs at the level of one-loop amplitudes for
certain kinematic configurations, namely the phase-space points specified in
Table~\ref{tab:SM_ew_comparison:psp}.
In order to be independent of the subtraction scheme used to cancel IR divergences,
we found it useful to compare the coefficients of a Laurent expansion of these
matrix elements, which are regularized in \mbox{$D=4-2\epsilon$} space-time dimensions
in the 't~Hooft--Veltman scheme.
The virtual renormalized matrix element, summed over colour and helicity configurations,
including averaging factors for initial states, can be written as
%
\begin{equation}
  2\Re\left\{\overline{\mathcal{M}^\ast_V\cdot\mathcal{M}_{\rm{Born}}}\right\}=\dfrac{(4\pi)^\epsilon}{\Gamma(1-\epsilon)}
  \left(\dfrac{V_2}{\epsilon^2} + \dfrac{V_1(\mu)}{\epsilon} + V_{\rm{finite}}(\mu)+
  \mathcal{O}(\epsilon)\right)\;.
\end{equation}
\\
The dependence on both $\epsilon$ and the regularization scale $\mu$ will cancel only upon combination
with the real-emission corrections, \emph{e.g.}\ by the corresponding dependencies of the $I$-operator
in a Catani--Seymour dipole-subtraction approach \cite{Catani:1996vz}.
In order to be independent of subtraction schemes,
we shall compare the relevant coefficients of the Laurent expansion in $\epsilon$,
$V_{\rm{finite}}(\mu)$, $V_1(\mu)$, and $V_2$ for a given regularization scale,
\begin{equation}
\mu=M_{\rm{Z}/\rm{W}}\;,
\end{equation}
for off-shell $\rm{ZZ}$ and $\rm{WW}$ production, respectively.
We start with the squared Born matrix elements, summed over colour and helicity configurations,
including averaging factors for initial states, denoted as 
\begin{equation}
B=\overline{\left|\mathcal{M}_{\rm{Born}}\right|^2}\;,
\end{equation} 
in order to validate that all input parameters have been set correctly.
As we shall see, it is interesting to perform the comparison both on
far off-shell and on-shell phase-space points
for each of the two different partonic channels considered,
namely the respective $\rm{u}\bar{\rm{u}}$-induced
and $\gamma\gamma$-induced channels of each of the processes.

In Table~\ref{tab:SM_ew_comparison:psp} we first list the two far off-shell
kinematic configurations labelled PSP~1 and PSP~2 that apply to all processes.
We then consider the points PSP~3 and PSP~4 for the $\rm{e}^+\rm{e}^-\mu^+\mu^-$
production processes a) and c) where the invariant mass of
the muon--anti-muon pair is exactly at the pole mass of the $\rm{Z}$ boson,
and finally the points PSP~5 and PSP~6 for the $\rm{e}^+\nu_{\rm{e}}\mu^-\bar{\nu}_\mu$
processes b) and d) where the invariant mass of the positron and its neutrino
sits at exactly the pole mass of the $\rm{W}$ boson.

The numerical results for $B$, $V_{\rm finite}\,$, $V_1\,$, and $V_2$ obtained with the different involved OLPs
(\textsc{MadLoop}, \textsc{Recola}, \textsc{OpenLoops}, \textsc{GoSam}, and \textsc{NLOX})
are displayed in Tables~\ref{tab:SM_ew_comparison:uux_emexmx}--\ref{tab:SM_ew_comparison:aa_emxnmnex}.
In general, an agreement among all OLPs of more than $10$ digits for off-shell kinematics
(typically a bit worse for on-shell kinematics) is found for the quantities $B$, $V_1\,$, and $V_2\,$, 
whereas $V_{\rm finite}$ obtained with \textsc{MadLoop} shows differences of up to the percent level compared to
the corresponding results from the other programs, which show reasonable agreement among themselves.
Such small deviations are to be expected due to differences in the details of the implementation of the complex-mass scheme renormalization.
In order to avoid complications in choosing the correct Riemann sheet when evaluating
two-point functions of self-energies with complex arguments, it was proposed in Ref.~\cite{Denner:2006ic}
to expand these functions in the EW coupling (see Eq.~(8) in Ref.~\cite{Denner:2006ic}), and
to truncate the series at the NLO EW level. This scheme has been adopted by
\textsc{Recola}, \textsc{OpenLoops}, \textsc{GoSam} and \textsc{NLOX}.
Conversely, \textsc{MadLoop} chooses not to perform such an expansion, and instead considers the
\emph{full} two-point function evaluated on the appropriate Riemann sheet. 
The details of this procedure will be documented in a forthcoming publication~\cite{MG5aMC_EW}. 

\begin{figure}[t!]
\centering
\includegraphics[width=1.\textwidth]{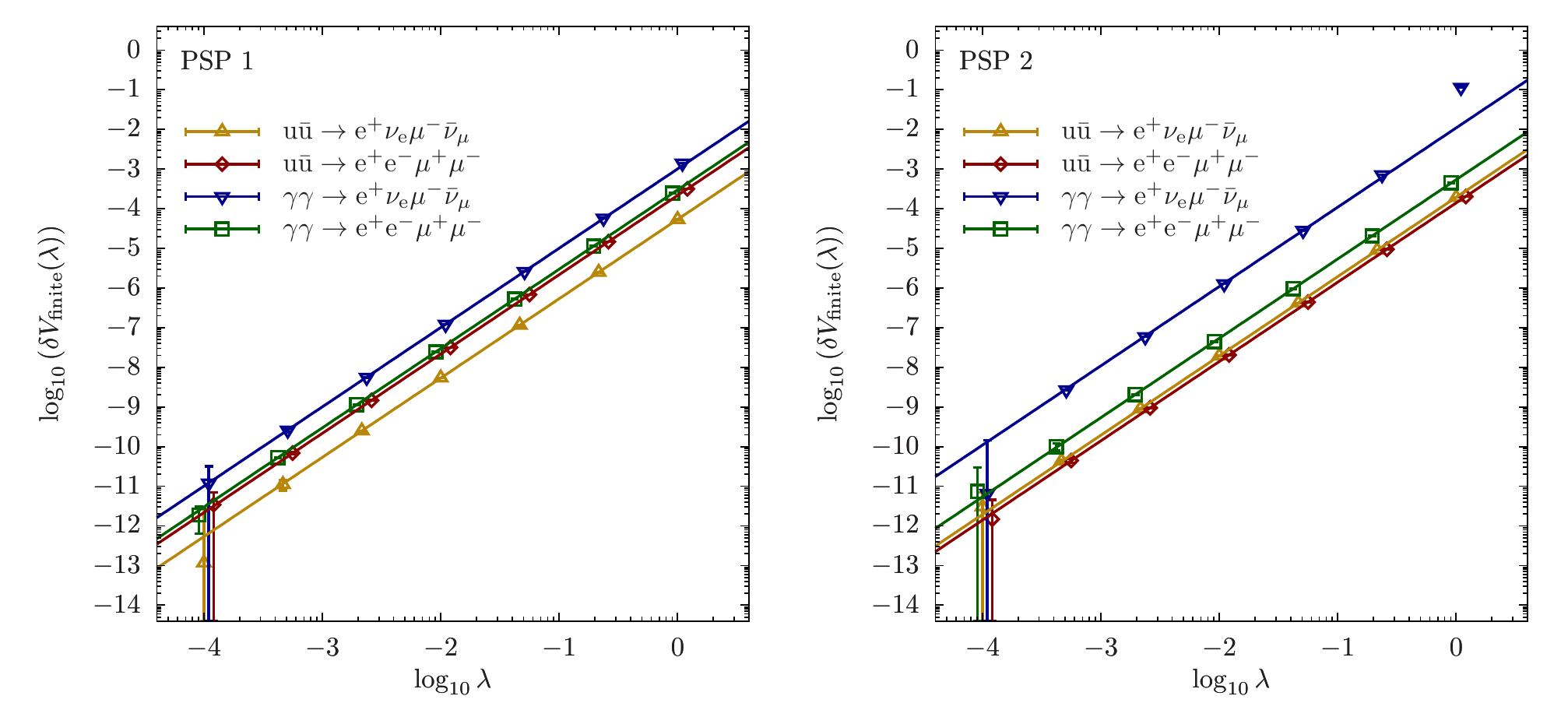}
\caption{Relative differences of $V_{\rm finite}(\lambda)$ between \textsc{MadLoop} and the
  average of \textsc{OpenLoops} and \textsc{Recola}, $\delta V_{\rm finite}(\lambda)$,
  for the two far off-shell points PSP~1 and PSP~2,
  accompanied by lines corresponding to a quadratic $\lambda$ dependence of
  $\delta V_{\rm finite}(\lambda)$. See main text for more details.}
\label{fig:SM_ew_comparison:madloopvsRecolaOpenLoops}
\end{figure}

\begin{figure}[t!]
\centering
\includegraphics[width=1.\textwidth]{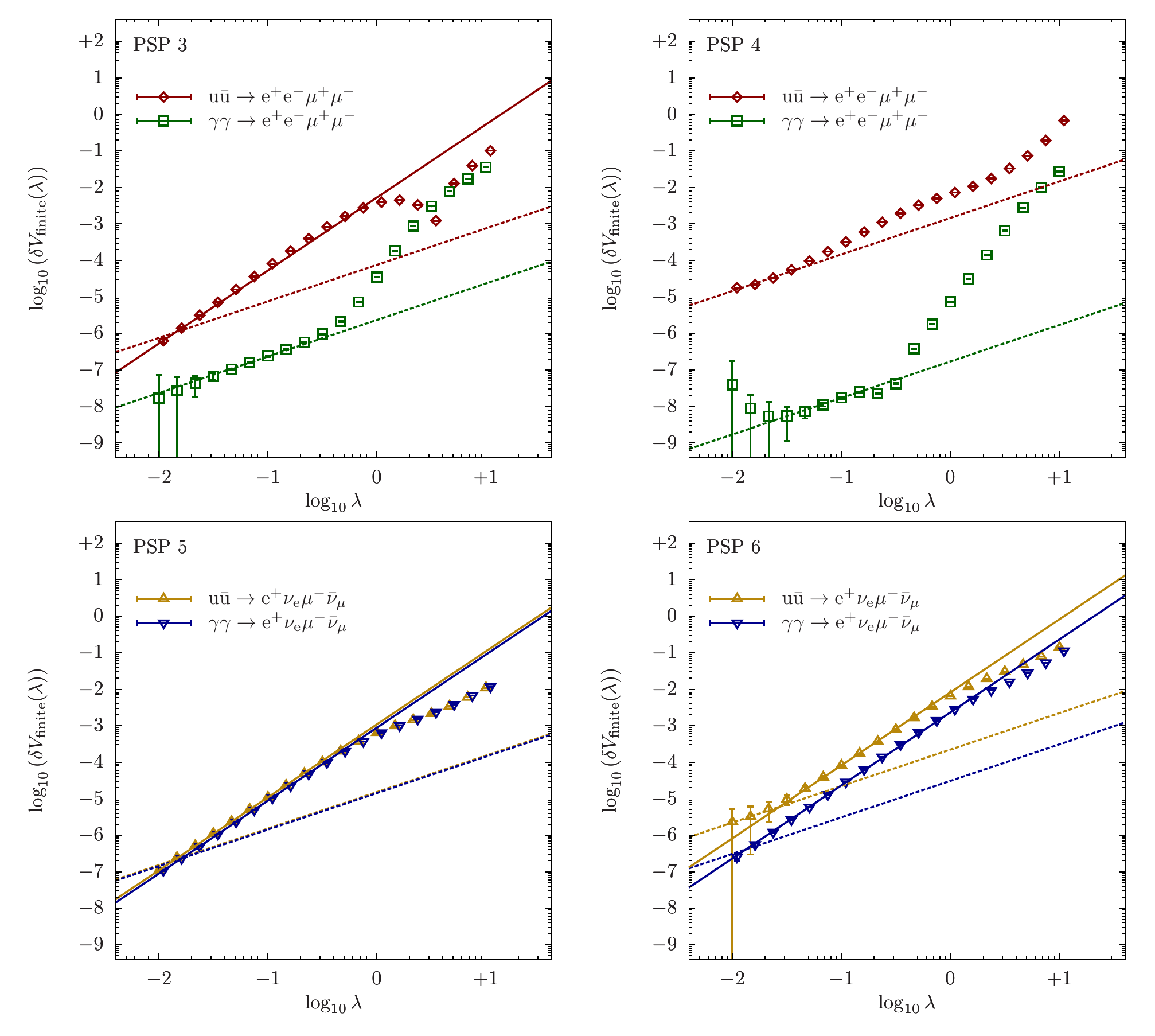}
\caption{Relative differences of $V_{\rm finite}(\lambda)$ between \textsc{MadLoop} and the
  average of \textsc{OpenLoops} and \textsc{RECOLA}, $\delta V_{\rm finite}(\lambda)$,
  for the on-shell configurations PSP~3+4 ($\rm{ZZ}$, upper plots)
  and PSP~5+6 ($\rm{WW}$, lower plots), accompanied by solid or dashed lines
  corresponding to a quadratic or linear $\lambda$ dependence of
  $\delta V_{\rm finite}(\lambda)$, respectively. See main text for more details.}
\label{fig:SM_ew_comparison:madloopvsRecolaOpenLoopsonshell}
\end{figure}

We stress that both approaches to handle the evaluation of self-energies in presence of
the complex-mass scheme
renormalization conditions are correct as they lead to identical results up to NLO EW accuracy.
It is important, however, to verify that the differences between \textsc{MadLoop} and the other involved OLPs
are indeed formally beyond the NLO EW accuracy targeted. To this end we propose here to
numerically investigate the asymptotic behaviour of $V_{\rm finite}$ in the limit of asymptotically
vanishing EW couplings.
This can be done numerically by rescaling the electromagnetic coupling constant and the related
parameters (\emph{i.e.}\ $G_\mu,\alpha, \Gamma_{\rm Z}, \Gamma_{\rm W}$) by a parameter $\lambda$,
\begin{equation}
\label{eq:SM_ew_comparison:lambdascaling}
G_{\mu} \rightarrow \lambda G_{\mu}\;,\quad
\alpha \rightarrow \lambda \alpha\;,\quad
\Gamma_{\rm Z} \rightarrow \lambda \Gamma_{\rm Z}\;,\quad
\Gamma_{\rm W} \rightarrow \lambda \Gamma_{\rm W}\;,
\end{equation}
while other model parameters such as ($M_{\rm Z},M_{\rm W},M_{\rm H},m_{\rm t}$) remain fixed.
Within this setup, we then compute the relative difference
of $V_{\rm finite}(\lambda)$ between 
\textsc{Recola}/\textsc{OpenLoops}\footnote{Given that the implementation in all involved
  OLPs except for \textsc{MadLoop} is identical, we restrict ourselves to the results of
  \textsc{Recola} and \textsc{OpenLoops} in this comparison:
  We use their point-wise average for the central values, and the difference between them
  as an estimate for the numerical precision.} and \textsc{MadLoop},
\begin{equation}
\label{eq:SM_ew_comparison:deltaVfinite}
\delta V_{\rm finite}(\lambda)=2\left|
\frac{V_{\rm finite}^{\textsc{MadLoop}}(\lambda)-V_{\rm finite}^{\textsc{Recola}/\textsc{OpenLoops}}(\lambda)}
{V_{\rm finite}^{\textsc{MadLoop}}(\lambda)+V_{\rm finite}^{\textsc{Recola}/\textsc{OpenLoops}}(\lambda)}\right|
\end{equation}
at different values of $\lambda$, as displayed in Figs.~\ref{fig:SM_ew_comparison:madloopvsRecolaOpenLoops} and \ref{fig:SM_ew_comparison:madloopvsRecolaOpenLoopsonshell}.
The expected scaling of the relative difference $\delta V_{\rm finite}(\lambda)$
with the parameter $\lambda$ is dictated by the expected contribution of the
UV mass counterterms $\delta m$, which we can schematically write as follows,
\begin{equation}
\label{eq:SM_ew_comparison:diffUVmassCT}
V_{\rm finite}^{\delta m} \propto \frac{1}{p_V^2-m_V^2 + i m_V \Gamma_V}\, \delta m\, \frac{1}{p_V^2-m_V^2 + i m_V \Gamma_V}\;.
\end{equation}
The assumption is that the UV mass counterterms $\delta m$ in \textsc{MadLoop}
only differs from the ones of the other OLPs by terms of order
$\mathcal{O}(\alpha\Gamma^2,\alpha^2\Gamma)$.
We can substitute this functional form of the difference in
Eqs.~\eqref{eq:SM_ew_comparison:diffUVmassCT} and \eqref{eq:SM_ew_comparison:deltaVfinite}
to deduce the expected dependence of $\delta V_{\rm finite}(\lambda)$
on the scaling parameter $\lambda$ for $\lambda \rightarrow 0$,
\begin{equation}
\label{eq:SM_ew_comparison:scaling}
\delta V_{\rm finite}(\lambda) = \lambda^\kappa \delta V_{\rm finite}^{(\kappa)} + \mathcal{O}(\lambda^{\kappa+1})\;,
\end{equation}
with $\kappa=1$ for exactly on-shell kinematic configurations ($p_V^2=m_V^2$) and $\kappa=2$ otherwise.
The coefficients $\delta V_{\rm finite}^{(\kappa)}$ are by definition independent of $\lambda$.

Figure~\ref{fig:SM_ew_comparison:madloopvsRecolaOpenLoops} shows the results obtained for $\delta V_{\rm finite}(\lambda)$
at progressively smaller values of $\lambda$ for the two off-shell points PSP~1+2.
Figure~\ref{fig:SM_ew_comparison:madloopvsRecolaOpenLoopsonshell} deals with the
on-shell points PSP~3+4 for $\rm{ZZ}$ production and PSP~5+6 for $\rm{WW}$ production, respectively.
The chosen $\lambda$ values are identical for all partonic processes in each frame, but slighly
shifted horizontally in Figs.~\ref{fig:SM_ew_comparison:madloopvsRecolaOpenLoops} and \ref{fig:SM_ew_comparison:madloopvsRecolaOpenLoopsonshell}
to improve readability.
The error bars on the relative difference $\delta V_{\rm finite}(\lambda)$ are extracted from the
internal numerical estimate of \textsc{MadLoop} on the one side,
and as the difference between \textsc{OpenLoops} and \textsc{Recola}, which should be entirely
due to numerical inaccuracies, on the other side,
resulting in an overall error estimate on $\delta V_{\rm finite}(\lambda)$ itself.
The numerical investigation of the on-shell phase-space points is particularly challenging
because the propagators involved in both the numerator and denominator of
Eq.~\eqref{eq:SM_ew_comparison:deltaVfinite} diverge when $\lambda \rightarrow 0$
in a way that crucially depends on the numerical accuracy at which $p_V^2=m_V^2$ is realised.
Moreover, the self-energy evaluation can also be numerically less stable in that regime (in the case of the
$\rm{W}$ boson \emph{e.g.}\ it involves terms of order $\mathcal{O}(\alpha \log(\Gamma))$).
The above-mentioned estimates of the numerical stability point towards 
$\log_{10}\lambda\gtrsim-4$ in the off-shell case and $\log_{10}\lambda\gtrsim-2$ in the on-shell case
being appropriate ranges for studying $\delta V_{\rm finite}(\lambda)$.

Figures~\ref{fig:SM_ew_comparison:madloopvsRecolaOpenLoops} and \ref{fig:SM_ew_comparison:madloopvsRecolaOpenLoopsonshell}
demonstrate that $\delta V_{\rm finite}(\lambda)$ between \textsc{MadLoop} and \textsc{OpenLoops} indeed follow
the expected scaling of Eq.~\eqref{eq:SM_ew_comparison:scaling}
(at some phase-space points with even greater values of $\kappa$),
confirming that they stand beyond the target NLO EW accuracy in \emph{all} kinematic regimes
and even beyond NNLO EW accuracy in the off-shell regions. 
We stress that within the complex-mass scheme, virtual matrix elements of NLO EW corrections always have
contributions beyond $\mathcal{O}(\alpha)$, and there is \emph{a priori} no argument for
deciding which incomplete set of beyond NLO contributions better approximates the all-order
result. The difference in the finite part of the virtual matrix element between
\textsc{MadLoop} and the other OLPs should therefore be considered, if anything, as a
systematic theoretical uncertainty, similarly to renormalization scale variations in QCD.
We note, however, that these small differences in the virtual matrix element do not lead
to resolvable effects in any of the differential observables presented in the context
of this comparison (see Sec.~\ref{sec:SM_ew_comparison:differential}).

\subsubsection{Integrated cross sections}
\label{sec:SM_ew_comparison:integrated}
The comparison of amplitudes at selected phase-space points in Sec.~\ref{sec:SM_ew_comparison:pointwise}
essentially validates the matrix-element generators (apart from their correct implementation
and initialisation within the corresponding integrators).
In contrast, integrated cross sections focus on the validation of the
phase-space integrators and the performance of their implementations of IR-subtraction
techniques. Naturally, such comparisons on integrated level --- even more so at the level of
distributions (discussed in Sec.~\ref{sec:SM_ew_comparison:differential}) --- might also uncover issues in
the stable evaluation of matrix elements in certain kinematical regions. However, to keep control
of such issues, most amplitude generators introduced in Sec.~\ref{sec:SM_ew_comparison:amplitudecodes} employ internal
stability checks for each evaluated matrix element to guarantee that their numerical accuracy
is sufficiently high such that phenomenological results are not affected by possible numerical
instabilities.

In this spirit, the results obtained from \textsc{Recola}, \textsc{OpenLoops}, and \textsc{GoSam} when interfaced to \textsc{Sherpa} have been combined, and are presented as a single result in the following presentation.
This is justified by the fact that all contributions apart from the virtual correction
within \textsc{Sherpa} are completely independent of the respective matrix-element generator,
and their comparison would essentially not go beyond an internal check of \textsc{Sherpa}.
Before combination, the statistical compatibility of the integrated virtual contributions
obtained with different OLPs was verified.
Consequently, only a single result labelled \textsc{Sherpa+GoSam/OpenLoops/Recola} is shown.
On the other side, it does make sense to involve integrated results obtained from different
integration frameworks that employ the same one-loop matrix elements in the comparison.

\begin{table}[t]
\setlength{\tabcolsep}{4pt}
  \centering
  \small
  \begin{tabular}{lllrrrr}
  $pp\to \rm{e}^+\rm{e}^-\mu^+\mu^-$ & \multicolumn{1}{c}{$\sigma^{\mathrm{LO}}$} & \multicolumn{1}{c}{$\sigma^{\mathrm{NLO}}_{\mathrm{EW}}$} & \multicolumn{2}{c}{$\Delta\sigma^{\rm{LO}}$} & \multicolumn{2}{c}{$\Delta\sigma_{\mathrm{EW}}^{\rm{NLO}}$} \\
  & \multicolumn{1}{c}{[fb]} & \multicolumn{1}{c}{[fb]} & [$\sigma$] & [\textperthousand] & [$\sigma$] & [\textperthousand] \\[2ex]
  {average}                              & $11.49675[8]$  & $10.88697[15]$ &  &  &  & \\[2ex]
  \textsc{MCBB+Recola}                   & $11.49648[12]$ & $10.88669[22]$ & $-2.9$ & $-0.02$ & $-1.7$ & $-0.03$ \\
  \textsc{Munich+OpenLoops}              & $11.49702[11]$ & $10.88720[25]$ & $+3.2$ & $+0.02$ & $+1.2$ & $+0.02$ \\
  \textsc{MoCaNLO+Recola}                & $11.49666[26]$ & $10.88734[56]$ & $-0.3$ & $-0.01$ & $+0.7$ & $+0.03$ \\
  \textsc{Sherpa+GoSam/OpenLoops/Recola} & $11.49670[34]$ & $10.88737[77]$ & $-0.1$ & $-0.00$ & $+0.5$ & $+0.04$ \\
  \textsc{MadGraph5\_\-aMC@NLO+MadLoop}  & $11.4956[22]$  & $10.8860[63]$  & $-0.5$ & $-0.10$ & $-0.1$ & $-0.09$ \\
  \end{tabular}
  \vspace*{2ex}
  
  \caption{Comparison of integrated cross sections for hadronic $\rm{e}^+\rm{e}^-\mu^+\mu^-$ (off-shell $ZZ$) production,
  obtained with different integration frameworks.}
  \label{tab:SM_ew_comparison:integrated_ZZ}
\end{table}
\begin{table}[t]
\setlength{\tabcolsep}{4pt}
  \centering
  \small
  \begin{tabular}{lllrrrr}
    $pp\to \rm{e}^+\nu_{\rm{e}} \mu^-\bar\nu_\mu$ & \multicolumn{1}{c}{$\sigma^{\mathrm{LO}}$} & \multicolumn{1}{c}{$\sigma^{\mathrm{NLO}}_{\mathrm{EW}}$} & \multicolumn{2}{c}{$\Delta\sigma^{\rm{LO}}$} & \multicolumn{2}{c}{$\Delta\sigma_{\mathrm{EW}}^{\rm{NLO}}$} \\
    & \multicolumn{1}{c}{[fb]} & \multicolumn{1}{c}{[fb]} & [$\sigma$] & [\textperthousand] & [$\sigma$] & [\textperthousand] \\[2ex]
    {average}                              & $448.5414[31]$ & $438.1902[56]$ &  &  & & \\[2ex]
    \textsc{Munich+OpenLoops}              & $448.5468[45]$ & $438.1920[75]$ & $+1.6$ & $+0.01$ & $+0.4$ & $+0.00$ \\
    \textsc{MoCaNLO+Recola}                & $448.538[10]$  & $438.193[13]$  & $-0.4$ & $-0.01$ & $+0.2$ & $+0.01$ \\
    \textsc{Sherpa+GoSam/OpenLoops/Recola} & $448.5364[46]$ & $438.186[11]$  & $-1.4$ & $-0.01$ & $-0.4$ & $-0.01$ \\
    \textsc{MadGraph5\_\-aMC@NLO}          & $448.541[40]$  & $438.113[70]$  & $-0.0$ & $-0.00$ & $-1.1$ & $-0.18$ \\
  \end{tabular}
  \vspace*{2ex}

\caption{Comparison of integrated cross sections for hadronic $\rm{e}^+\nu_{\rm{e}}\mu^-\bar{\nu}_\mu$ (off-shell $WW$) production,
  obtained with different integration frameworks.}
  \label{tab:SM_ew_comparison:integrated_WW}
\end{table}
We apply the input parameters and phase-space cuts specified in Sec.~\ref{sec:SM_ew_comparison:setup}.
The integrated results for off-shell $\rm{ZZ}$ production, which also involve the result of
Refs.~\cite{Biedermann:2016lvg,Biedermann:2017yoi}\footnote{We note again that those results, denoted
  here as \textsc{MCBB+Recola}, have been converted from the DIS to the $\overline{\rm{MS}}$
  factorization scheme. Besides, we recall that 
  in Ref.~\cite{Biedermann:2016lvg} a different treatment of the electromagnetic coupling $\alpha$
  was chosen for real photon radiation. These effects have been compensated for,
  which explains the numerical difference to the cross sections presented here.
  The \textsc{Sherpa+Recola} results from Ref.~\cite{Biedermann:2017yoi}
  have not been used here, but were generated anew for the present comparison.},
are collected in Table~\ref{tab:SM_ew_comparison:integrated_ZZ},
and those for off-shell $\rm{WW}$ production in Table~\ref{tab:SM_ew_comparison:integrated_WW}.
For each of the applied frameworks we state integrated cross sections
at LO~($\sigma^{\mathrm{LO}}$) and NLO EW~($\sigma^{\mathrm{NLO}}_{\mathrm{EW}}$) accuracy
with the corresponding statistical errors $\delta\sigma^{\mathrm{LO}}$ and
$\delta\sigma^{\mathrm{NLO}}_{\mathrm{EW}}$ indicated in square brackets,
referring to the last stated digits of the respective results.
Moreover, we provide a weighted average,
defined from the individual cross sections $\sigma^x_i$ and
their standard deviations $\delta\sigma^x_i$ via
\begin{equation}
  \label{eq:SM_ew_comparison:averagingprocedure}
  \bar\sigma^x=(\delta\bar\sigma^x)^2\,\sum_i\dfrac{\sigma^x_i}{(\delta\sigma^x_i)^2}\;,\qquad
  \delta\bar\sigma^x=\left(\sum_i\dfrac{1}{(\delta\sigma^x_i)^2}\right)^{-\frac12}\;\qquad
  x=\textnormal{LO, NLO EW}\;.
\end{equation}
The deviations for both LO~($\Delta\sigma^{\rm{LO}}$) and NLO EW~($\Delta\sigma_{\mathrm{EW}}^{\rm{NLO}}$)
cross sections with respect to the averaged cross sections are quantified both in terms of
standard deviations~$[\sigma]$\footnote{To compensate for the fact that this average depends
on each single calculation $i$, this comparison is performed in units of 
\begin{equation}
  \sqrt{(\delta\sigma^x_i)^2-(\delta\bar\sigma^x)^2},
\end{equation}
which corresponds to comparing
  each single calculation with the average of the remaining ones, with result and standard deviation
  of this average defined according to Eq.~\eqref{eq:SM_ew_comparison:averagingprocedure}.}
and as relative deviations in permille~[\textperthousand]
in order to validate both the statistical agreement and the level of precision on which this agreement
could be achieved.

For both processes we find good statistical agreement between all compared codes on the sub-permille level.
For \textsc{MCBB+Recola} (only for off-shell $\rm{ZZ}$), \textsc{Munich+OpenLoops}, \textsc{MoCaNLO+Recola}, and \textsc{Sherpa+GoSam/OpenLoops/Recola}
we are able to validate integrated results even on the remarkable level of better than a
tenth of a permille throughout, which corresponds to a relative agreement of the integrated
EW corrections at least at the level of a permille.

\subsubsection{Differential cross sections}
\label{sec:SM_ew_comparison:differential}
In order to validate the different calculations not only in
the resonance regions that dominate the integrated cross sections discussed in
the previous section, we perform also a comparison at the level of differential
distributions.
We have selected (pseudo-)observables that investigate different regions of
phase-space with certain peculiarities of the respective EW corrections:
invariant-mass distributions involving possible resonances exhibit large
corrections of QED type, which are driven by photon radiation off the
involved leptons; high-energy tails of invariant-mass and 
transverse-momentum distributions undergo large negative corrections due
to EW Sudakov logarithms, which are genuine weak effects and
dominated by the virtual contributions; finally, we also investigate
azimuthal-angle distributions as examples for observables whose EW
corrections are only weakly phase-space dependent.

\begin{figure}[p]
  \includegraphics[width=1.\textwidth]{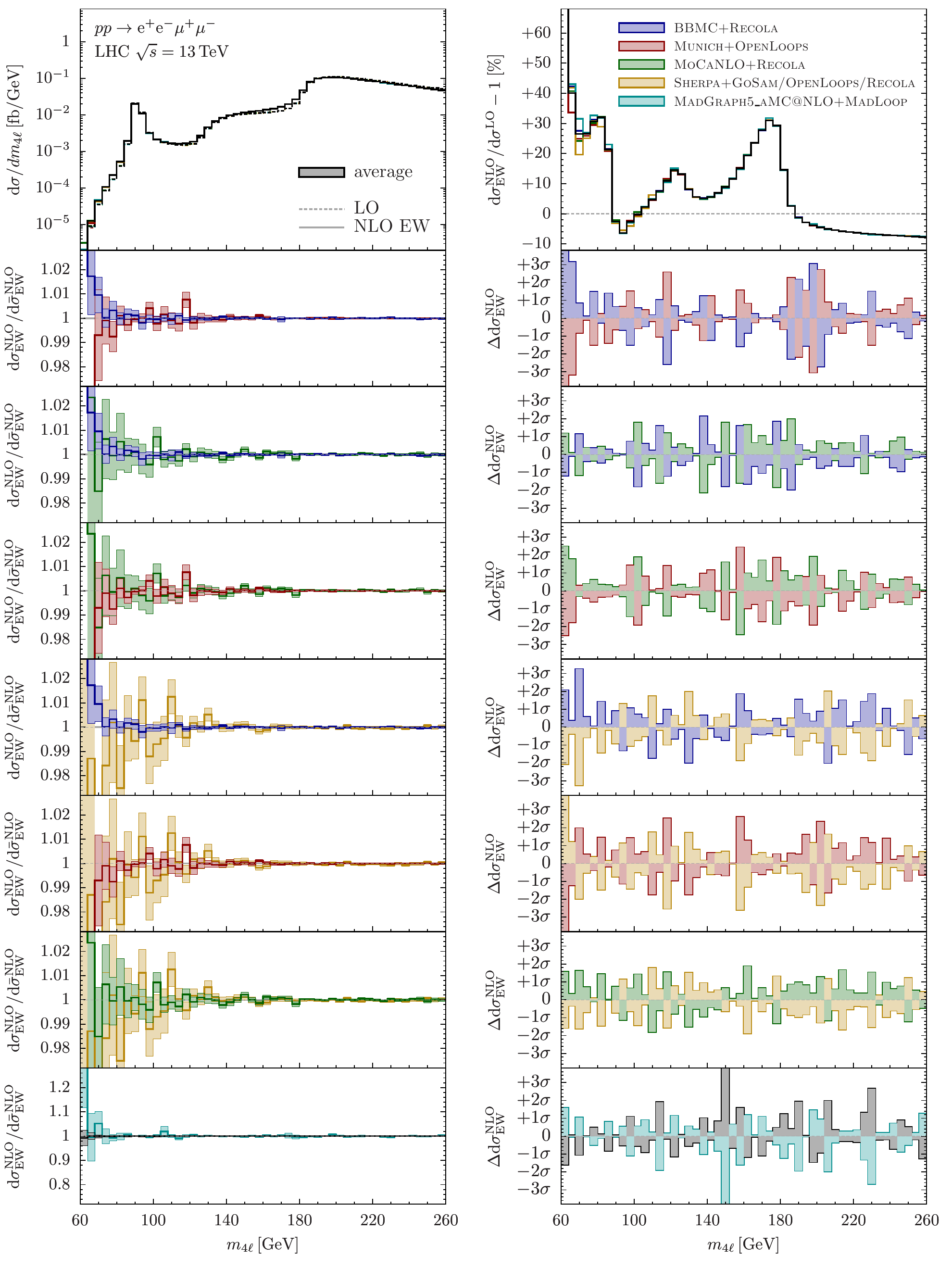}\\[-4ex]
  \caption{Technical comparison of NLO EW corrections to the distribution
    in the invariant mass of the $4\ell$ system (resonance/threshold region)
    for hadronic $\rm{e}^+\rm{e}^-\mu^+\mu^-$ (off-shell $ZZ$) production.
    See main text of Sec.~\ref{sec:SM_ew_comparison:differential} for details.}
  \label{fig:SM_ew_comparison:ZZ_m_4l_res}
\end{figure}

\begin{figure}[p]
  \includegraphics[width=1.\textwidth]{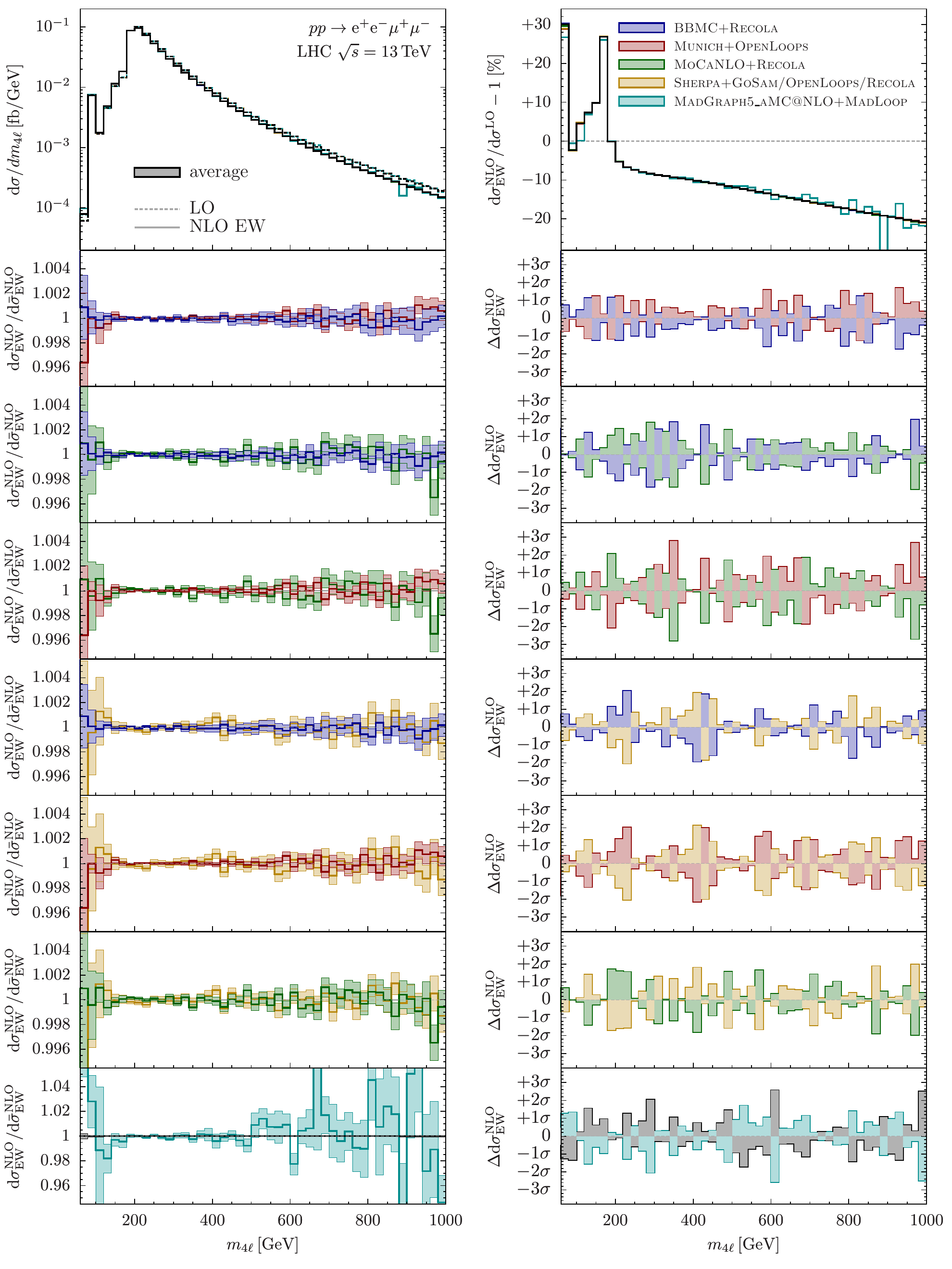}\\[-4ex]
  \caption{Technical comparison of NLO EW corrections to the distribution
    in the invariant mass of the $4\ell$ system (high-energy region)
    for hadronic $\rm{e}^+\rm{e}^-\mu^+\mu^-$ (off-shell $ZZ$) production.
    See main text of Sec.~\ref{sec:SM_ew_comparison:differential} for details.}
   \label{fig:SM_ew_comparison:ZZ_m_4l_high}
\end{figure}

\begin{figure}[p]
  \includegraphics[width=1.\textwidth]{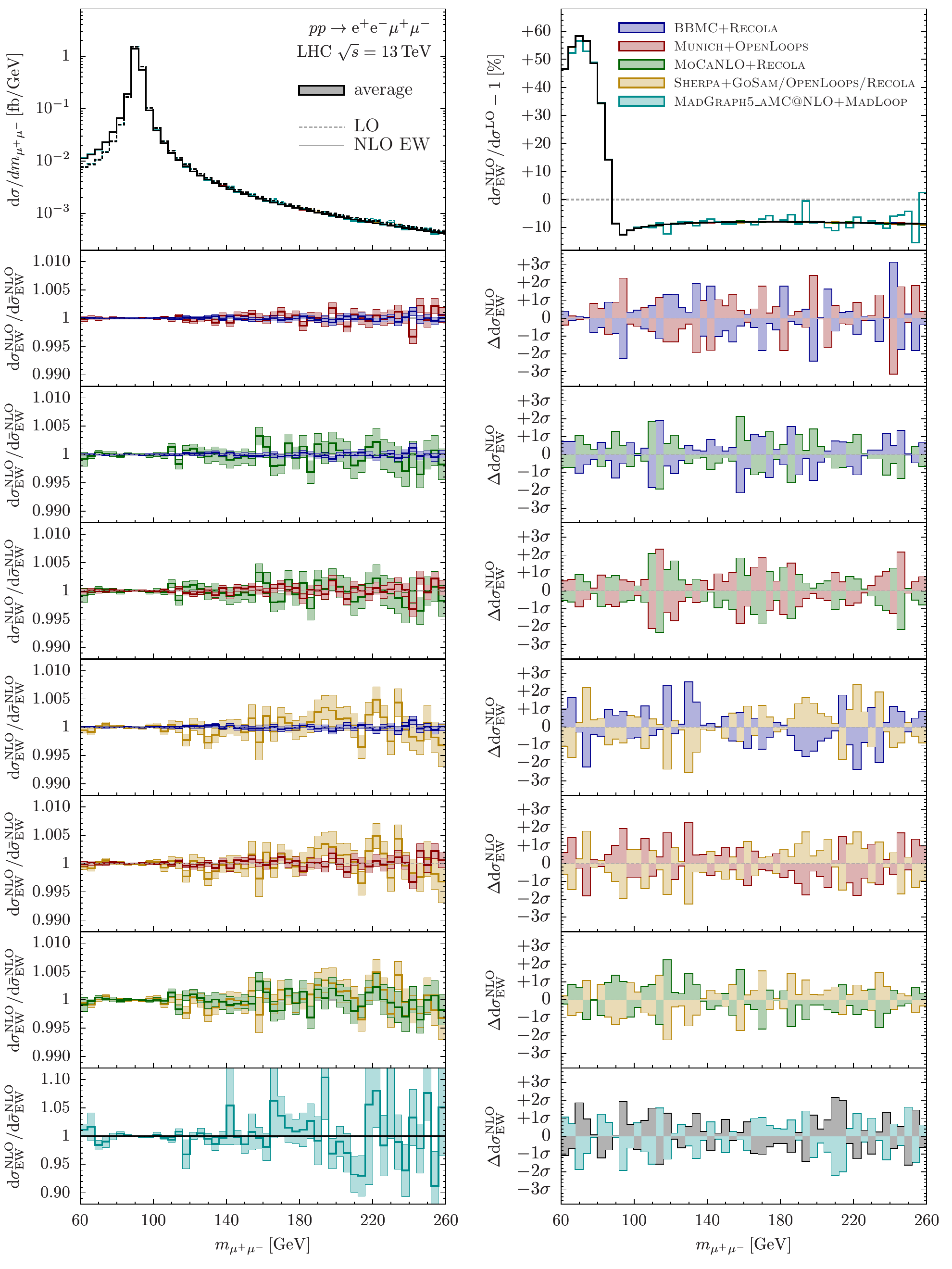}\\[-4ex]
  \caption{Technical comparison of NLO EW corrections to the distribution
    in the invariant mass of the $\mu^+\mu^-$ pair
    for hadronic $\rm{e}^+\rm{e}^-\mu^+\mu^-$ (off-shell $ZZ$) production.
    See main text of Sec.~\ref{sec:SM_ew_comparison:differential} for details.}
  \label{fig:SM_ew_comparison:ZZ_m_mmx_res}
\end{figure}

\begin{figure}[p]
  \includegraphics[width=1.\textwidth]{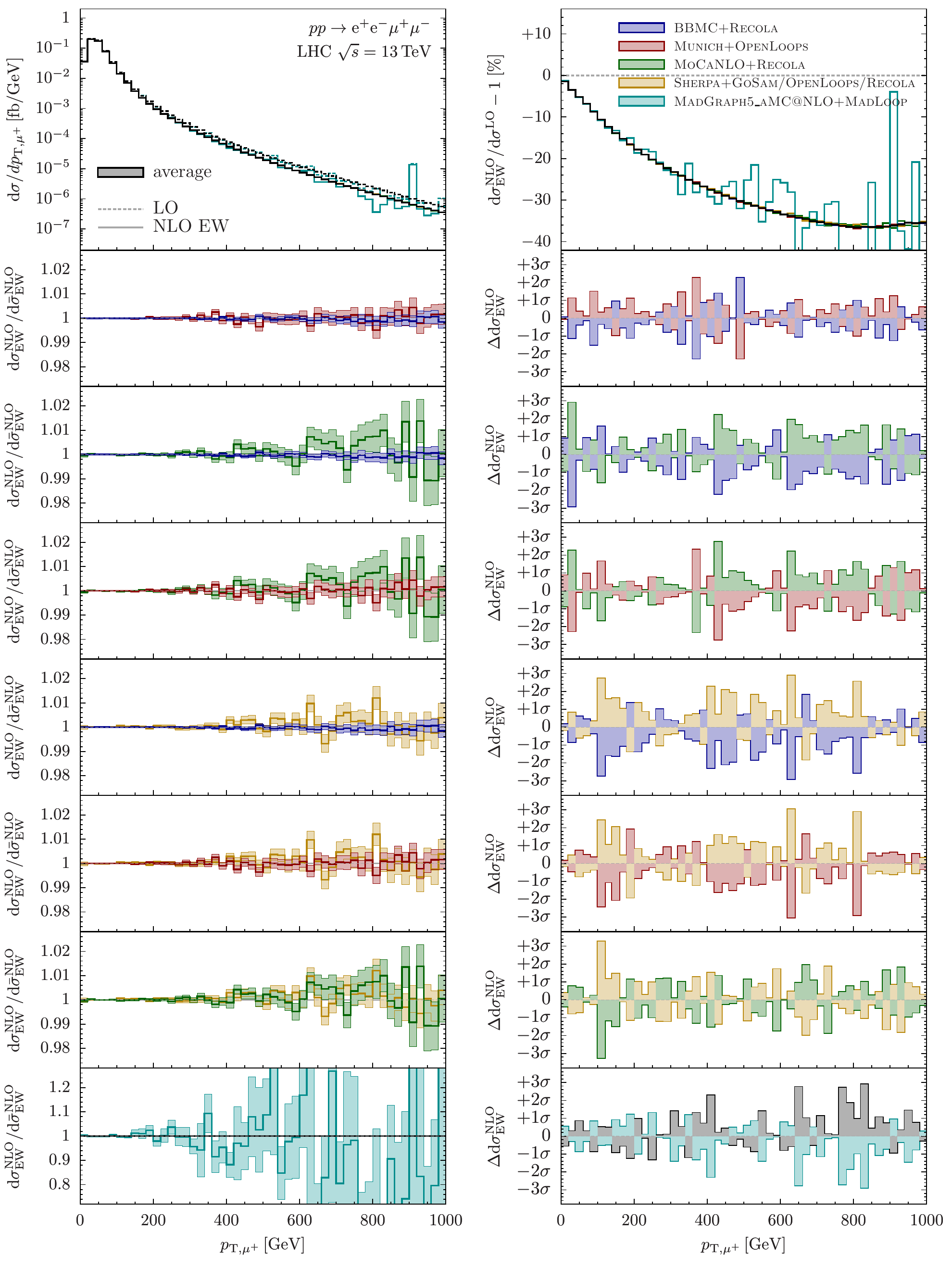}\\[-4ex]
  \caption{Technical comparison of NLO EW corrections to the distribution
    in the transverse momentum of the anti-muon, $p_{\mathrm{T,\mu^+}}$,
    for hadronic $\rm{e}^+\rm{e}^-\mu^+\mu^-$ (off-shell $ZZ$) production.
    See main text of Sec.~\ref{sec:SM_ew_comparison:differential} for details.}
  \label{fig:SM_ew_comparison:ZZ_pT_mx_20}
\end{figure}

\begin{figure}[p]
  \includegraphics[width=1.\textwidth]{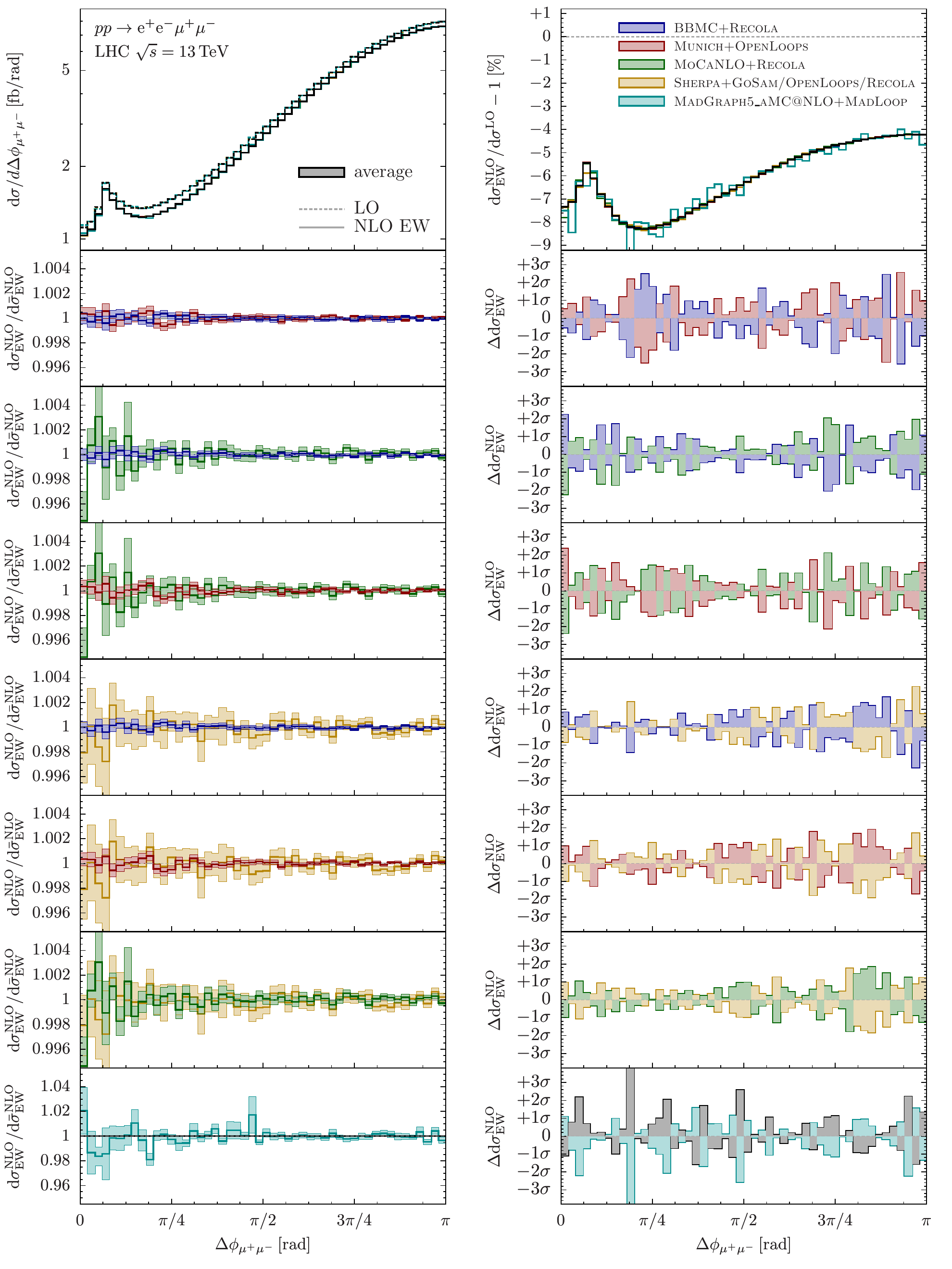}\\[-4ex]
  \caption{Technical comparison of NLO EW corrections to the distribution
    in the azimuthal angle between the anti-muon and the electron, $\Delta\phi_{\mu^+\mu^-}$,
    for hadronic $\rm{e}^+\rm{e}^-\mu^+\mu^-$ (off-shell $ZZ$) production.
    See main text of Sec.~\ref{sec:SM_ew_comparison:differential} for details.}
  \label{fig:SM_ew_comparison:ZZ_dphi_mmx}
\end{figure}
We start with a selection of distributions for off-shell $\rm{ZZ}$ production,
which are shown in
Figs.~\mbox{\ref{fig:SM_ew_comparison:ZZ_m_4l_res}--\ref{fig:SM_ew_comparison:ZZ_dphi_mmx}}.
All plots are organised as follows: The top--left frame depicts the absolute predictions
to give an idea of the overall size of the cross section in the respective region of
phase space, both at LO and NLO EW accuracy; we note here that we consider
observables whose cross sections cover a range of up
to five orders of magnitude.
The top--right frame shows the relative EW corrections, which vary between
about $-40\%$ in the high-$p_{\rm{T}}$ tails of lepton transverse-momentum distributions
and about $+60\%$ in invariant-mass distributions below resonance regions.
In both frames, all five individual predictions for the respective differential distribution are plotted, namely
\textsc{MCBB+Recola} (blue),
\textsc{Munich+OpenLoops} (red),
\textsc{MoCaNLO+Recola} (green),
\textsc{Sherpa+GoSam/OpenLoops/Recola} (yellow), and
\textsc{MadGraph5\_\-aMC@NLO+MadLoop} (cyan).
The average of the five calculations, defined binwise in analogy to the integrated
cross sections in the previous section, is plotted on top (black).
Absolute predictions are shown both at LO (dashed) and NLO EW (solid) accuracy.
To illustrate the relative as well as the statistical agreement,
we perform pairwise comparisons between the individual predictions in the frames below:
On the left-hand side, in each frame the ratio of two predictions $i$ and $j$ over the overall
average is plotted, accompanied by their respective $1\sigma$ error bands,
\emph{i.e.}\ 
\begin{equation}
(d\sigma_{i/j}\pm\delta d\sigma_{i/j})/d\bar\sigma\;.
\end{equation}
On the right-hand side, we show the deviation between the same two individual predictions $i$ and $j$
in units of their combined standard deviation,
\begin{equation}
  \delta d\sigma_{ij}=\sqrt{\delta d\sigma^2_i+\delta d\sigma^2_j}\;,
\end{equation}
\emph{i.e.}\ the plots are by construction symmetric. Since the available results from
     \textsc{MCBB+Recola}, \textsc{Munich+OpenLoops}, \textsc{MoCaNLO+Recola}, and
     \textsc{Sherpa+GoSam/OpenLoops/Recola} are typically about one order
     of magnitude more precise than the results provided by \textsc{Mad\-Graph5\_aMC@NLO+MadLoop},
     we start with their pairwise comparisons
in frames 2--7. 
In frame 8, we compare their average to
the results from \textsc{MadGraph5\_\-aMC@NLO+MadLoop}.

Figures~\mbox{\ref{fig:SM_ew_comparison:ZZ_m_4l_res} and \ref{fig:SM_ew_comparison:ZZ_m_4l_high}} depict
the invariant mass of the 4-lepton system, which
exhibits the \mbox{$Z\to4\ell$} peak around $m_{\rm{Z}}$ and the $\rm{ZZ}$ production threshold at
\mbox{$m_{4\ell}\gtrsim2m_{\rm{Z}}$}, with the corresponding photon-radiation dominated EW corrections
around the resonances, detailed in Fig.~\ref{fig:SM_ew_comparison:ZZ_m_4l_res}. In the high-energy region, shown in Fig.~\ref{fig:SM_ew_comparison:ZZ_m_4l_high},
the NLO EW corrections are driven by EW Sudakov
logarithms, and grow negatively to $-20\%$ at \mbox{$m_{4\ell}\approx1\,\rm{TeV}$}.
In the full range, all five predictions agree very well on a statistical level. For
the upper four calculations, this corresponds to a permille-level agreement throughout, apart
from the region below the $Z\to4\ell$
resonance where the cross section is suppressed by about four orders of magnitude compared to the
peak region.

Figure~\ref{fig:SM_ew_comparison:ZZ_m_mmx_res} shows the invariant mass of the $\mu^+\mu^-$
system, which is typically produced via an intermediate $\rm{Z}$ boson. The chosen
range \mbox{$m_{\mu^+\mu^-}\in[60;260]\,\rm{GeV}$} details both the resonance region with its QED-dominated
NLO EW corrections and an intermediate phase-space region where the pair is far off shell, but
the dominance of EW Sudakov logarithms does not set in yet.
We find agreement on a statistical level between all individual predictions, which again
corresponds to a permille level agreement in the full considered range
for \textsc{MCBB+Recola}, \textsc{Munich+OpenLoops}, \textsc{MoCaNLO+Recola} and \textsc{Sherpa+GoSam/OpenLoops/Recola}.

As an example of a Sudakov-dominated observable, Fig.~\ref{fig:SM_ew_comparison:ZZ_pT_mx_20}
depicts the distribution in the transverse momentum of the produced anti-muon, with the
typical large negative NLO EW corrections of $-35\%$ at \mbox{$p_{\rm{T},\mu^+}\approx1\,\rm{TeV}$}.
Statistically, the agreement is again good for all predictions in the full range.
However, in the highly suppressed tail, a relative numerical precision better than
a percent is only achieved by the most precise predictions.

The azimuthal-angle distribution of the $\mu^+\mu^-$ pair in
Fig.~\ref{fig:SM_ew_comparison:ZZ_dphi_mmx} exhibits a preference for back-to-back
configurations, while the collinear region dominated by \mbox{$\gamma^\ast\to\mu^+\mu^-$} splittings is
suppressed by the applied lepton--lepton isolation cut. Overall, the distribution varies by less than
one order of magnitude, and also the NLO EW corrections do not undergo particular enhancements and
show variations of less than $5\%$. All individual predictions agree again very well on a statistical
level, which corresponds to permille level agreement \textsc{Munich+OpenLoops}, \textsc{MCBB+Recola},
\textsc{MoCaNLO+Recola} and \textsc{Sherpa+GoSam/OpenLoops/Recola}, and at least percent level
agreement for the \textsc{MadGraph5\_\-aMC@NLO+MadLoop} prediction.

\begin{figure}[t!]
  \includegraphics[width=1.\textwidth]{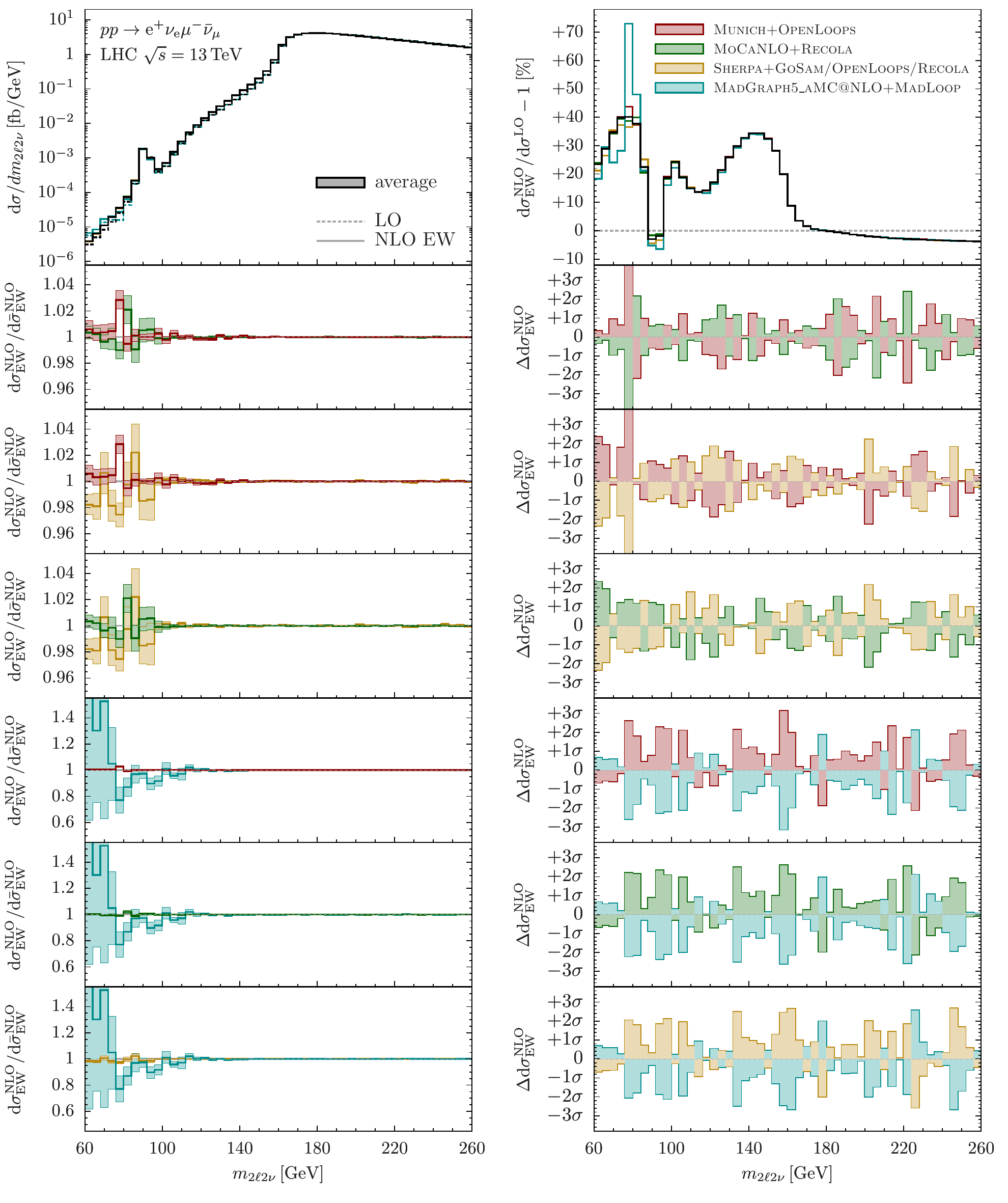}\\[-4ex]
  \caption{Technical comparison of NLO EW corrections to the distribution
    in the invariant mass of the $2\ell2\nu$ system (pseudo-observable)
    for hadronic $\rm{e}^+\nu_{\rm{e}}\mu^-\bar{\nu}_\mu$ (off-shell $WW$) production.
    See main text of Sec.~\ref{sec:SM_ew_comparison:differential} for details.}
  \label{fig:SM_ew_comparison:WW_m_2l2n_res}
\end{figure}

\begin{figure}[t!]
  \includegraphics[width=1.\textwidth]{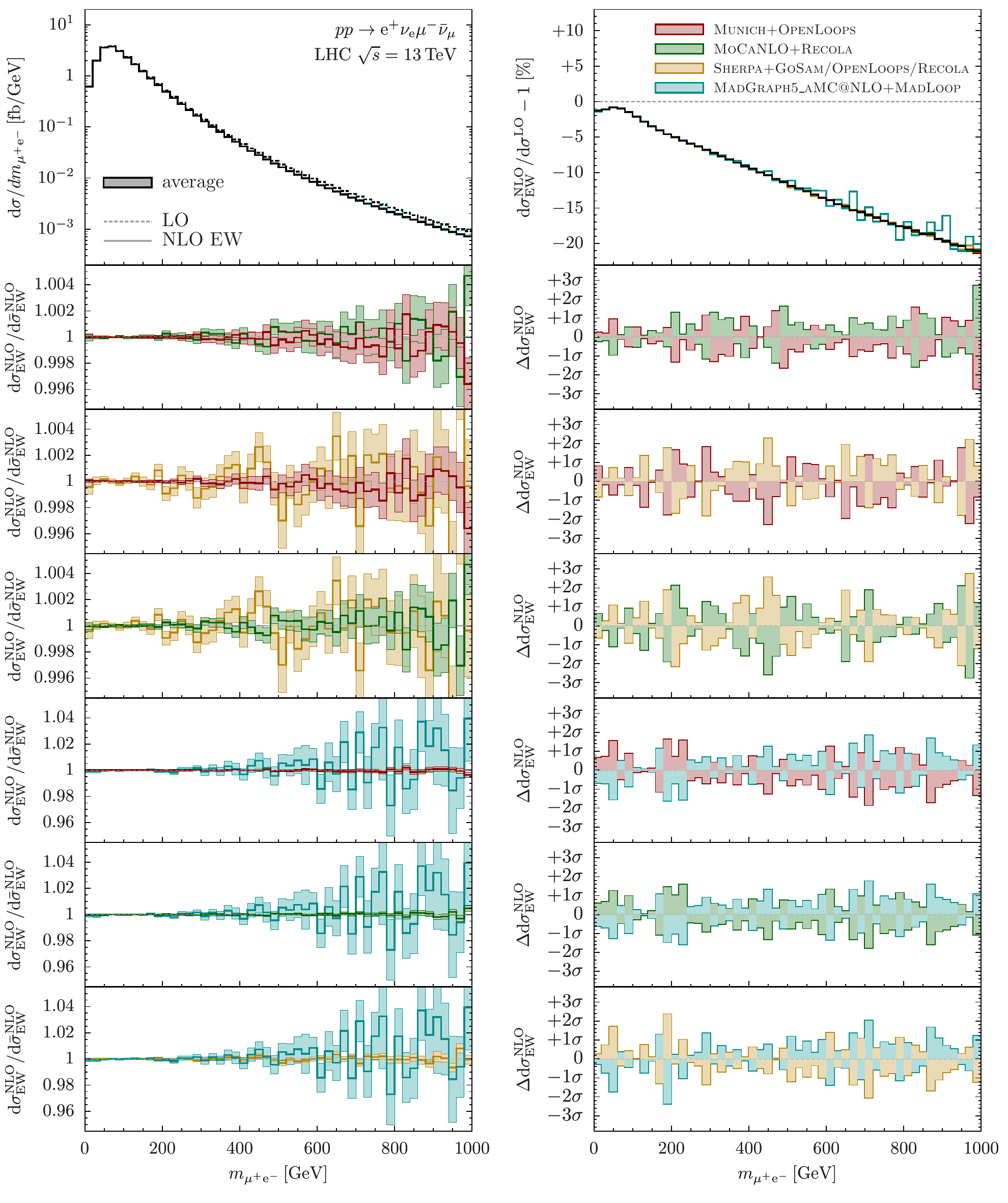}\\[-4ex]
  \caption{Technical comparison of NLO EW corrections to the distribution
    in the invariant mass of the $\mu^+e^-$ system,
    for hadronic $\rm{e}^+\nu_{\rm{e}}\mu^-\bar{\nu}_\mu$ (off-shell $WW$) production.
    See main text of Sec.~\ref{sec:SM_ew_comparison:differential} for details.}
  \label{fig:SM_ew_comparison:WW_m_llx_high}
\end{figure}

\begin{figure}[t!]
  \includegraphics[width=1.\textwidth]{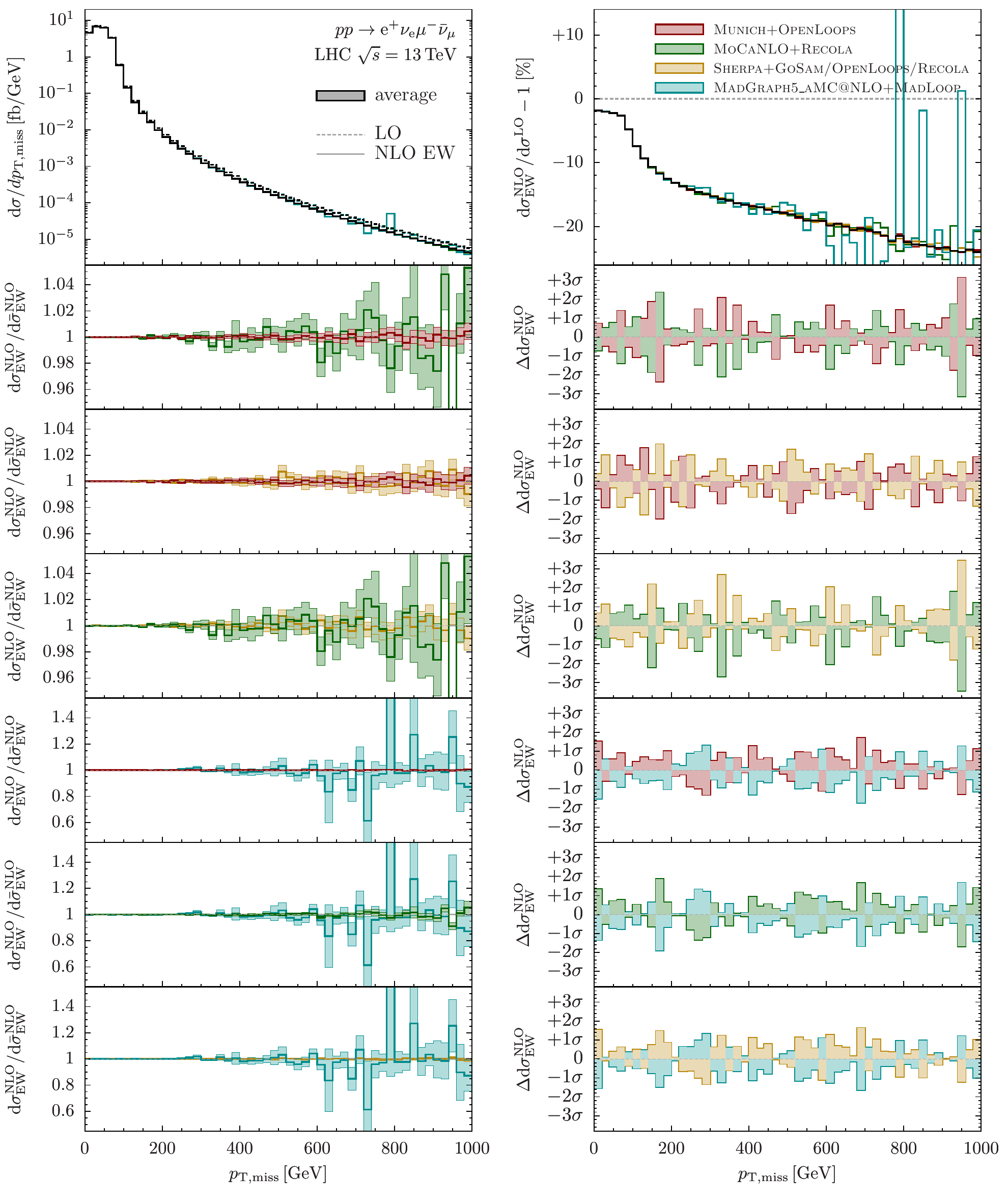}\\[-4ex]
  \caption{Technical comparison of NLO EW corrections to the distribution
    in the missing transverse momentum,
    for hadronic $\rm{e}^+\nu_{\rm{e}}\mu^-\bar{\nu}_\mu$ (off-shell $WW$) production.
    See main text of Sec.~\ref{sec:SM_ew_comparison:differential} for details.}
  \label{fig:SM_ew_comparison:WW_pT_2n_20}
\end{figure}

\begin{figure}[t!]
  \includegraphics[width=1.\textwidth]{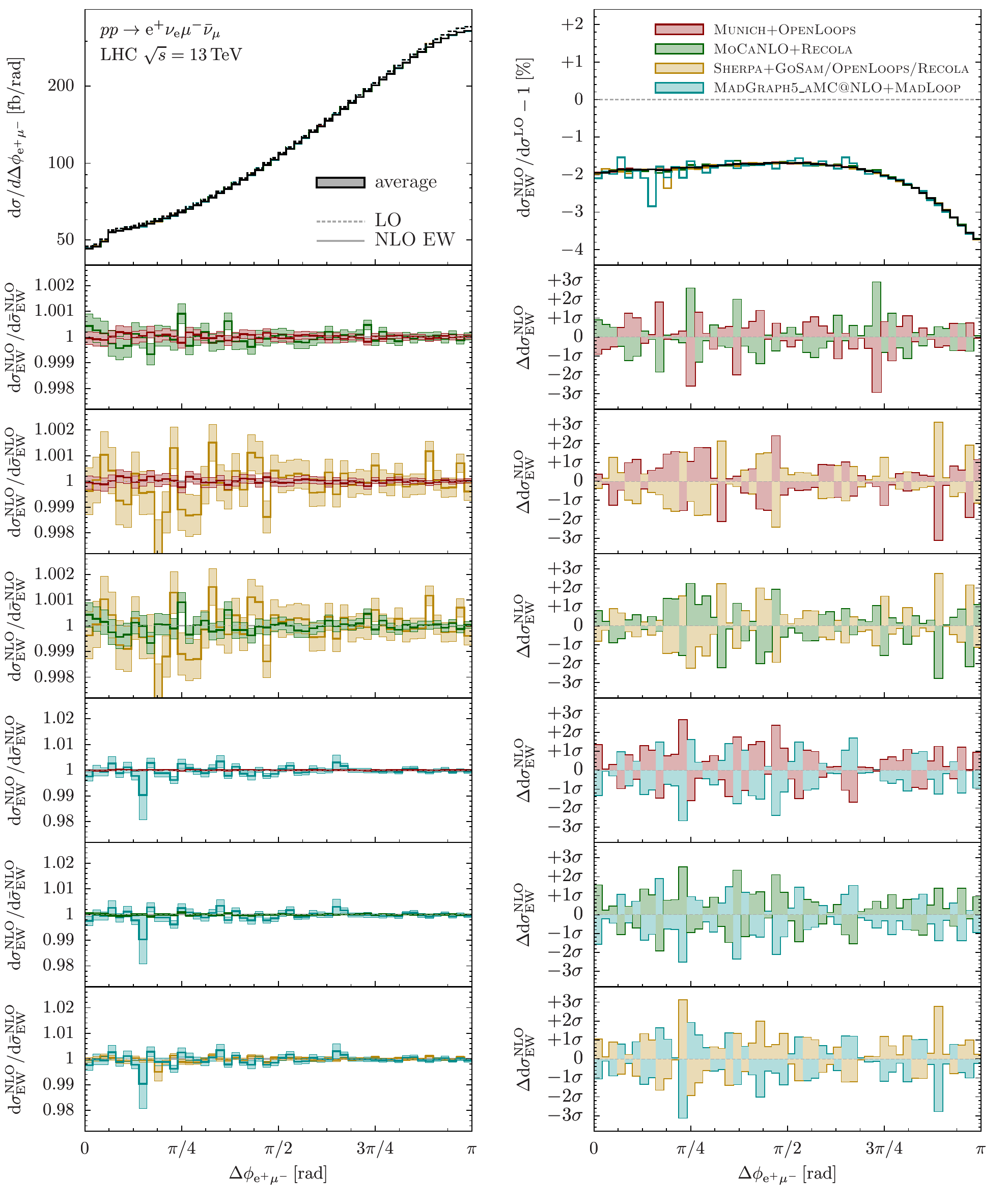}\\[-4ex]
  \caption{Technical comparison of NLO EW corrections to the distribution
    in the azimuthal angle between the anti-muon and the electron, $\Delta\phi_{\mu^+e^-}$,
    for hadronic $\rm{e}^+\nu_{\rm{e}}\mu^-\bar{\nu}_\mu$ (off-shell $WW$) production.
    See main text of Sec.~\ref{sec:SM_ew_comparison:differential} for details.}
  \label{fig:SM_ew_comparison:WW_dphi_llx}
\end{figure}

For off-shell $\rm{WW}$ production, we present a selection of distributions 
in Figs.~\ref{fig:SM_ew_comparison:WW_m_2l2n_res}--\ref{fig:SM_ew_comparison:WW_dphi_llx}.
The plots are organised analogously to those of $\rm{ZZ}$ production, but since with
\textsc{Munich+OpenLoops} (red),
\textsc{MoCaNLO+Recola} (green),
\textsc{Sherpa+GoSam/OpenLoops/Recola} (yellow), and
\textsc{MadGraph5\_\-aMC@NLO+MadLoop} (cyan)
there are only four predictions,
pairwise comparisons are shown for all of them in frames 2--7.
As for $\rm{ZZ}$ production, we find very good statistical agreement among the
individual predictions for all considered observables.

We start with the pseudo-observable $m_{2\ell2\nu}$, the invariant mass of the
$2\ell2\nu$ system, in Fig.~\ref{fig:SM_ew_comparison:WW_m_2l2n_res}. Above the $\rm{WW}$ threshold
where the bulk of the cross section is, all individual predictions agree on the permille level,
whereas the highly suppressed region below the threshold is by far less precise.

The distribution in the invariant mass of the $\mu^+e^-$ system, shown in
Fig.~\ref{fig:SM_ew_comparison:WW_m_llx_high}, exhibits the typical Sudakov behaviour with
negative NLO EW corrections of $\approx-20\%$ at \mbox{$m_{\mu^+ \rm{e}^-}\approx1\,\rm{TeV}$}.
The statistical agreement corresponds to a permille-level agreement for
\textsc{Munich+OpenLoops}, \textsc{MoCaNLO+Recola} and \textsc{Sherpa+GoSam/OpenLoops/Recola}
in the full range. under consideration. 

The missing transverse-momentum distribution, depicted in Fig.~\ref{fig:SM_ew_comparison:WW_pT_2n_20},
falls particularly steeply since the double-resonant $\rm{WW}$ configuration is suppressed on
the Born phase space for \mbox{$p_{\rm{T,miss}}\gtrsim M_{\rm{W}}$}. Correspondingly, the precision is
restricted to the level of few percent in the tail of the distribution.

Finally, the distribution in the azimuthal-angle between the anti-muon $\mu^+$ and the electron $\rm{e}^-$
exhibits no strong phase-space dependence, and the NLO EW corrections are quite flat.
For \textsc{Munich+OpenLoops}, \textsc{MoCaNLO+Recola} and \textsc{Sherpa+GoSam/OpenLoops/Recola},
the good statistical agreement corresponds to a permille agreement, while
deviations are well below one percent throughout for
\textsc{MadGraph5\_\-aMC@NLO+MadLoop} .

Beside the kinematic distributions discussed above, a number of observables was investigated in this
comparison. The result is a reasonable statistical agreement for all distributions under
consideration. Moreover, the observed deviations reflect the behaviour expected from a
purely statistical distribution.

\subsection{Conclusions}
\label{sec:SM_ew_comparison:conclusions}
In this contribution we have performed a detailed comparison of calculations of NLO EW corrections
between different automated fixed-order codes. As benchmark cases we considered off-shell
$WW$ and $ZZ$ production, which offer a rich resonance structure.
This allowed us to cross-validate many technical subtleties in the different implementations,
in particular the required renormalization in the complex-mass scheme.
First, we compared results obtained with the tools
{\sc MadLoop}, {\sc Recola}, {\sc OpenLoops}, {\sc GoSam} and {\sc NLOX} at the amplitude level
and found very good agreement at NLO EW accuracy.
Second, in order to compare results at the level of integrated cross sections,
these amplitude providers have been interfaced in different combinations with the Monte Carlo
frameworks {\sc BBMC}, {\sc MoCaNLO}, {\sc Munich}, {\sc Sherpa} and {\sc MadGraph\_aMC\@NLO},
which entail the subtraction of QED singularities.
Again very good agreement has been found in various differential distributions
including resonance peaks and high-energy tails.
This study therefore strengthens our confidence in the correctness of existing and future
predictions including NLO EW corrections from the tools considered here.

\subsection*{Acknowledgements}
BB, AD, and MP acknowledge financial support by the
German Federal Ministry for Education and Research (BMBF) under
contract no.~05H15WWCA1 and the German Science Foundation (DFG) under
reference number DE 623/6-1. The work of SB and SS has received
funding from BMBF under contract 05H15MGCAA and from the European Union's
Horizon 2020 research and innovation programme as part of the Marie Sklodowska-Curie
Innovative Training Network MCnetITN3 (grant agreement no. 722104).
NG was supported by the Swiss National Science Foundation
under contract PZ00P2\_154829. HSS is supported by the ILP Labex (ANR-11-IDEX-0004-02, ANR-10-LABX-63).
The work of SQ and CR is supported by the U.S. Department of Energy under grant DE-SC0010102.



\chapter{Parton distribution functions}
\label{cha:pdf}
\def\lsim{\mathrel{\rlap{\lower4pt\hbox{\hskip1pt$\sim$}}
    \raise1pt\hbox{$<$}}}  


\section{Theoretical uncertainties and dataset dependence of parton distributions~\protect\footnote{
      S.~Forte,
      Z.~Kassabov,
      J.~Rojo,
      L.~Rottoli}{}}
\label{sec:SM_globalPDFfits}


We study theoretical uncertainties on parton distributions (PDFs) due
to missing higher order (MHO) corrections by determining the change of PDFs
when going from  next-to-leading (NLO) to next-to-next-to-leading
order (NNLO) theory.
Based on the NNPDF3.1 framework,
we  compare PDF determinations obtained from different
datasets, specifically a global, a proton-only,  and a collider-only
dataset.
We show that PDF determinations obtained from a 
wider input dataset exhibit greater perturbative stability, and thus are
likely to be affected by smaller theoretical uncertainties from MHOs.
We also show that the effect of including deuterium nuclear corrections is smaller than
that of excluding the deuterium data altogether.

\subsection{Parton distribution uncertainties}
The accurate determination of the uncertainties on parton distribution functions
(PDFs) of the proton~\cite{Gao:2017yyd}
has been the main challenge in PDF studies at the LHC.
For instance, PDFs
represent one
of the main sources of uncertainty in Higgs
physics~\cite{Andersen:2014efa,deFlorian:2016spz}; they also have a
significant impact on searches of physics beyond  the standard model
(see e.g.~\cite{Beenakker:2015rna}), and on 
precision measurements such as the determination of the $W$ boson
mass~\cite{Aaboud:2017svj}. 
Uncertainties on
the current combined PDF4LHC15 PDF set~\cite{Butterworth:2015oua} are
typically of order 3-5\% in the region covered by data, but the more
recent NNPDF3.1 PDF set~\cite{Ball:2017nwa}, which includes a wide array
of LHC data, has PDF uncertainties typically between 1\% and 3\% in
the data region (not including, in either case, the uncertainty on
$\alpha_s$).

This uncertainty --- indeed, what is usually
referred to as ``PDF uncertainty'' --- includes the propagated
uncertainty on the data used for PDF determinations, as well as
further uncertainties due to the fitting methodology, but it does not include
any theory error. Specifically, the current PDF uncertainty does not
include a contribution accounting for the fact  
that PDFs are determined using fixed-order
perturbative QCD, and thus surely one has to account for a missing
higher order uncertainty (MHOU).
As the uncertainty due to the data
and methodology keeps decreasing, this theory error
is bound to stop being
negligible, and  eventually become dominant.

So far, there has been a broad consensus that the use of the widest
possible dataset for PDF determination --- leading to so-called global
PDF fits --- is advantageous. Indeed, on the one hand, more data contain
more information and thus allow for the accurate determination of the
widest set of PDF in the most extended kinematic region. On the other
hand, the use of multiple datasets provides a cross-check on both the
theory and methodology.
One may however ask whether the use of a wider
dataset is also advantageous --- or indeed not --- in terms of
theoretical uncertainties, specifically the MHOUs.
In particular, it is be important to understand if MHOUs
are likely to be smaller with a global dataset or with a reduced
dataset.

\subsection{Missing higher order corrections and dataset dependence}

\subsubsection{Comparing NLO and NNLO PDFs}
\label{sec:SM_globalPDFfits:settings}
Quite in general, there is currently no way to reliably
estimate the MHOUs. However, when several perturbative orders are known, we
may at least study the behaviour of the perturbative
expansion.
Assuming reasonable convergence, the shift between, say,
known NLO and NNLO results then provides a reasonable estimate of the
MHOU on the NLO result. 

We will thus address the problem of MHOU on
PDFs by studying the way PDF change from NLO to NNLO.
To this purpose, we have produced PDF determinations based essentially on
the same dataset as in the NNPDF3.1 global
analysis~\cite{Ball:2017nwa}.
The only difference is that, while in NNPDF3.1 some jet data for which NNLO
corrections were not yet available were treated approximately, here
we only include both at NLO and NNLO jet data for which exact NNLO
theory is available (see Sect.~4.4 of Ref.~\cite{Ball:2017nwa}), and
using  the exact NNLO corrections~\cite{Currie:2016bfm}. This ensures
perturbative consistency.

\begin{figure}[t]
\begin{center}
\includegraphics[width=\textwidth]{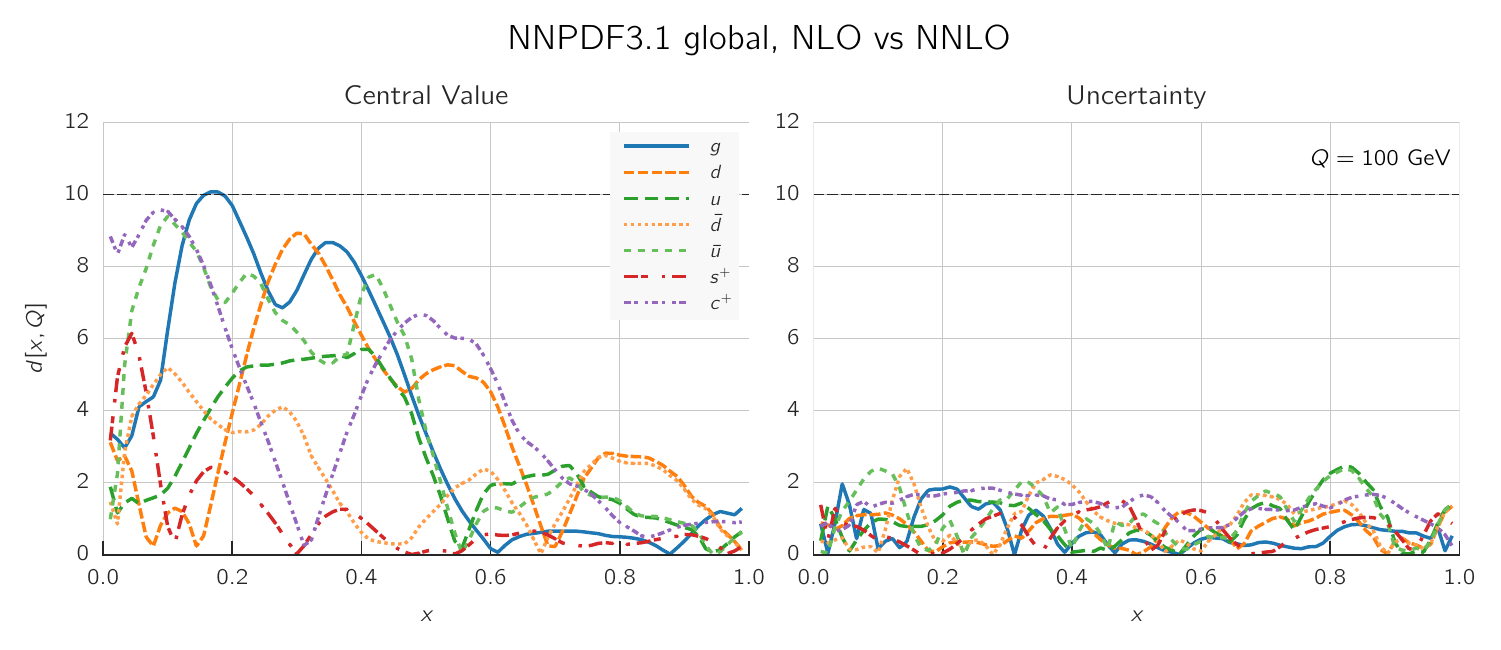}\\
\caption{\small Distance between the central values (left)
  and the uncertainties (right) of NLO and NNLO global PDFs.
}
\label{fig:SM_globalPDFfits:globdist}
\end{center}
\end{figure}

In Fig.~\ref{fig:SM_globalPDFfits:globdist}
we show the distance between PDFs determined at NLO and at NNLO.
Recall that the distance $d$ is defined as the difference in units
of standard deviation, normalized so that $d=10$ corresponds to
an one-$\sigma$ shift (see Ref.~\cite{Ball:2017nwa} and references
therein for a more detailed discussion).
It is clear from Fig.~\ref{fig:SM_globalPDFfits:globdist} that, whereas no distance between central values is
greater than one-$\sigma$ in 
the data region, several distances  (in particular for the gluon and light
quarks) are of order one-$\sigma$.
But a one-$\sigma$ distance means that NLO-NNLO shift is the same size
as the PDF uncertainty, so we conclude that, at NLO,
MHOUs on central values are comparable to  PDF uncertainties.
Distances between
uncertainties are instead of order one, {\it i.e.} comparable to a
statistical fluctuation.
This means that PDF uncertainties at NLO and
NNLO do not differ by a statistically significant amount, consistent with the
expectation that they reflect the uncertainty on the data
and methodology, and thus, by and large, do not systematically  depend on the
perturbative order.

\subsubsection{Dataset dependence}

In order to study the issue that we set out in the introduction,
namely, the dataset dependence of the MHOUs associated
to a fit of parton distributions,
we have repeated the previous PDF
determination now based on two reduced datasets.
First, we have produced
a  proton-only determination, in which we excluded
all data with nuclear and deuterium targets (specifically fixed-target
deep-inelastic and Drell-Yan production).
Then, we have produced a
 collider-only determination, in which we have excluded all
fixed-target DIS and DY data altogether.

These PDF determinations from smaller
datasets were already discussed in Ref.~\cite{Ball:2017nwa} (as well as in
previous NNPDF studies~\cite{Ball:2014uwa,Ball:2012cx})
where it was argued that, even though these
smaller datasets are in principle more consistent, the increased theoretical
reliability does not make up for the great loss in accuracy: the PDF
uncertainty increases monotonically when reducing the dataset, and
there is no evidence of inconsistency between the data entering the
global fit. With this motivation,
the use of the more global dataset for the baseline
PDF determination was advocated.
These two
PDF determinations have now been redone starting from the global 
dataset described in Sec.~\ref{sec:SM_globalPDFfits:settings}.

\begin{table}[t]
  \centering
  \begin{tabular}{ ccccc }
 \toprule
   &  NLO   &  NNLO & NLO/NNLO & $\Delta$  \\
 \midrule
global   &1.279 & 1.253 & 1.02& 1.16\\
proton  &1.248 & 1.193 & 1.05& 1.97\\
collider  &1.181 & 1.114 & 1.06& 2.07\\
\bottomrule
  \end{tabular}
  \caption{\small Value of $\chi^2$ per data point for the global,
    proton-only, and 
    collider-only fits at NLO and NNLO. The NLO/NNLO ratio, and
    difference normalized to the standard deviation (see text) are
    also given.
    Note that the total number of data points $N_{\rm dat}$
    in each of the three fits is different. 
    \label{tab:SM_globalPDFfits:chi2}
}\end{table}

The values of the total $\chi^2$ per data point for these NNPDF3.1-based
PDF determinations are collected in Table~\ref{tab:SM_globalPDFfits:chi2}. In each
case, we show the
$\chi^2$ per data point at NLO and NNLO, their ratio, and the
difference $\Delta=\frac{\chi^2_{\rm NLO}- \chi^2_{\rm
    NNLO}}{\sqrt{2 N_{\rm dat}}}$ which is a measure of the improvement
  of the $\chi^2$ in units of its standard deviation.
Note that the results in Table~\ref{tab:SM_globalPDFfits:chi2} only consider the
dataset that was included in the fit in each case, and consequently
the total number of data points $N_{\rm dat}$
    in each of the three fits is different.
Clearly, the PDF fits based on
smaller datasets lead to a better $\chi^2$, due to the greater
consistency of the dataset.
Interestingly,
however, the deterioration of the total $\chi^2$ from NNLO to NLO, as
measured both by the $\chi^2$ ratios, and the difference in units of
the standard deviation, is more
severe for the fits based on a  smaller datasets,
and thus largest in the case of
the collider-only fit.
This provides a first indication that the use
of a wider dataset may lead to greater perturbative stability.

\begin{figure}[t]
\begin{center}
\includegraphics[width=\textwidth]{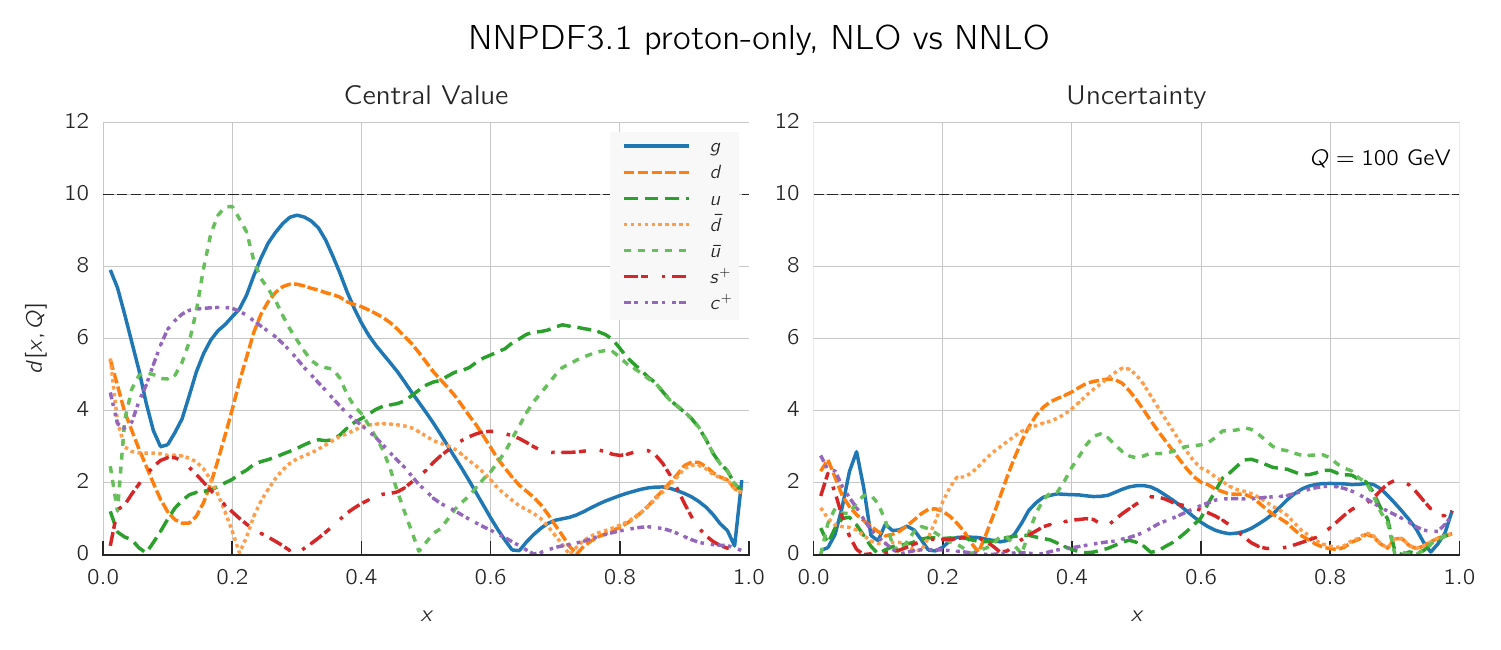}
\includegraphics[width=\textwidth]{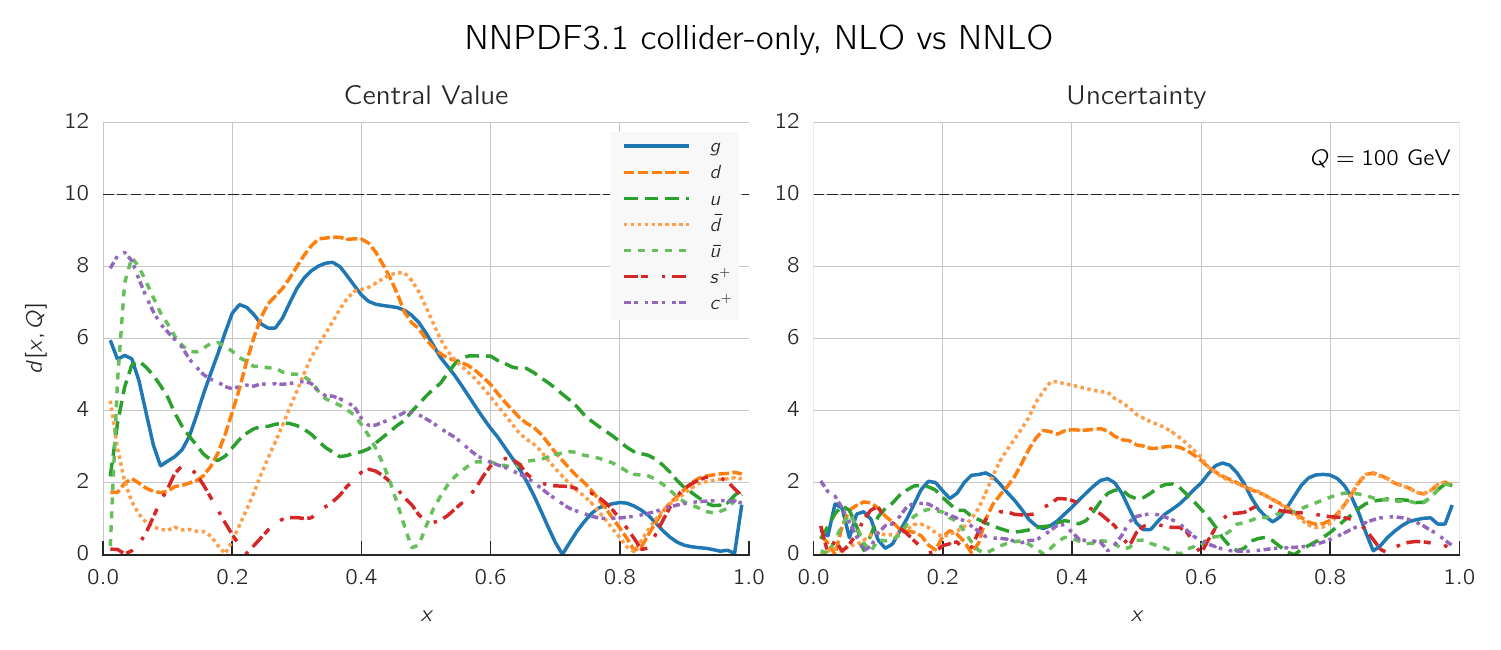}\\
 \caption{\small Same as Fig.~\ref{fig:SM_globalPDFfits:globdist}, but for the proton-only  (top)
   and collider-only (bottom) PDF determinations.
}
\label{fig:SM_globalPDFfits:pcolldist}
\end{center}
\end{figure}

To investigate this issue further, we have computed again the distance
between the NLO and NNLO fits, as shown in Fig.~\ref{fig:SM_globalPDFfits:globdist}, but now for the proton-only
and collider-only PDF sets.
Results are shown in
Fig.~\ref{fig:SM_globalPDFfits:pcolldist}.
%
It is clear that while distances which were
already sizable in the global fit (specifically for the gluon and down
quark) are still big, now also  the light quark PDFs (in particular
also up and anti-up) display sizable NLO-to-NNLO shifts.

\begin{figure}[h!]
\begin{center}
\includegraphics[width=0.48\textwidth]{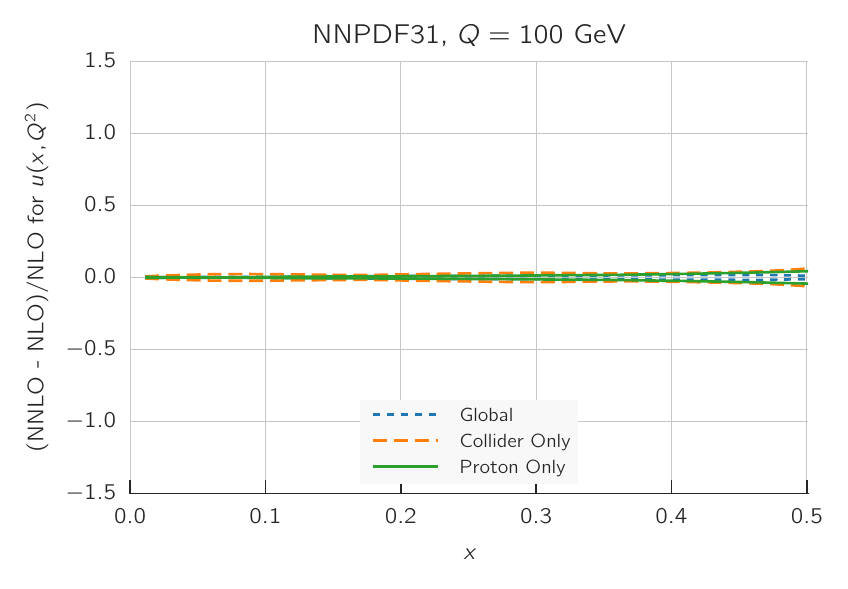}\hfill
\includegraphics[width=0.48\textwidth]{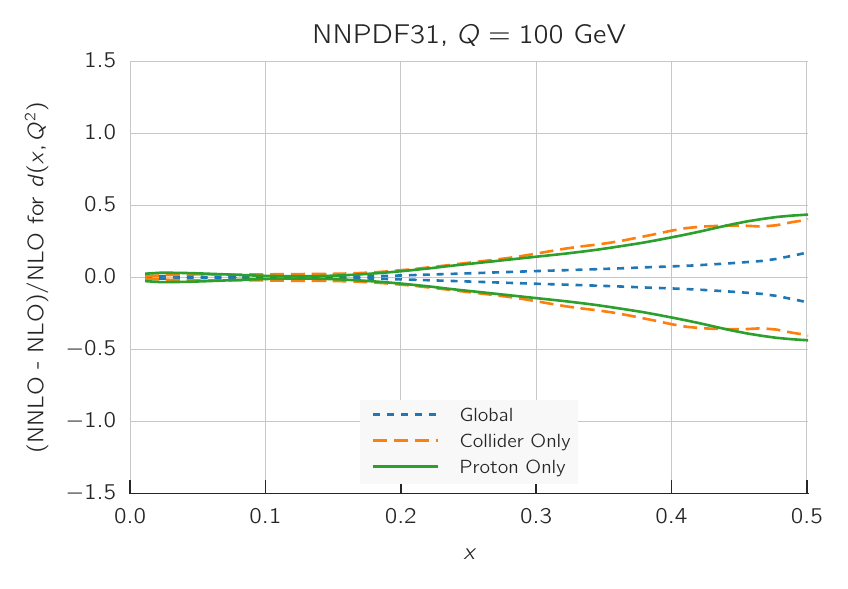}\\[1ex]
\includegraphics[width=0.48\textwidth]{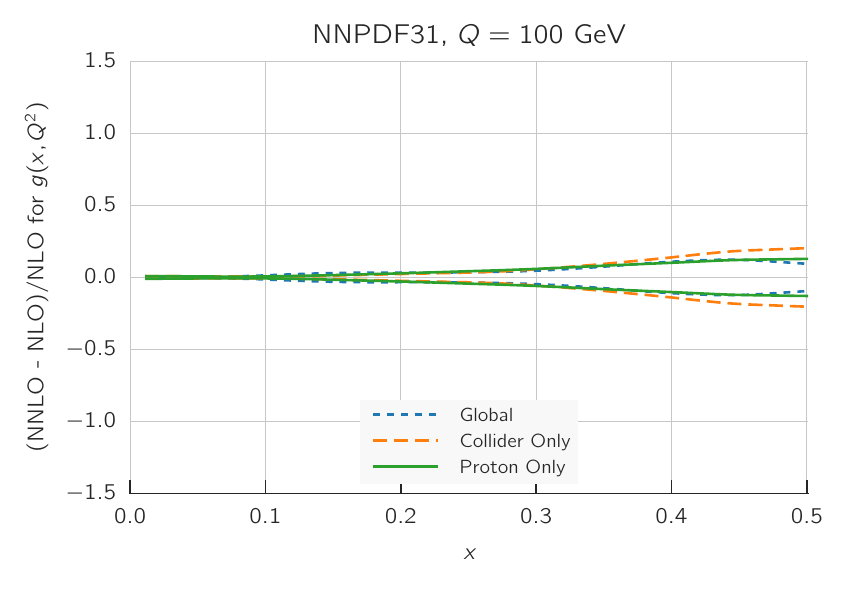}\hfill
\includegraphics[width=0.48\textwidth]{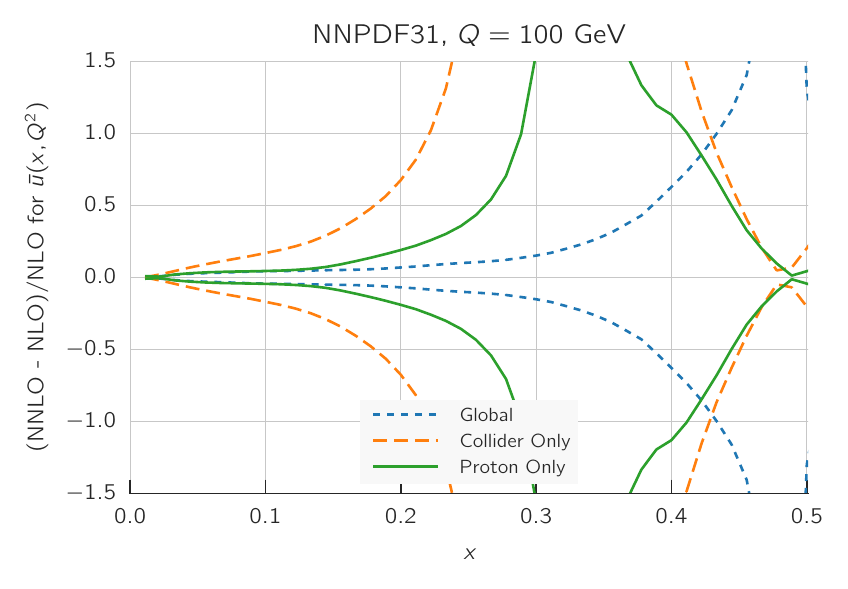}\\[1ex]
\includegraphics[width=0.48\textwidth]{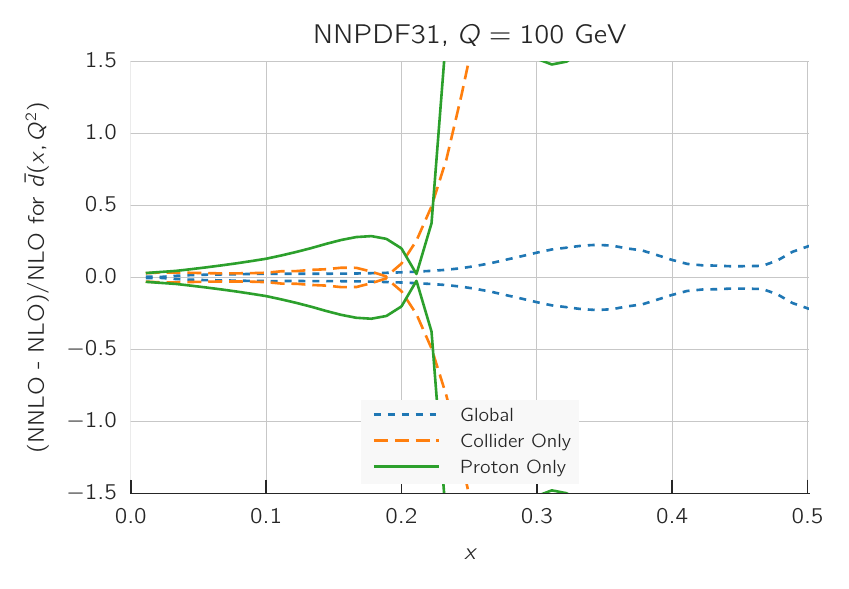}\hfill
\includegraphics[width=0.48\textwidth]{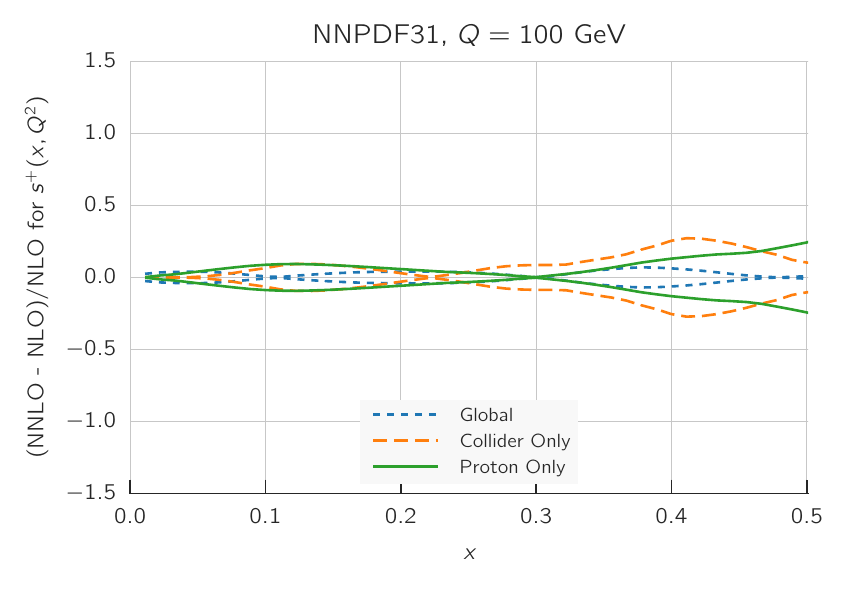}
\caption{\small The relative shift between
the central values of the
  NLO to NNLO PDFs normalized to the NLO, 
for the global, collider-only, and proton-only
  PDF determinations.
  Results are shown for, from left to right and
   from top to bottom, the
   gluon, up, down, anti-up, anti-down and
   total strange+antistrange PDFs.
   To facilitate visualization, the shifts are symmetrized about  the $x$ axis.
}
\label{fig:SM_globalPDFfits:nnshift}
\end{center}
\end{figure}

In order to achieve a fully quantitative comparison, in Fig.~\ref{fig:SM_globalPDFfits:nnshift}
we display the shift of PDF central values between the
NLO and NNLO fits, normalized to the NLO, for the
gluon and light quarks, comparing the three PDF determinations.
In order to facilitate visualization, the
shifts are symmetrized about the $x$ axis.
It is clear that while for the gluon the shift is of similar size (and
quite small) in the three
PDF determinations, for the quarks there is a uniform hierarchy: the 
smallest dataset, {\it i.e.} the collider-only PDF set, nearly always
displays the largest shifts, with very few localized exceptions.

%

A simple explanation of the greater perturbative stability of the more
global fit seen in Table~\ref{tab:SM_globalPDFfits:chi2} could be that hadron collider
processes, which have larger perturbative corrections than
deep-inelastic scattering, carry a greater weight in the fits to a
reduced dataset:  the global fit improves less because it is less consistent. 
But in this case,   we would expect the global
fit, due to its poorer consistency,
 not to show a significant improvement in PDF uncertainties, or to
 display a sizable  change in results when going
from NLO to NNLO.
Instead, the opposite is the case. As extensively discussed in
Ref.~\cite{Ball:2017nwa} the more global fit has
significantly smaller PDF uncertainty, and as shown in
Fig.~\ref{fig:SM_globalPDFfits:nnshift} it also changes less from NLO to NNLO.
Hence the global fit is both less uncertain and more perturbatively stable. 

An alternative explanation of the observed perturbative stability then
seems more likely. Namely, that it is  a
consequence of the fact that missing higher order terms for different processes
distort PDFs  randomly by pulling them in different directions.
Therefore, in a
more global dataset in which the same PDF combination is determined from
constraints by different
processes, these uncertainties tend to average out.

\subsection{Deuterium nuclear corrections}

Perhaps the main advantage of the  PDF fits based on smaller datasets is their greater consistency not
only from the experimental, but also from the theoretical point of
view.
In particular,  proton-only PDFs do not make use of any data that are
affected by the poorly known nuclear corrections.
It is then interesting to ask how the
size of nuclear corrections compares to the uncertainties that we have
discussed so far.
Whereas existing determinations of heavy nuclear
corrections are  affected by large uncertainties~\cite{Eskola:2016oht},
we may at least compare PDF determinations in which a
model of nuclear effects for deuterium is included. 

In order to isolate this effect, we have thus produced a PDF
determination in which data using heavier  nuclear targets have been
excluded (hence specifically neutrino deep-inelastic scattering data)
but data with deuterium targets (both deep-inelastic and fixed-target
Drell-Yan) are kept.
We then compare this fit either to the
proton-only fit, or to itself with deuterium  corrections included
using the MMHT14 best-fit model of Ref.~\cite{Harland-Lang:2014zoa} (with default
settings).

\begin{figure}[t]
\begin{center}
\includegraphics[width=\textwidth]{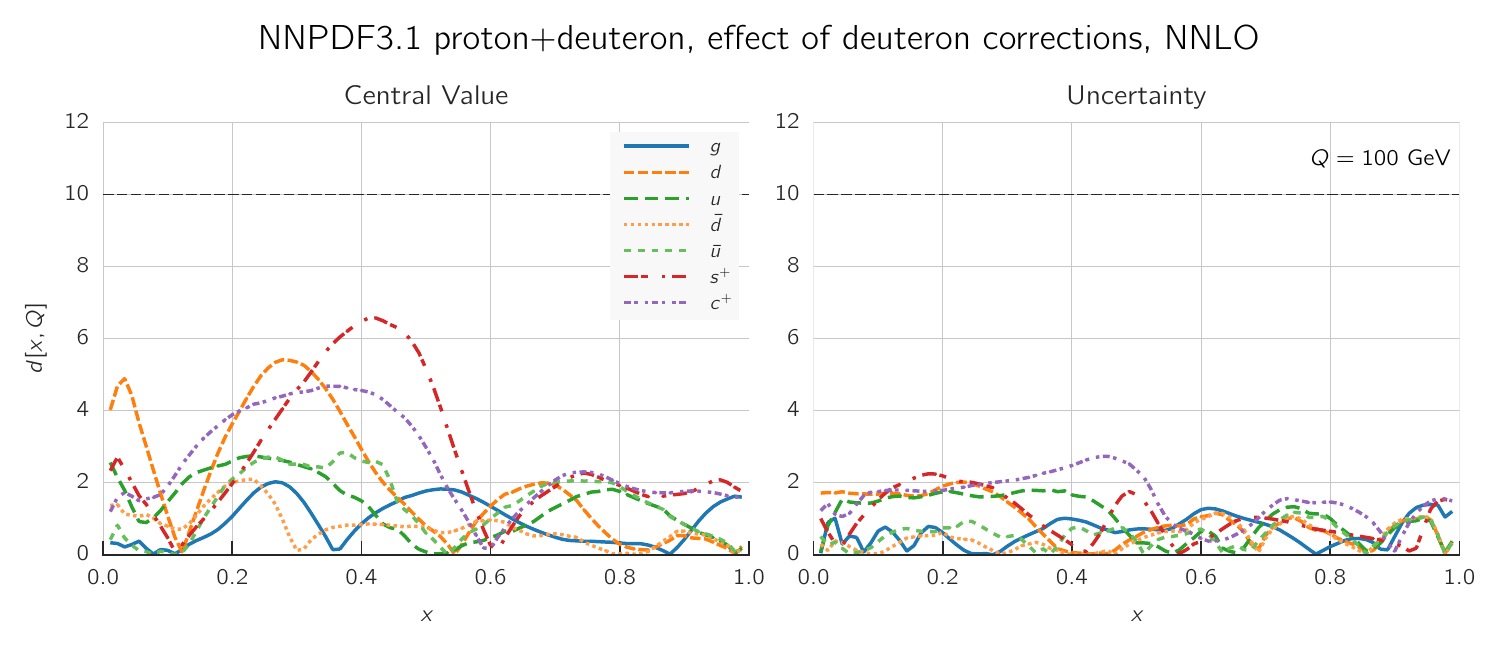}\\
\includegraphics[width=\textwidth]{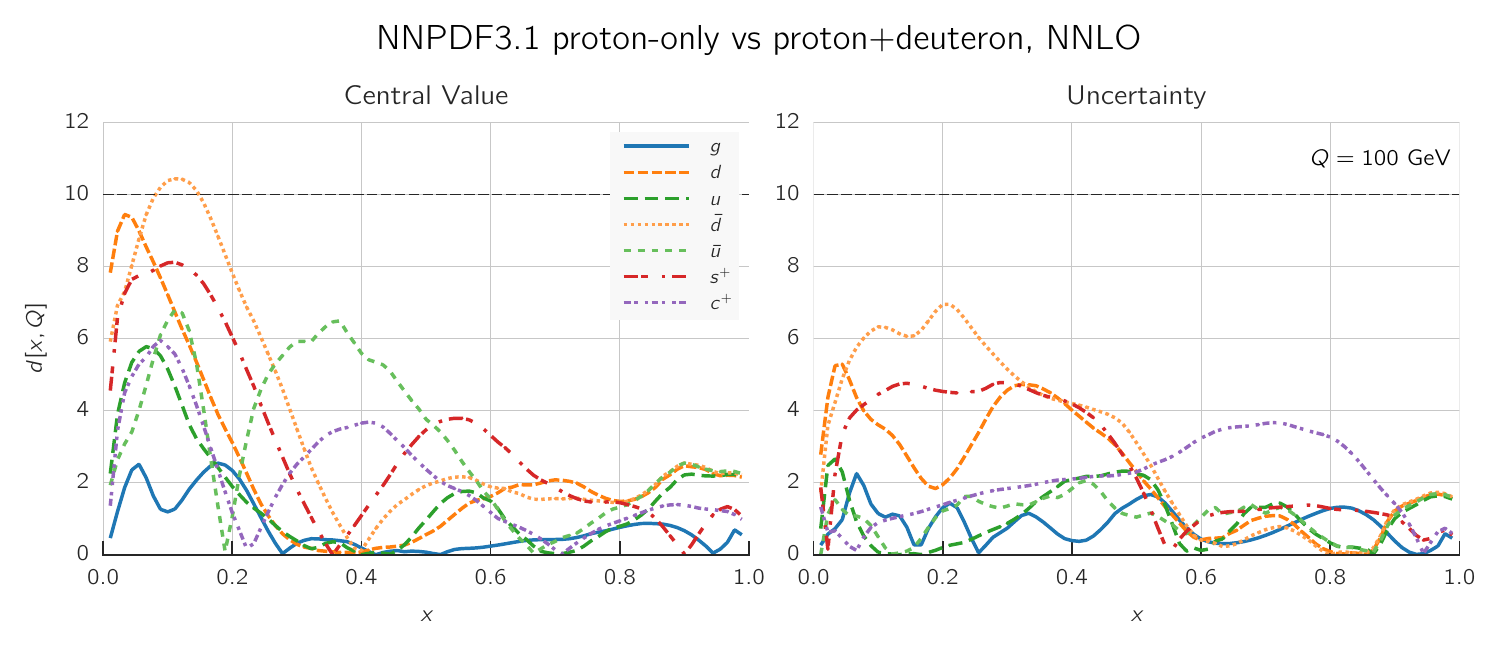}
 \caption{\small Distances between a fit with proton and deuterium target
   data and the proton-only fit (bottom) compared to the distances
   between a fit with proton and deuterium target data with and
   without linear corrections. 
}
\label{fig:SM_globalPDFfits:nucldist}
\end{center}
\end{figure}

The corresponding distances between these fits are shown in Fig.~\ref{fig:SM_globalPDFfits:nucldist}.
It is clear that the
effect of including the deuterium nuclear corrections is rather smaller than that of the deuterium
data itself --- whose impact is instead comparable to that of the NNLO
corrections.
We can thus conclude that the inclusion of deuterium data in the global PDF fit appears
to be currently advantageous, in the sense that the impact of this
data on the
PDFs is greater than the likely size of their uncertainty due to
missing nuclear corrections.
On the other hand, if these data are included,
some estimate of the associated theoretical
uncertainty, arising both from deuterium nuclear corrections
and from MHOs, should be performed, in view of the small size of the ensuing  PDF uncertainty.

\subsection{Conclusions}

We have demonstrated that PDF determinations based on a wider dataset are
characterized by greater perturbative stability, and thus
are most likely to exhibit
smaller theoretical uncertainties related to missing higher
orders.
In addition, we have shown that
the inclusion of data taken with nuclear target, and specifically 
with deuterium ones, appears to be advantageous at present, in that the
uncertainty related to the modeling of
deuterium nuclear corrections appears to be rather
smaller than both the MHOU and of the impact of this data on PDF
uncertainties.

These results provide further evidence
that the inclusion 
of theoretical uncertainties in PDF uncertainties, 
specifically those related to missing higher orders, but also
to nuclear corrections, is now one of the highest priorities.
An interesting observation in this
respect is that a possible way to approach the determination of MHOU
on PDFs
might be to study the way they vary between different
sets of experimental measurements.
Indeed, our study suggests that these uncertainties tend
to compensate when combining different classes of datasets.
This, in turn, hints at the fact that the size of the
MHOUs might be estimated by
performing dataset variations and studying the ensuing distribution of the
best-fit PDFs.
These topics are the subject of ongoing and forthcoming investigations. 

\subsection*{Acknowledgements}

Stefano Forte thanks Daniel Maitre for interesting discussions on PDF
uncertainties during the workshop.
Stefano Forte is supported by the
European Research Council under the European Union's 
Horizon 2020 research and innovation Programme (grant agreement ERC-AdG-740006).
Juan Rojo and Luca Rottoli are supported by an European Research Council Starting
Grant ``PDF4BSM''.
The research of Juan Rojo is also partially supported 
by the Netherlands Organization for Scientific Research (NWO).



\chapter{Jet substructure studies}
\label{cha:jets}




\section{Towards extracting the strong coupling constant from jet substructure at the LHC~\protect\footnote{
    I.~Moult,
    B.~Nachman,
    G.~Soyez,
    J.~Thaler (section coordinators); 
    S.~Chatterjee,
    F.~Dreyer,
    M.~V.~Garzelli,
    P.~Gras,
    A.~Larkoski,
    S.~Marzani,
    A.~Si\'{o}dmok,
    A.~Papaefstathiou,
    P.~Richardson,
    T.~Samui}{}}
\label{sec:SM_jetsub_alphas}

Recent advances in jet substructure have led to a new class of jet observables that are amenable to systematically-improvable calculations and are robust to the complex environment of the Large Hadron Collider (LHC).
These observables exploit grooming to reduce non-perturbative contributions and simplify perturbative calculations.
With these recent advances in both theory and experiment, we believe it is the appropriate time to begin investigating the possibility of extracting the strong coupling constant $\alpha_s$ from jet substructure.
In this section, we perform a proof-of-principle sensitivity study to demonstrate the suitability of such measurements to add useful information to the existing precision extractions of $\alpha_s$.
We highlight a number of theoretical and experimental advantages of using groomed observables for $\alpha_s$ extraction, and we discuss several difficulties of the LHC environment.
Using a simplified approach, we show that a measurement of $\alpha_s$ with an approximate 10\% uncertainty should be feasible at the LHC.
This result motivates a complete analysis with a full set of theoretical and experimental considerations, and we present a number of directions where improvements on both the theory and experiment sides could be made in the near future.

\subsection{Introduction}

In quantum chromodynamics (QCD), the strong coupling constant ($\alpha_s$) is responsible for the strength of interactions between quarks and gluons.
Governed by this fundamental parameter, a plethora of strong-force phenomena emerge, including the binding of partons into massive hadrons that are responsible for most of the visible energy-density in the universe, and the creation of collimated sprays of hadrons known as jets that are ubiquitous at high-energy particle colliders.
The internal structure of jets (jet substructure) has been extensively exploited to search for new particles at the Large Hadron Collider (LHC).
Now that both the theoretical and experimental tools of jet substructure have reached a high degree of maturity~\cite{Abdesselam:2010pt,Altheimer:2012mn,Altheimer:2013yza,Adams:2015hiv,Larkoski:2017jix}, it is time to ask if the radiation patterns inside jets can be used to extract fundamental parameters of the Standard Model, such as $\alpha_s$.

There is a strong motivation (pun intended) for making a precise measurement of $\alpha_s$.
The uncertainty in $\alpha_s$ is a limiting factor in our predictions for the stability of the universe~\cite{Andreassen:2017rzq}, and the
uncertainty in $\alpha_s$ at a variety of scales sets a model-independent sensitivity to strongly-interacting particles beyond the Standard Model~\cite{Kaplan:2008pt,Becciolini:2014lya}.
Now that many scattering processes have been calculated to a high perturbative order, the uncertainty on $\alpha_s$ ($\sigma_{\alpha_s}$) can be limiting in the overall accuracy; numerically $\alpha_s^3\sim \sigma_{\alpha_s}$, so higher-order terms can be smaller than the leading-order correction uncertainty (see e.g.\ $gg\rightarrow H$ at N$^3$LO~\cite{Anastasiou:2015ema}).
Determinations of $\alpha_s$ probe a wide variety of physical phenomena, and the consistency between methods is both a crucial test of the theory and an important ingredient to make accurate predictions for the LHC and beyond.

The world-average value for $\alpha_s$ at the $Z$ boson mass ($m_Z$) is $0.118\pm 0.0013$, an impressive $1.1\%$ total uncertainty~\cite{Olive:2016xmw}.%
\footnote{Unless otherwise specified, we always report $\alpha_s$ at the $Z$ boson mass in the $\overline{\mathrm{MS}}$ scheme.} 
There have been significant discussions in the community about the challenges and validity of various methods to extract $\alpha_s$ (see for instance \cite{Bethke:2011tr,Altarelli:2013bpa,Pich:2013sqa,Moch:2014tta,dEnterria:2015kmd,Olive:2016xmw,Salam:2017qdl}).
The most precise (and dominant) input to the world average is the $\alpha_s$ value from lattice QCD calculations combined with measurements of $B$-hadron mass differences, with an uncertainty that is less than $1\%$. 
After the lattice, the most precise determination is from measurements and calculations of thrust and the $C$-parameter in $e^+e^-$ collisions~\cite{Abbate:2010xh,Hoang:2015hka,Heister:2003aj,Abdallah:2004xe,Abreu:1996mk,Abreu:1999rc,Biebel:1999zt,Adeva:1992gv,Abbiendi:2004qz,Abe:1994mf}.
These methods are sensitive to very different regimes of QCD and, interestingly, differ from each other at about the 5\% level, corresponding to a more than $3\sigma$ tension.  

\begin{figure}[t]
\begin{center}
\includegraphics[width = 0.5\columnwidth]{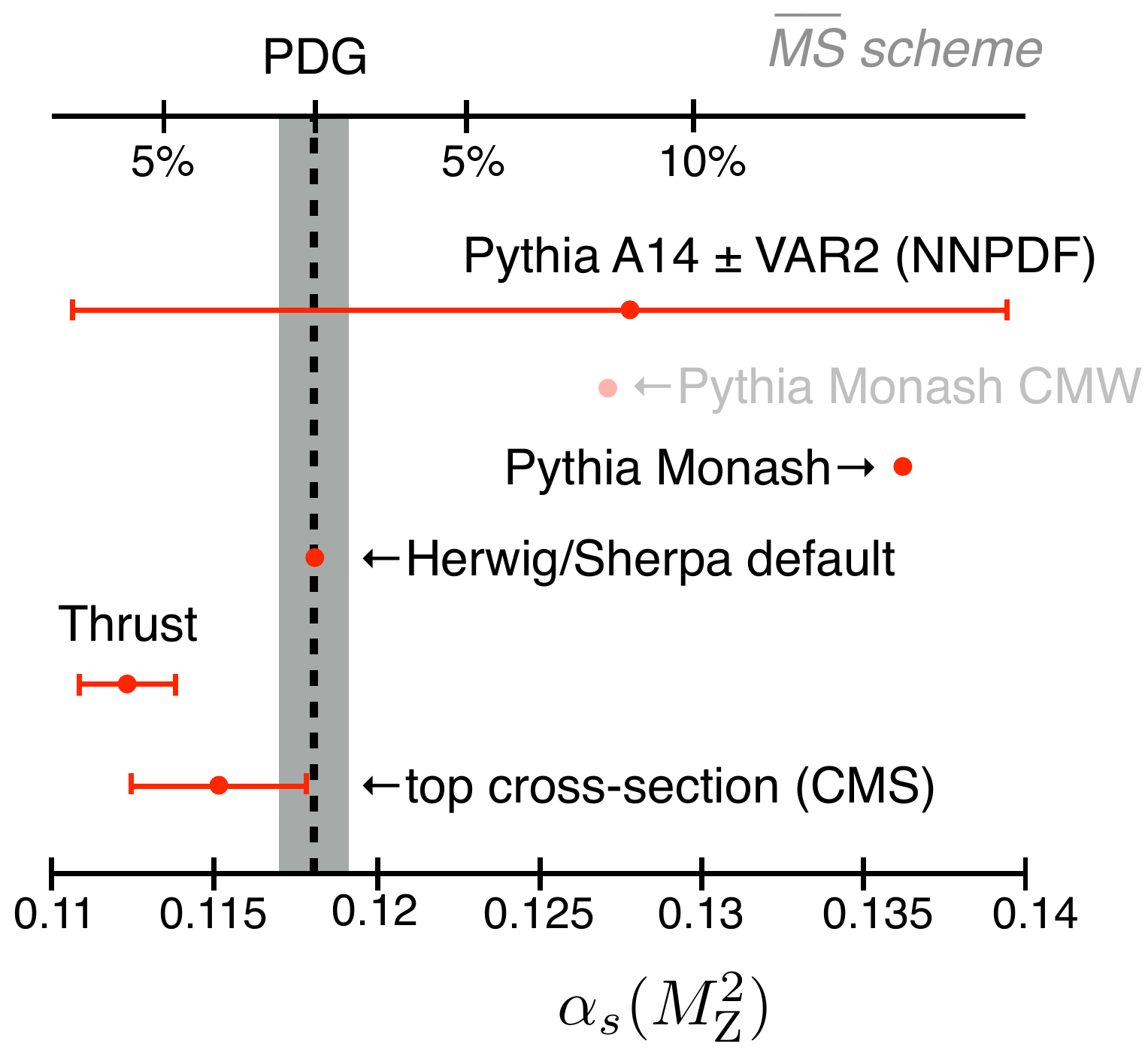}
\end{center}
\caption{Various values of $\alpha_s$ in the $\overline{\mathrm{MS}}$ scheme, including the world-average shown as a black dashed line with a grey uncertainty band~\cite{Olive:2016xmw}.
The point labeled \textit{Thrust} is from LEP data~\cite{Abbate:2010xh,Hoang:2015hka,Heister:2003aj,Abdallah:2004xe,Abreu:1996mk,Abreu:1999rc,Biebel:1999zt,Adeva:1992gv,Abbiendi:2004qz,Abe:1994mf}.
The measurement using the CMS top cross-section measurement is the first NNLO extraction at the LHC~\cite{Chatrchyan:2013haa}.
The other points are the $\alpha_s$ parameter value used in the final state shower in various PS generaters: \textsc{Herwig~7}~\cite{Bellm:2015jjp} and \textsc{Sherpa}~\cite{Gleisberg:2008ta} use the world-average by default, while \textsc{Pythia~8}~\cite{Sjostrand:2006za,Sjostrand:2014zea} employs a higher value in both the Monash default tune~\cite{Skands:2014pea} and the A14 tunes (with VAR2 as an A14 uncertainty variation)~\cite{ATL-PHYS-PUB-2014-021}.
The order at which $\alpha_s$ is utilized is not the same in all cases.}
\label{fig:SM_jetsub_alphas:propaganda}
\end{figure}

Extractions based on $e^+e^-$ event shapes are sensitive to soft and collinear regions of phase space, which are modeled with precision
using higher-order resummation.
Figure~\ref{fig:SM_jetsub_alphas:propaganda} shows various values of $\alpha_s$ extracted with next-to-next-to-leading order (NNLO) calculations including various levels of resummation, such as next-to-next-to-next-to-leading logarithmic (N$^3$LL) for the thrust-based extraction.
Also shown are the values of $\alpha_s$ used as inputs to various Parton Shower (PS) generators.%
\footnote{Note that these programs implement much more than just the QCD shower.  For example, radiated photons are also generated, though the impact of those electroweak processes is quite negligible here (as opposed to photons from $\pi^0\rightarrow\gamma\gamma$).}
The value of $\alpha_s$ in the final-state PS is also sensitive to the soft and collinear regime of QCD (albeit in different ways); interestingly, the values used in the \textsc{Pythia~8} program~\cite{Sjostrand:2006za,Sjostrand:2007gs}, which are fit to $e^+e^-$ event shape data, suggest a higher value of $\alpha_s$ than the lattice result by about 15\%.
Another challenge with the event-shape extraction is that non-perturbative (NP) corrections are nearly degenerate with changes to $\alpha_s$ \cite{Abbate:2010xh}.
This is in part because techniques to parametrically separate NP effects from perturbative effects (i.e.~jet grooming techniques to be discussed below) were not mature at the time of the Large Electron Positron (LEP) collider, and because the beam energies available at LEP did not allow for a large lever-arm between energy scales.
The sensitivity to low-energy scales where QCD is in the NP regime is also a key challenge for the lattice determination.
It would therefore be timely for a precision extraction of $\alpha_s$ using jets at the LHC.

For most analyses at the LHC so far, jets have been used mainly as proxies for quark and gluon four-vectors.
The multiplicity and kinematics of jets in purely hadronic final states can be predicted to high $\alpha_s$ orders in perturbation theory with e.g.\ NNLOJET~\cite{Currie:2016bfm,Currie:2017ctp} and NLOJet++~\cite{Nagy:2001fj,Nagy:2003tz}.
As a result, measurements of jet multiplicities, energies, and angles can be used to extract $\alpha_s$~\cite{ATLAS:2015yaa,Aaboud:2017fml,Khachatryan:2014waa,CMS:2014mna,Chatrchyan:2013txa}.
Even though these measurements have achieved an uncertainty of around $5\%$, they have not yet been included in the Particle Data Group (PDG) combination~\cite{Olive:2016xmw} since they are using only next-to-leading-order (NLO) theory calculations.
With recent progress in higher-order calculations, this will likely change soon, and it is noteworthy that an extraction using the recent NNLO $t\bar{t}$ cross-section \cite{Czakon:2013goa} is now included (see Fig.~\ref{fig:SM_jetsub_alphas:propaganda}).

In extractions based on fixed-order perturbation theory, the collinear region defining the internal
structure of jets is avoided in part due to the lack of precision calculations.%
\footnote{Logarithms of the jet radius $R$ may be important when the jet radius is small, and there are techniques to incorporate these corrections into fixed-order calculations~\cite{Dasgupta:2016bnd,Dasgupta:2014yra}.}
It is important to remember that jets are not in one-to-one correspondence with individual quarks and gluons, but rather to clusters of hadrons resulting from many quarks and gluons; for this reason, the substructure of jets is also governed by $\alpha_s$.
Recent theoretical and experimental advances in jet substructure have shown that the radiation pattern inside jets has great physics potential~\cite{Abdesselam:2010pt,Altheimer:2012mn,Altheimer:2013yza,Adams:2015hiv,Larkoski:2017jix}.
To date, the focus of jet substructures studies has been mostly on tagging the origin of jets and searching for physics beyond the Standard Model.
However, present and future precision may be sufficient to make a useful measurement of $\alpha_s$.

A key tool to facilitate a substructure extraction of $\alpha_s$ are jet grooming techniques, which systematically removes soft and wide-angle radiation from jets \cite{Butterworth:2008iy,Ellis:2009su,Ellis:2009me,Krohn:2009th,Dasgupta:2013ihk,Larkoski:2014wba}.
Jet grooming can parametrically separate the NP radiation in the jet from the hard perturbatively-described components.
At a hadron collider, grooming also mitigates the contribution from the underlying event and additional nearly simultaneous interactions (pileup).
Theoretical calculations for groomed observables have been performed at NNLL~\cite{Frye:2016okc,Frye:2016aiz} and NLL~\cite{Marzani:2017mva,Marzani:2017kqd} accuracy and will be extended to higher orders as calculations become available.
Recently, ATLAS~\cite{Aaboud:2017qwh} and CMS~\cite{CMS-PAS-SMP-16-010} demonstrated 5-10\% measurement uncertainties of these calculated quantities using existing technologies.

The purpose of this section is to study the feasibility of a measurement of $\alpha_s$ using jet substructure at the LHC, using jet shapes that are infrared and collinear (IRC) safe.
Similar techniques would of course also be interesting at both a low- and high-energy $e^+e^-$ collider; we focus here on $pp$ since high-quality data are now pouring out of the LHC.
We view this work as part of a broader program to apply jet grooming techniques for precision QCD (see also Ref.~\cite{Hoang:2017kmk} for applications to the top quark mass).
There are significant experimental and theoretical challenges to achieve success in this program, but we believe it is possible with community synergy.
An $\alpha_s$ extraction represents a concrete goal to push the accuracy and understanding of jet substructure calculations in particular and QCD calculations more generally.

This remainder of this work is organized as follows.
In Sec.~\ref{sec:SM_jetsub_alphas:definitions}, we define the observables and grooming strategies that will be considered in this study.
In Sec.~\ref{sec:SM_jetsub_alphas:softcomplications}, we highlight a number of theoretical and experimental benefits of jet grooming, which make groomed observables particularly interesting for extractions of $\alpha_s$.
In Sec.~\ref{sec:SM_jetsub_alphas:jetmass}, we discuss the sensitivity of groomed jet mass to the value of $\alpha_s$ using both analytic expressions and PS generators.
In Sec.~\ref{sec:SM_jetsub_alphas:ben_study}, an idealized setup is used to illustrate how an extraction of $\alpha_s$ might work at the LHC, using realistic estimates of both the theoretical and experimental uncertainties.
We conclude in Sec.~\ref{sec:SM_jetsub_alphas:future} and discuss future directions for improving both the theoretical and experimental uncertainties for $\alpha_s$ extractions from jet substructure.

\subsection{Observable and Algorithm Definitions}
\label{sec:SM_jetsub_alphas:definitions}

In this subsection, we briefly review the definitions of the observables and grooming procedures that we will focus on in this study.

\subsubsection{Two-point Correlators}\label{sec:SM_jetsub_alphas:shape_def}

A simple class of jet shape observables that have sensitivity to the value of $\alpha_s$ are two-point correlation functions~\cite{Banfi:2004yd,Larkoski:2013eya}.
There has been significant theoretical study of these objects, and their perturbative behavior is understood to relatively high accuracy.
For a set of constituents $\{i\}$ in a jet $J$, the two-point correlation functions in $pp$ collisions are defined as
\begin{equation}
\label{eq:SM_jetsub_alphas:ppe2}
\left.e_2^{(\alpha)}\right|_{pp}=\frac{1}{p_{TJ}^2}\sum_{i<j\in J} p_{Ti} \, p_{Tj} \, \left(\frac{R_{ij}}{R}\right)^\alpha\,, 
\end{equation}
where $p_{Ti}$ is the transverse momentum of particle $i$ with respect to the beam axis, $p_{TJ}$ is the transverse momentum of the jet, $R_{ij}$ is the distance between particles $i$ and $j$ in the rapidity-azimuthal angle plane, and $R$ is the jet radius.
Note that some authors define the correlation functions without the $R^\alpha$ factor in the denominator.
The angular exponent $\alpha$ is a parameter that controls the sensitivity to wide-angle emissions.
If all emissions in the jet are nearly collinear, then the two-point energy correlation function reduces to a function of the jet mass ($m$) when $\alpha=2$:
\begin{equation}
e_2^{(2)}\sim \frac{m^2}{R^2\, p_{T}^2}\,.
\end{equation} 

In this study, we primarily restrict ourselves to the case of $\alpha=2$ as this has been the focus so far from both the theoretical~\cite{Frye:2016okc,Frye:2016aiz,Marzani:2017mva,Marzani:2017kqd} and experimental~\cite{Aaboud:2017qwh,CMS-PAS-SMP-16-010} communities.
However, it is interesting to study if other values of $\alpha$ could provide improved sensitivity to $\alpha_s$.
Another benchmark value is $\alpha=1$, which corresponds to $k_T$ (or broadening) instead of mass.
The two-point correlation functions are closely related to the jet angularities~\cite{Berger:2003iw,Almeida:2008yp,Ellis:2010rwa,Larkoski:2014pca}, with the latter being defined with respect to a jet axis.

\subsubsection{Grooming Techniques}\label{sec:SM_jetsub_alphas:groom_tech}

There are a variety of jet grooming techniques to mitigate soft and wide-angle jet contamination~\cite{Butterworth:2008iy,Ellis:2009su,Ellis:2009me,Krohn:2009th,Dasgupta:2013ihk,Larkoski:2014wba}.
We focus here on the SoftDrop algorithm~\cite{Larkoski:2014wba}, defined using Cambridge/Aachen (C/A) reclustering \cite{Dokshitzer:1997in,Wobisch:1998wt,Wobisch:2000dk}.
Starting from a jet identified with an IRC-safe jet algorithm (such as anti-$k_t$ \cite{Cacciari:2008gp}), SoftDrop proceeds as follow:
\begin{enumerate}
\item Recluster the jet using the C/A clustering algorithm, producing an angular-ordered branching history for the jet.
\item Step through the branching history of the reclustered jet.  At each step, check the SoftDrop condition
\begin{equation}\label{eq:SM_jetsub_alphas:sd_cut}
\frac{\min\left[ p_{Ti}, p_{Tj}  \right]}{p_{Ti}+p_{Tj}}> z_{\mathrm{cut}} \left(   \frac{R_{ij}}{R}\right)^\beta \,,
\end{equation}
where $z_{\mathrm{cut}} $ is a parameter defining the scale below which soft radiation is removed, and $\beta$ is an exponent that controls the angular scale for removal.
If the SoftDrop condition is not satisfied, then the softer of the two branches is removed from the jet.  This process is then iterated on the harder branch.
\item The SoftDrop procedure terminates once the SoftDrop condition is satisfied.
\end{enumerate}
This procedure generalizes the modified Mass Drop Tagger (mMDT)~\cite{Dasgupta:2013ihk}, which corresponds to $\beta=0$ in the SoftDrop procedure described above.
Note that non-global logarithms~\cite{Dasgupta:2001sh} are formally removed from the SoftDropped mass distribution, even when $\beta>0$ (though clearly they are restored as $\beta\rightarrow\infty$).

\subsection{Grooming Away Soft Complications}
\label{sec:SM_jetsub_alphas:softcomplications}

Having defined the observables and grooming procedures of interest, in this subsection we provide a general discussion of the potential advantages of using groomed jet observables for extractions of $\alpha_s$.
The grooming procedure significantly simplifies a number of theoretical and experimental issues related to precision calculations and measurements in a $pp$ environment.
In particular, grooming provides:
\begin{itemize}
\item Mitigation of NP effects;
\item Perturbative simplicity; and
\item Improved detector resolution, due in large part to a reduced pileup sensitivity.
\end{itemize}
Each of these will be discussed in more detail below.
Combined, we believe that they provide a strong motivation for the measurement of  groomed jet observables, and the possibility of performing a precision extraction of $\alpha_s$ using jet substructure at the LHC.

\subsubsection{Mitigating Non-perturbative Effects}

\begin{figure}[t]
\begin{center}
\includegraphics[width = 0.5\columnwidth]{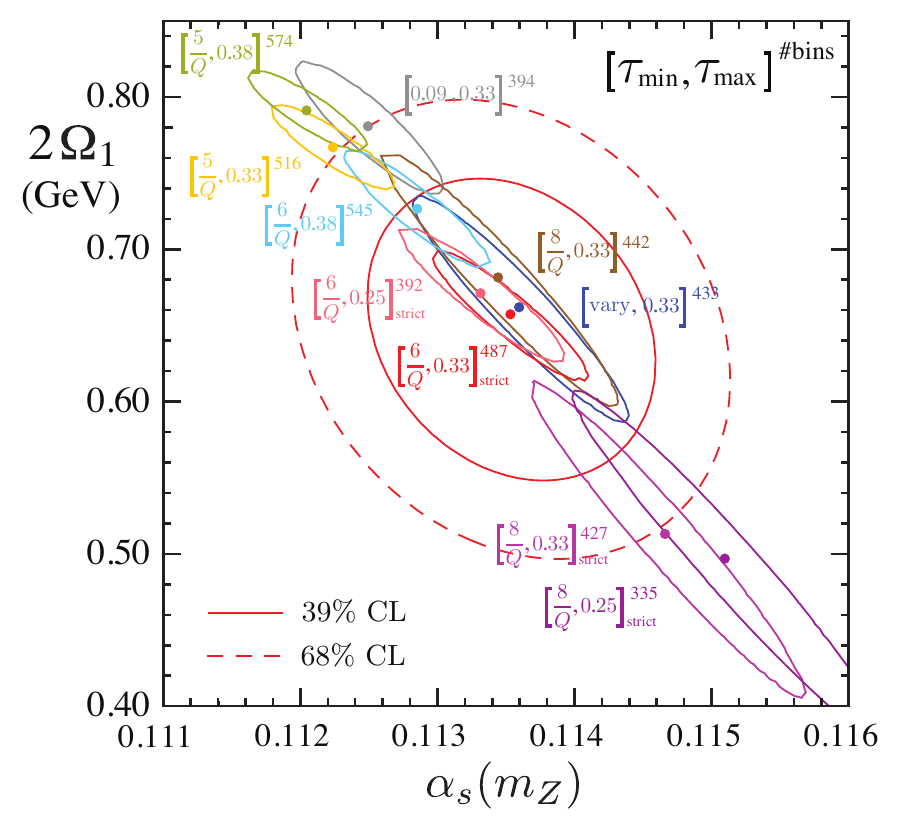}
\end{center}
\caption{An illustration of the correlations between the first NP moment $\Omega_1$ and $\alpha_s$ for the thrust observable.
NP effects for groomed observables have a significantly different structure, providing the possibility for complementary information to the extraction from (ungroomed) event shapes in $e^+e^-$.
Figure from Ref.~\cite{Abbate:2010xh}.}
\label{fig:SM_jetsub_alphas:correlation_firstmoment}
\end{figure}

One of the primary complications of $\alpha_s$ extractions using event shapes is NP corrections, which cannot currently be calculated from first principles.
Since event shape observables probe singular regions of the phase space, they are necessarily sensitive to NP corrections.  
For $e^+e^-$ dijet event shapes, such as thrust, these NP corrections can be given operator definitions in the dijet limit.
Although these operators cannot be calculated, they can be modeled using \textit{shape functions}~\cite{Korchemsky:1999kt,Korchemsky:2000kp,Hoang:2007vb,Ligeti:2008ac}.%
\footnote{There are other approaches to analytically describe and fit NP effects that have similar features to shape functions.  Another example is the dispersive model~\cite{Dokshitzer:1995qm,Dokshitzer:1995zt} which has also been used to make extractions of $\alpha_s$ using $e^+e^-$ event shapes~\cite{Gehrmann:2012sc}.}
These shape functions can be systematically expanded in moments, and for moderately large values of the observable, only the first moment ($\Omega_1$) is required.
This moment can be shown to be universal for a wide range of observables \cite{Lee:2006fn,Lee:2007jr}, and can therefore be extracted from data along with $\alpha_s$~\cite{Abbate:2010xh,Abbate:2012jh,Hoang:2015hka}.\footnote{See Ref.~\cite{Salam:2001bd,Mateu:2012nk} for subtleties related to hadron masses and the breakdown of universality.}
This moment, however, is highly correlated with $\alpha_s$ as shown in Fig.~\ref{fig:SM_jetsub_alphas:correlation_firstmoment}.
Therefore, observables with different, and preferably suppressed, NP effects could provide valuable complementary information for $\alpha_s$ extractions.

Using a scaling argument, it is a straightforward exercise to compute the value of the groomed two-point correlators at which NP effects are expected to be important.
This was considered in Ref.~\cite{Dasgupta:2013ihk,Frye:2016aiz} where it was shown that NP effects become important when (for $\alpha \geq 1$),
\begin{equation}
\label{eq:SM_jetsub_alphas:np}
\left. e_2^{(\alpha)}\right |_{\mathrm{NP}} \simeq  \left( \frac{\Lambda_{\mathrm{QCD}}}{z_{\mathrm{cut}}  Q}  \right)^{\frac{\alpha-1}{1+\beta}}  \frac{\Lambda_{\mathrm{QCD}}}{Q}\,,
\end{equation}
where $Q=p_TR$ is the starting scale of jet fragmentation and $\Lambda_{\rm QCD} \simeq 1$ GeV is a typical NP scale.
When $\alpha< 1$, the value is instead $(\Lambda_{\mathrm{QCD}}/(p_TR))^\alpha$.
If we consider for concreteness the jet mass ($\alpha=2$) we can learn a number of interesting lessons.
First, by taking $\beta\to \infty$, we obtain the ungroomed result
\begin{equation}
\left. e_2^{(2)} \right |_{\mathrm{NP}} \simeq  \frac{\Lambda_{\mathrm{QCD}}}{p_TR}\,,
\end{equation} 
while the groomed result gives 
\begin{equation}
\left. e_2^{(2)} \right |_{\mathrm{NP}} \simeq  \left( \frac{\Lambda_{\mathrm{QCD}}}{z_{\mathrm{cut}}  p_TR}  \right)^{\frac{1}{1+\beta}}  \frac{\Lambda_{\mathrm{QCD}}}{p_TR}\,.
\end{equation}
For $z_{\mathrm{cut}}  p_T R \gg \Lambda_{\mathrm{QCD}}$, we see that the grooming significantly suppresses the scale of the NP physics, extending the range of perturbative validity by a factor of $({\Lambda_{\mathrm{QCD}}} / {z_{\mathrm{cut}}  p_TR})^{1/(1+\beta)}$.
For a $1$ TeV jet with $z_{\mathrm{cut}}  =0.1$ and $\beta=0$, this is an extension by a factor of $\sim 100$.
Furthermore, we observe that, under the assumption that $\beta \geq 0$, this suppression is maximized for $\beta=0$, motivating this choice in later studies.%
\footnote{Negative values of $\beta$ are also possible and could be interesting to study.  However, one would need to dedicate specific attention in the vicinity of the endpoint of the distribution, which is $e_2^{(\alpha)}=z_{\mathrm{cut}}^{-\alpha\beta}$ at LL.}

It is important to emphasize that just because the value of the mass at which NP physics enters is suppressed, this does not by itself imply that there is a larger region over which one has perturbative control.
In principle the whole distribution could be shifted to lower values.
We know, however, that for $e_2^{(2)}\geq z_{\mathrm{cut}} $, the distribution is parametrically unaffected by the grooming procedure, and is therefore identical to the ungroomed jet mass.

\begin{figure}[t]
\begin{center}
\includegraphics[width = 0.5\columnwidth]{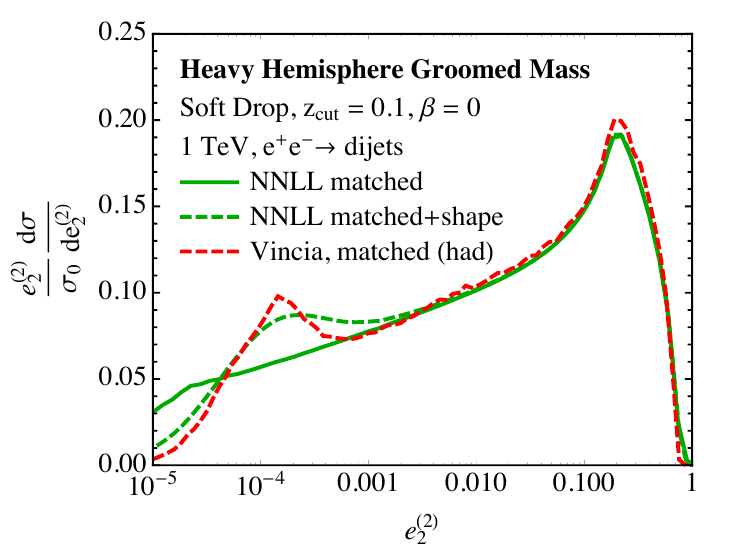}
\end{center}
\caption{A plot showing the effects of NP corrections to the groomed jet mass. This illustrates first that the NP corrections are suppressed over a large range, and second that they take a very different form than for double-logarithmic observables. Figure from Ref.~\cite{Frye:2016aiz}.}
\label{fig:SM_jetsub_alphas:shape_function}
\end{figure}

In addition to being suppressed, the structure of the NP corrections is quite different for groomed jet mass, which may provide complementary information compared to standard event shapes.
For event shape observables, NP corrections are typically large already at the Sudakov peak.
In the case of a groomed observable, since the NP corrections are suppressed and the observable is predominantly single logarithmic (exactly true when $\beta=0$), we instead see that the NP corrections appear as a bump on the falling distribution.
This is shown in Fig.~\ref{fig:SM_jetsub_alphas:shape_function} along with a model using a shape function.
Due to the fact that this is a completely different behavior, it may have different correlations as compared with standard event shapes, though a more complete understanding of this effect is required. 

Finally, while we have so far focused on the effect of the grooming procedure on hadronization corrections and a comparison to $e^+e^-$ event shapes, we must also emphasize that the grooming plays an important role in suppressing NP corrections from the underlying event.
This is crucial for achieving a precision measurement in the $pp$ environment, since there is a limited theoretical understanding of the underlying event.
Here, we must rely purely on Monte Carlo model implementations of the underlying event.
Monte Carlo studies have shown that in the regime where one has perturbative control, underlying event is highly suppressed after mMDT/SoftDrop~\cite{Dasgupta:2013ihk,Larkoski:2014wba}.
This is again a significant benefit of groomed observables.

\subsubsection{Perturbative Simplicity}
\label{sec:SM_jetsub_alphas:pertsimplicity}

By studying the analytic behavior of grooming on the jet mass, the authors of Ref.~\cite{Dasgupta:2013ihk} showed that grooming could be designed to enhance the simplicity and calculability of various jet shapes.
In particular, grooming significantly simplifies the perturbative resummation in the $pp$ environment in addition to suppressing NP corrections.
Calculations of the standard jet mass in $pp$ typically suffer from the following two perturbative difficulties:
\begin{itemize}
\item Global color correlations, which complicate the structure of higher-order soft functions; and
\item Non-global logarithms \cite{Dasgupta:2001sh}, which complicate the perturbative structure for non-global measurements.
\end{itemize}
The jet mass was calculated in \cite{Jouttenus:2013hs} (\cite{arXiv:1207.1640}) for $H+$ jet with(out) a veto on additional radiation, supplemented with arguments that non-global effects are small.
These above two features highlight that the ungroomed jet mass depends significantly on the particular process that created the jet, significantly reducing its universality, and complicating its structure in a busy $pp$ environment.
By contrast, the use of grooming can make the jet mass nearly universal.

In Ref.~\cite{Frye:2016aiz}, it was shown that for $e_2^{(\alpha)}\ll z_{\mathrm{cut}}  \ll 1$, one can express the cross-section for the groomed two-point correlator as
\begin{equation}
\label{eq:SM_jetsub_alphas:fac_pp_e2}
\frac{d\sigma^{pp}}{de_2^{(\alpha)}}=\sum\limits_{k=q,\bar q, g}D_k(p_T^{\mathrm{min}}, y_\mathrm{max}, z_{\mathrm{cut}} , R) \, S_{C,k}(z_{\mathrm{cut}} , e_2^{(\alpha)})\otimes J_k (e_2^{(\alpha)})\,,
\end{equation}
where $S_{C,k}$ is the soft function and $J_k$ is the jet function.
The function $D_k$ depends on the parton flavor, and can be interpreted as the quark and gluon fractions.
While the quark and gluon fractions are generally not IRC-safe quantities, in this limit the quark fraction can be defined jet-by-jet as 
\begin{equation}
f_J=\sum\limits_{i\in J_{\mathrm{SD}}} f_i\,,
\end{equation}
with $f_q=1$, $f_{\bar q}=-1$, $f_g=0$, and $J_{\mathrm{SD}}$ are the parton-level constituents of the groomed jet.
The grooming procedure makes this definition IRC safe at leading power (see related jet flavor approach in Ref.~\cite{Banfi:2006hf}).
These fractions are independent of the value of the jet mass observable and can be extracted from fixed-order generators, for example MCFM \cite{Campbell:1999ah,Campbell:2010ff,Campbell:2011bn}.

The formula in Eq.~\eqref{eq:SM_jetsub_alphas:fac_pp_e2} has a number of remarkable consequences.
In the resummation region, we see that there are no non-global logarithms and there are no global color correlations.
The collinear soft and jet functions depend only on the jet flavor, so they are therefore identical to in the case of groomed jets in $e^+e^-$ collisions.
In the resummation region, the grooming algorithm has therefore allowed the jet to be isolated from its $pp$ environment, and the only required input from the hard scattering process are the quark and gluon fractions. 

It is important to emphasize that this simplification does not occur throughout the entire distribution.
In particular, it does not hold for $m \gg Qz_{\mathrm{cut}} $, though this is the fixed order region, where we can match to standard fixed order perturbation theory.
This matching has been performed to (N)NLO for dijets ($V$+jets) \cite{Frye:2016aiz,Marzani:2017kqd,Marzani:2017mva}.
Ideally this would be done using $2\to 3$ matrix elements at NNLO, which are just now becoming available~\cite{Gehrmann:2015bfy,Dunbar:2016aux,Badger:2013yda,Badger:2017jhb,Abreu:2017hqn}.

\subsubsection{Pileup Resilience}

Arguably the biggest experimental challenge for precision physics at a high-luminosity $pp$ collider is noise from \textit{pileup}: multiple nearly simultaneous proton-proton collisions.
Jet shapes are particularly sensitive to pileup; for example, the jet mass scales as $\mathcal{O}(A^2)$~\cite{Salam:2009jx} for the jet catchment area $A$~\cite{Cacciari:2008gn} (whereas the jet $p_T$ scales linearly with $A$).
The jet-area subtraction that works well for $p_T$ has been extended to event shapes~\cite{Soyez:2012hv}, but must be re-calibrated per observable.
Constituent-based pileup subtraction schemes~\cite{Cacciari:2014gra,Krohn:2013lba,Bertolini:2014bba,Berta:2014eza,Komiske:2017ubm} show great promise and are actively being studied and adapted to the actual experimental settings~\cite{CMS-PAS-JME-14-001,CMS-DP-2015-034,ATLAS-CONF-2017-065,ATL-PHYS-PUB-2017-020,Aad:2015ina}.
Even without constituent-based subtraction techniques, though, there is a large reduction in pileup sensitivity to jet substructure from grooming~\cite{CMS-PAS-JME-14-001,Aad:2015rpa,Aad:2015ina,Altheimer:2013yza}.
Grooming systematically removes soft and wide-angle radiation, which is exactly the profile characteristic of pileup.
Even with extreme levels of pileup (up to 300 collisions), grooming can preserve the distribution of the jet mass distribution~\cite{JetSubstructureECFA2014}. 

Despite the power of grooming for pileup suppression, there is still a residual degradation of resolution with increased levels of pileup which makes precision jet substructure measurements challenging at high instantaneous luminosity.
Track-based observables are robust to pileup because their vertex of origin can be well-distinguished from pileup vertices.
Precision track-based substructure observables have been calculated~\cite{Krohn:2012fg,Waalewijn:2012sv,Chang:2013rca,Elder:2017bkd}, but typically require universal NP input.
It may be interesting to do a track- and jet-substructure-based extraction of $\alpha_s$, but this is left as a possibility for future work.

\subsection{Observable Sensitivity to $\alpha_s$}
\label{sec:SM_jetsub_alphas:jetmass}

In this subsection, we study the sensitivity of the groomed jet mass to variations in the value of $\alpha_s$.
We begin with a discussion based on the analytic formulae at LL accuracy.
We then perform a PS study, highlighting the interplay between the sensitivity of different parts of the distribution to variations in the value of $\alpha_s$ and NP effects.
Finally, we discuss the issue of Casimir scaling and the related issue of using normalized versus unnormalized distributions.

\subsubsection{Analytic Understanding}
\label{sec:SM_jetsub_alphas:analytic}

To get an understanding of the sensitivity of the groomed mass distribution both to the value of $\alpha_s$ as well as to the quark and gluon composition, it is enlightening to study the LL distribution.
Here, for simplicity, we consider only the leading logs in the observable, in the resummation region; complete expressions can be found in Refs.~\cite{Larkoski:2014wba,Frye:2016aiz,Marzani:2017mva,Marzani:2017kqd}.
For $\beta=0$, the LL result at fixed coupling for the cumulative distribution in the resummation region takes the schematic form
\begin{equation}
\Sigma(e_2^{(2)})=\exp\left[ - \frac{\alpha_s C_i}{\pi} [\log(z_{\mathrm{cut}})-B_i ] \log (e_2^{(2)}) \right]\,,
\end{equation}
where $B_i=-3/4$ for quarks and $B_g=-\frac{11}{12}+\frac{n_f}{6C_A}$ for gluons ($n_f$ is the number of active quark flavors).  This highlights that for $\beta=0$, the groomed jet mass is a single-logarithmic observable, contrasting with the standard double-logarithmic behavior of plain jet mass.
Differentiating the cumulative distribution, we obtain the spectrum
\begin{equation}
\label{eq:SM_jetsub_alphas:ecf_ll_dsitribution}
\frac{e_2^{(2)} }{\sigma}\frac{d\sigma}{d e_2^{(2)}}=   - \frac{\alpha_s C_i}{\pi} [\log(z_{\mathrm{cut}})-B_i ] \exp\left[ - \frac{\alpha_s C_i}{\pi} [\log(z_{\mathrm{cut}})-B_i ] \log (e_2^{(2)}) \right].
\end{equation}
Here, we immediately see several interesting consequences.
In the resummation region, the slope of the distribution when plotted against $\log e_2^{(2)}$ is set by the product $\alpha_s C_i$, where $C_i$ is the Casimir factor, namely $C_F = 4/3$ for quarks and $C_A = 3$ for gluons.
We therefore see that the groomed mass is indeed sensitive to the value of $\alpha_s$.
Due to the larger color charge of gluons, we expect that samples of pure gluon jets would have a significantly higher sensitivity to the value of $\alpha_s$; this expectation will be born out in our PS studies below.
Because $\alpha_s$ is always multiplied by a color factor, though, knowing the precise quark/gluon composition of a sample is essential, as discussed in Sec.~\ref{sec:SM_jetsub_alphas:casimir}.
In practice, the PS studies and the analytic studies that follow (see Sec.~\ref{sec:SM_jetsub_alphas:ben_study}) include higher-order effects, such as subleading terms in the splitting functions, that violate Casimir scaling.

\subsubsection{Parton Shower Study}

From the point of view of fitting for $\alpha_s$, a good observable is one whose probability distribution changes significantly with variations in $\alpha_s$.
However, many observables that significantly change with $\alpha_s$ are also very sensitive to NP effects, such as the constituent multiplicity inside jets.
Here, we study many two-point correlators to quantify the tradeoff between the sensitivity to $\alpha_s$ and the robustness to NP effects.

\begin{figure}[t]
\begin{center}
\includegraphics[width = 0.4\columnwidth]{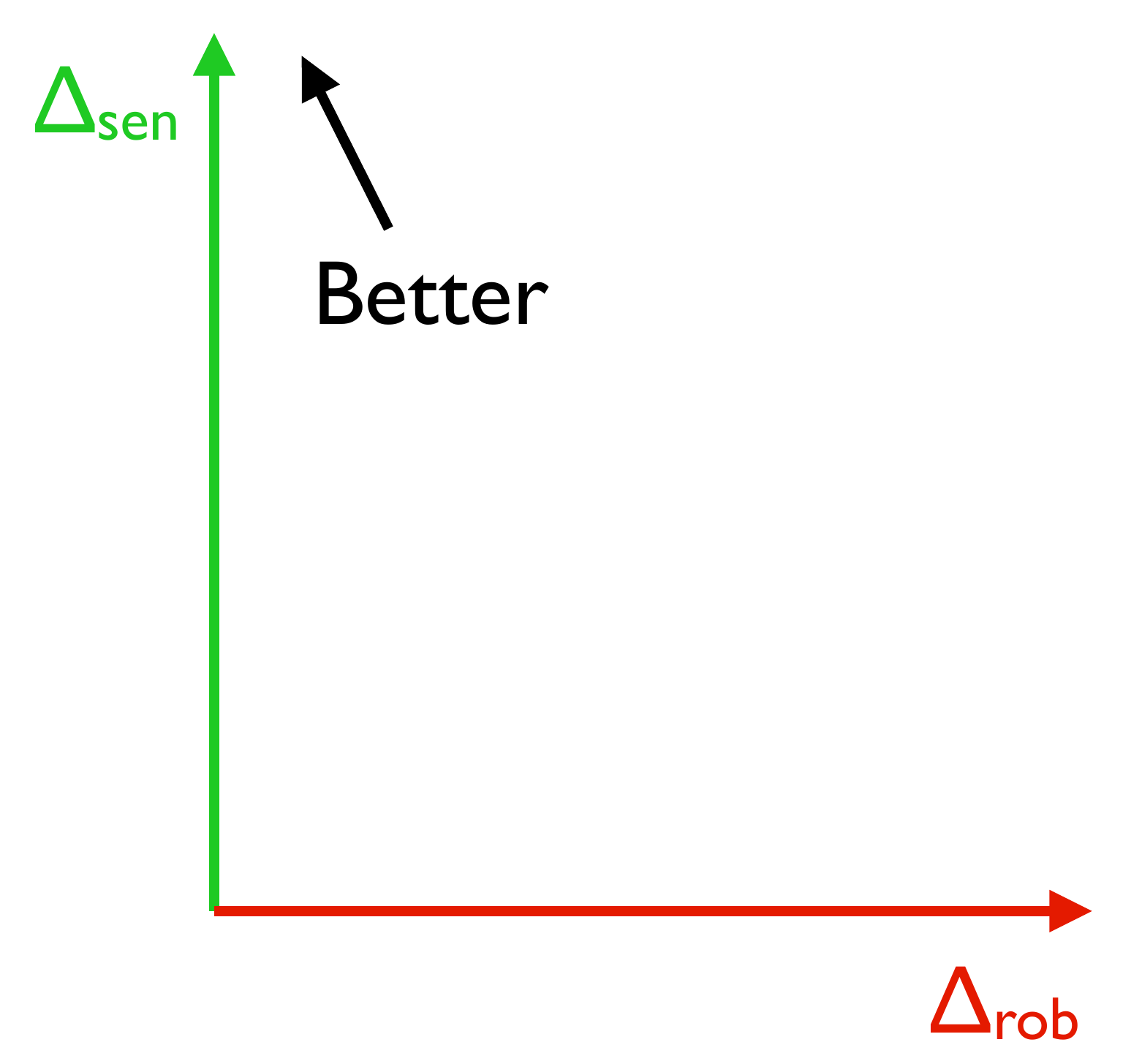}
\end{center}
\caption{A schematic diagram to illustrate the sensitivity-robustness plane defined by the separation power in Eq.~\eqref{eq:SM_jetsub_alphas:seppower}.}
\label{fig:SM_jetsub_alphas:robustnessschematic}
\end{figure}

Given two probability distributions $f$ and $g$ in some observable $\lambda$, we define the separation power $\Delta(f,g)$~\cite{Harrison:1998yr} as
\begin{equation}
\label{eq:SM_jetsub_alphas:seppower}
\Delta(f,g)=\frac{1}{2}\int d\lambda \, \frac{(f(\lambda)-g(\lambda))^2}{f(\lambda)+g(\lambda)}.
\end{equation}
The separation power is a number in $[0,1]$, where $\Delta=0$ if and only if $f=g$.
If $f$ is the nominal probability distribution of some observable and $g$ is the distribution of the same observable with a different value of $\alpha_s$, we would like $\Delta(f,g)$ close to 1 (sensitivity).
In contrast, if $g$ is the same as $f$ with some variation in the NP effects, then we would like $\Delta(f,g)$ to be close to $0$ (robustness).
The plane used to study the tradeoff between sensitivity and robustness is shown in Fig.~\ref{fig:SM_jetsub_alphas:robustnessschematic}.
Not all information about $\alpha_s$ sensitivity is captured by a single point in Fig.~\ref{fig:SM_jetsub_alphas:robustnessschematic}, because sensitivity to NP effects could be in regions of low $\alpha_s$ sensitivity and vice versa.
Therefore, it is also useful to study the integrand of Eq.~\eqref{eq:SM_jetsub_alphas:seppower} as a function of different observables $\lambda$.

This sensitivity-robustness tradeoff is studied using two PS generators: \textsc{Herwig~7}.1.0~\cite{Bellm:2015jjp,Bellm:2017bvx,Reichelt:2017hts} with its default settings%
\footnote{This uses the MMHT 2014 PDFs sets~\cite{Harland-Lang:2014zoa},
the angular-ordered parton shower~\cite{Gieseke:2003rz}, default settings
for the cluster hadronization model~\cite{Webber:1983if}, a Multiple Parton Interaction model~\cite{Bahr:2008wk,Bahr:2008dy,Gieseke:2016fpz}, and plain color reconnections~\cite{Gieseke:2012ft}.  The factorization and renormalization scale is set to $\mu_R^2=\mu_F^2=(\sum_i p_T^i)^2$ where $p_T^i$ is the transverse momentum of the $i$-th jet.}
and \textsc{Pythia~8}.223~\cite{Sjostrand:2006za,Sjostrand:2014zea} with the tune 4C~\cite{Corke:2010yf}.  
These programs are formally LL accurate, though they include effects beyond Eq.~\eqref{eq:SM_jetsub_alphas:ecf_ll_dsitribution}.
Results are presented separately for quark and gluon jets, using the $Z+q$ and $Z+g$ hard-scattering processes.
We select jets with transverse momentum $p_T>500$ GeV and rapidity $|y|<2.5$, clustered with the anti-$k_t$ algorithm~\cite{Cacciari:2008gp} algorithm using a jet radius $R=0.8$.
A variety of two-point correlators (as defined in Sec.~\ref{sec:SM_jetsub_alphas:shape_def}) are studied, with $\alpha\in\{0.5,1.0, 2.0\}$ corresponding to the Les Houches Angularity~\cite{Gras:2017jty}, width, and mass, respectively. 
Additionally, various SoftDrop grooming parameters are studied by varying $\beta\in\{0,1,2\}$ and $z_\mathrm{cut}\in \{0.05,0.1,0.2\}$.   The plots in this section are shown for \textsc{Herwig~7}.1 but the same qualitative conclusions hold for \textsc{Pythia~8}.2.

\begin{figure}[p]
\begin{center}
\includegraphics[width = 0.49\columnwidth]{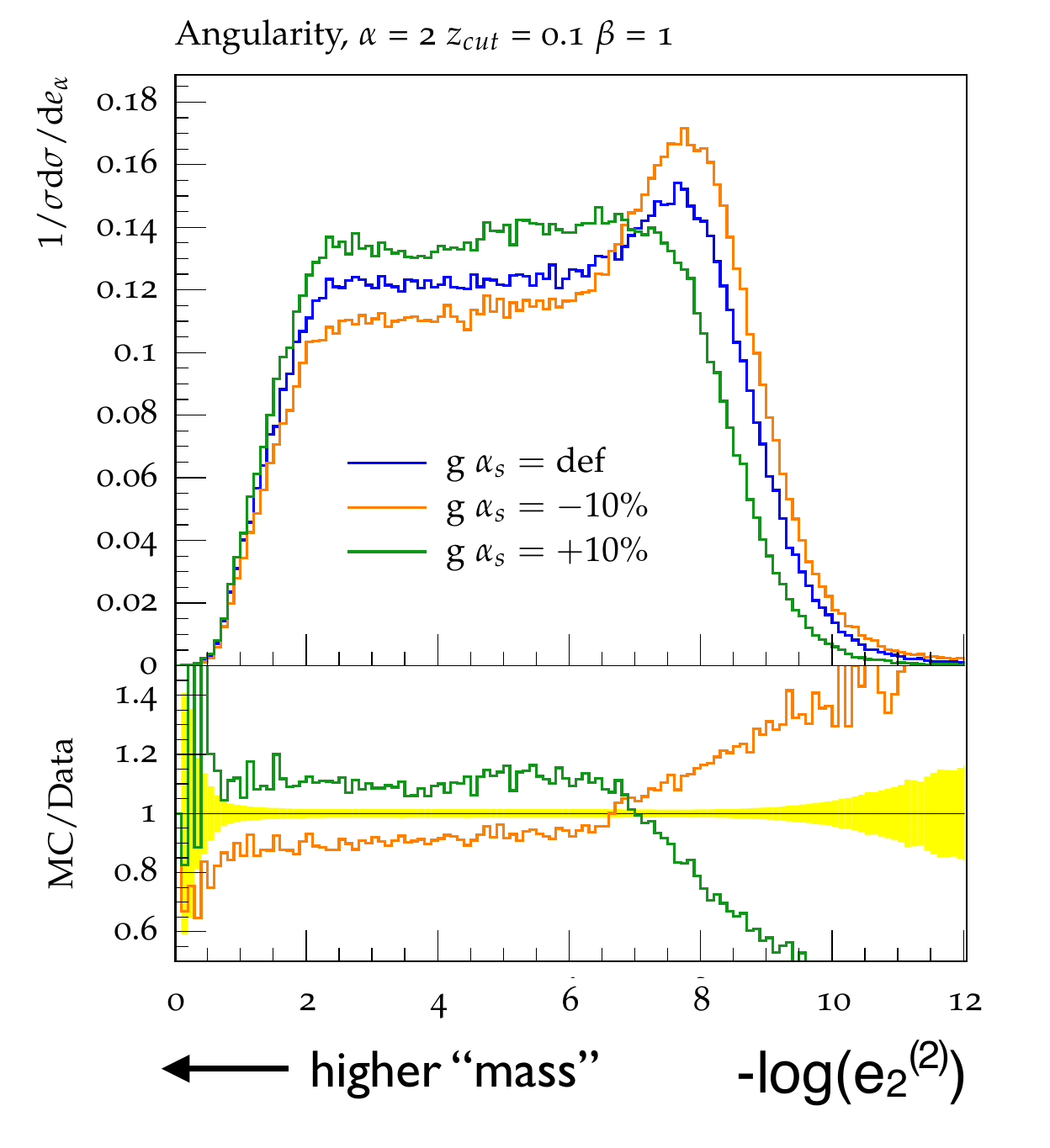}
\includegraphics[width = 0.48\columnwidth]{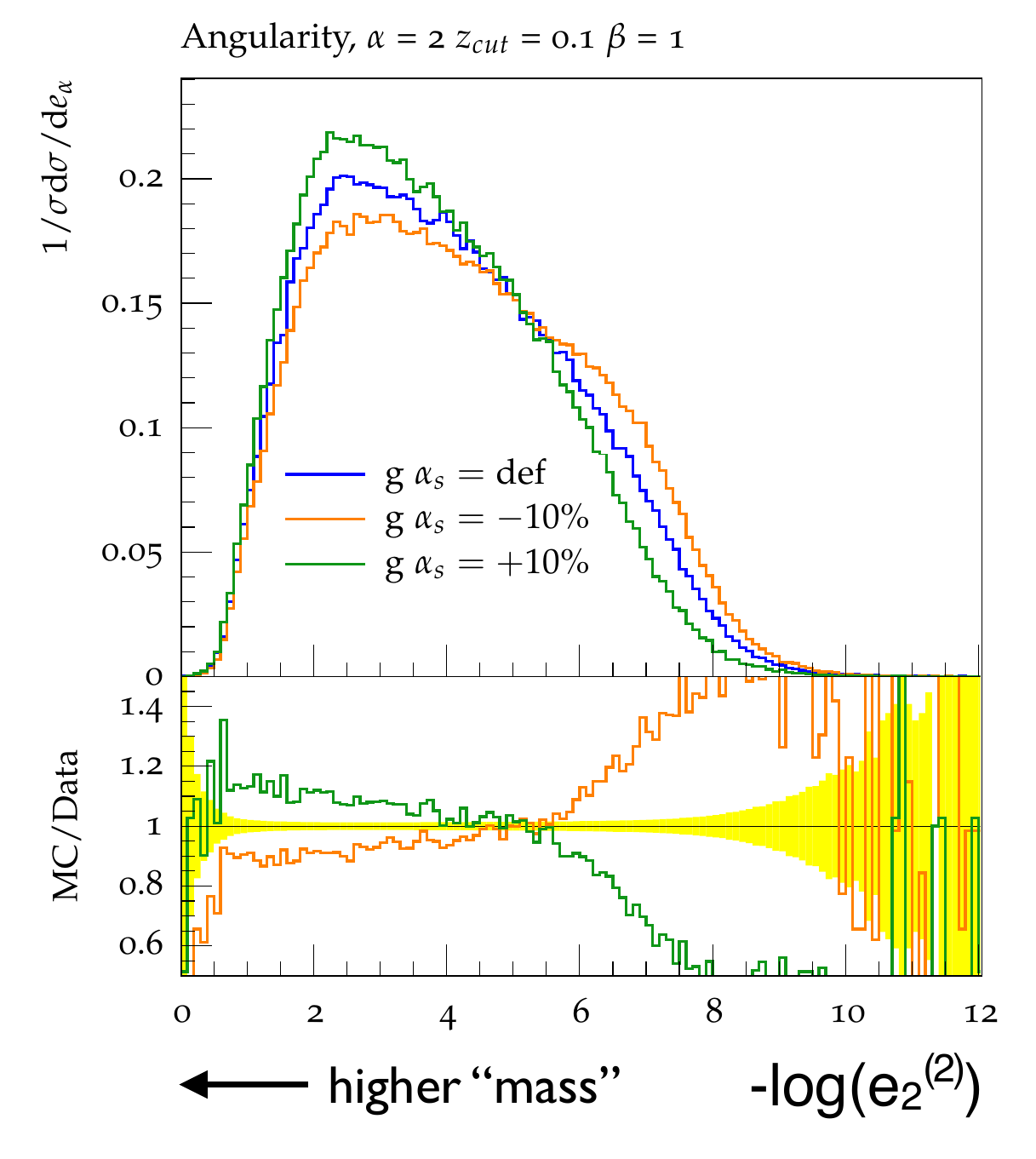}
\end{center}
\caption{The sensitivity to $\alpha_s$ in \textsc{Herwig~7}.  Shown is the distribution of the normalized squared jet mass ($e_2^{(2)}$) for quark jets (left) and gluon jets (right), with higher values of the mass are on the left.  The blue line uses $\alpha_s=0.118$ while the green and orange lines have the value of $\alpha_s$ varied by $10\%$.  The lower panels show the ratio with respect to the $\alpha_s=0.118$ curve.}
\label{fig:SM_jetsub_alphas:sensitivity}
\end{figure}

\begin{figure}[p]
\begin{center}
\includegraphics[width = 0.48\columnwidth]{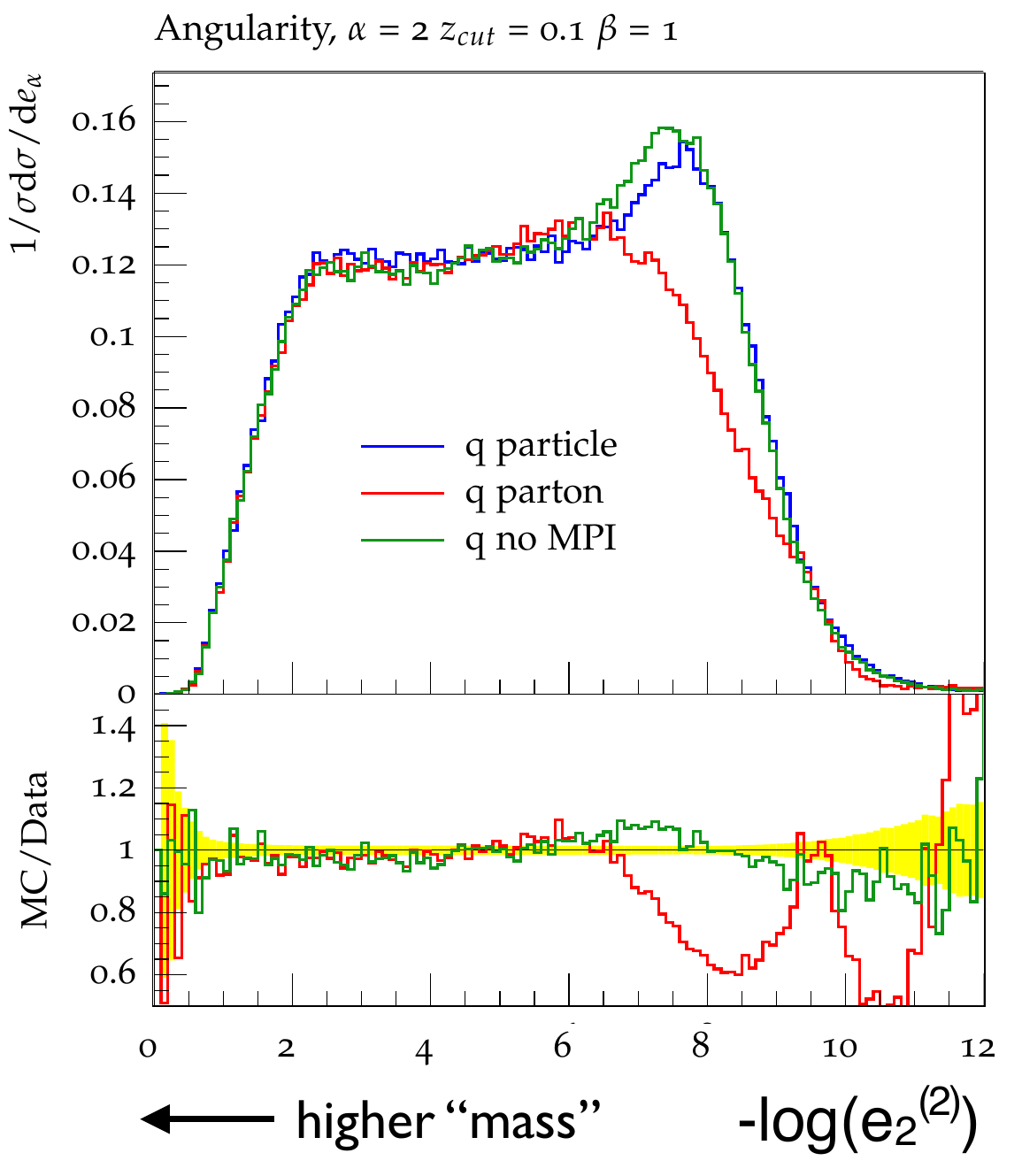}
\includegraphics[width = 0.49\columnwidth]{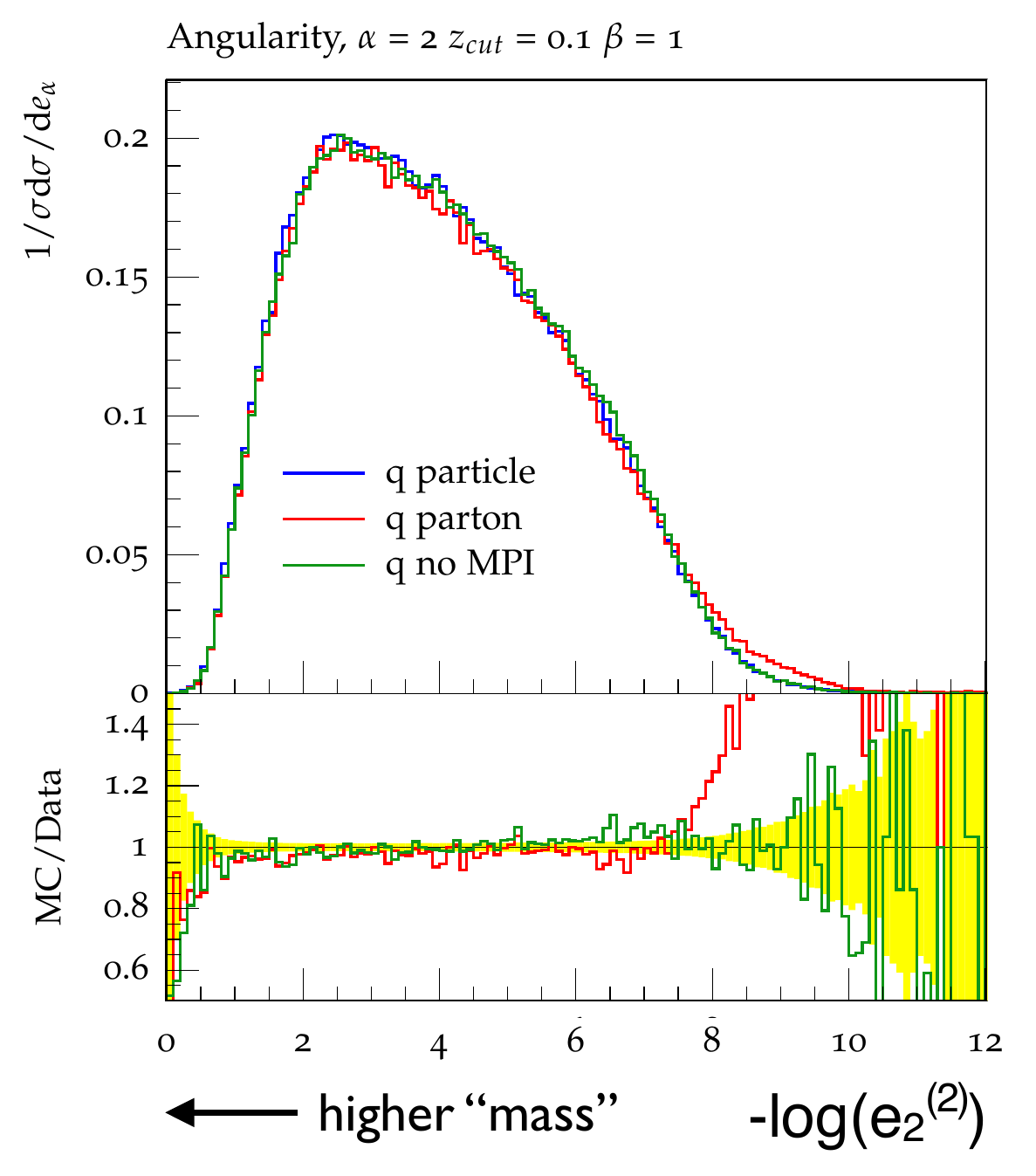}
\end{center}
\caption{The robustness to NP effects in \textsc{Herwig~7}, with the same distributions as Fig.~\ref{fig:SM_jetsub_alphas:sensitivity}.
The blue line shows the default particle-level simulation that includes the standard cluster hadronization model.
The red curve has hadronization turned off and the green curve is the same as the blue, but with the \textsc{Herwig~7} model for multiple parton interactions (MPI) turned off.
MPI is also off for the red curve.}
\label{fig:SM_jetsub_alphas:robustness}
\end{figure}

\begin{figure}[t]
\begin{center}
\includegraphics[width = 0.99\columnwidth]{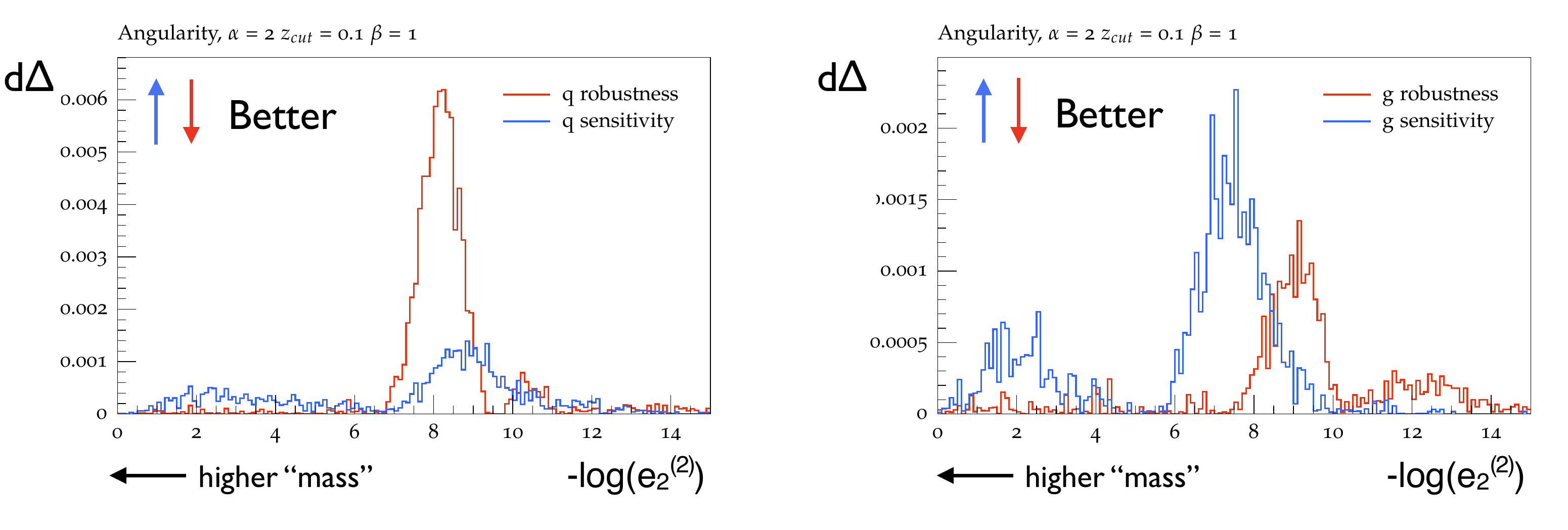}
\end{center}
\caption{The integrand of Eq.~\eqref{eq:SM_jetsub_alphas:seppower} in \textsc{Herwig~7}.  Shown is the normalized squared jet mass ($e_2^{(2)}$)  for quark jets (left) and gluon jets (right), with higher values of the mass are on the left.  The baseline $f$ function from Eq.~\eqref{eq:SM_jetsub_alphas:seppower} is the same for the red and blue curves.  For blue (sensitivity) the $g$ function is from varying $\alpha_s$ by 10\%, whereas for red (robustness) the $g$ function hadronization is turned off.  Note the different vertical scale in the left and right plots. }
\label{fig:SM_jetsub_alphas:differentialseparation}
\end{figure}

The sensitivity to $\alpha_s$ is shown in Fig.~\ref{fig:SM_jetsub_alphas:sensitivity}, in the case of normalized squared jet mass ($e_2^{(2)}$) for quark and gluon jets.%
\footnote{Here, we have varied the $\alpha_s$ value used in the PS only.  For future work, it would be interesting to test the sensitivity to $\alpha_s$ in the hard matrix element and PDFs as well.}
To the left of the low-mass NP peak, the shape of the $e_2^{(2)}$ distribution is nearly flat for quarks and nearly linear (in the log-space) for gluons, as expected from Eq.~\eqref{eq:SM_jetsub_alphas:ecf_ll_dsitribution}.
Increasing $\alpha_s$ shifts the quark distribution up but has nearly no impact on the shape of the distribution below the peak.
The size of the NP peak is significantly impacted by the value of $\alpha_s$, with smaller $\alpha_s$ values implying that NP effects are more important to the groomed mass distribution.
In contrast, the slope of the distribution for gluons does change with the variations in $\alpha_s$, but there is no low-mass NP peak.

The robustness to NP effects is shown in Fig.~\ref{fig:SM_jetsub_alphas:robustness}, using the same $e_2^{(2)}$ distribution but now with variations in the modeling of hadronization and MPI effects.
Overall, there is excellent stability in the $e_2^{(2)}$ distribution.
The low-mass peak for quarks is almost entirely due to hadronization effects.
According to \textsc{Herwig~7}, the impact of hadronization is much smaller for gluon jets, as expected since perturbative effects push the distribution away from the NP-sensitive region in Eq.~\eqref{eq:SM_jetsub_alphas:np}.

Figure~\ref{fig:SM_jetsub_alphas:differentialseparation} shows the distribution of the differential separation power (integrand of Eq.~\eqref{eq:SM_jetsub_alphas:seppower}), using the variation with $\alpha_s$ as the sensitivity and the change from turning off hadronization as the robustness.
For quark jets, Figs.~\ref{fig:SM_jetsub_alphas:sensitivity} and \ref{fig:SM_jetsub_alphas:robustness} showed that the biggest variations with $\alpha_s$ occurred at low mass which is also where NP effects are largest.
This corresponds to the peak in the blue and red distributions in Fig.~\ref{fig:SM_jetsub_alphas:differentialseparation} occurring in nearly the same location.
In contrast, the peaks are more well-separated in Fig.~\ref{fig:SM_jetsub_alphas:differentialseparation} for gluon jets and the blue is shifted to higher mass values where there is more perturbative control.
This suggests that gluon-enriched samples are going to play an important role for $\alpha_s$ determination from jet substructure.

\begin{figure}[t]
\begin{center}
\includegraphics[width = 0.6\columnwidth]{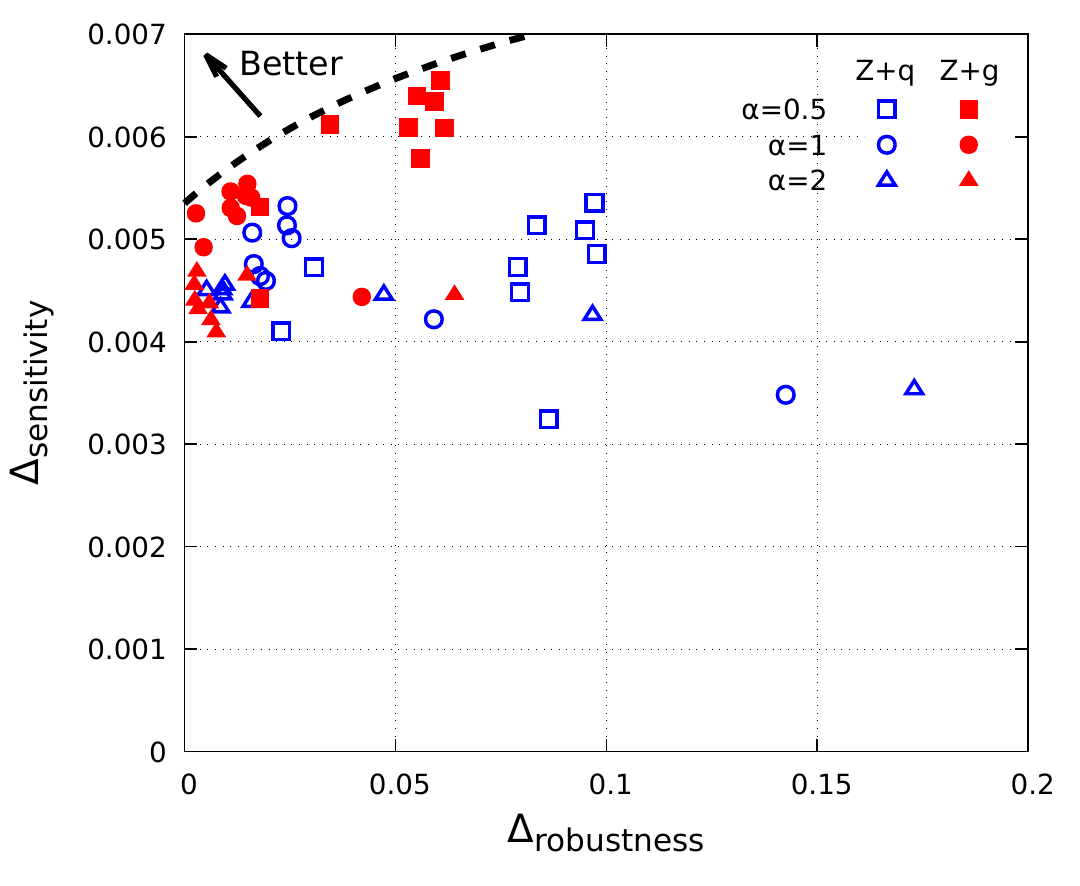}
\end{center}
\caption{The tradeoff between sensitivity and robustness for 27 two-point correlator/jet grooming combinations.  Open blue symbols represent quark jets and closed red symbols represent gluon jets.}
\label{fig:SM_jetsub_alphas:robseptradeoff}
\end{figure}

A summary of the sensitivity-robustness tradeoff for many two-point correlators is presented in Fig.~\ref{fig:SM_jetsub_alphas:robseptradeoff}.  As already observed for the jet mass, gluons tend to have superior sensitivity and robustness compared with quarks.
This is not surprising, as gluons have more perturbative radiation than quarks ($C_A>C_F$).
The jet mass has $(\Delta_\mathrm{sensitivity},\Delta_\mathrm{robustness})=(0.097,0.0043)$, $(0.015,0.0046)$ for quarks and gluons, respectively, using $\beta=0$ and $z_\mathrm{cut}=0.1$.
The groomed two-point correlators with the best quark and gluon sensitivity and robustness have $\alpha=1$, $z_\mathrm{cut}=0.05$ and $\beta=1$ or $\beta=2$.
The choice of $\alpha=1$ correspond to $k_t$ (a.k.a.~width) instead of mass, which may not be so surprising since the scale of $\alpha_s$ is set by $k_t$ and not mass.
Interestingly, $k_t$ with $\beta=0$ has significantly worse sensitivity than $\beta=1$ or $\beta=2$, highlighting the importance of having some double-logarithmic information. 
This study suggests that $k_t$ observables are important to include in future experimental and theoretical studies. The three-loop anomalous dimensions for the Energy-Energy Correlator (EEC) in $e^+e^-$, which is a $k_t$ sensitive observable, were recently derived \cite{Moult:2018jzp}. This will allow for a study of $k_t$ sensitive observables at N$^3$LL in $e^+e^-$, and a comparison with mass-type observables. 
It would also be interesting to study how the robustness versus sensitivity picture changes when considering only subsets of the available observable range.

\subsubsection{The Issue of Casimir Scaling}
\label{sec:SM_jetsub_alphas:casimir}

While we have illustrated that groomed jet mass provides excellent sensitivity to $\alpha_s$, particularly for gluon jets, one problem that is immediately clear from Sec.~\ref{sec:SM_jetsub_alphas:analytic} is that the leading behavior of the distributions is always dominated by the product $\alpha_s C_i$.
While this is broken at higher perturbative orders, it implies that at lowest order there is a complete degeneracy of the value of $\alpha_s$ and the quark versus gluon fraction of jets.
This problem is not faced for dijet event shapes at $e^+e^-$ colliders, which are almost entirely quark dominated.

There are a variety of different approaches to overcome this problem, each with their own advantages and disadvantages.
First, the quark and gluon fractions are perturbatively calculable given the parton distribution functions (PDFs).
Therefore, perturbatively calculating the quark and gluon fractions inputs the most possible information, and should correspondingly lead to the best sensitivity for $\alpha_s$.
This has the downside, however, of also introducing sensitivity to the PDFs, which in principle should be fitted along with $\alpha_s$~\cite{Accardi:2016ndt}.
This difficulty also enters into other $\alpha_s$ extractions at the LHC, for example the $3$-$2$ jet rate~\cite{Chatrchyan:2013txa} or the energy-energy correlator~\cite{ATLAS:2015yaa,Aaboud:2017fml}.
One of the key hopes of using jet substructure was that the sensitivity to the PDFs could be minimized, but that is not the case if the quark and gluon fractions cannot be determined independently.

\begin{figure}[t]
\begin{center}
\includegraphics[width = 0.4\columnwidth]{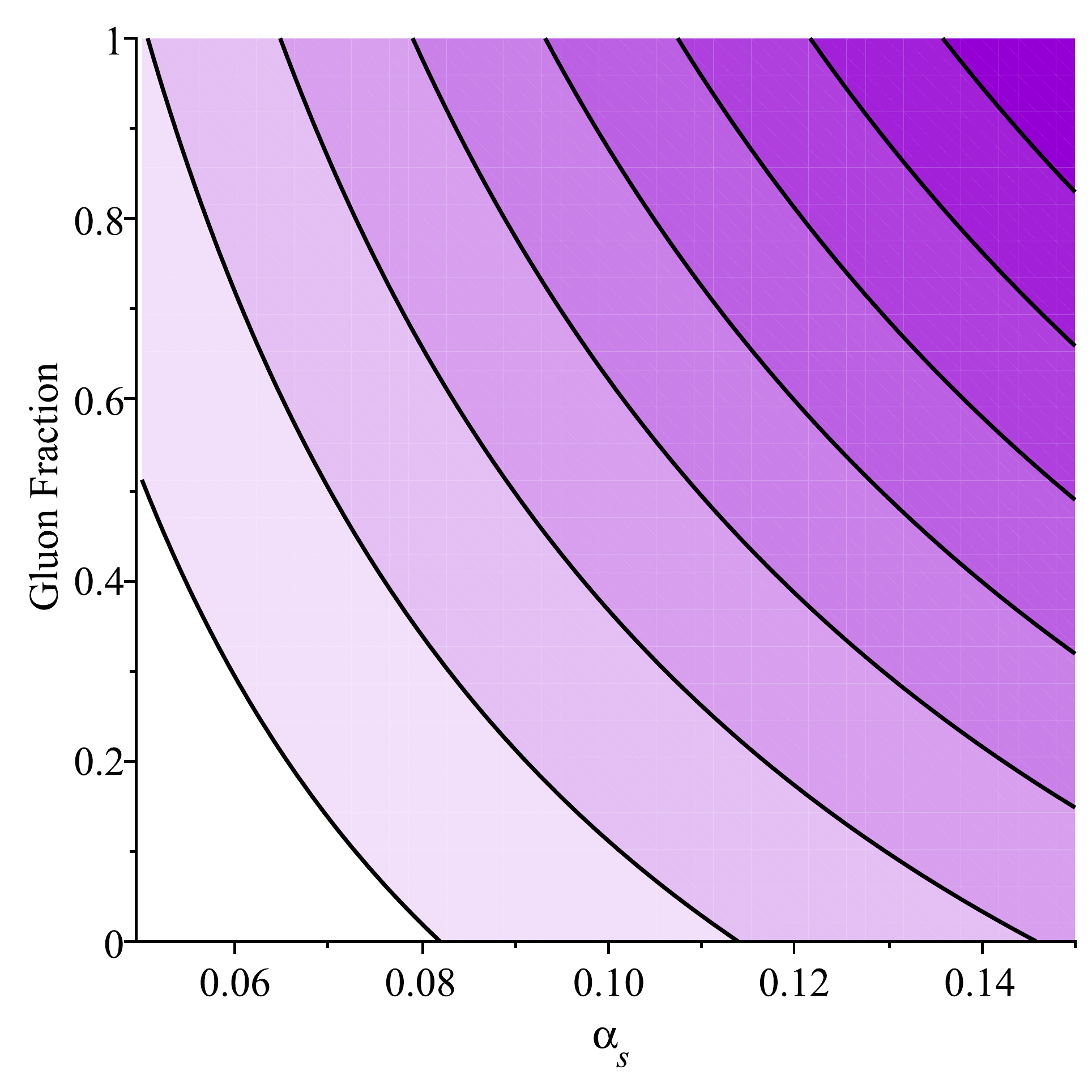}
\end{center}
\caption{The slope of the probability distribution of $\log(e_2^{(2)})$ is proportional to $\alpha_s(C_Af_g+C_F(1-f_g))$, which is plotted above.  The degeneracy between $\alpha_s$ and $f_g$ (gluon fraction) are represented by the banana-shaped isocontours. }
\label{fig:SM_jetsub_alphas:analyticbanana}
\end{figure}

A second approach is to simultaneously fit for $\alpha_s$ and the quark/gluon fraction.
In the resummation regime, the jet mass distribution only depends on $\alpha_s$ and the quark/gluon fraction.\footnote{At high mass, in the fixed-order regime, there is still residual explicit dependence on PDFs as is the case for the total cross-section.  This regime also suffers from sample-dependent effects where the quark and gluon substructure distributions have residual dependence on the hard process.}
Due to the fact that the two fractions must add to unity, this introduces a single additional parameter in the fit (see Fig.~\ref{fig:SM_jetsub_alphas:analyticbanana}).
The degeneracy between the quark/gluon fraction and $\alpha_s$ is broken by higher-order effects.
Furthermore, different $e_2^{(\alpha)}$ have different dependence on $\alpha_s$ and $C_i$ at higher orders.
Therefore, the measurement of multiple two-point correlators would allow the degeneracy to be broken.
From a theoretical perspective, this significantly complicates the analysis, since it would require precise
predictions to be made for the joint distribution of multiple two-point correlators.

In our fitting study in Sec.~\ref{sec:SM_jetsub_alphas:ben_study}, we consider both of the above approaches.
It would be interesting to develop other approaches to disentangling the quark and gluon fractions and $\alpha_s$.
Without some kind of conceptual breakthrough, though, we expect that the quark/gluon fraction will be a limiting aspect of $\alpha_s$ extractions from jet substructure at the LHC.

\subsubsection{Normalized vs.\ Unnormalized Distributions}
\label{sec:SM_jetsub_alphas:norm}

In addition to the complication of quark and gluon fractions, another issue which appears for the extraction of $\alpha_s$ from jet substructure is the issue of normalization.\footnote{We thank Gavin Salam for interesting discussions on this topic. }
Unlike for $e^+e^-$ event shapes, the Born dijet cross-section in $pp$ is sensitive to the value of $\alpha_s$.
This implies that the rate itself, in particular the absolute quark and gluon jet rates, carries information regarding $\alpha_s$. 

In our studies, we have decided to focus on using normalized distributions.
At a conceptual level, this is because we want to perform a true measurement of $\alpha_s$ from jet substructure, which is not dominated by the overall jet rate.
But there are two other practical concerns that favor using normalized distributions.

First, it is currently only possible to perform precise measurements of the groomed mass using normalized distributions.
Experimentally, the absolute rate is determined by the acceptance from kinematic requirements on the jet $p_T$.
The sophisticated in-situ jet energy scale and resolution program carried out for small-radius jets~\cite{Aad:2014bia,Aaboud:2017jcu,Khachatryan:2016kdb,CMS-DP-2016-020} has not yet reached the same level of maturity for groomed large-radius jets.
That said, this is simply a matter of time, and experimental efforts have already started in this direction~\cite{ATLAS-CONF-2017-063}.
A key challenge that still remains is to fully understand the correlations in the calibrations and uncertainties between the jet energy and the jet substructure observables.

\begin{figure}[t]
\begin{center}
\includegraphics[width = 0.5\columnwidth]{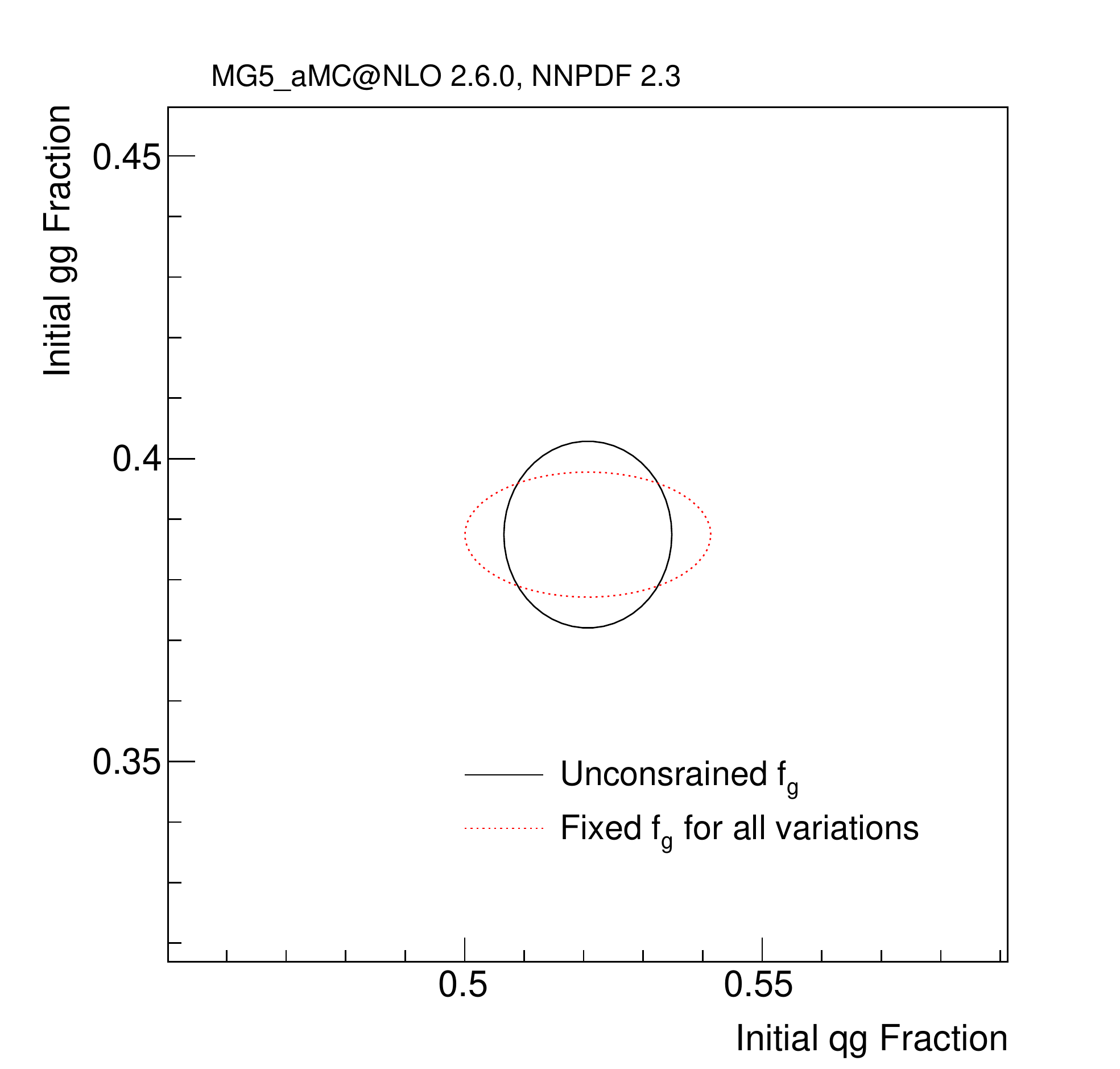}
\end{center}
\caption{The fraction of $qg$ and $gg$ initial states for leading-order dijet production with $p_T>200$ GeV, simulated at leading order with MG5\_aMC 2.6.0~\cite{Alwall:2014hca} using NNPDF 2.3~\cite{Ball:2012cx}.  The ellipses correspond to the uncertainty from the 100 error PDF sets.  For the red dashed line, the fraction of outgoing gluon jets $f_g$ is constrained to be the same for all variations (and equal to the nominal PDF set).}
\label{fig:SM_jetsub_alphas:pdf}
\end{figure}

Second, the use of normalized distributions minimizes the sensitivity to the PDFs.
As discussed in Sec.~\ref{sec:SM_jetsub_alphas:pertsimplicity}, grooming renders the quark and gluon two-point-correlator distributions universal (in the resummation regime).
Therefore, the measured distribution only depends on the fraction of gluon jets $f_g$ that pass the event selection.
In contrast, the total cross-section depends on the relative proportions of all possible partonic initial states, which introduces a source of uncertainty that is not present for the normalized cross-section.
For example, the initial state could be one of $qq$, $qg$ or $gg$.
The relative proportions can be parameterized by two numbers $f_{qg}$ and $f_{gg}$ where $f_{qg}+f_{gg}=1-f_{qq}$.
Figure~\ref{fig:SM_jetsub_alphas:pdf} shows the uncertainty in $f_{qg}$ and $f_{gg}$ from leading-order PDFs.
Fixing $f_g$, which carries all of the PDF sensitivity for the shape measurement, has little effect on the uncertainty in $f_{qg}$ and $f_{gg}$.
Thus, using unnormalized distributions results in additional PDF sensitivity from also measuring the total cross-section in addition to the shape of the substructure distribution. 

From the perspective of perturbative accuracy, there is an important issue with using normalized distributions, which is that jet-shape observables start at $\mathcal{O}(\alpha_s)$.
Specifically, the slope of the groomed mass distribution is $\mathcal{O}(\alpha_s)$.
This implies that to have an $\mathcal{O}(\alpha_s^2)$ uncertainty on the slope (as required to gain entry into the PDG world avarage), one needs to have $\mathcal{O}(\alpha_s^3)$ control, namely the NNLO $2 \to 3$ process for a jet with two constituents.
For the case of $e^+e^-$, this level of accuracy has been achieved, where the NNLO corrections to $e^+e^-\to 3$ jets are known and indeed used in extractions of $\alpha_s$.
Due to recent progress in the calculation of the relevant amplitudes, we believe this is a realistic target for jet substructure, but clearly requires substantial further work.

\subsection{Idealized Performance at the LHC}
\label{sec:SM_jetsub_alphas:ben_study}

The purpose of this subsection is to make some numerical estimates regarding $\alpha_s$ extraction from jet substructure at the LHC.
Many simplifying assumptions are made, with the goal of motivating a more complete effort within the context of ATLAS and CMS in collaboration with theorists.
First, we illustrate how $\alpha_s$ and the gluon fraction can be simultaneously extracted from the distribution of various two-point correlators.
Next, we estimate the needed experimental precision required to make a useful measurement of $\alpha_s$.
While both the theory and experimental precision will continue to improve over the next years, the community has already demonstrated that the work can begin with the first round of groomed jet mass results~\cite{Frye:2016okc,Frye:2016aiz,Marzani:2017mva,Marzani:2017kqd,Aaboud:2017qwh,CMS-PAS-SMP-16-010}.

\subsubsection{Extraction of Theory Templates}
\label{sec:SM_jetsub_alphas:templates}

\begin{figure}[t]
\begin{center}
\includegraphics[width = 0.45\columnwidth]{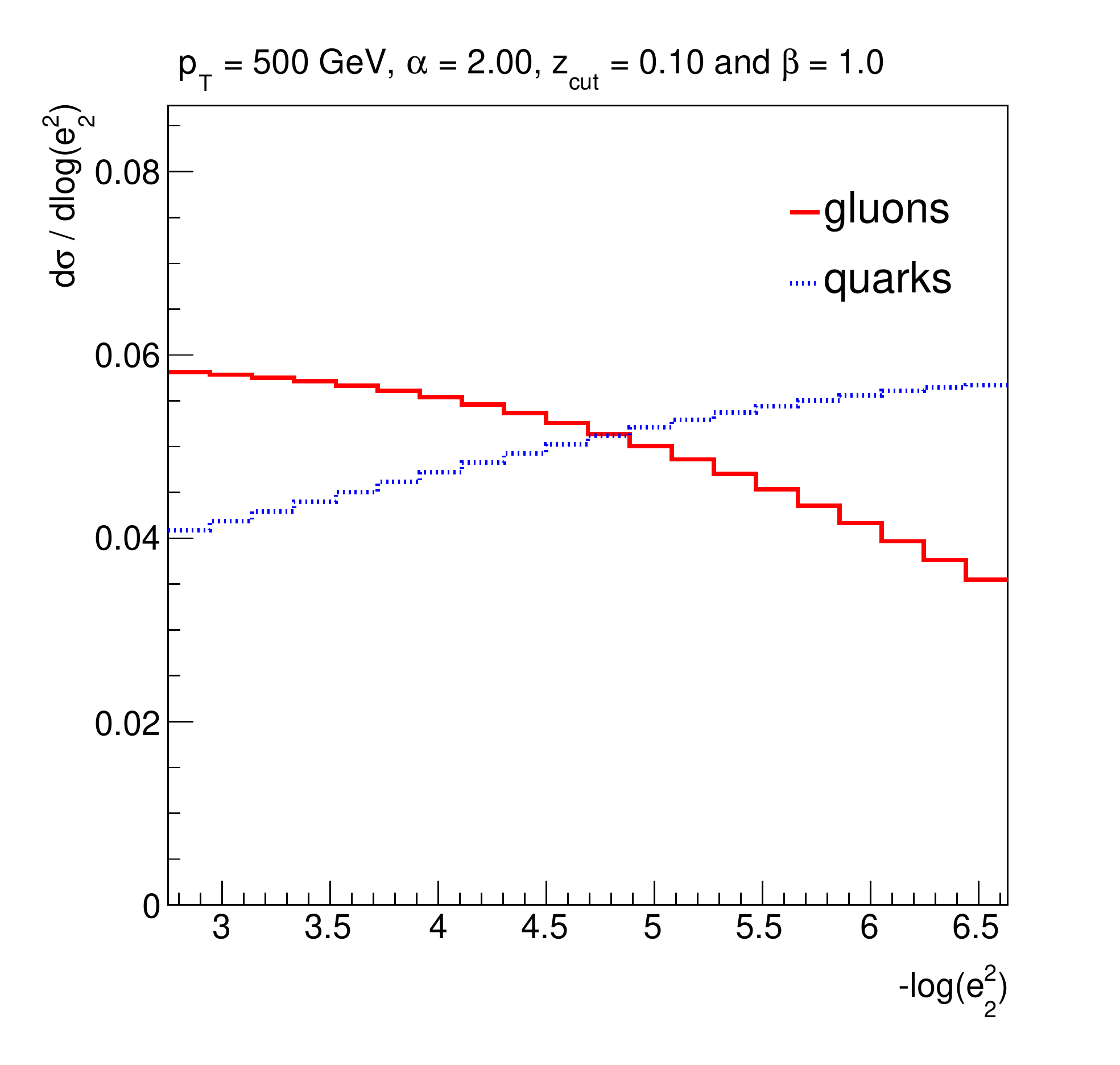}
\includegraphics[width = 0.45\columnwidth]{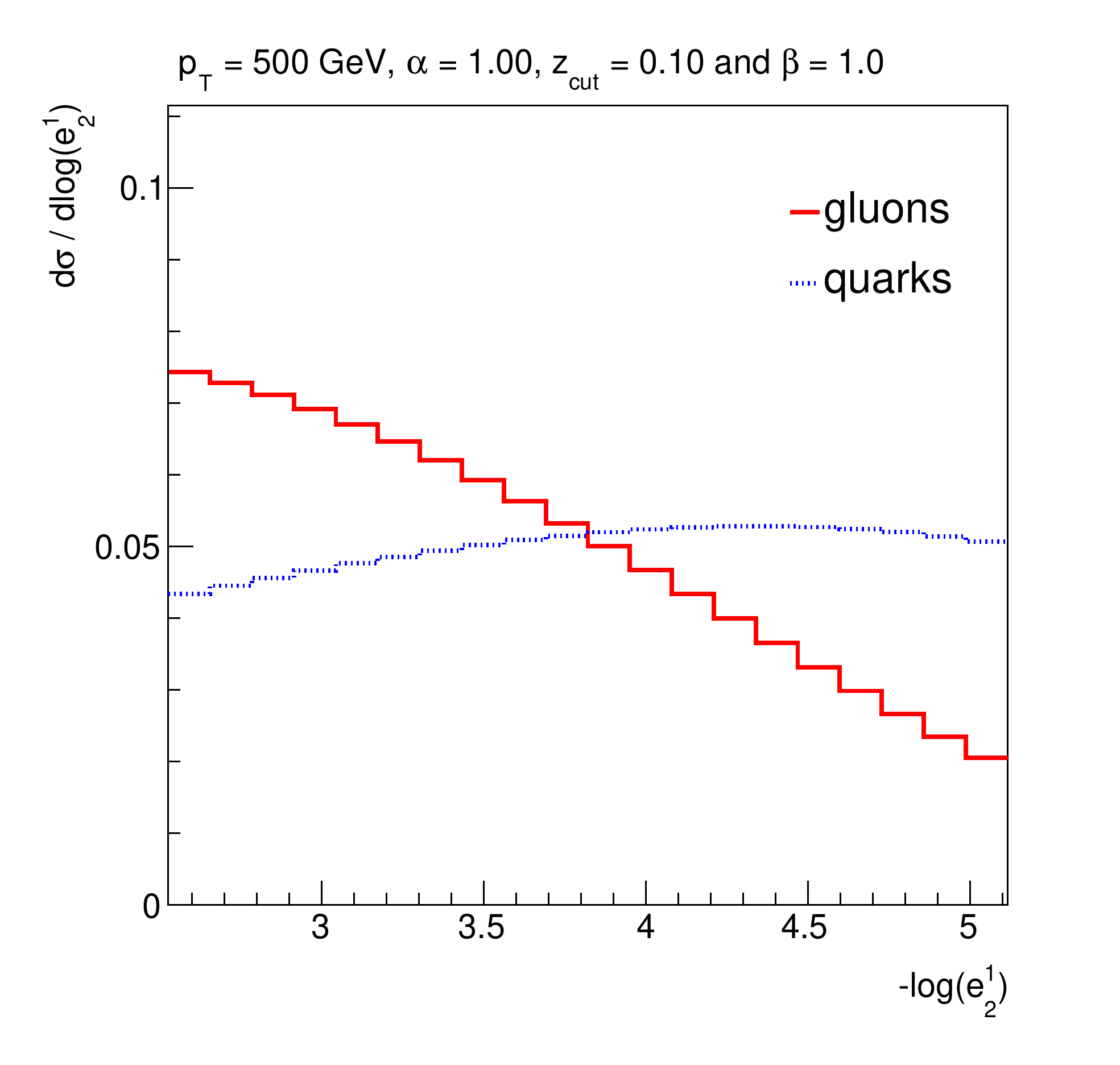}\\
\includegraphics[width = 0.45\columnwidth]{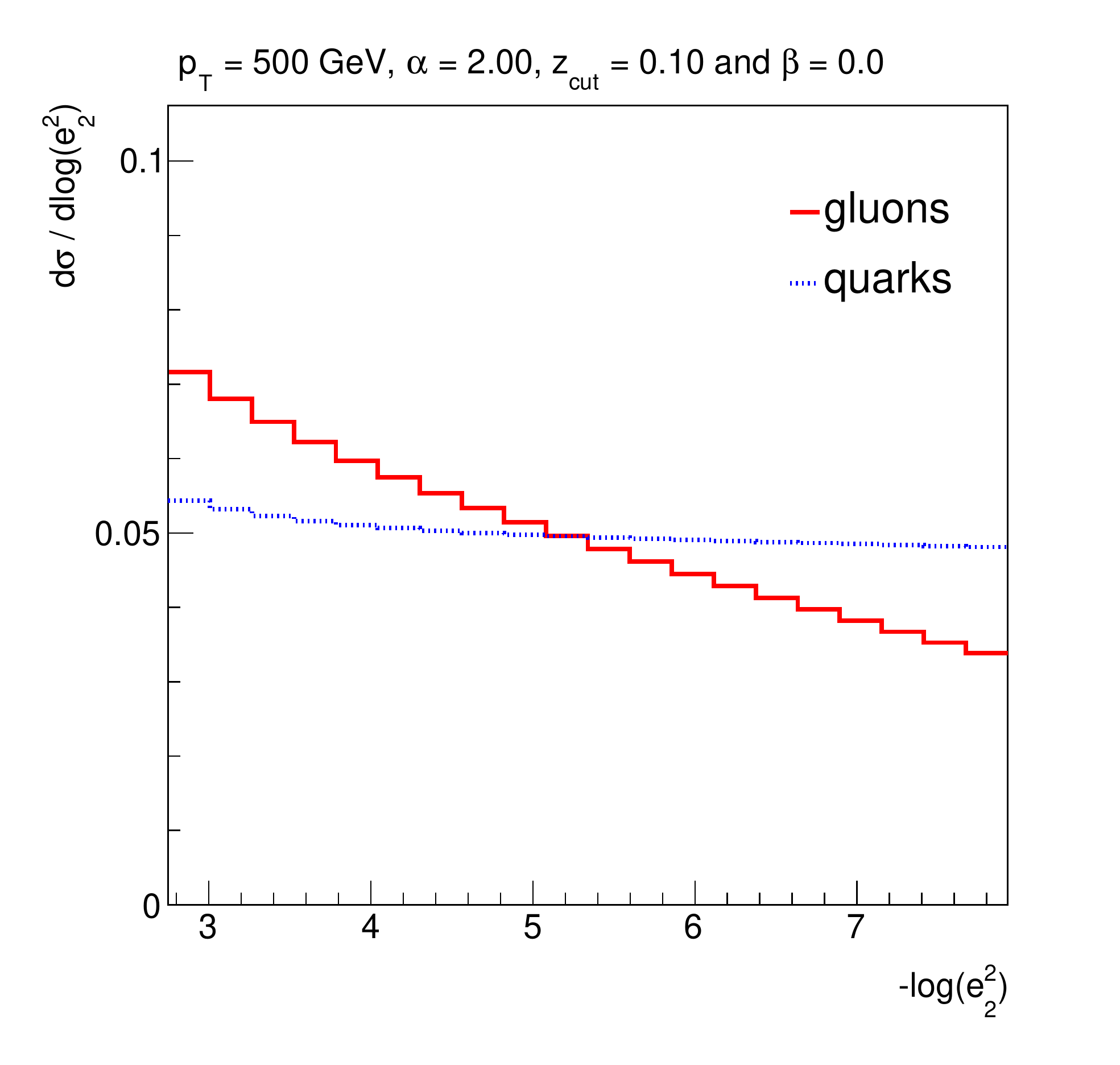}
\includegraphics[width = 0.45\columnwidth]{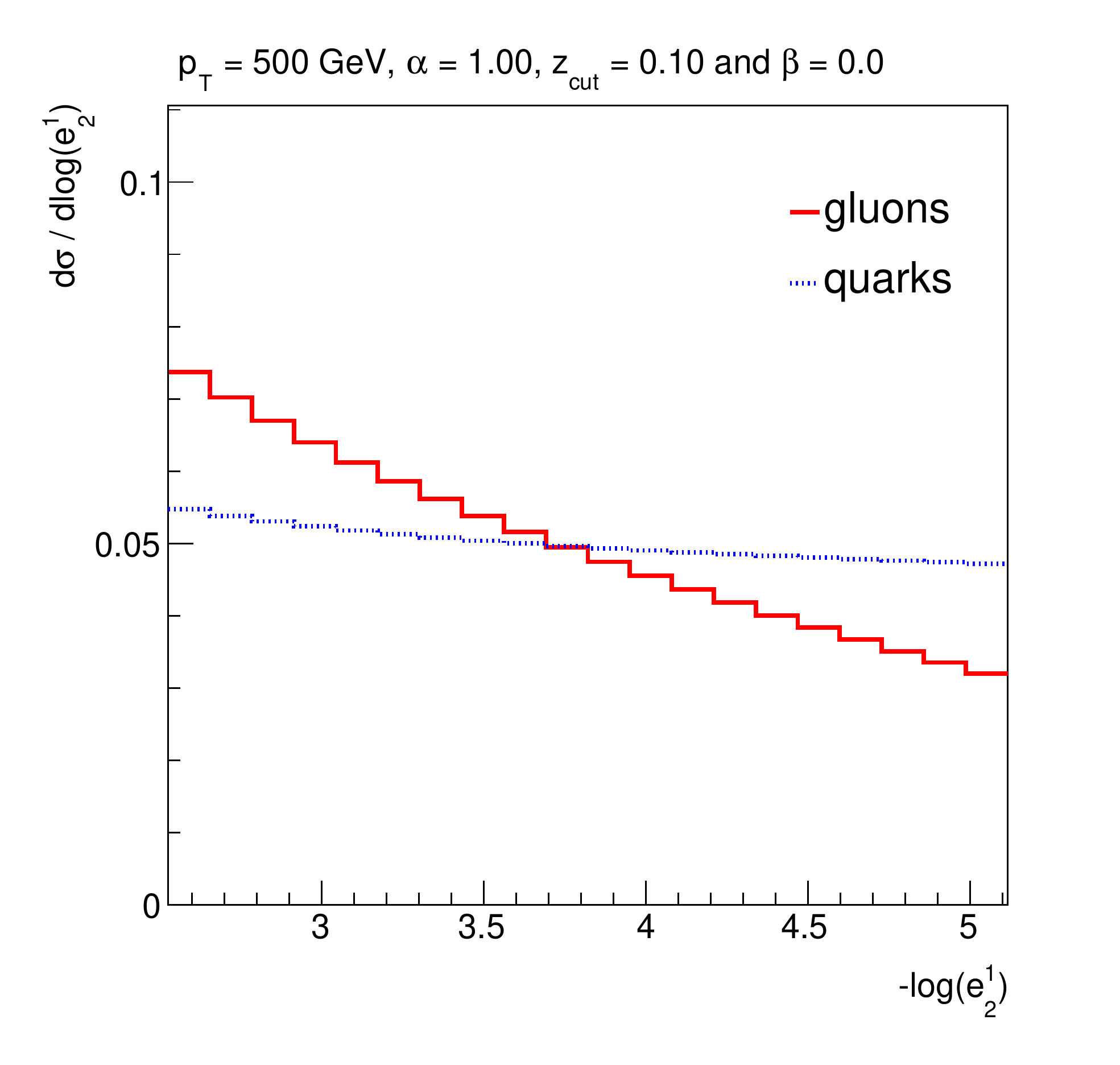}
\end{center}

\caption{The quark and gluon templates, computed at NLL~\cite{Marzani:2017mva,Marzani:2017kqd} for $\alpha=2$ (left column) and 
  $\alpha=1$ (right column), with grooming parameters $z_{\mathrm{cut}}  = 0.1$ and $\beta = 1$ (top row) and $\beta = 0$ (bottom row).  Note that larger masses are on the left so that the NP regime is on the
  right and the fixed-order regime is on the left.}
\label{fig:SM_jetsub_alphas:templates}
\end{figure}

A complete extraction of $\alpha_s$ will require matching resummed results to high fixed order and also estimating NP effects.
The two sets of predictions for dijets thus far have been matched to LO~\cite{Frye:2016okc,Frye:2016aiz} and NLO~\cite{Marzani:2017mva,Marzani:2017kqd} and have used hadronization models to study NP corrections~\cite{Marzani:2017mva,Marzani:2017kqd}.
Performing high-order fixed-order matching is conceptually straightforward but computationally expensive; while this will be required eventually, we focus here on a demonstration without matching.
Therefore, we isolate the resummation regime,
\begin{equation}
\label{eq:SM_jetsub_alphas:e2truncation}
\left. e_2^{(\alpha)}\right |_{\mathrm{NP}} \leq e_2^{(\alpha)}\leq z_\mathrm{cut}R^\alpha,
\end{equation}
where $\left. e_2^{(\alpha)}\right |_{\mathrm{NP}}$ is given in Eq.~\eqref{eq:SM_jetsub_alphas:np}, such that regions of phase space that are highly sensitive to NP or fixed-order effects are removed.
In this range, NLL calculations exist in analytic formulae that can be varied on-the-fly~\cite{Marzani:2017mva,Marzani:2017kqd}.
Figure~\ref{fig:SM_jetsub_alphas:templates} shows the quark and gluon templates for $\alpha=\{1,2\}$ and $\beta=\{0,1\}$, truncated according to Eq.~\eqref{eq:SM_jetsub_alphas:e2truncation}.

From these NLL distributions, pseudo-data are then generated from the binned analytic probability distribution $t(\alpha_s,f_g)$.
These distributions are a superposition of the quark and gluon distributions and depend only on $\alpha_s$ and $f_g$.
Each pseudo-dataset has $n$ events and its binned representation is denoted by $h(\alpha_s,f_g,n)$.
For a given pseudo-dataset, the fitted values of $\alpha_s$ and $f_g$ are determined from a $\chi^2$-like fit:
\begin{equation}
\label{eq:SM_jetsub_alphas:chi2fit}
\alpha_s,f_g=\mathrm{argmin} \sum_i \frac{\left(h_i(\alpha_s,f_g,n)-t_i(\alpha_s,f_g)\right)^2}{\sigma(h_i(\alpha_s,f_g,n))^2},
\end{equation}
where $t_i, h_i$ are the bin content of histograms $t$ and $h$, and $\sigma(h_i)$ is the statistical uncertainty in bin $i$ of histogram $h$.
In practice, there would also be systematic uncertainties (see Sec.~\ref{sec:SM_jetsub_alphas:resolution}), but the purpose of this study is to simply illustrate the sensitivity to $\alpha_s$ and $f_g$ for a given number of events.

\begin{figure}[t]
\begin{center}
\includegraphics[width = 0.49\columnwidth]{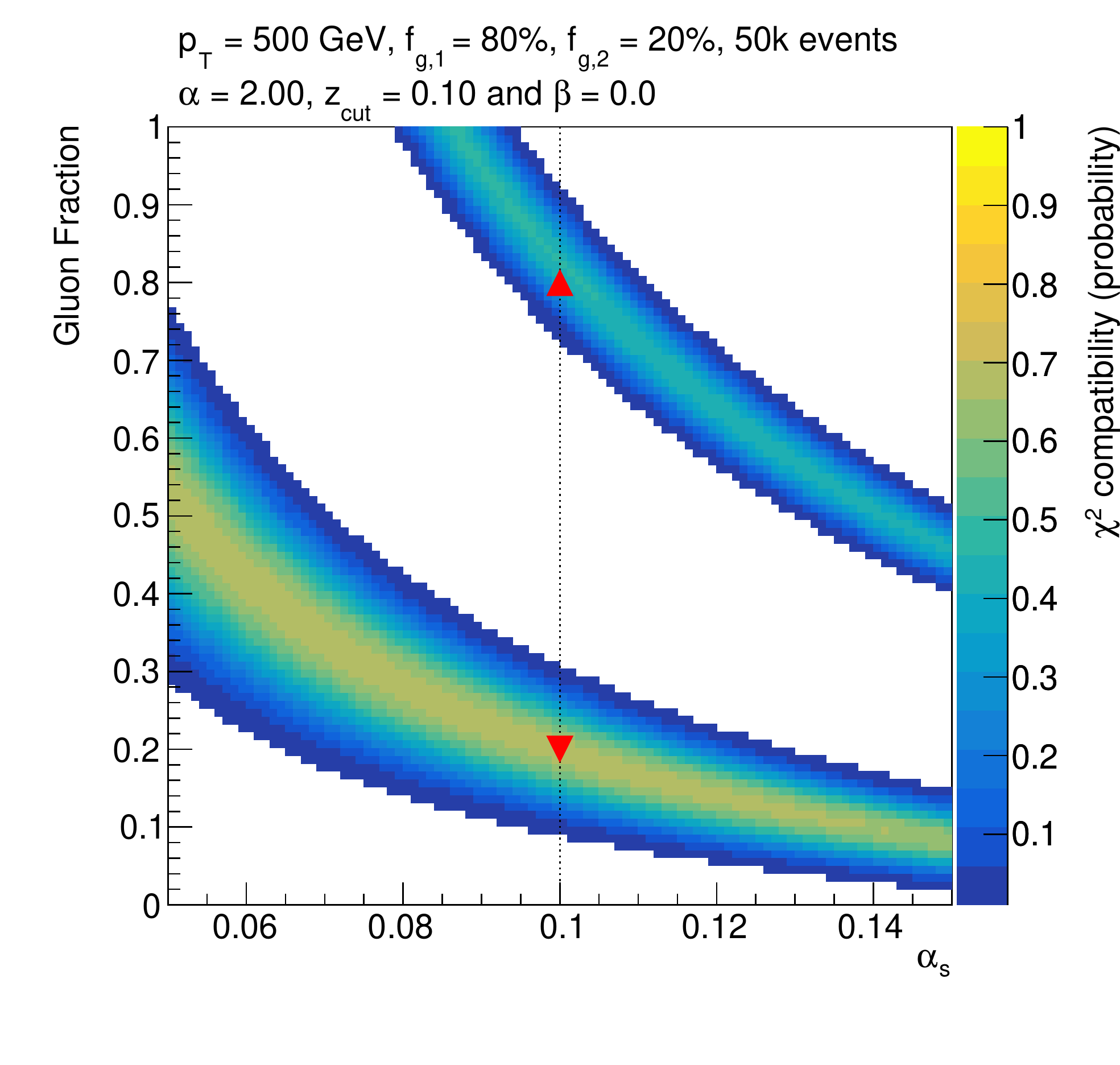}
\includegraphics[width = 0.49\columnwidth]{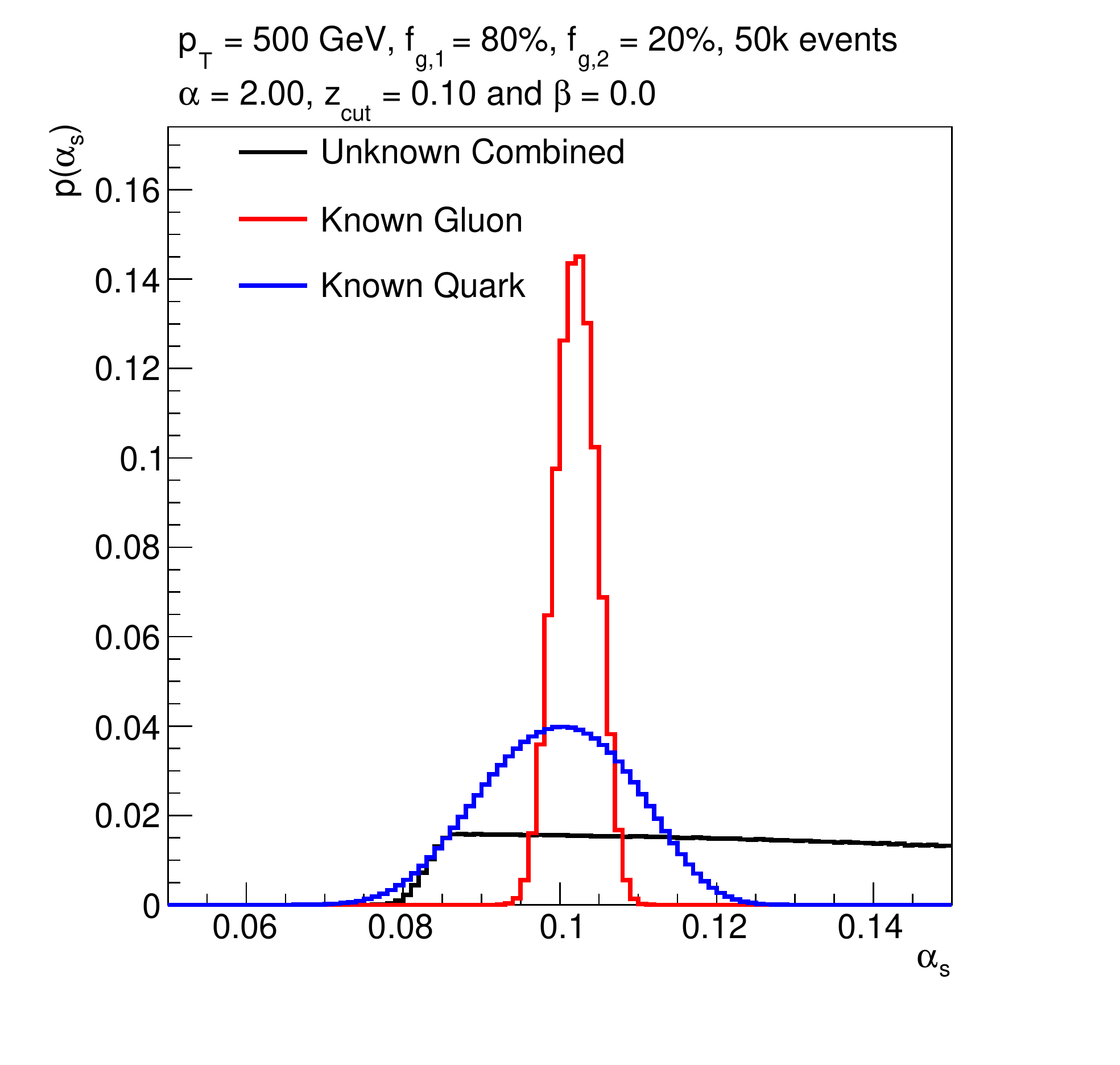}
\end{center}
\caption{Left: the probability of the minimized $\chi^2$ (assuming $n_\mathrm{bins}-1$ degrees of freedom) from Eq.~\eqref{eq:SM_jetsub_alphas:chi2fit} as a
  function of $f_g$ and $\alpha_s$ for one sample with 80\% gluons and another sample with 80\% quarks.  The true value of $\alpha_s$ is 0.1, as indicated by triangle markers.  Right: The right plot marginalized over $f_g$ and normalized to unity.  The three lines correspond to the fit performed on a pure sample of quarks, a pure sample of gluons, or a mixed sample of ($f_g\in\{0.2,0.8\}$) where the fractions are not known a priori.  This is the result from one pseudo-experiment with 100k events.}
\label{fig:SM_jetsub_alphas:alpha2fit}
\end{figure}

An example fit is demonstrated in Fig.~\ref{fig:SM_jetsub_alphas:alpha2fit} for the case of $\alpha=2$ and $\beta = 0$.
The left plot of Fig.~\ref{fig:SM_jetsub_alphas:alpha2fit} shows the $\chi^2$ from Eq.~\eqref{eq:SM_jetsub_alphas:chi2fit} for two samples, one with 20\% gluons and one with 80\% gluons.
The true value is taken to be $\alpha_s=0.1$, and as expected, the $\chi^2$ probability is high for $f_g=0.2$ and $f_g=0.8$.%
\footnote{It is not necessarily peaked at this value because this is the result of one pseudo-experiment.  Averaging over many pseudo-experiments results in peaks at $f_g=0.2$ and $0.8$.}
The banana shapes of the curves are a consequence of the degeneracy due to Casimir scaling discussed in Sec.~\ref{sec:SM_jetsub_alphas:casimir}.
From one sample alone, there is essentially no ability to distinguish between a larger $\alpha_s$ and a smaller $f_g$; the only constraint comes from the fact $0\leq f_g\leq 1$ which results in a crude bound on $\alpha_s$.
This is shown in the right plot of Fig.~\ref{fig:SM_jetsub_alphas:alpha2fit} where the distribution is marginalized over $f_g$ and normalized to unity.
One can view this as the posterior probability of the fitted $\alpha_s$: the peak is the fitted value of $\alpha_s$ and the width is the uncertainty.
When $f_g$ is known, the uncertainty in $\alpha_s$ is significantly reduced; this is illustrated with pure quark and gluon samples.
Due to the larger color factor, the measurement with pure gluon jets is more sensitive to $\alpha_s$ than the fit using pure quark jets, as anticipated in Sec.~\ref{sec:SM_jetsub_alphas:analytic}.
Using both the $f_g=0.2$ and $f_g=0.8$ samples to fit for $\alpha_s$, one can extract a $\sim 30\%$ measurement of $\alpha_s$, but there is no clear peak at the correct value of $\alpha_s$ due to the Casimir degeneracy.  

\begin{figure}[t]
\begin{center}
\includegraphics[width = 0.32\columnwidth]{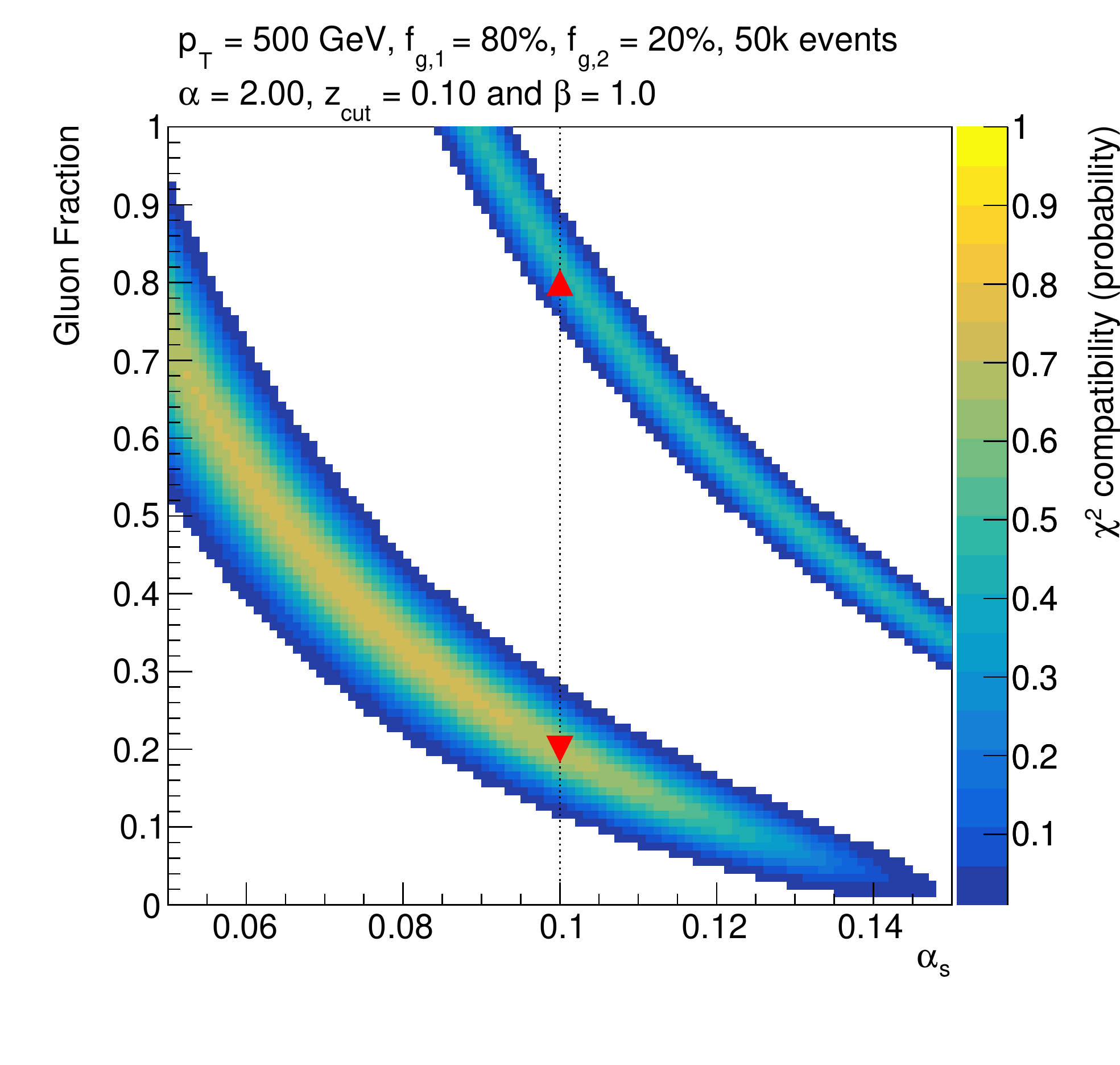}
\includegraphics[width = 0.32\columnwidth]{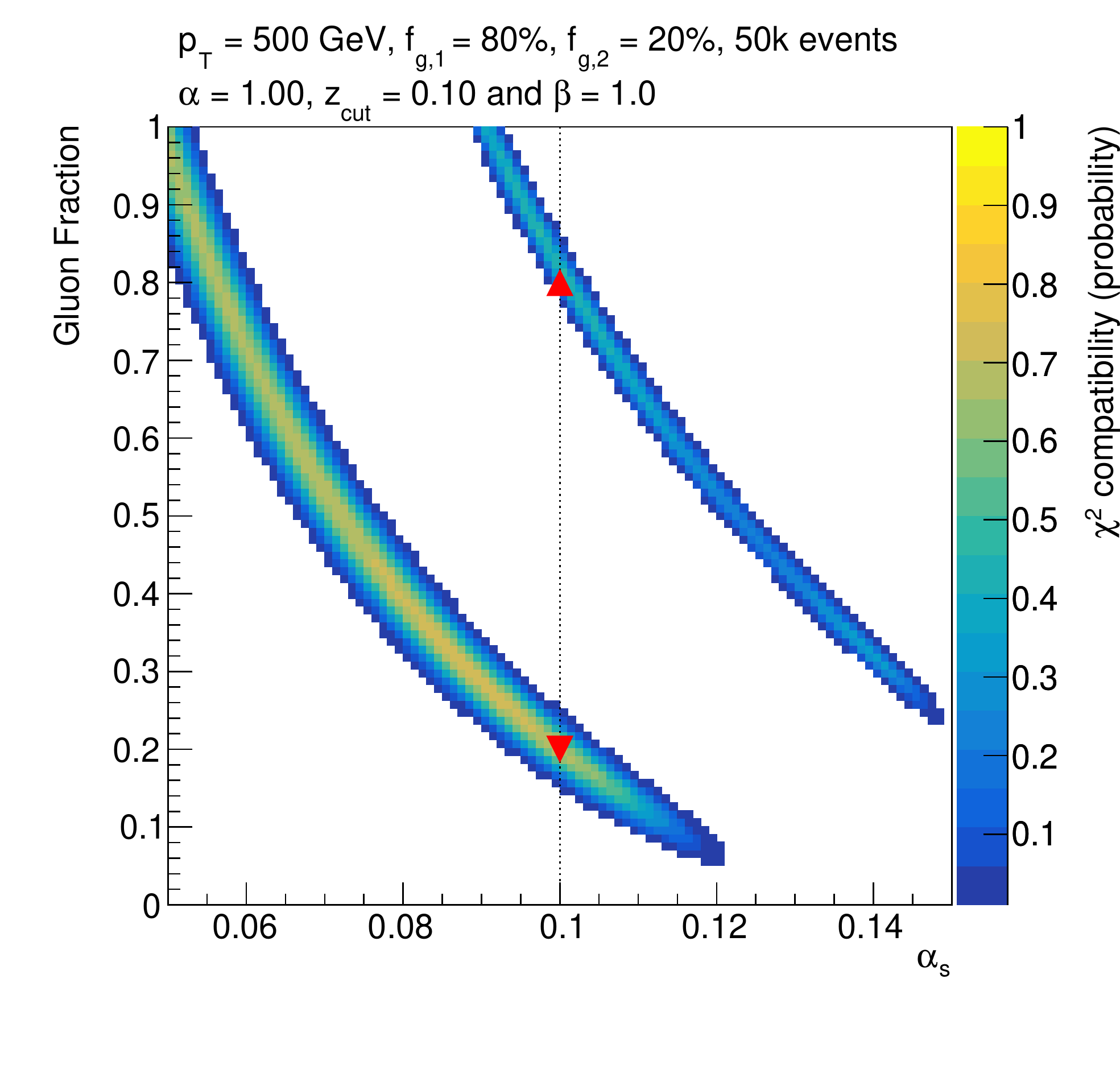}
\includegraphics[width = 0.32\columnwidth]{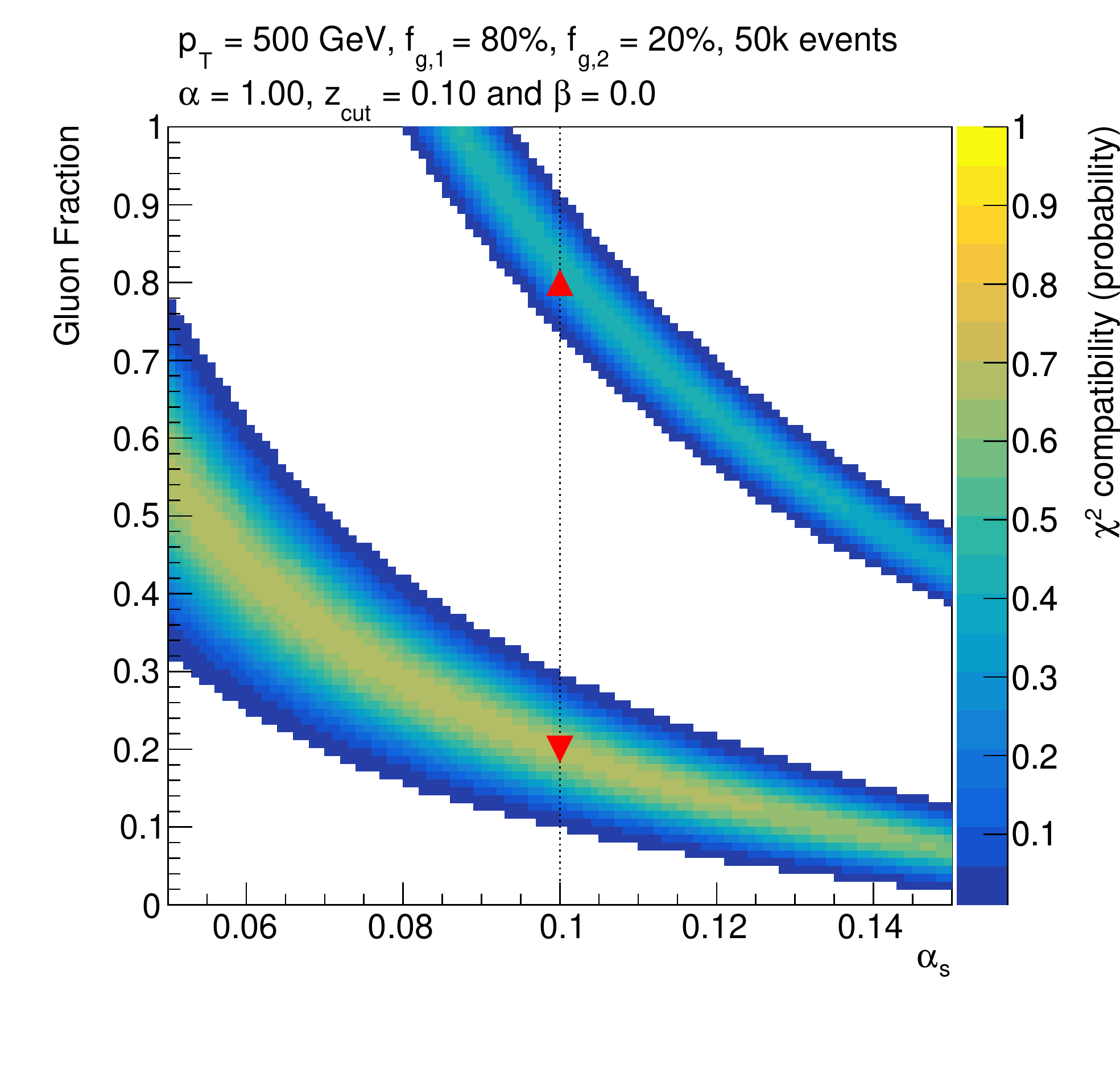}
\end{center}
\caption{The same as the left plot of Fig.~\ref{fig:SM_jetsub_alphas:alpha2fit}, but for the remaining three observables from Fig.~\ref{fig:SM_jetsub_alphas:templates}.}
\label{fig:SM_jetsub_alphas:morebananas}
\end{figure}

One way to improve the situation is to combine multiple $\alpha$, $\beta$, and $z_\mathrm{cut}$ values (only $\alpha$ and $\beta$ are varied here).
Figure~\ref{fig:SM_jetsub_alphas:morebananas} shows the $f_g,\alpha_s$ fit for all of the $\alpha,\beta$ values from Fig.~\ref{fig:SM_jetsub_alphas:templates} that were not shown in Fig.~\ref{fig:SM_jetsub_alphas:alpha2fit}.
As in Fig.~\ref{fig:SM_jetsub_alphas:alpha2fit}, there are two banana-shaped regions that correspond to the $f_g=20\%$ and $f_g=80\%$ samples.
The tilt of the bananas is slightly different than the $\alpha=2$, $\beta=0$ case and so there is a possibility to gain from combining the information in the observables.
In practice, a challenge with a multi-observable extraction strategy is the need to understand correlations between observables.
At the moment, joint distributions of two-point correlators are only known to NLL accuracy without grooming~\cite{Larkoski:2014pca}.

\begin{figure}[t]
\begin{center}
\includegraphics[width = 0.5\columnwidth]{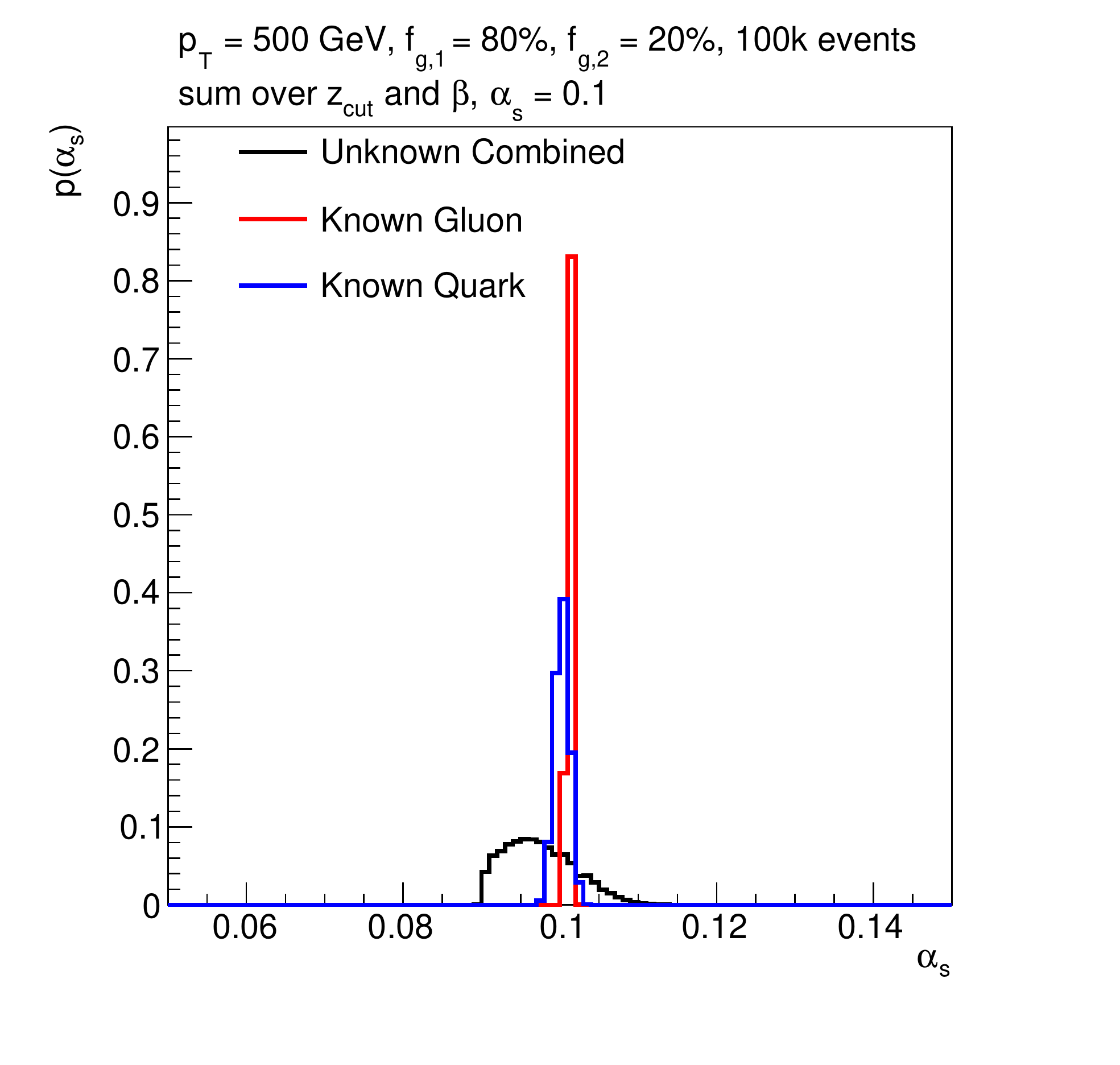}
\end{center}
\caption{Marginalizing the $f_g,\alpha_s$ fit over the gluon fraction for a fit that combines the four observables from Fig.~\ref{fig:SM_jetsub_alphas:templates}.  The unknown combined mixture uses two samples with $f_g=20\%$ and $f_g=80\%$.}
\label{fig:SM_jetsub_alphas:combo}
\end{figure}

With that caveat in mind, Fig.~\ref{fig:SM_jetsub_alphas:combo} shows the result of a combined fit assuming that all four distributions are statistically (and systematically) independent.
This is a rather strong assumption that is unlikely to be even approximately true in practice.
However, the benefit of having different tilts in the $f_g,\alpha_s$ plane is clearly shown and would be a generic feature of a multi-observable fit, even if the size of the gain is not as significant as shown here.
It is important to emphasize that the black curve in Fig.~\ref{fig:SM_jetsub_alphas:combo} assumes no prior knowledge of the gluon fraction of the event samples.
The fit can of course be improved by using some knowledge of the gluon fractions from the hard scattering process convolved with PDFs, as discussed in Sec.~\ref{sec:SM_jetsub_alphas:norm}.

\subsubsection{Estimate of Experimental Resolution}
\label{sec:SM_jetsub_alphas:resolution}

To keep pace with precise theory predictions, the experimental resolution must be well-under\-stood in order to ensure both precision and accuracy of jet substructure measurements.
To estimate the impact of detector resolution on $\alpha_s$ fit illustrated in Sec.~\ref{sec:SM_jetsub_alphas:templates}, a fast simulation from \textsc{Delphes} 3.4.1~\cite{deFavereau:2013fsa} was studied using particle-level input from \textsc{Pythia~8}.210.
This setup uses a CMS-like detector with jets built from particle-flow objects.
There are generically two regimes for determining the experimental resolution.
At high mass, there are well-resolved hard prongs in the jet, so the resolution is set by the jet energy resolution of those ``sub-jets''.
At low mass, the groomed jet is defined by nearly collinear splittings at which point the angular resolution can dominate.

\begin{figure}[t]
\begin{center}
\includegraphics[width = 0.5\columnwidth]{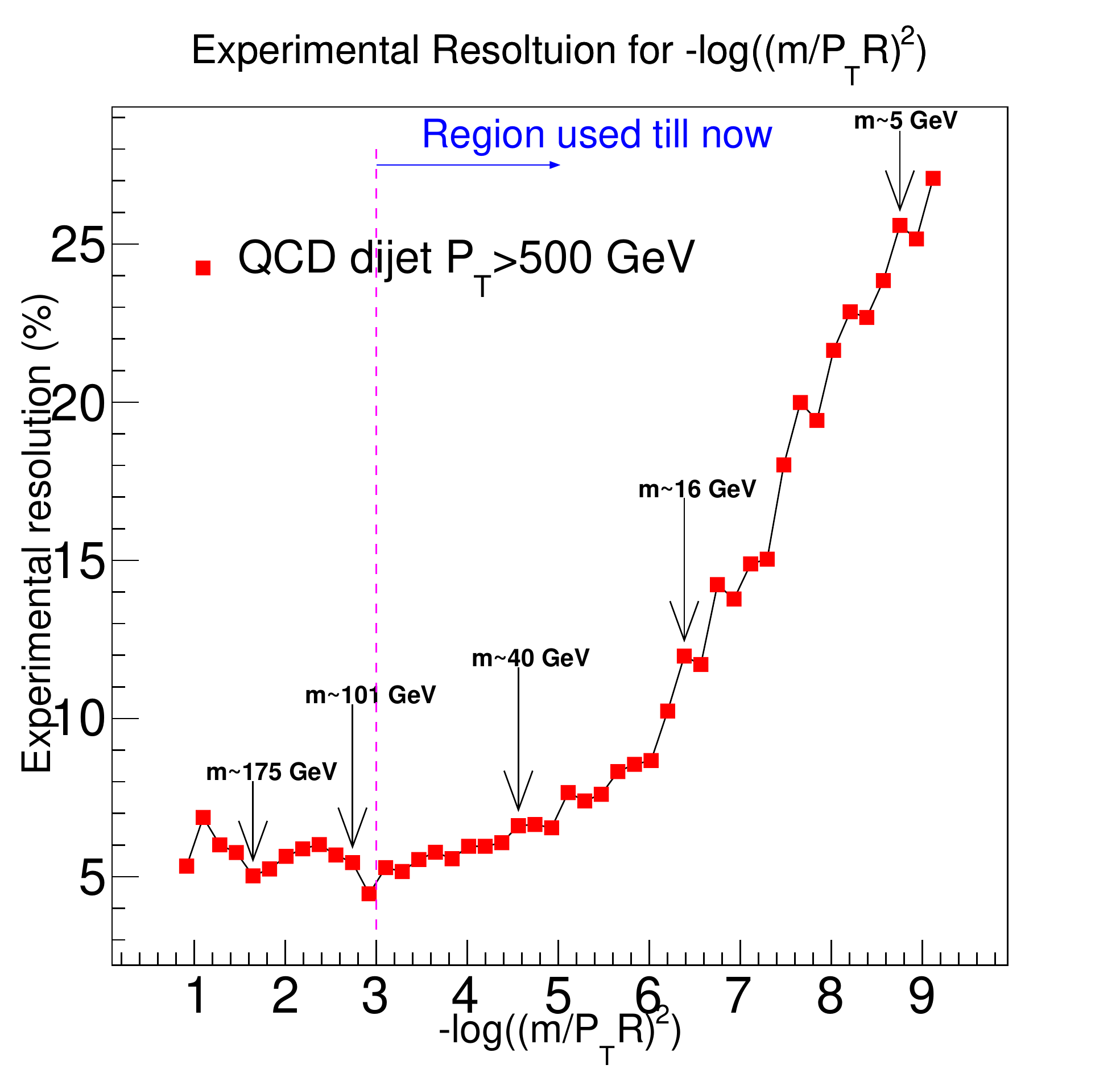}
\end{center}
\caption{The fractional $e_2^{(2)}$ distribution determined from dijet events simulated with \textsc{Pythia~8} plus \textsc{Delphes}.  Arrows indicate the mass at given values of the two-point correlator.  The upper bound of the resummation regime is indicated by a dashed line.  Larger masses are to the left.}
\label{fig:SM_jetsub_alphas:resolution}
\end{figure}

\begin{figure}[t]
\begin{center}
\includegraphics[width = 0.49\columnwidth]{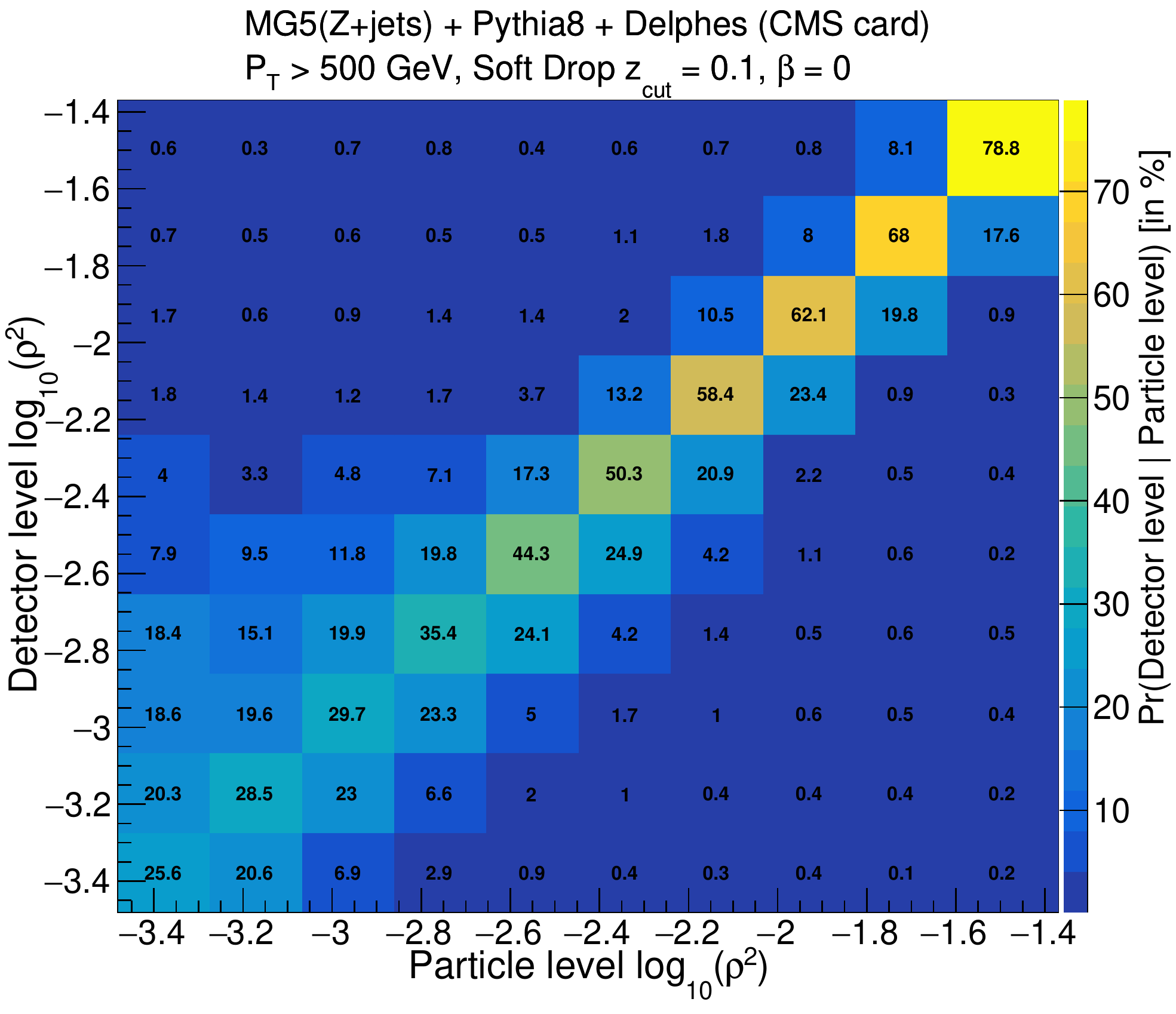}
\includegraphics[width = 0.49\columnwidth]{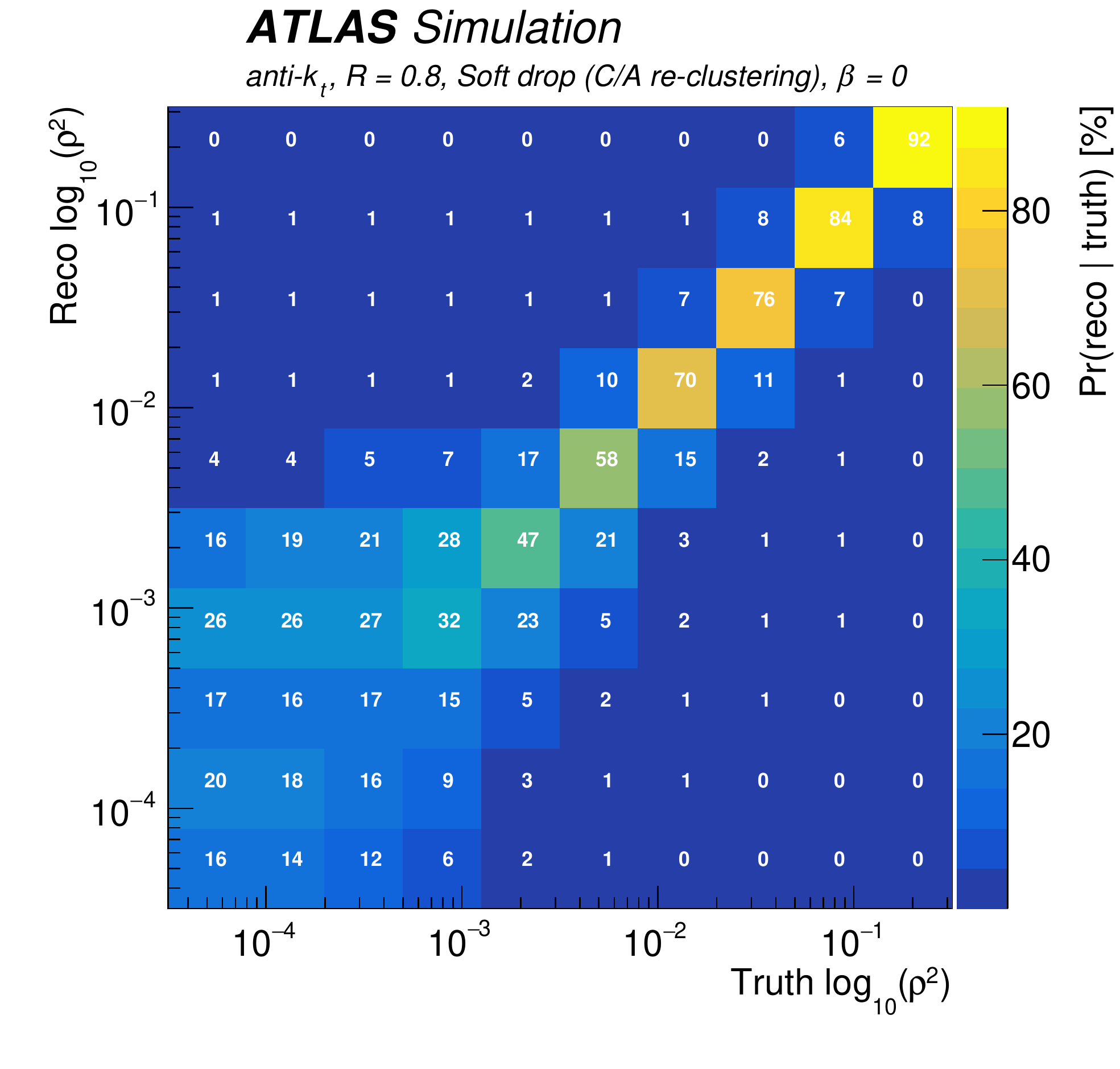}
\end{center}
\caption{Left: The migration matrix between particle-level and detector-level using the Delphes simulation (left) and from the ATLAS measurement (right; reproduced from Ref.~\cite{Aaboud:2017qwh}).  Unlike previous plots, larger masses are on the right (shown this way to match the ATLAS result). }
\label{fig:SM_jetsub_alphas:expres}
\end{figure}

Figure~\ref{fig:SM_jetsub_alphas:resolution} shows the fractional $e_2^{(2)}$ resolution estimated from \textsc{Delphes}.
The resolution is smallest at high mass due to the excellent energy resolution of the CMS-like detector.
The resolution at low mass is about 10\% near 15 GeV and reaches 30\% near the limit of $\mathcal{O}(\mathrm{few\hspace{1mm}GeV})$.
Encouragingly, the fits in Sec.~\ref{sec:SM_jetsub_alphas:templates} relied only on the regime where $\sim 10\%$ resolution seems to be achievable, which gives an indication that a 10\% extraction of $\alpha_s$ should be feasible.
As a check that this estimated resolution is sensible, Fig.~\ref{fig:SM_jetsub_alphas:expres} compares the migration matrix extracted from Delphes to the one published in the recent ATLAS measurement~\cite{Aaboud:2017qwh}.
Here, the comparison is between the particle-level and detector-level groomed mass values. 
The migration is qualitatively the same, with an excellent diagonal behavior (low migrations) at high mass and a worse resolution at low mass.

\begin{figure}[t]
\begin{center}
\includegraphics[width = 0.5\columnwidth]{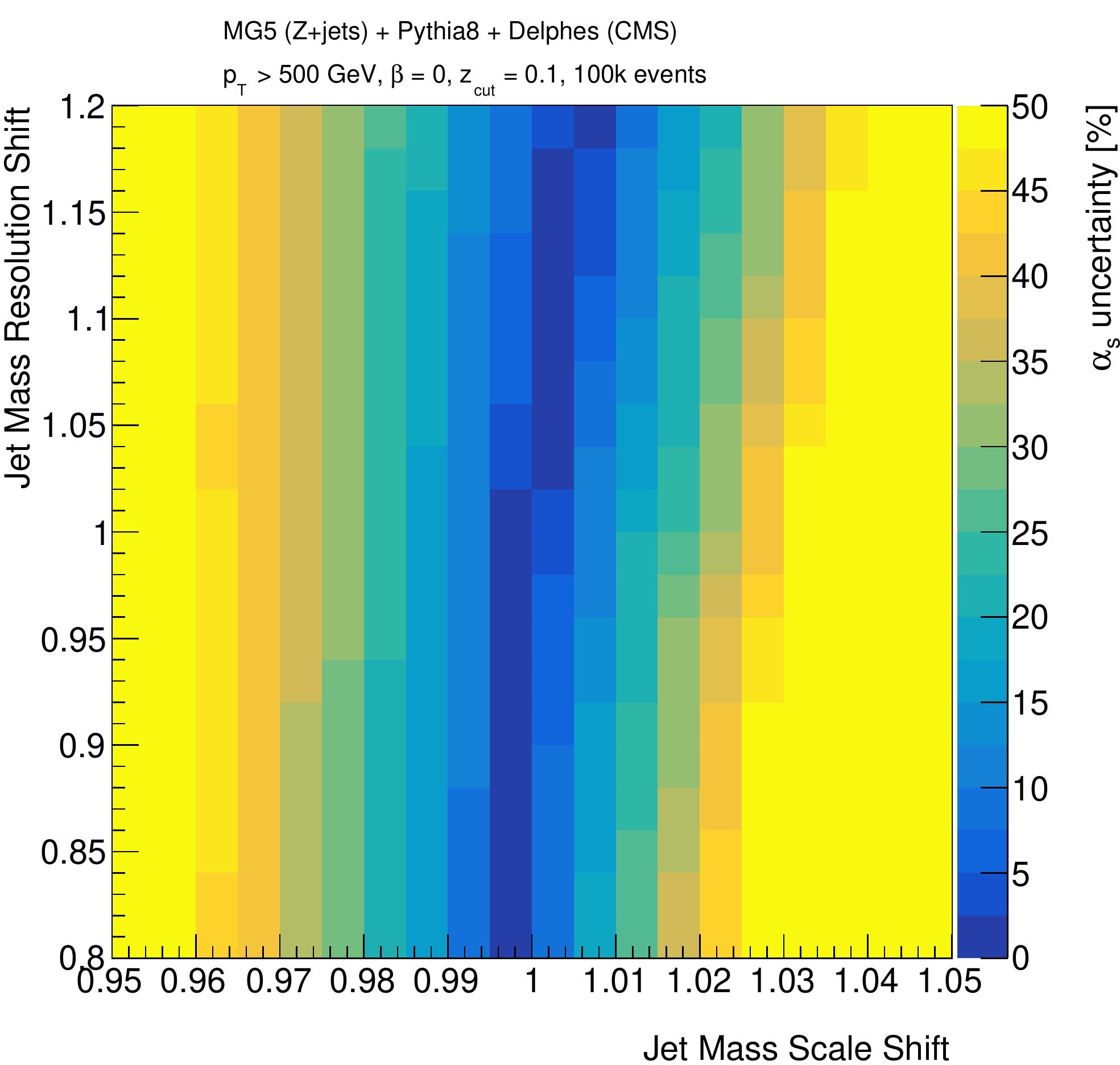}
\end{center}
\caption{The impact of jet mass scale and jet mass resolution uncertainties on the uncertainty in the measured value of $\alpha_s$.  See the text for details.}
\label{fig:SM_jetsub_alphas:expfit}
\end{figure}

To understand the sensitivity to jet mass scale and resolution uncertainties, we conduct pseudo-experiments by sampling from the templates described in Sec.~\ref{sec:SM_jetsub_alphas:templates} and smearing with the migration matrix shown in the left plot of Fig.~\ref{fig:SM_jetsub_alphas:expres}.
We then unfold with another migration matrix~\cite{DAgostini:1994fjx,Adye:2011gm} that has the jet mass scale shifted or smeared by a fixed amount, and extract $\alpha_s$ via Eq.~\eqref{eq:SM_jetsub_alphas:chi2fit} but for fixed and known $f_g=0$.
The results of this procedure are shown in Fig.~\ref{fig:SM_jetsub_alphas:expfit}.
There is little sensitivity to the jet mass resolution while there is a large uncertainty in the extracted $\alpha_s$ if the uncertainty in the jet mass scale exceeds a few percent.
Current jet mass scale uncertainties are below $5\%$ and resolution uncertainties are below 20\%~\cite{ATLAS-CONF-2017-063,CMS-PAS-JME-16-003}.
The mass scale uncertainty is as small as $2\%$ in various regions of phase space.
This again suggests that a $\sim 10\%$ measurement of $\alpha_s$ is feasible, as the uncertainty on the jet mass scale reaches the same level of maturity as the jet energy scale over the next few years.

\subsection{Conclusions and Future Outlook}
\label{sec:SM_jetsub_alphas:future}

In this section, we performed a preliminary study assessing a possible extraction of the strong coupling constant $\alpha_s$ from groomed jet substructure measurements at the LHC.
This has been made possible by recent advances in the calculation of groomed event shapes, which allow them to be resummed to NNLL accuracy, as well as advances in the understanding of the experimental uncertainties for groomed jet observables.
 
We have highlighted a number of features that an extraction of $\alpha_s$ from groomed substructure would offer.
In particular, the grooming procedure suppresses NP hadronization corrections to the mass distribution, extending the range of perturbative control by up to a factor of $\sim 100$ as compared to the ungroomed case, where hadronization corrections are a dominant uncertainty.
Furthermore, the nature of the NP corrections is completely different, and therefore a groomed extraction could provide complementary information to event shape measurements at LEP and other $e^+e^-$ colliders. 

Using the groomed jet mass as a concrete example, we performed a feasibility study for the extraction of $\alpha_s$ from groomed jet substructure.
Considered separately, the groomed quark and gluon distributions both exhibit good sensitivity to $\alpha_s$.
However, this study also highlighted a key issue relevant at the LHC.
The leading behavior of the shape of the distribution is sensitive to the product $\alpha_s C_i$, where $C_i$ is the color Casimir factor, namely $C_A$ for gluon jets and $C_F$ for quark jets.
This immediately implies that there is a degeneracy between the value of $\alpha_s$ and the quark/gluon fraction of the sample.
We highlighted two possible ways of overcoming this degeneracy, either by a fixed-order calculation of the quark and gluon fractions, or by a simultaneous measurement of different substructure observables.
This would ideally be combined with multiple samples with different (and relatively pure~\cite{Gallicchio:2011xc}) quark/gluon fractions.
While we expect significantly better precision to be achievable by the explicit calculation, this would introduce a stronger dependence on the PDFs which then should ideally be fitted simultaneously.

With currently achievable experimental and theoretical uncertainties, we have shown that an extraction of $\alpha_s$ at the $10\%$ level is realistic using the currently available data at the LHC.
We believe that this study motivates a serious effort to extract the strong coupling constant $\alpha_s$ from jet substructure at the LHC using groomed jet shapes.
A serious analysis will require advances on both the theory and experiment sides, so we conclude by discussing some of the major obstacles that must be overcome.

On the theory side, we can separately discuss three primary aspects of the calculation:
\begin{itemize}
\item {\bf Resummation Accuracy:} For $e^+e^-$ event shapes, the current state of the art is N$^3$LL.
This has been achieved only for a few select observables using soft-collinear effective theory.
The extension to N$^3$LL accuracy for the groomed jet mass would require only the calculation of the anomalous dimensions for the groomed soft function to three loops, which is possible using currently available techniques.
At this level of accuracy, it will also be important to assess perturbative power corrections in $z_{\mathrm{cut}}$.
While Refs.~\cite{Marzani:2017kqd,Marzani:2017mva} have shown that these corrections are numerically small, they may become important if multiple $z_{\mathrm{cut}}$ values are used that are not much less than one.
\item {\bf Fixed-Order Matching:} A key aspect, and a potential complication to achieve theoretical accuracy competitive with $e^+e^-$ determinations, is fixed-order matching.
For $e^+e^-$, the perturbative corrections to $e^+e^- \rightarrow$ 3 jets are known to NNLO \cite{GehrmannDeRidder:2007hr,Gehrmann-DeRidder:2007nzq,Weinzierl:2008iv,Weinzierl:2009ms}.
To achieve a similar perturbative accuracy for the matching for the groomed jet mass will require $2\rightarrow 3$ matrix elements at NNLO.
While results for the amplitudes are just becoming available \cite{Gehrmann:2015bfy,Dunbar:2016aux,Badger:2013yda,Badger:2017jhb,Abreu:2017hqn}, it will be a while before numerically-efficient evaluations of the relevant cross-sections are available.
We believe that this is currently the theoretically most difficult ingredient.
\item {\bf Non-Perturbative Corrections:} Finally, in addition to improving our understanding of the perturbative accuracy of the observable, it will also be crucial to improve our understanding of the NP aspects of groomed observables (mass and others).
While it has been shown through PS studies that NP effects are suppressed throughout a large component of the distribution, it will be important to quantify this further.
In the case of $e^+e^-$ event shapes, operator definitions of soft matrix elements allow for field-theoretic definitions of the NP parameters.
Ideally, this could also be performed for groomed observables, placing the groomed jet mass on a firmer theoretical footing.
\end{itemize}

\vspace{5mm}

Experimentally, there are three broad topics that need to be addressed to achieve a precision extraction of $\alpha_s$:
\begin{itemize}
\item {\bf Mass Scale Uncertainties:} At the moment, ATLAS~\cite{Aaboud:2017qwh} and CMS~\cite{CMS-PAS-SMP-16-010} have very different approaches for determining the jet mass scale uncertainty, both with known limitations.
CMS performs a fit to the hadronic $W$ boson mass peak in one-lepton $t\bar{t}$ events and takes the shift in the peak position as the uncertainty in the jet mass scale (which is negligible and thus ignored)~\cite{Sirunyan:2016cao}.
Two challenges with this approach are that (a) the peak position is a convolution of particle-level and detector-level effects and (b) it is not clear that uncertainties derived for boosted $W$ bosons should be the same as for generic quark and gluon jets at all masses.
One can overcome (a) with a technique like forward-folding~\cite{ATLAS-CONF-2016-008,ATLAS-CONF-2016-035}.
Various PS generators can be studied, but it is likely not sufficient to have one global model comparison.
The impact on the jet mass resolution in one topology may be completely different than the impact of the scale, resolution, or acceptance in another topology.

By contrast, the ATLAS measurement propagates constituent-based uncertainties through to the groomed mass.
These uncertainties are derived from matching tracks to calorimeter-cell clusters and studying the energy and angular matching.
Studies have shown that this ``bottom-up'' approach works well for reproducing the jet energy scale~\cite{Aaboud:2016hwh}, which has been validated also for groomed jets in Ref.~\cite{Aaboud:2017qwh}.
However, this does not hold exactly for the mass, which is not linear in the constituent energies~\cite{Nachman:2016qyc}.
The uncertainties are validated using the standard ATLAS approach using track-jets~\cite{Aad:2013gja,ATLAS-CONF-2017-063}, but to achieve higher precision, a more detailed understanding of the impact of energy thresholds, fluctuation correlations, and calorimeter cluster merging will be required.
\item {\bf Mass Resolution Uncertainties:} ATLAS and CMS use their same respective approaches for the mass resolution as for the mass scale (i.e.\ bottom-up and $W$ mass peak).
ATLAS validates their approach in a similar manner as CMS, by using the $W$ mass peak from $t\bar{t}$ events.%
\footnote{This validation does not have the same complete forward-folding machinery as was used by ATLAS for trimmed jets.}
From Fig.~\ref{fig:SM_jetsub_alphas:expfit}, it seems that the mass resolution will not be the limiting factor, but it could be if the situation is not improved as the jet mass scale precision improves.
\item {\bf Pileup Modeling and Mitigation:} Grooming significantly reduces the impact of pileup, but if the Run 3+ data are to be used (pileup levels of 80+), then a significantly better and more detailed understanding of the degradation due to pileup will be required.
Statistical uncertainties are currently not dominant, so it is conceivable that the higher instantaneous luminosity data will not be required for the precision $\alpha_s$ extraction.
This may change if one wants to exploit the largest lever-arm possible; to access the highest $p_{T}$ jets, we will need more data.
\end{itemize}

We are optimistic that these difficulties on both the theory and experiment side can be overcome, enabling a precision probe of the strong coupling constant from jet substructure in the LHC environment.
We also hope that this study motivates additional investigations to broaden the potential toolbox for $\alpha_s$ extraction.
At minimum, it would be interesting to study in more detail the use of other observables beyond the two-point correlators as well as alternative grooming techniques beyond mMDT/SoftDrop.
More ambitiously, one potentially interesting direction is the use of track-based measurements, since it may be the case that it is ultimately experimental uncertainties that are the limiting factor.
From the experimental perspective, this significantly reduces uncertainties, particularly in a high pile-up environment.
From the theoretical side, however, it introduces the need for NP track functions.
That said, only certain moments of the track functions are required for special observables, and it may be possible to perform a combined fits for the moments of the track functions and $\alpha_s$, much like how fits are performed for $e^+e^-$ event shapes.
Significantly more theoretical work is required to see if this is truly a viable possibility, although we believe that this is well motivated.

\subsection*{Acknowledgments}

We thank the participants of Les Houches 2017 for a lively environment and useful discussions.
The work of GS is supported in part by the French Agence Nationale de la Recherche,
under grant ANR-15-CE31-0016, and by the ERC Advanced Grant Higgs@LHC
(No.\ 321133).
The work of FD and JT is supported by the DOE under grant contract numbers DE-SC-00012567 and DE-SC-00015476.
IM and BN are supported in part by the Office of High Energy Physics of the U.S. Department of Energy under Contract No. DE-AC02-05CH11231, and the LDRD Program of LBNL.
AS acknowledges support from the National Science Centre, Poland Grant No. 2016/23/D/ST2/02605, the MCnetITN3 H2020 Marie Curie Initial Training Network, contract 722104 and COST Action CA15213 THOR.

\section{Performance versus robustness: Two-prong substructure taggers for the LHC~\protect\footnote{
    P.~Loch,
    I.~Moult,
    B.~Nachman,
    G.~Soyez,
    J.~Thaler (section coordinators); 
    D.~Bhatia,
    R.~Camacho,
    G.~Chachamis,
    S.~Chatterjee,
    F.~Dreyer,
    D.~Kar,
    A.~Papaefstathiou,
    T.~Samui,
    A.~Si\'{o}dmok}{}}
\label{sec:SM_jetsub_2prong}

The ability to robustly identify, or ``tag", boosted hadronically-decaying resonances plays a central role at the LHC, both in
  searches for new physics, as well as for probing the Standard Model
  in extreme regions of phase space.
  While a variety of powerful and
  theoretically well-understood tagging approaches exist, their behavior in a realistic experimental environment is complicated by a number of issues including hadronization, underlying event, pileup, and detector effects.
  In this section, we perform a
  systematic study contrasting the robustness and performance of
  different theoretical approaches to designing jet substructure
  observables.
  These include standard jet shape observables as well as various grooming
  strategies currently used by the LHC experiments.
  We also introduce a number of new observables, based on the idea of
  ``dichroic ratios'', which are designed to simultaneously maximize both
  robustness and performance.
  We discuss
  the different choices used by ATLAS and CMS, and we introduce
  reliable metrics for quantifying robustness and performance for
  substructure observables.
  Additionally, we study the dependence of taggers on the polarization of hadronically decaying $W$ bosons, and identify strategies to perform polarimetry using the hadronic decay products.
We conclude by making recommendations for future
  tagging strategies to ensure robust procedures based on sound
  theoretical organizing principles.

\subsection{Introduction}\label{sec:SM_jetsub_2prong:intro}

With the ever-increasing dataset from the Large Hadron Collider (LHC), we are able to probe the Standard Model and search for physics beyond the Standard Model in increasingly extreme regions of phase space.
Theoretically well-understood observables that are sensitive to phase space extremes are therefore playing an important role at the LHC.
One of the most exciting new approaches which has emerged at the LHC are tools from jet substructure, which allow for the identification of boosted hadronically-decaying particles within jets through a detailed study of their radiation patterns.
Techniques from jet substructure have now been widely used for Standard Model measurements \cite{Chatrchyan:2012sn,CMS:2013cda,Aad:2015cua,Aad:2015lxa,ATLAS-CONF-2015-035,Aad:2015rpa,Aad:2015hna,ATLAS-CONF-2016-002,ATLAS-CONF-2016-039,ATLAS-CONF-2016-034,CMS-PAS-TOP-16-013,CMS-PAS-HIG-16-004} as well as for searches for new physics  \cite{CMS:2011bqa,Fleischmann:2013woa,Pilot:2013bla,TheATLAScollaboration:2013qia,Chatrchyan:2012ku,CMS-PAS-B2G-14-001,CMS-PAS-B2G-14-002,Khachatryan:2015axa,Khachatryan:2015bma,Aad:2015owa,Aaboud:2016okv,Aaboud:2016trl,Aaboud:2016qgg,ATLAS-CONF-2016-055,ATLAS-CONF-2015-071,ATLAS-CONF-2015-068,CMS-PAS-EXO-16-037,CMS-PAS-EXO-16-040,Khachatryan:2016mdm,CMS-PAS-HIG-16-016,CMS-PAS-B2G-15-003,CMS-PAS-EXO-16-017}.%
\footnote{More LHC studies using jet substructure can be found at \url{https://twiki.cern.ch/twiki/bin/view/AtlasPublic} and \url{http://cms-results.web.cern.ch/cms-results/public-results/publications/}.} 

With the growing importance of jet substructure techniques, there has been a significant effort by both the theory and experimental communities to develop a more detailed understanding of the theoretical and experimental behavior of jet substructure observables.
On the theory side, this has been pursued through explicit calculations~\cite{Feige:2012vc,Field:2012rw,Dasgupta:2013ihk,Dasgupta:2013via,Larkoski:2014pca,Dasgupta:2015yua,Seymour:1997kj,Li:2011hy,Larkoski:2012eh,Jankowiak:2012na,Chien:2014nsa,Chien:2014zna,Isaacson:2015fra,Krohn:2012fg,Waalewijn:2012sv,Larkoski:2014tva,Procura:2014cba,Bertolini:2015pka,Bhattacherjee:2015psa,Larkoski:2015kga,Dasgupta:2015lxh,Frye:2016okc,Frye:2016aiz,Kang:2016ehg,Hornig:2016ahz,Marzani:2017mva,Marzani:2017kqd,Hoang:2017kmk,Larkoski:2017iuy,Larkoski:2017cqq}, scaling arguments~\cite{Walsh:2011fz,Larkoski:2014gra,Larkoski:2014zma}, and machine learning \cite{Cogan:2014oua,deOliveira:2015xxd,Almeida:2015jua,Baldi:2016fql,Guest:2016iqz,Conway:2016caq,Barnard:2016qma} approaches.
On the experimental side, there have been detailed studies of the behavior of substructure observables in data, and their interplay with experimental realities, such as detector resolution and pileup.
Summaries can be found in Refs.~\cite{Abdesselam:2010pt,Altheimer:2012mn,Altheimer:2013yza,Adams:2015hiv}, and Ref.~\cite{Larkoski:2017jix} provides a review of recent theoretical and machine learning developments in jet substructure. 

As a result of these efforts, there now exist a number of theoretically well-motivated jet substructure tools.
For tagging hadronically-decaying $W/Z/H$ bosons, which decay primarily to jets with two well-resolved prongs (also referred to as subjets), a variety of powerful two-prong taggers have been developed.
Modern two-prong taggers typically consist of two or three
ingredients: a groomer which removes low-energy contamination, a
two-prong finder which identifies two (or more) hard subjets, and a jet
shape observable which constrains radiation patterns in the
jet.
Often, the groomer and the two-prong finder are taken identical.
For jet shapes, it is well understood how to organize and study their
behavior using power counting \cite{Larkoski:2014gra}.
A variety of
different classes of observables exist, for example the energy
correlation functions \cite{Larkoski:2013eya,Moult:2016cvt,Komiske:2017aww} and $N$-subjettiness
observables \cite{Thaler:2010tr,Thaler:2011gf}, and their relation is
understood.
The behavior of groomers is also now well understood, and
a number of groomers with favorable experimental and theoretical properties have been introduced
\cite{Dasgupta:2013ihk,Larkoski:2014wba}.

The
theoretical understanding of the behavior of tagging observables is primarily based on perturbation theory, and it is therefore
not always clear how this translates to experimental reality, due to
the presence of hadronization, underlying event, detector effects, and
pileup.
Indeed, the different LHC experiments have settled on different
tagging combinations.
For grooming and prong finding, ATLAS using trimming \cite{Krohn:2009th} uses while CMS uses the modified
  Mass Drop Tagger (mMDT)~\cite{Butterworth:2008iy,Dasgupta:2013ihk} and its generalization, SoftDrop \cite{Larkoski:2014wba}.
  For jet shape observable, ATLAS uses $D_2$ \cite{Larkoski:2014gra,Larkoski:2015kga}, while CMS uses $N$-subjettiness ratio $\tau_{2,1}$ \cite{Thaler:2010tr,Thaler:2011gf} or $N_2$ \cite{Moult:2016cvt} with DDT \cite{Dolen:2016kst}.
  It is not clear whether these choices are driven by differences in the detectors, or an optimization with respect to different criteria, samples, or modeling. While detailed optimization must ultimately be performed by the experiments themselves, we believe that there is still much to be understood about the general organization and design of jet substructure observables.

In this section, we perform a comprehensive study of performance and robustness for two-prong tagging techniques.
To frame the study, we use a theoretical organization into different tagging strategies based on the idea of dichroic observables \cite{Salam:2016yht}, which generalize ratio observables to allow hybrid combinations of groomed and ungroomed shapes.
 These contain as special instances all the familiar observables used by the experiments, as well as new observables, such as dichroic versions of the $N_2$ and $D_2$ observables.
 We therefore place the ATLAS and CMS strategies as specific examples of broader classes of theoretical approaches for tagging two-prong substructure, about which we can draw general conclusions regarding robustness and performance.

The goal of this study is to highlight the interplay between performance and robustness, and assess the choices made by the different LHC experiments.
Here, ``performance'' refers to the tagging efficiency (for a given background rejection) in the absence of systematic uncertainties.
This has been the primary way to assess jet substructure observables in the past, but it does not capture the full set of considerations needed to apply jet substructure techniques in practice.
By contrast, ``robustness'' refers to modifications of the substructure observables as different physics aspects are added.
In particular, we consider robustness to non-perturbative effects, namely hadronization and underlying event, robustness to detector effects, and robustness to pileup radiation.
In this study, we refine the metic introduced
in Refs.~\cite{Dasgupta:2016ktv,Salam:2016yht} for quantifying
robustness.
We believe that this dual assessment of performance and
robustness will be useful in future studies of jet substructure
observables.
This allows us to study each tagging strategy in general, and the CMS
and ATLAS approaches in particular.
In all cases, we are able to identify observables with improved robustness and performance as compared with those currently used by the experiments.

As an additional aspect of this study, we also consider the robustness of the signal tagging efficiency to the polarization of the decaying boson.
We show that while jet shape observables themselves are fairly robust to polarization, groomed mass cuts are not, so that the tagging efficiency depends strongly on the polarization.
Furthermore, we propose that the momentum  asymmetry of the subjets is a good discriminant between longitudinal and transverse polarizations, and can be used to perform polarimetry for boosted hadronic decays. 

An outline of this study is as follows.
In Sec.~\ref{sec:SM_jetsub_2prong:pres}, we define our metrics for studying robustness.
We discuss the key physics issues that we would like to assess robustness to, both theoretical and experimental, and describe a chain of different steps in our simulation process such that each physics issue can be isolated and studied.
In Sec.~\ref{sec:SM_jetsub_2prong:obs_def}, we define all jet substructure observables that will be used throughout this study, and  provide details of our sample generation.
In Sec.~\ref{sec:SM_jetsub_2prong:hybrid_ratio}, we discuss theoretical approaches to designing robust two-prong taggers.
We extend the approach of Ref.~\cite{Salam:2016yht} and define several new dichroic observables formed from the energy correlation functions.  
In Sec.~\ref{sec:SM_jetsub_2prong:np}, we study the robustness of the observables to non-perturbative radiation both from hadronization and underlying event.
In Sec.~\ref{sec:SM_jetsub_2prong:exp}, we study robustness to detector and pileup.
In Sec.~\ref{sec:SM_jetsub_2prong:polar}, we study the robustness of substructure observables to the polarization of a decaying $W/Z$ boson, and introduce observables to distinguish polarized samples.
We summarize our results in Sec.~\ref{sec:SM_jetsub_2prong:conc} and make a number of recommendations for future jet substructure studies.

\subsection{Quantifying Performance and Robustness}\label{sec:SM_jetsub_2prong:pres}

The goal of this study is to study the interplay between tagging performance and robustness for two-prong taggers.
This requires a precise definition of the physics effects to which we are (or are not) robust, as well as a metric to quantify both performance and robustness.

Since we are able to generate pure signal and background samples, the tagging performance is straightforward to define using the signal and background efficiencies, $\epsilon_S$ and $\epsilon_B$.
In principle, we could evaluate the full receiver operating characteristic (ROC) curve relating $\epsilon_B$ to $\epsilon_S$.
For simplicity, we use the ``tagging significance'' as our metric for performance,
\begin{align}\label{eq:SM_jetsub_2prong:significance}
\epsilon=\frac{\epsilon_S}{\sqrt{\epsilon_B}}\,,
\end{align}
evaluated at a fixed signal efficiency, typically $\epsilon_S = 0.4$.

\begin{figure}[t]
\begin{center}
\includegraphics[width=0.75\columnwidth]{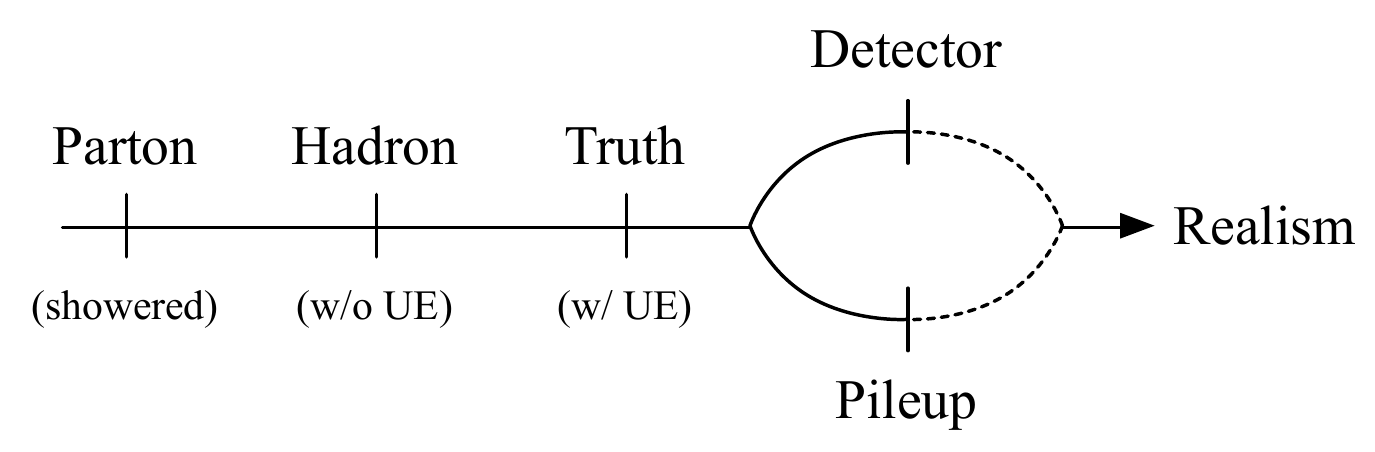}
\end{center}
\caption{A summary of the different stages of physics considered in
  this study, from idealized parton-level events to fully realistic
  events including detector simulation and pileup.
  This allows us to
  address robustness to physics at each stage.
  Detailed definitions of
  each stage, and the physics probes used, are given in the text.
   }
\label{fig:SM_jetsub_2prong:realism}
\end{figure}

We will approach robustness by moving from an idealized partonic description to a complete detector simulation including pileup; a``realistic" scenario representative of the LHC environment.
This chain of realism is shown in Fig.~\ref{fig:SM_jetsub_2prong:realism}, which illustrates the following stages:
\begin{itemize}
\item {\bf Parton Level}: We define the parton level result as the
  perturbative distribution for the active-active scattering (i.e.\ we
  do not include possible perturbative contributions to the underlying
  event).
  While this can be defined in an analytic calculation, it is
  more difficult in the context of parton shower generators, since
  there is necessarily a cut-off that must be imposed between
  perturbative and non-perturbative physics.
  Only the complete result
  is physical, and intermediate results should be interpreted with
  care.
  Nevertheless, to have some measure of the impact of
  non-perturbative effects, we will define parton level as generated
  by a parton shower generator with all hadronization effects turned
  off.
  We use \textsc{Pythia 8} \cite{Sjostrand:2006za,Sjostrand:2007gs} as our baseline generator, leaving a study of additional generators to future work.
\item {\bf Hadron Level}: We define the hadron level result as including hadronization in the shower, but not including any effects from the underlying event.
\item {\bf Truth Level}:  We define truth level as the hadronized result including the underlying event as implemented in an event generator.
Truth level therefore represents a complete hard scattering process in a hadron-hadron collider in isolation.
\item {\bf Detector Level}: Detector-level results are defined as truth level events passed through a detector simulation as implemented by TowerGrid.  See Sec.~\ref{sec:SM_jetsub_2prong:det_model} for details of the detector simulation.
\item {\bf Pileup Level}: Due to the high pileup environment of the LHC, we include in our study also the effects of uncorrelated proton interactions. We have done this separately from detector effects to be able to isolate and study the physics effects arising from pileup and pileup subtraction schemes. Our pileup subtraction scheme is described in Sec.~\ref{sec:SM_jetsub_2prong:pu_tech}.
\item {\bf Full Realism}: In the final stage of realism, we should consider events with pileup at detector
  level.
  These represent, to the level that we can consider in this
  study, realistic events as seen by the experiments at the
  LHC.
  Since our detector and pileup assessment is still fairly basic, we
  have left this final stage for future work.
\end{itemize}

Comparing the differences as we progress step by step through this sequence allows us to address at each stage the robustness to distinct physics issues, and we hope that our segmentation is sufficiently fine that we have a comprehensive view of robustness.
In particular, the different steps in the chain allow us to study robustness to the following physics: 
\begin{itemize}
\item {\bf Parton $\to$ Hadron}: Changes in the distribution from parton
  level to hadron level probe non-perturbative physics associated with
  hadronization.
  For many event shapes, hadronization corrections can
  be given a field-theoretic definition in terms of a matrix element
  whose symmetry properties can be used to prove basic
  results.
  Ultimately, such corrections cannot be computed
  from first principles and must be included through models, such as
  those included in parton shower generators, or through dispersive approaches or shape
  functions in analytic calculations
  \cite{Dokshitzer:1995qm,Dokshitzer:1995zt,Korchemsky:1999kt,Korchemsky:2000kp,Bosch:2004th,Hoang:2007vb,Ligeti:2008ac}.
    To have the best theoretical
  control and understanding of jet substructure observables, it is
  therefore desirable that their performance is robust to the effects
  of hadronization.
\item {\bf Hadron $\to$ Truth}: Changes in the distribution from hadron level to truth level probe the impact of the underlying event, namely the physics associated with interactions of the colliding protons and their remnants.
Such contributions are in principle both perturbative and non-perturbative.
They are poorly understood theoretically, and it is currently not known how to treat them systematically, or define them field theoretically.
It has been found empirically that the effects of underlying event are well modeled by a shape function \cite{Stewart:2014nna}, although the theoretical justification for this is not clear. Other simple theoretical models from which intuition can be gained have been proposed in Ref.~\cite{Cacciari:2009dp}.
The underlying event is implemented in parton shower event generators using models which are tuned to data, and we take these models as our definition of the underlying event.
Due to this lack of theoretical understanding, as well as the fact that radiation from the underlying event is typically not associated with the physics that we are interested in probing, it is desirable that jet substructure observables be robust to the underlying event.
\item {\bf Truth $\to$ Detector}: Since we are ultimately interested in using jet substructure in experiments, the behavior of the detectors plays an essential role.
The finite energy and spatial resolution of the detectors ultimately degrades the behavior of the observables.
Furthermore, the detector response must be unfolded, and is therefore difficult to compute analytically, or to include to higher accuracy. Therefore, both for performance and calculability, it is desirable that jet substructure observables are robust to detector effects.
\item {\bf Truth $\to$ Pileup}: Finally, due to the high pileup environment of the LHC, significant soft radiation can contaminate jet substructure observables.
Since this radiation is not correlated with the underlying hard scattering process, it is not associated with the physics of interest, and therefore can only act to degrade the performance.
Furthermore, it is difficult to model in an analytic calculation.
While techniques exist to mitigate pileup, as reviewed in Sec.~\ref{sec:SM_jetsub_2prong:pu_tech}, it is desirable that the substructure observables used are as robust as possible to pileup contamination.
\end{itemize}
We will classify the first two of these as ``Theory" issues, which will be discussed in Sec.~\ref{sec:SM_jetsub_2prong:np}, while the second two are classified as ``Experimental" issues, and will be discussed in Sec.~\ref{sec:SM_jetsub_2prong:exp}. This decomposition is of course somewhat arbitrary, since a coherent understanding involving the complete chain is required. However, this decomposition was chosen such that the ``Theory" issues cover an idealized hadronic collision in isolation. 

\begin{figure}[t]
\begin{center}
\includegraphics[width=0.9\columnwidth]{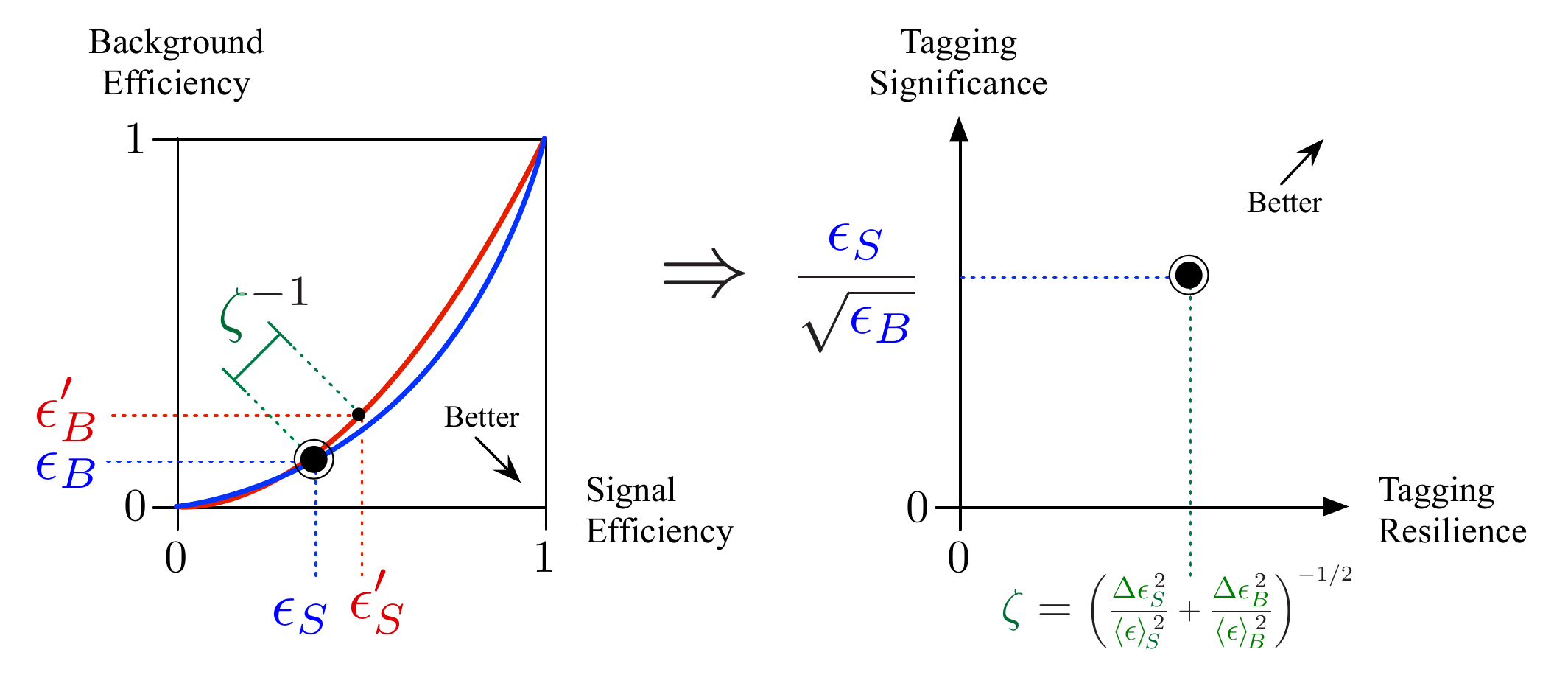}
\end{center}
\caption{An illustration of the resilience metric $\zeta$ used throughout the text to quantify robustness. In the left panel, we illustrate graphically $\zeta$ as the change in ROC curve to a particular aspect of the underlying physics. In the right panel, we illustrate the tagging resilience vs.\ tagging performance plane which we will use to graphically illustrate our results.  A simultaneously robust and performant tagger lives in the upper right hand corner of this space}
\label{fig:SM_jetsub_2prong:zeta_def}
\end{figure}

To compare the robustness to a
particular step in this chain for different observables, we must
introduce a metric.
There is of course a high degree of arbitrariness
in the definition of the metric.
For example, one could base the metric on the shape of the signal or background distribution.
Since we are ultimately interested in the performance of the observable, however, we introduce a metric which is based on the modification of the ROC curve.
Consider a reference stage (unprimed) and a modified stage (primed); in the case of hadronization, the reference stage would be the hadron level and the modified stage would be the parton level.
We first calculate the cut on the jet shape $v_{\text{cut}}$ that yields a fixed signal efficiency $\epsilon_S$ for the reference stage.
We then compute the reference stage background efficiency $\epsilon_B$ from that $v_{\text{cut}}$, as well as the modified stage signal and background efficiencies $\epsilon'_S$ and $\epsilon'_B$.
We can then defined a measure of robustness, which we refer to as resilience, $\zeta$, as
\begin{align}
\zeta=\left(  \frac{\Delta \epsilon_S^2}{ \langle \epsilon \rangle_S^2}  +\frac{\Delta \epsilon_B^2}{ \langle \epsilon \rangle_B^2}  \right)^{-1/2}\,,
\end{align}
where
\begin{align}
\Delta \epsilon_{S,B} & =\epsilon_{S,B}-\epsilon_{S,B}',\\
\langle \epsilon \rangle_{S,B} & = \frac{1}{2} \left(\epsilon_{S,B} + \epsilon_{S,B}'\right).
\end{align}
This approach gives an estimate of how much our signal and background efficiencies have changed, for a given set of cuts, when going from one stage to another.

The meaning of $\zeta$ is illustrated in Fig.~\ref{fig:SM_jetsub_2prong:zeta_def}, where larger values of $\zeta$ correspond to improved robustness.
Because we anchor to a fixed $v_{\text{cut}}$, this method can even detect a uniform shift in both the signal and background distributions (even though such a shift would not change the ROC curve).
We therefore believe that $\zeta$ provides a reliable metric for assessing the robustness of the tagger.
When presenting our results, we characterize observables
in the plane of $\epsilon$ (see Eq.~\eqref{eq:SM_jetsub_2prong:significance}) versus $\zeta$, with better observables being in
the upper-right corner.
We find that this allows us to synthesize the
information about a large number of observables in a compact manner.
A
similar metric and presentation style was used in Ref.~\cite{Dasgupta:2016ktv,Salam:2016yht}
to study robustness to non-perturbative effects.
In addition to
showing plots of the $\epsilon$-$\zeta$ plane, we will occasionally also show the modification of the distribution
itself to provide additional insight into the robustness of the
observables.

It is important to emphasize that it is impossible to completely
characterize an optimal observable, particularly as jet substructure
observables are being used for increasingly specific
purposes.
We hope that the issues that we have chosen to focus on
are representative of the issues that will be important for a broad
range of applications.
Other aspects, such as the robustness of the substructure observable distributions to changes in the jet mass or $p_T$ cuts, which are important for certain recent applications of jet substructure \cite{Sirunyan:2017dgc,CMS-PAS-HIG-17-010,CMS-PAS-EXO-17-001,Sirunyan:2017dnz,Sirunyan:2017nvi,Aaboud:2018zba}, and have received recent interest \cite{Shimmin:2017mfk,Aguilar-Saavedra:2017rzt,Moult:2017okx}, are
beyond the scope of the current project.
However, it would be interesting to investigate them
using similar techniques.

\subsection{Observable and Sample Definitions}\label{sec:SM_jetsub_2prong:obs_def}

In this subsection, we define all the observables that will be studied throughout this study.
This includes both the jet substructure shape observables and the grooming procedures.  We also present the details of our sample generation.

\subsubsection{Jet Shape Observables}\label{sec:SM_jetsub_2prong:shape_def}

The jet shape observables that we will consider are formed from ratios of the energy correlation functions \cite{Larkoski:2013eya,Moult:2016cvt} or the $N$-subjettiness observables \cite{Thaler:2010tr,Thaler:2011gf}.
The $N$-subjettiness observables are defined as~\cite{Stewart:2010tn,Thaler:2010tr,Thaler:2011gf}\footnote{For this
  study, we have used the un-normalized definition of $N$-subjettiness.}
\begin{equation}\label{eq:SM_jetsub_2prong:nsubdef}
  \tau_{N}^{(\beta)} =
  \sum_{1\leq i \leq n_J} p_{Ti}\min\left\{
R_{i1}^\beta,\dotsc,R_{iN}^\beta
\right\} \ .
\end{equation}
Here $p_{Ti}$ is the $p_T$ of particle $i$, and the sum is over all
particles in the jet.
The minimum is over the longitundinally-boost-invariant angle
\begin{align}
R_{iJ}^2 = (\phi_i-\phi_J)^2+(y_i-y_J)^2\,,
\end{align}
between the particle $i$ and the axis $J$.

Implicit in the definition of the $N$-subjettiness observable in
Eq.~\eqref{eq:SM_jetsub_2prong:nsubdef} is the definition of the axes $n_i$.
While their
placement is unambiguous (up to power corrections) in the limit of resolved substructure, an algorithmic definition is required to
determine their behavior in the unresolved limit.
Two main approaches
have been used for defining the axes.
The first approach is to define
the $N$-subjettiness axes as the axes found using an exclusive jet
clustering algorithm. The second approach is to minimize the sum in
Eq.~\eqref{eq:SM_jetsub_2prong:nsubdef} over possible light-like axes $n_i$.
In this study, we defined the axes using subjets obtained by reclustering the jet, with choices motivated by the studies in Refs.~\cite{Stewart:2015waa,Dasgupta:2015lxh}.
For $\beta = 1$, we recluster using the $k_t$ algorithm~\cite{Catani:1993hr} with the
  winner-take-all recombination scheme~\cite{Larkoski:2014uqa}.
For $\beta = 2$, we use the generalized $k_t$ algorithm~\cite{Cacciari:2011ma} with $p=1/2$.

For two-prong tagging, the relevant observable is the ratio \cite{Thaler:2010tr}
\begin{align}
\tau_{21}^{(\beta)}\equiv \frac{\tau_{2}^{(\beta)}}{\tau_{1}^{(\beta)}}\,.
\end{align}
For a jet with a well resolved two-prong structure, we have $\tau_{21}^{(\beta)}\ll 1$, while for a jet without a well resolved substructure, we have $\tau_{21}^{(\beta)}\sim 1$.
This observable has been extensively used at the LHC.
It has been calculated to LL accuracy \cite{Dasgupta:2015lxh}, and the effects of the axis definition on the perturbative behavior have been studied at NLO \cite{Larkoski:2015uaa}.

The second class of observables that we will consider are based on the energy correlation functions \cite{Larkoski:2013eya,Moult:2016cvt}.
Instead of correlating particles with axes, as is done for $N$-subjettiness, the idea of the energy correlation functions is to correlate $n$-tuples of particles.
In discussing the energy correlation functions, it is convenient to
work with dimensionless observables, written in terms of the angular
variable, $R_{ij}$ and the energy fraction variable $z_i$:
\begin{align}\label{eq:SM_jetsub_2prong:ptratio}  
z_i\equiv\frac{p_{Ti}}{\sum_{j \in \text{jet}} p_{Tj}}\,,
\end{align}
where $p_{Ti}$ is the transverse momentum of particle $i=1,\dots,n_J$. 
%
%
The generalized energy correlation function is defined as
\begin{equation}\label{eq:SM_jetsub_2prong:ecf_gen}
_v e_n^{(\beta)} = \sum_{1 \leq i_1 < i_2 < \dots < i_n \leq n_J} z_{i_1} z_{i_2} \dots z_{i_n} \prod_{m = 1}^{v} \min^{(m)}_{s < t \in \{i_1, i_2 , \dots, i_n \}} \left\{ R_{st}^{\beta} \right\},
\end{equation}
where $\min^{(m)}$ denotes the $m$-th smallest element in the list.  For a jet consisting of fewer than $n$ particles, $_v e_n$ is defined to be zero.  More explicitly, the three arguments of the generalized energy correlation functions are as follows:
\begin{itemize}
\item The subscript $n$, which appears to the right of the observable, denotes the number of particles to be correlated.   
\item The subscript $v$, which appears to the left of the observable, denotes the number of pairwise angles entering the product.  By definition, we take $v \leq \binom{n}{2}$, and the minimum then isolates the product of the $v$ smallest pairwise angles.
\item The angular exponent $\beta>0$ can be used to adjust the
  weighting of the pairwise angles as for $N$-subjettiness.
\end{itemize}

In this study, we use the 2-point energy correlation function,
\begin{align}\label{eq:SM_jetsub_2prong:explicit_twopointvar}
_1e_2^{(\beta)}&\equiv e_2^{(\beta)}=\sum_{1\leq i<j\leq n_J} z_{i}z_{j} \, R_{ij}^\beta\ ,
\end{align}
as well as the 3-point correlators,
\begin{align}\label{eq:SM_jetsub_2prong:explicit_ecfvar}
_1e_{3}^{(\beta)}&=\sum_{1\leq i<j<k\leq n_J} z_{i}z_{j}z_{k} \min \left\{ R_{ij}^\beta\,,  R_{ik}^\beta\,, R_{jk}^\beta  \right\} \ , \nonumber \\
_2e_{3}^{(\beta)}&=\sum_{1\leq i<j<k\leq n_J} z_{i}z_{j}z_{k} \min \left\{R_{ij}^\beta R_{ik}^\beta\,, R_{ij}^\beta  R_{jk}^\beta\,,     R_{ik}^\beta R_{jk}^\beta    \right\}  \ , \nonumber \\
e_{3}^{(\beta)}\equiv ~_3e_{3}^{(\beta)}&=\sum_{1\leq i<j<k\leq n_J} z_{i}z_{j}z_{k} \, R_{ij}^\beta R_{ik}^\beta R_{jk}^\beta \,.
\end{align}
A number of 2-prong discriminants have been formed from the energy correlation functions~\cite{Larkoski:2013eya,Larkoski:2014gra,Larkoski:2014zma,Moult:2016cvt}.  Here, we will focus on the observables
\begin{align}
 M_2^{(\beta)} = \frac{_1e_{3}^{(\beta)}}{e_{2}^{(\beta)}}, \qquad  N_2^{(\beta)} = \frac{_2e_{3}^{(\beta)}}{(e_{2}^{(\beta)})^2}\,, \qquad  D_{2}^{(\beta)}=\frac{e_{3}^{(\beta)}}{(e_{2}^{(\beta)})^{3}}\,, 
\end{align}
each of which probes the correlations between particles within the jet in a slightly different manner.
For a detailed discussion, see Ref.~\cite{Moult:2016cvt}.
The $N_2$ and $D_2$ observables are powerful discriminants and have
been used by CMS and ATLAS, respectively.
The $M_2$ observable is expected to have worse performance, except in particular scenarios, but we include it since it provides an example of a remarkably robust observable.

Beyond their discrimination power, these observables have nice theoretical properties.
First, since they can be written as a sum over particles in the jet without reference to external axes, they are automatically ``recoil-free'' \cite{Catani:1992jc,Dokshitzer:1998kz,Banfi:2004yd,Larkoski:2013eya,Larkoski:2014uqa}.
Second, since they have well-defined behavior in various soft and collinear limits, they are amenable to resummed calculations;  in Ref.~\cite{Larkoski:2015kga}, $D_2$ was calculated to next-to-leading-logarithmic (NLL) accuracy in $e^+e^-$ for both signal (boosted $Z$) and background (QCD) jets, and this was extended in Refs.~\cite{Larkoski:2017iuy,Larkoski:2017cqq} to a hadron-collider environment by exploiting the simplifying properties of grooming.

\subsubsection{Grooming Techniques}\label{sec:SM_jetsub_2prong:groom_tech}

Groomers, which remove wide-angle soft radiation and contamination from a jet, also play an important role in two-prong tagging.
While a variety of different grooming approaches have been defined~\cite{Butterworth:2008iy,Ellis:2009su,Ellis:2009me,Krohn:2009th,Dasgupta:2013via,Dasgupta:2013ihk}, we will focus on the mMDT/SoftDrop family, which is the most theoretically well understood, as well as trimming \cite{Krohn:2009th}, which is used by the ATLAS experiment.
In this subsection, we review the definition of the mMDT/SoftDrop and trimming algorithms and their parameters.

Starting from a jet identified with an IRC safe jet algorithm (such as
anti-$k_t$~\cite{Cacciari:2008gp}), the SoftDrop algorithm proceeds as follows:
\begin{enumerate}
\item Recluster the jet using the Camridge/Aachen (C/A) clustering
  algorithm~\cite{Dokshitzer:1997in,Wobisch:1998wt,Wobisch:2000dk},
  producing an angular-ordered branching history for the jet.
\item Step through the branching history of the reclustered jet.  At each step, check the SoftDrop condition
\begin{align}\label{eq:SM_jetsub_2prong:sd_cut}
\frac{\min\left[ p_{Ti}, p_{Tj}  \right]}{p_{Ti}+p_{Tj}}> {z_{\rm cut}} \left(   \frac{R_{ij}}{R}\right)^\beta \,.
\end{align}
Here, ${z_{\rm cut}}$ is a parameter defining the scale below which soft radiation is removed.  If the SoftDrop condition is not satisfied, then the softer of the two branches is removed from the jet.  This process is then iterated on the harder branch.
\item The procedure terminates once the SoftDrop condition is satisfied.
\end{enumerate}
SoftDrop generalises the mMDT procedure~\cite{Dasgupta:2013ihk} and
the two are equivalent for $\beta=0$.
For this reason, we will often use
the two names interchangeably.
Any IRC-safe observable can be measured on a jet groomed with the
SoftDrop procedure, without loosing IRC safety (for $\beta > 0$).
The aggressivity of the SoftDrop grooming can be adjusted by
tuning the parameters ${z_{\rm cut}}$ and $\beta$.
Larger values of ${z_{\rm cut}}$ groom away more radiation within the jet for a fixed value of $\beta$.
On the other hand, as $\beta$ is increased, the grooming becomes less
severe.
Typical values of ${z_{\rm cut}}$ are around $0.1$, while typical
values of $\beta$ are between $0$ and $2$.

Associated with the SoftDrop algorithm, in Sec.~\ref{sec:SM_jetsub_2prong:polar} we will study the observable $z_g$ as a means for performing polarimetry.
The $z_g$ observable is also referred to as the groomed momentum fraction, and is defined as
\begin{align}
z_g=\frac{\min\left[ p_{Ti}, p_{Tj}  \right]}{p_{Ti}+p_{Tj}}
\end{align}
for the first declustering that satisfies the SoftDrop criteria.
This observable probes the momentum sharing between the two prongs in the
jet. For $\beta \ge 0$, this observable is Sudakov safe \cite{Larkoski:2013paa,Larkoski:2015lea} on QCD
jets.

In addition to mMDT/SoftDrop, we also consider the trimming algorithm, since it is used by the ATLAS collaboration.
Starting from a jet of radius $R$ identified with an IRC-safe jet algorithm, trimming is defined by the following procedure:
\begin{enumerate}
\item Recluster the jet into subjets of radius $R_{\text{sub}}$.
\item Eliminate from the jet all particles in subjets that satisfy
  $p_{T,\text{subjet}} > z_{\text{cut}} \, p_{T,\text{jet}}$.
\item The trimmed jet is then defined to consist of the remaining particles.
\end{enumerate}
Trimming has been experimentally shown to be a powerful grooming algorithm, and it exhibits excellent mass resolution.
That said, trimming is known to suffer from non-global logarithms \cite{Dasgupta:2001sh}, and does not have a smooth spectrum as a function of the trimmed mass.
The trimmed mass was analytically calculated in Ref.~\cite{Dasgupta:2013ihk}.
The trimming parameters used by ATLAS are $R_{\text{sub}}=0.2$,  $ {z_{\rm cut}}=0.05$, and the $k_T$ algorithm is used to perform the reclustering.

\subsubsection{Parton Shower Samples}\label{sec:SM_jetsub_2prong:samples_sub}

For our QCD background jet samples we generated $pp\to$ dijets in
\textsc{Pythia} 8.230~\cite{Sjostrand:2006za,Sjostrand:2007gs} with
the Monash13 tune~\cite{Skands:2014pea}. 
An explicit comparison to the \textsc{Herwig}
generator~\cite{Bahr:2008pv,Bellm:2015jjp} is left for future work.
Samples were generated with hadronization off (parton level), with
hadronization on but underlying event off (hadron level), and with
hadronization and underlying event on (truth level). Pileup was
included by superimposing uncorrelated minimum bias events, which are
also generated in \textsc{Pythia} 8, this time using the 4C
tune. Details of our pileup removal strategies will be described in
Sec.~\ref{sec:SM_jetsub_2prong:pu_tech}.

For our polarized $W$ samples, we considered a $gg$ produced resonance, $X$, that decays to a pair of polarized $W$ bosons.
This kind of resonance decaying to longitudinally-polarized $W$s appear in warped extra-dimensional models, where the Standard Model fields propagate in the bulk.
On the other hand, models with graviton-like tensors with minimal couplings yield only transversely-polarized $W$ bosons.
They were produced with the \textsc{JHUGEN} 3.1.8~\cite{Gao:2010qx,Bolognesi:2012mm} generator, interfaced with \textsc{Pythia} 8 \cite{Sjostrand:2007gs} for parton showering including the effect of hard gluon radiation.
A resonance width of 1\% was chosen.
Table~\ref{tab:SM_jetsub_2prong:polarisedSamples} shows the coupling values used to generate the polarised $W$ samples (see also Ref.~\cite{Gao:2010qx} for more information). 

\begin{table}[t]
\centering
\begin{tabular}{|c|c|c|c|}
\hline
Model	&Production couplings	&Decay couplings	&Decay helicity amplitudes 	\\
\hline
$2_b^+$	& $g_1=1$		& $g_5=1$		& $f_{00}=0.98$			\\
$2_m^+$	& $g_1=1$		& $g_1=g_5=1$		& $f_{00}=0.08,f_{+-}=f_{-+}=0.46$\\	
\hline
\end{tabular}
\caption{A description of the production and decay constants used to produced the polarized $W$ samples used in this studies.}
\label{tab:SM_jetsub_2prong:polarisedSamples}
\end{table}

\subsection{Theory Approaches to Robust Ratio Observables}\label{sec:SM_jetsub_2prong:hybrid_ratio}

The observables defined in Sec.~\ref{sec:SM_jetsub_2prong:shape_def} are designed to
distinguish boosted bosons from QCD jets using the detailed structure
of the radiation within the jets.
From this perspective, it is then
immediately clear that there will be an interplay between performance
and robustness.\
By using a maximal amount of radiation within the jet,
the maximal information is available to identify the origin of the
jet.
However, an increased sensitivity to radiation also means an
increased sensitivity to contamination, both theoretically from
non-perturbative hadronization effects and underlying event, as well
as experimentally from pileup.
It also introduces sensitivity to the
experimental reconstruction of soft momenta in the event.
On the other
hand, a tight grooming procedure, which removes radiation from within
the jet, is expected to reduce the ability to distinguish boosted
bosons from QCD jets.
In that context, it is interesting to see that both ATLAS and CMS
have currently adopted a tight grooming strategy.

In this subsection, we introduce a classification of different tagging strategies which will be an organizing framework for our study of robustness and performance.
This organization will be based on the idea of dichroic observables \cite{Salam:2016yht}, which are ratio observables constructed from combinations of groomed and ungroomed observables.
In Sec.~\ref{sec:SM_jetsub_2prong:dichroic}, we review the physics of the dichroic approach for the example of the $\tau_{21}$ observable.
In Sec.~\ref{sec:SM_jetsub_2prong:dichroic_new}, we generalize the dichroic construction to energy correlation function based observables and give definitions of dichroic $N_2$, $D_2$, and $M_2$ observables.
This allows us to have multiple concrete examples of observables for each of our general tagging strategies.
Then, in Sec.~\ref{sec:SM_jetsub_2prong:dichroic_sum} we present the complete set of tagging strategies that we will consider in this work, including the different parameters we will scan within each general class.

\subsubsection{Review of the Dichroic Approach}\label{sec:SM_jetsub_2prong:dichroic}

In Ref.~\cite{Salam:2016yht}, it was proposed that in addition to considering the standard $\tau_{21}$ observable and its groomed counterpart, one should also consider a mixed version, with a groomed denominator, and an ungroomed numerator.
This observable was termed ``dichroic'', since different levels of grooming are sensitive to different color structures.
More explicitly, Ref.~\cite{Salam:2016yht} considered the three observables:
\begin{align}
  \tau_{21}^{\text{large}} =\frac{\tau_2^{\text{large}}}{\tau_1^{\text{large}}}\,,
  \qquad
  \tau_{21}^{\text{small}} =\frac{\tau_2^{\text{small}}}{\tau_1^{\text{small}}}\,,
  \qquad
  \tau_{21}^{\text{dichroic}} =\frac{\tau_2^{\text{large}}}{\tau_1^{\text{small}}}\,,
\end{align}
where ``large'' refers either to the plain jet of a
lightly/loosely groomed jet and ``small'' denotes a more aggressively/tightly
groomed jet.
It was then argued that the dichroic combination was optimal for
performance, while also increasing the robustness to non-perturbative
contamination.
Here we briefly summarize the motivation for the dichroic observable.
Readers interested in a more detailed discussion of the dichroic approach, including perturbative calculations, are referred to Ref.~\cite{Salam:2016yht}.

\begin{figure}[t]
\begin{center}
\includegraphics[width=0.5\columnwidth]{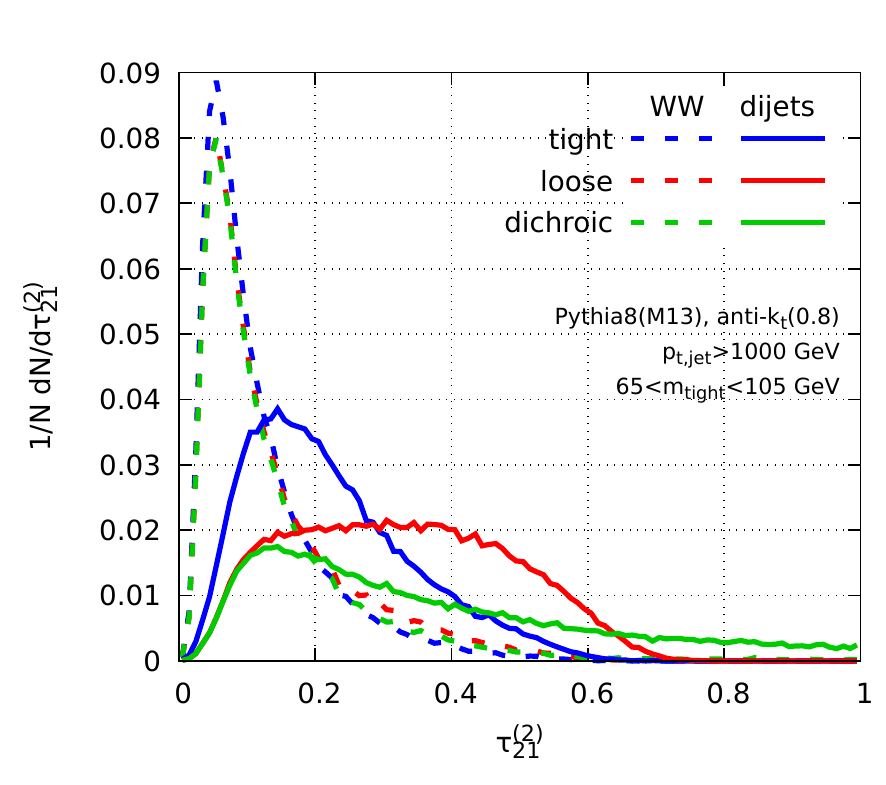}
\end{center}
\caption{Distributions of the tight, loose, and dichroic $\tau_{21}$ observables with a cut on the mMDT/SoftDropped mass. The dichroic approach offers considerably improved performance as compared with the tight grooming, almost preserving the performance of the loose grooming at high signal purity.  Figure taken from Ref.~\cite{Salam:2016yht}.}
\label{fig:SM_jetsub_2prong:dichroic_distribution}
\end{figure}

The benefits of the dichroic approach can simply be illustrated by
considering the distributions of the observables after aggressive
grooming (tight), moderate grooming (loose), and in the dichroic
approach.
This is shown in Fig.~\ref{fig:SM_jetsub_2prong:dichroic_distribution} for the particular case of $\tau_{21}$.
Considerable performance is lost in going from loose to tight grooming, since the background distribution is pushed to lower values of $\tau_{21}$.
For the dichroic observable, by constrast, the behavior of the loose
and dichroic observables is identical at small values of $\tau_{21}$,
but for larger values of $\tau_{21}$, the dichroic distribution is
pushed to yet larger values than the loose distribution, leading to
improved performance.
Since the dichroic ratio observable uses a partially-groomed
observable, it is expected to be less sensitive to non-perturbative effects due to hadronization.
It therefore represents an interesting new class of observables to consider. 

\subsubsection{New Dichroic Observables}\label{sec:SM_jetsub_2prong:dichroic_new}

It is almost straightforward to extend the dichroic definition from $N$-subjettiness ratios to observables formed from the energy correlation functions.
For $\tau_{21}$ it is immediately clear what the numerator and denominator of the observable are, however, this is initially less obvious for the energy-correlation-function-based observables that have a more complicated structure.
Here, we define the dichroic variants of the $M_2$, $N_2$, and $D_2$ by isolating a single factor of a mass like observable ($e_2$) as the denominator, and defining the remainder of the observable as the numerator.
The definitions of the dichroic variants of the $M_2$, $N_2$, and $D_2$
observables are then
\begin{align}
  M_2^{\text{dichroic}}&= \frac{( _1e_{3})^{\text{large}}  }{e_{2}^{\text{small}}}\,, \\
 N_2^{\text{dichroic}}&= \frac{\left( _{2}e_{3} / e_{2} \right)^{\text{large}} }{e_{2}^{\text{small}}}\,,\\
  D_2^{\text{dichroic}}&=\frac{\left( e_{3} / e_{2}^2 \right)^{\text{large}}}{ e_{2}^{\text{small}}}\,.
\end{align}
The above prescription is most easy to see for the $N_2$ observable. In the two-prong limit, the combination $ _{2}e_{3} / e_{2} $ reduces to $\tau_2$, and therefore the dichroic $N_2$ ratio behaves similarly to the dichroic $\tau_{21}$ ratio.

\subsubsection{Summary of Tagging Strategies}\label{sec:SM_jetsub_2prong:dichroic_sum}

\begin{table}
\begin{center}
\begin{tabular}{| c | c | c |c |c|c|c |c|r| }
  \hline                       
  Observable &  Numerator & Denominator \\
  \hline
  $M_2$ &   $_{1}e_{3}$ & $ e_{2}$ \\
  $N_2$ &   $_{2}e_{3} / e_{2} $ & $ e_{2}$ \\
  $D_2$ &   $e_{3} / e_{2}^2 $ & $ e_{2}$ \\
  $\tau_{21}$ &   $\tau_2$ & $\tau_1$ \\
  \hline  
\end{tabular}
\end{center}
\caption{
Definitions of the numerators and denominators for the different jet substructure observables.
}
\label{tab:SM_jetsub_2prong:dn}
\end{table}

\begin{table}[t!]
\begin{center}
\begin{tabular}{| c | c | c |c |c|c|c |c|c | }
  \hline                       
  Notation: $m \otimes \frac{n}{d}$ & $m$ (mass) & $n$ (numerator) & $d$ (denominator)\\
  \hline
  $p \otimes \frac{p}{p}$ & plain  &  plain & plain \\
  $\ell \otimes \frac{p}{p}$ & loose  &  plain & plain \\
  $\ell \otimes \frac{p}{\ell}$ & loose  &  plain & loose \\
  $\ell \otimes \frac{\ell}{\ell}$ & loose  &  loose & loose \\
  $t \otimes \frac{p}{p}$ & tight  &  plain & plain \\
  $t \otimes \frac{p}{\ell}$ & tight  &  plain & loose \\
  $t \otimes \frac{\ell}{\ell}$ & tight  &  loose & loose \\
  $t \otimes \frac{p}{t}$ & tight  &  plain & tight \\
  $t \otimes \frac{\ell}{t}$ & tight  &  loose & tight \\
  $t \otimes \frac{t}{t}$ & tight  &  tight & tight \\
  \hline
  $\text{trim}$ & trim &  trim & trim \\
  \hline  
\end{tabular}
\end{center}
\caption{ A summary of the different tagging strategies considered,
  including the notation, and the degree of grooming for the mass, and
  numerator and denominator of the shape observable. For simplicity,
  we have suppressed the jet radius, $R$. The definitions of plain,
  loose, tight and trim are given in the text.}
\label{tab:SM_jetsub_2prong:tag_summary}
\end{table}

The tagging strategies we consider can be put into the general form of a (groomed) mass cut ($m$) followed by a cut on a two-prong tagging observable, which takes the form
\begin{align}
\mathcal{O}=\frac{\text{3-particle observable}}{\text{2-particle (mass) observable}} \equiv \frac{n}{d}\,.
\end{align}
The explicit numerators ($n$) and denominators ($d$) for the different observables are summarized in Table~\ref{tab:SM_jetsub_2prong:dn}.

As our organizing principle for classes of  jet substructure taggers, we use the type of grooming applied to the initial mass cut, and the type of grooming applied to the numerator and denominator of the two-prong observable.
We use the notation 
\begin{align}
m \otimes \frac{n}{d}
\end{align}
to denote the grooming applied to a particular observable, where $m$, $n$, and $d$ can take the values
\begin{itemize}
\item plain ($p$): no grooming applied;
\item loose ($\ell$): SoftDrop with ${z_{\rm cut}}=0.05$, $\beta=2$;
\item tight ($t$): mMDT with ${z_{\rm cut}}=0.1$.
\item trim: trimming with $R_{\text{sub}}=0.2$,  $ {z_{\rm cut}}=0.05$ and the $k_T$ algorithm to perform the reclustering, as used by ATLAS.
\end{itemize}
In general, to have a good 2-prong tagging strategy, we want to have
\begin{align}
m \geq d \geq n\,,
\end{align}
where $\geq$ refers to the aggressiveness of the groomer, with $t > \ell > p$.

The complete set of configurations that we consider is given in Table~\ref{tab:SM_jetsub_2prong:tag_summary}.
These constitute generic grooming strategies, and will be studied for each of $N_2$, $D_2$, and $\tau_{21}$.
They include the dichroic ratios, as well as the current ATLAS and CMS approaches as specific examples.
This will allow us to draw general conclusions about the performance and robustness of jet substructure taggers.

\begin{table}
\begin{center}
\begin{tabular}{| c | c | c |c |c|c|c |c|r| }
  \hline                       
  Parameter &  Values Scanned \\
  \hline
  Jet Radius $R$ &   $0.8$ \\
  Jet Shape  &   $D_2$, $N_2$, $M_2$, $\tau_{21}$  \\
  Jet Shape Angular Exponent $\beta$ &   $1$, $2$ \\
  Jet $p_T$ &   $500$ GeV, $1000$ GeV, $2000$ GeV  \\
  \hline  
\end{tabular}
\end{center}
\caption{
Jet shapes and parameters scanned for each of the different strategies proposed in Table~\ref{tab:SM_jetsub_2prong:tag_summary}. 
}
\label{tab:SM_jetsub_2prong:params}
\end{table}

While our general approach is based on studying the different classes of grooming strategies in Table~\ref{tab:SM_jetsub_2prong:tag_summary}, for each of these different classes of strategies we will also scan different parameters.
In particular, we will scan the jet shape, jet shape angular exponent, and jet $p_T$.
The values scanned are summarized in Table~\ref{tab:SM_jetsub_2prong:params}.
This allows us to understand if the conclusions drawn are associated with specific observables within a given strategy, as well as to optimize over these parameters.
Due to the large number of physics and detector parameters that can be
varied in our study, only a subset of plots will be presented in the
study. More specifically, we will primarily show results for the
following 3 benchmark points
\begin{itemize}
\item an {\bf ATLAS-like tagger} putting cuts on the trimmed mass and
on $D_2^{(1)}$ computed on the trimmed jet: $D_2^{(1)}\text{[trim]}$;
\item a {\bf CMS-like tagger} putting cuts on the mMDT mass and on
$N_2^{(1)}$ computed on the tightly-groomed jet: $N_2^{(1)}[t\otimes t/t]$;
\item the {\bf Les-Houches Dichroic Taggger (LHDT)} putting cuts on
  the mMDT mass and on the dichroic $D_2^{(2)}$:
  $D_2^{(2)}[t\otimes l/t]$.
\end{itemize}
To study variations, we will either fix the grooming strategy to one
of the three benchmark points and vary the jet shape, or conversely,
fix the type of jet shape to one of our benchmark points and vary the
grooming strategy.
A more global summary of the observables scanned, highlighting those
which we find to be most promising, will be given in
Fig.~\ref{fig:SM_jetsub_2prong:phasespace}.

\subsection{Theory Robustness}\label{sec:SM_jetsub_2prong:np}

In this subsection, we study robustness to hadronization and underlying event, which we categorize as ``Theory" robustness. Robustness to the detector and to pileup are treated in Sec.~\ref{sec:SM_jetsub_2prong:exp}.

\subsubsection{Hadronization}\label{sec:SM_jetsub_2prong:hadr}

We begin by studying the robustness of different tagging techniques to hadronization.
As discussed in Sec.~\ref{sec:SM_jetsub_2prong:pres}, this must be interpreted with some care, since the unhadronized events are not themselves physical.
Nevertheless, the comparison of hadronized and unhadronized distributions is the best proxy for understand the impact of hadronization short of performing an analytic calculation.
To ensure that our conclusions are robust, ideally we would consider parton shower generators which implement different hadronization models.
For example, the \textsc{Pythia} shower uses the string model \cite{Andersson:1983ia,Andersson:1998tv}, while \textsc{Herwig}$++$ uses the cluster model \cite{Webber:1983if,Marchesini:1987cf}.
See for example Refs.~\cite{Buckley:2011ms,Skands:2011pf,Skands:2012ts} for a more detailed discussion.
Due to the restricted scope of this report, here we only consider \textsc{Pythia}.
The effect of hadronization on two-prong substructure observables has been studied in Refs.~\cite{Larkoski:2015kga,Salam:2016yht,Larkoski:2017iuy,Larkoski:2017cqq}.

\begin{figure}
  \centering{\includegraphics[width=0.4\textwidth]{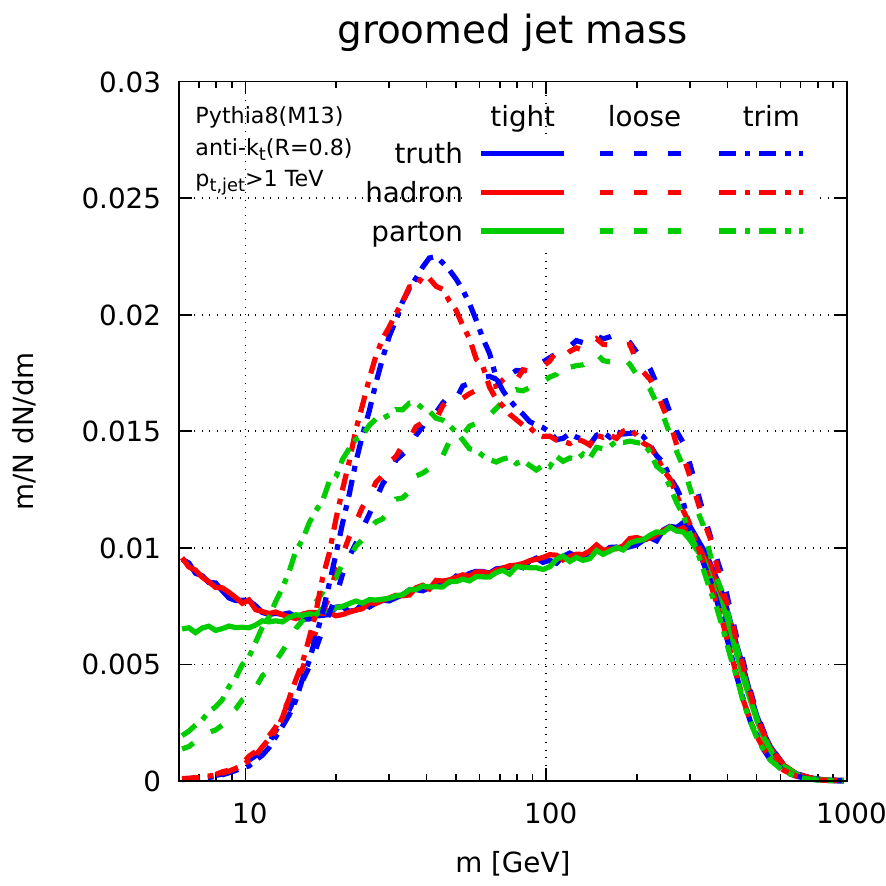}}
  \caption{Mass distribution after grooming for the three groomers
    considered in this paper. The distributions are shown at parton
    level, at hadron level, and at truth level (i.e.\ including both
    hadronization and the underlying event).}\label{fig:SM_jetsub_2prong:mass-distribution}
\end{figure}

Before studying robustness under hadronization quantitatively, we begin by showing several distributions with and without hadronization.
This will help to introduce the different observables, as well as to give the reader a feeling for the robustness at the level of the shape of the distribution, and how this compares to our resilience measure. 

In Fig.~\ref{fig:SM_jetsub_2prong:mass-distribution}, we show the jet-mass distribution for the three grooming strategies considered in this paper, for parton, hadron, and truth levels.
Although we will not focus directly on the mass distribution in this paper, it plays an important role since all of our studies will be performed with jet mass cuts.
Here we see two primary features.
First, all three groomers give rise to significantly different mass distributions.
This has been discussed in detail in Refs.~\cite{Dasgupta:2013ihk,Larkoski:2014wba}.
Second, with both tight and loose grooming, the distributions are robust to hadronization.
This is particularly true for tight grooming where hadronization has almost no effect, except at extremely small values of the observable.
On the other hand, the trimmed mass distribution is less robust to hadronization effects.

\begin{figure}
\subfloat[]{
  \includegraphics[width=0.315\textwidth,page=1]{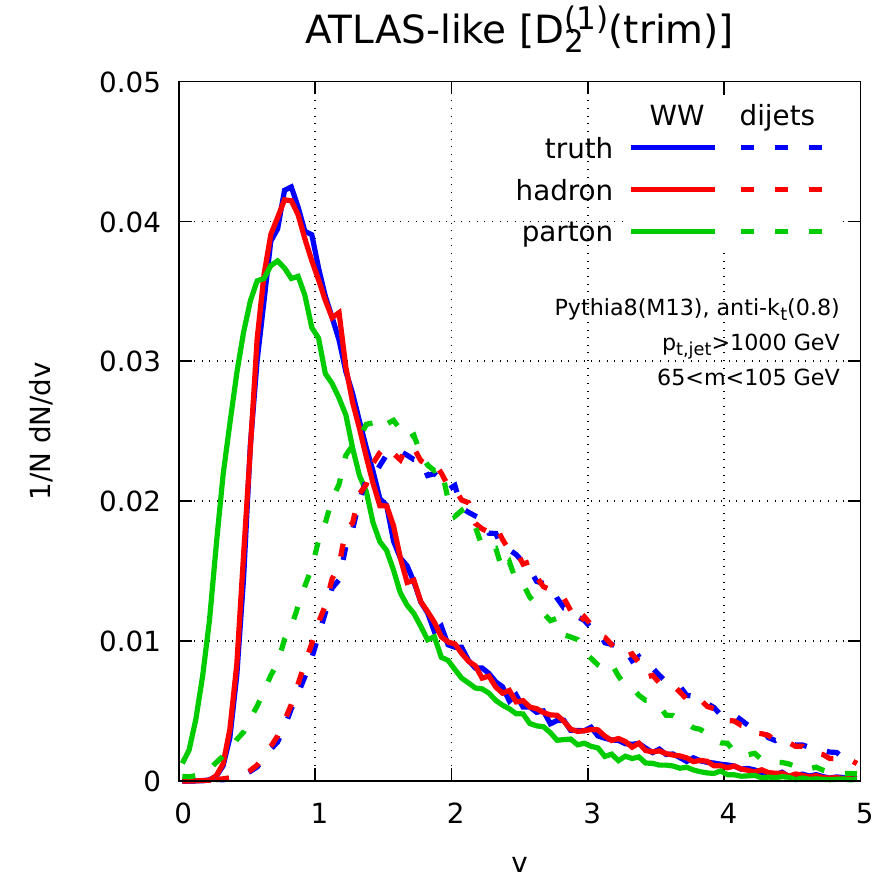}
}
\subfloat[]{
  \includegraphics[width=0.315\textwidth,page=2]{plots/jetsub_2prong_shape-distribs.pdf}
}
\subfloat[]{
  \includegraphics[width=0.315\textwidth,page=3]{plots/jetsub_2prong_shape-distribs.pdf}
}
  \caption{Distribution of three benchmark shapes: $D_2^{(1)}$
    computed on the trimmed jet (ATLAS-like), $N_2^{(1)}$ computed on
    the tight (mMDT) jet (CMS-like), and the dichroic $D_2^{(2)}$ with numerator
    computed on the loose jet and denominator computed on the tight
    jet (LHDT). The distributions are shown at parton level, at hadron level,
    and at ``truth'' level (i.e.\ including both hadronization and the
    underlying event), for both $WW$ (solid) and dijet (dashed)
    events.}\label{fig:SM_jetsub_2prong:shape-distribution}
\end{figure}

In Fig.~\ref{fig:SM_jetsub_2prong:shape-distribution}, we show
distributions for our benchmark observables, namely $D_2^{(1)}$, $N_2^{(1)}$, and a dichroic version of $D_2^{(2)}$, measured on both background and signal jets.
In all cases, we see that hadronization has a sizable effect on the shape of the distribution, pushing it to larger values.
For the $D_2$ observable, hadronization is mostly isolated to small values of the observable, and at larger values reduces simply to a shift of the distribution.
This has been discussed in detail for the case of $D_2$ in Refs.~\cite{Larkoski:2015kga,Larkoski:2017cqq,Larkoski:2017iuy}.
For the $N_2$ observable, hadronization effects are larger and are significant throughout the entire distribution.
Indeed, when we study performance and robustness quantitatively, we will find that while $N_2$-type observables tend to be more performant, they are also less resilient to hadronization effects. 

\begin{figure}
  \centering{\includegraphics[width=0.4\textwidth]{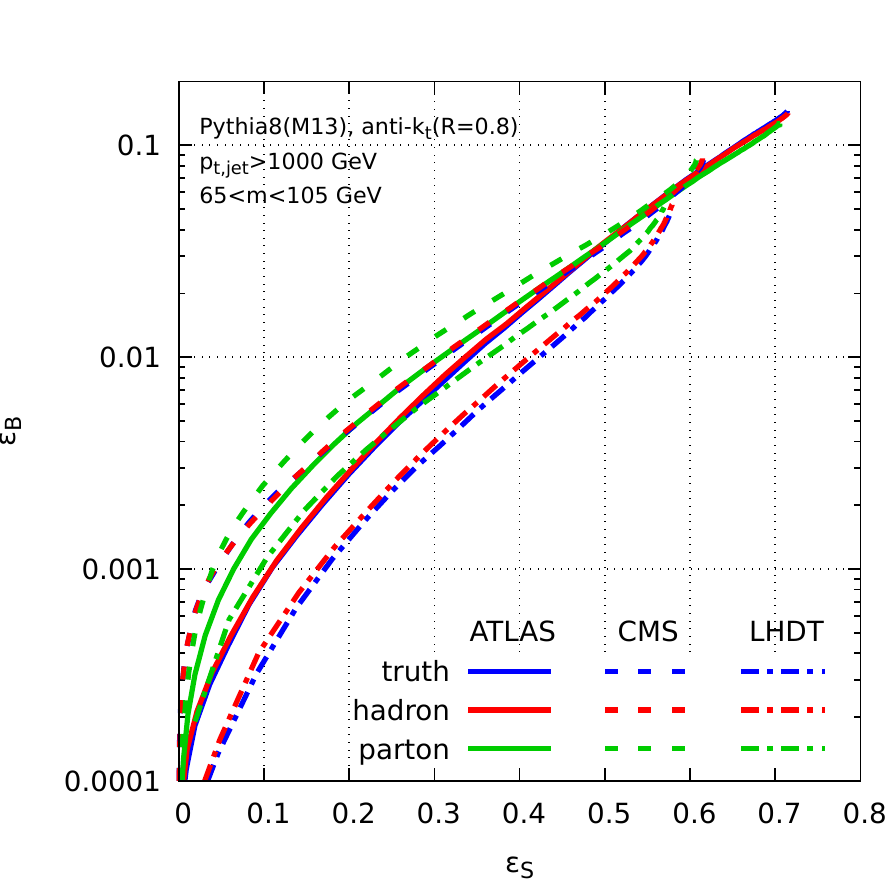}}
  \caption{ROC curves corresponding to the shapes plotted on
    Fig.~\ref{fig:SM_jetsub_2prong:shape-distribution}. The three line types correspond
    to the three benchmark points.}\label{fig:SM_jetsub_2prong:rocs}
\end{figure}

Since the primary role of hadronization is to push the distributions to larger values at small values of the observable, the performance of the observables is typically highly sensitive to hadronization, particularly at high signal purity.
In Fig.~\ref{fig:SM_jetsub_2prong:rocs} we illustrate this (lack of) robustness to hadronization at the level of the ROC curves for the different shape choices.
In all cases, we see that hadronization considerably improves the performance of the observables.
This is particularly true at high signal purity, and decreases as the signal purity is decreased.
Since the region of high signal purity is typically that of interest for jet substructure studies, this emphasizes the importance of understanding the robustness of observables to hadronization effects.

\begin{figure}
\begin{center}
\includegraphics[width=0.4\columnwidth]{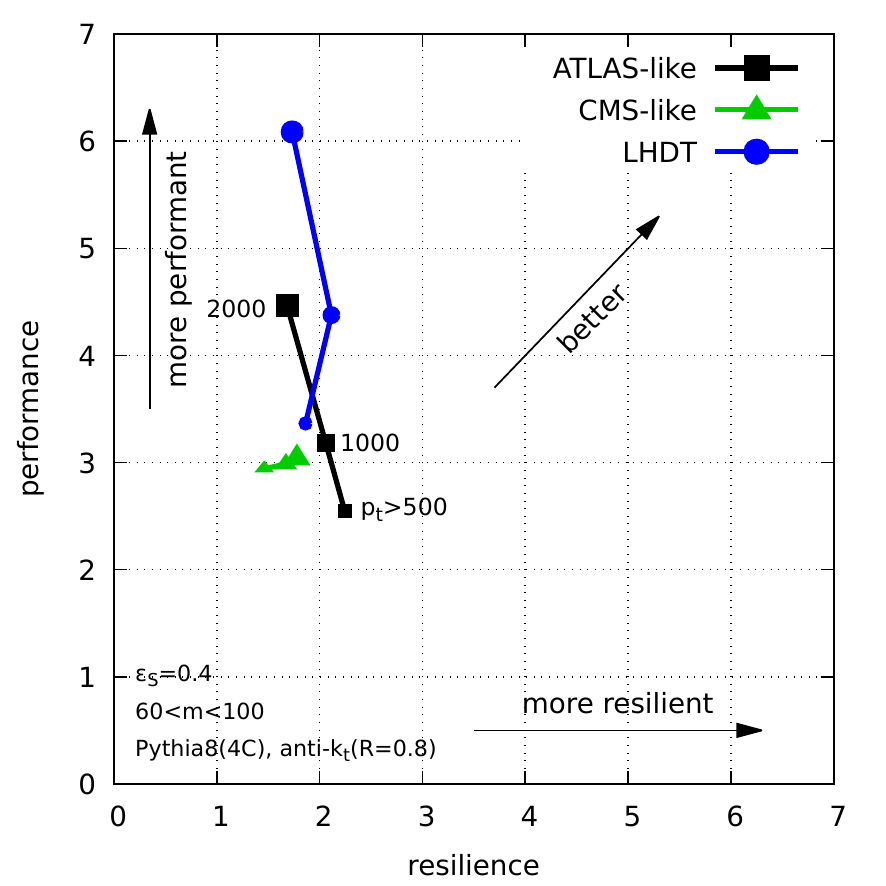}
\end{center}
\caption{An illustration of the performance-resilience plane that will be used to illustrate our results. More performant observables lie to the top, more resilient observables to the right, and better observables in the upper-right corner.}
\label{fig:SM_jetsub_2prong:sweep_pt}
\end{figure}

Having given a feel for the modifications due to hadronization at the level of both distributions and ROC curves, we now use our performance and resilience measures to perform a quantitative study.
Since our visualization method allows a considerable amount of information to be condensed into a single plot, we first briefly review our presentation method with a sample plot.
In Fig.~\ref{fig:SM_jetsub_2prong:sweep_pt}, we show a plot in the performance-resilience plane, in which we will display our results.
More performant observables appear higher on the $y$-axis (towards the top), while more robust (resilient) observables appear higher on the $x$-axis (to the right), as indicated by the arrows.
A performant and resilient observable will appear in the upper right corner.
For each observable, we also perform a scan of $p_T$, from $500-2000$ GeV, which are illustrated by the three connected points.
This will be the default format in which we display our results throughout the rest of this study.

\begin{figure}
\subfloat[]{
  \includegraphics[width=0.315\textwidth,page=3]{plots/jetsub_2prong_grooming-scan.pdf}
}
\subfloat[]{
  \includegraphics[width=0.315\textwidth,page=1]{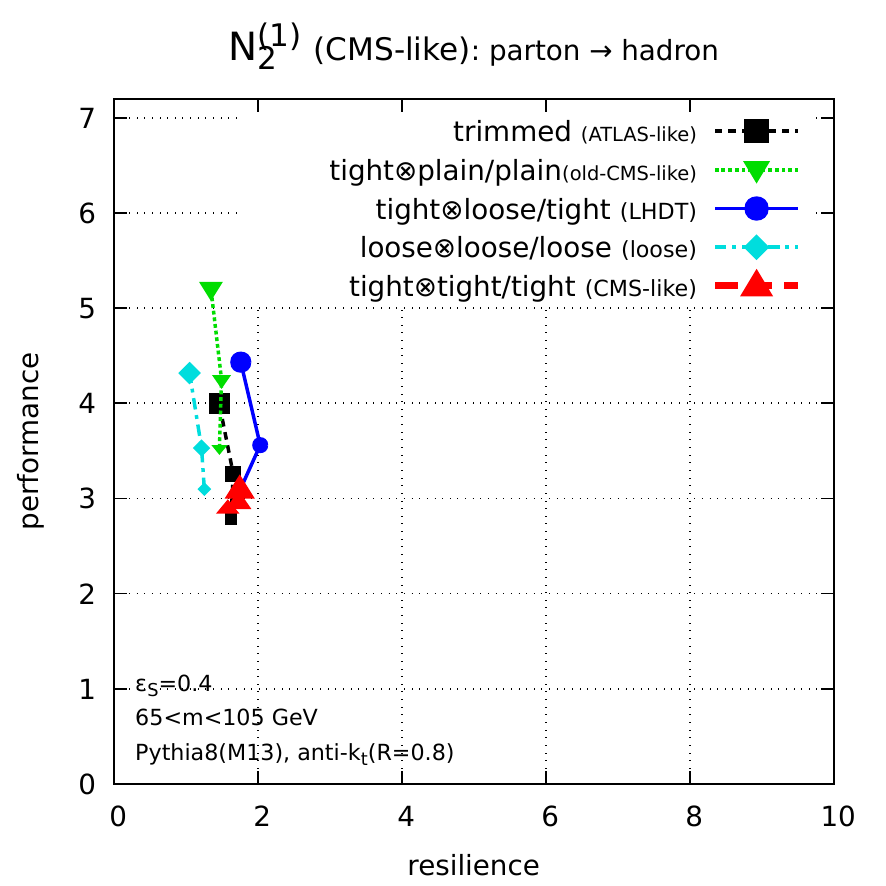}
}
\subfloat[]{\label{fig:SM_jetsub_2prong:grooming-hadronisation:c}
  \includegraphics[width=0.315\textwidth,page=5]{plots/jetsub_2prong_grooming-scan.pdf}
}
  \caption{Plots of the peformance-resilience plane under the addition of hadronization effects for the standard jet shape observables with different grooming strategies.}\label{fig:SM_jetsub_2prong:grooming-hadronisation}
\end{figure}

\begin{figure}
\subfloat[]{
  \includegraphics[width=0.315\textwidth,page=5]{plots/jetsub_2prong_shape-scan.pdf}
}
\subfloat[]{
  \includegraphics[width=0.315\textwidth,page=3]{plots/jetsub_2prong_shape-scan.pdf}
}
\subfloat[]{
  \includegraphics[width=0.315\textwidth,page=1]{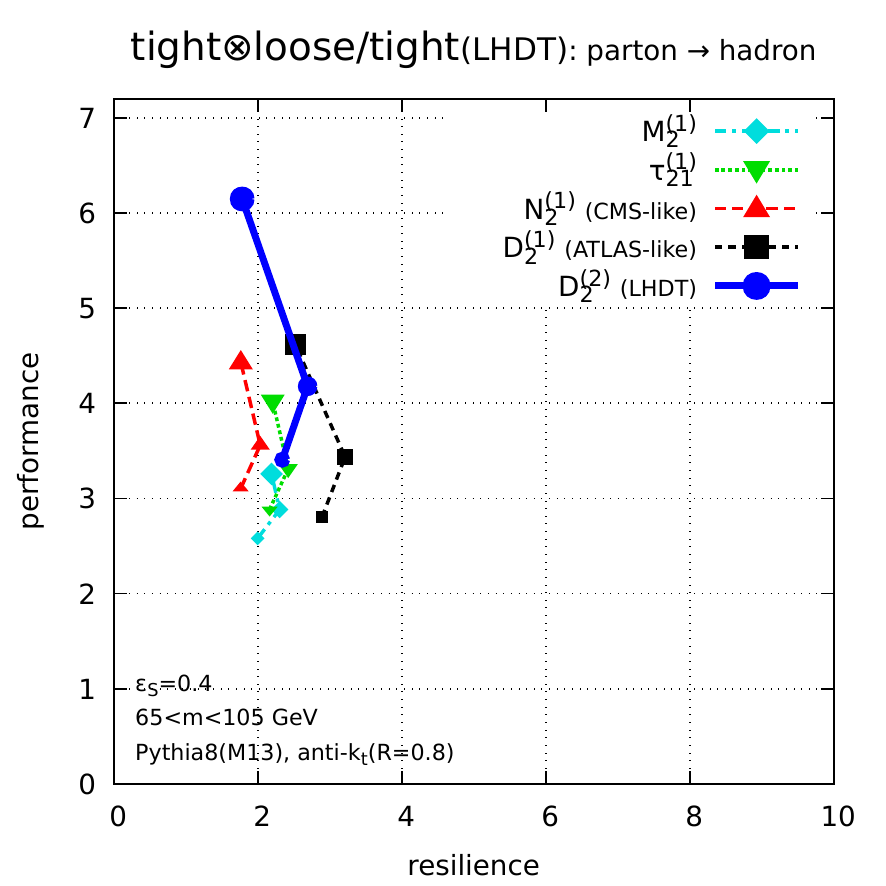}
}
  \caption{Plots of the peformance-resilience plane under the addition of hadronization effects for different jet shape observables, with fixed grooming strategies.}\label{fig:SM_jetsub_2prong:shapes-hadronisation}
\end{figure}

In Figs.~\ref{fig:SM_jetsub_2prong:grooming-hadronisation} and
\ref{fig:SM_jetsub_2prong:shapes-hadronisation}, we show the
performance-resilience plots for the effects of hadronization for our
benchmark observables.
Figure~\ref{fig:SM_jetsub_2prong:grooming-hadronisation} shows $D_2$, $N_2$, and dichroic $D_2$ for the different grooming strategies, while in Fig.~\ref{fig:SM_jetsub_2prong:shapes-hadronisation} we consider fixed grooming strategies in each plot, but different jet shape observables.
We first notice that, in almost all cases, a dichroic form of the observable can be used to improve resilience while maintaining a similar level of performance.
Among the shapes, we find that $D_2$ tends to be the most robust and
that $N_2$ tends to be slightly more performant, although $D_2$
becomes more performant than $N_2$ at larger $p_T$.
This agrees with what was seen by studying the distributions in Fig.~\ref{fig:SM_jetsub_2prong:shape-distribution} by eye, however, we are now able to quantify this.
We will see that this conclusion remains true under a larger scan of observables in Sec.~\ref{sec:SM_jetsub_2prong:exp_compare}.
We also notice that for almost all the observables, the trends with $p_T$ are similar.

\subsubsection{Underlying Event}\label{sec:SM_jetsub_2prong:UE}

\begin{figure}
\subfloat[]{
  \includegraphics[width=0.315\textwidth,page=4]{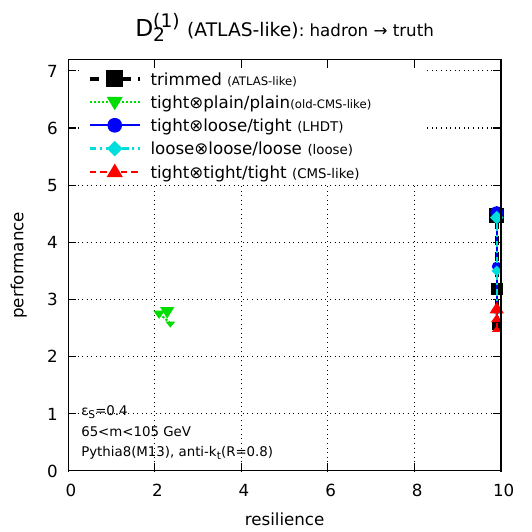}
}
\subfloat[]{
  \includegraphics[width=0.315\textwidth,page=2]{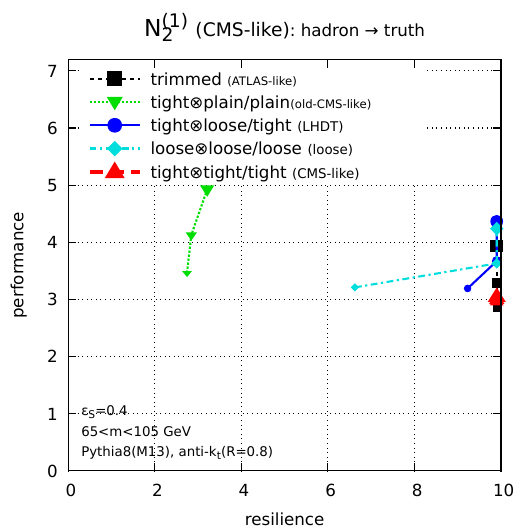}
}
\subfloat[]{\label{fig:SM_jetsub_2prong:grooming-UE:c}
  \includegraphics[width=0.315\textwidth,page=6]{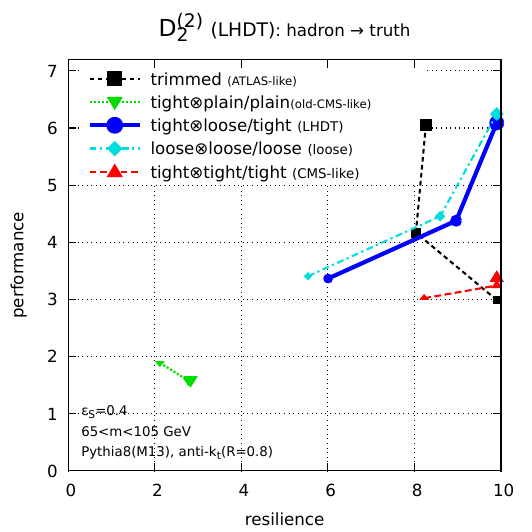}
}
  \caption{Plots of the peformance-resilience plane going from hadron
    level to truth level (inclusion of underlying event) for the
    standard jet shape observables with different grooming strategies.
    In this and the next figure, resilience values have been cut at
    $\zeta=10$ for easier comparison with the corresponding
    hadronisation plots,
    Figs.~\ref{fig:SM_jetsub_2prong:grooming-hadronisation}
    and~\ref{fig:SM_jetsub_2prong:shapes-hadronisation}.}\label{fig:SM_jetsub_2prong:grooming-UE}
\end{figure}

\begin{figure}
\subfloat[]{
  \includegraphics[width=0.315\textwidth,page=6]{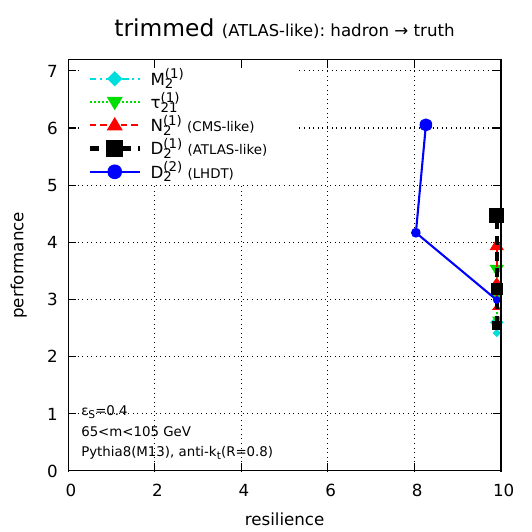}
 }
\subfloat[]{
  \includegraphics[width=0.315\textwidth,page=4]{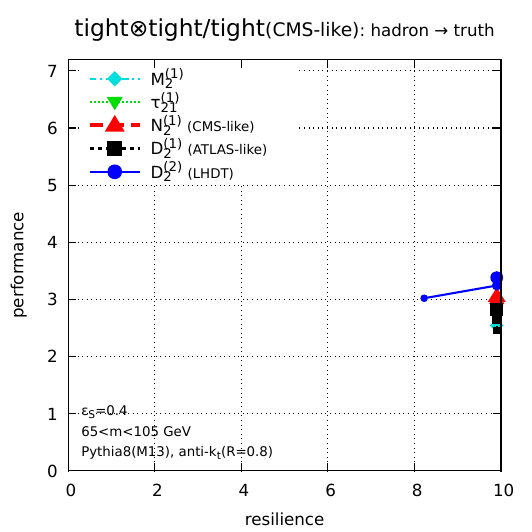}
}
\subfloat[]{
  \includegraphics[width=0.315\textwidth,page=2]{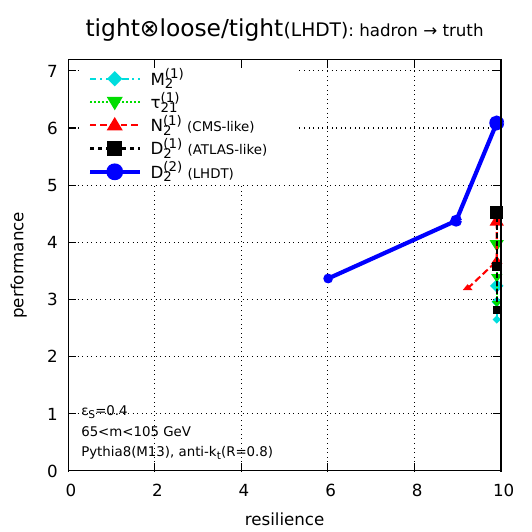}
}
  \caption{Plots of the peformance-resilience plane going from hadron level to truth level (inclusion of underlying event) for different jet shape observables, with fixed grooming strategies.}\label{fig:SM_jetsub_2prong:shapes-UE}
\end{figure}

We can now repeat the same exercise performed for hadronization to study the robustness to underlying event.
The results in the performance-resilience plane are shown in Figs.~\ref{fig:SM_jetsub_2prong:grooming-UE} and \ref{fig:SM_jetsub_2prong:shapes-UE}.
These are identical to Figs.~\ref{fig:SM_jetsub_2prong:grooming-hadronisation} and \ref{fig:SM_jetsub_2prong:shapes-hadronisation} but measure the robustness to underlying event instead of hadronization.
The first thing that is clear from comparing Figs.~\ref{fig:SM_jetsub_2prong:grooming-hadronisation} and \ref{fig:SM_jetsub_2prong:shapes-hadronisation} with  Figs.~\ref{fig:SM_jetsub_2prong:grooming-UE} and \ref{fig:SM_jetsub_2prong:shapes-UE} is that with modern grooming techniques, we are comparatively much less sensitive to underlying event than to hadronization effects.
With the exception of the tight$\otimes$plain/plain grooming strategy
(which indeed does not groom the observable), all the standard
observables are robust to underlying event for all the different
grooming strategies. Similarly, all the different jet shape
observables in this study are also robust to Underlying Event
effects. The dichroid $D_2^{(2)}$ observable
(Figs.~\ref{fig:SM_jetsub_2prong:grooming-hadronisation:c}
and~\ref{fig:SM_jetsub_2prong:grooming-UE:c}) shows a smaller resilience
against the Underlying Event although it remains much larger than the
corresponding resilience to hadronisation effects.
We believe that this should be viewed as a success of modern grooming tools.
We also believe that it is desirable, since underlying event effects are much less under theoretical control than hadronization effects.

\subsubsection{Towards Improved Performance and Robustness for ATLAS and CMS}\label{sec:SM_jetsub_2prong:exp_compare}

\begin{figure}
\begin{center}
\subfloat[]{\label{fig:SM_jetsub_2prong:unsoftdropped}
\includegraphics[width=7cm]{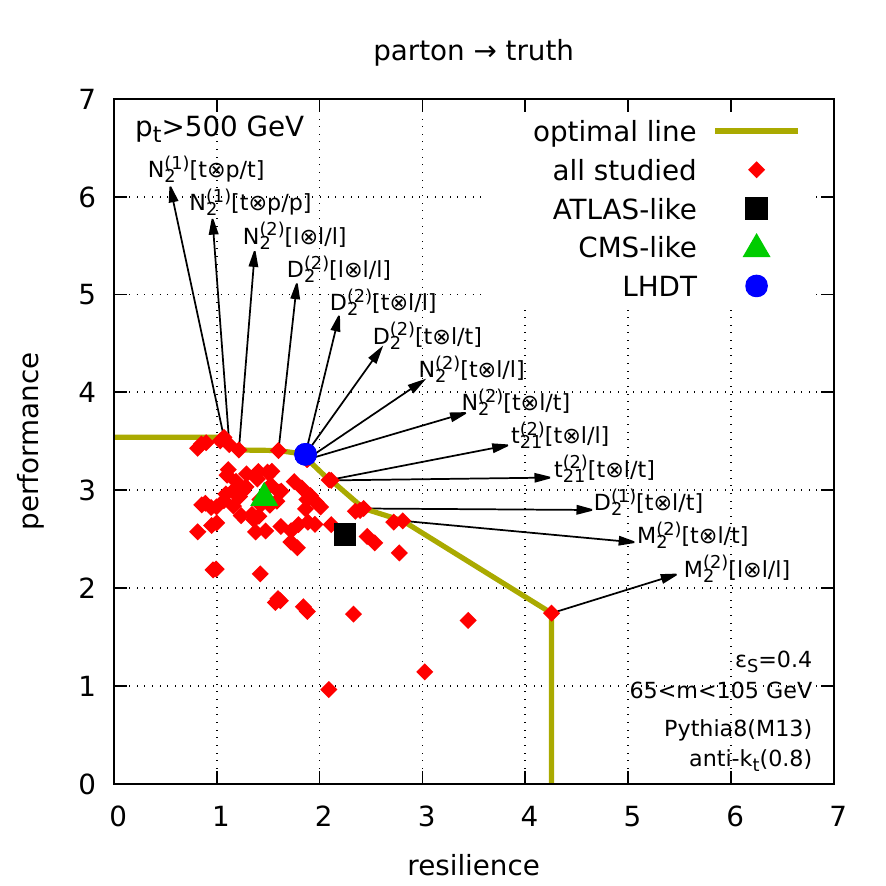}    
}\qquad
\subfloat[]{\label{fig:SM_jetsub_2prong:softdropped}
\includegraphics[width=7cm]{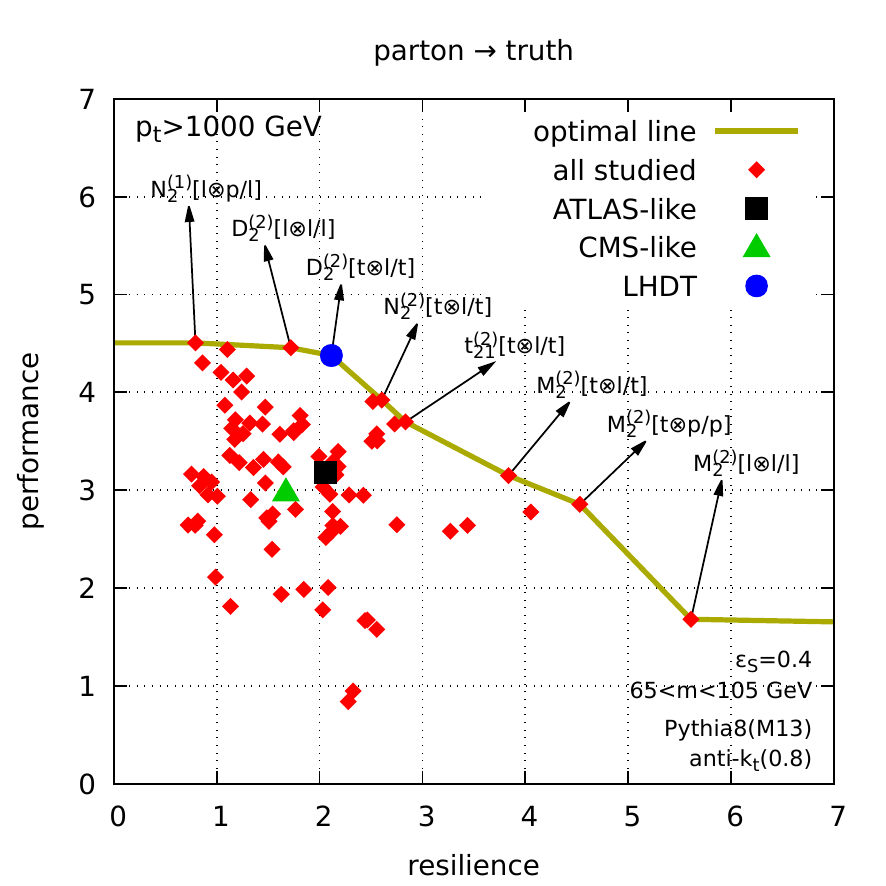}
}
\end{center}
\caption{The performance-resilience plane for the different observables scanned in our study at (a) $500$ GeV and (b) $1000$ GeV. The ATLAS- and CMS-like observables are marked in black and green, respectively. In both cases, more robust, and more performant observables can be selected, and a number of such observables are marked.
}
\label{fig:SM_jetsub_2prong:phasespace}
\end{figure}

The strategies used by ATLAS and CMS, namely trimmed $D_2$ \cite{Larkoski:2015kga,Larkoski:2014gra} and SoftDropped $N_2$ \cite{Moult:2016cvt} with DDT \cite{Dolen:2016kst} are specific examples of the broader approaches to two-prong tagging discussed above, and therefore our study allows us to gain insight into the different tagging strategies used by the collaborations,\footnote{And perhaps even into the sociology of the different experiments! However, such conclusions should be taken with a grain of salt.} as well as to suggest improved observables.

An overall scan of all the observables used in this study in the performance-resilience plane is shown in Fig.~\ref{fig:SM_jetsub_2prong:phasespace}.
The current CMS and ATLAS observables are highlighted with the black and green markers, respectively. We can draw two interesting conclusions from this plot.
First, for $p_T> 500$ GeV, it appears as though ATLAS is using a more robust, but less performant observable, while CMS is using a more performant, but less robust observable.
This must be caveated by the fact that we have not performed a full detector study, though it is suggestive.
Second, we can choose from the observables considered in our study observables that are simultaneously more performant, and more resilient than those that are currently being used.
At $500$ GeV the gains in performance are fairly minimal, but at $1000$ GeV, it seems that there are considerable gains in performance and robustness to be made.
The names of a large variety of those observables which lie on the upper boundary of the performance-resilience space are marked in the figure.

Looking more closely at the observables along this upper boundary, we see that there is in fact considerable structure, and a number of general lessons can be learned.
First, as we move along this boundary from least resilient to most resilient, we transition from $N_2^{(1,2)}$ observables which are the most performant, but less robust, through $D_2^{(2)}$, to $M_2^{(2)}$, which is more robust, but less performant.
This was also clearly observed in the distributions of Fig.~\ref{fig:SM_jetsub_2prong:shape-distribution}.
We believe that this is due to the fact that $N_2$ has a hard phase space boundary, and therefore non-perturbative effects are not isolated at small values of the observable, although it would be interesting to understand this behavior in more detail. 

A second pattern that is observed is that, in almost all cases, dichroic variants of the observables of the form $t\otimes \frac{p}{t}$ or $t\otimes\frac{\ell}{t}$ exhibit improved performance without significant loss in resilience.
We believe that it is worthwhile for the experiments to consider some of the dichroic observables that were newly introduced in Sec.~\ref{sec:SM_jetsub_2prong:dichroic_new} with simultaneous performance and resilience in mind.
We have highlighted in Fig.~\ref{fig:SM_jetsub_2prong:phasespace} that a whole interesting phase space of such observables exist, which map out the performance-resilience plane.
Different observables could be chosen depending on the particular
needs of a given study.
Note finally, that the above study has been carried using a specific
choices of grooming strategies (loose and tight). There is therefore a
potential additional gain that can be achieved by studying alternative
(more or less aggressive) options.

\subsection{Experimental Robustness}\label{sec:SM_jetsub_2prong:exp}

Having discussed robustness to theoretical issues, we now continue through our chain of realism of Fig.~\ref{fig:SM_jetsub_2prong:realism} and consider robustness to detector effects.
In Sec.~\ref{sec:SM_jetsub_2prong:det_model} we describe our detector model.
In Sec.~\ref{sec:SM_jetsub_2prong:pu_tech} we describe pileup removal.
In Sec.~\ref{sec:SM_jetsub_2prong:detector_robust} we study the robustness of jet mass to detector effects.
A more comprehensive study is left to a dedicated publication.

\subsubsection{Detector Models}\label{sec:SM_jetsub_2prong:det_model}

\begin{table}\centering
\renewcommand{\arraystretch}{1.25}
\begin{tabular}{|l|c|c|c|r@{\ =\ }l|}
\hline
Signal  & \multicolumn{5}{c|}{Model parameters}    \\ \cline{2-6}                      
        & Coverage
        & \ensuremath{p_{T}^{\text{min}}}
        & \ensuremath{p_{T}^{\text{max}}} 
        & \multicolumn{2}{c|}{Resolution function parameters} \\
\hline
\multicolumn{6}{|c|}{\textbf{Configuration A} ($R_{\text{calo}} = 1150$ mm, $B_{\text{solenoid}} = 2$ T)} \\
\hline
Towers     & $|\eta| < 2.5$ & 500 MeV &  10 TeV   & \ensuremath{a_{\text{calo}}} & $10\%\times\sqrt{\text{GeV}}$                  \\ 
(EM)       &                &         &           & \ensuremath{b_{\text{calo}}} & 0                                        \\
           &                &         &           & \ensuremath{c_{\text{calo}}} & 0.7\%                                    \\
Towers     & $|\eta| < 4.9$ & 500 MeV &  10 TeV   & \ensuremath{a_{\text{calo}}} & $50\%\times\sqrt{\text{GeV}}$                  \\
(HAD)      &                &         &           & \ensuremath{b_{\text{calo}}} & 0                                        \\
           &                &         &           & \ensuremath{c_{\text{calo}}} & 3\%                                      \\
\hline 
\multicolumn{6}{|c|}{\textbf{Configuration C} ($R_{\text{calo}} = 1290$ mm, $B_{\text{solenoid}} = 4$ T)} \\
\hline
Towers     & $|\eta| < 2.5$ & 500 MeV &  10 TeV   & \ensuremath{a_{\text{calo}}} & $3\%\times\sqrt{\text{GeV}}$                   \\
(EM)       &                &         &           & \ensuremath{b_{\text{calo}}} & 0                                        \\
           &                &         &           & \ensuremath{c_{\text{calo}}} & 0.5\%                                    \\
Towers     & $|\eta| < 4.9$ & 500 MeV &  10 TeV   & \ensuremath{a_{\text{calo}}} & $100\%\times\sqrt{\text{GeV}}$                 \\
(HAD)      &                &         &           & \ensuremath{b_{\text{calo}}} & 0                                        \\
           &                &         &           & \ensuremath{c_{\text{calo}}} & 5\%                                      \\
Tracks     & $|\eta| < 2.5$ &  1 GeV  &  300 GeV  & \ensuremath{a_{\text{track}}}  & $0.015\%\times\text{GeV}^{-1}$                  \\
           &                &         &           & \ensuremath{c_{\text{track}}} & 0.5\%                                     \\
\hline
\end{tabular}
\caption{Principal properties and smearing parameters of the detector model configurations A and C. Electromagnetic (EM) towers have a tower size of $\Delta\eta\times\Delta\phi = 0.025\times0.025$, while hadronic (HAD) towers have $\Delta\eta\times\Delta\phi = 0.1\times0.1$. Tracks are not modeled in configuration A, while configuration C models both tracks and calorimeter signals to simulate the effect of a particle flow algorithm. The resolution function parameters \ensuremath{a_{\text{calo}}}, \ensuremath{b_{\text{calo}}}{}, and \ensuremath{c_{\text{calo}}}{} are used together with the resolution function in Eq.~\eqref{eq:SM_jetsub_2prong:caloreso} to determine the width of the Gaussian energy smearing.
Similarly, \ensuremath{a_{\text{track}}}{} and \ensuremath{c_{\text{track}}}{} are used in Eq.~\eqref{eq:SM_jetsub_2prong:trkreso} to determine the width of the track-\ensuremath{p_{T}}{} resolution smearing.}
\label{tab:SM_jetsub_2prong:detmodel}
\end{table}

\begin{figure}
\begin{center}
\subfloat[]{
\includegraphics[width=0.48\columnwidth]{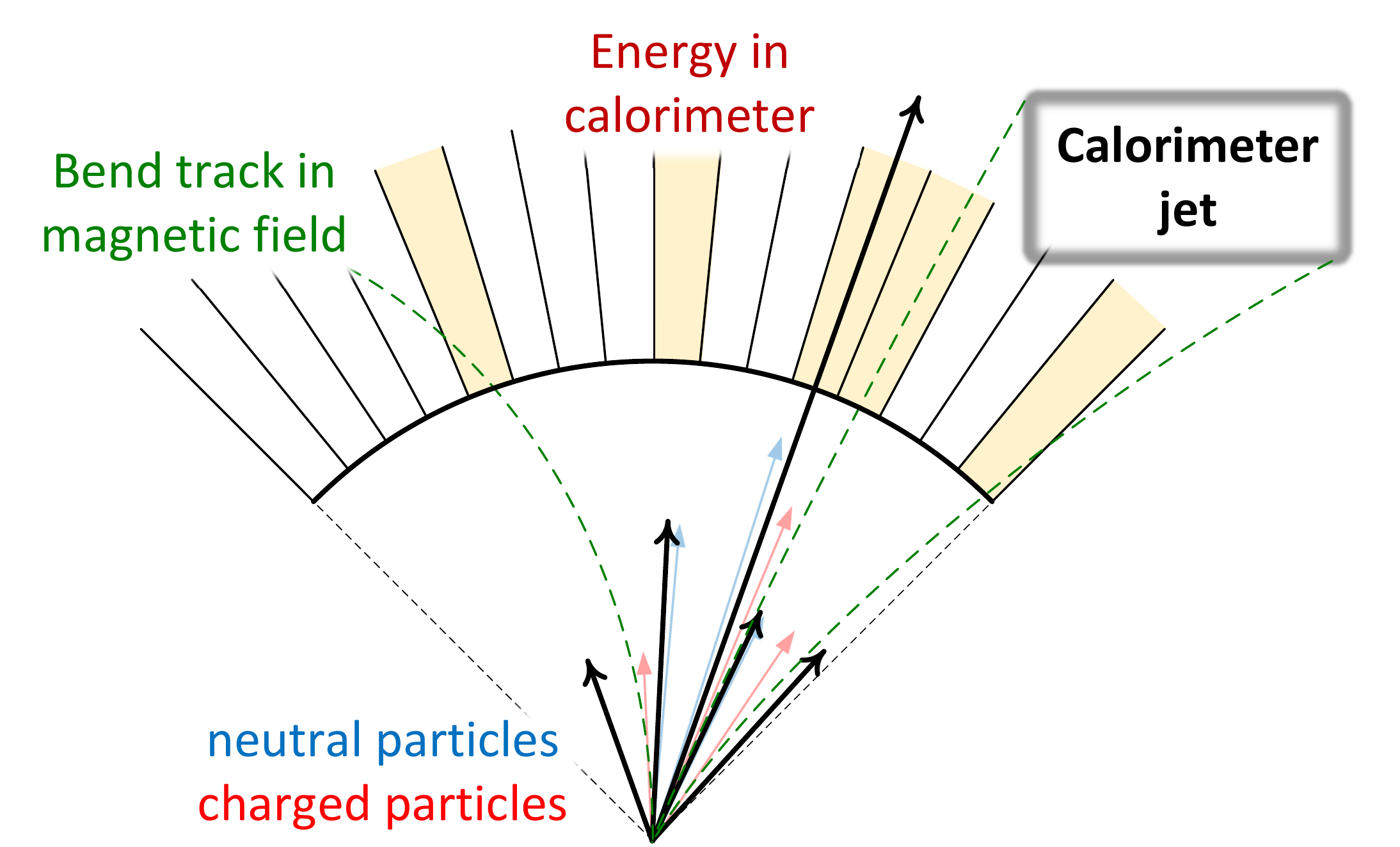}
}
\subfloat[]{
\includegraphics[width=0.48\columnwidth]{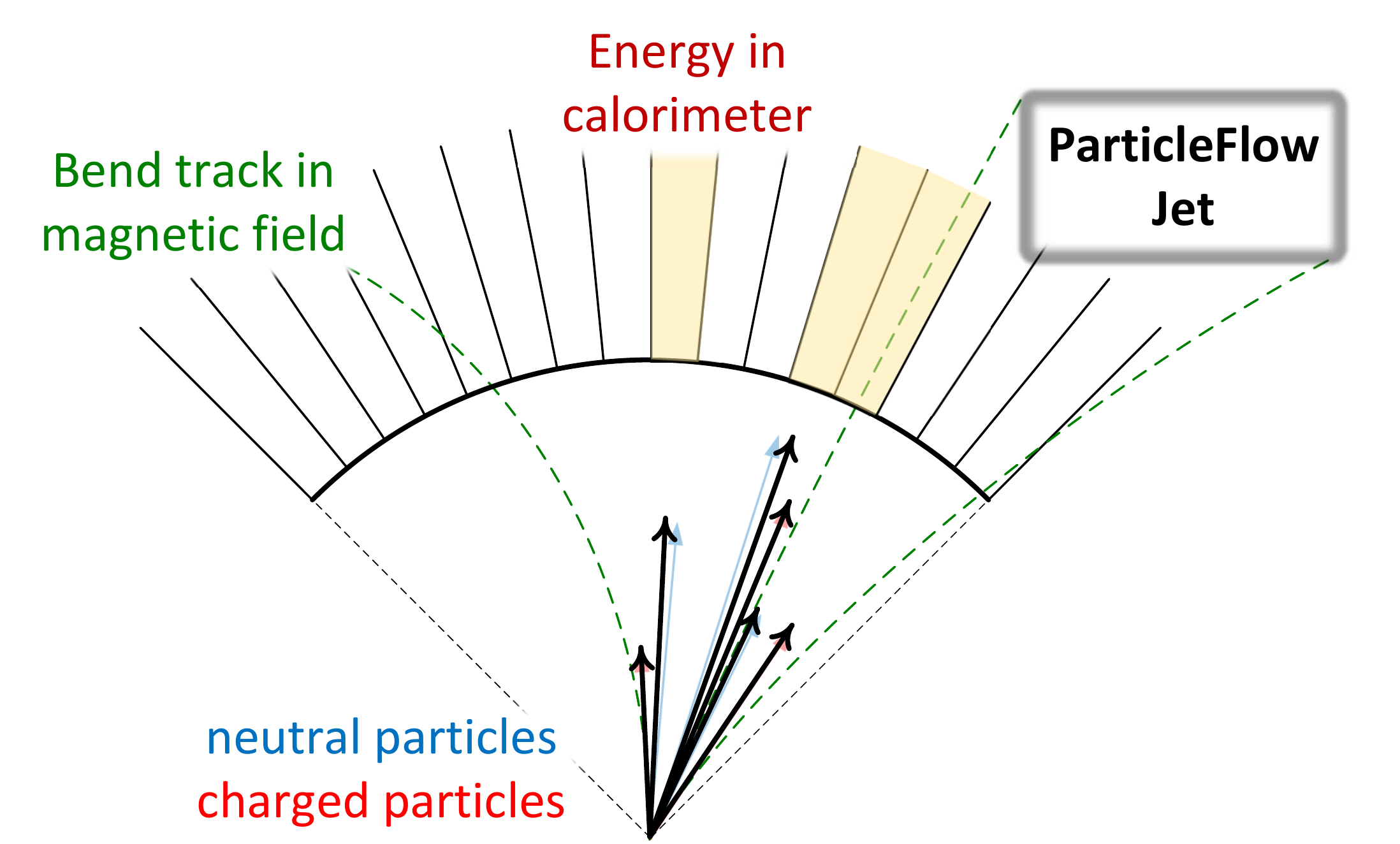}
}
\end{center}
\caption{An illustration of the basic jet features for the configuration A (ATLAS-like, calorimeter only) and configuration C (CMS-like, particle flow using tracks and calorimeter) detector signal models. The solid black arrows indicate the jet composition from representations of neutral (pale blue) and charged (pale red) particles by calorimeter towers or reconstructed tracks. The green dashed curves show charged particle tracks bend in the magnetic field. Both illustrations show the same truth-level jet.}
\label{fig:SM_jetsub_2prong:detmodel}
\end{figure}

The detector response is modeled by subjecting the particle-level objects\footnote{Those are stable particles produced by the generator that potentially reach a detector.  This principally requires a particle lifetime $\tau$ in the laboratory frame given by $c\tau > 10$ mm.}
to a simple acceptance and smearing model representing tracking detectors as well as calorimeters.
Basic detector feature descriptors are considered together with response features and signal reconstruction strategies similar to the ATLAS and CMS experiments at the LHC.
The respective model configurations are A (ATLAS{\emph{-like}} \cite{PERF-2007-01}) and C (CMS\emph{-like} \cite{CMS-TDR-08-001}).
Effects from particles showering in calorimeters are not modeled. 

Both configurations feature cylindrical detectors around the collision vertex with full azimuthal coverage $-\pi < \phi < \pi$.
The detector modeling produces a final state represented by a list of \emph{pseudoparticles} which are proxies for detector signals generated by stable particles.
Particles emitted at pseudorapidities $|\eta| > 4.9$ and non-detectable particles like neutrinos are excluded from this modeling and thus not part of this final state. 

In configuration A, the pseudoparticles are produced by a calorimeter-only detector model with regular projective readout in $(\eta,\phi)$ space. 
The energy of all stable particles generating hadronic shower when interacting with the detector material is collected 
in hadronic (HAD) \emph{calorimeter towers} of size $\Delta\eta\times\Delta\phi = 0.1\times0.1$. 
Electrons, positrons, and photons emitted within $|\eta| < 2.5$ fill electromagmetic (EM) towers of size $\Delta\eta\times\Delta\phi = 0.025\times0.025$ with their energy, mimicking the typical
coverage of the high granularity electromagnetic calorimeter. 
The direction of neutral particles is the generated direction at the vertex, while for all charged particles $\phi$ is changed to the azimuth at the entry point of the particle
into the calorimeter. 
This entry point is calculated using the particle trajectory in an axial uniform magnetic field of $B_{\text{solenoid}} = 2$~T and the radius $R_{\text{calo}} = 1150$ mm of
 the calorimeter front face.
If the transverse momentum of the charged particle  is too low to reach the calorimeter, i.e.\ its trajectory in the magnetic field does not exceed $R_{\text{calo}}$, 
the particle is considered invisible for the detector and thus ignored for further analysis.   
For particles reaching the calorimeter, the (bent) trajectory is not radial anymore at the front face.
This suggests a distribution of the particle energy into more than one tower.
In this model there is no energy sharing between towers, as this would require to at least model the longitudinal energy distribution in an electromagnetic or hadronic shower, with considerable computational effort.

After all energy is collected, the finite calorimeter resolution is modeled by smearing the tower energy \ensuremath{E_{\text{tower}}}{} following a Gaussian distribution with width 
$\sigma_{\ensuremath{E_{\text{tower}}}}$ given by the canonical calorimeter resolution function:
\begin{align}
  \frac{\sigma_{\ensuremath{E_{\text{tower}}}}(\ensuremath{E_{\text{tower}}})}{\ensuremath{E_{\text{tower}}}} = \sqrt{\frac{\ensuremath{a_{\text{calo}}}^{2}}{\ensuremath{E_{\text{tower}}}} + \frac{\ensuremath{b_{\text{calo}}}^2}{\ensuremath{E_{\text{tower}}}^{2}} + \ensuremath{c_{\text{calo}}}^{2}}.
  \label{eq:SM_jetsub_2prong:caloreso}
\end{align}
The three components of this function are the stochastic term \ensuremath{a_{\text{calo}}}{} reflecting sampling and intrinsic shower fluctuations, the noise term \ensuremath{b_{\text{calo}}}{} quantifying the detector noise, 
and the constant term \ensuremath{c_{\text{calo}}}{} capturing fluctuations introduced in the process of the detector signal extraction.  
The values for \ensuremath{a_{\text{calo}}}{} and \ensuremath{c_{\text{calo}}}{} are shown in Table~\ref{tab:SM_jetsub_2prong:detmodel}. 
The detector noise is not modeled, thus the noise term is $\ensuremath{b_{\text{calo}}}=0$ for both configurations.

The tower energy after smearing is required to pass $\ensuremath{p_{T}^{\text{tower}}} > \ensuremath{p_{T}^{\text{min}}}$. 
Towers passing this requirement are converted into massless pseudoparticles using the 
nominal tower center $(\ensuremath{\eta_{\text{tower}}},\ensuremath{\phi_{\text{tower}}})$ and the tower energy \ensuremath{E_{\text{tower}}}{} ($(\ensuremath{E_{\text{tower}}},\ensuremath{\eta_{\text{tower}}},\ensuremath{\phi_{\text{tower}}}) \mapsto (\ensuremath{E_{\text{tower}}},\vec{p}_{\text{tower}})$ with $|\vec{p}_{\text{tower}}| = \ensuremath{E_{\text{tower}}}$).

Configuration C models a particle flow signal from a tracking detector combined with a calorimeter. 
Charged particles are bent in an axial uniform magnetic field with $B_{\text{solenoid}} = 4$ T.
The transverse momentum of the charged particles emitted within a tracking detector acceptance of $|\eta|<2.5$
is smeared along a Gaussian distribution function with width $\sigma_{\ensuremath{p_{T}^{\text{track}}}}$ given by:
\begin{align}
  \frac{\sigma_{\ensuremath{p_{T}^{\text{track}}}}}{\ensuremath{p_{T}^{\text{track}}}} = \sqrt{(\ensuremath{a_{\text{track}}}\cdot\ensuremath{p_{T}^{\text{track}}})^{2} + \ensuremath{c_{\text{track}}}^{2}}\,.
  \label{eq:SM_jetsub_2prong:trkreso}
\end{align} 
If \ensuremath{p_{T}^{\text{track}}}{} after smearing is within $\ensuremath{p_{T}^{\text{min}}} < \ensuremath{p_{T}^{\text{track}}} < \ensuremath{p_{T}^{\text{max}}}$, the charged particle is added to the list of pseudoparticles with its direction at the interaction vertex. 
The trajectories of all charged particles within $|\eta| < 2.5$ and outside of this transverse momentum range, 
and all charged particles with $|\eta| > 2.5$, are extraploted to the front face of the calorimeter at $R_{\text{calo}} = 1290$ mm.
If the extrapolated particle trajectory reaches the calorimeter, the particle energy is added to a calorimeter tower at the extrapolated $\phi$, similar to the treatment of all charged particles in configuration A.

The energy of neutral particles is added to the calorimeter tower in the same way as in configuration A. 
The selection employing $\ensuremath{p_{T}^{\text{tower}}} > \ensuremath{p_{T}^{\text{min}}}$ is applied as well after the tower energy smearing with the parameters for configuration C given in Table~\ref{tab:SM_jetsub_2prong:detmodel}. 
Figure~\ref{fig:SM_jetsub_2prong:detmodel} shows the calorimeter-only composition of a given truth-level jet in configuration A together with the track-and-calorimeter jet composition of configuration C.
The two model configurations produce significantly different jet
compositions, in particular with respect to the energy flow from
low-\ensuremath{p_{T}}{} constituents.

The implementation of this detector model is available from the
{\tt{DetectorModel}} subdirectory of the {\tt{git}} repository at \url{https://github.com/gsoyez/lh2017-2prongs/}.

\subsubsection{Pileup Mitigation}\label{sec:SM_jetsub_2prong:pu_tech}

Besides the detector response discussed above, LHC collisions are also
contaminated by pileup.
While pileup multiplicities remained reasonably low, around 20, during Run~I of
the LHC, they already increased to 40-60 in Run~II and are expected to
increase even further, in the 140-200 range, for the high-luminosity
upgrade of the LHC.

To correct for the energy bias and smearing associated with the
pileup contamination, one uses a variety of pileup mitigation
techniques (see~Ref.~\cite{Soyez:2018opl} for a recent review). The
standard approach for most applications is the area--median
subtraction
method~\cite{Cacciari:2007fd,Cacciari:2008gn,AlcarazMaestre:2012vp,Soyez:2012hv}
(potentially using the ConstituentSubtractor~\cite{Berta:2014eza} for
subtracting pileup from jet shapes). Recently, more complex pileup
subtraction methods have been
proposed~\cite{Krohn:2013lba,Bertolini:2014bba,Cacciari:2014gra,Komiske:2017ubm} (see
also~\cite{Tseng:2013dva,Cacciari:2014jta}). In particular,
PUPPI~\cite{Bertolini:2014bba} and the
SoftKiller~\cite{Cacciari:2014gra} both provide event-wide pileup
mitigation techniques that show good performance in terms of
resolution, at the expense of requiring some degree of fine-tuning of
their free parameters.
PUPPI has already been used by the CMS collaboration in a series of
jet-substructure studies.

For these proceedings, we will concentrate on the SoftKiller approach.
This is motivated by its speed---for a typical LHC event with pileup
would be subtracted and clustered in 200-700~$\mu$s---by the fact
that a public implementation is available, and by the fact that the
SoftKiller and PUPPI have shown similar performance (see e.g.~\cite{puws14}).

The SoftKiller algorithm works by iteratively removing the softest particle in
the event until the area-median pileup density estimate $\rho$ is
zero. This is equivalent to breaking the event in patches and
iteratively removing the softest particle in the event until half of
the patches are empty, i.e. imposing a cut
\begin{align}
p_T^{\text{cut}}=\underset{p \in \text{patches}}{\text{median}}\Big\{
  \underset{\text{particle }i\in p}{\text{max}}\{ p_{Ti}\}\Big\}\,.
\end{align}
SoftKiller then removes particles below a cutoff $p_T^{\text{cut}}$, chosen such that $\rho=0$. We have
\begin{align}
p_T^{\text{cut}}=\text{median} \{ p_{Ti}^{\text{max}} \}\,.
\end{align}
SoftKiller has been shown to provide good performance for removing pileup contamination.

\subsubsection{Impact on Mass Resolution}\label{sec:SM_jetsub_2prong:detector_robust}

The impact of detector resolution and pileup is far more complex than
including the detector simulation discussed in
Sec.~\ref{sec:SM_jetsub_2prong:det_model}.
In particular, we have not performed any calibrations or in-situ
corrections.
We have therefore limited the scope of these proceedings to a
discussion of their effects on mass resolution.
This will highlight the need to perform a more sophisticated study
when considering detector effects, which will be left to a future
publication.

\begin{figure}
\subfloat[]{\label{fig:SM_jetsub_2prong:mass-detector:a}
  \includegraphics[width=0.45\columnwidth]{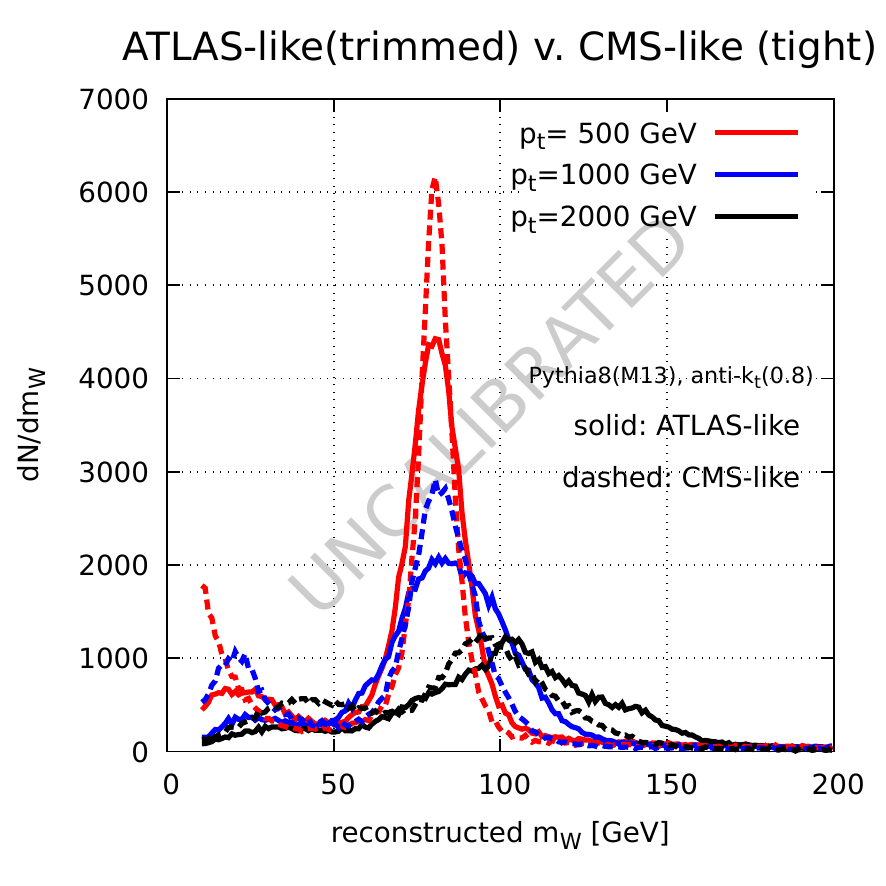}
}
\subfloat[]{\label{fig:SM_jetsub_2prong:mass-detector:b}
  \includegraphics[width=0.45\columnwidth]{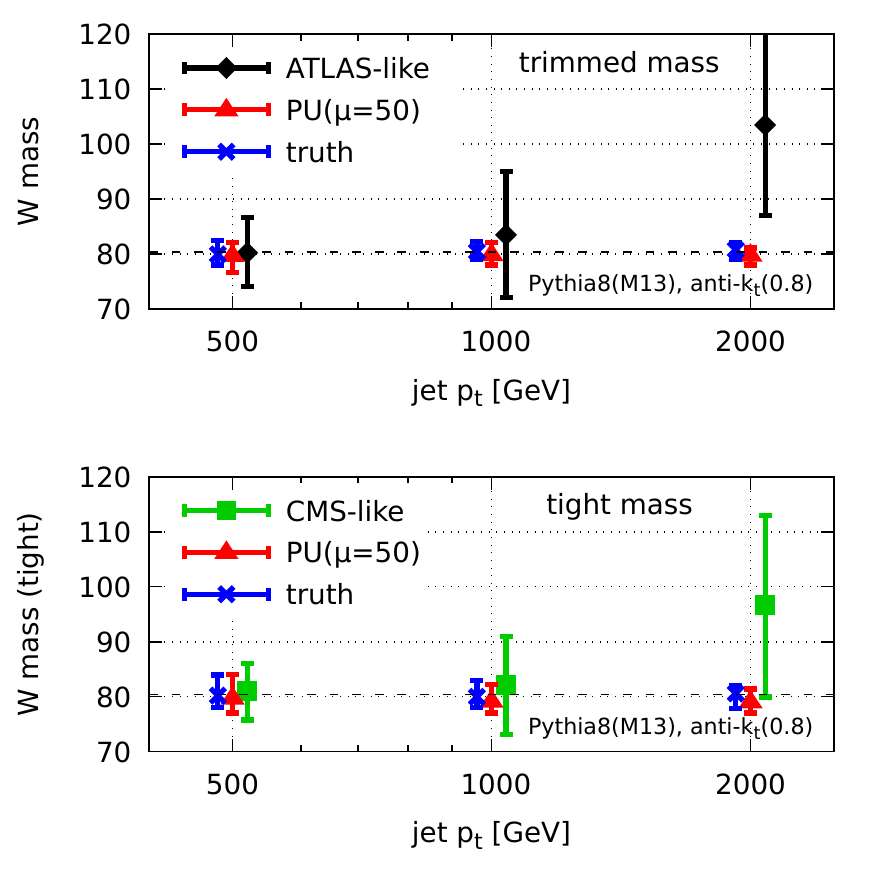}
}
  \caption{(a) Mass reconstruction with detector effects at different jet $p_T$.  (b) The peak position and width of the mass distribution, after pileup and, separately, after detector effects.}\label{fig:SM_jetsub_2prong:mass-detector}
\end{figure}

In Fig.~\ref{fig:SM_jetsub_2prong:mass-detector}, we show distributions
of the groomed mass for a hadronically-decaying boosted $W$ boson in
our CMS-like and ATLAS-like detectors at different values of the jet
$p_T$ (Fig.~\ref{fig:SM_jetsub_2prong:mass-detector:a}), together with
the measure peak position and width, for the trimmed mass with the
ATLAS-like detector simulation and the tight (mMDT) mass with the
CMS-like simulation (Fig.~\ref{fig:SM_jetsub_2prong:mass-detector:b}).%
\footnote{The position and width
  are defined as the median and width of the smallest mass window
  containing 40\% of the events.}
We see that the detector has a significant impact on the
distributions: while the $W$ peak remains decent for $500$-GeV jets,
typical of most applications today, the raw resolution considerably
degrades as the $p_T$ is increased.
We also see that both detectors show a similar trend, despite their
different approaches.
For $p_T$'s up to 1~TeV, it seems that the configuration C gives a
slightly better resolution thans the configuration A, an effect that
is tentatively attributed to particle flow.
However, at high $p_T$ there is a trade off between tracking and
calorimetry, which is better in configuration A, leading to comparable detector
resolution.
Furthermore, in the full experiments, the reconstruction algorithms
are complementary and adapted to the experiments' detector
technologies, so that the two experiments achieve similar performance.
Due to the fact that the end results of the two detectors are quite
similar, we find it unwise to draw any deeper conclusions, since there
are potentially other effects, not included in our simulation, that
could determine the final outcome.
We will however point out that, in our (over-simplified) preliminary
studies, the main patterns observed in
Sec.~\ref{sec:SM_jetsub_2prong:exp_compare}, i.e.\ the overall good
performance of $N_2^{(1,2)}$ and $D_2^{(2)}$ and the interest in
investigating dichroic ratios, seem to remain valid.

Figure~\ref{fig:SM_jetsub_2prong:mass-detector:b} also includes the mass
resolution in the presence of pileup, here Poisson-distributed with an
average multiplicity of 50, mitigated using the SoftKiller method
described in Sec.~\ref{sec:SM_jetsub_2prong:pu_tech}.
Based on this simple analysis, it
seems that the reconstruction of $W$ mass peak still behaves properly
in the presence of pileup.
It is important to note, however, that this pileup result is in the absence of detector effects, and there is a non-trivial interplay between detector corrections and pileup corrections.

This brief study highlights the essential importance
of understanding detector effects when designing jet substructure
observables.
Although we have found that the $N_2$ and $D_2$ observables emphasize,
respectively, performance and robustness, it would also be interesting
to understand if the different detectors played a role in the choice
of one observable as compared with the other.
As highlighted by the specific example of the groomed jet mass here, such a study requires considerable care to perform in a meaningful manner.
The detectors have a large impact, and the two detectors appear qualitatively similar in our analysis setup.
Differences may therefore be due to more subtle features, beyond those included in our study, and this is probably something that is best left to the experimental collaborations to study with realistic detector simulations.
We hope that this study motivates the experimental collaborations to further investigate the performance and robustness of different two-prong tagging observables at each of the different steps along the chain of realism in order to understand the different choices in observables.

\subsection{Polarization Dependence}\label{sec:SM_jetsub_2prong:polar}

For signal jets, there is another consideration related to the robustness of two-prong tagging, namely the specific nature of the decaying electroweak-scale resonance, which can also affect the substructure observables.
Assuming that this electroweak-scale resonance is a color-singlet decaying to quarks, it is completely
characterized by its spin structure.

All of our taggers consist of two conceptually distinct components, which will be affected in different ways by the polarization.
We begin by making a cut on the (groomed) mass, followed by a cut on a particular jet shape which is sensitive to the two-prong structure. These two steps are associated with very different physics. A jet from a $W\to q\bar q$ decay consists of two collimated sprays of radiation, which are proxies for the $q$ and $\bar q$, along with radiation emitted from the dipole.
Ideally, a groomer should terminate when it de-clusters the jet into two subjets corresponding to the jets initiated by the two quarks.
If there is a large fraction of decays where the momentum sharing between the two subjets is hierarchical, though, then the groomer can remove one of the subjets.
In this case, the jet will fail the groomed mass criteria.
Since the polarization controls the momentum sharing of the subjets, this introduces a sensitivity on the polarization into the tagging procedure.
On the other hand, the $2$-prong tagging observables that we consider are all formed as ratios of an observable which is sensitive to radiation from the prongs, divided by a mass-type observable, which (largely) removes the dependence on the momentum sharing of the subjets.

In this subsection, we consider two main issues, namely
the dependence of the standard jet substructure observables to the
polarization of the signals, and the ability to perform polarimetry
using jet substructure observables measured on the hadronic decay
products.

\subsubsection{Performance Impact of Polarization}\label{sec:SM_jetsub_2prong:polar_robust}

We begin by studying the sensitivity of two-prong taggers to polarization.
To do this, we consider samples of hadronically decaying $W$s, which are either purely transversely polarized, purely longitudinally polarized, or have the Standard Model fraction (mostly transverse).
The details of the sample generation were discussed in Sec.~\ref{sec:SM_jetsub_2prong:samples_sub}.

\begin{figure}
\begin{center}
\subfloat[]{\label{fig:SM_jetsub_2prong:polarization_dist}
\includegraphics[width=0.4\columnwidth]{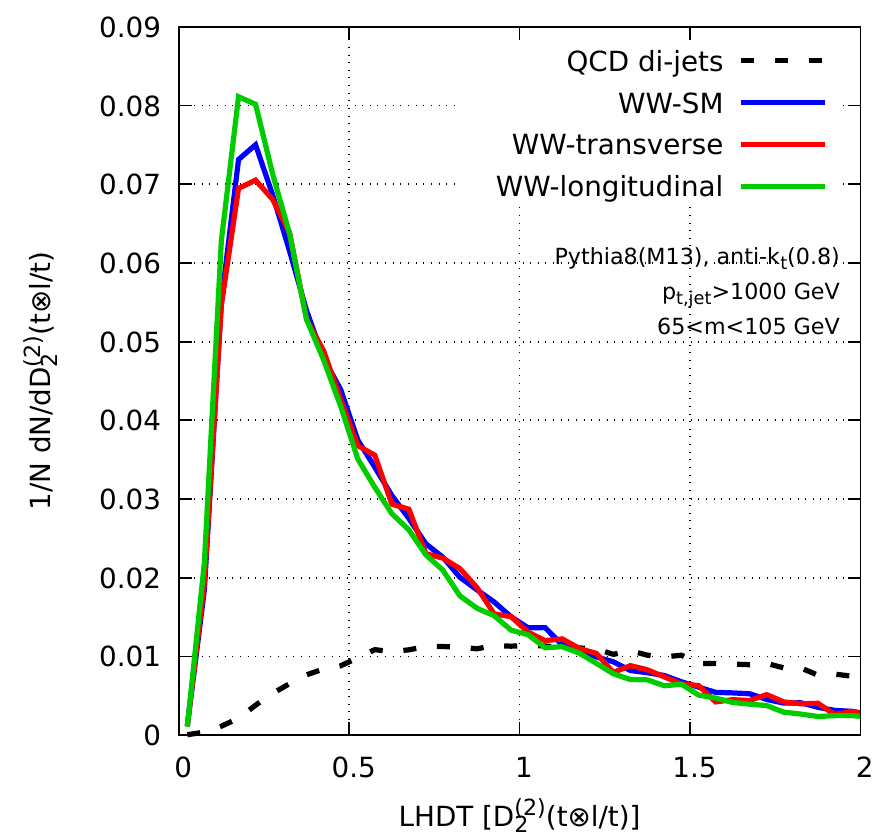}
}
\subfloat[]{\label{fig:SM_jetsub_2prong:polarization_roc}
\includegraphics[width=0.4\columnwidth]{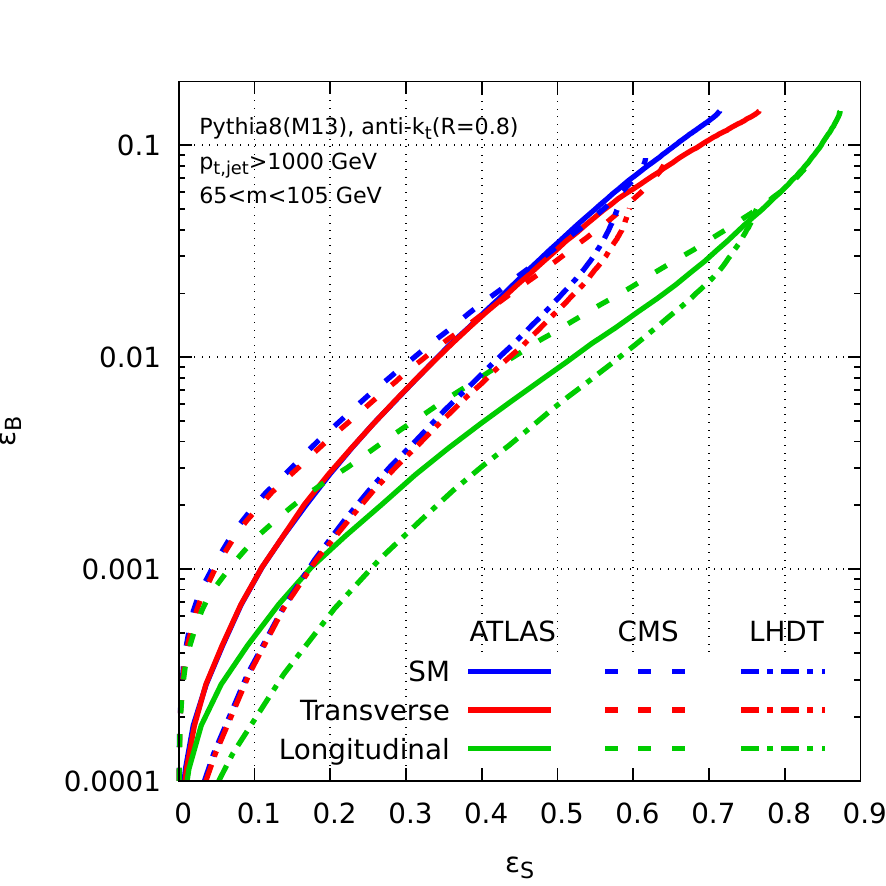}
}
\end{center}
\caption{a) Distributions of the $D_2$ observable as measured on the samples with different polarization compositions. These observables are found to be largely insensitive to the polarization. b) ROC curves comparing the different grooming strategies for different polarization samples. While the polarization has a limited effect on the jet shape observables, it modifies the groomed mass acceptance, and therefore the tagging efficiency, significantly.}
\end{figure}

In Fig.~\ref{fig:SM_jetsub_2prong:polarization_dist}, we show the $D_2$ observable as measured on longitudinal and transverse hadronically-decaying W bosons, as well as the Standard Model mixture, as a representative example of the dependence of a two-prong tagger on polarization.
From this figure, we see that the jet shape observable itself is remarkably insensitive to the polarization, due to its ratio nature.
However, it is important to emphasize that this does not imply that the tagging performance is also independent of the polarization.
As can be seen in Fig.~\ref{fig:SM_jetsub_2prong:polarization_roc}, the tagging performance is significantly worse for transversely-polarized $W$ bosons as compared with longitudinally-polarized $W$ bosons.
This is due to the fact that transversely-polarized $W$ bosons have a more asymmetric energy sharing, and it is therefore more likely that one of the subjet prongs is groomed away, leading to the jet failing the mass cut.
This difference between longitudinally and transversely-polarized bosons should be taken into account in studies at the LHC.
It would also be interesting to develop tagging or reconstruction schemes that are less sensitive to polarization.

\subsubsection{Tagging Longitudinal vs.\ Transverse Bosons}\label{sec:SM_jetsub_2prong:polar_tag}

Having understood the dependence of standard jet substructure observables on polarization, it is interesting to know whether jet substructure observables are able to tag distinct polarizations on hadronically-decaying particles, and with what efficiency this can be done.
Here, we do not perform a comprehensive study, but restrict ourselves to studying a particular example of an observable which is sensitive to the $W$ polarization, and we evaluate its performance for distinguishing transverse and longitudinal $W$ bosons. 

The sole impact of the polarization of the decaying object is to determine the kinematics of the decaying subjets.
Indeed, this can be made rigorous in the sense of a factorization theorem for boosted jets in the two-prong limit.
We are therefore interested in an observable that is sensitive to the kinematics of the two subjets.
While a variety of different observables could be considered, here we consider the $z_g$ observable \cite{Larkoski:2014wba,Larkoski:2014bia,Larkoski:2015lea}, which measures the momentum sharing of the subjets. The precise definition of $z_g$ was given in Sec.~\ref{sec:SM_jetsub_2prong:groom_tech}, which we recall for convenience
\begin{align}
z_g=\frac{\min\left[ p_{Ti}, p_{Tj}  \right]}{p_{Ti}+p_{Tj}}\,.
\end{align}
Here $p_{Ti}$ and $p_{Tj}$ are the momenta of the first set of subjets that pass the SoftDrop criteria.

Since we are focused on robustness in this paper, it is also worth commenting on the robustness of the $z_g$ observable. For signal jets, since this observable measures global energy properties of the subjets, it is stable.
Interestingly, it is also remarkably stable on the background, where it flows in the high-$p_T$ limit to the QCD splitting function \cite{Larkoski:2015lea}.

\begin{figure}
\begin{center}
\subfloat[]{\label{fig:SM_jetsub_2prong:z_g_dist}
\includegraphics[width=0.45\columnwidth]{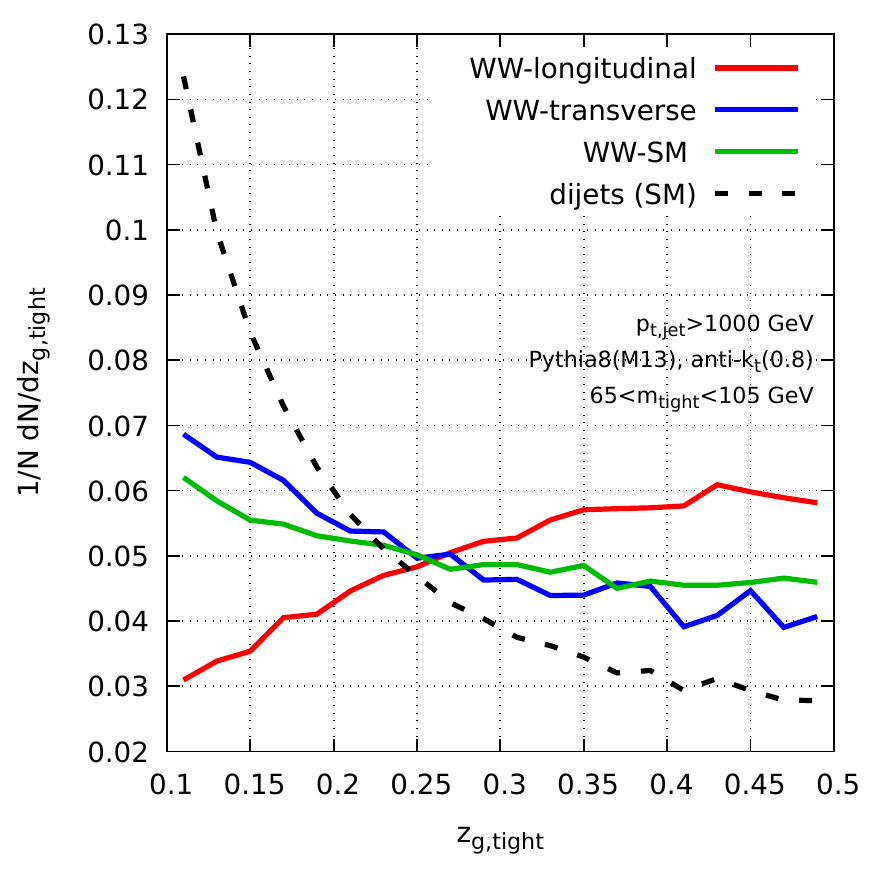}
}
\subfloat[]{\label{fig:SM_jetsub_2prong:z_g_roc}
\includegraphics[width=0.45\columnwidth]{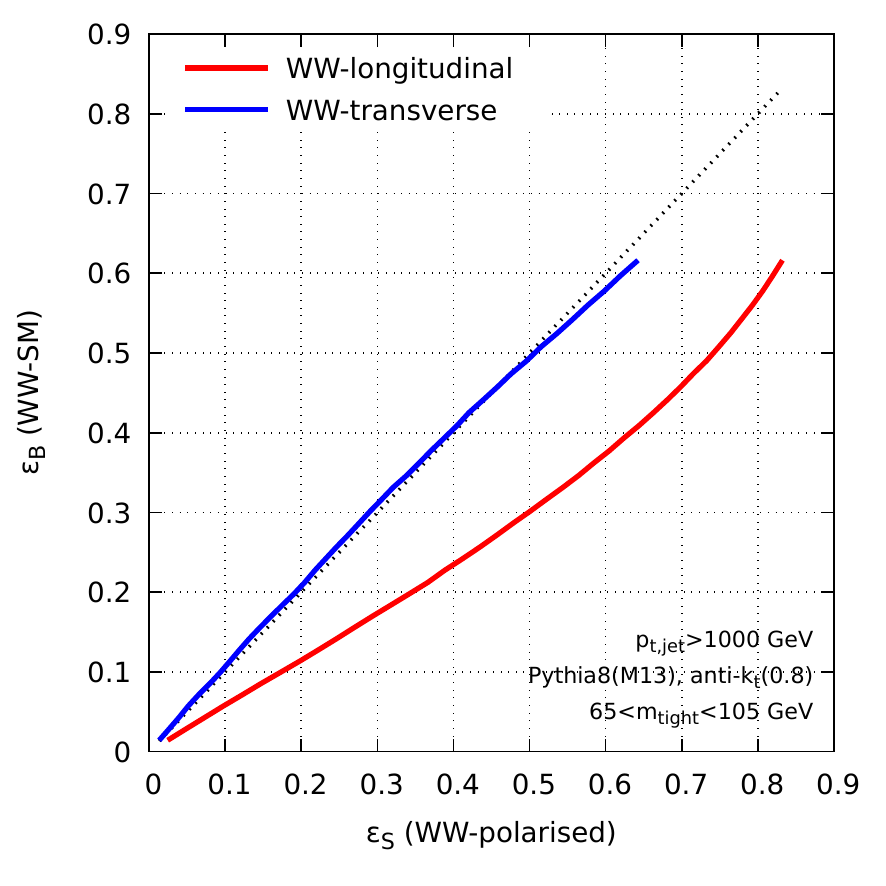}
}
\end{center}
\caption{a) The $z_g$ distribution as measured on boosted $W$ samples with different polarization compositions. Transversely-polarized $W$s behave similarly to the QCD background, while decent separation is observed between longitudinally- and transversely-polarized $W$s. b) A ROC curve showing the separation between transversely- and longitudinally-polarized $W$ bosons using the $z_g$ observable. Moderate separation is observed.}
\end{figure}

In Fig.~\ref{fig:SM_jetsub_2prong:z_g_dist}, we show the $z_g$ distribution for different polarized $W$ bosons.
For reference, the $z_g$ distribution for QCD dijets is also shown.
The $z_g$ distribution for transversely-polarized $W$ bosons follows most closely the QCD distribution, as expected, being peaked at small values of $z_g$.
On the other hand, for longitudinally-polarized $W$ bosons, the $z_g$ distribution is peaked at high values of $z_g$.
This illustrates that the $z_g$ observable is indeed behaving as expected, and is achieving sensitivity to the polarization of the decaying $W$ boson only from its hadronic decay products.
In Fig.~\ref{fig:SM_jetsub_2prong:z_g_roc}, we show  the ROC curve for separating longitudinal from transverse $W$ bosons.
Here we see that $z_g$ provides moderate separation power for tagging the polarization of the decaying $W$ bosons.
It would be interesting to investigate this further to see if more ideal observables could be found, however, we are not optimistic that significant improvement can be achieved, since the primary imprint of the polarization should be in the kinematics of the subjets.

\subsection{Summary and Recommendations}\label{sec:SM_jetsub_2prong:conc}

In this paper, we performed a comprehensive study of performance and robustness for two-prong tagging, and provided a unifying approach to understanding different classes of two-prong taggers based on dichroic observables, which allow for different amounts of grooming in the numerator and denominator of observables.
We introduced measures of robustness, in addition to the standard measures of performance, and we used these measures to study the robustness of two-prong taggers to theory issues, namely hadronization and underlying event.
We believe that these will be of more general utility in jet substructure, and could be applied also to study three-prong tagging, for example.

As a part of our study, we have also introduced a number of new dichroic observables, which generalize the dichroic $N$-subjettiness observables to observables formed from the energy correlation functions.
This offers a general and unifying approach to designing new jet substructure observables which are simultaneously performant and resilient.
We have shown that the $N_2$-style observables used by CMS tend to be more performant, but less resilient than the $D_2$-style observables used by ATLAS.
For a given observable, we have found that moving to a dichroic variant can typically improve performance without significantly decreasing its robustness to hadronization.

We have also studied the effect of polarization on two-prong taggers.
We found that while polarization has minimal effect on standard two-prong tagging observables, since they are typically defined as ratios, it has a large effect on their tagging efficiency due to applied mass cuts.
Significantly better tagging performance is observed for longitudinally-polarized bosons.
An interesting avenue beyond standard two-prong tagging is using jet substructure observables to tag the polarization of decaying Standard Model bosons using their hadronic decay products.
We proposed the observable $z_g$, which measures the momentum sharing asymmetry between subjets, as an effective polarization tagger.
We illustrated that separation between longitudinal and transverse bosons can be achieved.
It would be interesting to study this problem in more detail.

We have identified a  number of new two-prong taggers that outperform, in both robustness and tagging performance, those currently used by the ATLAS and CMS collaborations.
These findings are summarized in Fig.~\ref{fig:SM_jetsub_2prong:phasespace}, which also lists a number of promising new observables.
We therefore believe that further studies using more detailed simulations of the ATLAS and CMS detectors, and ultimately on real data, would be of significant interest.
More generally, we expect that the emphasis on a simultaneous evaluation of the performance and robustness, as well as the particular metrics introduced in this paper, will play a significant role in future studies of jet substructure techniques at the LHC.

\subsection*{Acknowledgments}

We thank the participants of Les Houches 2017 for a lively environment and useful discussions.
The work of GS is supported in part by the Paris-Saclay IDEX under the
IDEOPTIMALJE grant, by the French Agence Nationale de la Recherche,
under grant ANR-15-CE31-0016, and by the ERC Advanced Grant Higgs@LHC
(No.\ 321133).
The work of DK is supported by NRF CSUR and Incentive funding
The work of FD and JT is supported by the DOE under grant contract numbers DE-SC-00012567 and DE-SC-00015476.
The work of IM is supported by Office of High Energy Physics of the U.S. Department of Energy under Contract No. DE-AC02-05CH11231, and the LDRD Program of LBNL.
AS acknowledges support from the National Science Centre, Poland Grant No. 2016/23/D/ST2/02605, the MCnetITN3 H2020 Marie Curie Initial Training Network, contract 722104 and COST Action CA15213 THOR.




\chapter{Standard Model Higgs}
\label{cha:higgs}


\section{Transverse momentum resummation in Higgs boson plus two jet production via weak boson fusion process~\protect\footnote{
      P. Sun,
      C.-P. Yuan,
      F. Yuan}{}}
\label{sec:Higgs_2jet_VBF}

We study the soft gluon resummation effect in the Higgs boson plus two-jet production
at the LHC. By applying the transverse momentum
dependent factorization formalism, the large logarithms introduced by the small total transverse momentum
of the Higgs boson plus two-jet final state system, are resummed
to all orders in the expansion of the strong interaction coupling at the accuracy of Next-to-Leading Logarithm.
We also compare our result with the prediction of the Monte Carlo event generator Pythia8, and found
noticeable difference in the distributions of the total transverse momentum and the azimuthal angle
correlations of the final state Higgs boson and two-jet system.

\subsection{Introduction}
In this letter, we apply the transverse momentum
dependent (TMD) resummation method to study
the soft gluon resummation effect on the production of
Higgs boson plus two jets via weak boson fusion (VBF) at the LHC.
By applying the TMD factorization formalism, the large logarithms introduced by the small total transverse momentum ($q_\perp$)
of the Higgs boson plus two-jet final state system, are resummed
to all orders in the expansion of the strong interaction coupling at the accuracy of Next-to-Leading Logarithm.
In terms of our TMD factorization formalism, the $q_\perp$ differential cross section
is factorized into several individual factors, and each factor is calculated up to the one-loop order.
As of today, the Monte Carlo (MC) event generators
are the only available tools to predict the soft gluon shower effect for this channel.
Our calculation, for the first time, provides an important test on the validity of the commonly used
MC event generators.

\subsection{The Model}

Our TMD resummation formula can be written as:

\begin{eqnarray}
\frac{d^6\sigma}
{dy_H dy_{J1}dy_{J2} d P_{J1\perp}^2 d P_{J2\perp}^2
d^2\vec{q}_{\perp}}=\sum_{ab}\left[\int\frac{d^2\vec{b}}{(2\pi)^2}
e^{-i\vec{q}_\perp\cdot
\vec{b}}W_{ab\to Hcd}(x_1,x_2,b)+Y_{ab\to Hcd}\right] \ ,\label{eq:Higgs_2jet_VBF:label1}
\end{eqnarray}
where $y_H$, $y_{J1}$ and $y_{J2}$ denote the rapidities of the Higgs boson and the jets, respectively,
$P_{J1\perp}$ and $P_{J2\perp}$ are the jets transverse momentum,
 and $\vec{q}_\perp=\vec{P}_{H\perp}+\vec{P}_{J1\perp}+\vec{P}_{J2\perp}$ is the imbalance
transverse momentum of the Higgs boson and the two final state jets.
The first term ($W$) contains all order resummation effect
and the second term ($Y$) accounts for the difference between
the fixed order result and the so-called asymptotic
result which is given by expanding the resummation result
to the same order in $\alpha_s$ as the fixed order term.
$x_1$ and $x_2$ are the momentum fractions of
the incoming hadrons carried by the incoming partons.

\subsection{Numerical Results}
Following the study in Ref~\cite{Sun:2016kkh},
The uncertainty of our resummation calculation is estimated by varying the resummation scale $\hat \mu$ from $P^{lead}_{J\perp}$ to $P^{sub}_{J\perp}$, where the $P^{lead}_{J\perp}$ and $P^{sub}_{J\perp}$
are the transverse momenta of the final state leading jet and sub-leading jet, respectively.
In addition, we set the renormalization scale in the hard factor to be $\tilde{\mu}=m_{H}$,
and use the CT14 NNLO PDFs~\cite{Dulat:2015mca}, with the mass of the Higgs boson ($m_H$)
set to be 125 GeV.
Following the experimental analysis in Ref.~\cite{ATLAS:2016nke},
we require the rapidity of the observed jets to satisfy $|y_J|<4.4$.
We use the anti-$k_t$ algorithm to define the observed jets, and the jet size and the minimal transverse
momentum are set at $R=0.\textrm{4}$ and $P_{J\perp}>30$ GeV.
In our calculation we have applied the narrow jet approximation.
We have also constrained the two final state
jets to have a large rapidity separation with $|\Delta y_{JJ}|>2.6$, which is
the signal region for detecting the Higgs boson production via weak boson fusion process.
In the same figure we also compare to the predictions from Pythia8 which are based on the tree level scattering amplitudes with parton showers.
\begin{figure}
\begin{center}
\includegraphics[width=0.48\textwidth]{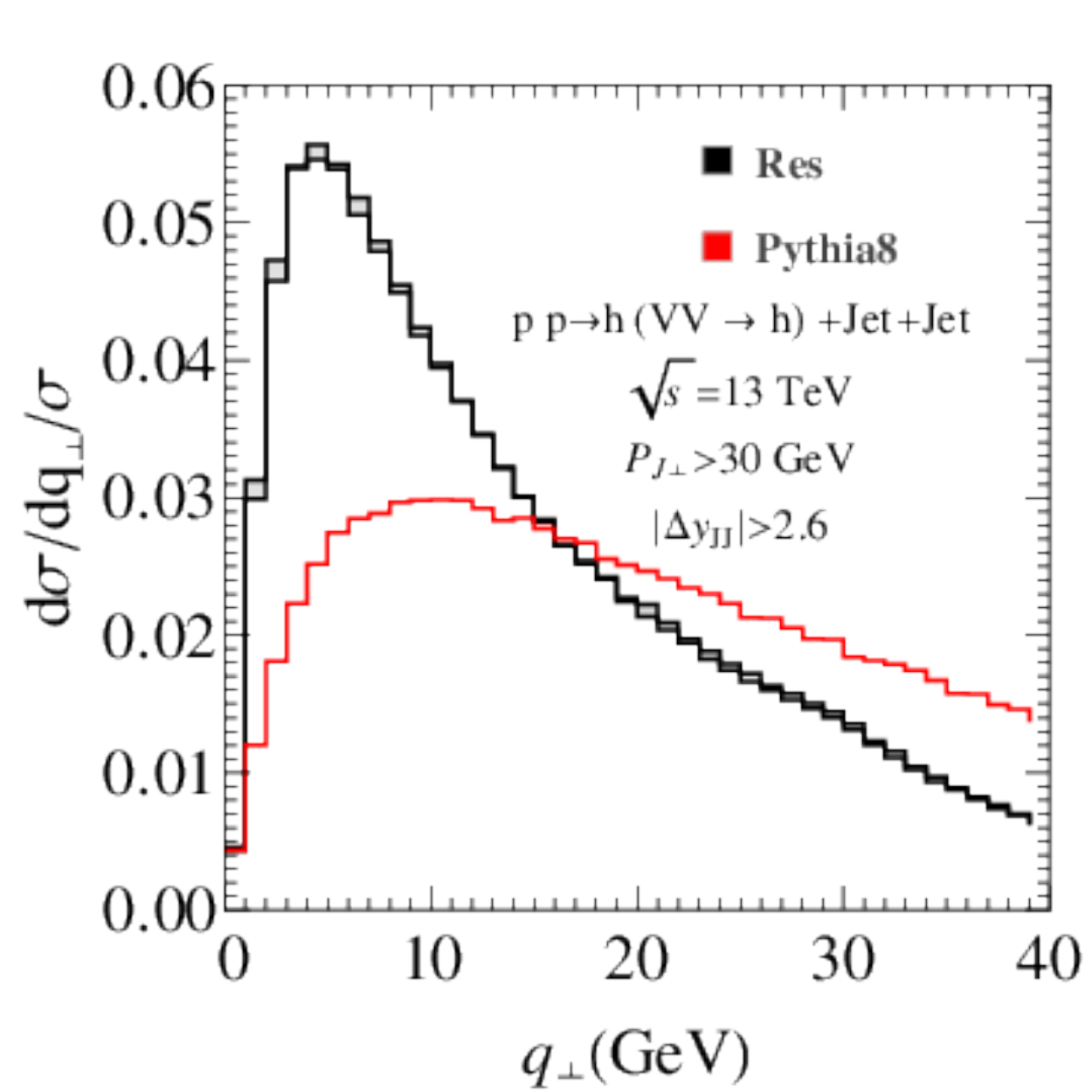}\hfill
\includegraphics[width=0.48\textwidth]{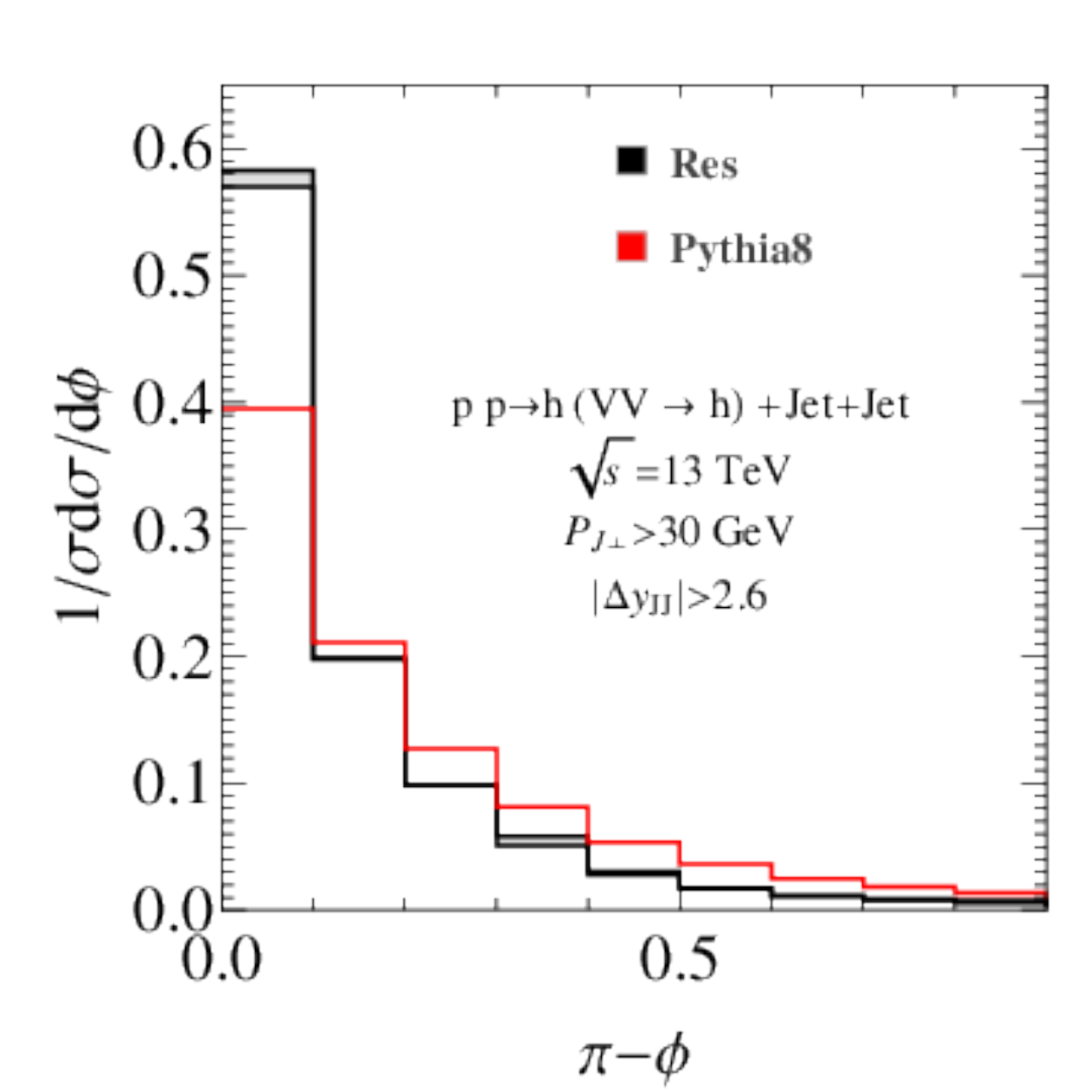}
\vspace*{3ex}

 \caption{The differential cross sections of Higgs boson plus two jet production
at the LHC as functions of $q_\perp$ and azimuthal angle $\phi$ between
Higgs boson and the final state two-jet system.
In these plots, the $\alpha_s$ order $Y$ pieces are included in the resummation curves.
The predictions from Pythia8 are based on the tree level scattering amplitudes with parton showers.
The uncertainty of our resummation calculation is estimated by varying the resummation scale $\hat \mu$ from $P^{lead}_{J\perp}$ to $P^{sub}_{J\perp}$.
}
\label{fig:Higgs_2jet_VBF:label2}
\end{center}
\end{figure}

As shown in Fig.~\ref{fig:Higgs_2jet_VBF:label2}, we find large difference between the Pythia8 and our predictions in the
distributions of the total transverse momentum ($q_\perp$) and the azimuthal angle
($\phi$) correlations of the final state Higgs boson and two-jet system,
 after imposing the kinematic cuts
used in the LHC data analysis.
Specifically, Pythia8 predicts a flatter shape, in $q_\perp$ distribution, than our resummation calculation. Another significant disagreement lies in the peak position of the $q_\perp$ distribution.
Pythia8 predicts a peak in $q_\perp$ around $10$ GeV, while ours is at about $5$ GeV.
Similarly, they also differ in predicting the $\phi$ distribution.
Pythia8 predicts a less back-to-back configuration, between Higgs boson and the final state two-jet system, than ours.
In Ref.~\cite{ATLAS:2016nke}, the ATLAS Collaboration required the azimuthal angle
separation ($\phi$) between the Higgs boson and the di-jet system to be
$\phi>2.6$, and compared the measured fiducial cross section with the Pythia8~prediction.
For $\phi>2.6$, Pythia8 and ours differ by about 8\%, and
our resummation calculation results in a larger
total fiducial cross section.
This implies a larger value in the coupling of Higgs boson to weak gauge bosons by about 4\%.
At the High-Luminosity LHC, with an integrated luminosity of up to
3000 fb$^{-1}$, the expected precision on the measurement of the production cross section of
the SM-like Higgs boson via VBF mechanism is around 10\%~\cite{CMS:2017cwx}.
Hence, the difference found in our resummation and Pythia8 calculations of the fiducial cross
sections could become important. Further comparisons on various event shapes between
the experimental data and our resummation predictions could also be carried out in order
to test the Standard Model and to search for New Physics.







\section{Higher order corrections in VBF Higgs production~\protect\footnote{
    F. Dreyer,
    A. Karlberg,
    M. Rauch}{}}
\label{sec:Higgs_vbf_nnlo}

Recently, a significant error was
discovered~\cite{Cruz-Martinez:2018rod} in the virtual matrix elements
of the VBF Higgs plus three jet NLO cross section calculated
in Ref.~\cite{Figy:2007kv} and implemented in the program packages
VBFNLO~\cite{Arnold:2008rz,Arnold:2011wj,Baglio:2014uba} and
Powheg-Box~\cite{Nason:2004rx,Frixione:2007vw,Alioli:2010xd,Jager:2014vna}.
Since this result was used as part of the calculation of the fully
differential NNLO QCD corrections~\cite{Cacciari:2015jma}, this bug
impacts the size of the second order corrections.
We report here on updated cross sections after implementation of the
correct virtual corrections.

Because the issue is in the $H+3j$ calculation, there is no impact on
the inclusive cross section.
However, after VBF cuts, the cross section is slightly increased.
We impose the following conditions on the tagging jets
\begin{eqnarray}
  \label{eq:Higgs_vbf_nnlo:cuts}
  m_{j_1,j_2}>600\text{GeV}\,,\quad \Delta y_{j_1,j_2} > 4.5\,,
\end{eqnarray}
where the jets $j_1,j_2$ are required to have transverse momentum
$p_{t,j}>25$ GeV and rapidity $|y_j|<4.5$, as well as being in
opposite hemispheres ($y_{j_1} y_{j_2}<0$).
The NNLO cross section after these cuts is
$\sigma = 0.844^{+0.008}_{-0.008}$ pb, which is a $\sim 2\%$ change to
the result originally reported in Ref.~\cite{Cacciari:2015jma}.

\begin{figure}[p]
  \centering
  \includegraphics[clip,height=0.45\textwidth,page=1,angle=0]{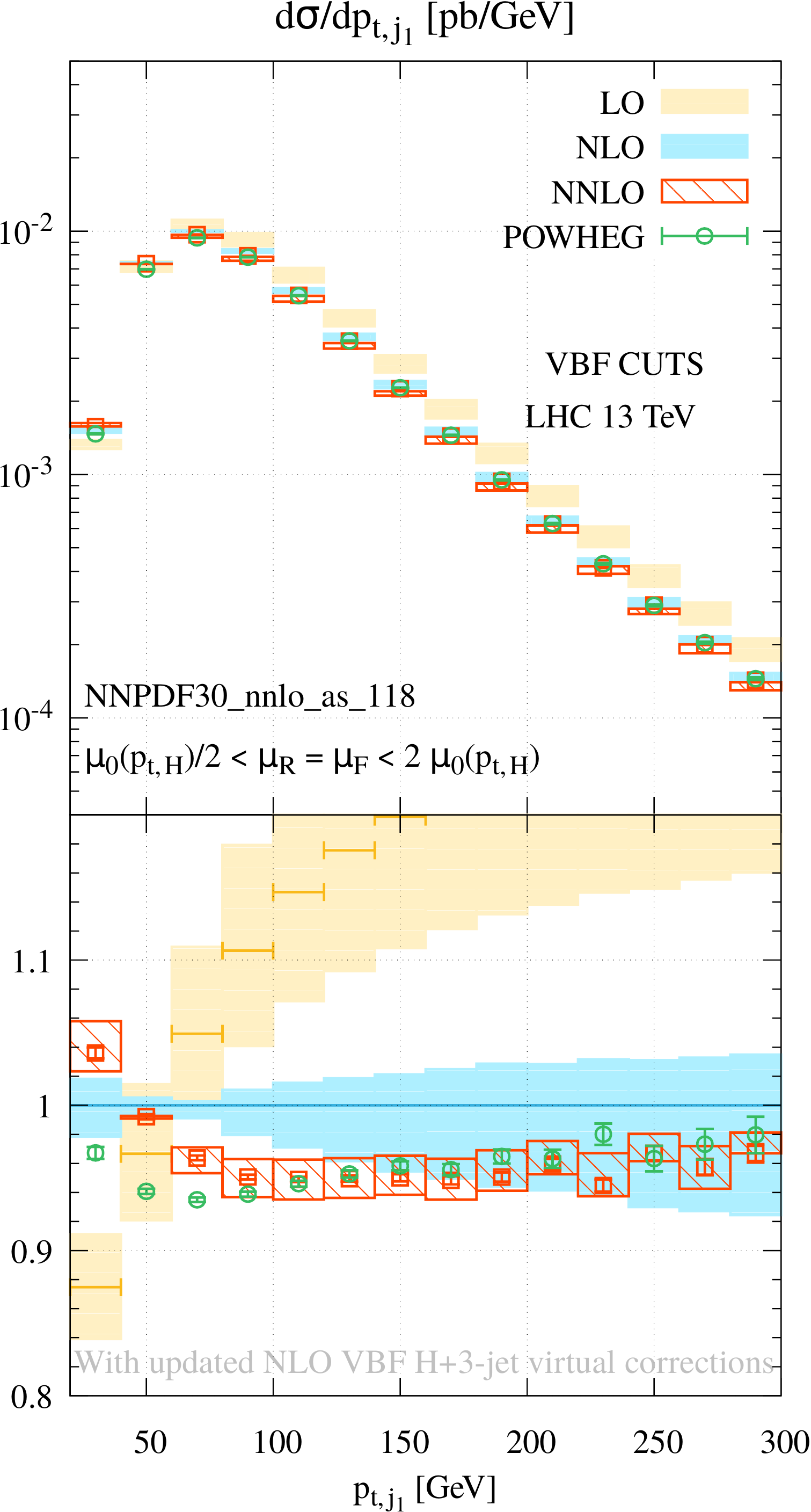}%
    \hfill\includegraphics[clip,height=0.45\textwidth,page=2,angle=0]{plots/bugfix-crop.pdf}%
    \hfill\includegraphics[clip,height=0.45\textwidth,page=3,angle=0]{plots/bugfix-crop.pdf}\hspace{0.8mm}%
    \hfill\includegraphics[clip,height=0.45\textwidth,page=4,angle=0]{plots/bugfix-crop.pdf}%
\vspace*{2ex}
  \caption{Differential cross sections for the transverse momentum of the tagging jets and the higgs boson, as well as for the rapidity separation between the jets.}
  \label{fig:Higgs_vbf_nnlo:nnlo-diff}
\vspace*{15ex}

  \centering\includegraphics[width=0.8\textwidth]{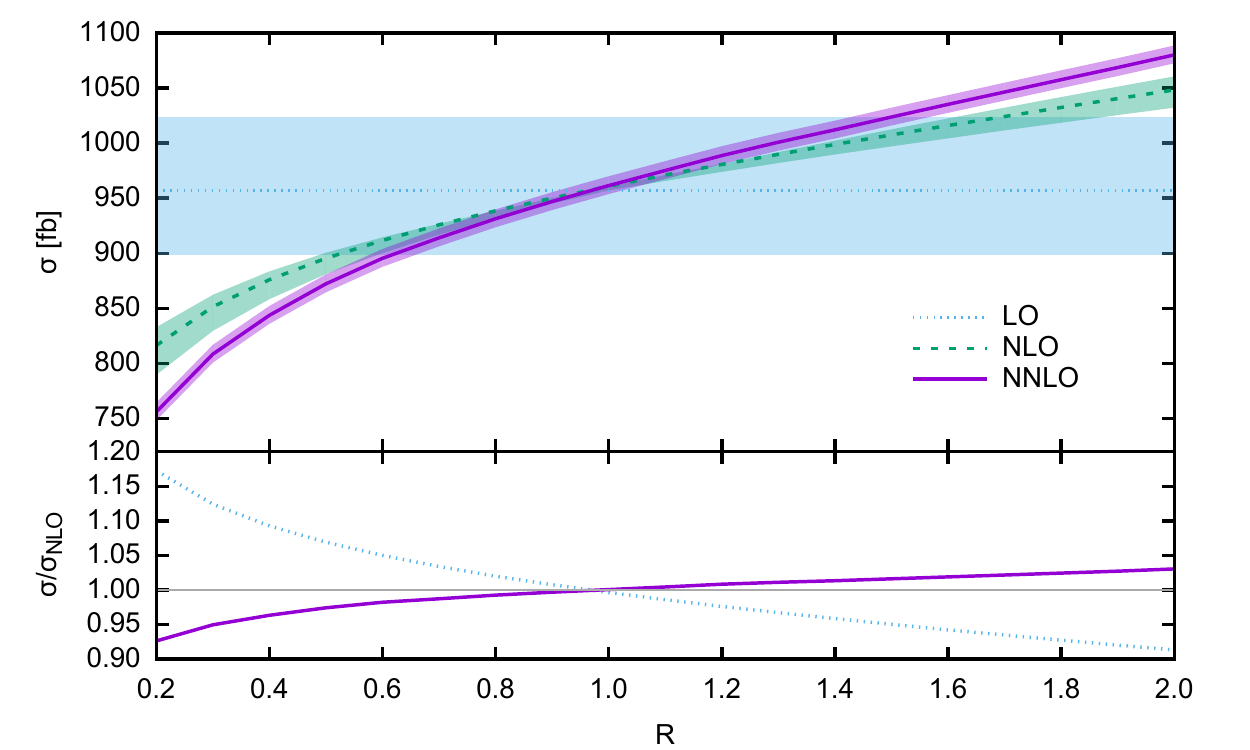}
\vspace*{2ex}
  \caption{Integrated cross section for VBF-$H$ production as function of
  the jet clustering radius $R$.}
  \label{fig:Higgs_vbf_nnlo:intRdep}
\end{figure}

In Fig.~\ref{fig:Higgs_vbf_nnlo:nnlo-diff}, we present the differential cross
sections of the transverse momentum of the two leading jets and the
Higgs boson, and the rapidity separation between the tagging jets.
The correction of the bug in Ref.~\cite{Figy:2007kv} leads to slightly
harder jets, with an increase of $~\sim 2\%$ of the NNLO cross section
at high transverse momentum.
The difference is most visible on the rapidity separation, where the
shape of the $K$-factor is changed for very widely separated jets.

\begin{figure}[p]
  \centering
  \includegraphics[width=\textwidth]{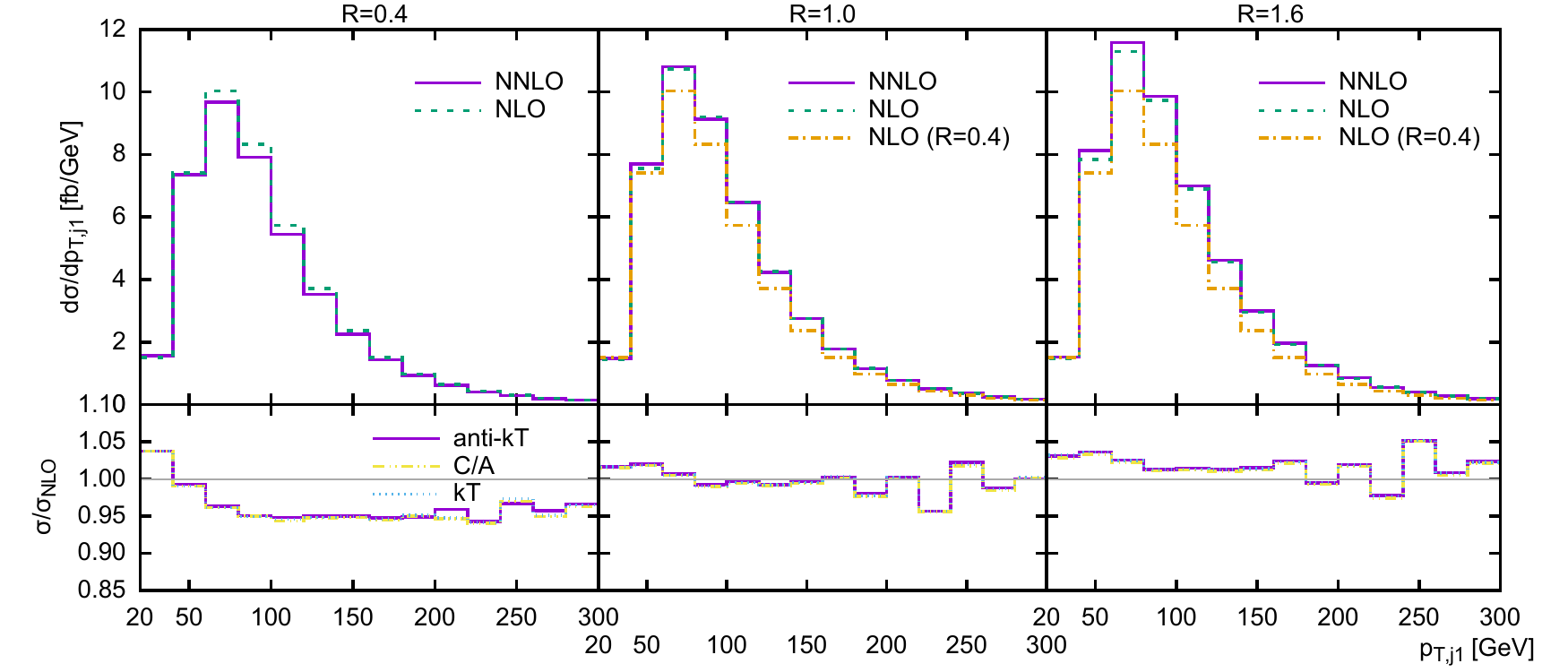} \\[1ex]
  \mbox{}\hfill%
  \includegraphics[width=0.42857\textwidth]{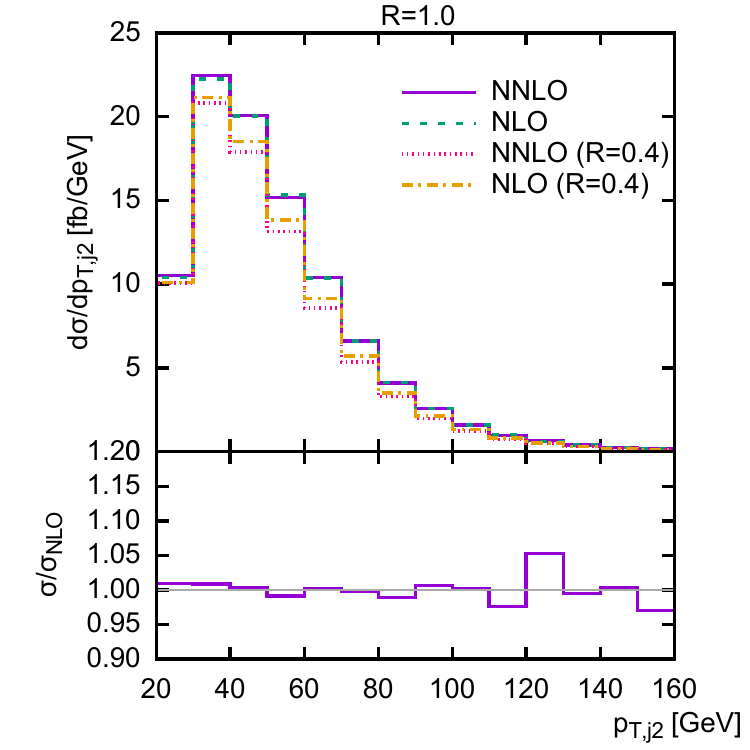}%
  \hfill
  \includegraphics[width=0.42857\textwidth]{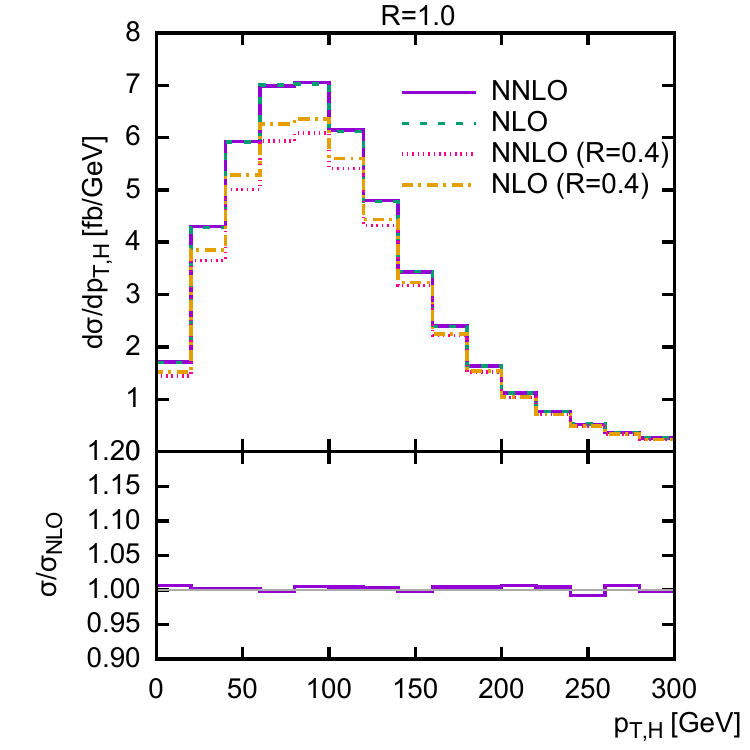}%
  \hfill\mbox{}\\[1ex]
  \includegraphics[width=\textwidth]{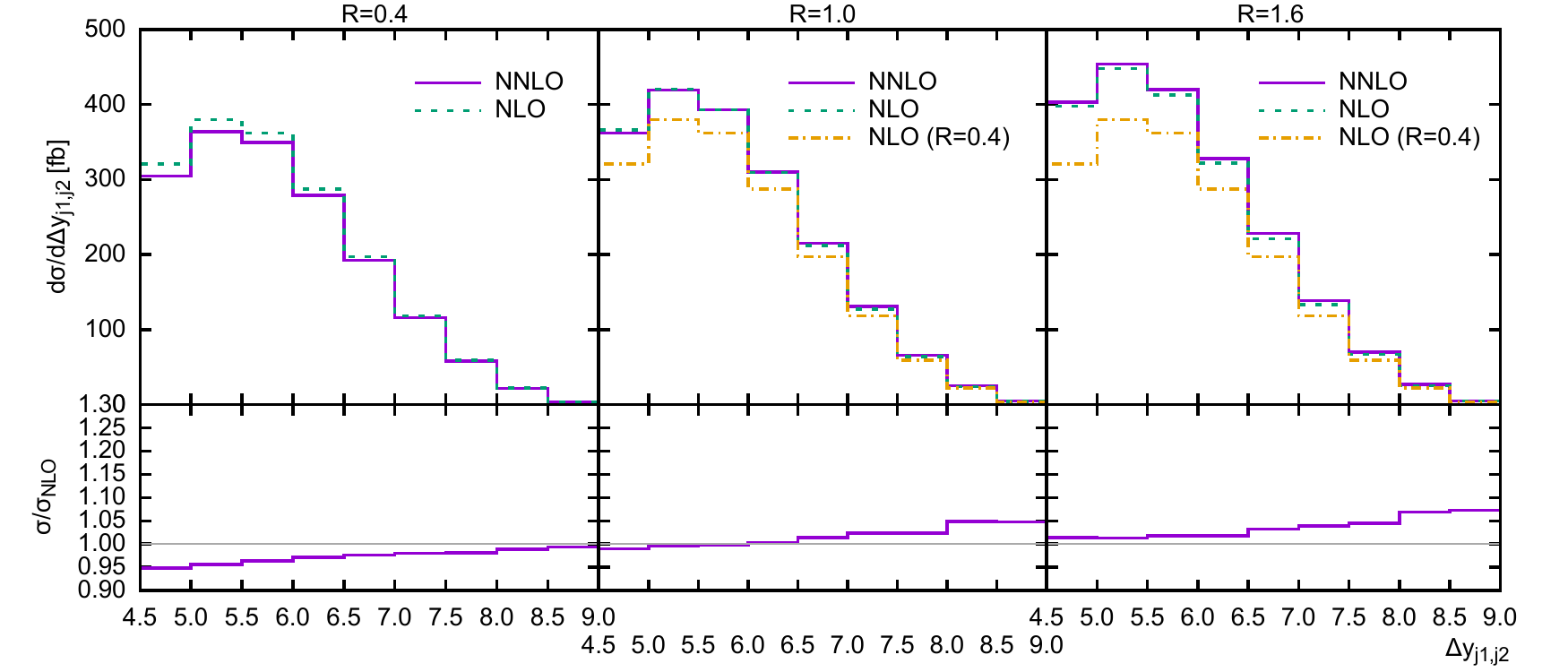}
  \caption{Transverse momentum distribution for the leading jet
  (\textit{top}), the subleading jet (\textit{centre left}) and the
  Higgs boson (\textit{centre right}) as well as the rapidity difference
  of the two tagging jets (\textit{bottom}). 
  }
  \label{fig:Higgs_vbf_nnlo:diffRdep}
\end{figure}

Both the VBF Higgs NNLO QCD and the VBF Higgs plus three jet NLO cross
sections have then been used to study the dependence of the NNLO result
on the jet definition in Ref.~\cite{Rauch:2017cfu}. For the dependence
of the integrated cross section on the jet clustering radius $R$, shown
in Fig.~\ref{fig:Higgs_vbf_nnlo:intRdep}, the matching value between LO, NLO and NNLO
results stays at approximately $R{=}1.0$ after the correction. The slope
of the NNLO cross section flattens slightly. In Fig.~\ref{fig:Higgs_vbf_nnlo:diffRdep}
we show the updated results on the differential distributions. The
smallness of the NNLO corrections in the transverse momentum
distributions of the subleading jet and the Higgs boson when using
$R{=}1.0$ is unchanged from our original result. In contrast, we had
seen relevant remaining NNLO effects when using the larger clustering
radius at small transverse momenta of the leading jet and at large
rapidity differences of the two tagging jets. In both cases, the
remaining effects are significantly reduced after the update. 






\section{The gluon-fusion component of Higgs boson-jet-jet production and its suppression in weak-boson fusion studies~\protect\footnote{
      J.~R.~Andersen,
      M.~Heil,
      A.~Maier,
      J.~M.~Smillie}{}}
\label{sec:MC_gg_vbf}



  We compare various perturbative predictions for the gluon-fusion component
  of Higgs-boson production in association with dijets (Hjj), in particular
  within the cuts designed to enhance the acceptance of Hjj from the
  production through weak-boson fusion.

  The first sub-leading corrections from the high-energy limit have
  recently been included in the description of \textsc{HEJ}. The
  predictions are changed only minimally by the inclusion of these
  sub-leading corrections, and are here contrasted with those obtained at
  next-to-leading order accuracy. We show that a moderate cut
  on further central hard jet activity can reduce the gluon-fusion
  component in weak-boson fusion measurements by about 10--15\%.

\subsection{Introduction}
The study of the process of Higgs boson production in association with dijets
is significant for several reasons. It can proceed through the Born level
fusion of two weak bosons exchanged between quarks from each proton to a
Higgs boson\cite{Cahn:1983ip} (WBF) -- or as a radiative correction to the
top-loop mediated fusion of two gluons\cite{DelDuca:2001eu,DelDuca:2001fn}
(GF). The quantum interference between the two processes is
negligible\cite{Andersen:2007mp}, so the processes can in theory be studied
independently. As such, the WBF process provides a direct and independent
measurement of the Higgs boson coupling to weak bosons. Meanwhile, the GF
process allows for tests of the $CP$-properties of the Higgs-boson couplings
to gluons\cite{Hankele:2006ma,Klamke:2007cu,Andersen:2010zx}; in particular
the small admixtures from extended sectors with direct $CP$-violations can be
measured by studying the azimuthal angle between the
jets\cite{Plehn:2001nj,Andersen:2010zx}.

The degree of success of all such studies will, however, depend not just on
the standard suppression of background events unrelated to Higgs-boson
production, but also on separating the WBF and GF component of the Hjj
production. We will here report on an updated study on possibilities of
suppressing the GF component during WBF studies. The GF component has the
largest cross section at the inclusive level of Hjj, but it is dominated by
the sub-processes involving two incoming gluons. This skews the partonic
processes towards smaller total invariant masses than that for WBF, since the
gluons predominantly carry less of the proton energy than the valence quarks
do. A minimal cut on the invariant mass between the two jets (anti-kt,
$p_t>30$~GeV) will therefore favour WBF over GF. We will consider cuts of
$m(j_1j_2)>400$~GeV, in line with current experimental studies. Similarly, a
minimal rapidity difference between the jets is often required,
e.g.~$|y_{j_1}-y_{j_2}|>2.8$. This constitutes the WBF cuts.

A reliable calculation of the GF component within these cuts obviously rely
on a successful description not just of the total rate, but also the spectrum
in $m_{j_1 j_2}$ and $|y_{j_1}-y_{j_2}|$. The description of QCD processes at
large dijet invariant masses is traditionally difficult, and has been studied
for both Wjj\cite{Aad:2014qxa,Aaboud:2017fye} and Zjj\cite{Aaboud:2017emo},
where also both a WBF and a QCD-dominated process contribute.

We will here be concerned with predictions from pure
NLO\cite{Campbell:2006xx,Campbell:2010cz} and \emph{High Energy Jets}
(HEJ)\cite{Andersen:2009nu,Andersen:2011hs,Andersen:2012gk,Andersen:2016vkp,Andersen:2017kfc},
which both give a good description of e.g.~the dijet invariant mass spectrum
for the QCD-dominated processes in
Wjj\cite{Aad:2014qxa,Aaboud:2017fye,Abazov:2013gpa,Aaboud:2017soa}. The
properties of the radiative corrections associated with the colour-octet
exchange between jets separated by a large invariant mass is similar between
various underlying processes, such that a test of the predictions for $Zjj$
or $Wjj$ can aid the investigation of $Hjj$. This is particularly useful,
when further discrimination between the WBF and GF processes is sought by
utilising the observation that the colour exchange associated with the GF
process leads to increased perturbative jet-activity in the rapidity range of
the colour-octet exchange compared to the WBF
process\cite{Dokshitzer:1991he}. The recent measurement by ATLAS utilised
this to isolate the WBF component of $Zjj$, but found very large variations
in the predictions of the spectrum of $m_{j_1 j_2}$ for the GF enhanced
sample where a third jet was required, and between these predictions and
data.

The reliability of the perturbative prediction for the jet veto can be
jeopardised by succumbing to the temptation of using a small transverse
momentum for the cut. In this study we will investigate how the suppression
of the GF component can be obtained with perturbatively more stable cut on
larger transverse momenta, but with the cut applied in a larger region of
rapidity. The difference in jet radiation pattern between GF and WBF is valid
not just for region in rapidity between the two hardest jets, but in the full
region of colour octet exchange, which predominantly is between the forward
($j_f$) and backward ($j_b$) jet. Ref.\cite{Binoth:2010nha} found stable
predictions for $d\sigma/dm_{j_f j_b}$ between shower, NLO and HEJ for a jet
cut of $p_\perp>40$GeV. Current analyses of LHC data frequently apply a jet veto cut 25GeV to veto
QCD contributions. In this contribution we will investigate the effect of
applying a harder jet cut (which will reduce the effect of the veto), but in
a larger region of rapidity (which will increase the effect of the veto).


\subsection{Results}
\begin{figure}[t]
\begin{center}
\includegraphics[width=0.495\textwidth]{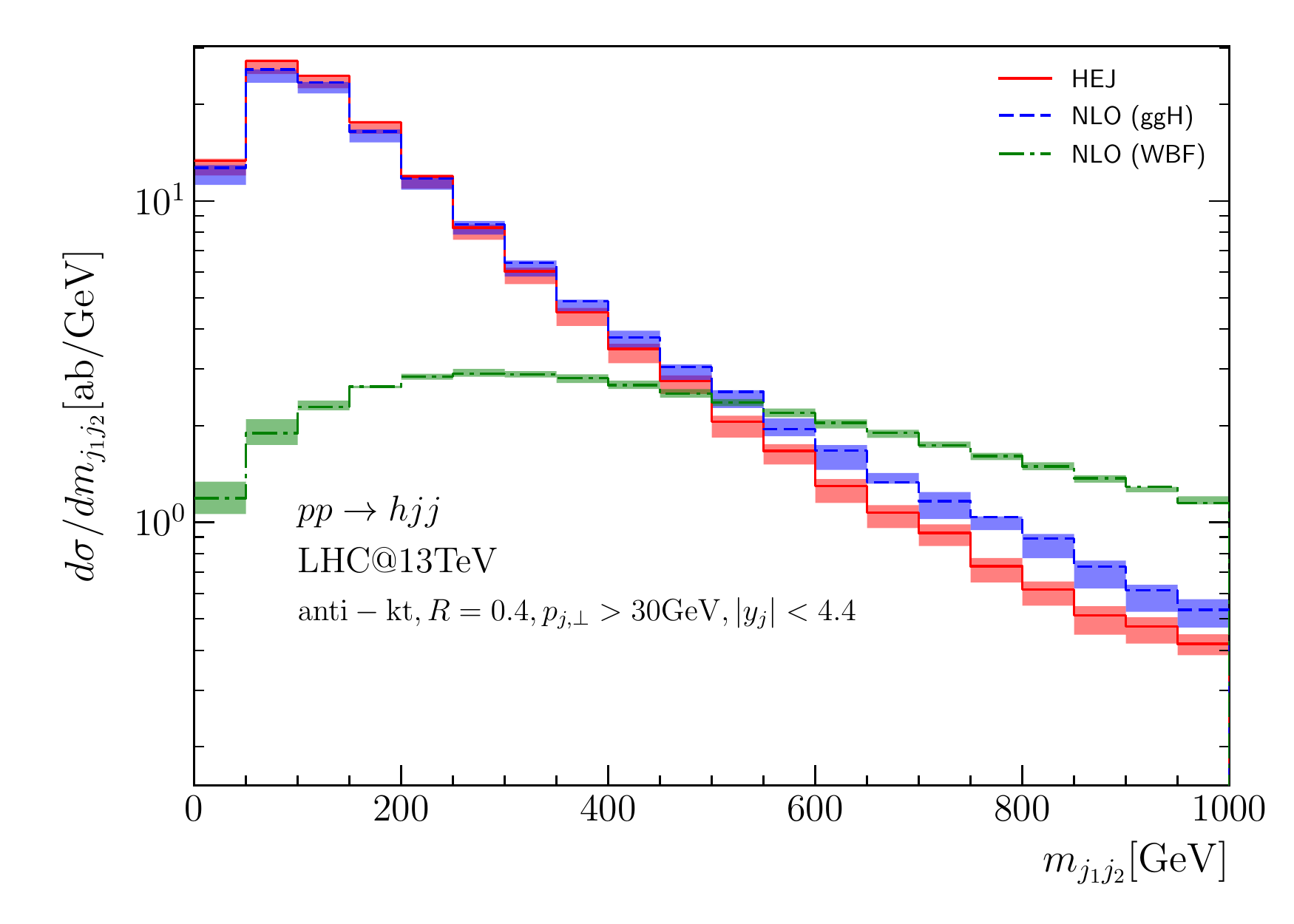}
\caption{$d\sigma/dm_{j_1 j_2}$: The spectrum of the invariant mass between
  the two hardest jets for WBF and GF HJJ calculated at NLO and with HEJ. The
  GF component dominates at small $m_{j_1 j_2}$. See text for further details.}
\label{fig:MC_gg_vbf:mjj}
\end{center}
\end{figure}

The predictions for the invariant mass spectrum between the two hardest jets
for the GF (obtained at NLO and with HEJ) and WBF component (at NLO) of $Hjj$
is displayed on Fig.~\ref{fig:MC_gg_vbf:mjj}. The cross sections for HEJ reported
here have been normalised to NLO accuracy at the level of the inclusive cross
section for $Hjj$. This decreases the scale variation of the results compared
to those reported in Ref.\cite{Andersen:2017kfc}.  Frequently applied WBF
cuts designed to enhance the WBF component over that of GF apply a cut on the
minimum invariant mass of the hardest two jets of 400GeV. The spectrum for
the GF component falls off slightly faster for HEJ than for NLO, which leads
to differences for the prediction of the GF component within the WBF
cuts. Predictions for the distribution in $m_{j_1 j_2}$ with similar jet cuts
as those applied in the studies of $Hjj$ have been checked against data for
$Wjj$\cite{Aaboud:2017fye,Aaboud:2017soa}, albeit for centre-of-mass
energies of 7TeV and 8TeV. A
measurement of the distribution for pure dijet, $Wjj$ or $Zjj$ for similar
jet cuts and parameters as those applied in $Hjj$ would help in determining
the shape at large invariant dijet mass, and thus reduce the uncertainty in
the determination of the GF component. We note that it was
found\cite{Aaboud:2017emo} for $Zjj$ at 13TeV that the predictions for the
QCD-dominated contributions vary by more than a factor of 5 at large
invariant masses.

The study of \cite{Aaboud:2017emo} discriminates further the GF and WBF
components by referring to the dominance of further jet radiation from the GF
component from the colour connection between the two
jets\cite{Dokshitzer:1991he}, and thus applies a slightly softer jet veto
in-between the two hardest jets. However, the discrimination need not apply
just to the rapidity region in-between the two hardest jets. We will here
investigate how a harder jet cut (thus perturbatively safer) can be applied
to a larger region in rapidity to get the same effect. Instead of applying a
jet veto to suppress the GF contribution only if a hard jet exists in-between
the two hardest jets of the event, we will investigate the following
procedure: Define the two tagging jets to be either the two hardest
($j_1, j_2$) or the two furthest apart in rapidity ($j_f, j_b$), and let
$y_0 = (y_{j_1} + y_{j_2} )/2$ or $y_0 = (y_{j_f} + y_{j_b} )/2$. The region
of jet veto is then characterised by a rapidity $y_c>0$, and an event is
removed by the veto if it contains a further jet (with transverse momentum at
least 30GeV) in-between the two tagging jets, and with a rapidity
$y_{j}$ with $|y_j -y_0 | < y_c$ (cf.\cite{Rainwater:1996ud}). This
definition corresponds to applying no central jet veto at all for $y_c =
0$, whilst in the case of the tagging jets being the forward/backward
pair the limit of large $y_c$ corresponds to vetoing all additional
jets, and $\sigma(y_c = \infty)$ is the cross section for the
\emph{exclusive} production of a Higgs boson with two jets.

\begin{figure}[t]
\begin{center}
\includegraphics[width=0.49\textwidth]{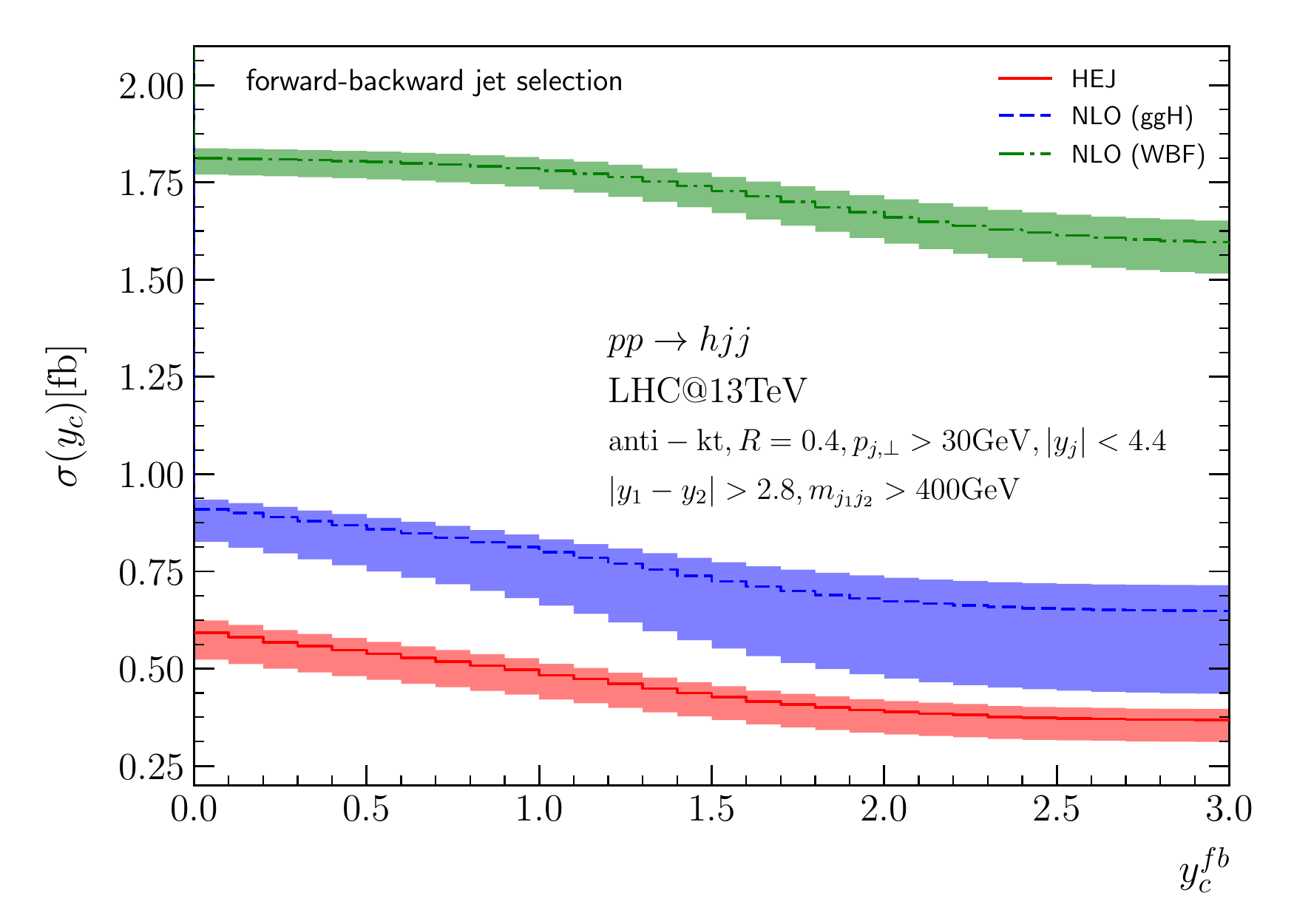}
\includegraphics[width=0.49\textwidth]{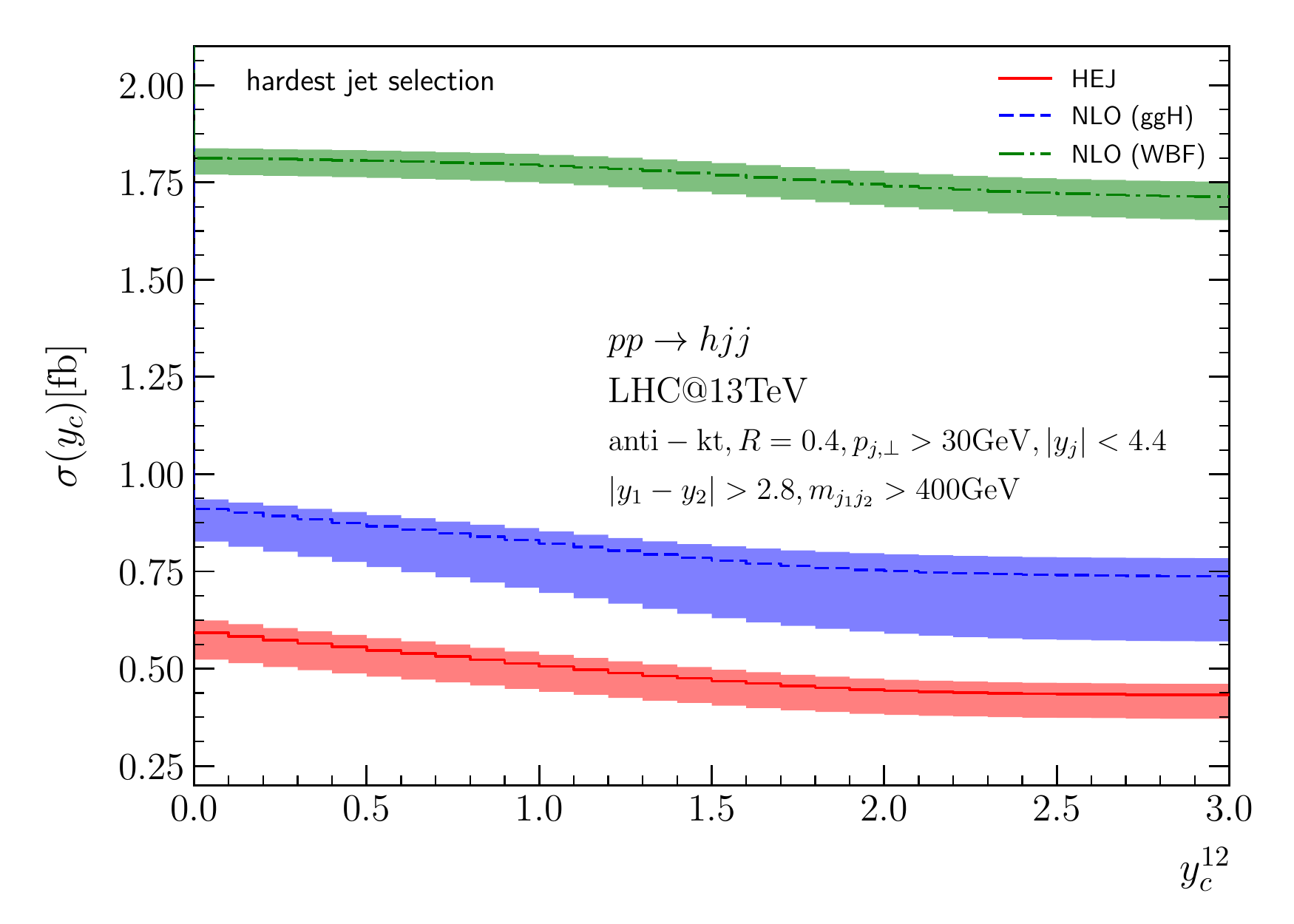}
\includegraphics[width=0.49\textwidth]{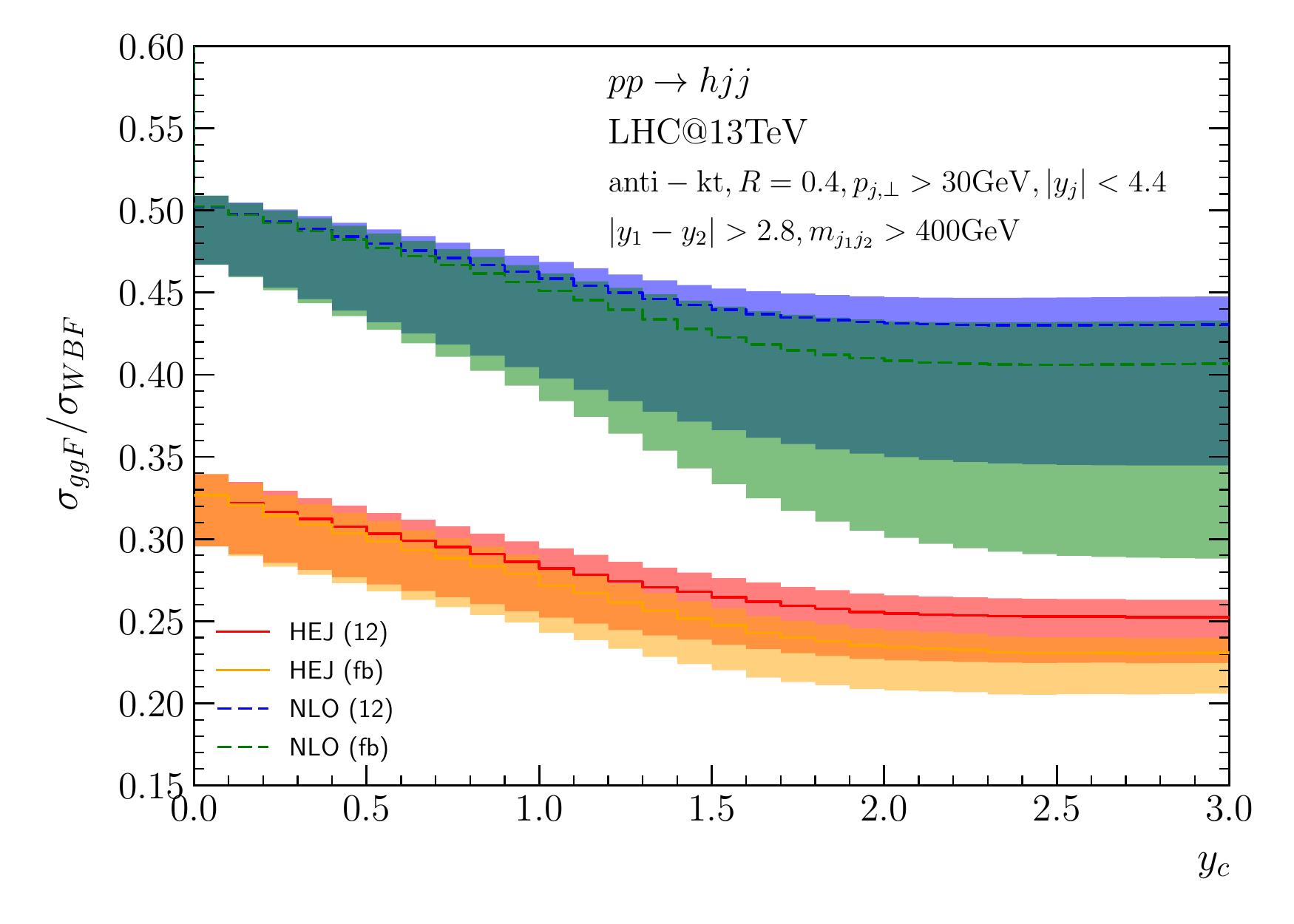}
\caption{The cross section passing the WBF cuts on the two hardest jets,
  and a further veto on central jet activity, either between the forward
  and backward jets (top left) or between the two hardest jets (top
  right). The former is rejecting more of the gluon-fusion
  component. The lower panel shows the relative size of the GF component
  compared to WBF.}
\label{fig:MC_gg_vbf:CJV}
\end{center}
\end{figure}
At the top of Fig.~\ref{fig:MC_gg_vbf:CJV} we plot the result for the
predictions of the GF and WBF components passing the cuts on a minimum
invariant $m_{j_1 j_2}$ and rapidity separation between the two hardest jets,
as a function of a further veto of jet activity in an increasing region of
rapidity described as mentioned above by $y_c$, either between the
forward/backward jet pair or just the two hardest jets (right). The
difference between vetoing jets in-between the forward/backward pair or the
hardest pair is modest, as illustrated on the bottom plot, which shows the
relative GF component of WBF+GF passing the cuts, calculated at both NLO and
with HEJ. We observe that for a veto characterised by $y_c=1$ there is almost
no drop in the WBF component, whereas the relative size of the GF component
compared to WBF can be lowered from 50\% to 45\% (NLO predictions, a 10\%
reduction in the GF contribution) or 33\% to 28\% (HEJ predictions, a 15\%
reduction in the GF contributions). We observe that the NLO scale
dependence is comparable to the promised reduction in the GF contribution by
such a central jet veto.

\subsection{Conclusions}
We find that while a jet veto certainly can serve as a useful tool for
further suppressing the GF contribution to HJJ in WBF studies, the GF
contribution at large dijet invariant mass in itself has an uncertainty
comparable to the promised effects of a moderate central jet veto on the
surviving cross section. The difference between the HEJ and NLO
predictions is larger than the suppression, which in turn is roughly of
the same size as the NLO scale dependence. It therefore seems prudent to perform detailed measurements of the dijet
invariant mass spectrum for related processes, and utilise these to
improve predictions compared to those contrasted to electroweak
production of $Zjj$\cite{Aaboud:2017emo}.




\section[The sensitivity of a $ZH/WH$ ratio measurement in the
$H\rightarrow b\bar{b}$ channel~\protect\footnote{
      R. Harlander,
      J. Klappert,
      C. Pandini,
      A. Papaefstathiou,
      L. Perrozzi}{}]
        {\boldmath The sensitivity of a $ZH/WH$
  ratio measurement in the $H\rightarrow b\bar{b}$
  channel~\protect\footnote{
      R. Harlander,
      J. Klappert,
      C. Pandini,
      A. Papaefstathiou,
      L. Perrozzi}{}}
\label{sec:SM_VHratio}

\subsection{Introduction}
Since no new fundamental particles beyond the SM have been found at the
LHC up to now, the focus is turning towards indirect signals for New
Physics. This requires the comparison of precise measurements with
higher-order calculations. At a hadron collider, however, effects such
as parton luminosities or jet tagging efficiencies influence both the
accuracy of the theory prediction as well as the measurement. The
identification of suitable observables where the sources of uncertainty
are minimised is thus essential for a successful high-precision program
at the LHC.

With this objective, we investigate the ratio of $ZH$ to $WH$
production. As suggested in \cite{Harlander:2013mla}, the close
similarity in the leading-order production mechanisms for both processes
should lead to a large cancellation of both the theoretical and the
experimental uncertainties in this ratio. However, due to qualitatively
different contributions to $ZH$ production which, in the SM, occur at
higher orders, the ratio reveals significant sensitivity to potential
new physics. In the SM, the dominant difference between $WH$ and $ZH$
production is due to the partonic process $gg\to ZH$, which is mostly
mediated by top- and bottom-quark loops. In extensions of the SM, this
process may receive contributions from virtual loops of new particles
which may alter its effect on the total cross section, as well as on its
kinematics\,\cite{Harlander:2018yio} (see also
\cite{Englert:2013vua}). In addition, new $s$-channel effects due to the
exchange of new Higgs bosons may occur\,\cite{Harlander:2013mla}.

A detailed analysis of such effects on various kinematical distributions
will be presented elsewhere~\cite{OurPaper}. In this contribution, we
study the potential of a dedicated measurement of the $ZH$/$WH$ ratio at
the LHC, based on recent results for the individual $WH$ and $ZH$ cross
sections and assuming various degrees of correlation among their
uncertainties.

\subsection{\boldmath The double ratio $R^{ZW}_R$}
We consider first the experimental quantity:
\begin{equation}
    R^{ZW}_\mathrm{exp} = \frac{ \mathrm{d} \sigma^{ZH} } { \mathrm{d}
      \sigma^{WH}} \;,
    \label{eq:SM_VHratio:Rzw}
\end{equation}
where $\mathrm{d} \sigma^{VH}$ represents the (differential) rate for
vector boson ($V = (Z,W)$) plus Higgs boson ($H$) production. Despite
the different final states in the numerator and the denominator, we
expect that a number of systematic experimental uncertainties
cancel. Here we derive rough estimates of the statistical and systematic
uncertainties.

We can also define the contributions of the ``gluon-fusion'' (ggF) and
``Drell-Yan''-like (DY) component of the $ZH$ process by:
\begin{equation}
    R_{gg}= \frac{ \mathrm{d} \sigma^{gg} } { \mathrm{d}
      \sigma^{ZH}_{\mathrm{DY}} } = R^{ZW}_R\;-1,
    \label{eq:SM_VHratio:RzwR}
\end{equation}
The ``double ratio'' $R^{ZW}_R$ can be constructed by dividing the
experimental measurement of ratio of $ZH$ and $WH$ events with the
theoretical prediction for the ratio between the DY $ZH$ and $WH$:
\begin{equation}\label{eq:SM_VHratio:doubleratio}
R^{ZW}_R = \frac{R^{ZW}_{\mathrm{exp}}} {R^{ZW}_{DY,\mathrm{th}}} \;.
\end{equation}
\subsubsection{Statistical uncertainty}
Error propagation on $R^{ZW}_R$ yields:
\begin{equation}\label{eq:SM_VHratio:doubleratioerror}
\left(\frac{ \delta R^{ZW}_R } { R^{ZW}_R } \right)^2 = \left(\frac{
  \delta {R^{ZW}_\mathrm{DY}} } { R^{ZW}_{\mathrm{DY}} }
\right)^2_{\mathrm{th}} + \left(\frac{ \delta R^{ZW}} { R^{ZW
}}\right)^2 _{\mathrm{exp}} \;,
\end{equation} 
where the second term on the r.h.s.\ contains the experimental
statistical and systematic uncertainties. In practice, the experimental
ratio $R^{ZW}_{\mathrm{exp}}$ can be evaluated either by performing a
fit, or explicitly, e.g.:
\begin{equation}
R_\mathrm{exp}^{ZW} = \frac{\mathrm{d} N^{ZH}}{ \mathrm{d} N^{WH} } = \frac{
  \mathrm{d} N^{\ell^+ \ell^- b\bar{b}} - \mathrm{d} N^{\ell^+ \ell^-
    b\bar{b}}_\mathrm{bkg} } {\mathrm{d} N^{\ell^\pm b\bar{b}} -
  \mathrm{d} N^{\ell^\pm b\bar{b}}_\mathrm{bkg}}\,,
\end{equation}
where $\mathrm{d} N^{X b\bar{b}}$ and $\mathrm{d} N^{X
  b\bar{b}}_\mathrm{bkg}$ represent the total and background events,
respectively, in a certain bin for a given observable, in the final
state $X b\bar{b}$, where $X$ is either one or two leptons. The
background subtraction induces a systematic uncertainty which we absorb
in the estimations of the next section. Assuming sufficiently large
event samples, the (data) statistical uncertainty on the experimental
ratio is given by:
\begin{eqnarray}
\left(\frac{ \delta R^{ZW}} { R^{ZW} }\right)^2 _{\mathrm{exp,~stat}}
&=& \left( \frac{ \delta (\mathrm{d} N^{\ell^+ \ell^- b\bar{b}}) } {
  \mathrm{d} N^{\ell^+ \ell^- b\bar{b}} - \mathrm{d} N^{\ell^+ \ell^-
    b\bar{b}}_\mathrm{bkg}} \right)^2 + \left( \frac{ \delta (\mathrm{d}
  N^{\ell^\pm b\bar{b}}) } { \mathrm{d} N^{\ell \pm b\bar{b}} -
  \mathrm{d} N^{\ell \pm b\bar{b}}_\mathrm{bkg} } \right)^2 \nonumber \\ &=&
\frac{ \mathrm{d}N^{\ell^+ \ell^- b\bar{b}} } { (\mathrm{d} N^{\ell^+
    \ell^- b\bar{b}} - \mathrm{d} N^{\ell^+ \ell^-
    b\bar{b}}_\mathrm{bkg})^2 } + \frac{ \mathrm{d} N^{\ell^\pm
    b\bar{b}} } { (\mathrm{d} N^{\ell \pm b\bar{b}} - \mathrm{d} N^{\ell
    \pm b\bar{b}}_\mathrm{bkg})^2 } \;,
    \end{eqnarray}
where we have assumed that the event sample sizes are sufficiently large. 

The theoretical ratio $R^{ZW}_\mathrm{DY, th}$ is affected by theoretical systematic uncertainties due to parton density functions and higher-order corrections. Due to the similarity between the DY $WH$ and $ZH$ processes, these are expected to be highly correlated. Under the assumption that the scale uncertainties are fully correlated, we have verified that the combined QCD scale and PDF uncertainties on this ratio are of the order of $\sim 1\%$ and we neglect them here for simplicity. The effect of electroweak corrections is left to future work~\cite{OurPaper}. In the next section we describe an estimation of the systematic experimental uncertainty on $R^{ZW}_\mathrm{exp}$.

\subsubsection{Systematic uncertainty}
The systematic uncertainty on the $R^{ZW}_\mathrm{exp}$ includes all
uncertainties contributing to an experimental measurement of this
quantity. While a precise determination of the systematics would require
a fully-fledged experimental analysis, we can nevertheless extract an
estimate of the uncertainty from the results presented
in~\cite{Aaboud:2017xsd}, for the separate $ZH$ and $WH$ signal
strengths:
\begin{eqnarray}
\mu_{ZH} &=&
1.12^{+0.34}_{-0.33}\mathrm{(stat.)}^{+0.37}_{-0.30}\mathrm{(syst.)}\,,\nonumber\\
\mu_{WH} &=&
1.35^{+0.40}_{-0.38}\mathrm{(stat.)}^{+0.55}_{-0.45}\mathrm{(syst.)}
\label{eq:SM_VHratio:signalstrengths}
\end{eqnarray}
The systematic term of the signal strength uncertainty includes all
sources of experimental nature, related to the background and signal
Monte Carlo simulation and data driven estimates, and to the finite size
of the simulated samples.

We assume that these systematic uncertainties can be propagated directly
on the experimental ratio defined by Eq.~\eqref{eq:SM_VHratio:Rzw}. For
simplicity, the uncertainties are symmetrized, and three different
correlation scenarios are explored: uncertainties on $\mu_{ZH}$ and
$\mu_{WH}$ completely uncorrelated, 50\% correlated or 100\% correlated,
which we denote as: $p_{WZ} = (0, 0.5, 1.0)$ respectively.

\subsection{Analysis and results}
We present here a proof-of-principle Monte-Carlo level analysis
demonstrating the usefulness of the double ratio $R^{ZW}_R$. The
$VH(bb)$ selection follows the experimental cuts applied
in~\cite{Aaboud:2017xsd} as closely as possible.

In the analysis, jets are reconstructed using the anti-$kt$ clustering
algorithm with a jet-radius parameter of $R=0.4$.  The jet transverse
momentum is required to be greater than $20$~GeV for `central jets'
($|\eta|<2.5$) and greater than $30$~GeV for `forward jets'
($2.5<|\eta|<5$). Selected central jets are labeled as `$b$-tagged' if a
$b$-hadron is found within the jet. A $b$-tagging efficiency of 70\% is
considered, flat over transverse momentum of the jets, to reproduce the
efficiency of the experimental $b$-tagging algorithm. The leading
$b$-jet is required to have a transverse momentum larger than
$45$~GeV. The missing transverse energy is reconstructed by taking the
transverse momentum of the negative sum of the four-momenta of all the
visible particles. Electrons and muons were subject to isolation
criteria, by requiring the scalar sum of the transverse momenta of
tracks in $R=0.2$ around them to be less than $0.1$ times their
transverse momentum: $\sum_{R < 0.2} p_{T, \mathrm{tracks}} < 0.1 \times
p_{T,\ell}$.

Three event selections were considered in~\cite{Aaboud:2017xsd}, corresponding to the $Z\rightarrow\nu\nu$, the $W\rightarrow l\nu$, and the $Z\rightarrow ll$ channels. Here we only consider the 1- and 2-lepton channels. 
All selections require at least 2 b-tagged central jets, used to define the invariant mass $m_{bb}$. For the $W\rightarrow l\nu$ selections events with more than 3 central and forward jets are discarded.

To calculate the reconstructed top quark mass, $m_{top}$, an estimate of
the four-momentum of the neutrino from the $W$ boson decay is obtained,
considering the neutrino's transverse momentum components identical as
the $E_{T}^{miss}$ vector component, and constraining $p^{\nu}_z$ from
the $W$ mass. Further details on 1-lepton and 2-lepton channels are as
follows:\\ \ \\ The $Z\rightarrow ll$-channel selection is further
defined by:
\begin{itemize}
  \item exactly 2 same-flavour (and opposite charge for muons) leptons $p_{T}>7$ GeV  and $|\eta|<2.5$, of which at least one with $p_{T}>25$ GeV
  \item lepton invariant mass $81$ GeV $< m_{ll}< 101$ GeV
\end{itemize}
\ \\
The $W\rightarrow l\nu$-channel selection is further defined by:
\begin{itemize}
  \item exactly 1 lepton with $p_{T}>25$ GeV  and $|\eta|<2.5$
  \item $p_{T}^{W} > 150$ GeV
  \item $E_{T}^{miss} > 30$ GeV in the electron sub-channel
  \item $m_{bb} > 75$ GeV or $m_{top} \leq 225$ GeV
\end{itemize}
~\\ Event samples at 13~TeV have been generated using
\texttt{MG5\_aMC@NLO}~\cite{Alwall:2011uj, Alwall:2014hca}, interfaced
to the general-purpose event generator \texttt{HERWIG
  7}~\cite{Gieseke:2011na, Arnold:2012fq, Bellm:2013hwb, Bellm:2015jjp,
  Bellm:2017bvx} for the parton shower and non-perturbative effects,
such as hadronization and multiple-parton interactions. No smearing was
applied to the momenta of jets or leptons entering the analysis and no
backgrounds originating from mis-tagging of charm jets or light jets
were considered.

For simplicity, we concentrate on the ``dijet-mass analysis''
of~\cite{Aaboud:2017xsd}, where the $BDT_{VH}$ discriminant is replaced
by the $m_{bb}$ variable. This results in 10 signal regions, shown in
the second and third rows of Table~12 in~\cite{Aaboud:2017xsd}. In the
present analysis, we have only included signal regions with $p_{T,V} >
150$~GeV. We have verified that the signal and background event yields
are in reasonable agreement at the ``selection'' level with those
of~\cite{Aaboud:2017xsd}. We have calculated the double ratio by
constraining the invariant mass of the two $b$-jets to lie within
$m_{bb} \in [110, 140]$~GeV.

\subsubsection{Results}
Table~\ref{tab:SM_VHratio:results} shows results for the double ratio
$R_R^{ZW}$ after applying the analysis, as well as rough estimates of
the statistical and systematic uncertainties. The ggF component of $ZH$
was rescaled by a flat $K$-factor of $K=2$ to account for higher-order
QCD corrections that are presently unavailable. The systematic
uncertainty is given assuming a correlation between the individual $ZH$
and $WH$ sample systematic uncertainties of $p_{WZ} = (0, 0.5, 1.0)$,
i.e.\ fully uncorrelated, moderately-correlated, and fully correlated,
respectively. We show results for integrated luminosities of
$\mathcal{L} = 36.1,~300,~3000~\mathrm{fb}^{-1}$, corresponding to
Ref.~\cite{Aaboud:2017xsd}, the LHC before LS3, and the HL-LHC,
respectively.

\begin{table}[t]
  \begin{center}
    \begin{tabular}{r|c|ccc|ccc}
      && \multicolumn{3}{c|}{stat.\ (${\cal L}/\text{fb}^{-1}$)}
      & \multicolumn{3}{c}{syst.\ ($p_{WZ}$)}\\ & $R_R^{ZW}$ & $36.1$ &
      $300$ & $3000$ & 0 & 0.5 & 1 \\ \hline all $m_{VH}$ & 1.32 & $\pm
      0.88$ & $\pm 0.31$ & $\pm 0.10$ & $\pm 0.79$ & $\pm 0.58$ & $\pm
      0.22$\\ high $m_{VH}$ &1.51 & $\pm 1.43$ & $\pm 0.50$ & $\pm 0.10$
      & $\pm 0.79$ & $\pm 0.58$ & $\pm 0.22$
    \end{tabular}
  \end{center}
  \caption{Results for the double ratio $R^{ZW}_R$ following the
    analysis of~\cite{Aaboud:2017xsd}. The statistical uncertainty is
    shown for integrated luminosities of $36.6$, 300, and
    3000\,fb$^{-1}$, and the systematic uncertainty for fully
    uncorrelated, moderately-correlated, and fully correlated
    uncertainties in $WH$ and $ZH$ production, respectively ($p_{WZ} =
    (0, 0.5, 1.0)$). In the second line, the $VH$ invariant mass was
    restricted to $m_{VH}\in(300,600)$\,GeV. The systematic
    uncertainties are assumed to be unchanged by this restriction.}
\label{tab:SM_VHratio:results}
\end{table}


It is evident from Table~\ref{tab:SM_VHratio:results} that a measurement
of $R^{ZW}_R \neq 1$, demonstrating the presence of a ggF component of
$ZH$, will eventually be dominated by systematic uncertainties. However,
the analysis of~\cite{Aaboud:2017xsd} has not been optimised to maximize
the performance of $R^{ZW}_R$. Indeed, if one restricts the range of the
invariant mass of the reconstructed vector boson and Higgs boson system
to $m_{VH} \in (300, 600)$~GeV, the ratio increases to $R^{ZW}_R \simeq
1.51$, with an associated increased statistical error, as shown in the
second line of Table~\ref{tab:SM_VHratio:results}. Such a cut is not
applied in \cite{Aaboud:2017xsd}, but we expect its effect on the
systematic uncertainty to be rather minor and leave it unchanged for
the current discussion. We also stress that the numbers of
Table~\ref{tab:SM_VHratio:results} do not take into account any
reduction of the systematic uncertainties to be expected from future
improved analysis methods.


\subsection{Conclusions}
We have presented rough results on the feasibility of measurement of the
double ratio $R^{ZW}_R$ in the $VH$ process, defined by
Eq.~\eqref{eq:SM_VHratio:doubleratio}. By exploiting this potential
measurement, we have examined the prospects of constraining the
gluon-fusion component of $ZH$ production. Our preliminary estimates of
statistical and systematic uncertainties, obtained through a
hadron-level Monte Carlo analysis, indicate that both an optimised
experimental analysis and a detailed investigation of systematic
uncertainties would facilitate extraction of ggF-induced $ZH$
production. This would allow constraints to be placed on several new
physics models affecting ggF $ZH$.

\subsection*{Acknowledgements}
AP acknowledges support by the ERC grant ERC-STG-2015-677323.  CP
acknowledges support by the University of Geneva. RH and JK are
supported by BMBF under contract 05H15PACC1.



\newcommand{\powheg}{{\tt POWHEG}\xspace}
\newcommand{\powhegbox}{{\tt POWHEG-BOX}\xspace}
\newcommand{\madgraph}{{\tt MG5\_aMC@NLO}\xspace}
\newcommand{\mcnlo}{{\tt MC@NLO}\xspace}
\newcommand{\sherpa}{{\tt Sherpa}\xspace}
\newcommand{\pythia}{{\tt Pythia\,8}\xspace}
\newcommand{\pythiaold}{{\tt Pythia\,6}\xspace}

\newcommand{\mhh}{\ensuremath{m_{\mathrm{hh}}}\xspace}
\newcommand{\pthh}{\ensuremath{p_{T}^{\mathrm{hh}}}\xspace}
\newcommand{\pt}{p_{T}}
\newcommand{\ptj}{p_{T}^{\mathrm{j}_1}}
\newcommand{\ptjj}{p_{T}^{\mathrm{j}_2}}
\newcommand{\pth}{p_{T}^{\mathrm{h}}}
\newcommand{\pthl}{p_{T}^{\mathrm{h}_1}}
\newcommand{\pths}{p_{T}^{\mathrm{h}_2}}
\newcommand{\dphihh}{\Delta\Phi^{\mathrm{hh}}}
\newcommand{\drhh}{\Delta R^{\mathrm{hh}}}

\section{Parton shower and top quark mass effects in Higgs boson pair production~\protect\footnote{
      G.~Heinrich,
      S.~P.~Jones,
      M.~Kerner,
      S.~Kuttimalai,
      G.~Luisoni}{}}
\label{sec:SM_nlops_ggHH}


We study the effects of various parton shower approaches on
distributions relevant for Higgs boson pair production measurements.
In particular, we investigate 
the behaviour of the Higgs boson pair transverse momentum spectrum
which is particularly sensitive to both parton shower  and top quark mass effects.

\subsection{Introduction}

In order to obtain fully exclusive descriptions of final states, 
and to improve the accuracy of theoretical predictions in regions of the
phase space where the convergence of the perturbative series is spoiled by
large soft and collinear logarithms,  matching
Next-to-Leading Order (NLO) predictions to parton shower (PS) Monte Carlo
programs is desirable and mostly represents the current state of the art. 
Parton shower Monte Carlo programs describe multiple soft and
collinear emissions
by resumming at least the leading tower of logarithms to all orders in
perturbation theory. There are several different schemes which allow
to match a fixed order calculation to a PS without spoiling the NLO accuracy
of the former. One of the aspects which distinguishes them from each other is
a different treatment of corrections which are formally of higher order, suppressed by
higher powers of the strong coupling. It is however well known that these formally subleading
terms can lead to large differences in the final prediction, especially
for processes with large higher order corrections. A typical example
which was already studied extensively in the past is the production of a
Higgs boson in gluon fusion~\cite{Alioli:2008tz,Bagnaschi:2015bop}.

More recently Higgs boson pair production in gluon fusion was
computed at NLO accuracy with full top quark mass dependence~\cite{Borowka:2016ehy,Borowka:2016ypz}. 
In a second step this was matched to different PS Monte Carlo programs~\cite{Heinrich:2017kxx,Jones:2017giv}
following the three
different matching schemes as implemented in the \powhegbox~\cite{Frixione:2007vw,Alioli:2010xd},  \madgraph~\cite{Alwall:2014hca,Hirschi:2015iia}
and \sherpa~\cite{Gleisberg:2008ta}. The aim of the present study is
to identify where the differences between the various PS Monte Carlo
descriptions come from and to assess how they depend on the matching
and shower parameter settings.

\subsection{MC@NLO Parton Shower Matching}
\label{sec:SM_nlops_ggHH:mcnlo}

In this report we will consider the two most widely used NLO parton
shower matching schemes, POWHEG \cite{Nason:2004rx} and MC@NLO
\cite{Frixione:2002ik}. In order to point out the most relevant
differences between them it is both sufficient and notationally
convenient to consider the special case where the splitting kernels
appearing in the fixed order subtraction terms $D$ are identical to
the splitting kernels used for parton shower evolution. In the MC@NLO
formalism, the fixed-order NLO cross section is split up in two parts,
one to be integrated over the born phase space $\phi_B$ and one to be
integrated over the real emission phase space
$\phi_R=\phi_B\times\phi_1$. In terms of the leading order
contributions $B$, the virtual corrections $V$, and the real emission
corrections $R$, the seed cross sections for the two contributions are
respectively given by
\begin{align}
  \bar B(\phi_B) &= B(\phi_B)+V(\phi_B)+\int D(\phi_R) \Theta(\mu^2_\text{PS}-t(\phi_R))\dif\phi_1\label{eq:SM_nlops_ggHH:sevent}\\
                   H(\phi_R) &= R(\phi_R) -
                               D(\phi_R)\Theta(\mu^2_\text{PS}-t(\phi_R))\label{eq:SM_nlops_ggHH:hevent}\ ,
\end{align}
where $t$ is the parton shower evolution variable and $\mu_\text{PS}$
is the parton shower starting scale. Events of both types are then
dressed with additional QCD radiation through a parton shower
algorithm. Making this explicit in the hardest emission off the $\bar B$
events, we thus have
\begin{align}
    &\sigma_\text{NLO+PS} = \nonumber\\
    &\int \bar B(\phi_B)
      \left[\Delta(t_0,\mu^2_\text{PS}) + \int
      \Delta(t,\mu^2_\text{PS})\frac{D(\phi_B,
      \phi_1)}{B(\phi_B)}\Theta(\mu_\text{PS}^2-t)\Theta(t-t_0)\dif\phi_1\right]\dif\phi_B\label{eq:SM_nlops_ggHH:sevent_ps}\\
    +&\int H(\phi_R)\dif\phi_R,
  \label{eq:SM_nlops_ggHH:mcnlo}
\end{align}
with the Sudakov form factor
$\Delta(t_0,t_1)=\exp\left[
  -\int^{t_1}_{t_0}\frac{D(\phi_R)}{B(\phi_B)}\dif\phi_1 \right]$ and
the infrared cutoff scale of the parton shower $t_0$.

It is useful to consider the above expression in the kinematic regimes
of soft and hard emissions separately. In the soft real emission phase
space region where $t\ll\mu_\text{PS}^2$ we have $D\approx R$ and thus
$H\approx 0$ by construction of the subtraction terms. The remaining
contribution is given by the term \eqref{eq:SM_nlops_ggHH:sevent_ps} and gives a
LO+PS like result that is rescaled by a local K-factor of $\bar B/B$.
Considering only hard emissions with $t\approx\mu_\text{ps}$, on the
other hand, we can set $\Delta\approx 1$ and also drop the first term
in the square bracket of \eqref{eq:SM_nlops_ggHH:sevent_ps}, where no emission
occurs at all. After some re-arrangements, this gives
\begin{align}
    \sigma_\text{NLO+PS} &= \int\left[\bar B(\phi_B)-B(\phi_B)\right]\frac{D(\phi_B,\phi_1)}{B(\phi_B)}
  \Theta(\mu_\text{PS}^2-t)\dif\phi_B\dif\phi_1\label{eq:SM_nlops_ggHH:ps-cancellation0}\\
  &+\int R(\phi_R)\dif\phi_R\,, \label{eq:SM_nlops_ggHH:ps-cancellation1}
\end{align}
where the term on the second line gives the fixed-order result and the
term on the first line cancels up to differences between $\bar B$ and
$B$, which are of higher order in $\alpha_s$.

In MC@NLO matching, the scale $\mu_\text{PS}$ therefore separates the real
emission phase space in a resummation region that is populated by the
parton shower through the $\bar B$ events and a region that is
populated mostly by the fixed-order real-emission contributions in
$H$. Variations of this scale can thus be used in order to assess
uncertainties associated with this separation.

\subsection{POWHEG Parton Shower Matching}

In the formulation presented above, the \powheg method can be
understood as the limit in which $\mu_\text{PS}\rightarrow \infty$ and
$D=R$. This leads to $H=0$. All real emission contributions are thus
generated by parton shower emission off $\bar B$ events. Taking
$\mu_\text{PS}\rightarrow\infty$ ensures that the full real-emission
phase space is covered while setting $D=R$ in the first emission
ensures that the fixed-order radiation pattern is recovered. This
method comes with the benefit that no negative event weights occur (in
MC@NLO, the event weight \eqref{eq:SM_nlops_ggHH:hevent} can become negative).
However, setting $D=R$ also leads to the exponentiation of the full
real-emission corrections in the Sudakov form factor. This is in
general not justified since $R$ contains hard, non-factorizing
contributions. Instead of setting $D=R$, it has therefore been
suggested in Ref.~\cite{Alioli:2008tz} to use
\begin{align}
D=\frac{h^2_\text{damp}}{p_T^2+h^2_\text{damp}}R\,,
\end{align}
where $p_T$ is the transverse momentum of the Born final state
($p_T=\pthh$ in our case). This implements a phase space separation
as it is achieved in terms of the parton shower evolution variable in
MC@NLO. Effectively, the damping term
$\frac{h^2_\text{damp}}{p_T^2+h^2_\text{damp}}$ assumes the role of
the Heaviside theta functions in Eqs.~\eqref{eq:SM_nlops_ggHH:sevent} and
\eqref{eq:SM_nlops_ggHH:hevent}.


\subsection{POWHEG and MG5\_aMC@NLO results}

\begin{figure}
\centering
\includegraphics[width=0.48\textwidth]{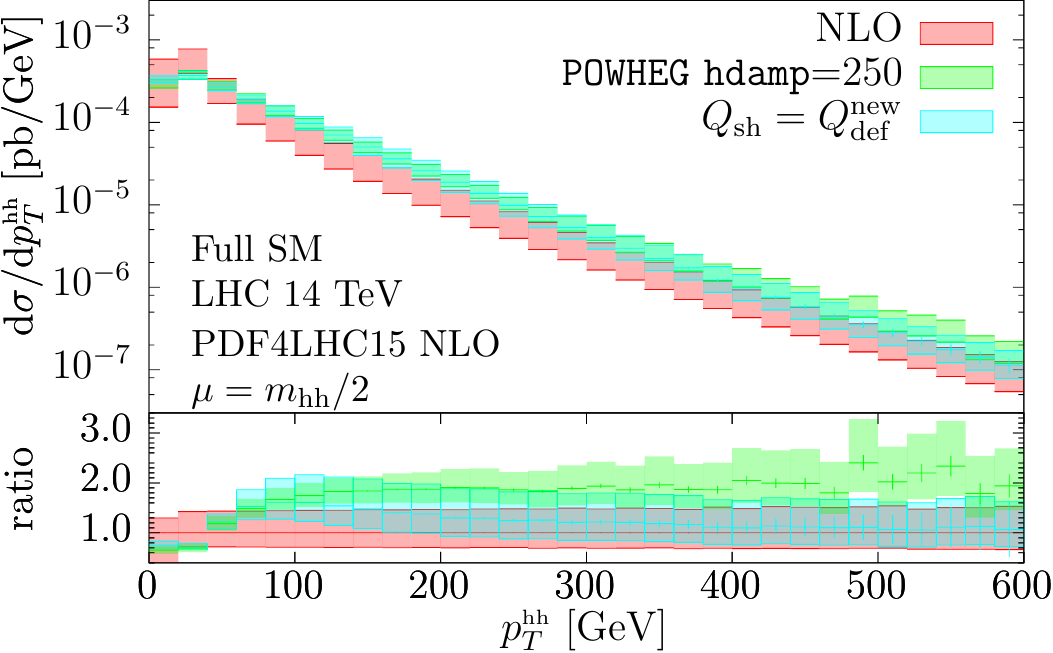}
\hfill
\includegraphics[width=0.48\textwidth]{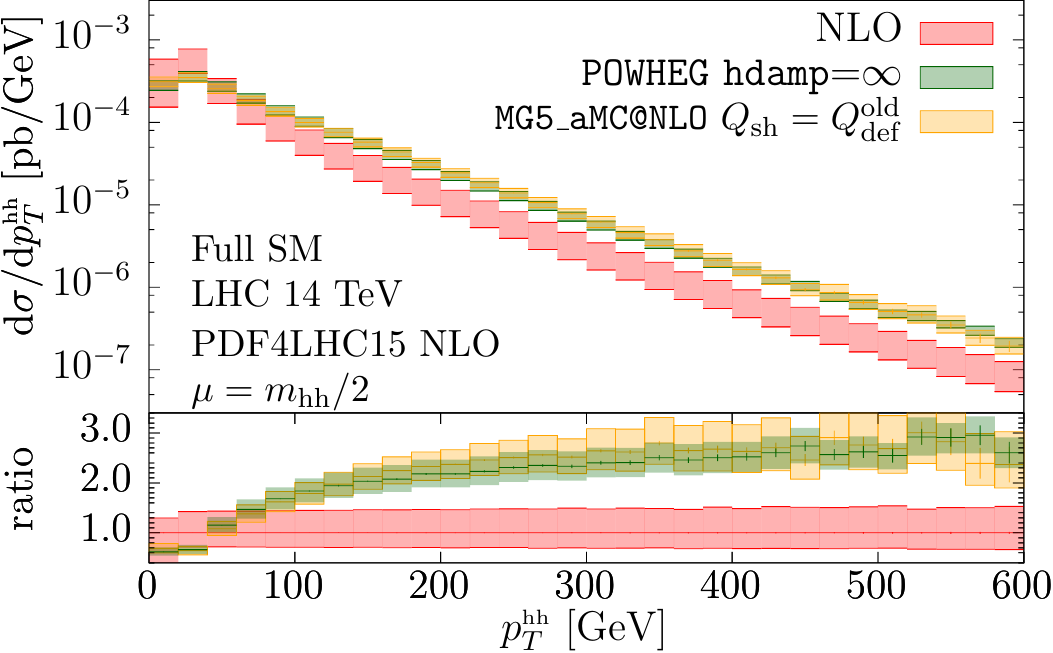}
\caption{%
Comparison between \powheg and \madgraph results Higgs boson pair
 transverse momentum distribution. Left: results based on the new
 definition of the shower starting scale $Q_{\rm{sh}}$ (\madgraph version
 $\geq$ 2.5.3) , compared to \powheg with \texttt{hdamp=250}. Right: results based on the shower starting scale
 $\mu_\text{PS}$ before version 2.5.3, compared to \powheg with \texttt{hdamp}=$\infty$. \label{fig:SM_nlops_ggHH:pwg_mg5}}
\end{figure}
\begin{figure}
\centering
\includegraphics[width=0.48\textwidth]{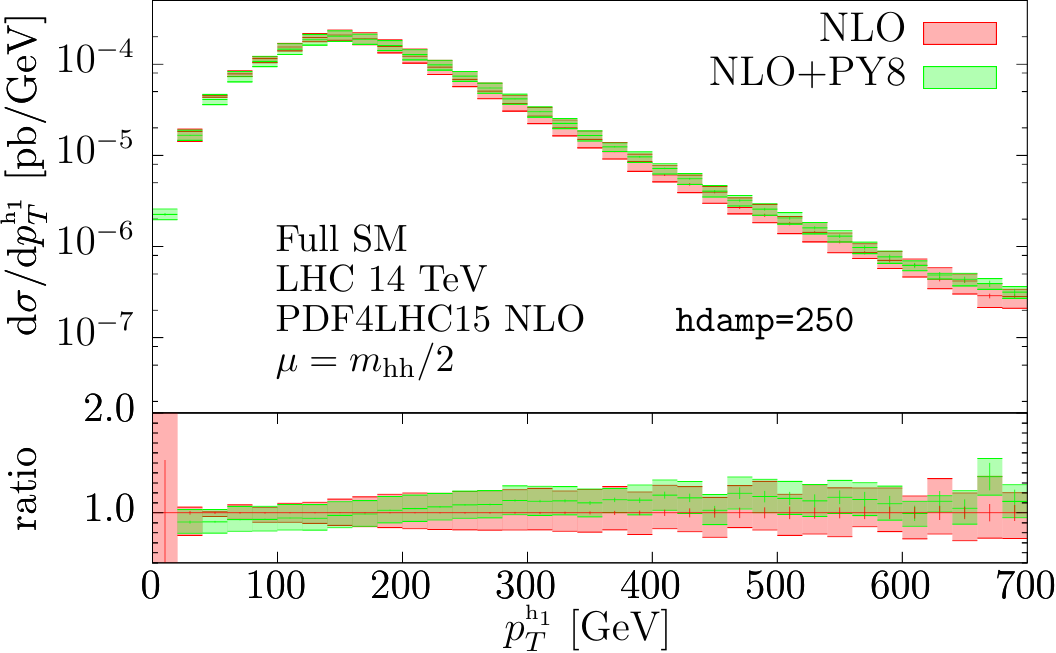}
\hfill
\includegraphics[width=0.48\textwidth]{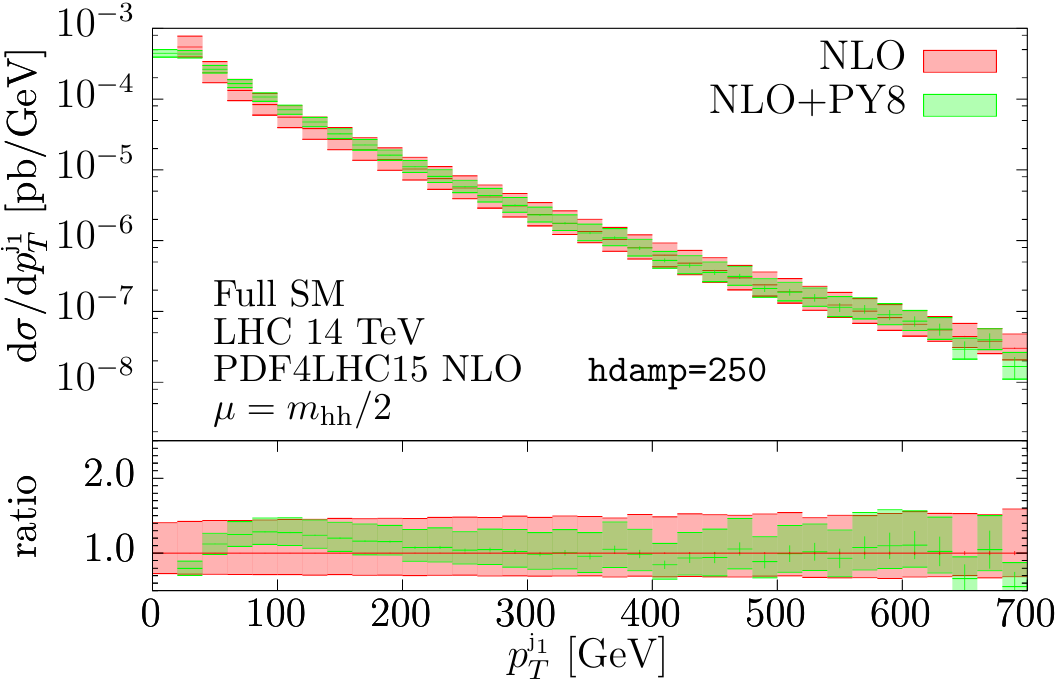}
\caption{%
 Comparison between fixed
 order and \powheg+\pythia results. Left:  
 transverse momentum distribution of the leading-$p_T$ Higgs boson. Right:  transverse
 momentum distribution of the leading jet.\label{fig:SM_nlops_ggHH:ptH1_ptj1}}
\end{figure}

In Fig.~\ref{fig:SM_nlops_ggHH:pwg_mg5} we clearly see the effect of damping the
radiation as explained above by using a value for \texttt{hdamp} different from
\texttt{hdamp}=$\infty$ in \powheg and by using a lower shower starting
scale $Q_{\rm{sh}}=\mu_\text{PS}$ in the \mcnlo approach\footnote{In version 2.5.3 of \madgraph,
the shower starting scale is picked with some probability distribution
to be in the interval {\tt shower\_scale\_factor}
$\times \,[0.1\, H_T/2,H_T/2]$ with $H_T$ computed with Born
kinematics, while previously it was
picked in the interval   {\tt shower\_scale\_factor}
$\times \,[0.1\,\sqrt{\hat{s}},\sqrt{\hat{s}}]$.}.
However, even with \texttt{hdamp=250}, the effects in the tail of the
$\pthh$ distribution are surprisingly large. This is why we made
further investigations in order to understand its origin.
First of all, we should mention that the $\pthh$ distribution is
particularly sensitive to extra radiation because NLO is the first
order where a non-zero $\pthh$ is generated at the level of the hard
matrix element. Therefore, it is not too surprising that the extra emission generated by the parton
shower strongly affects this distribution.  However, the large effects
are not expected to extend into a region where resummation should not
play a role. 
Figure~\ref{fig:SM_nlops_ggHH:ptH1_ptj1} shows that indeed the impact of the parton
shower is small for other distributions like the transverse momentum
of the leading Higgs boson, and even for the transverse momentum distribution
of the leading jet. As the Higgs boson pair recoils against this
jet in the fixed order calculation, while in the showered results more
jets can be present, we again conclude that the enhancement in the tail of
the $\pthh$ distribution must be related to the fact that the shower
generates a lot of radiation.
From Eq.~\eqref{eq:SM_nlops_ggHH:ps-cancellation0} we can at
least partly 
trace this back to the fact that $\bar{B}-B$ is large for Higgs boson
pair production, with an NLO K-factor of about 1.6.
However, this cannot be the only reason, as we will see in the following.

\begin{figure}
\centering
\includegraphics[width=0.48\textwidth]{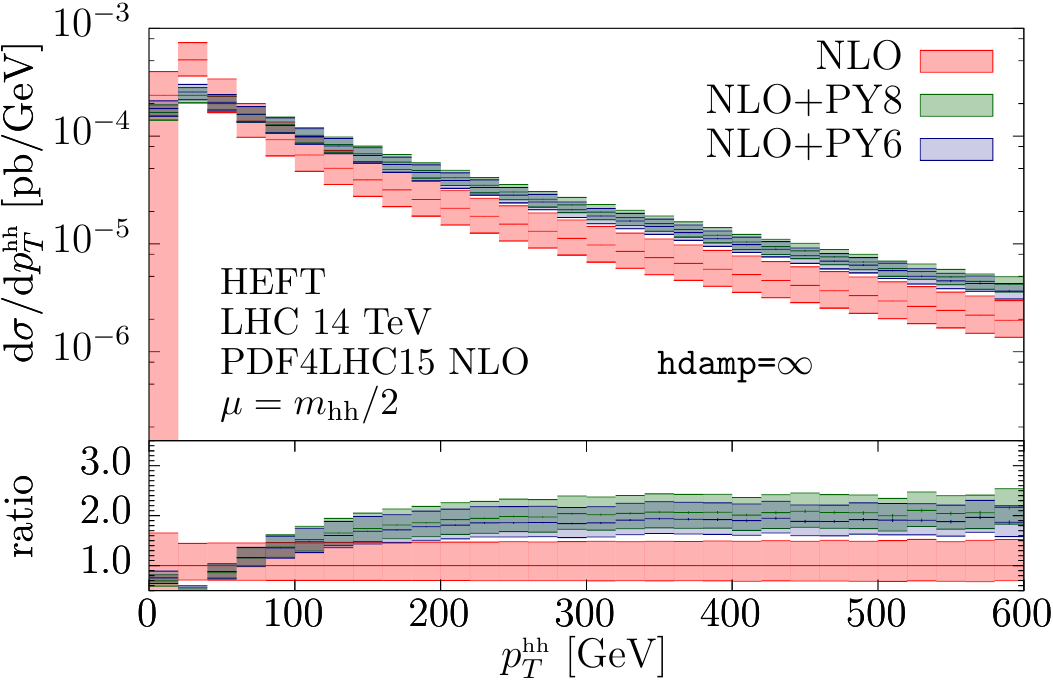}
\hfill
\includegraphics[width=0.48\textwidth]{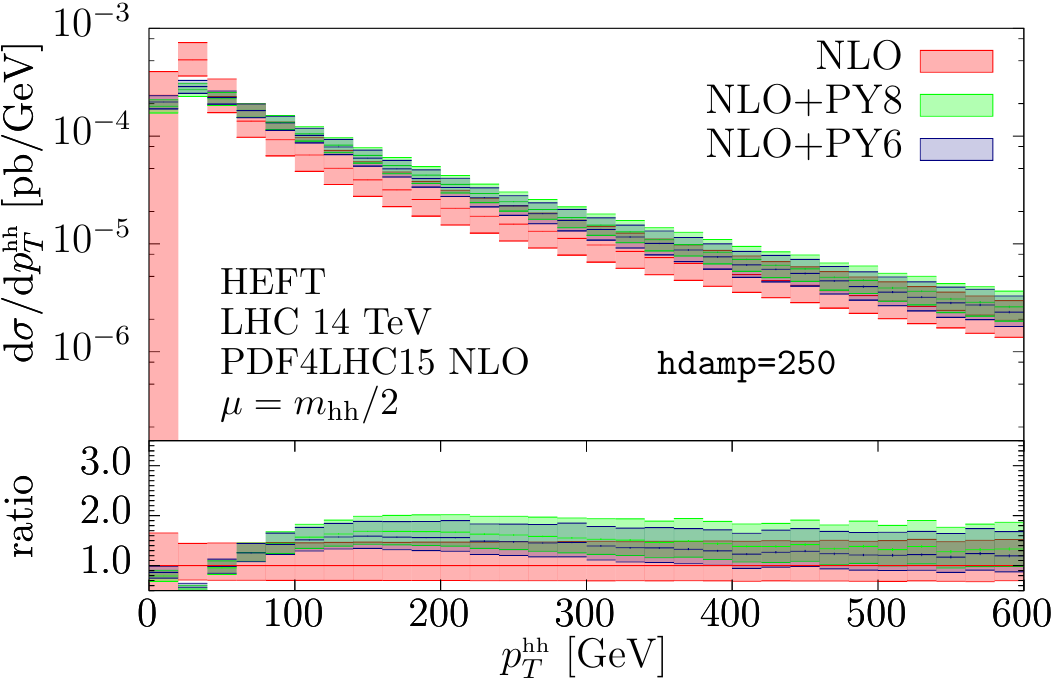}
\\
\includegraphics[width=0.48\textwidth]{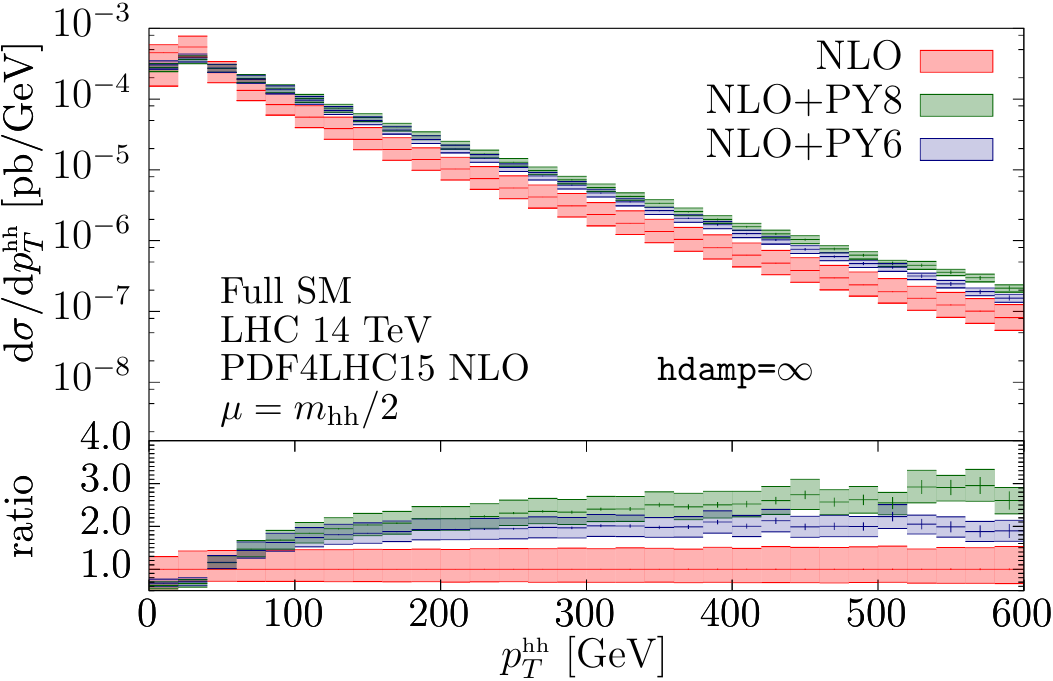}
\hfill
\includegraphics[width=0.48\textwidth]{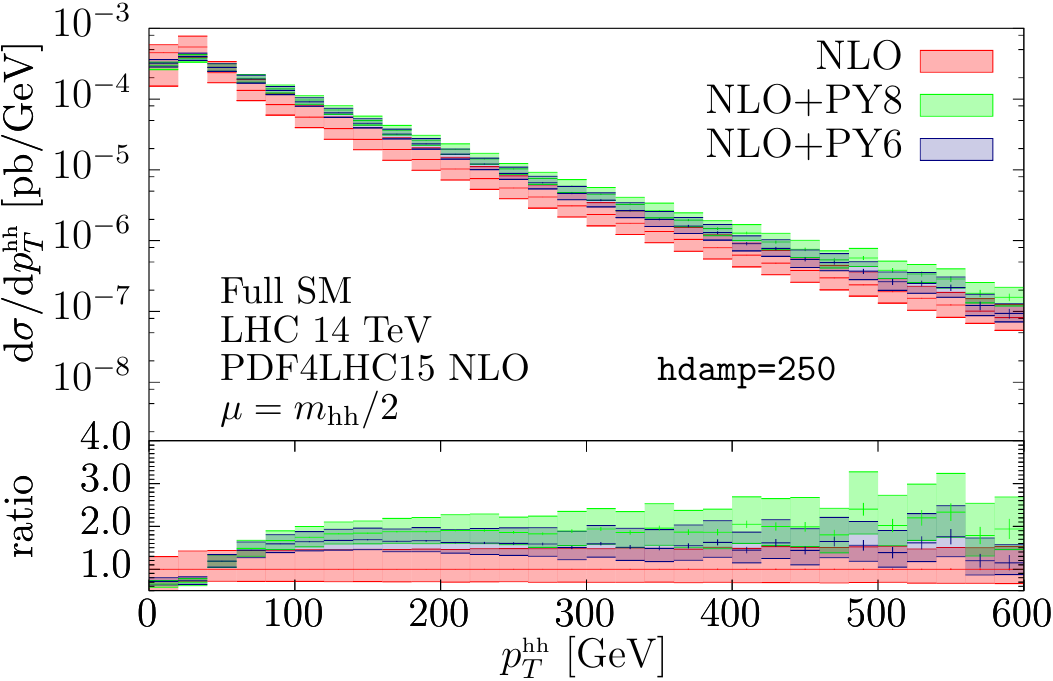}
\caption{%
  Higgs boson pair transverse momentum distribution $\pthh$\,: left column with \texttt{hdamp=$\infty$}, right column
  with \texttt{hdamp=250}. We compare the fixed order result with
  showered results from both \pythiaold and \pythia in the basic HEFT
  approximation (upper row) and in the full SM (lower
  row).\label{fig:SM_nlops_ggHH:pythia6}}
\end{figure}

\begin{figure}
\centering
\includegraphics[width=0.48\textwidth]{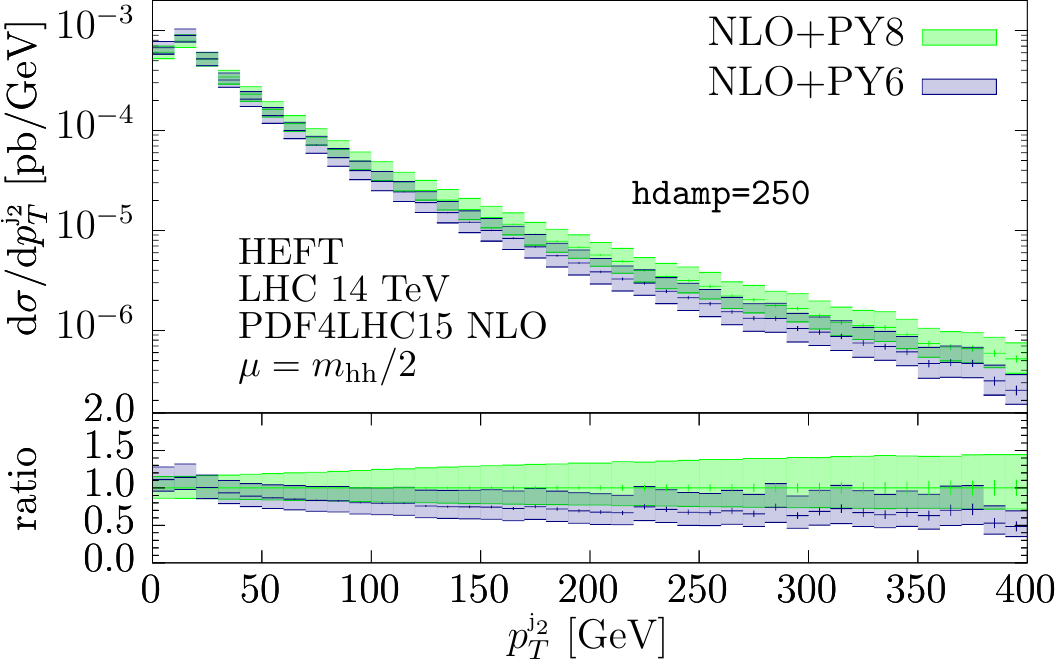}
\hfill
\includegraphics[width=0.48\textwidth]{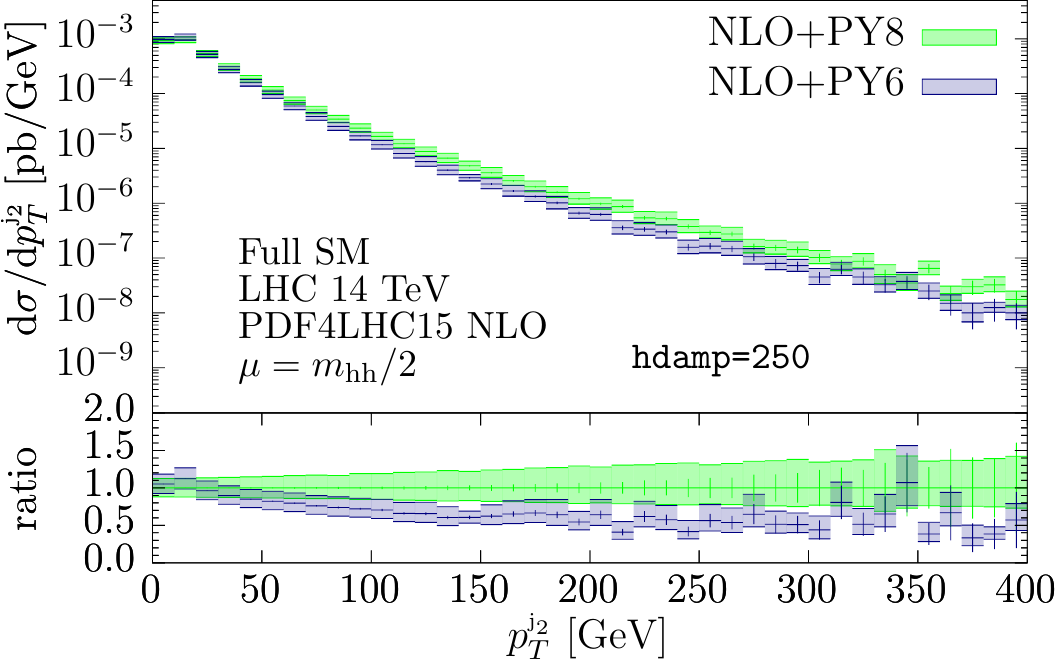}
\vspace*{1ex}
\caption{%
  Subleading jet transverse momentum distribution $\ptjj$ as obtained with \pythia and \pythiaold with \texttt{hdamp=$250$}. Left: basic HEFT
  approximation. Right: Full top quark mass dependence.\label{fig:SM_nlops_ggHH:ptj2pythia}}
\end{figure}

In Fig.~\ref{fig:SM_nlops_ggHH:pythia6} we compare the fixed order results to predictions obtained with
the \pythiaold shower and the \pythia shower both in the basic HEFT
approximation and in the full SM, for \texttt{hdamp=250} and
\texttt{hdamp=}$\infty$.
It is interesting to see that with \pythiaold, the enhancement in the
tail of the $\pthh$ distribution is much less pronounced, which means
that the hardness of the radiation recoiling against the Higgs-pair
system is higher in \pythia than in the older \pythiaold. Although
this behaviour is known~\cite{Corke:2010zj,Corke:2010yf}, 
it has a particularly large impact in the case
considered here\footnote{Very recent developments in
  \pythia~\cite{Cabouat:2017rzi} are likely to soften the observed behaviour in
the tail of the $\pthh$ distribution. We thank Stefan Prestel for
pointing this out to us.}. To exhibit very clearly the difference in the
hardness of the radiation generated by the two showers, 
in Fig.~\ref{fig:SM_nlops_ggHH:ptj2pythia} we compare the transverse momentum spectrum of the subleading jet $\ptjj$ in basic HEFT and in the full theory. This jet is of pure shower origin, since the matrix elements in our calculations can describe only one jet. Above $200$ GeV the difference between the two reaches a factor of $2$ for predictions in the full SM.
\begin{figure}
\centering
\includegraphics[width=0.48\textwidth]{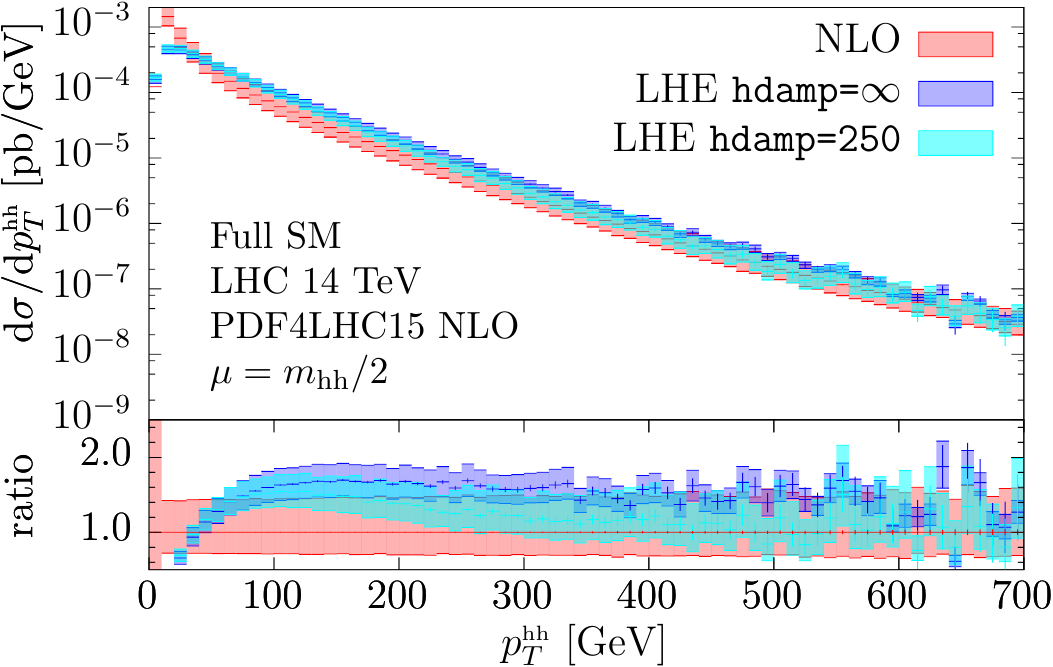}
\hfill
\includegraphics[width=0.48\textwidth]{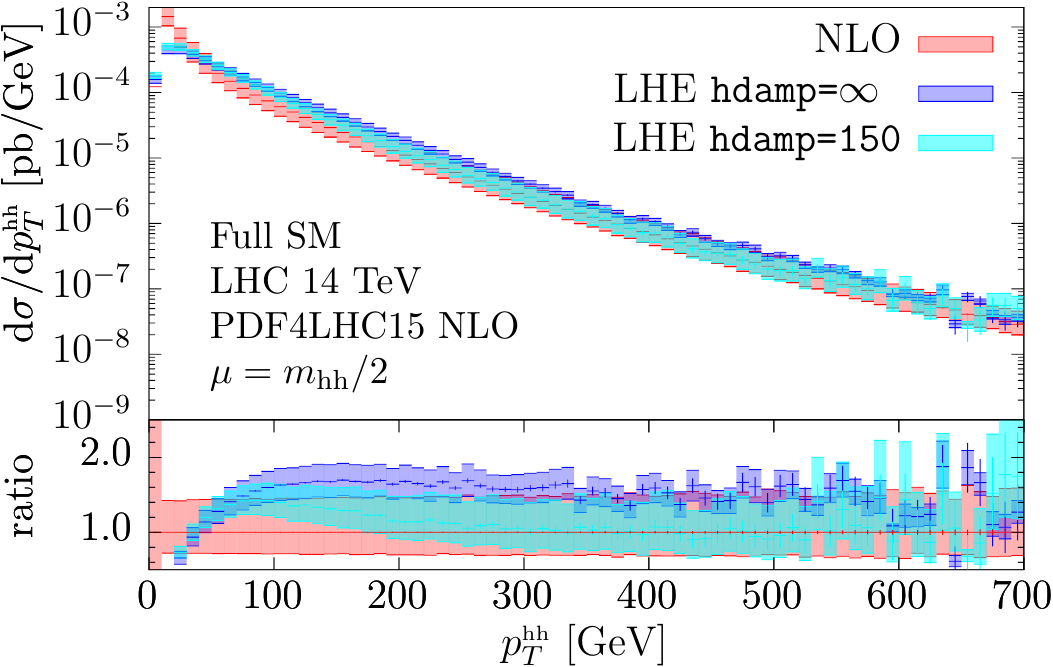}
\vspace*{1ex}
\caption{%
  Higgs boson pair transverse momentum distribution $\pthh$ at LHE
  level within the basic HEFT
  approximation: left column with \texttt{hdamp=250}, right column
  with \texttt{hdamp=150}.\label{fig:SM_nlops_ggHH:hdamp}}
\end{figure}
Even more evidence is given in Fig.~\ref{fig:SM_nlops_ggHH:hdamp}, which compares \powhegbox predictions at the Les Houches
event level (LHE) for two different values of $h_\text{damp}$. These plots corroborate the hypothesis that the
enhancement of the tail is not solely caused by the \powheg matching
procedure, since at the LHE level 
the tail of the distribution touches down on the fixed order result
for \texttt{hdamp=150}, \texttt{hdamp=250} and even for \texttt{hdamp=}$\infty$
in the shown $\pthh$ range. This is not the case after showering with
\pythia, as can be seen from Fig.~\ref{fig:SM_nlops_ggHH:pwg_mg5}.

\begin{figure}
\centering
\includegraphics[width=0.48\textwidth]{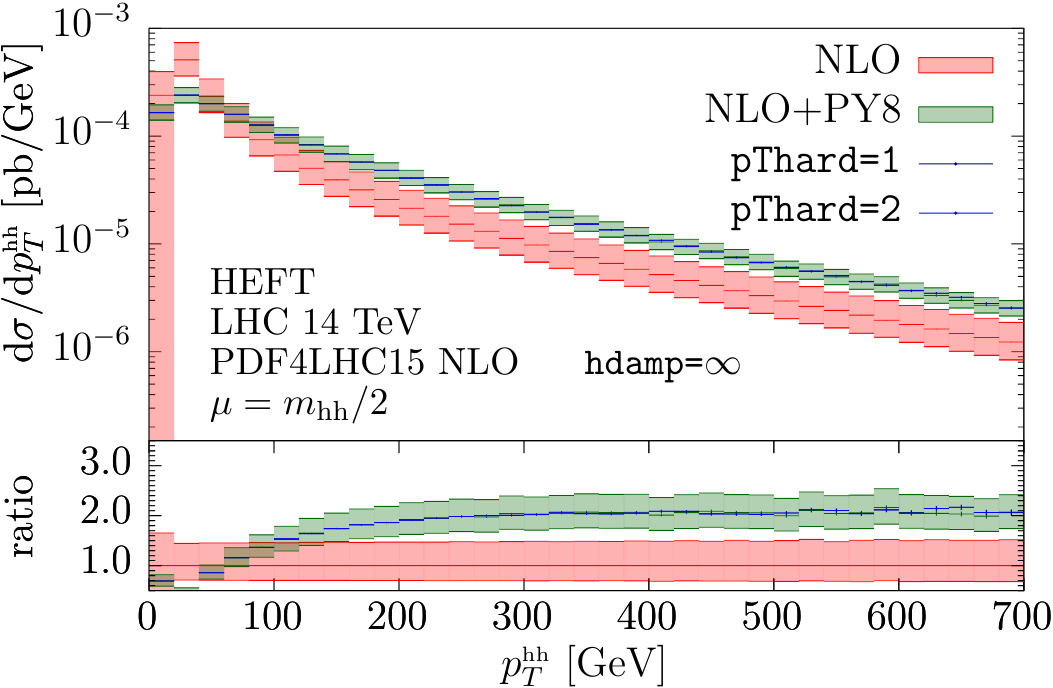}
\hfill
\includegraphics[width=0.48\textwidth]{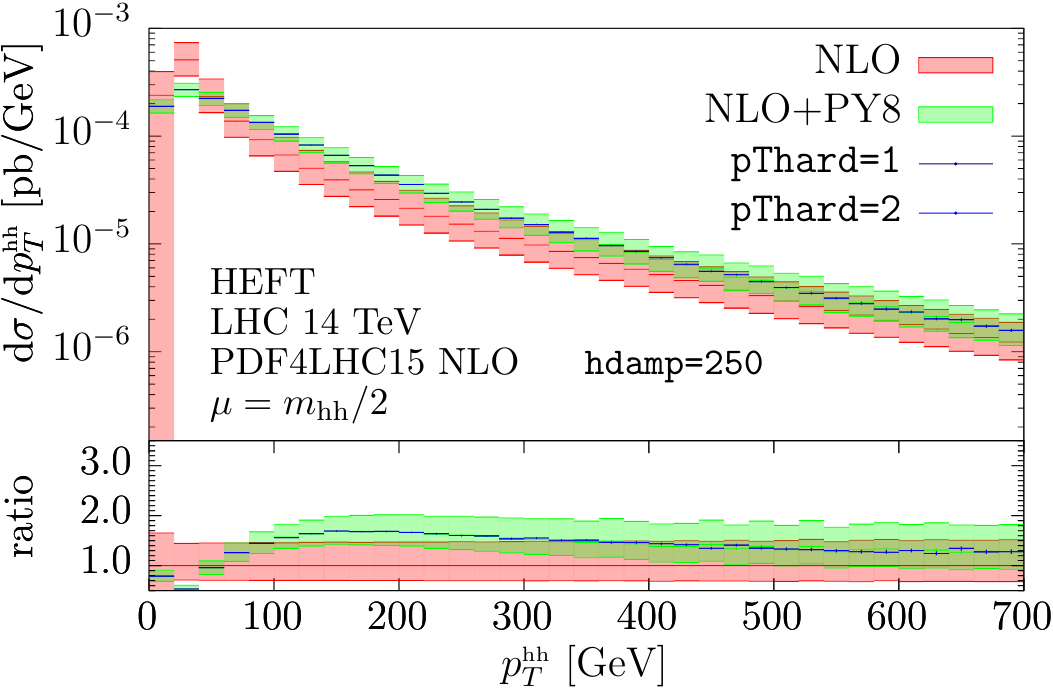}
\\
\includegraphics[width=0.48\textwidth]{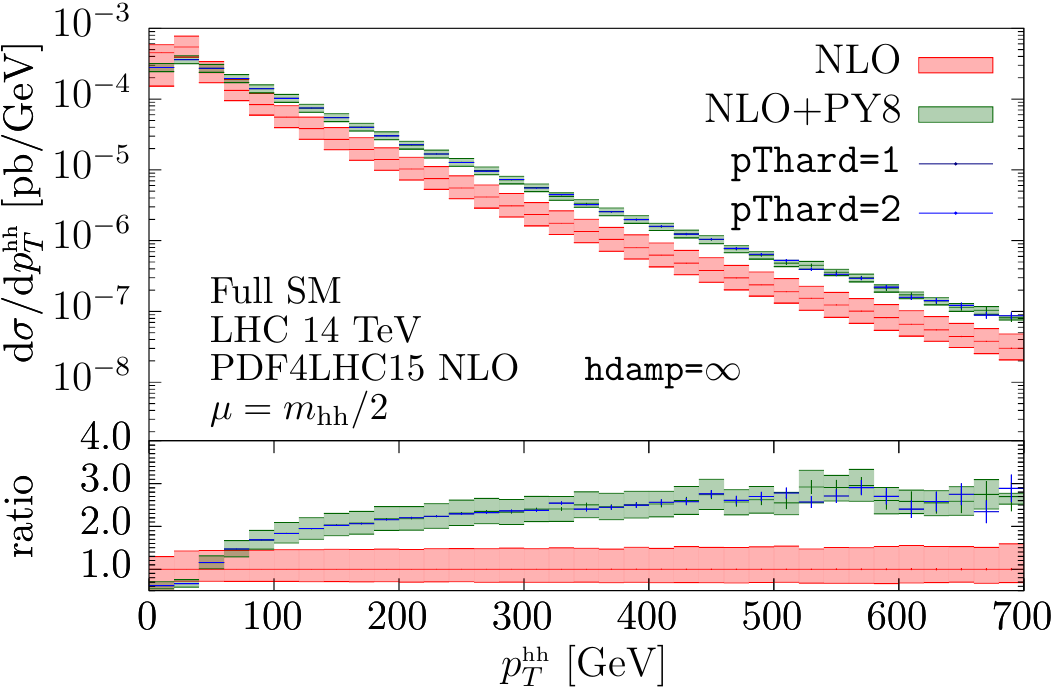}
\hfill
\includegraphics[width=0.48\textwidth]{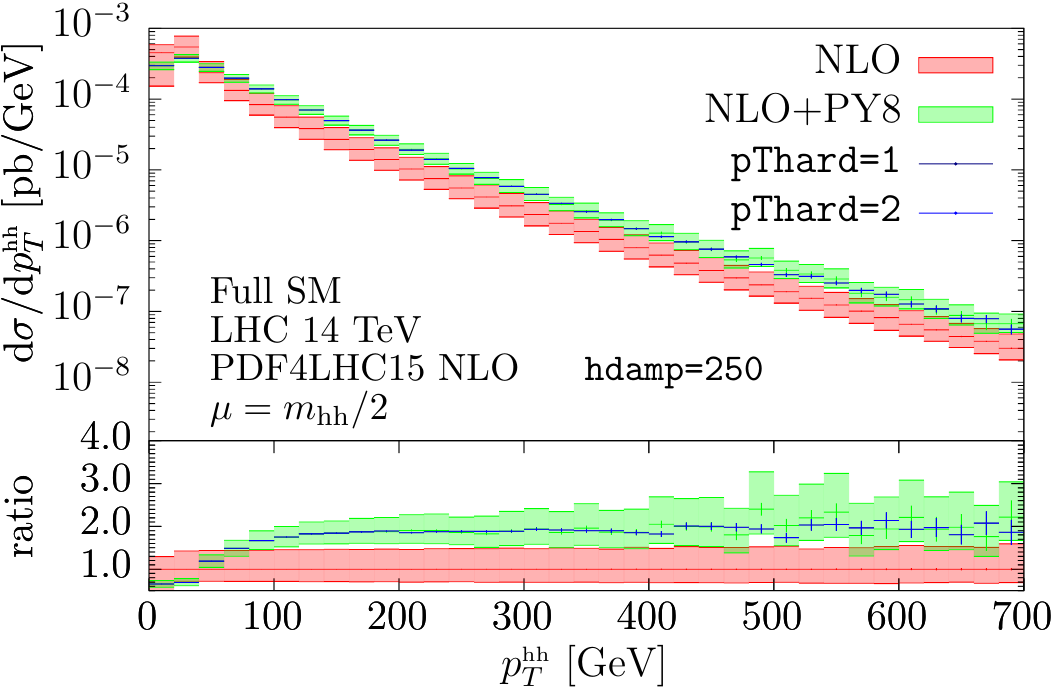}
\caption{%
  Higgs boson pair transverse momentum distribution $\pthh$ with
  variations of the parameter $p_T^{\rm{hard}}$. The default
  corresponds to  {\tt pThard=0}.\label{fig:SM_nlops_ggHH:pthard}}
\end{figure}
Therefore we investigated further the influence of the \pythia
settings by varying the parameters {\tt SCALUP}  and {\tt pThard}.
Both these parameters control the hardness criterion. By default the
\powhegbox transfers this value in the {\tt SCALUP} member of Les
Houches Events. When {\tt pThard=0} the default {\tt SCALUP} value is
used, leading to the results already discussed. In order to estimate
the uncertainty due this choice it is possible to change the value of
the {\tt pThard} flag to be {\tt pThard}=1 or {\tt pThard}=2.
In the former case the transverse momentum of the \powheg emission is
tested against all other incoming and outgoing partons, and the
minimal value is chosen, in the latter case the $\pt$ of all
final-state partons is tested against all other incoming and outgoing
partons, and the minimal value is selected. Since in Higgs boson pair
production there is only one coloured final state particle, and this
coincides with the \powheg emission, the two options lead to identical
results, as shown in Fig.~\ref{fig:SM_nlops_ggHH:pthard}. The two curves where {\tt
  pThard}$\neq0$ are exactly superimposed. Changing {\tt pThard} does
not lead to significant changes because, as already mentioned above,
in this process only one coloured final state parton is present, and
hence there is only a single choice for a $\pt$ reference value.

\begin{figure}
\centering
\includegraphics[width=0.48\textwidth]{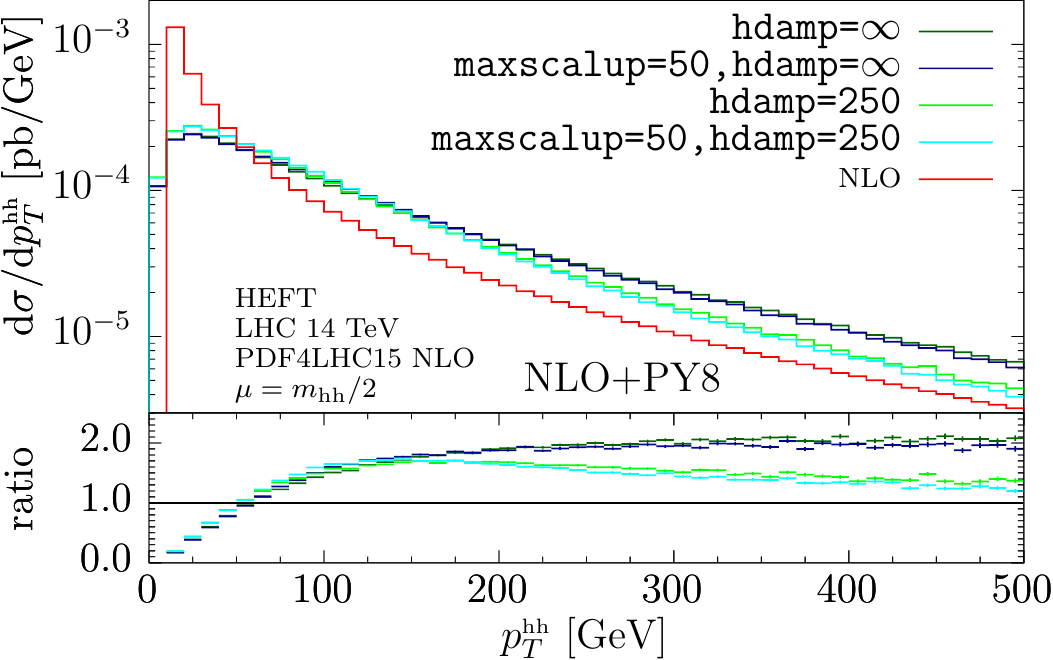}
\hfill
\includegraphics[width=0.48\textwidth]{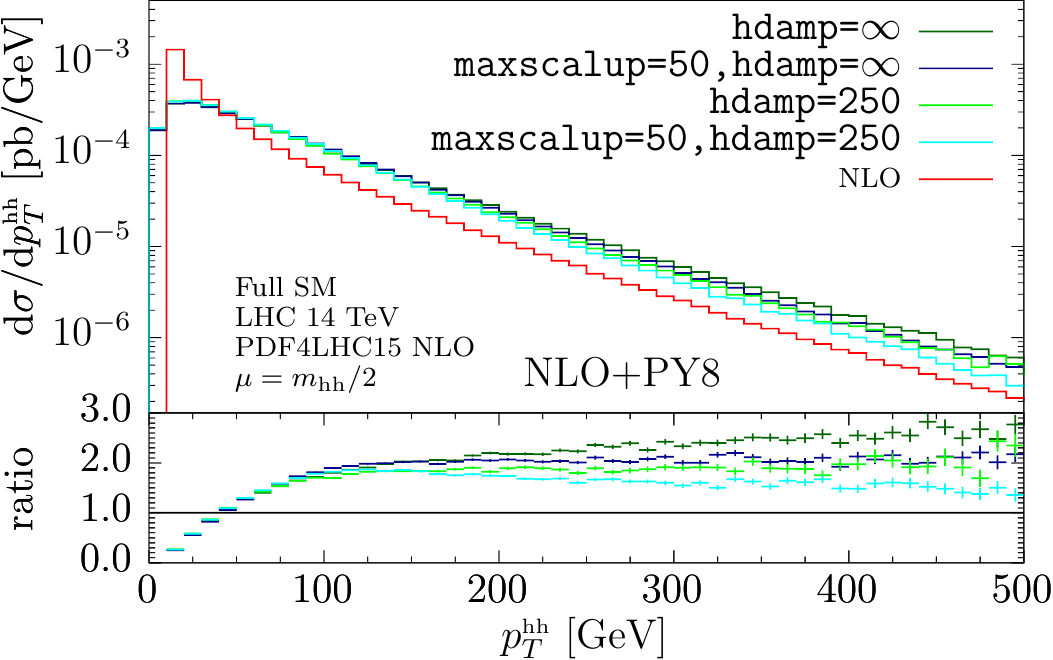}
\vspace*{1ex}
\caption{%
  Higgs boson pair transverse momentum distribution $\pthh$ with
  variations of the {\tt maxscalup} parameter in \pythia. Left: basic HEFT
  approximation, Right:
  Full top quark mass dependence.\label{fig:SM_nlops_ggHH:scalup}}
\end{figure}

\begin{figure}
\centering
\includegraphics[width=0.48\textwidth]{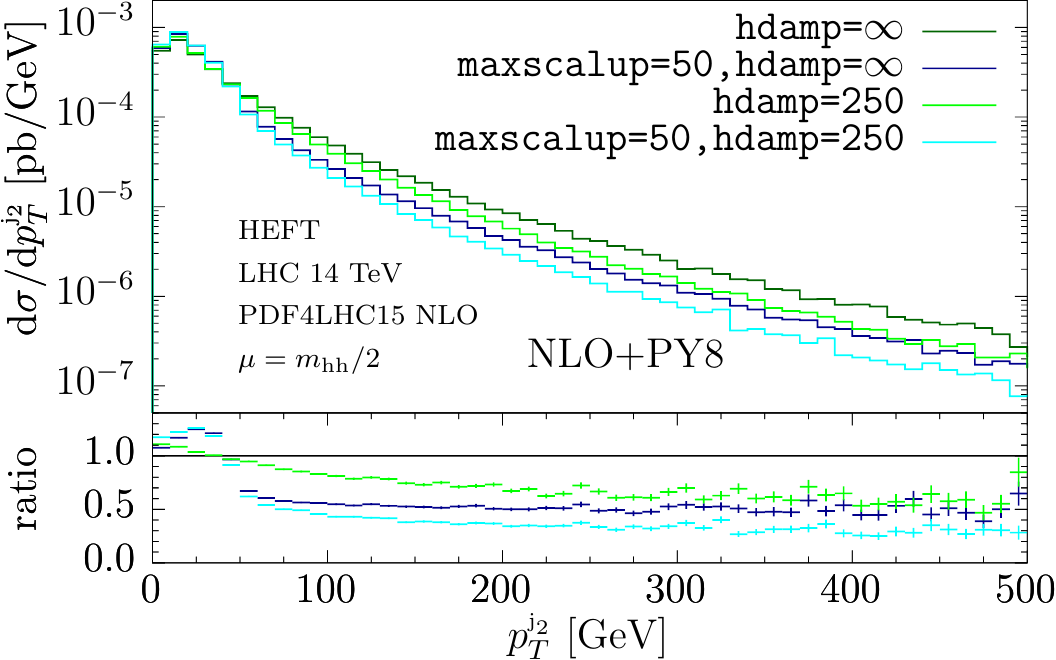}
\hfill
\includegraphics[width=0.48\textwidth]{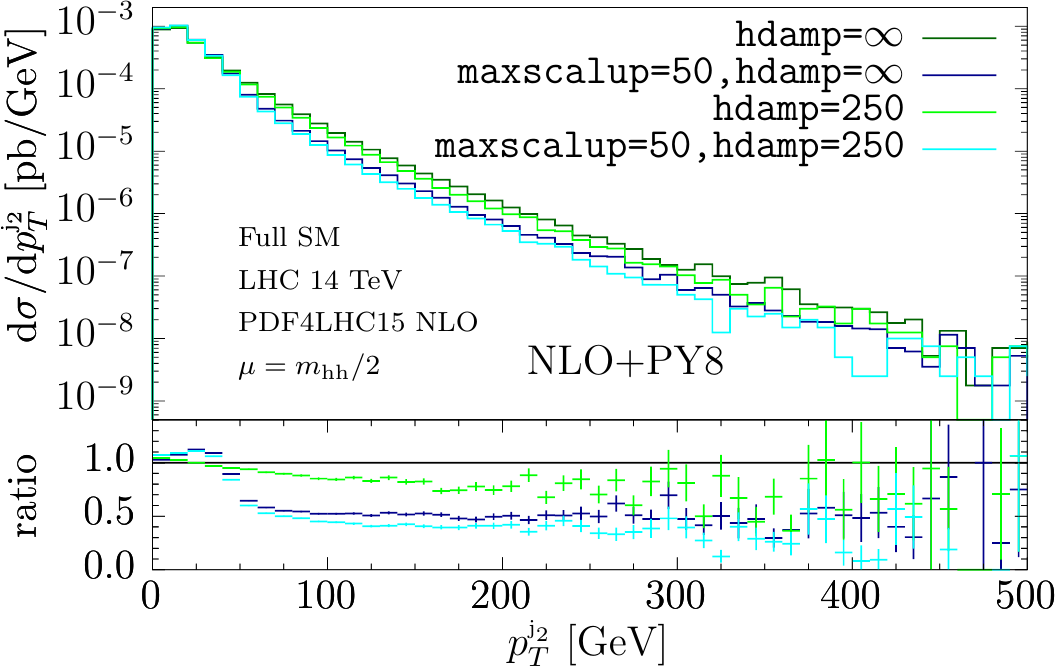}
\vspace*{1ex}
\caption{%
  Higgs boson pair transverse momentum distribution $\ptjj$ with
  variations of the {\tt maxscalup} parameter in \pythia. Left: HEFT
  approximation, Right:
  Full top quark mass dependence.\label{fig:SM_nlops_ggHH:scalup_j2}}
\end{figure}

Another possibility is instead to keep the
default value {\tt pThard}=0 and modify the {\tt SCALUP} value
within the \powhegbox using the {\tt changescalup} and {\tt maxscalup}
flags. In Fig.~\ref{fig:SM_nlops_ggHH:scalup} we show results where we set {\tt
  maxscalup=50}. This means that the shower starting scale in
\pythia is lowered to be less than 50~GeV, irrespective of the
hardness of the first \powheg emission. To verify this we can look at
the transverse momentum of the subleading jet, shown in
Fig.~\ref{fig:SM_nlops_ggHH:scalup_j2}, which is purely generated by shower
emissions. The plots clearly show a sharp decrease at
50~GeV. Comparing the corresponding transverse momentum spectra of
the Higgs-pair system we note a qualitative difference between basic
HEFT predictions, shown on the left in Fig.~\ref{fig:SM_nlops_ggHH:scalup}, and full
theory predictions, shown in the right plot in
Fig.~\ref{fig:SM_nlops_ggHH:scalup}. While in the effective theory predictions the
change in {\tt maxscalup} is barely visible for both values of {\tt
  hdamp}, in the full theory the curves start to deviate around 200~GeV. 
In both plots we also show fixed order NLO predictions, which we use as
a reference in the ratio plots displayed in the lower inset. Imposing a smaller
upper limit on the shower starting scale seems to cure the
problem partially, at least in the full theory predictions, bringing the NLO+PS
curves more in agreement with the fixed order NLO ones at large transverse
momenta.

\hfill

\subsection{SHERPA results}
Using an implementation in the \sherpa event generator it was shown in
\cite{Jones:2017giv} that the large parton shower effects in the tail
of the Higgs boson pair transverse momentum distribution can be traced
back to the formally subleading term \eqref{eq:SM_nlops_ggHH:ps-cancellation0}. Here
we will repeat the main arguments made in \cite{Jones:2017giv} and
show the corresponding results obtained from the \sherpa
implementation.

As stated above, in the full theory the term
\eqref{eq:SM_nlops_ggHH:ps-cancellation0} can become numerically large compared to
the fixed-order result that is given by \eqref{eq:SM_nlops_ggHH:ps-cancellation1}.
This is due to the fixed-order spectrum rapidly dropping above scales
of the order of the top quark mass. The impact of the uncanceled
parton shower contributions in \eqref{eq:SM_nlops_ggHH:ps-cancellation0} therefore
becomes large relative to the fixed-order prediction. In the HEFT
approximation the fixed-order spectrum is considerably harder and the
parton shower effects in the tail are thus of smaller significance.

\begin{figure}
  \centering
  \includegraphics[width=.48\textwidth]{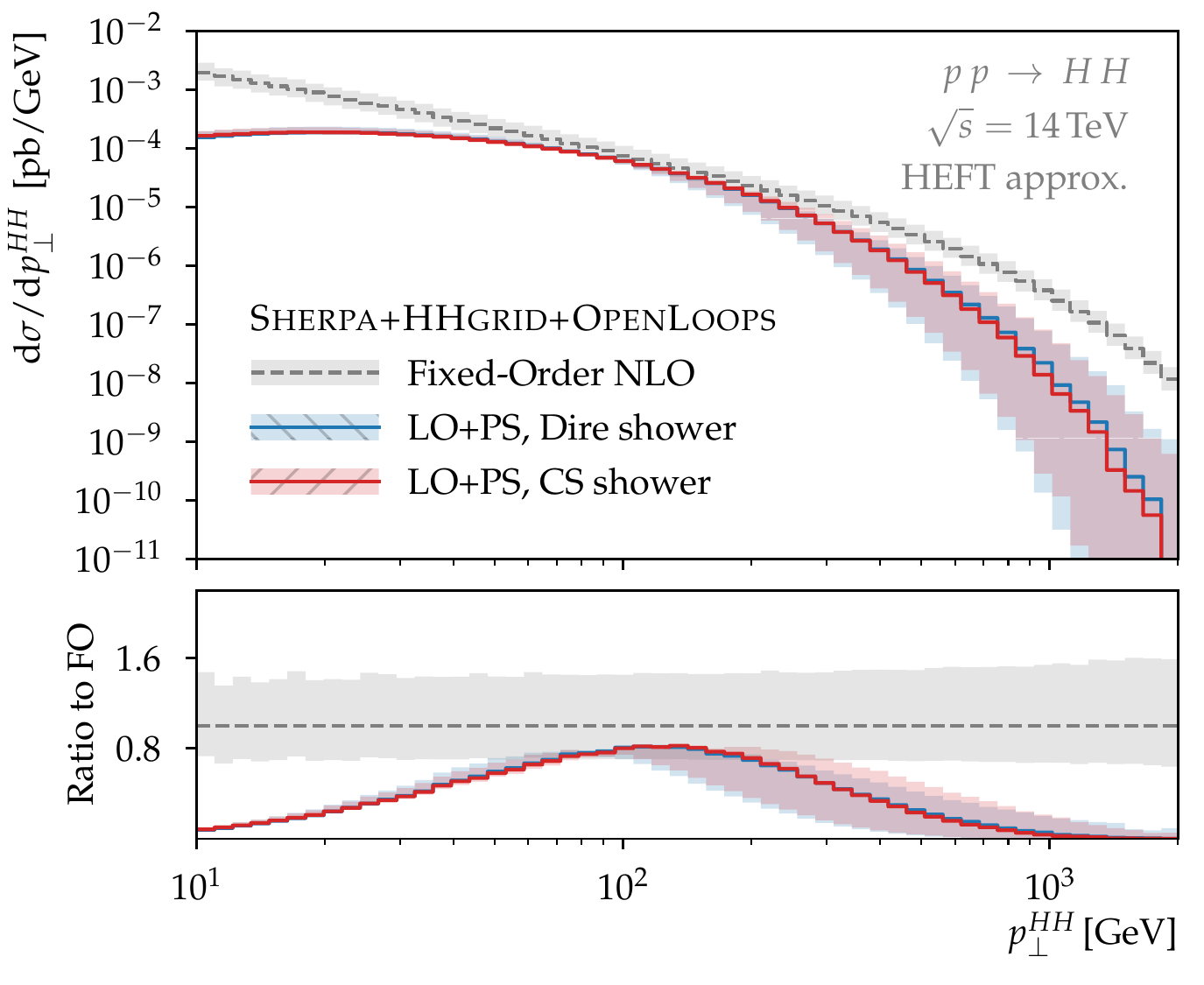}
  \hfill
  \includegraphics[width=.48\textwidth]{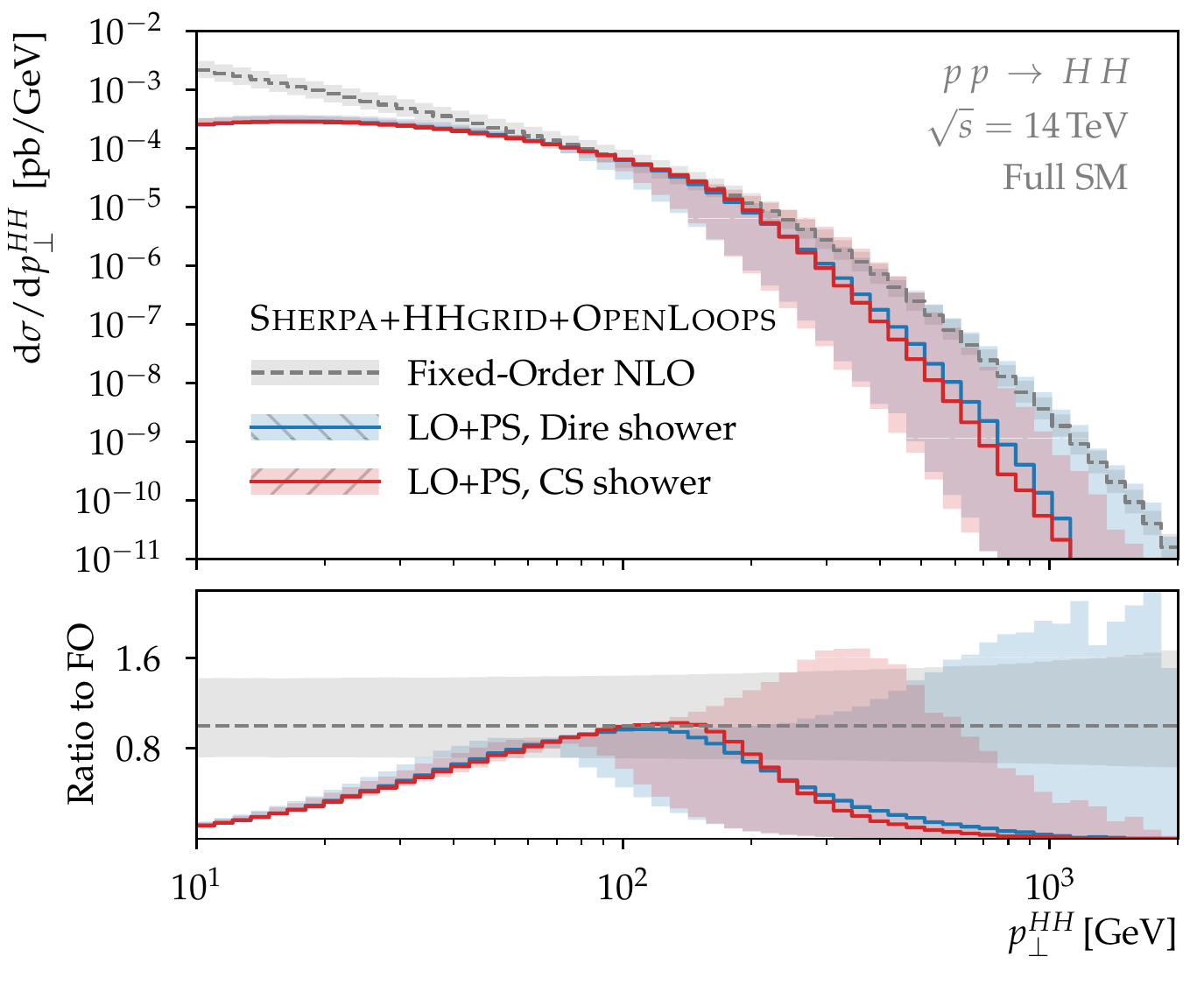}
  \caption{LO+PS results for the Higgs pair transverse momentum
    spectrum in the HEFT approximation (left panel) and in the full theory (right panel). The results are compared to the
    fixed-order prediction shown with an uncertainty band that
    corresponds to variations of $\mu_f$ and $\mu_r$ by factors of
    two. The bands around the LO+PS results are obtained through
    variations of the parton shower starting scale $\mu_\text{PS}$ by
    factors of two.}
  \label{fig:SM_nlops_ggHH:sherpa-lops}
\end{figure}

This effect is illustrated in Fig.~\ref{fig:SM_nlops_ggHH:sherpa-lops}, where the
fixed-order transverse momentum spectrum is compared to predictions
derived from LO+PS type simulations. The uncertainty bands on the
LO+PS simulation correspond to variations of the parton shower
starting scale $\mu_\text{PS}$ by factors of 2. The central scale
choice in case of the Dire shower \cite{Hoche:2015sya} is
$\mu_\text{PS}=\mhh/4$, whereas $\mu_\text{PS}=\mhh/2$ is set for the
CS shower \cite{Schumann:2007mg}. As expressed by the Heaviside theta
function in \eqref{eq:SM_nlops_ggHH:sevent_ps}, the parton shower starting scale
implements a phase space restriction that restricts parton shower
emissions to the phase space region where $t<\mu_\text{PS}^2$. If the
hard region of phase space is made accessible to the parton shower by
choosing a high parton shower starting scale (represented by the upper
edge of the uncertainty band), both parton showers overestimate the
fixed-order real-emission result in the tail of the distribution in
the full theory. This is not the case in the HEFT approximation. In
this approximation, the fixed-order spectrum is considerably harder
and the parton shower therefore underestimates the fixed-order
spectrum in the tail.

\begin{figure}
  \centering
  \includegraphics[width=.48\textwidth]{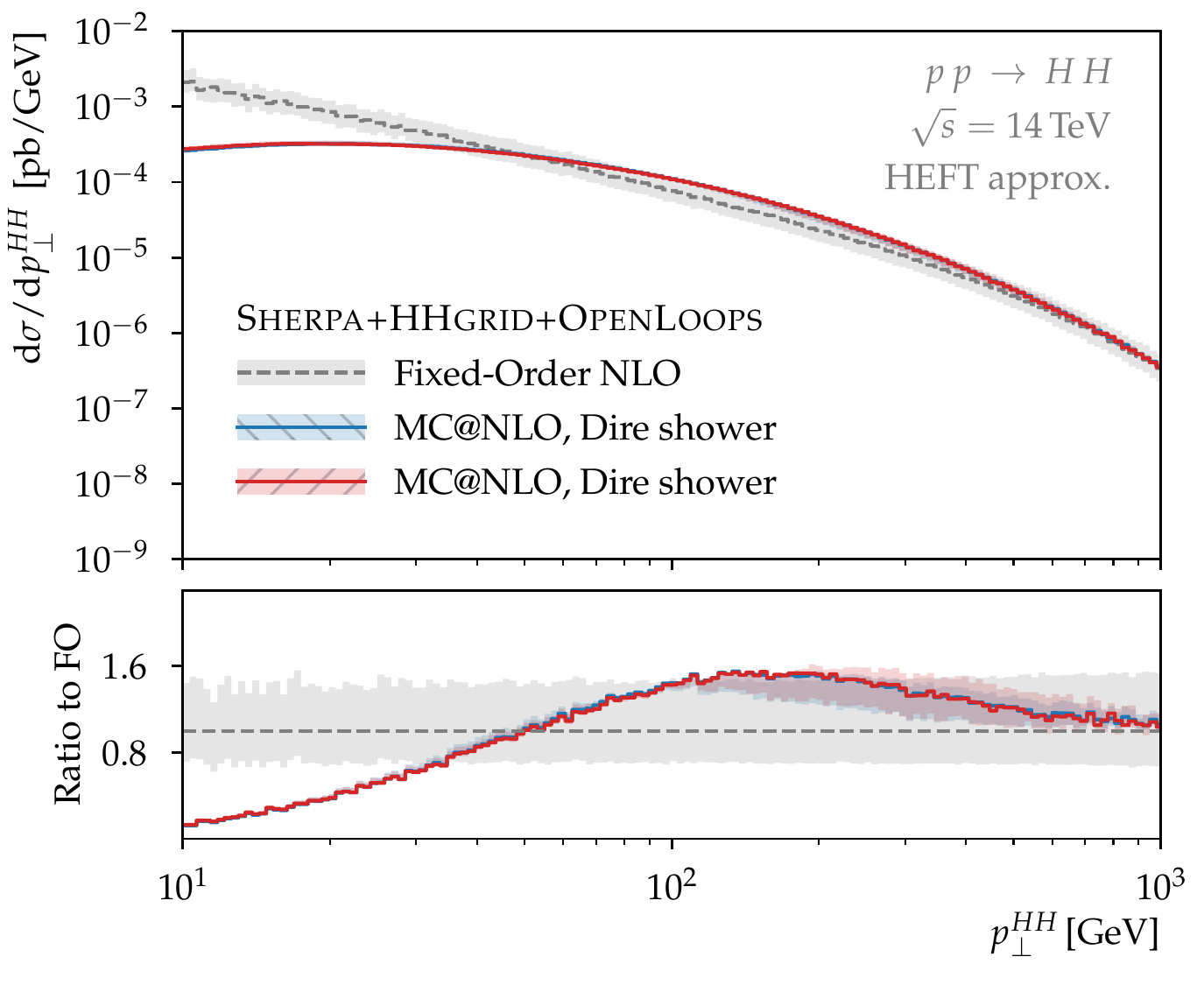}
  \hfill
  \includegraphics[width=.48\textwidth]{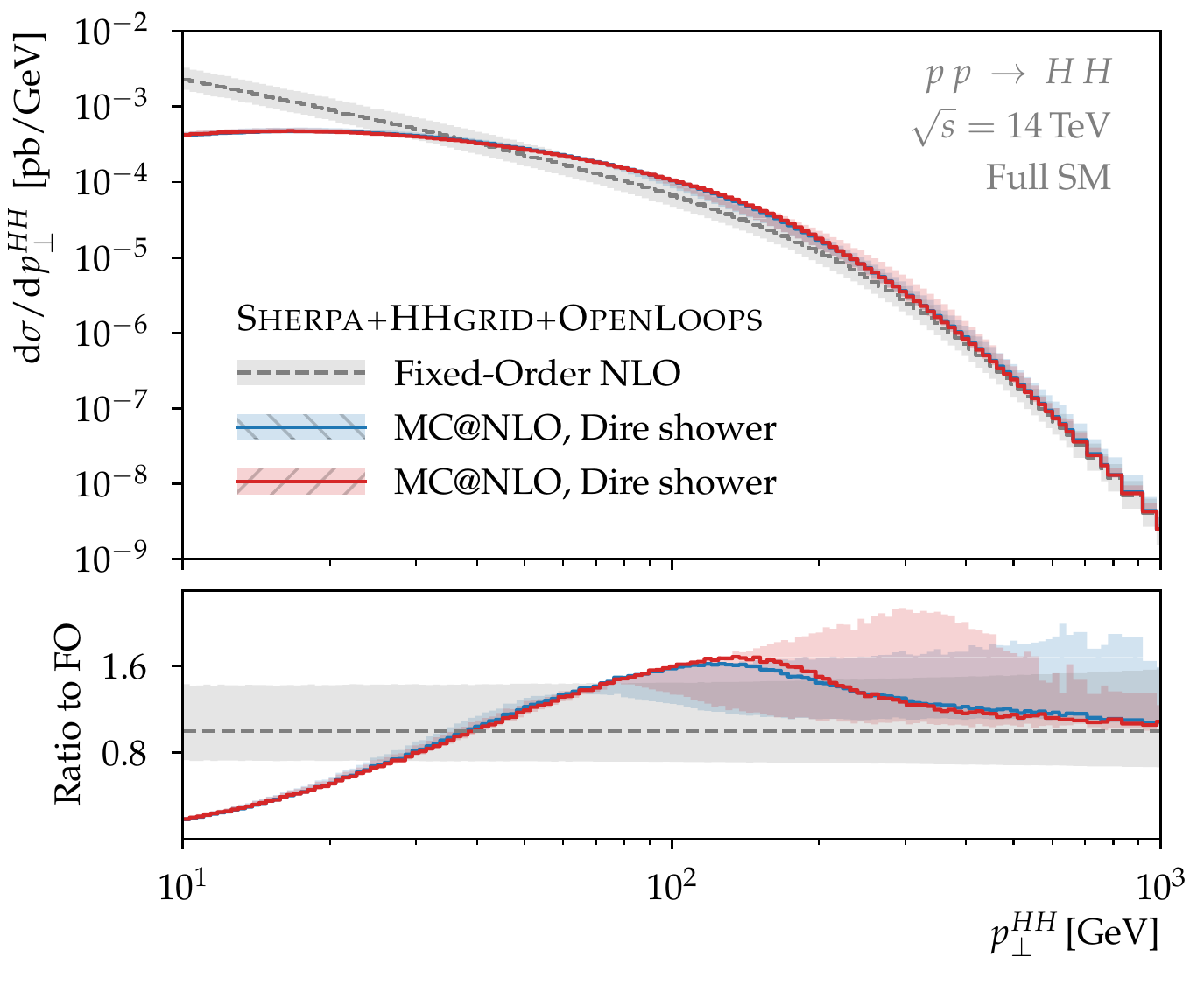}
  \caption{MC@NLO results for the Higgs pair transverse momentum
    spectrum in the HEFT approximation (left panel) and in the full theory (right panel). The results are compared to the
    fixed-order prediction shown with an uncertainty band that
    corresponds to variations of $\mu_f$ and $\mu_r$ by factors of
    two. The bands around the MC@NLO results are obtained through
    variations of the parton shower starting scale $\mu_\text{PS}$ by
    factors of two.}
  \label{fig:SM_nlops_ggHH:sherpa-mcnlo}
\end{figure}

In MC@NLO, the parton shower effects in the tail are removed at order
$\alpha_s$ relative to the born through the modified subtraction
prescription, i.e. by the term
$-D(\phi_R)\Theta(\mu^2_\text{PS}-t(\phi_R))$ in Eq.~\eqref{eq:SM_nlops_ggHH:hevent}.
The higher-order remainder \eqref{eq:SM_nlops_ggHH:ps-cancellation0} that is left
over is, however, relatively large due to the large numerical
difference between $\bar B$ and $B$. This is shown in
Fig.~\ref{fig:SM_nlops_ggHH:sherpa-mcnlo}. As indicated by the upper uncertainty
bands, the left-over parton shower effects in the tail can be of order
one if a high shower starting scale is chosen in the full theory. For
more moderate choices, however, the fixed-order result is reproduced
at large transverse momenta. In the HEFT approximation, parton shower
effects are much smaller. This is to be expected, since, as shown in
Fig.~\ref{fig:SM_nlops_ggHH:sherpa-lops}, the fixed-order spectrum is extremely
hard. Even though $\bar B- B$ is numerically large in the HEFT as
well, any parton shower effects thus appear small when compared to the
fixed-order spectrum.

\subsection{Conclusions}
We have studied parton shower effects on the Higgs boson
pair transverse momentum spectrum based on NLO calculations matched to
several parton shower Monte Carlo programs. 
Since these effects were found to be surprisingly large, 
we investigated in more detail the cause of the
enhancement in the tail of the $\pthh$ spectrum,
comparing results in basic HEFT and in the full SM,
obtained with three different matching schemes and various different
shower settings.

We identified the following reasons for the growth of the $\pthh$ tail in
NLO+PS matched predictions. 
As can be seen from Eq.~\eqref{eq:SM_nlops_ggHH:ps-cancellation0}, 
there are three factors which play a role: large K-factor
($B-\bar{B}$), large splitting kernel, and the shower starting scale
in the case of MC@NLO matching.
The large NLO K-factor is certainly given for the process under
consideration, further the contribution from the splitting kernel also
seems to be large relative to the hard real radiation.

Depending on the matching scheme, the largeness of the (formally
subleading) terms in Eq.~\eqref{eq:SM_nlops_ggHH:ps-cancellation0} can be controlled
by tuning the shower starting scale or by separating the hard
real radiation phase space from the Sudakov region.
On the other hand, we observed important differences between various
parton shower programs while using the same matching procedure. For example,
the radiation pattern produced by the \pythia and \pythiaold showers
seems to be quite different.
These differences are harder to control without an in-depth knowledge of the
parton shower programs. They are however a manifestation of the theoretical
uncertainty related to the showering, which should be taken into
account when estimating the total theoretical
uncertainty related to an NLO+PS prediction.

\subsection*{Acknowledgements}
We are grateful to Stefan Prestel for useful comments about {\tt Pythia}.
We also thank the organizers of the ``LH2017 Phyisics at TeV Colliders''
workshop for creating a productive and pleasant environment.
This research was
supported in part by the Research Executive Agency (REA) of the
European Union under the Grant Agreement PITN-GA2012316704
(HiggsTools) and by the U.S. Department of Energy
under contract number DE-AC02-76SF00515.




\let\powheg\undefined
\let\powhegbox\undefined
\let\madgraph\undefined
\let\mcnlo\undefined
\let\sherpa\undefined
\let\pythia\undefined
\let\pythiaold\undefined

\let\mhh\undefined
\let\pthh\undefined
\let\pt\undefined
\let\ptj\undefined
\let\ptjj\undefined
\let\pth\undefined
\let\pthl\undefined
\let\pths\undefined
\let\dphihh\undefined
\let\drhh\undefined








\section{Treatment of theory uncertainties for Simplified Template Cross Sections~\protect\footnote{
    N. Berger,
    F. J. Tackmann,
    K. Tackmann}{}}
\label{sec:Higgs_STXS}

\subsection{Introduction}
\label{sec:Higgs_STXS:intro}

Simplified Template Cross Sections (STXS) have been adopted
as an evolution of the signal strength measurements
performed during Run 1 of the LHC. Their goal is reduce the theoretical uncertainties that
are directly folded into
the measurements and provide more finely-grained measurements, while at the same time
allowing and benefiting from the combination of measurements in many decay channels.
For a detailed discussion, see Sec.~III.3 of Ref.~\cite{Badger:2016bpw} and
Sec. III.2 of Ref.~\cite{deFlorian:2016spz}.

The primary features to achieve these goals are as follows.
First, they are cross sections (instead of signal strengths) defined in mutually exclusive
regions of phase space. The bin definitions are abstracted and simplified compared
to the exact fiducial volumes of specific analyses in different Higgs decay channels.
The measurements are unfolded to these signal regions, the STXS ``bins'',
which are common for all analyses, allowing for a subsequent global combination
of different decay channels as well as measurements from ATLAS and CMS.
In particular, the STXS are defined inclusively in the Higgs boson decay,
to naturally combine the various decay modes.
While the bin definitions are simplified to allow for the combination of
different decay channels (and also for ease of use), they nevertheless try to be as close
as
possible to the typical experimental selections to avoid any unnecessary
extrapolations.  The goal is to allow the use of advanced analysis
techniques such as event categorization or multivariate techniques while
still keeping the unfolding uncertainties small.

Second, the STXS bins are defined for specific production modes,
with the SM production processes serving  as kinematic templates.
This separation into production modes is an essential aspect to reduce their model
dependence, i.e., to eliminate the dependence of the measurements on the relative
fractions of the production modes in the SM.

The number of separately measured bins can evolve with time, such that the
measurements can become more fine-grained as the size of the available dataset increases.
Several ``stages'' have been defined, and can be further extended.
The stage 0 bin definitions essentially correspond to the production mode measurements of Run 1,
while stage 1 has been defined as the target for the full Run 2 measurements.
At intermediate stages individual bins can be merged and only their sum measured
according to the sensitivity of each analysis and decay channel. While the full
stage 1 granularity should become possible in the combination of all decay
channels by the end of Run 2, individual decay channels will require
some bins to be merged for the foreseeable future.

The measured STXS together with the partial decay widths are meant to serve as
input for subsequent interpretations.
Such interpretations could for example be the determination of signal strengths
or coupling scale factors $\kappa$ (providing compatibility with earlier
results), EFT coefficients, tests of specific BSM models, and so forth.
The treatment of theory uncertainties when reporting experimental measurements 
will be discussed in Sec.~\ref{sec:Higgs_STXS:reportingresults}.

\subsection{Overview of Theory Uncertainties}

There are two places where theory uncertainties enter, which are important to
distinguish. First, there are residual theory uncertainties in the measurements
of the STXS bins, resulting from the assumed SM predictions of the kinematic
distribution within each bin that enter in the unfolding of the experimental
event categories to the STXS bins. Second, there are the theory uncertainties on
the SM and beyond predictions of the STXS bins which are required in any
subsequent interpretation.
One of the primary purposes of the STXS bin definitions is to move the dominant
theory uncertainties to the second interpretation step, since they are much
easier to deal with there. Here, were are primarily concerned with the theory
uncertainties at the interpretation level.

To combine the information from all measured bins in the interpretation, the
theoretical predictions and their uncertainties must be evaluated separately for
each bin, since different bins will in general contribute with different relative
weights to the final result.
In this context, the correlations of the theory uncertainties for different bins
must be taken into account. This is particularly important whenever a certain
binning cut induces an important additional source of perturbative uncertainties
that affects each bin but should cancel in their sum. (A well-known example is
the case of jet binning, where it is essential to treat the uncertainties
induced by the jet-binning cut as anticorrelated between the jet
bins~\cite{Berger:2010xi, Stewart:2011cf}.)

Hence, to enable such interpretations in a flexible way, it is important to have
a common treatment of theory uncertainties across all bins and production modes.
The goal is to provide a general parameterization of theory uncertainties, which
is physically motivated, in particular allowing to take into account the
possible theory correlations between different bins, and is flexible enough to
accommodate different (types of) theory predictions. In this way, the underlying
uncertainty parameterization can be implemented once and used to easily test
different predictions or include future theory improvements. For the
gluon-fusion production process, such a parameterization has already been worked
out~\cite{WG1indico}. 
Here we will discuss similar parameterizations for the vector-boson fusion
(VBF) and associated (VH) production processes.

In general, to properly treat the theoretical uncertainties one should try to
identify and distinguish different sources of uncertainties and take into
account the correlation implied by each common uncertainty source. For the
practical implementation, an essential requirement is to parametrize the
uncertainties such that they can be implemented in terms of fully correlated and
uncorrelated nuisance parameters. This is facilitated by parameterizing the
uncertainties in terms of uncertainty sources, each of which is considered
common and hence fully correlated among all bins, while the different sources
are considered mutually independent and hence uncorrelated. The uncertainties
can then be implemented via a single nuisance parameter for each source.
For this purpose, we follow the generic parameterization strategy for
uncertainties in kinematic bins discussed in Sec.~I.4.2.a of
Ref.~\cite{deFlorian:2016spz}, and which is briefly reviewed in
Sec.~\ref{sec:Higgs_STXS:thunc_in_kinematic_bins} below.

It should be stressed that the theory uncertainty parameterization discussed
here is not meant and not necessarily suitable to assess the residual theory
uncertainties in the measurement step. However, whenever two bins are merged in
the measurement, this in general reintroduces the dependence on the SM
prediction for the (ratio of the) cross sections of the two merged bins with the
resulting theory uncertainty. In this context, a common parameterization of the
theory uncertainties between measurement and interpretation is essential to
allow for a consistent treatment in the combination of measurements with
differently merged bins as well as between measurements and interpretation.

Finally, we note that here we are only concerned with the perturbative theory
uncertainties due to missing higher-order corrections. Uncertainties due to the
imprecise knowledge of input parameters, like parton distribution functions or
quark masses, are not considered here, since their correct treatment including
correlations is straightforward.

\subsection{Theory Uncertainties in Kinematic Bins}
\label{sec:Higgs_STXS:thunc_in_kinematic_bins}

In this subsection, we review a general parameterization strategy for
uncertainties in kinematic bins. For a more detailed discussion we refer to
Sec.~I.4.2.a of Ref.~\cite{deFlorian:2016spz}.
First, consider a single bin boundary $a/b$ that splits the cross section $\sigma_{ab} = \sigma_a + \sigma_b$ into two bins with cross sections $\sigma_a$ and $\sigma_b$.
In general, the binning cut has a nontrivial influence on the perturbative structure of
$\sigma_a$ and $\sigma_b$, for example, it can introduce sensitivity to an additional energy scale or separate different jet multiplicities.
This implies that the binning cut corresponds to an additional and a priori nonnegligible source of uncertainty that is not present in $\sigma_{ab}$.

The uncertainty matrix for $\{\sigma_a, \sigma_b\}$ can be parametrized in terms
of fully correlated and fully anticorrelated components as
\begin{equation} \label{eq:Higgs_STXS:Cgeneral}
C(\{\sigma_a, \sigma_b\}) =\!
\begin{pmatrix}
(\Delta^{\rm y}_a)^2 &  \Delta^{\rm y}_a\,\Delta^{\rm y}_b  \\
\Delta^{\rm y}_a\,\Delta^{\rm y}_b & (\Delta^{\rm y}_b)^2
\end{pmatrix}
\!+
\begin{pmatrix}
 \Delta_{a/b}^2 &  - \Delta_{a/b}^2 \\
-\Delta_{a/b}^2 & \Delta_{a/b}^2
\end{pmatrix}
\!.\end{equation}
This is a general parameterization of a $2\times2$ symmetric matrix, and hence the uncertainties obtained with any prescription can always be written in this form, provided sufficient information or assumptions on the correlations are available.
This parameterization is convenient for two reasons:
First, the separation into independent components that are $\pm 100\%$ correlated
between the different bins allows for a straightforward implementation in terms of independent nuisance parameters for each component. That is, we can use two nuisance parameters $\theta^{\rm y}$ and $\theta_\mathrm{cut}$,
whose absolute uncertainty amplitudes for $\{\sigma_{ab}, \sigma_a, \sigma_b\}$ are
\begin{equation} \label{eq:Higgs_STXS:simple}
\theta^{\rm y}: \quad \{ \Delta^{\rm y}_{ab},\, \Delta^{\rm y}_a,\, \Delta^{\rm y}_b \}
\qquad\qquad
\theta_{a/b}: \quad \{ 0,\, \Delta_{a/b}, -\Delta_{a/b} \}
\,,\end{equation}
where $\Delta_{ab}^{\rm y} = \Delta^{\rm y}_a + \Delta^{\rm y}_b$.
Second, this parameterization admits a simple physical interpretation: The first correlated component with superscript ``y'' can be interpreted as an overall yield uncertainty of a common source for all bins. The second anticorrelated component can be interpreted as a migration uncertainty between the two bins $\sigma_a$ and $\sigma_b$, which is introduced by the $a/b$ binning cut and must drop out in their sum. Having such a physical interpretation
is very useful to identify and estimate each component in the actual theory calculation.
For example, for jet-binning this interpretation has been explicitly exploited to construct uncertainty estimates for both fixed-order and resummed predictions~\cite{Berger:2010xi, Stewart:2011cf, Stewart:2013faa}.

Consider now the general case of multiple bins, where
each bin can have more than one boundary and each boundary can be shared by different bins.
To make this tractable in a systematic fashion, the idea is to use Eq.~\eqref{eq:Higgs_STXS:simple}
for any given single bin boundary $a/b$ with all additional subdivisions removed. Then
given the bin boundary $a/b$ separating the cross section as $\sigma_{ab} = \sigma_a + \sigma_b$, we can include additional binning cuts that further subdivide $\sigma_a$ and $\sigma_b$ as $\sigma_a = \sum_i \sigma_{ai}$ and $\sigma_b = \sum_j \sigma_{bj}$.
Since we interpret the $a/b$ boundary as a common uncertainty source, we can consider it as fully correlated among each set of sub-bins and implement it via a single nuisance parameter $\theta_{a/b}$. The corresponding uncertainty amplitudes for all the bins are now given as
\begin{equation}
\theta_{a/b}: \quad \Delta_{a/b} \times \bigl\{\{ x_{ai} \}, - \{ x_{bj} \} \bigr\}
\qquad\text{with}\qquad
\sum_i x_{ai} = \sum_j x_{bj} = 1
\,,\end{equation}
where the parameters $x_{ai}$ and $x_{bj}$ specify how the absolute uncertainty
$\Delta_{a/b}$ gets distributed among the sub-bins. That is, the impact on
$\sigma_{ai}$ is $x_{ai} \Delta_{a/b}$ and on $\sigma_{bj}$ it is $-x_{bj}
\Delta_{a/b}$. The $x_{ai}$ and $x_{bj}$ parameters are specific to each
nuisance parameter $\theta_{a/b}$.

In this way, we can consider each bin boundary as a potential source of an uncertainty with an associated nuisance parameter.
Of course, in practice with sufficiently complicated bin boundaries, we have to apply some theoretical judgment in how to choose the relevant independent binning cuts.
In addition, we can have one or more overall yield uncertainties correlated among all bins.

Mathematically speaking, this way of parameterizing the uncertainties has some redundancy.
This is desired and makes it flexible enough to accommodate different scenarios while maintaining the simple physical interpretation in terms of the underlying uncertainty sources.
On the other hand, for the case of $N$ bins, it is also desirable to have at least $N$ independent nuisance parameters, such that in principle any generic $N\times N$ uncertainty matrix can be reexpressed in this form.

\subsection{Parameterization of Theory Uncertainties}

To parametrize the perturbative uncertainties, we consider both QCD and EW sources.
We give the physical interpretation of each source that should be followed when
estimating the uncertainty in practice.
While providing explicit uncertainty estimates is beyond the scope here, we will
try to indicate in which cases estimates from fixed-order calculations are likely to be appropriate and where resummation (or parton-shower) uncertainties are expected to be
relevant. We also point out where additional studies are needed to decide on whether certain
sources should be treated as correlated or uncorrelated, or where additional sources
might be needed in the future.

We hope that the parameterizations below will provide a sufficiently flexible and
somewhat future-proof baseline for the experimental measurements and interpretations
of STXS bins at stage 1.

\subsubsection{VBF Production}

\begin{figure}
\centering
\includegraphics[scale=0.5]{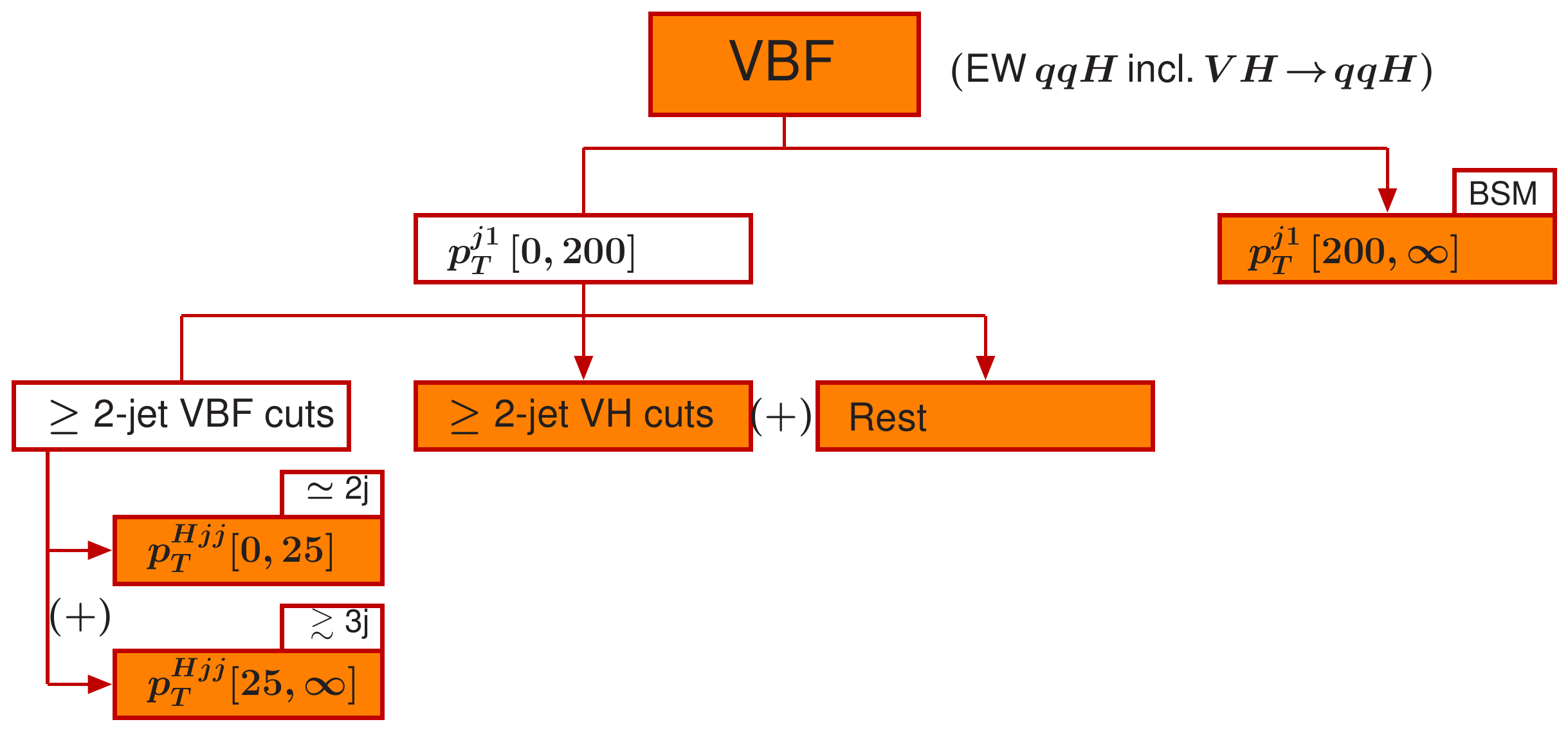}
\caption{Stage 1 bins for VBF production.}
\label{fig:Higgs_STXS:VBFstage1}
\end{figure}

The VBF template process is defined as electroweak $qqH$ production, which includes
the usual VBF topology but also $q\bar q\to VH$ production with hadronic $V\to q\bar q$ decays.
Since they lead to the same final state, they essentially represent the $t$-channel
and $s$-channel contributions to the same physical process,
and can only be distinguished by enriching one or the other type of contribution
via kinematic cuts.

The full stage 1 bins for vector-boson fusion are depicted in Fig.~\ref{fig:Higgs_STXS:VBFstage1}.
They include a high-$p_T$ bin and low-$p_T$ bin, defined by a cut on the leading
jet to be above or below $p_T^{j1} = 200\,\mathrm{GeV}$.
The low-$p_T$ bin is further split into three bins targeting the typical VBF and VH topologies and a bin for the rest.
The VBF topology cuts are $m_{jj} > 400\,\mathrm{GeV}$ and $\Delta\eta_{jj} > 2.8$
with both signal jets required to have $p_T^j > 30\,\mathrm{GeV}$.
The VH topology cuts are $60\,\mathrm{GeV} < m_{jj} < 120\,\mathrm{GeV}$.
For our purposes we can essentially consider these as splitting the $m_{jj}$ spectrum into three regions. (Technically, the Rest bin also includes events where one or zero jets
pass the jet selection cuts.)
Finally, the VBF topology bin is split into an exclusive 2-jet-like and inclusive 3-jet-like
bin using  a cut on $p_T^{Hjj}$ to be above or below $25\,\mathrm{GeV}$.

\paragraph{QCD Uncertainties} For the QCD uncertainties, we identify the following sources/nuisance parameters:
\begin{itemize}
\item {\boldmath $\theta^{\rm y}_{\rm VBF}$}:
The primary yield uncertainty for the high $m_{jj} > 400\,\mathrm{GeV}$ region.

\item {\boldmath $\theta^{\rm y}_{\rm Rest}$}:
The primary yield uncertainty for the low $m_{jj} < 400\,\mathrm{GeV}$ region.

\item {\boldmath $\theta^{\rm y}_{\rm VH}$}:
The overall yield uncertainty for the underlying VH production process.

\item {\boldmath $\theta_{200}$}:
The migration uncertainty related to the $p_T^{j1} = 200\,\mathrm{GeV}$ bin boundary.

\item {\boldmath $\theta_{25}$}:
The migration uncertainty related to the $p_T^{Hjj} = 25\,\mathrm{GeV}$ bin boundary.
\end{itemize}

For the VBF process, an important question that should be studied is how the uncertainties are
correlated between the low and high $m_{jj}$ regions, i.e., between the ``VBF cuts''
and the ``Rest'' bin. Here, we took the approach of considering them as primarily
uncorrelated with each having their own nuisance parameter
$\theta_{\rm Rest}$ and $\theta_{\rm VBF}$.
In principle, correlations between the bins can be taken into account by
allowing these to also impact the corresponding other regions.
An alternative would be to use an overall yield uncertainty and a migration
uncertainty across the $m_{jj} = 400\,\mathrm{GeV}$ boundary. However, since the
bins differ by more cuts, this seems less appropriate in this case.

For completeness, we have included the $\theta_{\rm VH}^{\rm y}$ nuisance
parameter here. It is the same parameter entering for the
VH process below, such that the uncertainties for the hadronic and leptonic VH
process are correlated. It will primarily impact the VH cuts bin, providing the dominant
uncertainty there. In principle, it could also have a small impact on the other bins as well.
In general, all the VH nuisance parameters discussed below can have some impact
here. For example, the high $p_T^j$ bin can have contributions from the boosted regime
of VH production, where the hadronically decaying boosted vector boson produces
a high-$p_T$ jet.

The overall impacts of the yield uncertainties ($\Delta^{\rm y}_{\rm VBF}$, $\Delta^{\rm y}_{\rm Rest}$) and of the $\theta_{200}$ migration uncertainty ($\Delta_{200}$) can be evaluated
based on fixed-order uncertainties. Their separation into the individual bins
for most bins could be evaluated from fixed-order predictions or based on Monte-Carlo predictions. (If different predictions are utilized to estimate the overall size and the separation into the bins, one must be careful to check that both predictions are
consistent with each other.) The $p_T^{Hjj}$ cut is in principle sensitive to resummation
effects. Therefore, the size of the $\theta_{25}$ migration uncertainty ($\Delta_{25}$) as well as the separation of the other uncertainties within the $p_T^{Hjj}$ bins should preferably be evaluated (or at least cross checked) based on parton-shower Monte Carlos or resummed calculations.

\begin{table}
\centering
\renewcommand{\arraystretch}{1.5}
\renewcommand{\tabcolsep}{1.5ex}
\begin{tabular}{l || c | c | c | c | c || c | c }
& \multicolumn{5}{c||}{QCD uncertainties} & \multicolumn{2}{c}{EW uncertainties}
\\
& $\Delta^{\rm y}_{\rm VBF}$ & $\Delta^{\rm y}_{\rm Rest}$ & $\Delta^{\rm y}_{\rm VH}$ & $\Delta_{200}$ & $\Delta_{25}$ & $\Delta_\mathrm{Sud}$ & $\Delta_{\rm hard}$
\\\hline\hline
$p_T^{j1}$ [0,200] &$\approx 1$   &$\approx 1$ & $\approx 1$  & $-1$ &        &$y$ & $y$
\\\hline\hline
$\,\, \geq$ 2-jet VBF cuts & $\approx 1$ &$\approx 0$ & $\approx 0$ &$-x_1$ & $0$  &$x_1 y$ & $x_1 y$
\\\hline
$\quad p_T^{Hjj}\, [0,25]$   & $(\approx 1)z$ & $\cdots$ & $\cdots$ &$-x_1 z$ & $+1$ & $\cdots$ & $\cdots$
\\
$\quad p_T^{Hjj}\, [25,\infty]$ & $(\approx 1)(1\!-\!z)$ & $\cdots$ & $\cdots$ & $-x_1(1\!-\!z)$  & $-1$ & $\cdots$ & $\cdots$
\\\hline
$\,\, \geq$ 2-jet VH cuts &$\approx 0$ & $\approx 0$ & $\approx 1$ &$-x_2$ & & $x_2 y$ & $x_2 y$
\\
$\,\,$ Rest            &$\approx 0$ & $\approx 1$ & $\approx 0$ & $-x_3$  & & $x_3 y$ & $x_3 y$
\\\hline\hline
$p_T^{j1}$ $[200,\infty]$ & $\approx 0$ & $\approx 0$ & $\approx 0$ &$+1$  &        &$1-y$ & $1-y$
\end{tabular}
\caption{Structure and expected impact of different uncertainty sources for VBF production.
Each column correspond to one independent nuisance parameter.
The $\Delta_i$ denote the absolute impact, which are multiplied by the
corresponding parameters for each bin.
Empty entries mean that a source has by definition no impact on a bin.
The $x_i$, $y$, $z$ parameters can be different in each column, and $\sum_i x_i = 1$.}
\label{tab:Higgs_STXS:VBF}
\end{table}

\paragraph{EW Uncertainties}
For the EW uncertainties, we identify the following sources/nuisance parameters:
\begin{itemize}
\item {\boldmath $\theta_{\rm Sud}$}:
The uncertainty related to EW Sudakov effects.

\item {\boldmath $\theta_{\rm hard}$}:
The uncertainty related to hard EW (non-Sudakov) effects.
\end{itemize}
The reason to separate these two sources is that they correspond to structurally different
types of EW corrections, and hence can be considered as largely uncorrelated.
The Sudakov effects refer to the appearance of EW logarithms
from the virtual exchange of EW gauge bosons. They
are expected to be most relevant in the high-$p_T^j$ bin,
while the non-Sudakov effects are likely to affect all bins.
Since the EW uncertainties can be expected to be less relevant than the QCD ones,
it should not be necessary to separate them further based on the individual bins,
and hence they effectively act as overall yield uncertainties correlated among all bins.
Their evaluation should be based on the corresponding EW calculations.

The structure and expected impact of the different sources are illustrated
in Table~\ref{tab:Higgs_STXS:VBF}. Here, the $\Delta_i$ denote the absolute uncertainties, and
the $x_i$, $y$, $z$ parameters in the different columns are
different parameters, i.e., they are specific to each source (to each column).
For notational simplicity, we did not adorn them with additional subscripts.
The $x_i$ satisfy $\sum_i x_i = 1$.

\subsubsection{VH Production}

\begin{figure}
\centering
\includegraphics[scale=0.5]{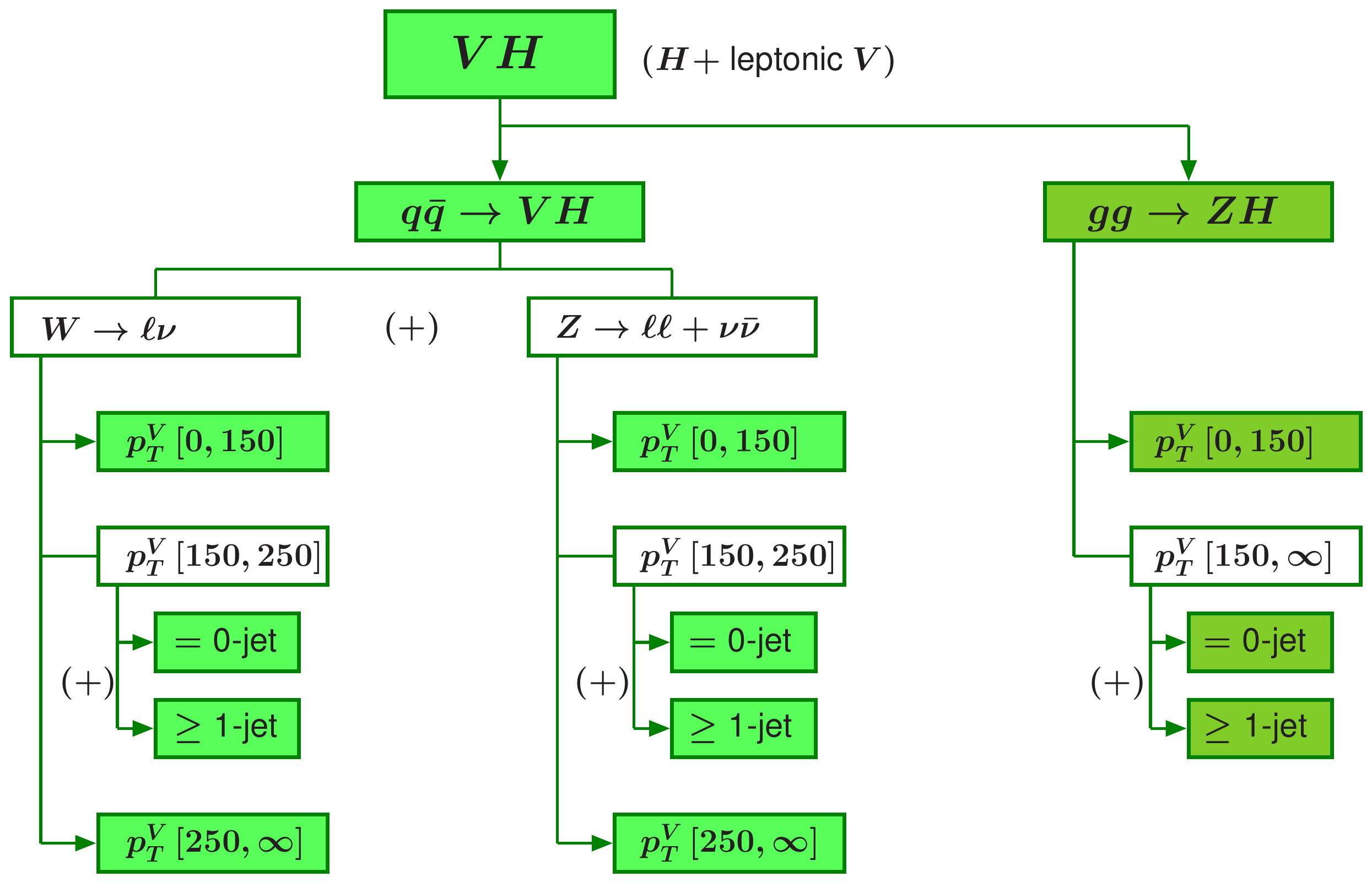}
\caption{Stage 1 bins for VH production.}
\label{fig:Higgs_STXS:VHstage1}
\end{figure}

The VH template process is formally defined as Higgs production in association
with a leptonically decaying vector boson. It is separated into the three
underlying processes $q\bar q'\to WH$, $q\bar q\to ZH$, and $gg\to ZH$. The full
stage 1 bins are depicted in Fig.~\ref{fig:Higgs_STXS:VHstage1}. They include three $p_T^V$
bins (two for $gg\to ZH$) with binning cuts at $p_T^V = 150\,\mathrm{GeV}$ and $p_T^V =
250\,\mathrm{GeV}$. The second $p_T^V$ bin is further separated into an
exclusive 0-jet bin and an inclusive $\geq 1$-jet bin, with a cut at $p_T^j =
30\,\mathrm{GeV}$.

The hadronic VH processes $q\bar q\to V(\to q\bar q) H$ are included as part of
the VBF template process. Similarly, the gluon-induced $gg\to Z(\to q\bar q) H$
is included as part of the $gg\to H$ template process (where it corresponds to
an electroweak real-emission correction).

\paragraph{QCD Uncertainties} For the QCD uncertainties, we identify the following sources/nuisance parameters:
\begin{itemize}
\item {\boldmath $\theta^{\rm y}_{\rm VH}$, $\theta^{\rm y}_{gg\to ZH}$}:
The overall yield uncertainty for the underlying VH production process.

\item {\boldmath $\theta_{150}^{\rm VH}$, $\theta_{150}^{gg\to ZH}$}:
The migration uncertainty related to the $p_T^V = 150\,\mathrm{GeV}$ bin boundary.

\item {\boldmath $\theta_{250}^{\rm VH}$}:
The migration uncertainty related to the $p_T^V = 250\,\mathrm{GeV}$ bin boundary.

\item {\boldmath $\theta_{0/1}^{\rm VH}$, $\theta_{0/1}^{gg\to ZH}$}:
The migration uncertainty related to the $0/1$-jet bin boundary.
\end{itemize}
These parameters are the same for the $q\bar q'\to WH$ and $q\bar q\to ZH$
subprocesses, whose QCD uncertainties are thus treated as correlated.
The overall impacts ($\Delta$s) for the two subprocesses however do not have to be the same.
The corresponding nuisance parameters for $gg\to ZH$ are separate,
such that the quark-induced and gluon-induced processes are treated as uncorrelated.

The overall impacts of the yield and $p_T^V$ migration uncertainties
($\Delta^{\rm y}_{\rm WH}$, $\Delta^{\rm y}_{\rm ZH}$, $\Delta^{\rm y}_{gg\to ZH}$,
$\Delta^{\rm WH}_{150}$, $\Delta^{\rm ZH}_{150}$, $\Delta^{\rm WH}_{250}$, $\Delta^{\rm ZH}_{250}$) can be evaluated
based on fixed-order uncertainties. Their separation into the individual bins
for most bins could be evaluated from fixed-order predictions or based on Monte-Carlo predictions. The $0/1$-jet binning cut is likely to be sensitive to resummation
effects. Therefore, the size of the $\theta_{0/1}$ migration uncertainty ($\Delta_{0/1}^{\rm ZH}$, $\Delta_{0/1}^{\rm WH}$, $\Delta_{0/1}^{gg\to ZH}$) as well as the separation of the other uncertainties within the jet bins should preferably be evaluated
based on parton-shower Monte Carlos or resummed calculations.

For the $gg\to ZH$ process it should be studied to what extent the uncertainties
should be correlated with corresponding uncertainties in the $gg\to H$ process.
For example, most likely it would be most appropriate to identify
the $\theta_{0/1}$ for $gg\to H$ and $gg\to ZH$, since the relevant resummation
effects for both processes are the same.

\paragraph{EW Uncertainties}
For the EW uncertainties, we identify the following sources/nuisance parameters:
\begin{itemize}
\item {\boldmath $\theta_{\rm Sud}^{\rm VH}$}:
The uncertainty related to EW Sudakov effects.

\item {\boldmath $\theta_{\rm hard}^{\rm WH}$, $\theta_{\rm hard}^{\rm ZH}$}:
The uncertainty related to hard EW (non-Sudakov) effects.
\end{itemize}
As for the VBF process, we can distinguish EW Sudakov and hard (non-Sudakov) uncertainties.
The Sudakov corrections are universal and are thus treated as correlated for WH and ZH
production. On the other hand, the hard EW corrections are different with separate nuisance parameters for WH and ZH.
One aspect to be studied is to what extent the EW uncertainties should be treated
as correlated or uncorrelated between the VBF and VH processes. The naive expectation
is that it would be most appropriate to treat the EW Sudakov uncertainties as correlated
and the others as uncorrelated.

We do not yet consider EW uncertainties for $gg\to ZH$, since they are not yet relevant
(due to the fact that the contribution of the subprocess itself is small with relatively large QCD uncertainties). Once they become relevant, they can easily be added as for the
dominant quark-induced subprocesses.
We also do not separately distinguish uncertainties due to mixed QCD-EW corrections,
which for now should be included as part of the (hard) EW uncertainties. If it becomes
necessary, one could add a separate source for these in the future.
Uncertainties from photon-induced corrections are also not yet separately considered.

The structure and expected impact of the different sources are illustrated
in Table~\ref{tab:Higgs_STXS:VH}. As before, the $\Delta_i$ denote the absolute uncertainties, and
the $x_i$, $y$, $z$ parameters in the different columns are
different parameters, i.e., they are specific to each source (to each column).
The $x_i$ satisfy $\sum_i x_i = 1$.
The uncertainties for $gg\to ZH$ production are not shown. They have the same structure
but with an independent set of nuisance parameters.

\begin{table}
\centering
\renewcommand{\arraystretch}{1.5}
\renewcommand{\tabcolsep}{1.3ex}
\begin{tabular}{l || c | c | c | c || c | c | c}
& \multicolumn{4}{c||}{QCD uncertainties} & \multicolumn{3}{c}{EW uncertainties}
\\
$q \bar q'\to W$ & $\Delta^{\rm y}_{\rm WH}$ & $\Delta^{\rm WH}_{150}$ & $\Delta^{\rm WH}_{250}$ & $\Delta^{\rm WH}_{0/1}$ & $\Delta^{\rm WH}_\mathrm{Sud}$ & $\Delta^{\rm WH}_{\rm hard}$ &
\\\hline\hline
$\,\, p_T^V$ [0,150]      & $x_1$   & $-1$  & $-y$   &      & $x_1$    & $\cdots$ & \\\hline
$\,\, p_T^V$ [150,250]    & $x_2$   & $+1-y$  & $-(1-y)$  & $0$  & $x_2$    & $\cdots$
\\\hline
$\quad =$ 0-jet           & $x_2 z$ & $+(1\!-\!y)z$ & $-(1\!-\!y)z$ & $+1$ & $\cdots$ & $\cdots$ &
\\
$\quad \geq$ 1-jet        & $x_2 (1\!-\!z)$ & $+(1\!-\!y)(1\!-\!z)$ & $-(1\!-\!y)(1\!-\!z)$ & $-1$ & $\cdots$ & $\cdots$ &
\\\hline
$\,\,p_T^V$ [250,$\infty$]& $x_3$   & $y$   & $+1$  &      & $x_3$    & $\cdots$ &
\\ \hline\hline
$q \bar q \to Z$& $\Delta^{\rm y}_{\rm ZH}$ & $\Delta^{\rm ZH}_{150}$ & $\Delta^{\rm ZH}_{250}$ & $\Delta^{\rm ZH}_{0/1}$ & $\Delta_\mathrm{Sud}^{\rm ZH}$ & & $\Delta_{\rm hard}^{\rm ZH}$
\\ \hline\hline
$\,\, p_T^V$ [0,150]      & $x_1$   & $-1$  & $-y$   &      & $x_1$    & & $\cdots$
\\\hline
$\,\, p_T^V$ [150,250]    & $x_2$   & $+1-y$  & $-(1-y)$  & $0$  & $x_2$    & & $\cdots$
\\\hline
$\quad =$ 0-jet           & $x_2 z$ & $+(1\!-\!y)z$ & $-(1\!-\!y)z$ & $+1$ & $\cdots$ & & $\cdots$
\\
$\quad \geq$ 1-jet        & $x_2 (1\!-\!z)$ & $+(1\!-\!y)(1\!-\!z)$ & $-(1\!-\!y)(1\!-\!z)$ & $-1$ & $\cdots$ & & $\cdots$
\\\hline
$\,\,p_T^V$ [250,$\infty$]& $x_3$   & $y$   & $+1$  &      & $x_3$    & & $\cdots$

\end{tabular}
\caption{Structure and expected impact of different uncertainty sources for VH production.
Each column correspond to one independent nuisance parameter.
The $\Delta_i$ denote the absolute impact, which are multiplied by the
corresponding parameters for each bin.
Empty entries mean that a source has by definition no impact on a bin.
The $x_i$, $y$, $z$ parameters can be different in each column, and $\sum_i x_i = 1$.
In addition, $gg\to ZH$ production has an independent set of nuisance parameters
with the analogous structure of uncertainties.}
\label{tab:Higgs_STXS:VH}
\end{table}

\subsection{Treatment of Theory Uncertainties in Reporting Experimental Results}
\label{sec:Higgs_STXS:reportingresults}

The STXS measurements are performed with the SM production processes serving 
as kinematic templates. They can be interpreted in the context of the SM, but 
can also be recast in the context of beyond-the-SM models to help set 
constraints on new physics.
In order to make optimal use of the measurement information, the interpretation should be performed using the full experimental likelihood. 

This is however difficult in practice: full likelihoods are typically not made public by the experimental collaborations, and also involve a large number of parameters including both the parameters of interest (POIs) and a large number of nuisance parameters (NPs) representing systematic uncertainties. Experimental results are usually reported in terms of the profile likelihood ratio (PLR), in which all nuisance parameters have been profiled. A Gaussian form is further assumed for the PLR, and results are reported in terms of the best-fit values of the POIs and their covariance matrix.

This presentation of results suffers from two limitations: firstly, non-Gaussian effects in the measurements are neglected; and secondly it is not possible to correlate systematic uncertainties in the interpretation stage with the same uncertainties in the experimental measurements, since the NPs representing the uncertainties in the latter case have been profiled in the reported results. 

In this section, we present possible directions of improvement for the reporting of experimental results to address the second point.

\subsubsection{Gaussian measurements with linear parameter response}

We start by defining a simplified model in which the measurement is described as $N$ independent Gaussian observations $n_i$, which can represent for instance event counts in $N$ mutually exclusive signal regions. The corresponding likelihood is
\begin{equation}
L_n(\mathbf{n}) = \exp \left[-\frac{1}{2} (\mathbf{n} - \hat{\mathbf{n}})^T H_n (\mathbf{n} - \hat{\mathbf{n}}) \right]
\end{equation}
where $\hat{\mathbf{n}}$ is the best-fit value of $\mathbf{n}$ and $H_n$ is the Hessian matrix, defined as the inverse of the covariance matrix
\begin{equation}
\renewcommand{\arraystretch}{1.5}
C = \left( \begin{array}{ccc}
\sigma_1^2 & 0 & \cdots \\
0 & \ddots & \ddots \\
\vdots & \ddots & \sigma_N^2 \\
\end{array}\right)
\end{equation}
We further assume that the expected values of the $n_i$ are linear functions $n_i = \sum\limits_{\alpha} k_{i\alpha} x_{\alpha}$ of the model parameters $x_{\alpha}$. The measurement of the $x_{\alpha}$ is then described by the Gaussian likelihood
\begin{equation}
L_x(\mathbf{n}) = \exp \left[-\frac{1}{2} (\mathbf{x} - \hat{\mathbf{x}})^T H_{\mathbf{x}} (\mathbf{x} - \hat{\mathbf{x}}) \right]
\label{eq:Higgs_STXS:L}
\end{equation}
where $\hat{x}$ is the maximum-likelihood estimator of $x$, and
\begin{equation}
H_{\mathbf{x}} = K^T H_n K,
\end{equation}
$K$ denoting the matrix of the $k_{i\alpha}$.
We furthermore explicitly split $\mathbf{x}$ into POI and NP components, $\mathbf{\mu}$ and $\mathbf{\theta}$ respectively, 
$\mathbf{x}^T = \left(\begin{array}{c|c} \mathbf{\mu} & \mathbf{\theta} \end{array}  \right)$, and assume that the $\mathbf{n}$ similarly decompose into true experimental measurements and auxiliary measurements~\cite{ATLAS_CMS_Stats} providing constraints on the $\mathbf{\theta}$. We also follow the usual convention~\cite{ATLAS_CMS_Stats} of orthogonalizing and normalizing the $\mathbf{\theta}$ so that the auxiliary measurements are all represented by standard Gaussians of unit width. We then have
\begin{equation}
\renewcommand{\arraystretch}{1.5}
H_n = \left( \begin{array}{c|c}
\Sigma^{-1} & 0 \\
\hline
0 & I \\
\end{array} \right) \qquad
K = \left( \begin{array}{c|c}
R & \Delta \\
\hline
0 & I \\
\end{array} \right)
\end{equation}
where $\Sigma^{-1}$ is the diagonal matrix of the uncertainties of the experimental measurements; $R$ is the response matrix giving the impact of the POIs on the measurements; and $\Delta$ provides the impacts of the NPs on the measurements, representing systematic uncertainties. It follows that
\begin{equation}
\renewcommand{\arraystretch}{1.5}
H_{\mathbf{x}} = \left( \begin{array}{c|c}
R^T \Sigma^{-1} R & R^T \Sigma^{-1} \Delta \\
\hline
\Delta^T \Sigma^{-1} R & I + \Delta^T \Sigma^{-1} \Delta \\
\end{array} \right)
 = \left( \begin{array}{c|c}
C_{\text{stat}}^{-1} & \Lambda \\
\hline
\Lambda^T & I + K_{\theta} \\
\end{array} \right).
\label{eq:Higgs_STXS:Hmes}
\end{equation}
The $C_{\text{stat}}$ matrix represents the statistical component of the measurement of $\mathbf{\mu}$; the bottom-right block describes the uncertainties of the $\mathbf{\theta}$, which can be reduced from the initial unit values due to constraints from the measurements; and the off-diagonal blocks $\Lambda$ represent the correlation between the measurements of $\mathbf{\mu}$ and $\mathbf{\theta}$ through systematic effects.

The measurement of the POIs $\mathbf{\mu}$ is performed using the profile likelihood ratio
\begin{equation}
\lambda(\mathbf{\mu}) = \frac{L \left(\mathbf{\mu}, \hat{\hat{\mathbf{\theta}}}(\mathbf{\mu})\right)}{L \left(\hat{\mathbf{\mu}}, \hat{\mathbf{\theta}}\right)}
= \exp\left[-\frac{1}{2} (\mathbf{\mu} - \hat{\mathbf{\mu}})^T H'_{\mu} (\mathbf{\mu} - \hat{\mathbf{\mu}}) \right].
\end{equation}
In terms of a general Hessian matrix
\begin{equation}
\renewcommand{\arraystretch}{1.5}
H_{\mathbf{x}} = \left( \begin{array}{c|c}
H_{\mathbf{\mu}\mathbf{\mu}} & H_{\mathbf{\mu}\mathbf{\theta}} \\
\hline
H_{\mathbf{\mu}\mathbf{\theta}}^T & H_{\mathbf{\theta}\mathbf{\theta}} \\
\end{array} \right),
\label{eq:Higgs_STXS:Hx}
\end{equation}
the profiled values of the nuisance parameters are
\begin{equation}
\hat{\hat{\mathbf{\theta}}}(\mathbf{\mu}) = \hat{\mathbf{\theta}} + H_{\mathbf{\theta}\mathbf{\theta}}^{-1} H_{\mathbf{\mu}\mathbf{\theta}}^T \left(\mathbf{\mu} - \hat{\mathbf{\mu}} \right)
\label{eq:Higgs_STXS:prof}
\end{equation}
and the profile likelihood ratio has a Gaussian form with the reduced Hessian matrix
\begin{equation}
H'_{\mathbf{\mu}} = H_{\mathbf{\mu}\mathbf{\mu}} - H_{\mathbf{\mu}\mathbf{\theta}} H_{\mathbf{\theta}\mathbf{\theta}}^{-1} H_{\mathbf{\mu}\mathbf{\theta}}^T.
\label{eq:Higgs_STXS:HmuGen}
\end{equation}
which for the Hessian of Eq.~\eqref{eq:Higgs_STXS:Hmes} gives
\begin{equation}
H'_{\mathbf{\mu}} = C_{\text{stat}}^{-1} - \Lambda \left( I + K_{\theta} \right)^{-1} \Lambda^T.
\label{eq:Higgs_STXS:Hmu}
\end{equation}
In terms of the covariance matrix 
\begin{equation}
\renewcommand{\arraystretch}{1.5}
C_{\mathbf{x}} = H_{\mathbf{x}}^{-1} = \left( \begin{array}{c|c}
C_{\mathbf{\mu}} & C_{\mathbf{\mu}\mathbf{\theta}} \\  \hline
C_{\mathbf{\mu}\mathbf{\theta}}^T & C_{\mathbf{\theta}}
\end{array} \right)
\label{eq:Higgs_STXS:Cx}
\end{equation}
of the full measurement, one has $H'_{\mathbf{\mu}} = C_{\mathbf{\mu}}^{-1}$ : the sector of $C_{\mathbf{x}}$ corresponding to the POI measurement is the inverse of $H'_{\mathbf{\mu}}$ computed above after profiling the NPs and thus already includes their impacts. Assuming these are small compared to statistical uncertainties ($\Delta\Delta^T \ll \Sigma$), this can be written to linear order as
\begin{align}
C_{\mathbf{\mu}} &=  C_{\text{stat}} + C_{\text{stat}} \Lambda \Lambda^T C_{\text{stat}} \\
& \equiv C_{\text{stat}} + C_{\text{syst}},
\label{eq:Higgs_STXS:Cmu}
\end{align}
where the first term represents statistical uncertainties and the second one systematics.

\subsubsection{Link with experimental results}

Higgs couplings measurements typically report uncertainties using the profile likelihood ratio, computed from  the full experimental likelihood. They also often provide the covariance matrix $C_{\mathbf{\mu}}$, obtained from the Hessian matrix $H_{\mathbf{x}}$ of second derivatives of the likelihood evaluated at the best-fit position and truncated to POIs only.  This allows the experimental results to be recast under specific theoretical models, accounting for correlation between the measured parameters. This suffers from the two limitations mentioned in the introduction: first the Gaussian assumption on the likelihood form, which is assumed to hold here; and secondly the non-inclusion of the nuisance parameters describing systematic uncertainties. If the same parameters also play a role at the level of the reinterpretation, $C_{\mathbf{\mu}}$ alone does not provide sufficient information to properly correlate the two effects.

Such a treatment can be performed using the full covariance matrix $C_{\mathbf{x}} = H_{\mathbf{x}}^{-1}$. To limit the size of the matrix, one can still truncate from $C$ the uncertainties that are not relevant to the interpretation stage (for instance experimental uncertainties), leading to a reduced parameter set $\tilde{\mathbf{x}} = (\mathbf{\mu}, \theta_1 \cdots \theta_p)$. This approach allows to rebuild an experimental likelihood as in Eq.~\eqref{eq:Higgs_STXS:L}, assuming Gaussian assumptions. 

A third option, inspired by Eq.~\eqref{eq:Higgs_STXS:Cmu}, is to report $C_{\mathbf{\mu}}$ together with $C_{\text{stat}}$, which allows to extract the effect of the relevant systematic uncertainties. $C_{\text{stat}}$ can be obtained from the full likelihood in the same way as $C_{\mathbf{\mu}}$, but keeping the systematics NPs fixed to their best-fit value. The method is most straightforwardly applied by reporting the matrices
\begin{equation}
C_p = C_{\mathbf{\mu}}, C_{p-1} = C_{\mathbf{\mu}|\theta_p}, C_{p-2} = C_{\mathbf{\mu}|\theta_p, \theta_{p-1}} \cdots, C_0 = C_{\mathbf{\mu}|\theta_p, \cdots \theta_1}
\end{equation}
where the NPs after the vertical bar are fixed to their best-fit value when computing the correlation matrix. $C_0$ therefore corresponds to the covariance matrix of the $\mathbf{\mu}$ without the effect of any of the $\theta_1 \cdots \theta_p$ included, $C_1$ corresponds to the covariance matrix including the effect of $\theta_1$ only, and so on until all the $\theta_1 \cdots \theta_p$ are included in $C_p$.
Under the same assumptions as those leading to Eq.~\eqref{eq:Higgs_STXS:Cmu}, one then has
\begin{equation}
\delta C_i = C_i - C_{i - 1} = C_{i - 1} \Lambda_i \Lambda_i^T C_{i-1}
\end{equation}
which allows to extract the impact $\Lambda_i$ of the NP $\theta_i$. The sign of $\Lambda_i$ is however ambiguous and needs to be provided in addition to the covariance matrices. The $\Lambda_i$ can also be obtained without simplifying assumptions by re-expressing Eq.~\eqref{eq:Higgs_STXS:HmuGen} as
\begin{equation}
C_i^{-1} = C_{i - 1}^{-1}  - \Lambda_i (1 + K_i)^{-1}\Lambda_i^T.
\end{equation}
The $\Lambda_i$ (the impact of $\theta_i$ on the POIs) and $1 + K_i$ (which determines the uncertainty on $\theta_i$) cannot be unambiguously determined since the covariance matrix is only sensitive to their combination. A possible convention is to require that the post-fit uncertainty on $\theta_i$ is $1$, as for the pre-fit case, by setting $K_i = \Lambda_i^T C_{i-1} \Lambda_i$. The components of $\Lambda_i$ are then given by 
\begin{equation}
[\Lambda_i]_{\alpha} = \pm\sqrt{(1 + K_i) [C_{i - 1}^{-1} - C_i^{-1}]_{\alpha\alpha}}, \qquad 1 + K_i =  [1 - \text{Tr}(I -  C_{i - 1} C_i^{-1})]^{-1}
\end{equation}
with the sign of the $[\Lambda_i]_{\alpha}$ again provided as external input. The full matrix $H_{\tilde{\mathbf{x}}}$ can then be reconstructed as
\begin{equation}
\renewcommand{\arraystretch}{1.5}
H_{\mathbf{x}} = \left( \begin{array}{c|c|c|c|c}
 C_0^{-1} & \Lambda_p & \Lambda_{p-1} & \cdots & \Lambda_1 \\
\hline
 \Lambda_p^T & I + K_p & 0  & \cdots & 0 \\
\hline
\Lambda_{p-1}^T & 0 & I + K_{p-1}  & \ddots & \vdots \\
\hline
\vdots  & \vdots & \ddots & \ddots & 0 \\  
\hline
\Lambda_1^T & 0 & \cdots & 0 &  I + K_1
\end{array} \right).
\label{eq:Higgs_STXS:Hxx}
\end{equation}
This expression of $H_{\mathbf{x}}$ can then be used to represent the experimental likelihood as above. It has the property that successively profiling away $\theta_1 \cdots \theta_p$ leads to reduced Hessian matrices with top-left sectors equal to $C_1^{-1} \cdots C_p^{-1}$, so that profiling recovers at each stage the covariance matrices obtained from the full profile likelihood computation\footnote{Note that this works only if the parameters are profiled in the same order as for the computation of the $C_i$, since correlation information between the NPs is not included.}. 

\subsubsection{Example}

\begin{figure}
\centering
\subfloat[]{\label{fig:Higgs_STXS:toy1}\includegraphics[width=.48\textwidth]{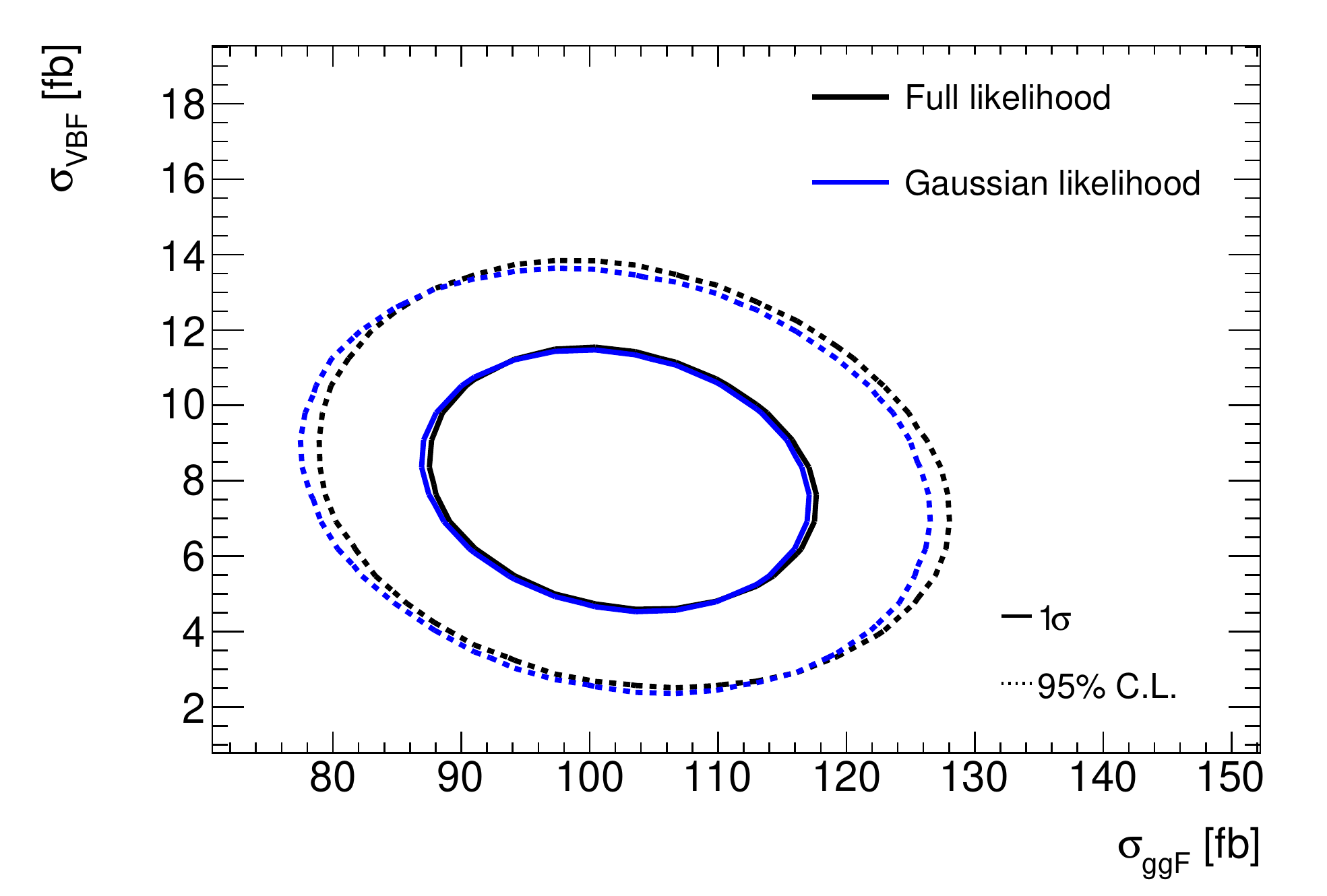}}
\subfloat[]{\label{fig:Higgs_STXS:toy2}\includegraphics[width=.48\textwidth]{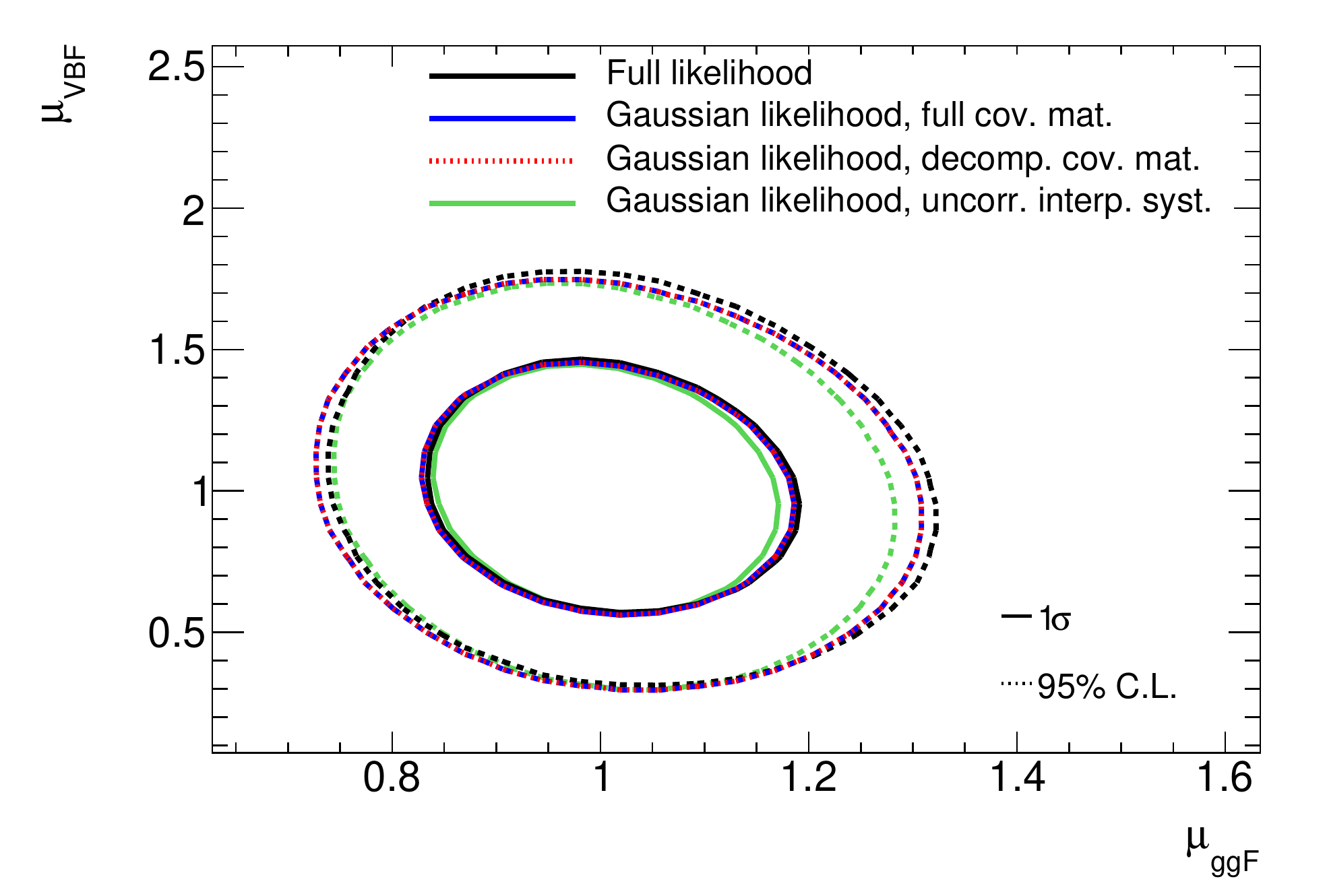}}
\caption{Confidence level contours at the $1\sigma$ (solid lines) and $95\%$ CL (dashed lines) levels, computed from the profile likelihood ratio in the asymptotic approximation. (a) Contours in the $(\sigma_{\text{ggF}}, \sigma_{\text{VBF}})$ plane of the original measurement, using the full likelihood (black) and the Gaussian likelihood built from the covariance matrix reported by HESSE at the best-fit point. (b) Contours in the  $(\mu_{\text{ggF}}, \mu_{\text{VBF}})$ plane computed from those in (a), using a reparameterization the full likelihood (black); the Gaussian likelihood built from the covariance matrix of the parameter set $(\sigma_{\text{ggF}}, \sigma_{\text{VBF}}, \theta_{\text{ggF}}, \theta_{\text{VBF}})$ reported by HESSE (blue); the Gaussian likelihood built from the covariance matrix recomposed as in Eq.~\eqref{eq:Higgs_STXS:Hxx} (red); and the Gaussian likelihood built from the covariance matrix of $(\sigma_{\text{ggF}}, \sigma_{\text{VBF}}$), with additional interpretation-stage uncertainties uncorrelated with those of the measurement.}
\end{figure}

As an illustration, we consider a simple example inspired by the measurement of the gluon-fusion and VBF cross-sections in the $H \rightarrow \gamma\gamma$ decay channel~\cite{hgg}. The measurement is modeled by counting experiments in two bins, denoted as \textit{0-jet} and \textit{2-jet}, and considering contributions from gluon-fusion Higgs production (ggF), VBF production and background. Theoretical and experimental uncertainties are considered on the experimental acceptance values, and implemented with log-normal profiles. The parameters of the model are summarized in Table~\ref{tab:Higgs_STXS:toy_params}.
A profile likelihood ratio (PLR) scan in the $(\sigma_{\text{ggF}}, \sigma_{\text{VBF}})$ plane for the SM hypothesis is performed, and the corresponding confidence level contours computed in the asymptotic approximation~\cite{Asimov} are shown in Fig.~\ref{fig:Higgs_STXS:toy1}. The numerical results are
\begin{align}
\sigma_{\text{ggF}} &= 102 \substack{+8.2 \\ -8.5} (\text{stat}) \substack{5.7 \\ -4.1} (\text{exp}) \substack{2.3 \\ -1.7} (\text{theo})\, \text{fb} \\
\sigma_{\text{VBF}} &= 8.0 \pm 2.1 (\text{stat}) \substack{+0.7 \\ -0.5} (\text{exp}) \substack{+0.7 \\ -0.4} (\text{theo})\, \text{fb}
\end{align}
with a correlation coefficient of $17\%$. The covariance matrix of the fit is used to build a Gaussian likelihood following Eq.~\eqref{eq:Higgs_STXS:L}. The PLR contours for this likelihood are also shown in Fig.~\ref{fig:Higgs_STXS:toy1}. Small differences are visible between the two due to the non-Gaussian form of the full likelihood. A similar comparison without systematic uncertainties included shows excellent agreement, which suggests the log-normal profiles of systematics is the main source of non-Gaussianity.

\begin{table}[]
\centering
  \begin{tabular}{lcc} 
  \hline\hline
  Parameters & \textit{0-jet} & \textit{2-jet} \\
  \hline
  $\sigma_{\text{ggF}}$ & $102~\text{fb} \times 96\%$  & $102~\text{fb} \times 4\%$ \\
  $\sigma_{\text{VBF}}$ & -  & $8~\text{fb}$ \\
  Signal $A \times \epsilon$ & \multicolumn{2}{c}{$40\%$} \\
  $\sigma_{\text{Bkg}}^{\text{vis}}$ & $1500~\text{fb}$  & $100~\text{fb}$ \\
  $\mathcal{L}$ & \multicolumn{2}{c}{$150~\text{fb}^{-1}$} \\
  \hline\hline
  Systematics & \textit{0-jet} & \textit{2-jet} \\
  \hline
  Experimental, \textit{0-jet} & $5\%$ & - \\ 
  Experimental, \textit{2-jet} & - & $5\%$ \\ 
  ggF theory on inclusive $A \times \epsilon$ & $2\%$ & $2\%$ \\ 
  VBF theory on inclusive $A \times \epsilon$ & - & $2\%$ \\ 
  ggF theory  on \textit{2-jet} $A \times \epsilon$ & - & $15\%$  \\
  \hline\hline
\end{tabular}
\caption{Summary of numerical parameters defining the example model. $\sigma_{\text{ggF}}$ and $\sigma_{\text{VBF}}$ are the total cross-sections for signal production in the ggF and VBF modes; $A \times \epsilon$ is the overall efficiency of the signal selection; $\mathcal{L}$ is the integrated luminosity of the data sample; and $\sigma_{\text{Bkg}}^{\text{vis}}$ is the visible cross-section the continuum background (the ratio of the number of observed background events after all selections to the integrated luminosity). In the lower half of the table, each line corresponds to an independent source of systematic uncertainty, associated with a separate nuisance parameter.}
\label{tab:Higgs_STXS:toy_params}
\end{table}

The measurement is reinterpreted in the $\mu$ parameterization of Higgs couplings, with parameters of interest defined as the ratios $\mu_{\text{ggF}} = \sigma_{\text{ggF}}/\sigma_{\text{ggF,SM}}$ and $\mu_{\text{VBF}} = \sigma_{\text{VBF}}/\sigma_{\text{VBF,SM}}$ of the measured cross-sections to their SM values. The parameterization includes uncertainties of $5\%$ on the SM ggF and VBF cross-sections with log-normal profiles. They are considered to be fully correlated with the corresponding inclusive acceptance systematics ($\theta_{\text{ggF}}$ and $\theta_{\text{VBF}}$ respectively), making it possible to compare the possibilities described in the preceding section. 

We consider the 3 cases of the model of Eq.~\eqref{eq:Higgs_STXS:Hx} in which the full covariance matrix is provided; the model of Eq.~\eqref{eq:Higgs_STXS:Hxx} in which the covariance matrix is provided in decomposed form; and the case where the covariance matrix is only reported for the POIs, and the systematics on the interpretation are added separately without accounting for their correlation with those of the measurement.

The results are shown in Fig.~\ref{fig:Higgs_STXS:toy2}. The contours from the two methods of reporting the full covariance matrix are rather close to those of the exact likelihood, differing by the non-Gaussian effects already visible on Fig.~\ref{fig:Higgs_STXS:toy1}. No difference between the two methods is visible on the plot, as expected due to the smallness of the impact of NP/NP correlations. Both results provide a better description than the case where the systematics for the interpretation stage are not correlated with those of the measurement. The uncertainties are somewhat smaller in this case, since to first approximation the uncertainties at each stage add in quadrature, whereas they add linearly when fully correlated. The magnitude of the difference would however be reduced for smaller values of the theoretical uncertainties on the measured values.

\subsection*{Acknowledgments}
This writeup is based on discussions in Les Houches 2017 as well as subsequent discussions,
with contributions and feedback from Josh Bendavid, Marco Delmastro, Fr\'ed\'eric Dreyer, Michael Duehrssen-Debling, Markus Ebert, Maria Vittoria Garzelli, Nicolas Greiner, Stefan Kallweit, Predrag Milenovic, Luca Perrozzi, Carlo Pandini, Marco Zaro, and many others.



\chapter{Phenomenological studies}
\label{cha:pheno}



\newcommand{\mr}[1]{{\ensuremath{\mathrm{#1}}}}
\newcommand{\mc}[1]{{\ensuremath{\mathcal{#1}}}}
\newcommand{\mb}[1]{{\ensuremath{\mathbf{#1}}}}
\newcommand{\done}{\ensuremath{\mr{d}}}
\newcommand{\nmaxnlo}{{\ensuremath{n_{\mathrm{max}}^{\text{NLO}}}}}
\newcommand{\Qcut}{{\ensuremath{Q_{\mathrm{cut}}}}}
\newcommand{\MEPSatNLO}{M\protect\scalebox{0.8}{E}P\protect\scalebox{0.8}{S}@N\protect\scalebox{0.8}{LO}\xspace}
\newcommand{\MCatNLO}{M\protect\scalebox{0.8}{C}@N\protect\scalebox{0.8}{LO}\xspace}
\newcommand{\QCD}{\text{QCD}\xspace}
\newcommand{\EWvirt}{\ensuremath{\EW_\text{virt}}\xspace}
\newcommand{\QCDpEW}{\text{\QCD{}+\EW}\xspace}
\newcommand{\QCDpEWvirt}{\text{\QCD{}+\EW{}$_\mathrm{virt}$}\xspace}
\newcommand{\Gmu}{{\ensuremath{G_\mu}}}
\newcommand{\shortequal}{{\ensuremath{\!\!\!=\!\!\!}}}
\newcommand{\ttbar}{{\ensuremath{t\bar{t}}}}
\newcommand{\ttbarj}{{\ensuremath{t\bar{t}+\text{jet}}}}
\newcommand{\core}{{\ensuremath{\text{core}}}}
\newcommand{\muR}{{\ensuremath{\mu_R}}}
\newcommand{\muF}{{\ensuremath{\mu_F}}}
\newcommand{\muQ}{{\ensuremath{\mu_Q}}}
\newcommand{\CKKW}{\text{CKKW}\xspace}
\newcommand{\muCKKW}{{\ensuremath{\mu_\CKKW}}}
\newcommand{\mucore}{{\ensuremath{\mu_\core}}}
\newcommand{\zthr}{{\ensuremath{z_{\mathrm{thr}}}}}
\newcommand{\alphaS}{{\ensuremath{\alpha_s}}}

\section{Top-pair production including multi-jet merging and EW corrections~\protect\footnote{
    C.~G\"utschow,
    J.~M.~Lindert,
    M.~Sch\"onherr}{}}
\label{sec:SM_ewmerging_ttbar}


\subsection{Introduction}
\label{sec:SM_ewmerging_ttbar:intro}

The detailed study of top-quark properties, in particular in the dominant
top-pair production mode, is amongst the main goals of the ongoing physics
program at the LHC. 
After initial measurement at the inclusive cross section
level~\cite{Chatrchyan:2012bra,Chatrchyan:2013faa,Aad:2014kva,Khachatryan:2015uqb,Aaboud:2016pbd,Nayak:2017aes}, where very good agreement with perturbative calculations at the
NNLO+NNLL level in QCD~\cite{Cacciari:2011hy,Beneke:2011mq,Czakon:2013goa} has
been observed, the attention in the study of top-pair production has shifted
towards differental measurements. In fact, the ATLAS and CMS collaborations
already performed various measurements in the different top-decay channels, 
at 7 TeV~\cite{Chatrchyan:2012saa,Aad:2014zka,Aad:2015eia}, 8 TeV~\cite{Khachatryan:2015oqa,Aad:2015hna} and 13 TeV~\cite{Khachatryan:2016mnb,Sirunyan:2017mzl,Aad:2015hna,Aaboud:2016syx,Aaboud:2018eqg}, presenting in particular results for the
reconstructed top transverse momentum distribution, whose reliable modelling is
of great importance for background estimates in a multitude of searches for
Physics Beyond the Standard Model. 
These measurements have mostly been compared against predictions obtained with the Monte Carlo
frameworks \textsc{MC@NLO}~\cite{Frixione:2003ei} (or more recently MadGraph\_aMC@NLO~\cite{Alwall:2014hca}) and \textsc{POWHEG}~\cite{Frixione:2007nw}. In these
frameworks top-pair production at NLO QCD is matched to parton showers from
\textsc{Pythia}~\cite{Sjostrand:2006za} or \textsc{Herwig}~\cite{Corcella:2000bw}. These measurements
consistently indicate that the top quark transverse momentum distribution at low
$\pT$ is well predicted by the Monte Carlo programs, both in normalisation and
shape, but these predictions exceed the data at high $\pT$. Comparing these
measurements at the unfolded parton level to differential NNLO QCD
predictions~\cite{Czakon:2015owf}, this excess has been alleviated. This
indicates the relevance of including higher jet
multiplicities~\cite{Dittmaier:2007wz,Kardos:2011qa,Alioli:2011as,Bredenstein:2009aj,Bredenstein:2010rs,
Bevilacqua:2009zn,Bevilacqua:2010ve,Bevilacqua:2011aa,Kardos:2013vxa,Cascioli:2013era,Hoeche:2013mua,
Hoeche:2014qda,Czakon:2015cla,Bevilacqua:2017cru,Jezo:2018yaf,Hoche:2016elu,Bevilacqua:2015qha}
in the modelling of the top quark transverse momentum distribution at high $\pT$. 
At the same time, it is well known that higher-order EW corrections alter the
shape of the top transverse momentum distribution at high $\pT$ due to the
appearance of EW Sudakov
logarithms~\cite{Kuhn:2006vh,Bernreuther:2006vg,Kuhn:2013zoa,Bernreuther:2010ny,Hollik:2011ps,Pagani:2016caq,
Denner:2016jyo,Czakon:2017wor} yielding corrections of about $-10\%$ at
$p_{T,{\rm top}}=1$~TeV.

In this contribution \cite{Gutschow:2018tuk}, we first present a calculation of $t\bar t$(+jet)
production at fixed-order NLO EW focussing on the top transverse momentum
distribution, however, also showing results for the invariant mass distribution
of the top-quark pair. Second, we present predictions based on the \MEPSatNLO
multi-jet merging in {\sc
Sherpa}~\cite{Hoeche:2009rj,Hoeche:2012yf,Gehrmann:2012yg}, incorporating the EW
corrections in an approximation that we show holds at the level of a few percent
up the TeV range. In this approximation, the dominant virtual NLO EW corrections
are incorporated exactly, while the NLO QED bremsstrahlung is first integrated
out and subsequently incorporated via YFS multi-photon
emission~\cite{Schonherr:2008av}. We compare the resulting \MEPSatNLO QCD+EW$_{\rm
virt}$ predictions for the top-quark transverse momentum distribution at
particle-level against a recent measurement performed by ATLAS in the lepton+jet
channel, based on a selection of top-quark candidates in the boosted
regime~\cite{Aad:2015hna}.

\subsection{From NLO EW to approximate EW corrections in the \MEPSatNLO framework}
\label{sec:SM_ewmerging_ttbar:ewvirt}

Based on the automated {\sc Sherpa+OpenLoops} framework fixed-order NLO EW
corrections to $t\bar t$ and $t\bar t+$jet production can readily be
computed~\cite{Cascioli:2011va,Kallweit:2014xda,Schonherr:2017qcj} . They
comprise virtual corrections that are enhanced at large energies due to the
appearance of EW Sudakov logarithms and QED bremsstrahlung corrections. For
$t\bar t+$jet production also mixed QCD-EW interference contributions arise in
4-quark channels and are included in our NLO EW predictions.

To be precise, at NLO EW, the cross section is defined as 
\begin{equation}\label{eq:SM_ewmerging_ttbar:nlo}
  \begin{split}
    \done\sigma^\text{NLO}
    =\;&\done\Phi_B\,
	  \left[
	    \mr{B}(\Phi_B)
	    +\tilde{\mr{V}}(\Phi_B)
	  \right]
	+\done\Phi_R\,
	  \mr{R}(\Phi_R)\;.
  \end{split}
\end{equation}
Therein, $\mr{B}$ is the Born matrix element including 
all PDF and symmetry/averaging factors and $\Phi_B$ is 
its accompanying phase space configuration. 
$\mr{R}$ and $\Phi_R$ are defined analogously for 
the $\mathcal{O}(\alpha)$ real correction and its phase space, while 
$\tilde{\mr{V}}$ contains the virtual $\mathcal{O}(\alpha)$ correction 
$\mr{V}$ as well as the corresponding collinear counterterm 
of the PDF mass factorisation.
At subleading orders of the strong coupling $\alpha_S$, further tree and
one-loop contributions and photon-induced contributions arise. However, as shown
in~\cite{Pagani:2016caq,Czakon:2017wor} for $t\bar t$ production these are
numerically unimportant, given for the latter a determination of the photon
content of the proton is based on the recent advances of~\cite{Manohar:2016nzj}.
In the following we consider the subleading tree contributions, but neglect any
subleading one-loop or photon-induced contributions.

In order to approximate the full expression for the NLO EW contribution in the
high-energy regime, by a form that is local in the Born phase space, and thus
suitable for straightforward incorporation in the current \MEPSatNLO framework in
Sherpa, we define the $\EWvirt$ approximation \cite{Kallweit:2015dum} as
\begin{equation}\label{eq:SM_ewmerging_ttbar:ewvirt}
  \begin{split}
    \done\sigma^\text{NLO \EWvirt}
    =\;&\done\Phi_B\,
	  \left[
	    \mr{B}(\Phi_B)
	    +\mr{V}(\Phi_B)
	    +\int_1\done\Phi_1\,\mr{R}_\text{approx}(\Phi_B\cdot\Phi_1)
	  \right]\;.
  \end{split}
\end{equation}
Here, we introduce the approximated real-emission contribution 
$\mr{R}_\text{approx}$ such that its integral over the 
real-emission phase space equals the standard Catani-Seymour 
$\mb{I}$-operator. 
This construction is both finite and correctly reproduces the 
exact NLO EW corrections in the Sudakov limit, but also 
contains important non-logarithmic terms extending its 
validity in practice.

This result can now be straightforwardly used to incorporate 
approximate electroweak corrections in the \MEPSatNLO 
method \cite{Hoeche:2009rj,Hoeche:2012yf,Gehrmann:2012yg} 
of merging multiple samples of successive multiplicities at
next-to-leading order accuracy in QCD.
For any such multiplicity $n$, $n<\nmaxnlo$,
its exclusive $n$-jet cross sections are defined as 
\begin{equation}\label{eq:SM_ewmerging_ttbar:mepsatnlo}
  \begin{split}
    \done\sigma_{n}^\text{(\MEPSatNLO)}
    \;=&\;\biggl[\done\Phi_{n}\,
          \bar{\mr{B}}_{n}(\Phi_n)\,
          \bar{\mc{F}}_{n}(\mu_Q^2\,;<\!\Qcut)
    \\
    &{}\;\;
    +\done\Phi_{n+1}\,
         \mr{H}_{n}(\Phi_{n+1})\,
         \Theta(\Qcut-Q_{n+1})\,
         \mc{F}_{n+1}(\mu_Q^2\,;<\!\Qcut)\biggr]
     \,\Theta(Q_{n}-\Qcut)\;,
  \end{split}
\end{equation}
with the familiar $\bar{\mr{B}}_n$ and $\mr{H}_n$ functions 
of the \MCatNLO matching method \cite{Hoeche:2011fd,
  Hoeche:2012ft,Hoche:2012wh} and their parton shower 
functionals $\bar{\mc{F}}_{n}$ and $\mc{F}_{n+1}$. 
To this end, we modify the standard NLO QCD 
$\bar{\mr{B}}_{n}$ function to also incorporate 
the approximate NLO EW corrections now local in $\Phi_n$, 
\begin{equation}\label{eq:SM_ewmerging_ttbar:mepsatnloewa}
  \begin{split}
    \bar{\mr{B}}_{n}(\Phi_{n})
    \;\longrightarrow\;
    \bar{\mr{B}}_{n,\QCDpEW}(\Phi_{n})
    =\;&
    \bar{\mr{B}}_{n}(\Phi_{n})
        +{\mr{V}}_{n,\EW}(\Phi_{n})
        +\mr{I}_{n,\EW}(\Phi_{n})
        +\mr{B}_{n,\rm sub}(\Phi_{n})\;.
  \end{split}
\end{equation}
Additional multiplicities can be merged at leading order 
accuracy. 
They receive a differential $K$-factor \cite{Hoche:2010kg,
  Gehrmann:2012yg,Hoeche:2014rya}, ensuring the 
continuity of the corrections applied throughout the 
spectrum. The term $\mr{B}_{n,\rm sub}$ allows for the inclusion
of additional subleading Born contributions in the merging.

\subsection{$\mathbf{t\bar t}$ and $\mathbf{t\bar t+}$jet production at NLO EW}
\label{sec:SM_ewmerging_ttbar:resultsfo}

\begin{figure*}[t]
  \includegraphics[width=.48\textwidth]{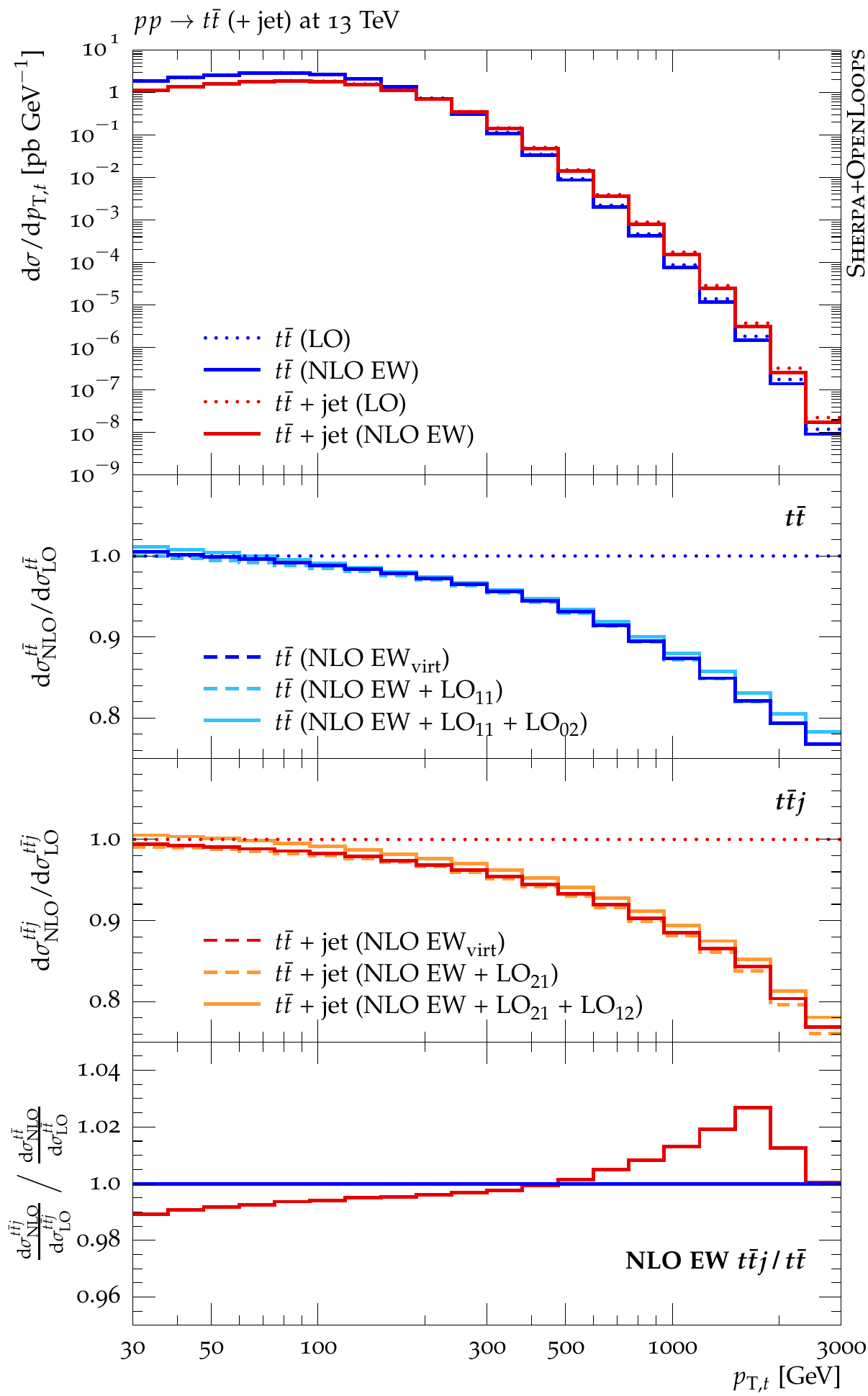}
  \hfill
  \includegraphics[width=.48\textwidth]{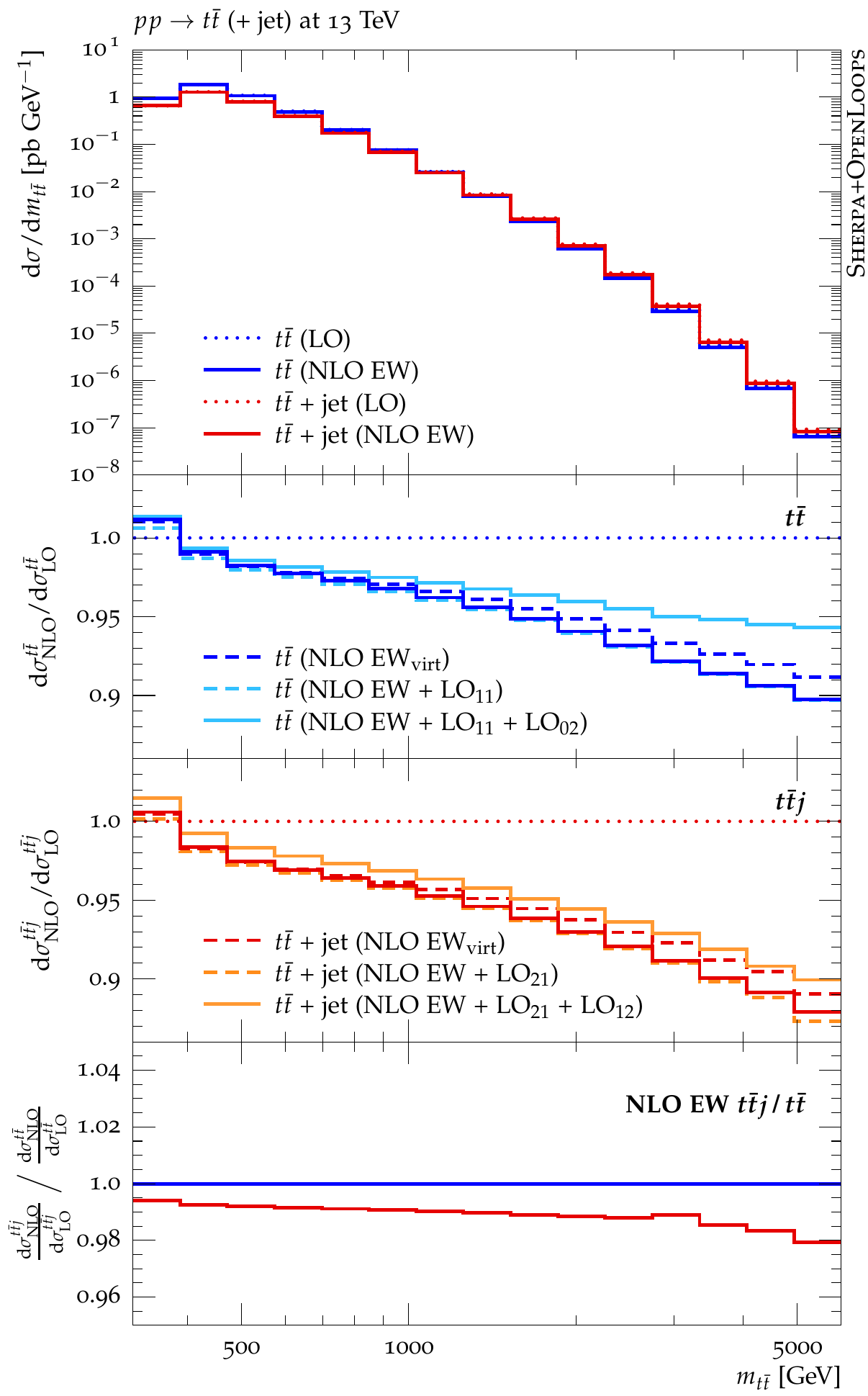}
  \caption{
    Top quark transverse momentum (left) and top-antitop 
    invariant mass (right) distributions in inclusive $\ttbar$ production 
    (blue) and $\ttbarj$ production (red) at 
    NLO EW at 13\,TeV at the LHC. 
    The top panel shows the differential cross section, 
    while the three lower panels show, from top to bottom, 
    the NLO EW corrections to inclusive $\ttbar$ production 
    and $\ttbarj$ production, respectively. 
    The lowest panel shows the ratio of both corrections.
    \label{fig:SM_ewmerging_ttbar:pTtop_mtt}
  }
\end{figure*}

In Fig.~\ref{fig:SM_ewmerging_ttbar:pTtop_mtt} we present predictions for the top-quark
transverse momentum distribution and top-pair invariant mass distribution in
top-pair and top-pair plus jet production at the LHC with $13~$TeV including NLO
EW corrections. For our predictions input parameters are chosen as listed in
Table~\ref{tab:SM_ewmerging_ttbar:inputs} and the electroweak coupling $\alpha$ is fixed and
renormalized according to the $G_\mu$-scheme, $\alpha = \frac{\sqrt{2}}{\pi}
G_{\mu} \left| \mu_{W}^2 \sin_{\theta_{\rm w}}^2 \right|$,
where $\mu_{W}$ denotes the complex-valued W mass, with $\mu_V^2=M_V^2-i\Gamma_V
M_V$ and $\theta_{\rm w}$ the also complex valued weak mixing angle, derived
from the ratio $\mu_W/\mu_Z$. The massive vector bosons and the Higgs are
renormalised in the complex-mass scheme, while the top-quark is kept stable and
correspondingly renormalised in the on-shell scheme. The introduction of finite
widths for the massive vector bosons is mandatory due to the appearance of
otherwise singular resonant internal propagators in the $\mathcal{O}(\alpha)$
bremsstrahlung. As renormalisation and factorisation scales for the strong
coupling $\alpha_S$ we use $\mu=\mu_R=\mu_F=\frac{1}{2}(E_{T,t} + E_{T,\bar
t})$, where $E_{T,t/\bar{t}}$ denotes the transverse energy of the top/anti-top.
In the predictions for top-pair plus jet production we cluster jets according to
the anti-$k_{\rm T}$ algorithm implemented in {\sc
FastJet}~\cite{Cacciari:2011ma} and require $p_{T,j} > 30$ GeV and $|\eta_{j}| <
4.5$.

\begin{table}[t]
  \begin{center}
    \begin{tabular}{rclrcl}
      $\Gmu$ & \shortequal & $1.1663787\cdot 10^{-5}~\text{GeV}^2$ 
        & \qquad & &\\
      $m_W$ & \shortequal & $80.385~\text{GeV}$  
        & $\Gamma_W$ & \shortequal & $2.0897~\text{GeV}$ \\
      $m_Z$ & \shortequal & $91.1876~\text{GeV}$ 
        & $\Gamma_Z$ & \shortequal & $2.4955~\text{GeV}$ \\
      $m_h$ & \shortequal & $125~\text{GeV}$     
        & $\Gamma_h$ & \shortequal & $4.07~\text{MeV}$ \\
      $m_t$ & \shortequal & $173.2~\text{GeV}$   
        & $\Gamma_t$ & \shortequal & $0$ 
    \end{tabular}
  \end{center}\vspace*{-5mm}
  \caption{
    Numerical values of all input parameters. While the masses are 
    taken from \cite{Agashe:2014kda}, the widths are obtained from 
    state-of the art calculations.
    \label{tab:SM_ewmerging_ttbar:inputs}
  }
\end{table}

The EW corrections to the top transverse momentum distribution shown
in Fig.~\ref{fig:SM_ewmerging_ttbar:pTtop_mtt}
display a typical Sudakov behaviour and reach -10(-20)\% at
1(2) TeV consistently in $t\bar t$ and $t\bar t+$jet production. In fact, the
difference in the relative NLO EW corrections between $t\bar t$ and $t\bar
t+$jet production is below 1(2)\% up to about $p_{T,t}=1(2)$ TeV. This indicates
a factorisation of the EW correction from QCD radiation and supports a
multiplicative combination of QCD and EW corrections for $t\bar t$ production.
The NLO EW$_{\rm virt}$ approximation introduced in Eq.~\eqref{eq:SM_ewmerging_ttbar:ewvirt}
reproduces the full NLO EW corrections in the transverse momentum of the top at
the percent level, both for $t\bar t$ and $t\bar t+$jet production. Subleading
Born contributions of $\mathcal{O}(\alpha_S\alpha)$ and $\mathcal{O}(\alpha^2)$
in $t\bar t$ production and $\mathcal{O}(\alpha_S^2\alpha)$ and
$\mathcal{O}(\alpha_S\alpha^2)$ in $t\bar t+$jet production only contribute at
the percent level.

In the invariant mass of the top-quark pair the corrections reach -5(-10)\% at
1(5) TeV, where again we observe a universality of the EW corrections between $t
\bar t$ and $t \bar t+$jet production. Here the accuracy of the NLO EW$_{\rm
virt}$ approximation is slightly worse compared to the transverse momentum
distribution. Still, the full NLO EW correction and the approximation agree at
the percent level. Subleading Born contributions are marginally relevant for $t
\bar t+$jet production, while for $t \bar t$ production the subleading
$\mathcal{O}(\alpha^2)$ contribution yields a relevant contribution at very
large top-pair invariant masses.

\subsection{Multi-jet merged predictions for $t\bar t$ production including EW corrections}
\label{sec:SM_ewmerging_ttbar:results}

\begin{figure*}[t]
  \includegraphics[width=.48\textwidth]{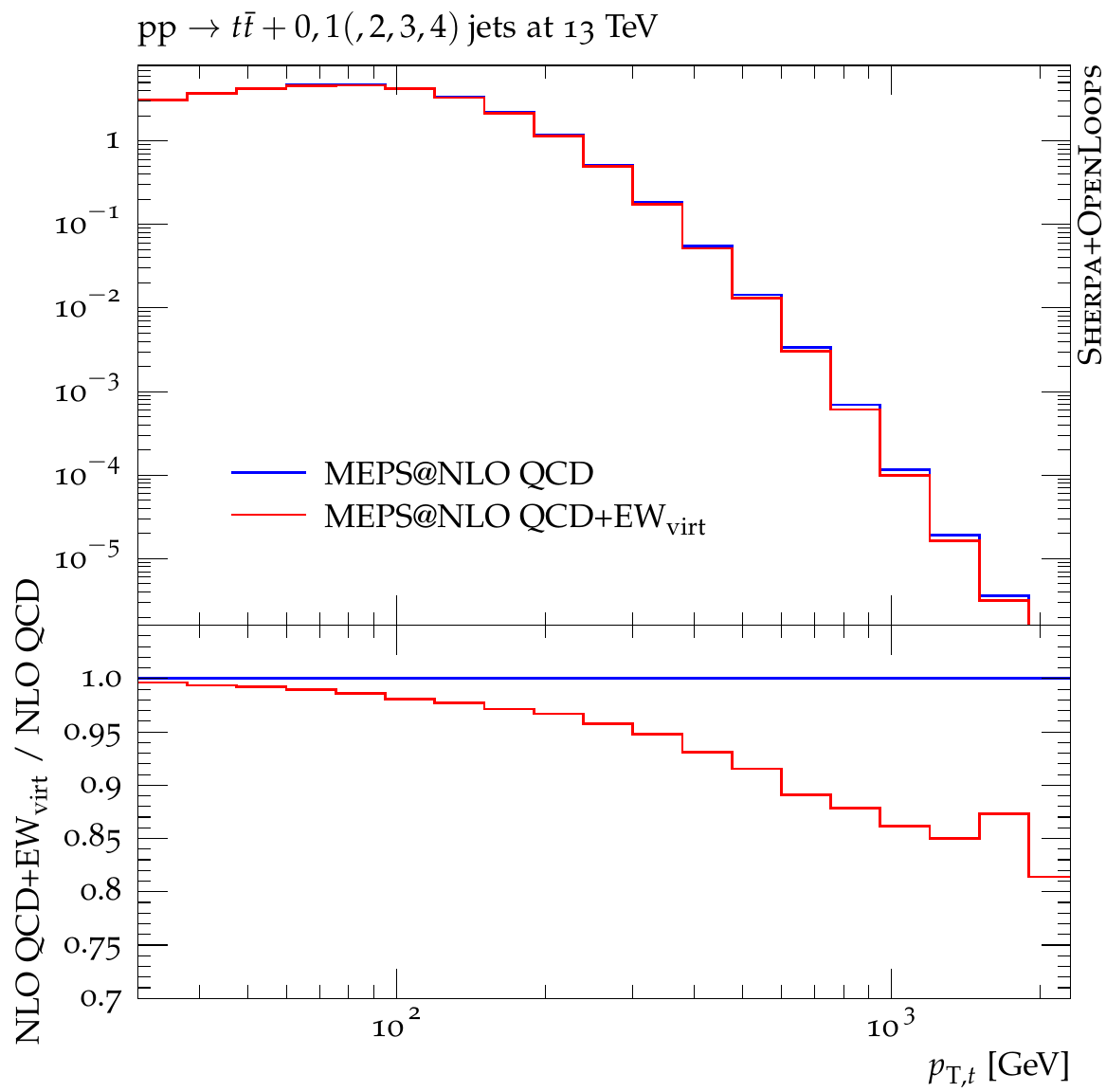}
  \hfill
  \includegraphics[width=.48\textwidth]{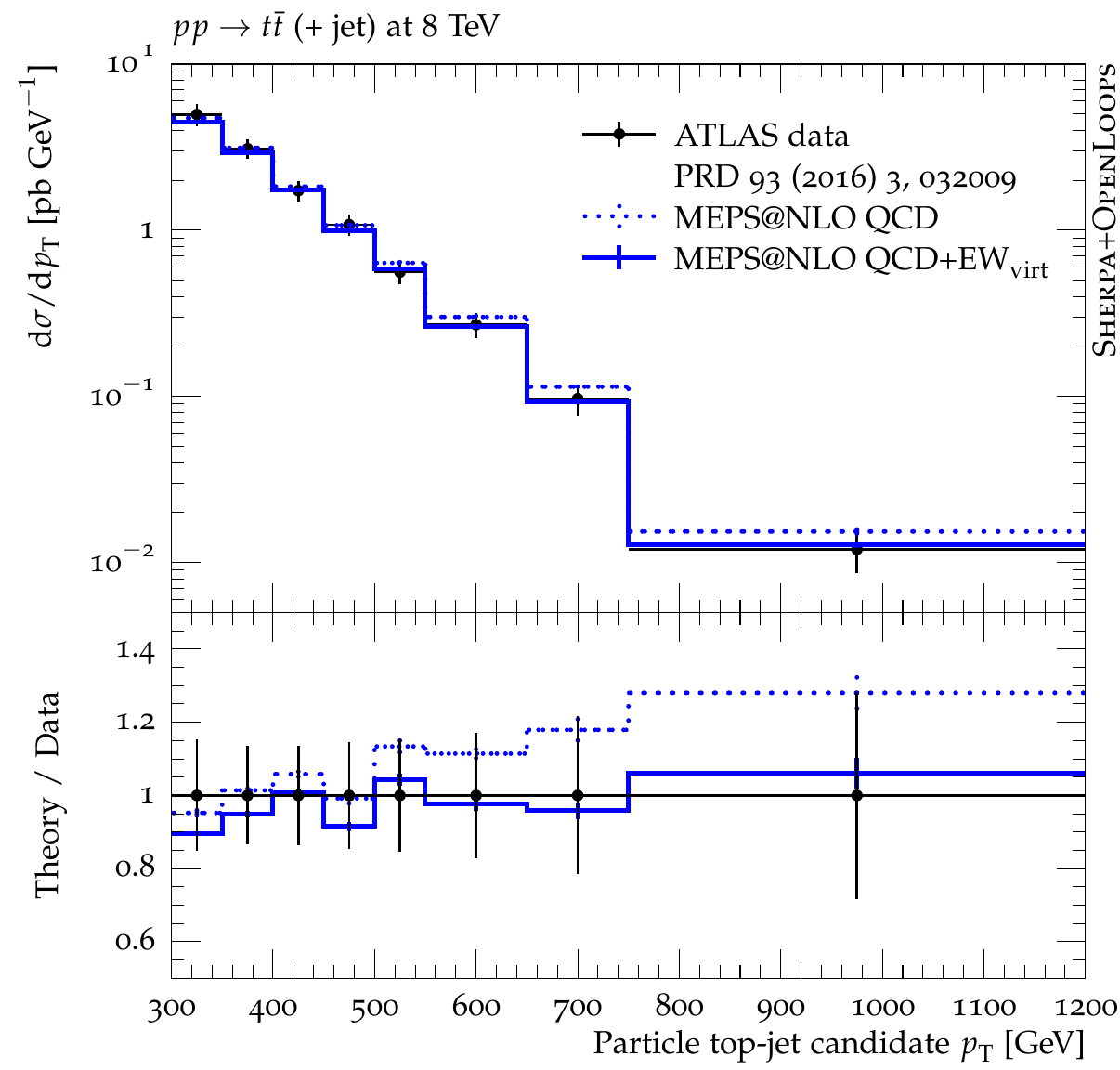}
  \caption{
   Top-quark transverse momentum distribution at parton-level (left) and particle-level (right), where for the latter hadronically decaying top candidates are reconstructed in a lepton+jets sample based on jet substructure techniques. Compared are \MEPSatNLO \QCD and 
    \MEPSatNLO \QCDpEWvirt predictions obtained with {\sc Sherpa+OpenLoops} . The particle-level predictions are also compared against a recent measurement performed by ATLAS~\cite{Aad:2015hna}.
    \label{fig:SM_ewmerging_ttbar:pTtop_mtt_merged}
  }
\end{figure*}

In Fig.~\ref{fig:SM_ewmerging_ttbar:pTtop_mtt_merged} we present on the left parton-level multi-jet 
merged \MEPSatNLO predictions  for the top-quark transverse momentum at the LHC with $13~$TeV. 
The input parameters and settings are chosen as detailed in Sec.~\ref{sec:SM_ewmerging_ttbar:resultsfo}. 
We merge $\ttbar$ plus zero and one jet production based on NLO matrix elements including 
EW corrections in the EW$_{\rm virt}$ approximation of Eq.~\eqref{eq:SM_ewmerging_ttbar:mepsatnloewa} 
and subleading Born contributions are included. 
Additional jet multiplicities up to $\ttbar+$4 jets are merged at LO accuracy. 
Comparing the \MEPSatNLO QCD predictions with the \MEPSatNLO QCD+EW$_{\rm virt}$
predictions we recover an EW correction factor consistent with the fixed-order results presented in Sec.~\ref{sec:SM_ewmerging_ttbar:resultsfo}.

In the right plot of Fig.~\ref{fig:SM_ewmerging_ttbar:pTtop_mtt_merged} we finally compare particle-level
\MEPSatNLO predictions for the reconstructed top-quark transverse momentum
distribution at the LHC with $8~$TeV against a recent ATLAS measurement of
differential top-pair production in the lepton+jets channel, based on a tagging
of hadronically decaying top-quark candidates in the boosted regime~\cite{Aad:2015hna}. In both the
QCD and the QCD+EW$_{\rm virt}$ predictions the top-quarks are decayed
preserving spin correlations \cite{Hoche:2014kca}. 
Non-perturbative effects, i.e. hadronization and MPI, are also included, as are 
higher order QED corrections \cite{Schonherr:2008av}. 
Here, scales are set using the CKKW procedure \cite{Catani:2001cc,Hoeche:2009rj}.
Thus, $\muR=\muCKKW$ is defined through 
\begin{equation}\label{eq:SM_ewmerging_ttbar:ckkwscale}
  \begin{split}
    \alphaS^{2+n}(\mu_\CKKW^2)
    =\alphaS^2(\mu_\core^2)\cdot\alphaS(t_1)\cdots\alphaS(t_n)\,,
  \end{split}
\end{equation}
and $\muF=\muQ=\mucore$, where as core scale was have \cite{Hoeche:2013mua,Hoeche:2014qda} 
\begin{equation}\label{eq:SM_ewmerging_ttbar:corescale}
  \begin{split}
    \mu_\core
    =\frac{1}{2} \left(\frac{1}{\hat s}+\frac{1}{m_t^2-\hat t}+\frac{1}{m_t^2-\hat u} \right)^{-\frac{1}{2}}\,,
  \end{split}
\end{equation}
based on a reclustered core $2\to2$ process. The remaining parameters are as discussed above.
Again zero and one jet multiplicities are merged based on NLO matrix elements.
In the high-$p_T$ tail the inclusion of the EW corrections results in a significantly improved agreement 
of the ATLAS data with the {\sc Sherpa+OpenLoops} Monte Carlo prediction.

\subsection{Conclusions}
\label{sec:SM_ewmerging_ttbar:conclusions}

In this contribution we have presented predictions for the top-quark transverse momentum distribution and top-pair invariant mass
distribution in top-pair production and top-pair plus jet production including NLO EW corrections. Subsequently, based on the
\MEPSatNLO multijet merging framework in Sherpa, we derived parton- and particle-level predictions for inclusive top-pair production
including higher-order EW corrections. The EW corrections are incorporated in an approximation, based on exact virtual NLO EW
contributions combined with integrated-out QED Bremsstrahlung. We showed that this approximation is able to reproduce the full NLO
EW result for $t\bar t$ and $t\bar t+$jet production at the percent level. Comparing our predictions against a recent measurement
for the top-quark $\pT$-spectrum performed by ATLAS in the lepton+jet channel we find very good agreement between Monte Carlo predictions 
and data when the EW corrections are included.

\subsection*{Acknowledgements}
We thank Frank Krauss and Stefano Pozzorini for useful conversations and support.
This work has received funding from the European Union's Horizon 2020 research and innovation programme as part of the Marie Sk\l{}odowska-Curie Innovative Training Network MCnetITN3 (grant agreement no. 722104).



\let\mc\undefined
\let\mr\undefined
\let\mb\undefined

\let\done\undefined
\let\nmaxnlo\undefined
\let\Qcut\undefined
\let\MEPSatNLO\undefined
\let\MCatNLO\undefined
\let\QCD\undefined
\let\EWvirt\undefined
\let\QCDpEW\undefined
\let\QCDpEWvirt\undefined
\let\Gmu\undefined
\let\shortequal\undefined
\let\ttbar\undefined
\let\ttbarj\undefined
\let\core\undefined
\let\muR\undefined
\let\muF\undefined
\let\muQ\undefined
\let\CKKW\undefined
\let\muCKKW\undefined
\let\mucore\undefined
\let\zthr\undefined
\let\alphaS\undefined



\newcommand{\powheg}{{\tt POWHEG}\xspace}
\newcommand{\powhegbox}{{\tt POWHEG-BOX\xspace}}
\newcommand{\sherpa}{{\sc Sherpa}\xspace}
\newcommand\TwoFigBottom{-2}

\newcommand{\mt      }{\ensuremath{m_{t}}\xspace}
\newcommand{\mtin    }{\ensuremath{m^{\mathrm{in}}_{t}}\xspace}
\newcommand{\mtou    }{\ensuremath{m^{\mathrm{out}}_{t}}\xspace}
\newcommand{\mlb     }{\ensuremath{m_{lb}}\xspace}
\newcommand{\mwb     }{\ensuremath{m_{Wb}}\xspace}
\newcommand{\mtwo     }{\ensuremath{m_{T2}}\xspace}
\newcommand{\mll     }{\ensuremath{m_{ll}}\xspace}
\newcommand{\etdr     }{\ensuremath{E_T^{\Delta R}}\xspace}
\newcommand{\bjet     }{\ensuremath{b}-jet}
\newcommand{\nlofull}{\mathrm{NLO_{full}}}
\newcommand{\nlodec}{\mathrm{NLO_{NWA}^{NLOdec}}}
\newcommand{\lodec}{\mathrm{NLO_{NWA}^{LOdec}}}
\newcommand{\nlops}{\mathrm{NLO_{PS}}}
\newcommand{\lops}{\mathrm{LO_{PS}}}
\newcommand{\lofull}{\mathrm{LO_{full}}}
\newcommand{\lolo}{\mathrm{LO_{NWA}^{LOdec}}}
\newcommand{\Chiq      }{\ensuremath{\chi^{2}}\xspace}
\newcommand{\Oi       }[2]{\ensuremath{{#1}_{#2}}\xspace}
\newcommand{\nprod}{n_{\mrm{prod}}}
\newcommand{\ndec}{n_{\mrm{dec}}}

\def\MSbar{$\overline{{\rm MS}}$}
\def\gev{\mathrm{\:GeV}}
\def\tev{\mathrm{\:TeV}}
\def\mrm{\mathrm}
\def\as {\ensuremath{\alpha_s}}

\section{Parton shower and off-shell effects in top quark mass determinations~\protect\footnote{
      G.~Heinrich,
      L.~Scyboz}{}}
\label{sec:MC_WWbb}

We  compare different theoretical descriptions of top quark pair
production in the di-lepton channel. 
The full NLO corrections to $pp\rightarrow W^+W^-b\bar b\rightarrow
(e^+ \nu_e)\,(\mu^- \bar{\nu}_{\mu})\,b\bar b$ production are compared
to calculations in the narrow width approximation, where the production
of a top quark pair is calculated at NLO and combined with 
different descriptions of the top quark decay: LO, NLO and via a parton shower.

\subsection{Introduction}

A precise knowledge of the top quark mass is very important for many
aspects of collider physics, ranging from electroweak precision tests
and constraints on New Physics to investigations about the vacuum
stability of the universe. 
Recent measurements at the LHC~\cite{Khachatryan:2015hba,ATLAS-CONF-2017-071,Pearson:2017jck,Castro:2017yxe}
reach an uncertainty below 1~GeV, which implies that theoretical work is demanded
to further reduce it.
As the top quark mass is not a physical observable, its definition is
 scheme dependent.  The most commonly used mass definitions are the pole mass
 and the \MSbar\ mass, where the different masses are related by a perturbative
 series, see e.g.~Refs.~\cite{Chetyrkin:1999ys,Melnikov:2000qh}.
Recent theoretical studies with regards to the definition and
extraction of the top quark mass can be found in Refs.~\cite{Beneke:2016cbu,Butenschoen:2016lpz,Kawabata:2016aya,Hoang:2017suc,Hoang:2017btd,Hoang:2017kmk,Bevilacqua:2017ipv}.

For stable top quarks, NNLO corrections to differential distributions are
known~\cite{Czakon:2015owf,Czakon:2016dgf,Czakon:2017dip} and have been combined with
NLO electroweak corrections recently~\cite{Czakon:2017wor}.
Due to their very high complexity, the NNLO fixed-order calculations have so far only
been combined with top quark decay in the narrow-width approximation
(NWA), which factorises the production and decay processes.
NLO corrections to top quark decays have been calculated in
Refs.~\cite{Bernreuther:2004jv,Melnikov:2009dn,Campbell:2012uf}, and 
NNLO QCD corrections to the decay are also known meanwhile~\cite{Brucherseifer:2013iv,Gao:2017goi}.
The combination of fixed-order NLO corrections to both production and
decay with a parton shower
in the narrow width approximation has been done in 
Ref.~\cite{Campbell:2014kua}  within
an extension of the {\tt PowHeg}~\cite{Frixione:2007vw,Alioli:2010xd} framework, called {\tt ttb\_NLO\_dec}
in the {\tt POWHEG-BOX-V2}.
Within the {\sc Sherpa}~\cite{Gleisberg:2008ta} framework, NLO QCD
predictions based on the NWA for top quark
pair production with up to three jets matched to a parton shower are
also available, see Refs.~\cite{Hoeche:2014qda,Hoche:2016elu}.

However, a  description of top quark production and decay which can
describe the {\em shapes} of distributions to an accuracy required for
improvements on current experimental precision needs to go beyond the
narrow-width approximation. 
NLO QCD calculations of $W^+W^- b\bar{b}$ production, including leptonic
decays of the $W$ bosons,  have been performed in
Refs.~\cite{Denner:2010jp,Denner:2012yc,Bevilacqua:2010qb,Heinrich:2013qaa}. 
These calculations use the 5-flavour scheme, where
the $b$-quarks are treated as massless partons.
NLO  electroweak corrections with complete off-shell effects have been
calculated in the di-lepton channel in Ref.~\cite{Denner:2016jyo}.
In the lepton+jets channel, NLO QCD corrections with full off-shell
effects are available since recently~\cite{Denner:2017kzu}.

The $b$-quark mass effects on observables like the invariant mass of a
lepton-$b$-quark pair  ($m_{lb}$)  are very small. 
However, the use of 
massive $b$-quarks (more precisely, the 4-flavour scheme, 4FNS)  has the
technically important feature that it avoids collinear singularities
due to $g\to b\bar{b}$ splittings. This implies that any phase space restrictions on
the $b$-quarks can be made without destroying infrared safety, and
thus allows  to consider 0, 1- and 2-jet bins for 
$pp\to e^+\nu_e\mu^-\bar{\nu}_\mu b\bar{b}$ in one and the same setup, 
which is important for cross sections defined by jet vetos.
In Refs.~\cite{Frederix:2013gra,Cascioli:2013wga}, NLO calculations in the 4FNS have been performed.

Based on an NLO calculation of $W^+W^- b\bar{b}$ production combined
with the {\tt Powheg} framework, first results of the $W^+W^-
b\bar{b}$ calculation in the 5-flavour scheme matched to a parton
shower have been presented in Ref.~\cite{Garzelli:2014dka}. 
However, it has been noticed subsequently that the matching of NLO matrix elements
involving resonances of coloured particles to parton showers poses
problems which can lead to artifacts in the top quark lineshape~\cite{Jezo:2015aia}.
As a consequence, an improvement of the resonance treatment has been
implemented in {\tt POWHEG-BOX-RES}, called ``resonance aware matching'', 
and combined with NLO matrix elements from {\sc OpenLoops}~\cite{Cascioli:2011va}, to arrive at 
the most complete description so far~\cite{Jezo:2016ujg}, based on the
framework developed in Ref.~\cite{Jezo:2015aia} and the 4FNS calculation of
Ref.~\cite{Cascioli:2013wga}.  An alternative algorithm to treat
radiation from heavy quarks in the {\tt Powheg} NLO+PS framework has been
presented in Ref.~\cite{Buonocore:2017lry}. 
An improved resonance treatment in the matching to parton showers for off-shell single top production at NLO
has been worked out in Ref.~\cite{Frederix:2016rdc}, and for off-shell
$t\bar{t}$ and $t\bar{t}H$ production in $e^+e^-$ collisions in Ref.~\cite{Nejad:2016bci}.
A recent study of top quark mass measurements at the LHC
using various NLO+PS approaches can be found in
Ref.~\cite{Ravasio:2018lzi}, and a detailed assessment of fragmentation
uncertainties in Ref.~\cite{Corcella:2017rpt}.

\medskip

Here  we investigate the impact of different theoretical 
approximations on distributions relevant to top quark mass
measurements, as described in more detail in
Ref.~\cite{Heinrich:2017bqp}.
We compare the NLO calculation of $W^+W^- b\bar{b}$
production of Ref.~\cite{Heinrich:2013qaa} with the calculation based
on the narrow-width approximation 
where both $t\bar{t}$ production {\it and} decay are calculated at
NLO~\cite{Melnikov:2009dn} and with a calculation 
 in the narrow-width approximation where the
 Sherpa~\cite{Gleisberg:2008ta} parton shower is combined with our
matrix elements of top quark pair production calculated at NLO.

\subsection{Theoretical descriptions}

We consider the following descriptions of the top quark pair production
cross section in the di-lepton channel:
\begin{description}
\item[~~~~~$\nlofull$:]
  full NLO corrections to $pp\to W^+W^- b\bar{b}$ with leptonic $W$-decays,
\item[~~~~~$\nlodec$:]
  NLO $t\bar{t}$ production $\otimes$ NLO decay,
\item[~~~~~$\lodec$:]
  NLO $t\bar{t}$ production $\otimes$ LO decay,
\item[~~~~~$\nlops$:]
  NLO $t\bar{t}$ production+shower $\otimes$ decay via parton showering.
\end{description}

The calculations $\nlofull$ and $\lodec$ follow the ones described in
Ref.~\cite{Heinrich:2013qaa}, but now at $\sqrt{s}=13\tev$ and with
cuts as described in Ref.~\cite{Heinrich:2017bqp}. The results for the $\nlodec$ calculation are obtained
as described in Ref.~\cite{Melnikov:2009dn}, relying on  the factorisation of the matrix elements according to
\begin{eqnarray}\label{eq:MC_WWbb:NWA1}
   \mathcal{M}^\mathrm{NWA}_{ij\to t\bar{t}\to b \bar{b} 2l 2\nu}
   \;=\;
   \mathcal{P}_{ij \to t\bar{t}} \otimes \mathcal{D}_{t\to b l^+ \nu}
   \otimes \mathcal{D}_{\bar{t}\to \bar{b} l^- \bar{\nu}}~,
\end{eqnarray}
where $\mathcal{P}_{ij \to t\bar{t}}$ describes the $t\bar{t}$ production process and $\mathcal{D}_{t\to b l \nu}$ the top quark decay dynamics.
Spin correlations are included as indicated by the symbol $\otimes$.

The $\nlops$ computations shown here are based on the
NLO plus parton-shower matching scheme as implemented in
{\sc Sherpa}~\cite{Hoeche:2011fd,Hoeche:2013mua}. 
Using this scheme, we obtain an NLO+PS accurate description of
$t\bar t$ production. The top quark decays are attached
afterwards in a way that spin correlations are preserved, and
supplemented by their respective decay showers following the same
procedure as proposed in \cite{Hoche:2014kca}.

In the following we will focus on the observable \mlb 
which is defined as the invariant mass
  \begin{equation}\label{eq:MC_WWbb:def:mlb}
    \mlb^2\;=\;(p_l+p_b)^2\;,
  \end{equation}
where $p_l$ denotes the four-momentum of the lepton and $p_b$ the
  four-momentum of the $b$-jet. As there are two top quarks, there are also
 two possible \mlb\ values per event. Since experimentally, it is not possible to reconstruct
 the $b$-quark charge on an event-by-event basis with sufficient accuracy, one also needs a
 criterion to assign a pair of a charged lepton and a $b$-jet
 as the one stemming from the same top quark decay.
Following~\cite{Aaboud:2016igd}, the algorithm applied here is to choose
that $(l^+b\textrm{-jet},\,l^-b\textrm{-jet}')$ pairing which
 minimises the sum of the two \mlb\ values per event.
 Finally, the \mlb\ observable used in the analysis is the mean of the two \mlb\
 values per event obtained when applying the above procedure.

The \mlb\ distribution has a kinematic edge at $m_{lb}^{\rm{edge}}=\sqrt{m_t^2-M_W^2}= 152.6$~GeV,
beyond which it is only populated by additional radiation, non-resonant contributions and incorrect $b$-lepton pairings.

\subsection{Phenomenological results}

We use the  PDF4LHC15\_nlo\_30\_pdfas
sets~\cite{Butterworth:2015oua,Dulat:2015mca,Harland-Lang:2014zoa,Ball:2014uwa}
and a centre-of-mass energy of $\sqrt{s}=13$~TeV.
Our default top quark mass is  $m_t=172.5\gev$.
LO top quark and $W$ boson widths are used in the LO calculations and
the NLO $t\bar{t}\, \otimes$ LO decay calculation, 
while NLO widths~\cite{Jezabek:1987nf} are used in the remaining NLO calculations.
NLO widths appearing in propagators are not expanded in $\alpha_s$.
The QCD coupling in the NLO widths is varied according to the chosen
scale. For more details on the input parameters an kinematic requirements we refer to Ref.~\cite{Heinrich:2017bqp}.
The $b$-quarks are treated as massless in all fixed-order calculations.

We chose $\mu_R=\mu_F=m_{t}$ as our central scale. The impact of
choosing $H_T/2$ (rather than $m_t$) as the central scale on the top
quark mass determined by our method has been shown to be very
small~\cite{Heinrich:2013qaa}. Further it would be difficult to
come up an $H_T$ definition for the $\nlops$ approach that matches
the one in the full $W^+W^-b\bar{b}$ calculation.

For the  $\nlops$ case, the standard $\mu_R$ and $\mu_F$ variations that we employ for our
fixed-order calculations are not fully sufficient to assess the
theory uncertainties, as the
showering depends on further scale and parameter settings. 
Our variation in the $\nlops$ case, denoted by
$\mu_F\mu_R\as^{\mathrm{PS}}$, is a combination of simultaneously
varying $\mu_F$, $\mu_R$ and $\mu_R^\mrm{PS}\sim p_T^\mrm{emit}$ by a
factor of two up and down. By evaluating $\as(\mu_R^\mrm{PS})$, the
latter variation determines the shower uncertainty which we abbreviate
by $\as^{\mathrm{PS}}$.
Other ways of uncertainty assessment include the variation of the
shower starting scale $\mu_Q$,
which regulates the overall size of the resummation regime primarily
affecting the $t\bar t$ production showers. The {\sc Sherpa} default
is to set this resummation scale equal to the factorisation scale,
i.e.~for our $\nlops$ results, we have set $\mu_Q=\mu_F=m_t$.

\begin{figure}[tbp!]
\centering
\includegraphics[width=0.7\textwidth]{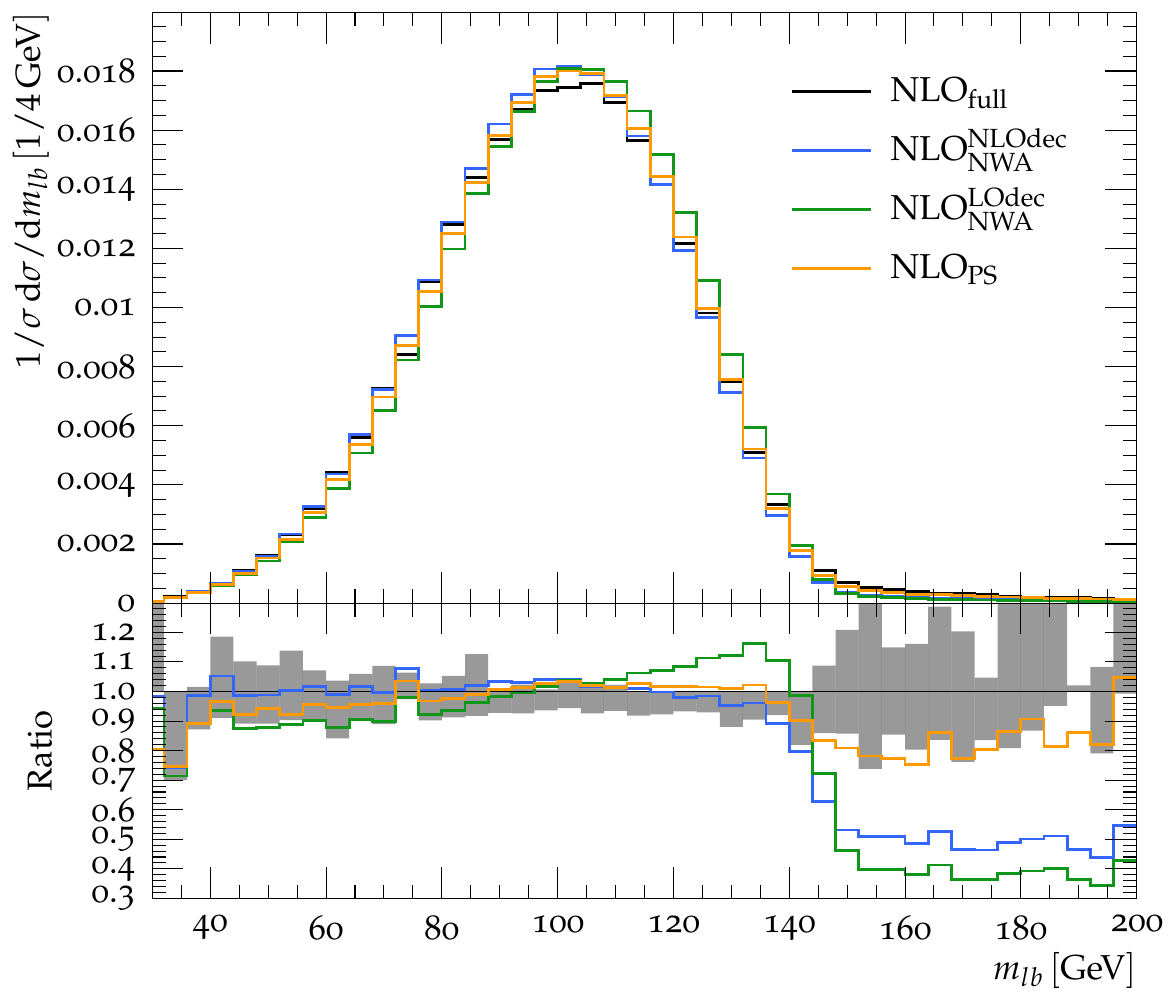}
\caption{\label{fig:MC_WWbb:scalevar_mlb}%
  Normalised differential cross sections for the invariant mass
  \mlb\ at the 13~TeV LHC for four different theoretical
  descriptions: full NLO calculation including off-shell and
  non-resonant contributions ($\nlofull$), NLO calculation in the NWA
  with and without NLO corrections to the top quark decay ($\nlodec$
  and $\lodec$, respectively) and NLO calculation for top quark pair
  production matched to a parton shower  ($\nlops$). The ratio
  of all descriptions to $\nlofull$ including its scale uncertainty
  band is also shown.}
\end{figure}

Experimentally, the top quark mass is measured most precisely by
fitting normalised differential cross sections to Monte Carlo
templates produced with different values for the top mass. 
This procedure is therefore sensitive to the modeling of the shape of the distributions entering the fit. 
Figure~\ref{fig:MC_WWbb:scalevar_mlb} shows the differential cross section for the \mlb\ observable based on the four theoretical predictions
$\lodec$, $\nlodec$, $\nlops$ and $\nlofull$, where the latter is the baseline of the ratio plot, and its
scale uncertainties are shown as grey bands. We observe that the simplest description, $\lodec$, differs substantially from the $\nlofull$ prediction, 
in the range $40 \gev \leq m_{lb} \leq 150\gev$, which comprises the bulk of the cross section, 
effectively moving the peak to greater values of $m_{lb}$.
The NLO corrections to the top quark decay, $\nlodec$, restore agreement with $\nlofull$ to within less than $10\%$, while
the $\nlops$ prediction mostly lies between $\lodec$ and $\nlodec$ in this region.
In contrast, the high-mass tail features substantial differences between the predictions, ranging from $70\%$ for $\lodec$, $50\%$ for $\nlodec$, 
to  $20\%$ for $\nlops$. 
Taking only the range $40 \gev \leq m_{lb} \leq 150\gev$ into consideration, 
which is also within the range used for fits determinng the top quark mass based on the \mlb\ distribution~\cite{Heinrich:2017bqp}, 
the NWA predictions $\nlodec$ and $\nlops$, which describe the top quark decay 
beyond the leading order, but do not contain non-resonant, singly-resonant and non-factorising contributions, 
come rather close to the full $W^+W^-b\bar{b}$ calculation. 
Thus we conclude that the non-resonant and off-shell contributions are less important than 
corrections to the decay for top quark mass determinations based on the \mlb\ observable.

\begin{figure}[tbp!]
\centering
\subfloat[]{\label{fig:MC_WWbb:mlb_scalevar}
\includegraphics[width=0.48\textwidth]{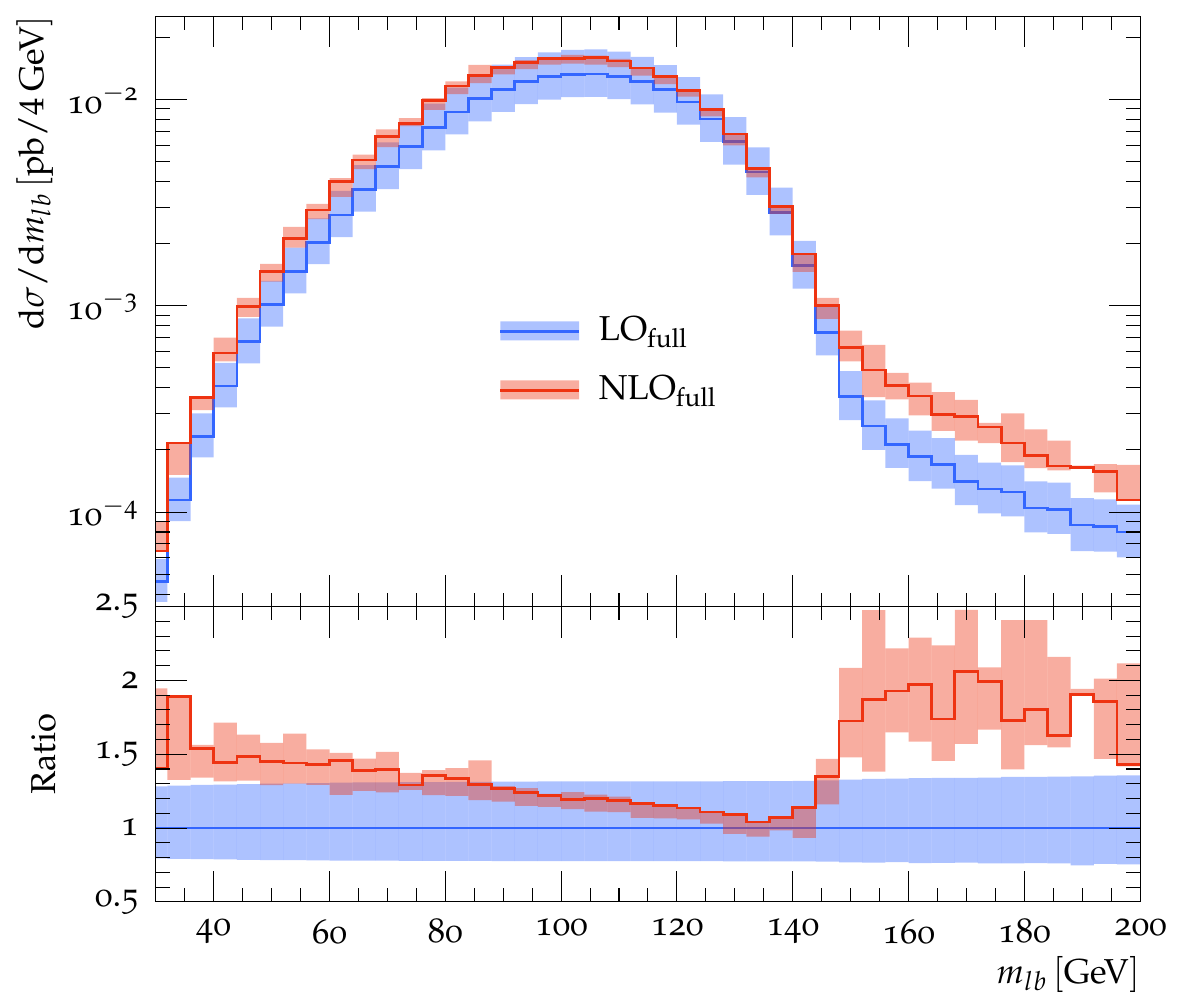}
}
\hfill
\subfloat[]{\label{fig:MC_WWbb:mlb_scalevar_nwa}
\includegraphics[width=0.48\textwidth]{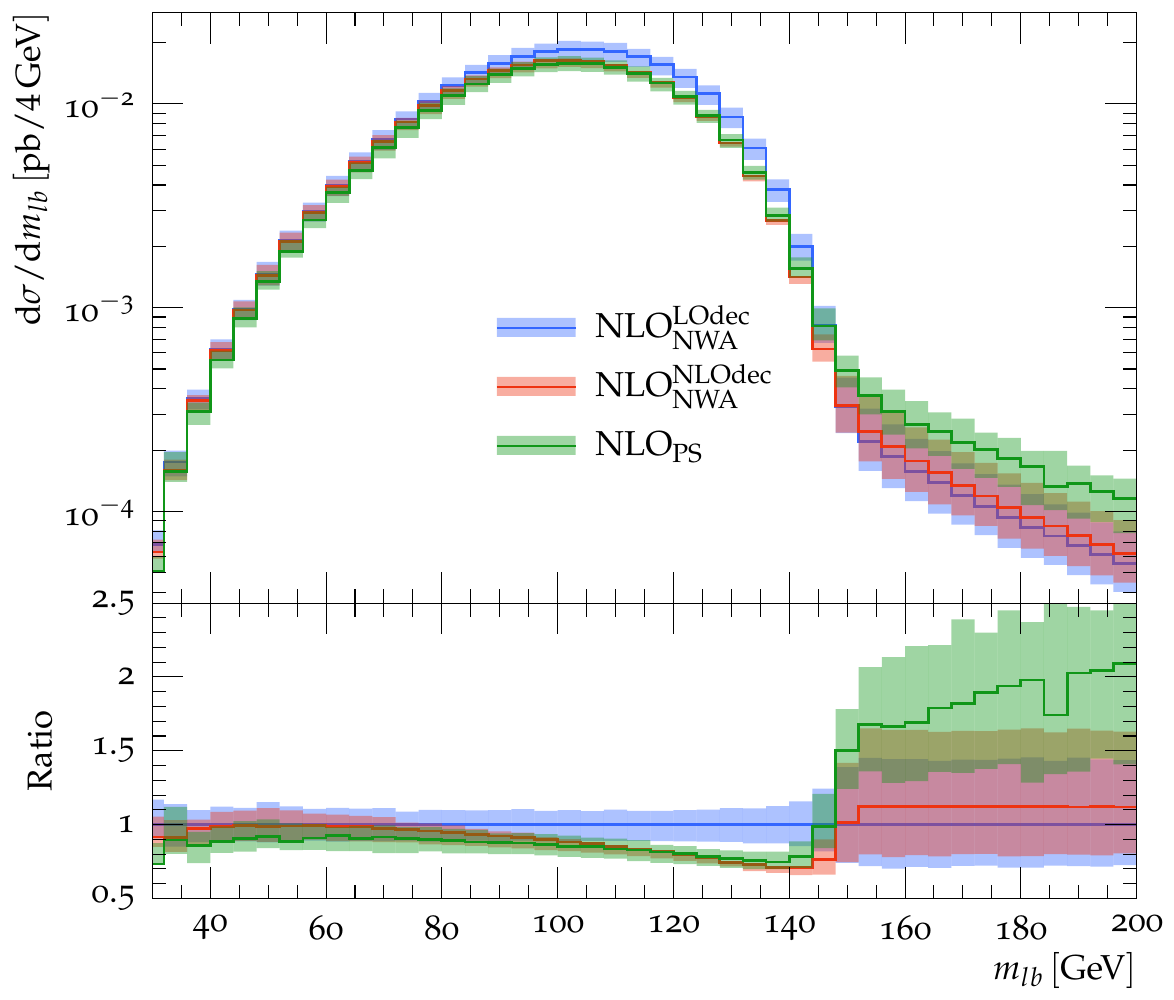}
}\\[2ex]
\caption{Results including scale variation bands for $m_{lb}$, for (a) the $\lofull$ and $\nlofull$ calculations, (b) the calculations based on the NWA.
The ratio plot in (a) is with respect to $\lofull$, while the ratio plot in (b) is with respect to $\lodec$.}
\end{figure}

In the case of the full $W^+W^-b\bar{b}$ calculation, the NLO corrections introduce important changes in the shape of the \mlb\ distribution, 
as shown in Fig.~\ref{fig:MC_WWbb:mlb_scalevar}. While this is expected for the tail of the distribution, it is remarkable that the result also gets corrections 
of up to $50\%$ for lower \mlb\ values. The various descriptions of the top quark decay in the NWA exhibit a different behaviour, as shown in 
Fig.~\ref{fig:MC_WWbb:mlb_scalevar_nwa}: when compared to the $\lodec$ central prediction, both the $\nlodec$ and the $\nlops$ predictions soften the 
distribution up to the kinematic edge, while the region beyond the kinematic edge gets substantially enhanced by the parton shower. 
Notice that incorporating higher-order corrections to the decay pushes the \mlb distribution outside of the $\lodec$ scale uncertainty bands, 
also in the peak region.

\begin{figure}[tbp!]
\centering
\subfloat[]{\label{fig:MC_WWbb:mlb_restricted}
\includegraphics[width=0.48\textwidth]{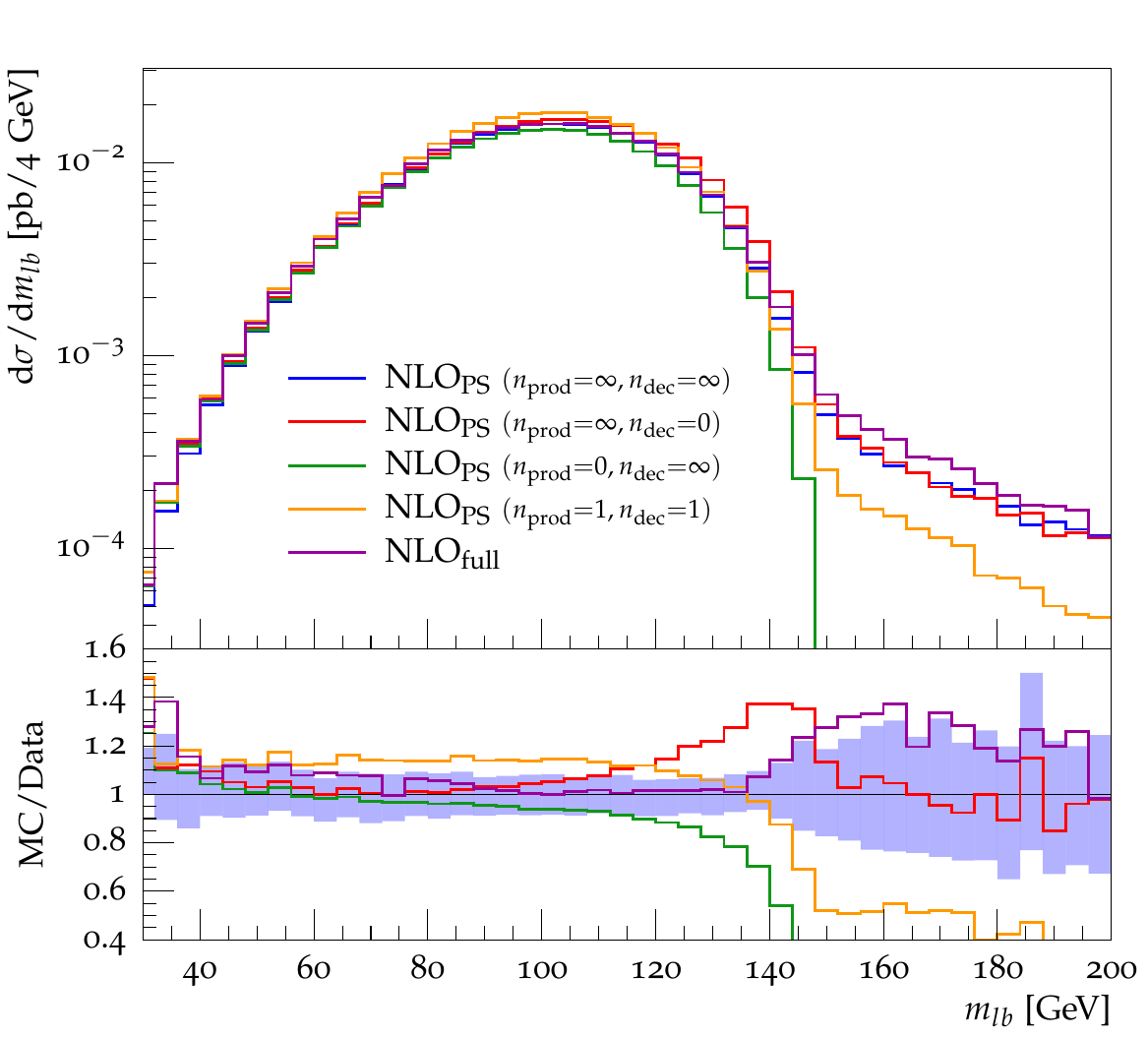}
}
\hfill
\subfloat[]{\label{fig:MC_WWbb:mwb_restricted}
\includegraphics[width=0.48\textwidth]{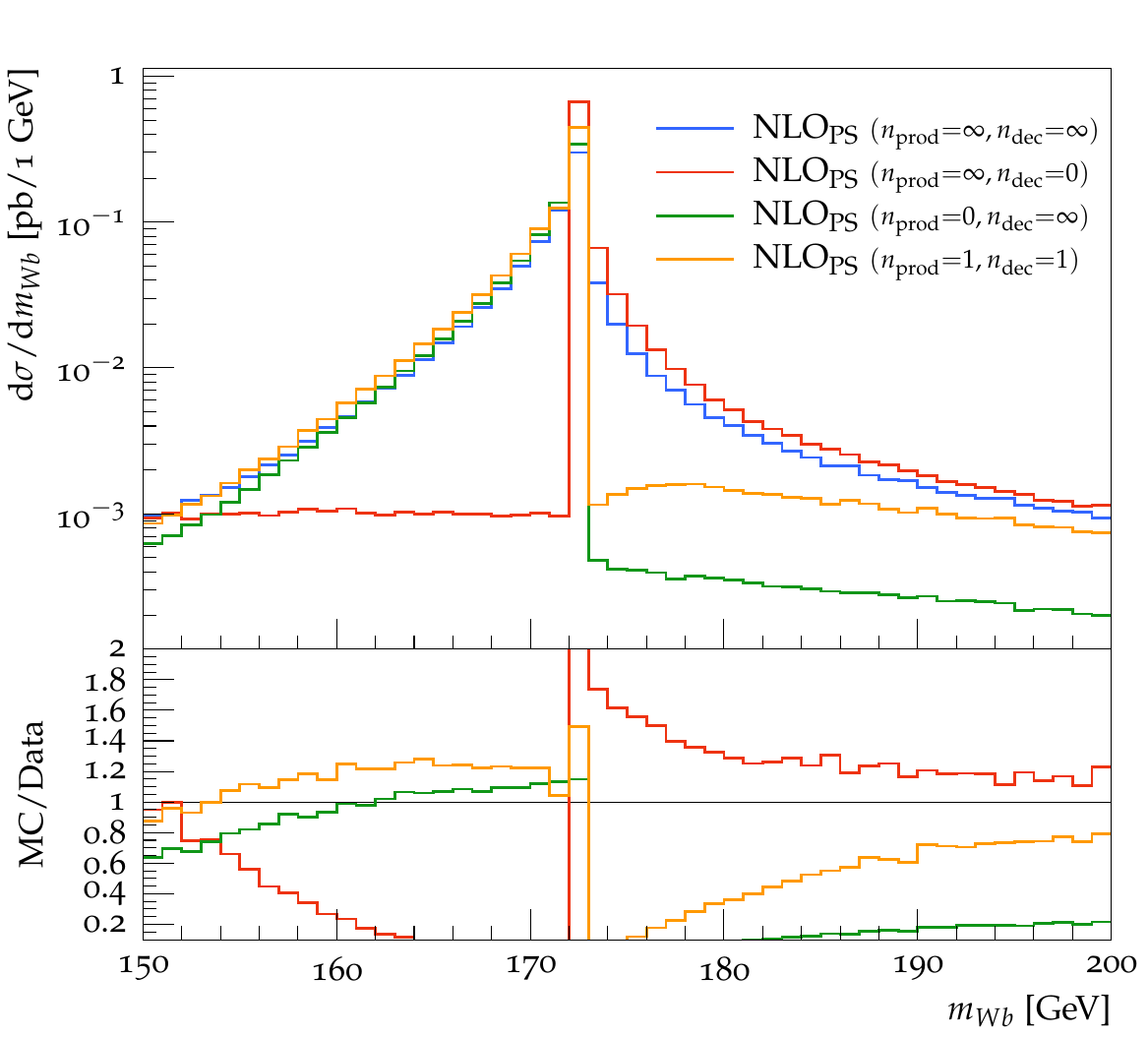}
}\\[2ex]
\caption{Truncated shower emissions for the observabls (a) \mlb, (b)
  \mwb. The shaded area in the ratio plot left denotes the scale
  uncertainty of the full $\nlops$ prediction.}
\label{fig:MC_WWbb:restricted_shower}
\end{figure}

As a complementary study, and in order to disentangle the effect of additional emissions in the parton shower before and after the decay, 
we now consider the $\nlops$ description in more detail. 
The parton shower can be artificially truncated after a given number of emissions $\nprod$ and $\ndec$ in the $t\bar{t}$ production, 
respectively in the top quark decay. In Fig.~\ref{fig:MC_WWbb:mlb_restricted}, the unnormalised differential cross section for the \mlb\ observable is shown 
for four combinations of $\nprod$ and $\ndec$, namely 
\begin{eqnarray*}
(A) & (\nprod=\infty,\ndec=\infty)\\
(B) & (\nprod=\infty,\ndec=0)\\
(C) & (\nprod=0,\ndec=\infty)\\
(D) & (\nprod=1,\ndec=1)~,
\end{eqnarray*}
where $n=\infty$ symbolically denotes the full shower emission (with default cutoff $\mu_Q$). 
Comparing (A) to (B), that is the full shower and the deactivated
shower in the top quark decay, we observe that the high-\mlb\ tail is
still described in a similar way in both cases, 
while the ratio (B)/(A) peaks sharply just before the turnover around the kinematic edge at $\sim 145 \gev$. 
On the other hand, comparing (A) and (C), this time deactivating the shower in the top quark pair production, 
the agreement between the full and truncated shower is fair in the
bulk of the distribution, while the ratio (C)/(A)  drops in the high-\mlb region. 
This is because there is no additional (mainly initial state) radiation in the production, 
which would further dilute the kinematic edge by radiating into the b-jet cone.
If we consider case (D), where one emission is allowed in both the production and the decay shower, we can anticipate that allowing more emissions would 
progressively flatten the truncated curves shown in the ratio plot to
the full shower baseline. 
We also observe that for \mlb\ values between $40\gev$ and $140\gev$,
the $\nlofull$ result is within the uncertainty band of the $\nlops$ result.

The underlying dynamics is exhibited more clearly if we look at a more theoretical observable, 
which does not suffer from the spread in phase space coming from the unknown neutrino four-momentum. 
We define the invariant mass of the $W$ and b-jet system as
  \begin{equation}\label{eq:MC_WWbb:def:mwb}
    \mwb^2\;=\;(m_{lb} + p_{\nu})^2\;=(p_l + p_b + p_{\nu})^2\;,
  \end{equation}
where the same assignment for lepton and b-jets is used as in Eq.~\eqref{eq:MC_WWbb:def:mlb}, and $p_{\nu}$ designates the four-momentum of the neutrino corresponding to the lepton $l$. Again, the value of \mwb used in the analysis is the mean of the two \mwb values per event. 

Figure~\ref{fig:MC_WWbb:mwb_restricted} describes the truncation of the parton shower for the \mwb observable. 
In the case of the full shower (A), it peaks at the top mass $m_t = 172.5\gev$, and is spread by the real radiation on both sides. Going to the case (B), 
which corresponds to switching off the shower for the top quark decay,
the spread of the peak drops to a constant baseline for \mwb\ values below the peak, 
while even exceeding the $\nprod=\infty$ result for \mwb\ values beyond the peak. 
This behaviour can be understood from the fact that, if there is no radiation off the top quark at all, larger $m_t$ values are preferred. 
In addition, initial state radiation being clustered into the b-jet increases the \mwb\ values.
Mirroring this situation, if the production shower is switched off as in case (C), the left side of the peak follows the fully-showered result (A) within $40\%$. 
The shower contribution above the top mass peak, in contrast, is reduced to almost zero, as 
additional radiation in the decay shower can only decrease the invariant mass of the $Wb$ system.
In case (D), we observe that one emission in the decay shower suffices to bring the curve below the peak to its final, fully-showered level. 
In contrast, the distribution only slowly builds up with each emission from the production shower in the high-mass tail.

These considerations show that the shape of observables sensitive to the top quark mass, like \mlb and \mwb, 
is considerably affected by higher order corrections to both, production and decay.

\subsection{Conclusions}

We have investigated how various theoretical descriptions of top
quark pair production translate into 
the behaviour of the distributions \mlb and \mwb
 in view of 
measurements of the top quark mass in the
di-lepton channel. In particular, we have compared the NLO QCD results
for $W^+W^- b\bar{b}$ production ($\nlofull$) to results based on the narrow-width
approximation, combining $t\bar{t}$ production at NLO with (i) LO top
quark decays ($\lodec$), (ii) NLO top quark decays
($\nlodec$) and (iii) a parton shower ($\nlops$). 
We have found that corrections beyond the leading order in the {\it decay} ($\nlodec$ and $\nlops$ descriptions) 
are important, being always closer to the $\nlofull$ result than the leading order $W^+W^- b\bar{b}$ result, 
even though the latter also contains non-resonant and non-factorising contributions.

We have further seen that the $\nlodec$ and $\nlops$ descriptions are pretty close to each other within an \mlb\ range that is usually taken into account for template fits to measure the top quark mass. 
We also considered  parton showers with restricted emissions, restricting to zero or one emission in production and/or decay.
The behaviour of the \mlb and \mwb\ distributions in these cases shows that both the production and the decay shower play an important role 
to arrive at the fact that the $\nlops$ result, based on the narrow-width approximation, comes quite close to the $\nlofull$ result.

 \subsection*{Acknowledgements}
 We would like to thank our collaborators on this project, Andreas
 A. Maier, Richard Nisius, Johannes Schlenk and Jan Winter.
 We also thank the organizers of the ``LH2017 Phyisics at TeV Colliders''
 workshop for creating a productive and pleasant environment.
 This research was
 supported in part by the Research Executive Agency (REA) of the
 European Union under the Grant Agreement PITN-GA2012316704
 (HiggsTools).



\let\powheg\undefined
\let\powhegbox\undefined
\let\sherpa\undefined
\let\TwoFigBottom\undefined

\let\mt\undefined
\let\mtin\undefined
\let\mtou\undefined
\let\mlb\undefined
\let\mwb\undefined
\let\mtwo\undefined
\let\mll\undefined
\let\etdr\undefined
\let\bjet\undefined
\let\nlofull\undefined
\let\nlodec\undefined
\let\lodec\undefined
\let\nlops\undefined
\let\lops\undefined
\let\lofull\undefined
\let\lolo\undefined
\let\Chiq\undefined
\let\Oi\undefined
\let\nprod\undefined
\let\ndec\undefined

\newcommand{\Sherpa}{S\protect\scalebox{0.8}{HERPA}\xspace}
\newcommand{\Herwig}{H\protect\scalebox{0.8}{ERWIG}7\xspace}
\newcommand{\Matchbox}{M\protect\scalebox{0.8}{ATCHBOX}\xspace}
\newcommand{\MGaMC}{M\protect\scalebox{0.8}{AD}G\protect\scalebox{0.8}{RAPH}5\_aMC@NLO\xspace}
\newcommand{\MoCaNLO}{M\protect\scalebox{0.8}{oCaNLO}\xspace}
\newcommand{\Recola}{R\protect\scalebox{0.8}{ecola}\xspace}
\newcommand{\VBFNLO}{V\protect\scalebox{0.8}{BFNLO}\xspace}

\section{Study of electroweak production of WZ in association with two jets at the LHC~\protect\footnote{
    K.~Long,
    M.~Pellen (section coordinators);
    S.~Br\"auer,
    V.~Ciulli,
    S.~Gieseke,
    M.~Herndon,
    M.~Mozer,
    S.~Pl{\"a}tzer,
    M.~Rauch,
    E.~Yazgan}{}}
\label{sec:MC_vbs}

\subsection{Introduction}
\label{sec:MC_vbs:intro}

The electroweak (EW) production of vector-boson pairs in association with two jets at the CERN Large Hadron Collider (LHC) forms an important class of processes both theoretically and experimentally.
This signature includes vector-boson scattering (VBS) contributions, which constitute the principle example of processes where the scattering of two massive gauge bosons can be observed at the LHC.
These processes provide a natural probe of the vector-boson quartic couplings, which arise due to the non-Abelian nature of the electroweak gauge group and are exactly predicted in the Standard Model (SM).
The Higgs boson plays a unique role in these interactions, preventing the cross section from diverging in the high-energy limit and preserving the unitarity of the associated scattering amplitudes.
Deviations in these channels could therefore indicate physics beyond the SM in the electroweak sector.

The measurements of these processes are particularly challenging due to their high multiplicities and small cross sections.
They have been observed and even measured only recently by the experimental collaborations at the LHC.
The most precise current measurement \cite{Aad:2014zda,Khachatryan:2014sta,Sirunyan:2017ret,Aaboud:2016ffv} concerns the scattering of two same sign W bosons (usually denoted ${\rm W}^\pm{\rm W}^\pm{\rm j}{\rm j}$).
This process has a unique signature due to the same-sign charged leptons in the final state.
The cross section is well within reach with the large data sets being collected in the LHC Run II, and the
signature is experimentally accessible due to the precise lepton identification, charge assignment, and momentum resolution of the LHC experiments.
The nature of the final state is also attractive due to the low rate of background processes
producing two prompt, same sign leptons associated with forward jets (referred to as the irreducible background).

Measurements of other VBS signatures such as ${\rm Z}{\rm Z}{\rm j}{\rm j}$, ${\rm W}^+{\rm W}^-{\rm j}{\rm j}$, or ${\rm W}^{\pm}{\rm Z}{\rm j}{\rm j}$ present additional challenges due to the lower cross section and larger irreducible backgrounds, which dominate over the VBS contribution in most regions of phase space.
Nonetheless, studies have already been made at the LHC of both the ${\rm Z}{\rm Z}{\rm j}{\rm j}$ \cite{Sirunyan:2017fvv} and ${\rm W}^{\pm}{\rm Z}{\rm j}{\rm j}$ \cite{Aad:2016ett} signature at $\sqrt{s} = 8$~TeV.
In these cases, separation of the electroweak component of the $VV{\rm j}{\rm j}$ state depends on exploiting the different characteristics of the EW and non-EW components through kinematic selections.
Accurate predictions for discriminating distributions and a detailed understanding of their associated uncertainties therefore directly impact such measurements.
A theory-agnostic measurement that does not attempt separation of states by production mode avoids such dependencies,
but cannot fully leverage the statistical tools and treatment of uncertainties available in an experimental analysis.
The approaches are therefore complementary, and are often presented together in experimental results.

From a theoretical point of view, one of the challenges for the predictions of the EW di-boson production in association with two jets is its high multiplicity.
This is why in the past, next-to-leading order (NLO) predictions have focused on VBS approximations.
Only recently, full NLO computations became available at NLO QCD and EW for both the EW component and the QCD-induced process for ${\rm W}^\pm{\rm W}^\pm{\rm j}{\rm j}$ \cite{Biedermann:2017bss}.
For this process, preliminary results \cite{Anders:2018gfr} of a comparison of different theoretical predictions have shown that differences between the full computation and VBS-approximated ones are not significant given the present experimental accuracy.
Qualitatively, one could expect a similar conclusion for other processes such as ${\rm W}^{\pm}{\rm Z}{\rm j}{\rm j}$ but a quantitative check is still needed.
Therefore, in these proceedings we aim at making a similar comparison of theoretical predictions implemented in Monte Carlo programs.
This allows to infer the quality of the VBS approximation at leading-order (LO).

As such there is strong motivation for an investigation of the theoretical predictions, tools, and uncertainties used in a typical VBS analysis. 
We focus on the ${\rm W}^{\pm}{\rm Z}{\rm j}{\rm j}$ \cite{Aad:2016ett} state, the measurement of which is strongly limited by the size of the data set, 
but which has not yet been studied at the 13 TeV LHC. Hence, a preliminary study on this process is of particular interest.

Due to its colour structure, the ${\rm W}^{\pm}{\rm Z}{\rm j}{\rm j}$ signature possesses three different contributions at LO.
The first, of order $\mathcal{O} (\alpha^6)$, is usually referred to as the EW component or even VBS component (even if it also possesses non-VBS contributions such as tri-boson production).
The two quark lines can also be connected via a gluon while the gauge bosons are radiated off the quark lines.
This contribution is of order $\mathcal{O} (\alpha_{\rm s}^2\alpha^4)$ and is called QCD contribution/background.
Finally, there exists a non-zero interference of order $\mathcal{O} (\alpha_{\rm s}\alpha^5)$.
For this signature (as opposed to \emph{e.g.}\ ${\rm W}^\pm{\rm W}^\pm{\rm j}{\rm j}$) the EW component is highly suppressed with respect to the QCD one.
This implies having very exclusive experimental cuts in order to enhance the EW contribution and a good control over the description of both the EW signal and QCD background.
To that end, the understanding of theoretical predictions and Monte Carlo programs is key.
In particular, it is vital to ensure that all different programs can provide equivalent physics result.
Such comparisons can also shed light on the true uncertainties of such predictions.
Typically, uncertainty treatments asses the impact of parton density function (PDF) uncertainties or of missing higher orders in perturbative QCD.
But in addition, Monte Carlo generators require many input parameters and the selection of which can have a significant impact on the predictions both in shape and normalisation.
In the present study we use a common set-up for all predictions at LO accuracy. We additionally provide results with
variations in some parameters and comment on their effects and motivations.

In these proceedings, we start with Sec.~\ref{sec:MC_vbs:theory} where a short review of the theoretical state-of-the-art predictions is given.
The programs used in the present work are also briefly described.
In Sec.~\ref{sec:MC_vbs:setup}, the set-up of the calculation is presented.
It amounts to give the input parameters as well as the event selection.
Section~\ref{sec:MC_vbs:results} is devoted to the results of the study.
It starts with LO predictions for the three different contributions to the ${\rm e}^+  \nu_{\rm e}  \mu^+ \mu^- {\rm j} {\rm j}$ final state at the LHC.
Then the various comparisons are reported at both fixed-order and with parton shower.
Finally, Sec.~\ref{sec:MC_vbs:concl} contains a summary, concluding remarks as well as recommendations.

\subsection{Theory and event generators}
\label{sec:MC_vbs:theory}

The study focuses on the ${\rm W}^{\pm}{\rm Z}{\rm j}{\rm j}$ signature and more precisely on the partonic processes

\begin{equation}
 {\rm p} {\rm p} \to {\rm e}^+  \nu_{\rm e}  \mu^+ \mu^- {\rm j} {\rm j} ,
\end{equation}
and
\begin{equation}
 {\rm p} {\rm p} \to {\rm e}^-  \bar \nu_{\rm e}  \mu^+ \mu^- {\rm j} {\rm j}.
\end{equation}
The predictions presented are all at LO but NLO QCD corrections to the EW contribution and its irreducible background are known since 10 years in the VBS approximation \cite{Bozzi:2007ur} while the QCD corrections to the QCD-induced process have been computed more recently \cite{Campanario:2013qba}.
The NLO EW corrections are currently unknown.
In Ref.~\cite{Biedermann:2016yds}, it has been argued that large NLO EW corrections to the EW contributions are an intrinsic feature of VBS at the LHC.
Therefore, they are expected to play a significant role for all VBS signatures.
In Ref.~\cite{Biedermann:2017bss}, which focuses on the computation of the full NLO corrections to the ${\rm W}^\pm{\rm W}^\pm{\rm j}{\rm j}$ process, it has been shown that the EW corrections to the EW process are the dominant NLO corrections.
This means that the EW corrections to the EW contributions are expected to be at least of the same order as the QCD corrections for other VBS signatures.
Parton-shower effects to VBS processes and the resulting uncertainties
induced by a variation of factorisation, renormalisation, and
shower-starting scale have also been studied
recently~\cite{Rauch:2016upa,Rauch:2016pai}, taking the ${\rm W}^+{\rm
W}^-{\rm j}{\rm j}$ VBS process as an example. Thereby, the standard LHC
jet definition with a cone radius of $R{=}0.4$ leads to migration
effects~\cite{Rauch:2017cfu} and a reduction of the cross section by about $10\%$ compared to
the fixed-order results after VBS cuts.

In the following, the codes used for the predictions are briefly described.

\subsubsection*{\protect\MGaMC}
\label{sec:MC_vbs:mgamc}

{\sc MadGraph5\_aMC@NLO}~\cite{Alwall:2014hca} is an automatic meta-code (a code that generates codes) which makes it possible to simulate any scattering process
      including NLO QCD corrections both at fixed-order and including matching to parton showers. 
      The commands that have been used for the present computation (for the W$^{-}$ case) are 
\begin{verbatim}
> set complex_mass_scheme
> set gauge Feynman
> generate p p > e- ve~ mu+ mu- j j QED=6 QCD=0
> output WmZJJToENu2MuJJ
\end{verbatim}
  The version used is 2.6.0. We use the complex mass scheme and Feynman gauge for consistency with \MoCaNLO+\Recola, but 
  verify that we obtain equivalent results if they are not explicitly specified.
  We note that the ``deltaeta'' variable in the MG5\_aMC run\_card also applies a requirement that the two jets
  have opposite rapidity sign. We do not make this requirement for our fiducial region definition, so we remove this condition for our comparisons.
  This is accomplished by modifying the cut behaviour in cuts.f in the ``SubProcesses'' folder, and validated by
  generating events without this condition and applying selections to the LHE-level partons using standalone code.
  
  \subsubsection*{\protect\Herwig}
  \label{sec:MC_vbs:herwig}

Based on extensions of the previously developed \Matchbox
module~\cite{Platzer:2011bc}, the \Herwig event generator~\cite{Bellm:2015jjp,Bahr:2008pv} facilitates the automated set-up of all ingredients necessary for a full NLO QCD calculation.
It relies on an implementation of the Catani--Seymour dipole
subtraction method~\cite{Catani:1996vz,Catani:2002hc}, as well as
interfaces to a list of external matrix element providers -- either
at the level of squared matrix elements, based on extensions of the
BLHA standard~\cite{Binoth:2010xt,Alioli:2013nda,Andersen:2014efa} or
at the level of colour-ordered sub-amplitudes.

For this study the relevant tree-level matrix elements have been
provided by an interface to VBFNLO \cite{Arnold:2008rz,Arnold:2011wj,Baglio:2014uba} using an extension of the BLHA
accord. We have matched these calculations to the angular ordered
shower using the subtractive matching and default settings of the
\Herwig~7.1 release. The PDF sets that have been used to tune the parton shower are MMHT2014lo68cl and
MMHT2014nlo68cl~\cite{Harland-Lang:2014zoa}.
These sets are not the one used for this computation (see below).
This can in particular impact initial-state radiations.
We consider variations of the hard shower scale as detailed in Ref.~\cite{Bellm:2016rhh}, using the 'resummation' profile scale choice.

\subsubsection*{\protect{\MoCaNLO\!+\Recola}}
\label{sec:MC_vbs:MoCaNLO_Recola}

The program {\sc MoCaNLO+Recola} is made of a flexible Monte Carlo program dubbed {\sc MoCaNLO} and of the general matrix element generator {\sc Recola}~\cite{Actis:2012qn,Actis:2016mpe}.
The program can compute arbitrary processes in the Standard Model with NLO QCD and EW accuracy.
The fast integration is ensured by using similar phase-space mappings to those of Refs.~\cite{Berends:1994pv,Denner:1999gp,Dittmaier:2002ap}.
The complex-mass scheme~\cite{Denner:1999gp,Denner:2005fg} to treat unstable particles is always used.
These tools have been successfully used for the computation of NLO corrections for high-multiplicity processes and in particular VBS processes \cite{Biedermann:2016yds,Biedermann:2017bss}.

\subsubsection*{\protect\Sherpa}
\label{sec:MC_vbs:sherpa}
\Sherpa~\cite{Gleisberg:2008ta,Gleisberg:2003xi} is a multipurpose event generator for high-energy particle collisions. 
It is built out of several algorithms and modules tailored to many different physics challenges of collider physics.
In the \Sherpa framework the infrared divergences appearing in the real-emission are treated by the Catani--Seymour dipole-subtraction method~\cite{Catani:1996vz,Catani:2002hc,Gleisberg:2007md}. 
The default parton-shower algorithm of \Sherpa is based on the Catani--Seymour factorisation~\cite{Schumann:2007mg,Hoeche:2009xc}.
For NLO computations, interfaces to several one-loop generators exist, with the interface to \textsc{Recola}~\cite{Actis:2012qn,Actis:2016mpe} being the latest addition~\cite{Biedermann:2017yoi}.
For correctly combining the matrix elements in multijet-production processes with the parton shower, \Sherpa has adapted the MEPS@LO method~\cite{Hoeche:2009rj} at LO and its generalisation MEPS@NLO~\cite{Hoche:2010kg} at NLO. 
In this study, the matrix elements are provided by \textsc{Comix}~\cite{Gleisberg:2008fv}, one of the two built-in generators next to \textsc{Amegic}~\cite{Krauss:2001iv}.
Both of the studied partonic processes are then showered with the default shower.

\subsubsection*{\protect\VBFNLO}
\label{sec:MC_vbs:VBFNLO}
\VBFNLO~\cite{Arnold:2008rz,Arnold:2011wj,Baglio:2014uba} is a flexible
Monte Carlo event generator for processes with electroweak bosons.
Besides the Standard Model, selected processes can also be calculated in
a variety of new-physics models, including effective field theories with
dimension-6 and dimension-8 operators.
The matrix elements for VBF and VBS processes are calculated in the VBS
approximation. The corresponding $s$-channel contribution, which can be
seen as triboson production where one vector boson decays hadronically,
is available as well, but its contribution not included in the studies
presented here.
Its use of leptonic tensors in the calculation of the matrix elements
can lead to a significant speed improvement compared to automatically
generated code.
For results with parton showers in this study, \VBFNLO can serve as the
matrix-element provider and phase-space generator for \Herwig. The
interface between the two programs is based on an extension of the BLHA
standard~\cite{Binoth:2010xt,Alioli:2013nda,Andersen:2014efa}. When
combining \VBFNLO results with showering performed by Pythia, event
files following the Les Houches LHE file
standard~\cite{Boos:2001cv,Alwall:2006yp} are used instead.

\subsection{Details of the set-up}
\label{sec:MC_vbs:setup}

In this section we describe default input parameters that have been used.
Also, the event selection used in the comparison is described.

\subsubsection*{Input parameters}

All simulations are performed for the LHC running with a center-of-mass energy $\sqrt s = 13 {\rm~TeV}$.
The default PDF used is the NNPDF~3.0 set~\cite{Ball:2014uwa} with four active flavour at LO and a strong coupling constant $\alpha_{\rm s}\left( M_{\rm Z} \right) = 0.130$.\footnote{Its {\tt lhaid} in LHAPDF6~\cite{Buckley:2014ana} is 263400.} 
If this default PDF is not employed, it is explicitly stated.
The masses and widths of the particle used in the simulations are
\begin{alignat}{2}
                M_{\rm t}   &=  173.21 {\rm~GeV},             & \quad \quad \quad \Gamma_{\rm t} &= 0 {\rm~GeV},  \nonumber \\
                M_{\rm Z}^{\rm OS} &=  91.1876{\rm~GeV},      & \quad \quad \quad \Gamma^{\rm OS}_{\rm Z} &= 2.4952{\rm~GeV},  \nonumber \\
                M_{\rm W}^{\rm OS} &=  80.385{\rm~GeV},       & \Gamma^{\rm OS}_{\rm W} &= 2.085{\rm~GeV},  \nonumber \\
                M_{\rm H} &=  125.0{\rm~GeV}, 		      & \Gamma_{\rm H}   &=  4.07 \times 10^{-3}{\rm~GeV}.
\end{alignat}
The value of the mass and width of the Higgs boson are those recommended by the Higgs cross section working group \cite{deFlorian:2016spz}.
The top quark does not appear at tree level in the simulations when the bottom-quarks in the initial state are neglected.
Therefore its width is set to zero.
The numerical values used in the simulation for the pole mass/width of the gauge bosons ($V={\rm W,Z}$) are obtained from the measured on-shell (OS) values for the masses and widths according to Ref.~\cite{Bardin:1988xt} as:
\begin{equation}
        M_V = M_{\rm V}^{\rm OS}/\sqrt{1+(\Gamma_{\rm V}^{\rm OS}/M_{\rm V}^{\rm OS})^2}\,,\qquad  \Gamma_V = \Gamma_{\rm V}^{\rm OS}/\sqrt{1+(\Gamma_{\rm V}^{\rm OS}/M_{\rm V}^{\rm OS})^2}.
\end{equation}
The EW coupling is renormalised in the $G_\mu$ scheme \cite{Denner:2000bj} where
\begin{equation}
    G_{\mu}    = 1.16637\times 10^{-5}{\rm~GeV}^{-2}.
\end{equation}
The input parameters above yield a numerical value for $\alpha$ of
\begin{equation}
 \alpha = 7.555310522369 \times 10^{-3}.
\end{equation}
Note that for the EW contribution of order $\mathcal{O} (\alpha^6)$, no dependence on the strong coupling appears.
For contributions (interference or QCD-induced contributions) where there is a dependence on $\alpha_{\rm s}$, the numerical value used is the one extracted from the PDF set.

For the renormalisation and factorisation scales, two choices have been adopted.
For the fixed scale, it is
\begin{equation}
 \mu = \mu_{\rm fix} = M_{\rm W},
\end{equation}
while the dynamical scale used is
\begin{equation}
 \mu = \mu_{\rm dyn} = {\rm Max}\left[p_{\rm T, j}\right].
\end{equation}
The latter should be understood as the maximum of the transverse momenta of the tagging jets (defined below).
This observable is closely connected to the momentum transfer through
the virtual vector boson, which has been shown to be a reasonable scale
choice, and is also a reasonable choice when used as the starting scale
of the parton shower~\cite{Rauch:2016upa}.
These two scales are the default fixed and dynamical scale, respectively.
If the scale used is different, this is explicitly stated in the text.

Photon-induced as well as bottom-induced contributions have been neglected.
The photon contributions are expected to be small \cite{Biedermann:2017bss} while the bottom-induced contribution can lead to single-top resonant contributions.
The later can in principle be isolated thanks to kinematic constraints.

\subsubsection*{Event selection}

Following experimental studies \cite{Aad:2016ett,CMS-PAS-SMP-14-008}, the event selection used for the present work is:

\begin{itemize}
\item All charged leptons are required to have
    \begin{align}
        \label{eq:MC_vbs:cut:1}
         p_{\rm T, \ell} >  20{\rm~GeV},\qquad |y_{\ell}| < 2.5.
    \end{align}
\item For the leptons of opposite charge and same flavour, an invariant mass cut to single out the Z-boson resonance is applied:
    \begin{align}
        \label{eq:MC_vbs:cut:2}
         76 {\rm~GeV} < m_{l_i^+ l_i^-} < 106 {\rm~GeV}.
    \end{align}

\item QCD jets are clustered thanks to the anti-$k_T$ algorithm~\cite{Cacciari:2008gp} with radius parameter $R=0.4$.
      At least two jets are required to have
        \begin{align}
        \label{eq:MC_vbs:cut:3}
         p_{\rm T, j} >  30{\rm~GeV}, \qquad |y_{\rm j}| < 4.7, \qquad \Delta R_{\rm j \ell} > 0.4,
        \end{align}
        and are called tagging jets.
\item On the two leading tagging jets, typical VBS cuts are applied:
        \begin{align}
        \label{eq:MC_vbs:cut:4}
         m_{\rm j j} >  500{\rm~GeV},\qquad |\Delta y_{\rm j j}| > 2.5.
        \end{align}
\end{itemize}

These cuts have been issued either directly in the Monte Carlo programs or using a {\sc Rivet} routine \cite{Buckley:2010ar}.
This file will be made public in order to make the present study easily reproducible.

\subsection{Results}
\label{sec:MC_vbs:results}

\subsubsection*{Several contributions for one process}

\begin{figure}[t]
\begin{center}
   \includegraphics[width=0.48\textwidth]{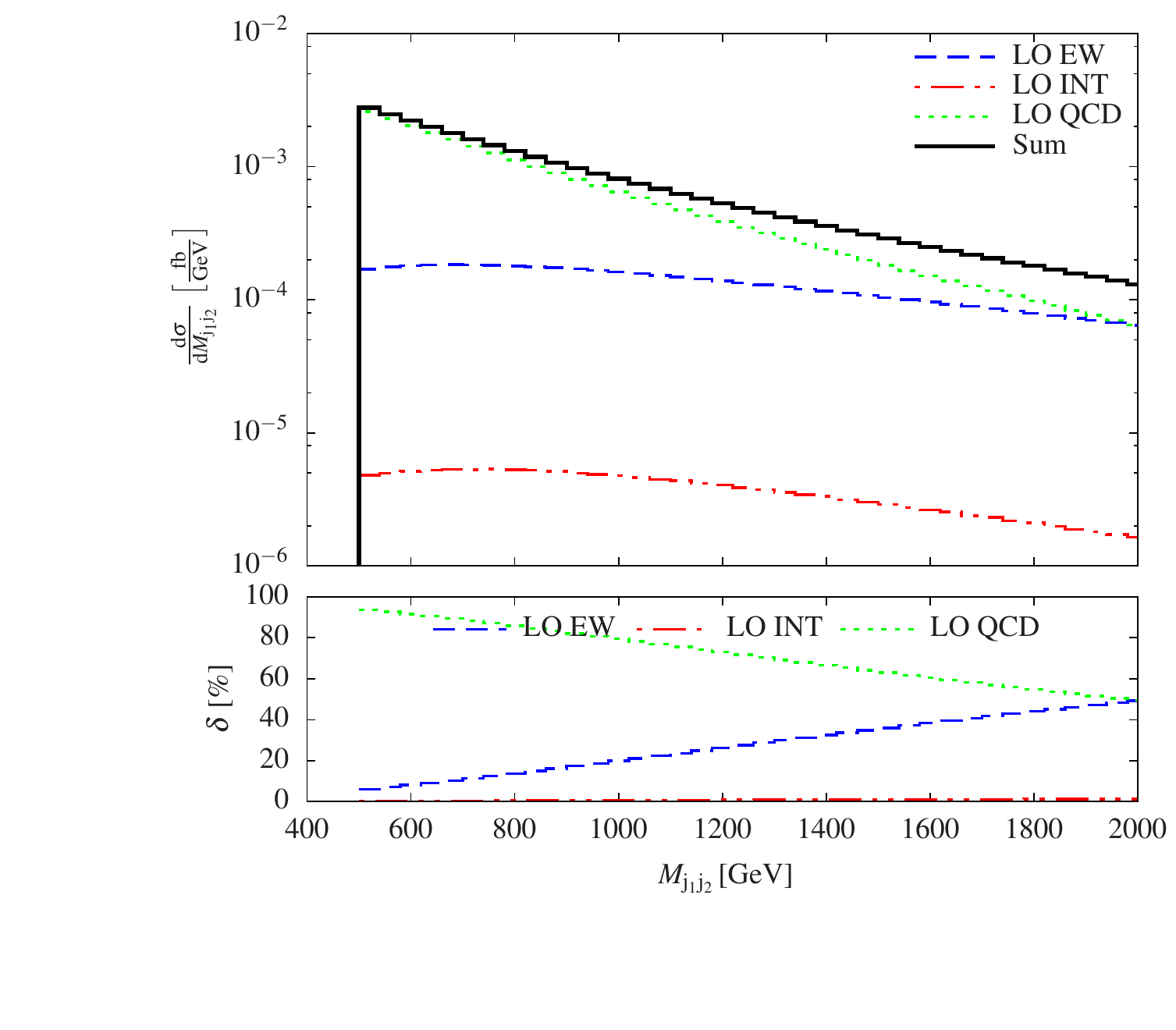}\hfill
   \includegraphics[width=0.48\textwidth]{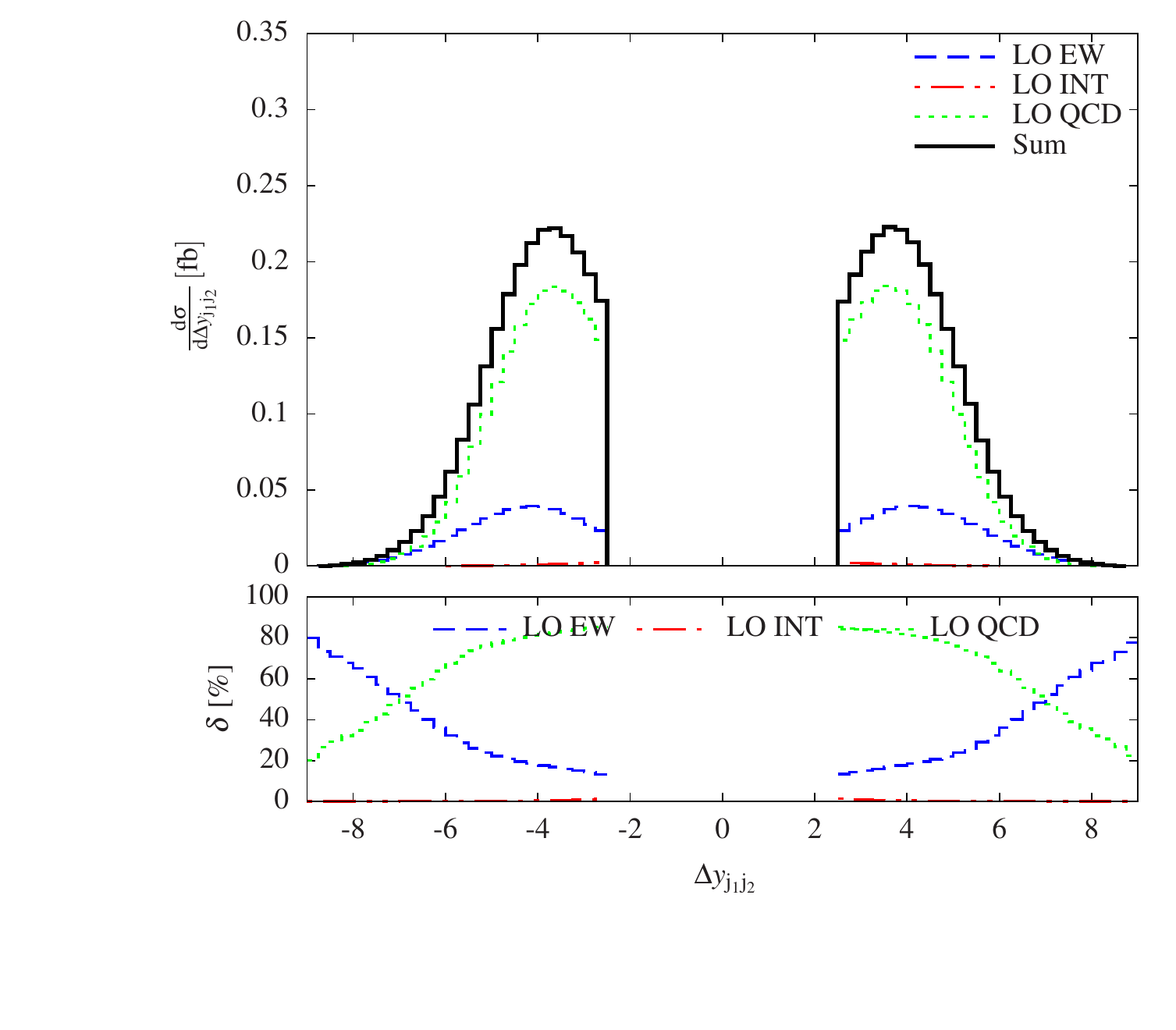}\\
   \includegraphics[width=0.48\textwidth]{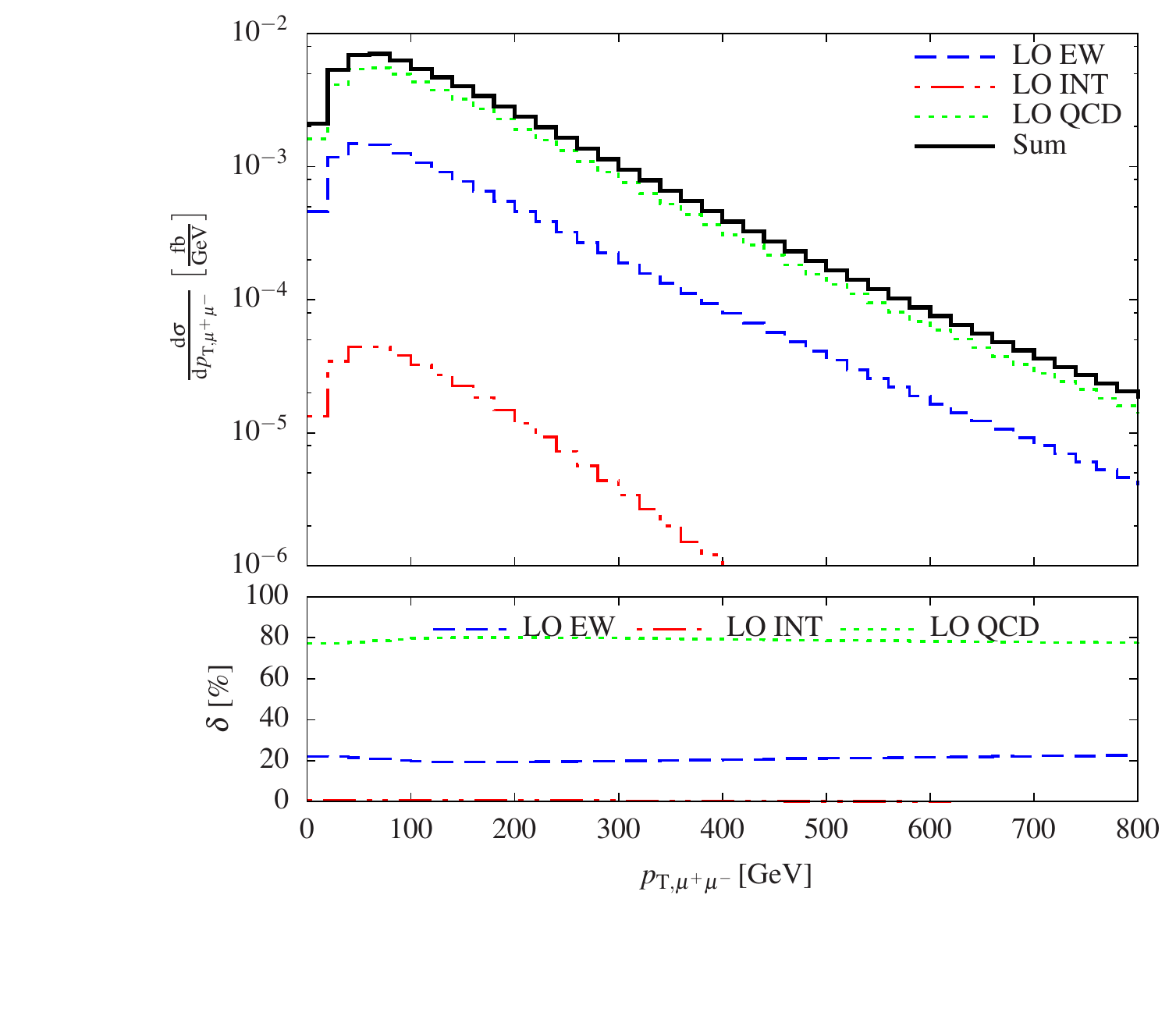}\hfill
   \includegraphics[width=0.48\textwidth]{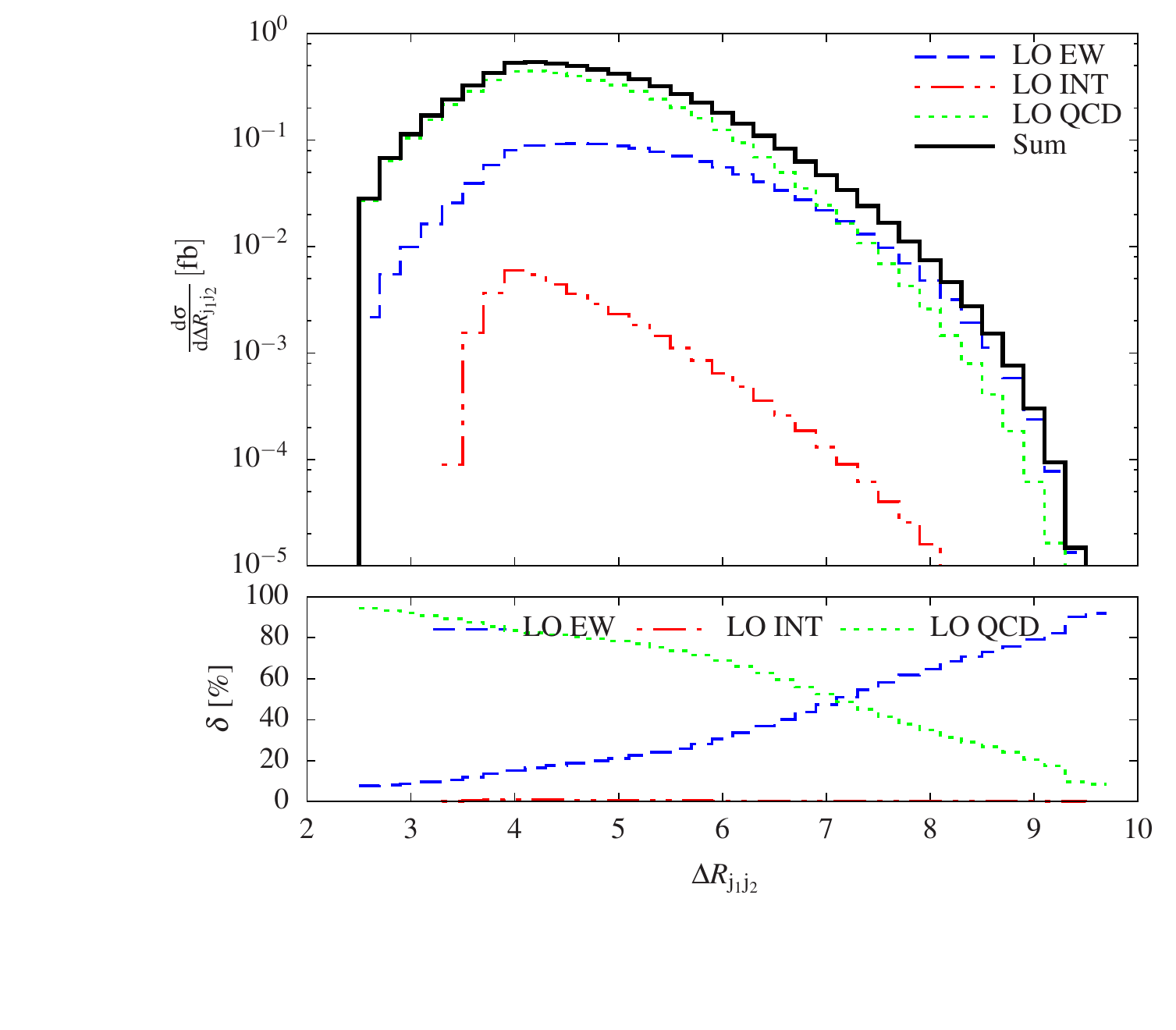}
\caption{Differential distributions at a centre-of-mass energy $\sqrt{s}=13{\rm~TeV}$ at the LHC for ${\rm p} {\rm p} \to {\rm e}^+  \nu_{\rm e}  \mu^+ \mu^- {\rm j} {\rm j}$ 
with the contributions at orders $\mathcal{O} (\alpha^6)$ (EW), $\mathcal{O} (\alpha_{\rm s}\alpha^5)$ (INT), and $\mathcal{O} (\alpha_{\rm s}^2\alpha^4)$ (QCD): 
                invariant mass of the two jets~(top left),
                rapidity separation between the two jets~(top right)
                transverse momentum of the anti-muon--muon system~(bottom left), and
                distance between the two jets~(bottom right).
                The upper panels show the LO predictions as well as their sum.
                The lower panels display the respective contributions normalised to their sum.}
\label{fig:MC_vbs:diffcontr}
\end{center}
\end{figure}

As explained previously, the processes ${\rm p} {\rm p} \to {\rm e}^+  \nu_{\rm e}  \mu^+ \mu^- {\rm j} {\rm j}$  and ${\rm p} {\rm p} \to {\rm e}^-  \bar \nu_{\rm e}  \mu^+ \mu^- {\rm j} {\rm j}$
possess at LO three contributions of orders $\mathcal{O} (\alpha^6)$, $\mathcal{O} (\alpha_{\rm s}\alpha^5)$, and $\mathcal{O} (\alpha_{\rm s}^2\alpha^4)$.
As it can be seen in Table~\ref{tab:MC_vbs:xsectallLOdyn}, the EW component represents only about $20\%$ of the total cross section\footnote{In this subsection and in particular for the results of Table~\ref{tab:MC_vbs:xsectallLOdyn} and Fig.~\ref{fig:MC_vbs:diffcontr}, the cut $|y_{\rm j}| < 4.5$ as been used instead of $4.7$ as everywhere else. This has a $1\%$ effect on the fiducial cross section.}.
This is in contrast with the ${\rm W}^\pm{\rm W}^\pm{\rm j}{\rm j}$ signature where the EW component represents almost $90\%$ of the cross section \cite{Biedermann:2017bss} in a comparable fiducial volume.
Hence measuring the EW component is much more challenging.
Therefore, inferring the shape of the signal and irreducible background is key.
Note that the interference contribution is about $0.5\%$ which is negligible with respect to current experimental accuracy.\footnote{For the ${\rm W}^\pm{\rm W}^\pm{\rm j}{\rm j}$ signature, the interference contribution has been shown to be around $3\%$ \cite{Biedermann:2017bss}.}

\begin{table}
\begin{center} 
\begin{tabular}{ c | c | c }
 $\mu = \mu_{\rm dyn}$ / $\sigma_{\rm LO}$ [fb] & ${\rm p} {\rm p} \to {\rm e}^+  \nu_{\rm e}  \mu^+ \mu^- {\rm j} {\rm j}$  & ${\rm p} {\rm p} \to {\rm e}^-  \bar \nu_{\rm e}  \mu^+ \mu^- {\rm j} {\rm j}$  \\
  \hline\hline
  $\mathcal{O} (\alpha^6)$                        & $0.25416(6)$  & $0.15003(3)$   \\
  $\mathcal{O} (\alpha_{\rm s}\alpha^5)$          & $0.006833(6)$ & $0.003977(3)$  \\
  $\mathcal{O} (\alpha_{\rm s}^2\alpha^4)$        & $0.9912(2)$   & $0.6306(6)$   \\
  \hline
\end{tabular}
\end{center}
\caption{
Fiducial cross sections at LO for the process ${\rm p}{\rm p}\to{\rm e}^+\nu_{\rm e}\mu^+\mu^-{\rm j}{\rm j}$ and ${\rm p}{\rm p}\to{\rm e}^-\bar\nu_{\rm e}\mu^+\mu^-{\rm j}{\rm j}$ at orders $\mathcal{O} (\alpha^6)$, $\mathcal{O} (\alpha_{\rm s}\alpha^5)$, and $\mathcal{O} (\alpha_{\rm s}^2\alpha^4)$.
The predictions are expressed in fb and are for the LHC running at a centre-of-mass energy of $\sqrt{s}=13 {\rm~TeV}$.
The scale used in the simulations is $\mu = \mu_{\rm dyn} = {\rm Max}\left[p_{\rm T, j}\right]$.
The integration errors of the last digits are given in parentheses.}
\label{tab:MC_vbs:xsectallLOdyn}
\end{table}

In Fig.~\ref{fig:MC_vbs:diffcontr}, several differential distributions are shown.
In the upper plot the absolute predictions for each component as well as their sum are displayed.
In the lower plot, each contribution is normalised to their sum and expressed in percent.
These distributions reflect the same conclusion as for the cross section namely that the processes are largely dominated by QCD-induced contribution.
The first two observables displayed (top) are the invariant mass and the rapidity difference of the two tagging jets.
These two observables are used as cuts [see Eq.~\eqref{eq:MC_vbs:cut:4}] in order to enhance the EW component over the QCD background.
The cuts are clearly visible on the plots and it is easily understandable why they are enhancing the EW contribution.
Toward high invariant-mass, the EW component becomes more and more important.
It even becomes of the same size as the QCD one for an invariant-mass of $2000{\rm~GeV}$.
The same holds true for high rapidity separation for the two jets.
On the other hand, the transverse invariant mass (bottom left) of muon--anti-muon pair does not display significant differences in the different contributions over the whole range.
Other transverse-momentum distributions display the same pattern.
Finally, we show the distance between the two jets, 
defined as $\Delta R_{jj} = \sqrt{\Delta\eta(\mathrm{j}_{1}, \mathrm{j}_{2})^2 + \Delta\phi(\mathrm{j}_{1}, \mathrm{j}_{2})^2}$.
This observable also seems to possess a good discriminating power.
In particular, for large distances, the EW component becomes dominant but with very low statistics.
The interference effects are extremely suppressed due to the smallness of the cross section.
Contribution of each of the studied observable does not exceed the per-cent level over the whole phase-space range.

\subsubsection*{Parton level comparisons}

We start the comparison of the various predictions by a comparison at the level of the cross section.
The fiducial volume is the one described in Eqs.~\eqref{eq:MC_vbs:cut:1}--\eqref{eq:MC_vbs:cut:4}.
The results are documented in Tables~\ref{tab:MC_vbs:xsectLOfix} and \ref{tab:MC_vbs:xsectLOdyn},
which give the cross sections for both processes ${\rm p} {\rm p} \to {\rm e}^+  \nu_{\rm e}  \mu^+ \mu^- {\rm j} {\rm j}$ and ${\rm p} {\rm p} \to {\rm e}^-  \bar \nu_{\rm e}  \mu^+ \mu^- {\rm j} {\rm j}$ at fixed and dynamical scales.
The predictions of \MoCaNLO\!+\Recola and \Sherpa are generally in perfect agreement, but those of \MGaMC show slight statistical disagreement for fixed scale.
In this case the difference between \MoCaNLO\!+\Recola/\Sherpa and \MGaMC is about one percent as it can be seen in Table~\ref{tab:MC_vbs:xsectLOfix}.
We note that the source of the difference is likely unchanged in the dynamic scale result, where the setup is otherwise unchanged,
but is not clear due to the higher statistical uncertainty.

The predictions of \VBFNLO are not expected to be in perfect agreement with the others as \VBFNLO is not using full matrix elements but VBS-approximated ones.
Nonetheless, the difference between the \VBFNLO predictions and the ones of \MoCaNLO\!+\Recola/\Sherpa amounts to $0.6\%$ for the fixed scale.

\begin{table}
\begin{center} 
\begin{tabular}{ c | c | c }
 $\mu = \mu_{\rm fix}$ / $\sigma_{\rm LO}^{\rm EW}$ [fb] & ${\rm p} {\rm p} \to {\rm e}^+  \nu_{\rm e}  \mu^+ \mu^- {\rm j} {\rm j}$  & ${\rm p} {\rm p} \to {\rm e}^-  \bar \nu_{\rm e}  \mu^+ \mu^- {\rm j} {\rm j}$  \\
  \hline\hline
  \MGaMC                  & $0.2857(8)$     & $0.1657(4)\phantom{0}$   \\
  {\sc MoCaNLO}+{\sc Recola}      & $0.2885(1)$     & $0.16718(3)$  \\
  {\sc Sherpa}                    & $0.2885(2)$     & $0.1670(2)\phantom{0}$   \\
  {\sc VBFNLO}                    & $0.2867(5)$     & $0.1661(3)\phantom{0}$   \\
  \hline
\end{tabular}
\end{center}
\caption{
Fiducial cross sections at LO for the process ${\rm p}{\rm p}\to{\rm e}^+\nu_{\rm e}\mu^+\mu^-{\rm j}{\rm j}$ and ${\rm p}{\rm p}\to{\rm e}^-\bar\nu_{\rm e}\mu^+\mu^-{\rm j}{\rm j}$ at order $\mathcal{O} (\alpha^6)$.
The predictions are expressed in fb and are for the LHC running at a centre-of-mass energy of $\sqrt{s}=13 {\rm~TeV}$.
The scale used in the simulations is $\mu = \mu_{\rm fix} = M_W$.
The integration errors of the last digits are given in parentheses.}
\label{tab:MC_vbs:xsectLOfix}
\end{table}

\begin{table}
\begin{center} 
\begin{tabular}{ c | c | c }
 $\mu = \mu_{\rm dyn}$ / $\sigma_{\rm LO}^{\rm EW}$ [fb] & ${\rm p} {\rm p} \to {\rm e}^+  \nu_{\rm e}  \mu^+ \mu^- {\rm j} {\rm j}$  & ${\rm p} {\rm p} \to {\rm e}^-  \bar \nu_{\rm e}  \mu^+ \mu^- {\rm j} {\rm j}$  \\
  \hline\hline
  \MGaMC                  & $0.2550(8)$  & $0.1492(5)\phantom{0}$ \\
  {\sc MoCaNLO}+{\sc Recola}      & $0.2574(2)$  & $0.15003(3)$  \\
  {\sc Sherpa}                    & $0.2574(2)$  & $0.14998(6)$   \\
  \hline
\end{tabular}
\end{center}
\caption{
Fiducial cross sections at LO for the process ${\rm p}{\rm p}\to{\rm e}^+\nu_{\rm e}\mu^+\mu^-{\rm j}{\rm j}$ and ${\rm p}{\rm p}\to{\rm e}^-\bar\nu_{\rm e}\mu^+\mu^-{\rm j}{\rm j}$ at order $\mathcal{O} (\alpha^6)$.
The predictions are expressed in fb and are for the LHC running at a centre-of-mass energy of $\sqrt{s}=13 {\rm~TeV}$.
The scale used in the simulations is $\mu = \mu_{\rm dyn} = {\rm Max}\left[p_{\rm T, j}\right]$.
The integration errors of the last digits are given in parentheses.}
\label{tab:MC_vbs:xsectLOdyn}
\end{table}

\begin{figure}[t]
\begin{center}
   \includegraphics[width=0.48\textwidth]{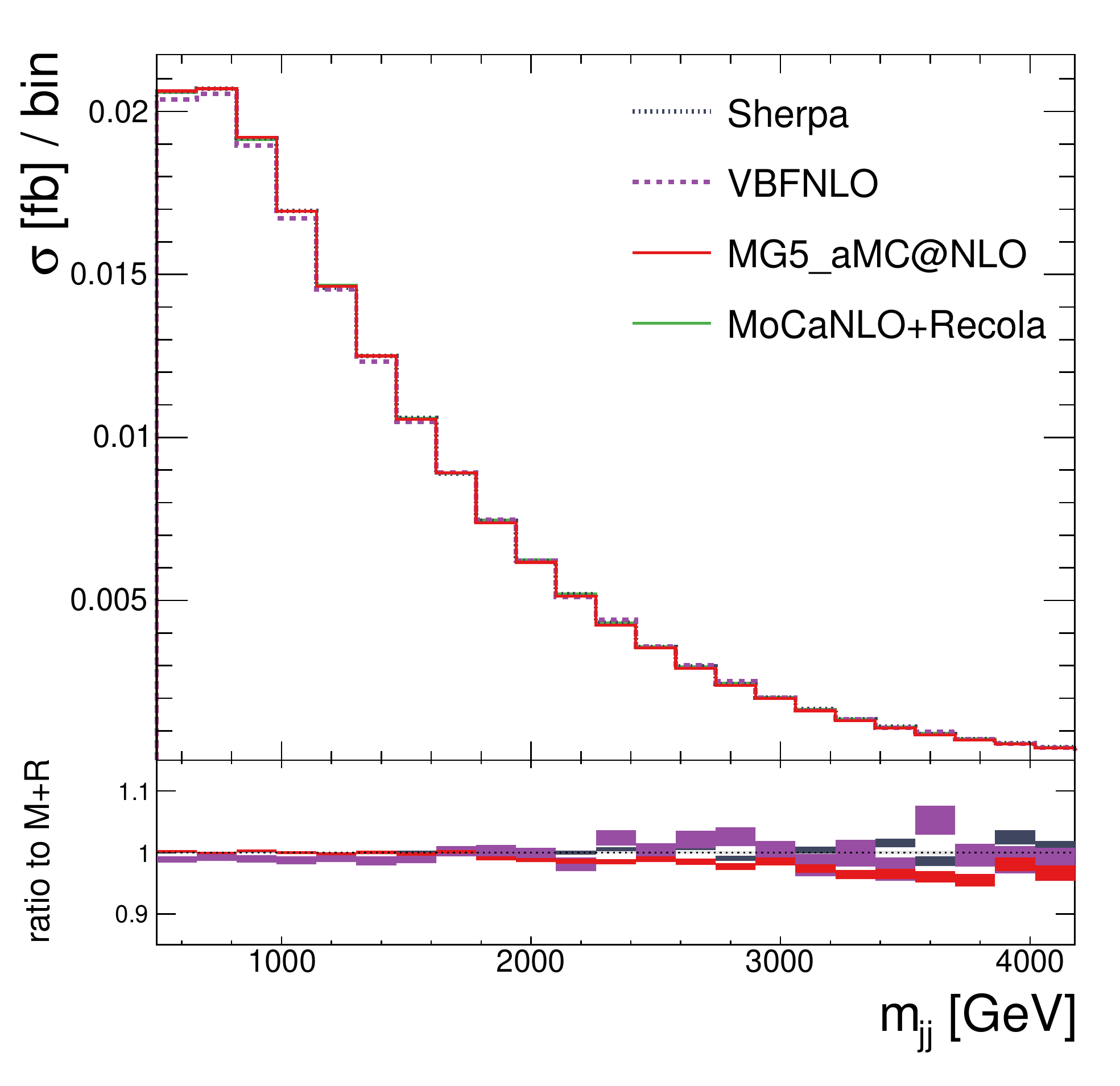}\hfill
   \includegraphics[width=0.48\textwidth]{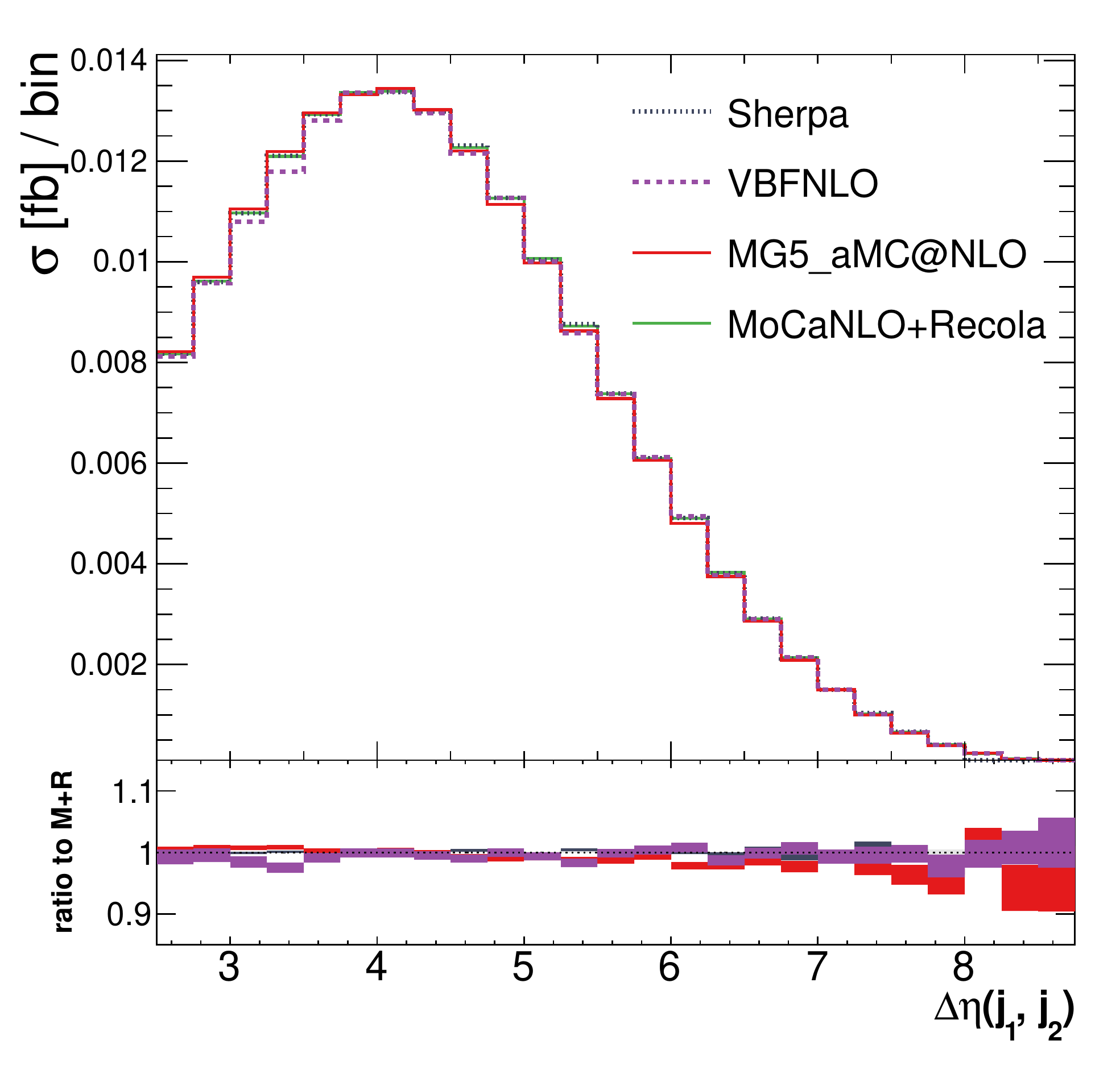}
\caption{Differential distributions computed at fixed-order with centre-of-mass energy $\sqrt{s}=13{\rm~TeV}$ at the LHC for ${\rm p} {\rm p}
  \to {\rm e}^-  \nu_{\rm e}  \mu^+ \mu^- {\rm j} {\rm j}$ at LO with fixed scaled $\mu = M_{\rm W}$: 
                invariant mass of the two jets~(left),
                rapidity separation between the two jets~(right).
                The predictions in the lower plot are normalised to the prediction of \MoCaNLO\!+\Recola.
		The shaded bands indicate the relative statistical uncertainty by bin for each sample.
 		The statistical uncertainty on the \MoCaNLO\!+\Recola predictions is shown in grey, other samples
 		are indicated in the ratio with the color indicated in the legend.
                }
\label{fig:MC_vbs:fixed_order}
\end{center}
\end{figure}

\begin{figure}[t]
\begin{center}
   \includegraphics[width=0.48\textwidth]{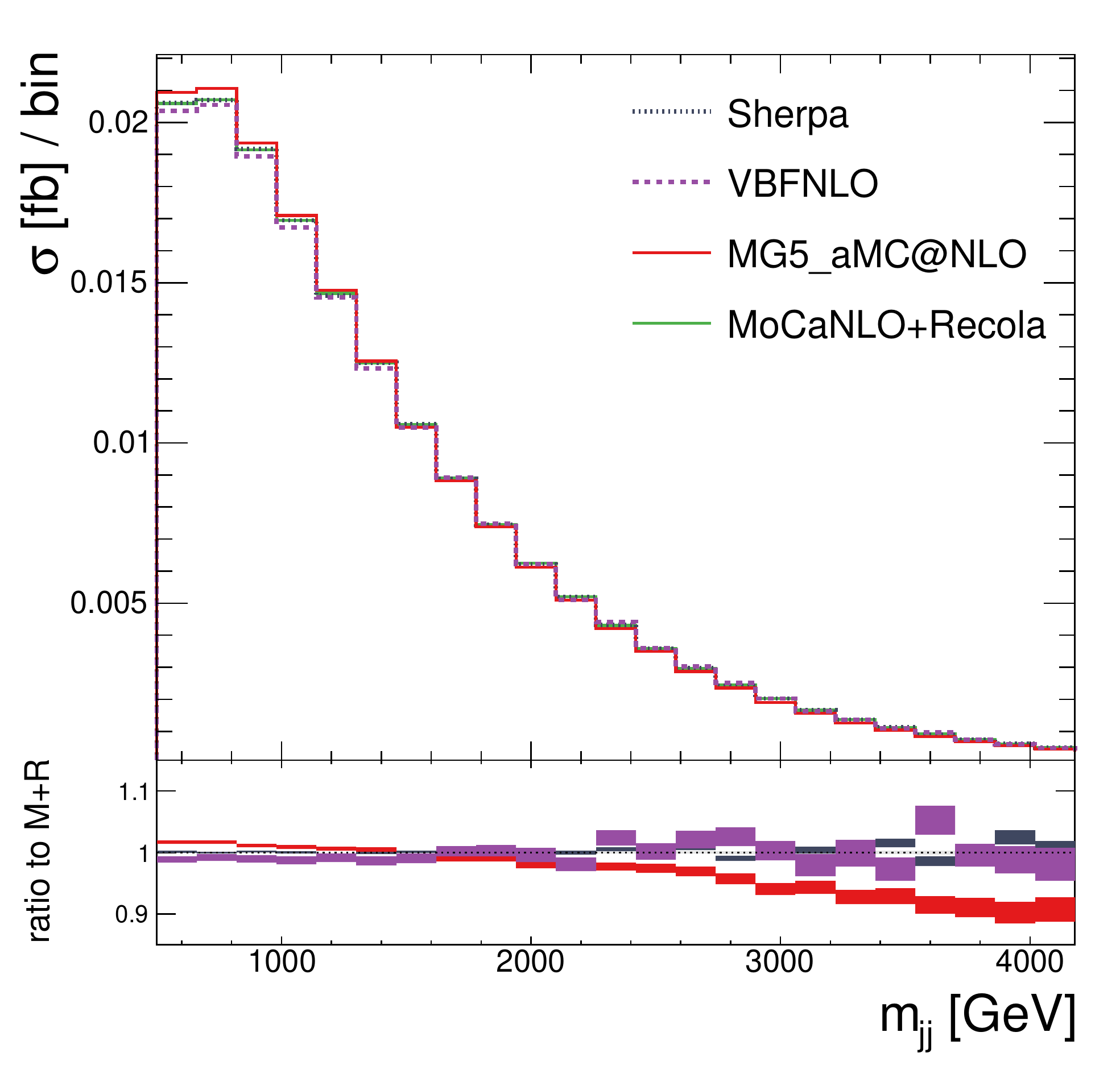}
\caption{Invariant mass of the two jets computed at fixed-order with centre-of-mass energy $\sqrt{s}=13{\rm~TeV}$ at the LHC for ${\rm p} {\rm p}
  \to {\rm e}^-  \nu_{\rm e}  \mu^+ \mu^- {\rm j} {\rm j}$ at LO with fixed scaled $\mu = M_{\rm W}$. 
                The predictions in the lower plot are normalised to the prediction of \MoCaNLO\!+\Recola.
		The shaded bands indicate the relative statistical uncertainty by bin for each sample.
 		The statistical uncertainty on the \MoCaNLO\!+\Recola predictions is shown in grey, other samples
 		are indicated in the ratio with the color indicated in the legend.
    \MGaMC events are generated using loose selections in the run card before applying the selection on the LHE
    events. The trend of disagreement in high $m_{jj}$ is accentuated for looser generation cuts, pointing 
    to a possible phase-space integration effect.
                }
\label{fig:MC_vbs:fixed_order_loose}
\end{center}
\end{figure}

As it can be seen in Fig.~\ref{fig:MC_vbs:fixed_order}, where the invariant mass and rapidity separation of the two tagging jets are displayed, the agreement between the predictions is good in both shape and normalisation.
This is particularly true for the jet rapidity separation, where all predictions agree within statistical uncertainties over the whole range.
On the other hand, for the di-jet invariant mass, subtle but statistically significant differences are more pronounced towards higher masses between \MGaMC's predictions and the others. 
We do not see such discrepancies in other distributions, however, we see that this discrepancy is accentuated when relaxing the cuts applied at generation
time through the run card and manually applying the selection to the LHE events, as shown in Fig.~\ref{fig:MC_vbs:fixed_order_loose}.

Concerning the validation of the VBS approximation ({\sc MoCaNLO}+{\sc Recola} vs.\ \linebreak{\sc VBFNLO} for example), both predictions are in very good agreement as at the level of the cross section.
This supports the findings of Ref.~\cite{Anders:2018gfr} where preliminary results for similar comparisons for ${\rm W}^\pm{\rm W}^\pm{\rm j}{\rm j}$ have been reported.
This means that the VBS approximation ({\sc VBFNLO}) approximates rather well the full computation ({\sc MoCaNLO}+{\sc Recola}) in the fiducial region chosen.

We stress that differences of configuration should be considered independently of typical estimates of theoretical uncertainties such as QCD scale and PDF uncertainties. 
To illustrate this, we compute the PDF uncertainty for the NNPDF3.0 set and the two scale choices considered here. 
The PDF uncertainty is evaluated to be 3--5\% using \MGaMC. 
Scale uncertainties are evaluated using the typical prescription of varying $\mu_{R}$ and $\mu_{F}$ 
subject to the constraint $1/2 \le \mu_{F}/\mu_{R} \le 2$, using \MGaMC and cross-checked with \MoCaNLO+\Recola, 
and are found to be between 7--10\% for the scale choices considered. The full results obtained with \MGaMC are shown in Table~\ref{tab:MC_vbs:xsecWithUnc}.

\begin{table}[htbp]
\begin{center} 
\begin{tabular}{ c | c | c }
 Scale choice & ${\rm p} {\rm p} \to {\rm e}^+  \nu_{\rm e}  \mu^+ \mu^- {\rm j} {\rm j}$  & ${\rm p} {\rm p} \to {\rm e}^-  \bar \nu_{\rm e}  \mu^+ \mu^- {\rm j} {\rm j}$  \\
  \hline\hline
  $\mu_{R} = \mu_{\rm fix} = m_{\mathrm{W}}$                     & $0.286^{+9.2\%}_{-7.8\%} \pm 3.7\%$  & $0.166^{+9.0\%}_{-7.7\%} \pm 4.3\%$ \\
  $\mu = \mu_{\rm dyn} = {\rm Max}\left[p_{\rm T, j}\right]$ & $0.255^{+8.0\%}_{-6.9\%} \pm 3.7\%$  & $0.149^{+9.0\%}_{-7.7\%} \pm 4.3\%$  \\
  \hline
\end{tabular}
\end{center}
\caption{
Fiducial cross sections at LO for the process ${\rm p}{\rm p}\to{\rm e}^+\nu_{\rm e}\mu^+\mu^-{\rm j}{\rm j}$ and ${\rm p}{\rm p}\to{\rm e}^-\bar\nu_{\rm e}\mu^+\mu^-{\rm j}{\rm j}$ at order $\mathcal{O} (\alpha^6)$ 
  via \MGaMC.
The predictions are expressed in fb and are for the LHC running at a centre-of-mass energy of $\sqrt{s}=13 {\rm~TeV}$.
  Uncertainties are expressed as $\sigma^{+\delta_{\mathrm{scale}}}_{-\delta{\mathrm{scale}}} \pm \mathrm \delta_{\mathrm{PDF}}$.
}
\label{tab:MC_vbs:xsecWithUnc}
\end{table}

We also compare the effect of configurations changes using the general-purpose generators {\sc Sherpa} and \MGaMC. 
These generators are able to compute arbitrary processes in the standard model, 
and therefore have default configurations designed to cover a wide range of processes. It is thus advisable
to explicitly configure settings most appropriate to the process considered. 

Using the same fiducial definition, but with the default parameter settings (defined in param\_card.dat) for 
\MGaMC, we obtain a cross section of $0.1561(5)$ fb for the ${\rm Max}\left[p_{\rm T, j}\right]$
scale, an increase of $4\%$ from the nominal value. The primary source of this difference is the 
settings of the boson masses and widths, which are set to their LO values by default in \MGaMC.
Such a setting can be considered appropriate in cases where the partial widths of many decay channels
must be considered together, but for an explicitly leptonic process where no gauge boson is set on-shell, 
we argue that the best measured values are more appropriate.
We do not believe that this $4\%$ difference constitutes an additional uncertainty, but warn that
careful configuration of such parameters should be considered.

We similarly study the default dynamical scale choices in \MGaMC and {\sc Sherpa}. In both cases,
this choice is motivated by a desire for broad application to a variety of processes. Sherpa uses the inverted
parton shower to cluster the matrix element onto a core $2 \to 2$ process. This procedure determines the scales
and is especially suited for truncated shower merging. For \MGaMC,
the default scale depends on the order of the process and shower settings. At LO without
merging of parton multiplicities, the scale is set to the central $m_{\rm T}$ scale after k$_{\rm T}$-clustering of the event.
If the last clustering is a $t$-channel colourless exchange, the scale is set to the last $m_{\rm T}$ values on either side \footnote{The procedure is described at https://cp3.irmp.ucl.ac.be/projects/madgraph/wiki/FAQ-General-13.}.
The resulting cross section is $5\%$ greater than the ${\rm Max}\left[p_{\rm T, j}\right]$ result.
Using both the default scale and parameter settings (note however, that we still specify the PDF)
we obtain a value of $\sigma_{\mathrm{default}} = 0.1638(5)$ fb, $9\%$ greater than our configured
$\mu = {\rm Max}\left[p_{\rm T, j}\right]$ value. We reiterate that this difference, which is
of comparable size to the phenomenological uncertainties, should not necessarily be considered on equal footing.
The fact that the functional form of the dynamical scale choice can have a significant impact compared to the
usual factor of two variation around a nominal value is well established. One should therefore take care
to select an appropriately motivated choice. Additional uncertainties may be appropriate,
but should be considered with care rather than taking a broad envelope of less motivated choices.
This comparison additionally highlights the advantage of the reduced scale dependency at NLO-QCD accuracy.

\subsubsection*{Comparisons after parton shower}

\begin{figure}[p]
\begin{center}
   \includegraphics[width=0.48\textwidth]{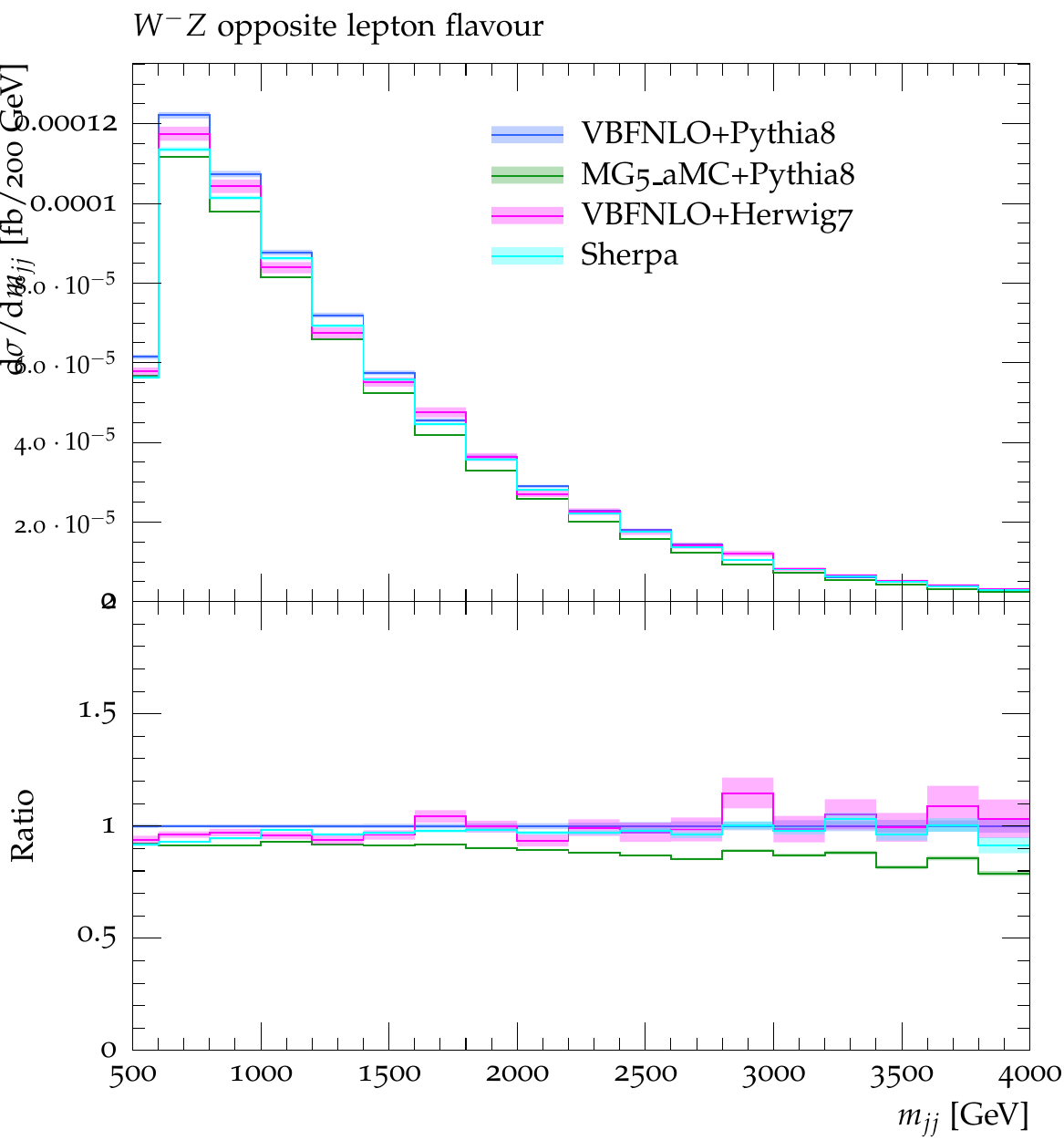}\hfill
   \includegraphics[width=0.48\textwidth]{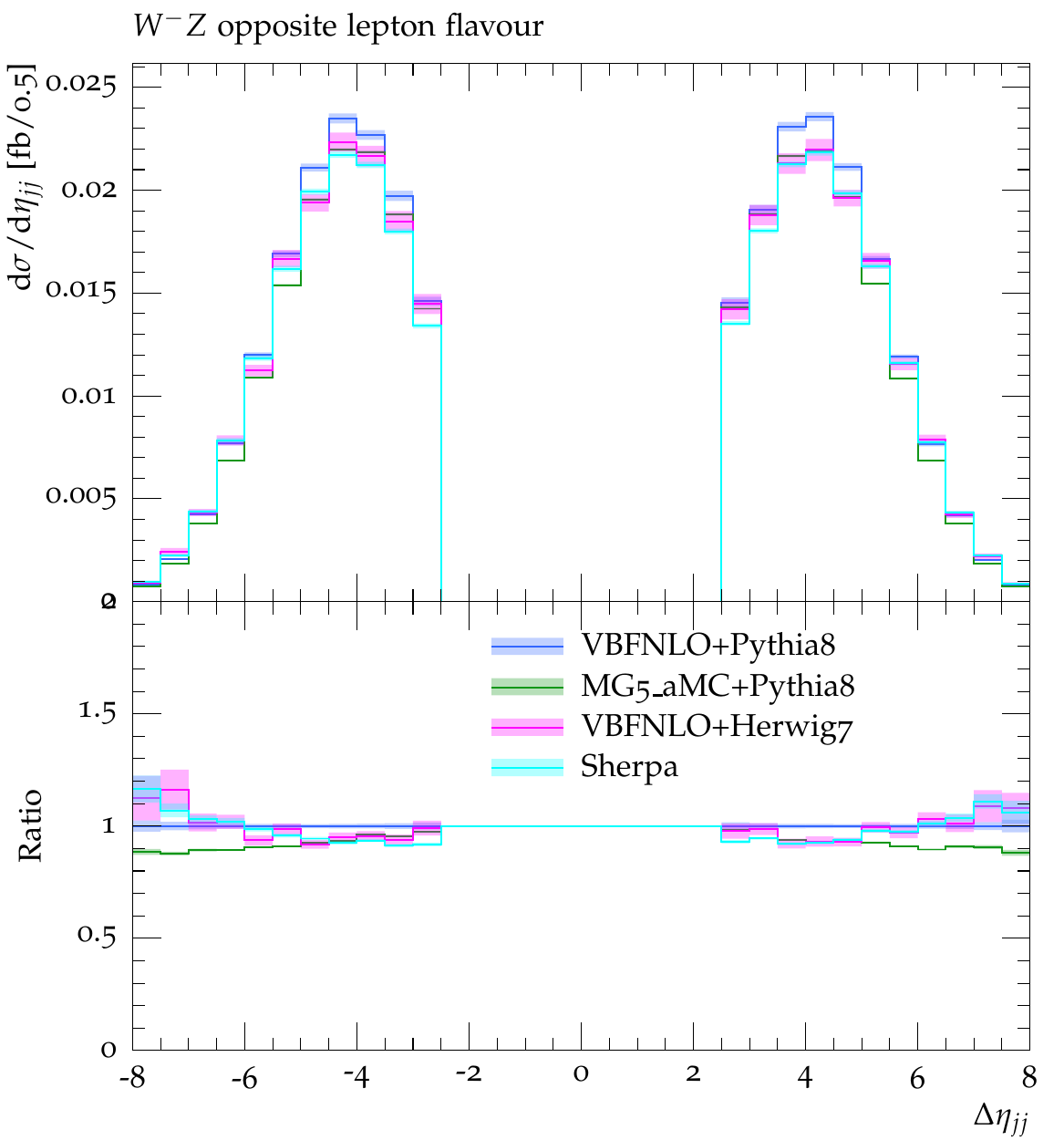}\\[2ex]
   \includegraphics[width=0.48\textwidth]{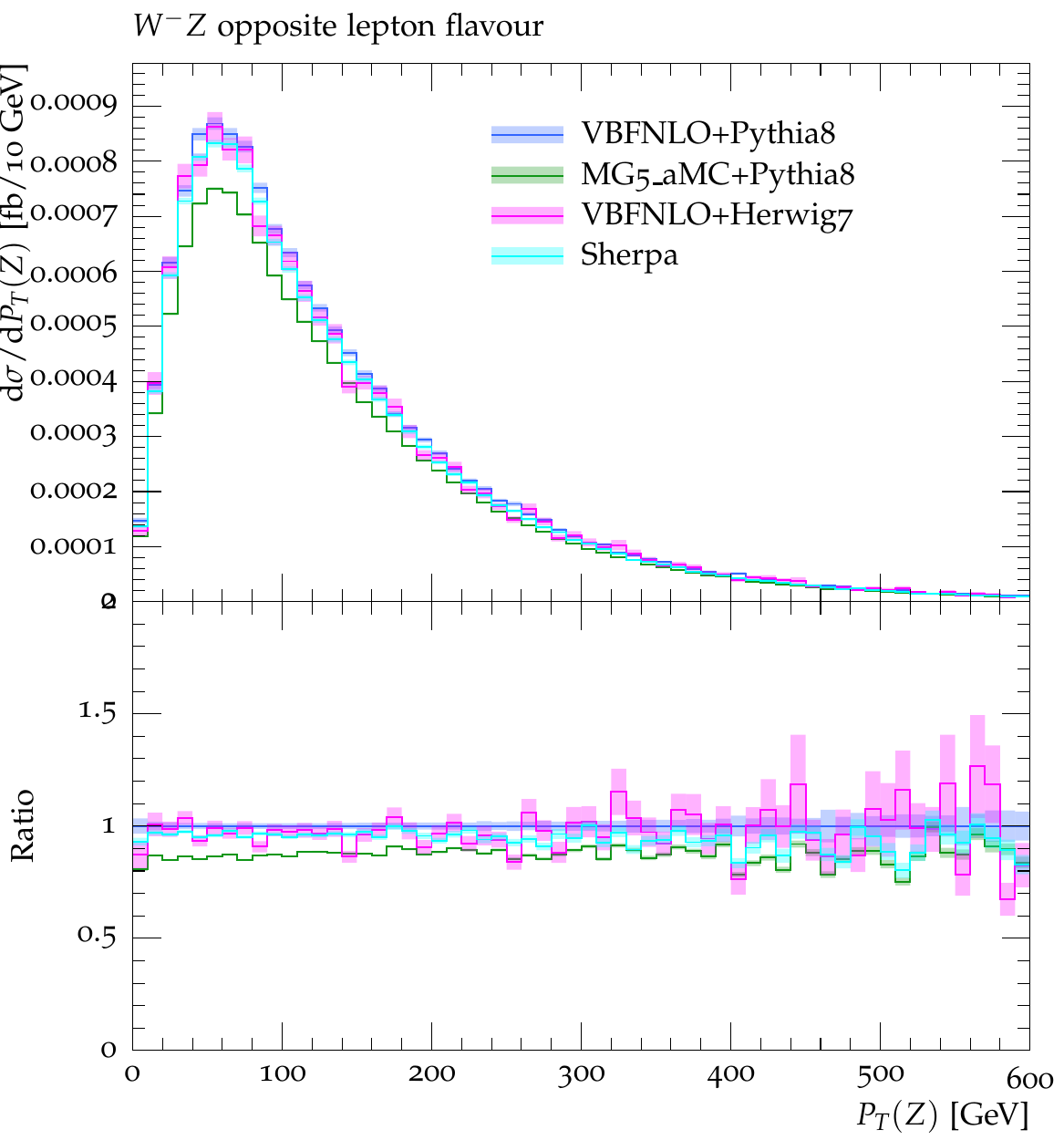}\hfill
   \includegraphics[width=0.48\textwidth]{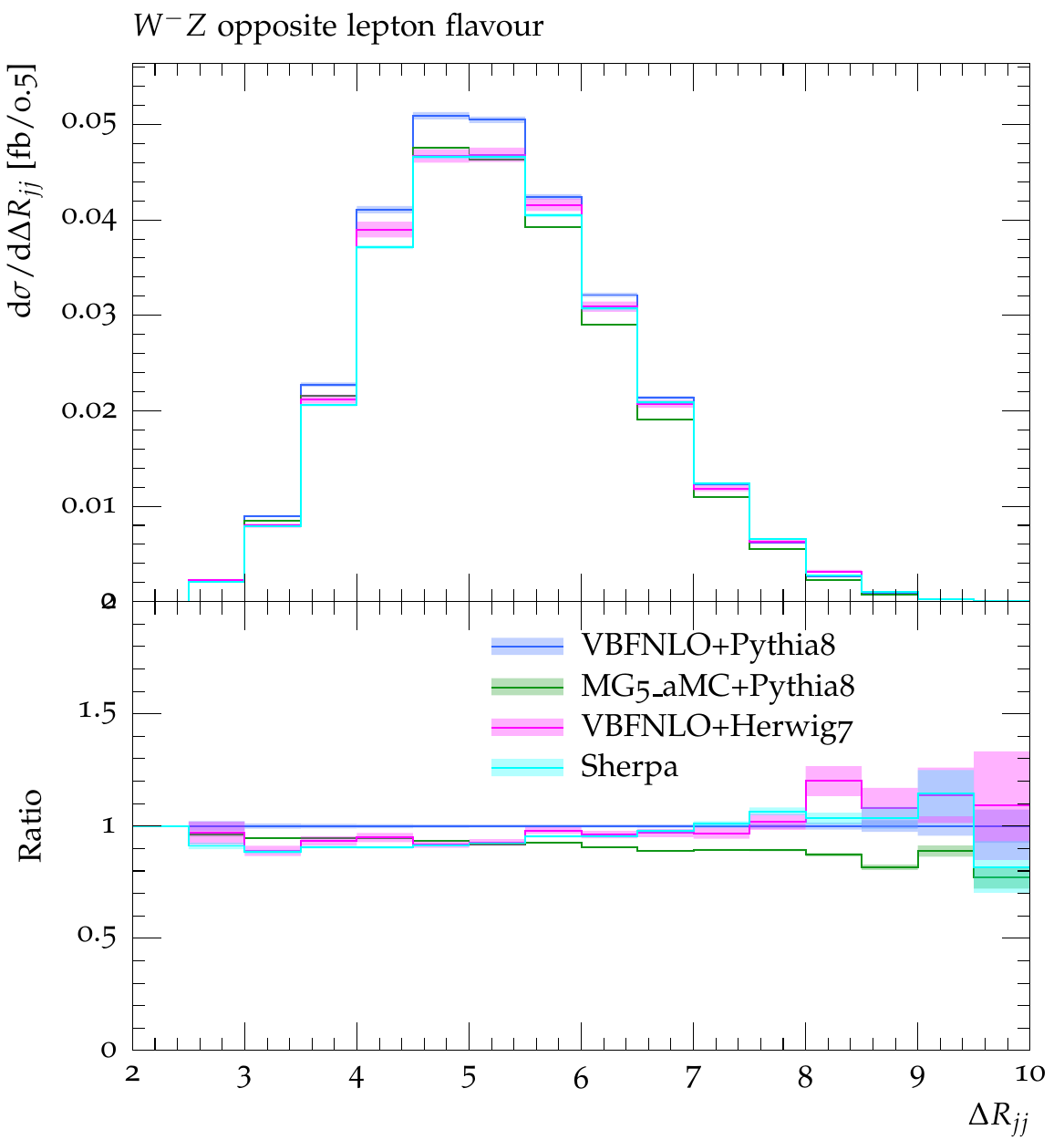}
   \vspace*{2ex}
\caption{Differential distributions at a centre-of-mass energy $\sqrt{s}=13{\rm TeV}$ at the LHC for ${\rm p} {\rm p}
  \to {\rm e}^-  \nu_{\rm e}  \mu^+ \mu^- {\rm j} {\rm j}$ at LO matched with parton shower with fixed scaled $\mu = M_{\rm W}$: 
                invariant mass of the two jets~(top left),
                rapidity separation between the two jets~(top right),
                transverse momentum of the anti-muon--muon system~(bottom left), and
                distance between the two jets~(bottom right). 
                In the lower plot, the normalisation is with respect to the \VBFNLO+{\sc PYTHIA8} predictions.
                The error band represents
                the statistical error of the Monte Carlo integrations.}
\label{fig:MC_vbs:shower_1a}
\end{center}
\end{figure}

\begin{figure}[p]
\begin{center}
   \includegraphics[width=0.48\textwidth]{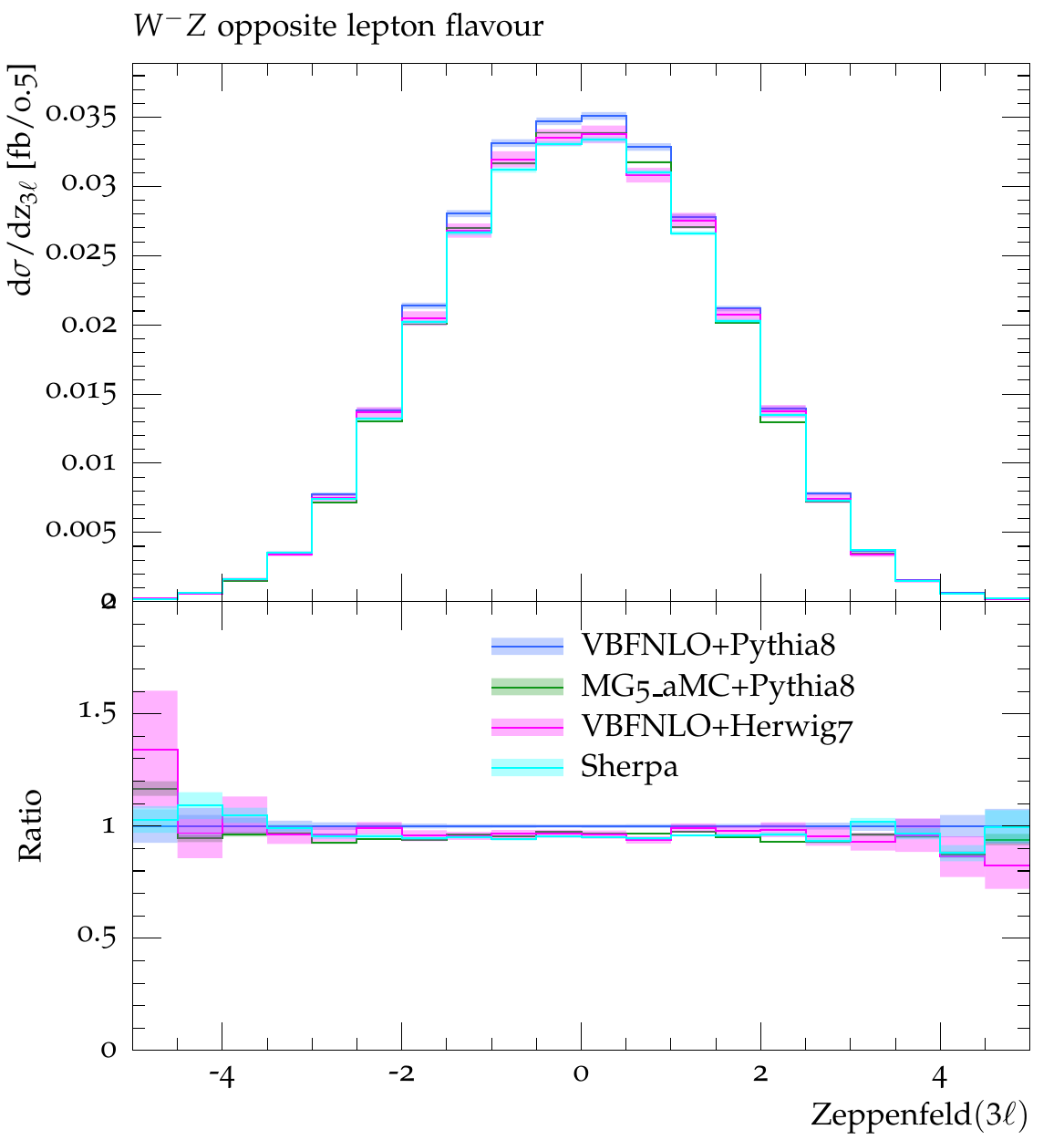}\hfill
   \includegraphics[width=0.48\textwidth]{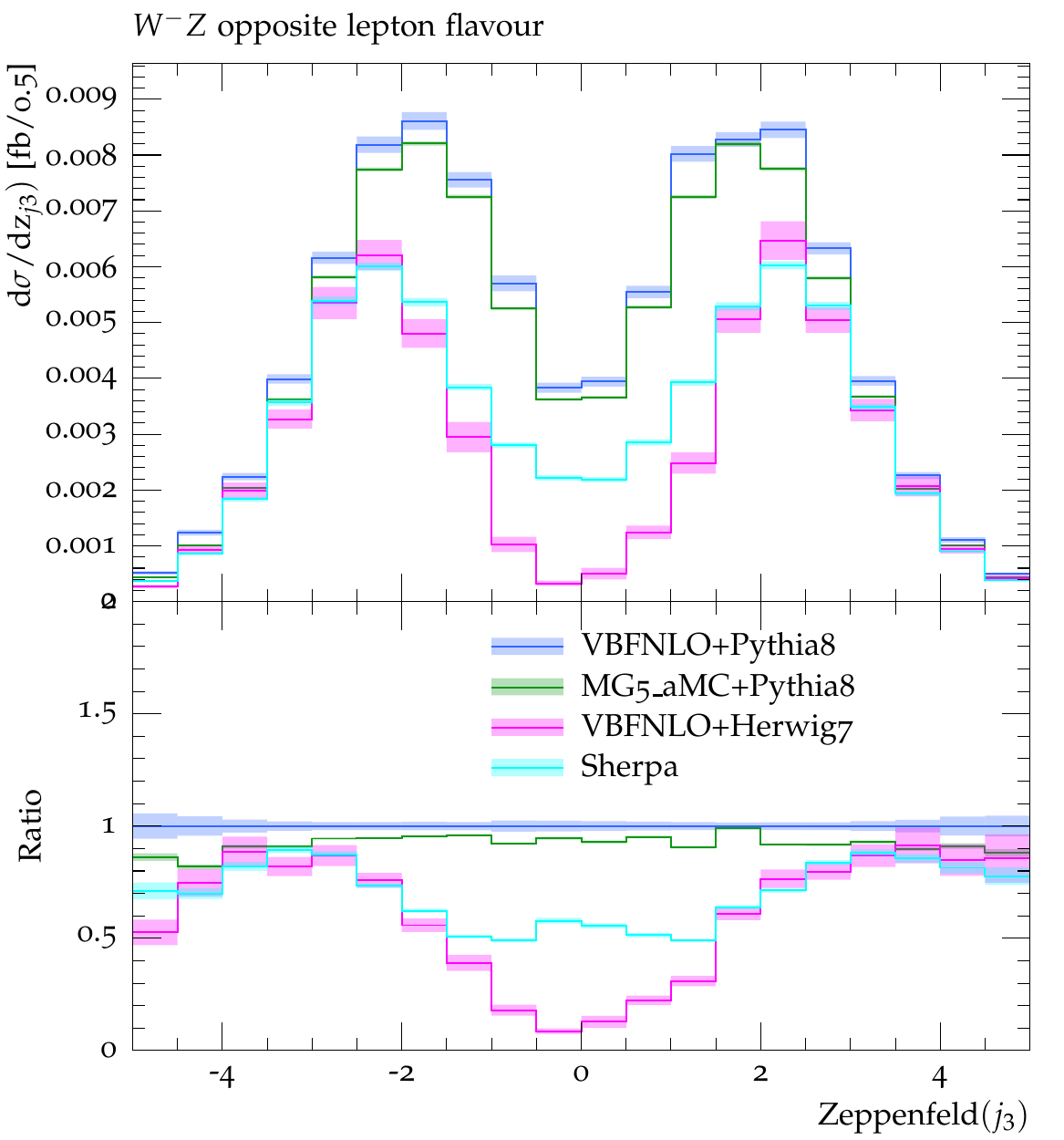}\\[2ex]
   \includegraphics[width=0.48\textwidth]{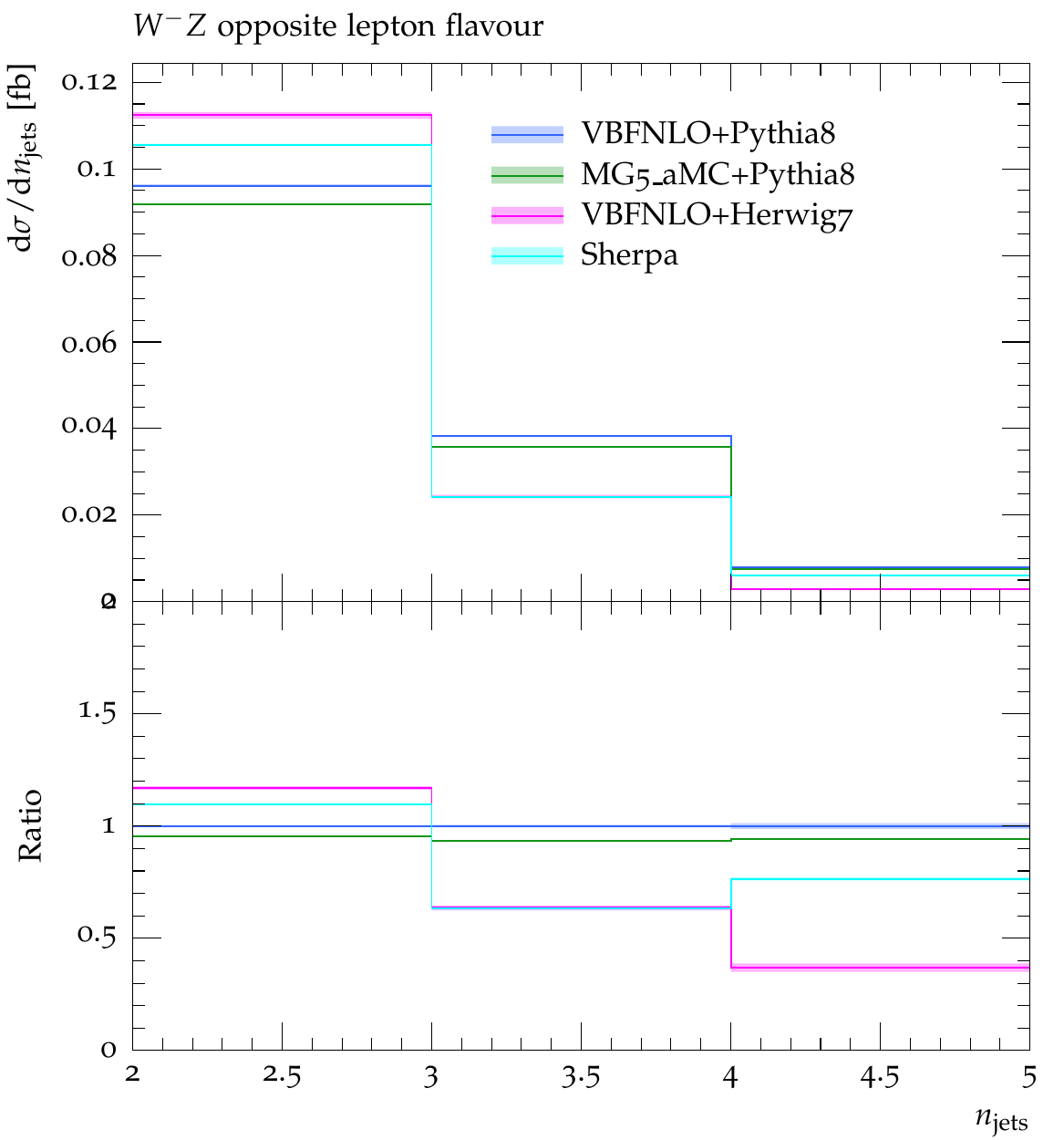}
   \vspace*{2ex}
\caption{Differential distributions at a centre-of-mass energy $\sqrt{s}=13{\rm TeV}$ at the LHC for ${\rm p} {\rm p}
  \to {\rm e}^-  \nu_{\rm e}  \mu^+ \mu^- {\rm j} {\rm j}$ at LO matched with parton shower with fixed scaled $\mu = M_{\rm W}$: 
                Zeppenfeld variable for the three leptons~(top left),
                Zeppenfeld variable for the third jet~(top right), and
                number of jets~(bottom). 
                In the lower plot, the normalisation is with respect to the \VBFNLO+{\sc PYTHIA8} predictions.
                The error band represents
                the statistical error of the Monte Carlo integrations.}
\label{fig:MC_vbs:shower_1b}
\end{center}
\end{figure}

In Figs.~\ref{fig:MC_vbs:shower_1a} and \ref{fig:MC_vbs:shower_1b}, a comparison of results obtained with different
generators for the process ${\rm p} {\rm p} \to {\rm e}^-  \nu_{\rm e}  \mu^+ \mu^- {\rm j} {\rm j}$ at LO supplemented with parton shower with fixed scaled $\mu =M_W$ is shown. 
The predictions have been obtained from \VBFNLO and \MGaMC in association with {\sc PYTHIA8}, \VBFNLO with {\sc HERWIG7}, and \Sherpa with its own parton shower.
In Fig.~\ref{fig:MC_vbs:shower_1a} several differential distributions are shown:
the invariant mass of the two jets, the rapidity separation between
the two jets, the transverse momentum of the anti-muon--muon system,
and the distance between the two jets. 
The error band in the plots represents the statistical error of the Monte Carlo integrations.  
The overall picture is that the predictions obtained from \Sherpa, \VBFNLO+{\sc PYTHIA8}, and \VBFNLO+{\sc HERWIG7} agree rather well over the whole kinematic range.
This is not the case for the predictions of \MGaMC+{\sc PYTHIA8} which differ by up to $20\%$ in certain phase-space regions.
Because of kinematic changes in jet and lepton kinematics due to the parton shower, it is necessary to generate events with a selection looser than the one 
used in the fixed-order fiducial definition and applied in the Rivet routine.
The showered events for \MGaMC are therefore generated with the conditions of Fig.~\ref{fig:MC_vbs:fixed_order_loose}.
The disagreement at high di-jet invariant mass, already seen at fixed order, is thus not an effect of the parton shower.
The normalisation differences are also consistent with a reduction in acceptance into the fiducial region
for the \MGaMC sample due to this trend.

Differences are more pronounced in Fig.~\ref{fig:MC_vbs:shower_1b} where the Zeppenfeld variable for the three charged leptons and the third jet as well as the number of jets are displayed.
The Zeppenfeld variable for a given particle $X$ is defined as

\begin{equation}
  z_{X} = \frac{y_{X}-\frac{y_{{\rm j}_1}+y_{{\rm j}_2}}2}{|y_{{\rm j}_1}-y_{{\rm j}_2}|} ,
\end{equation}
where $y_{{\rm j}_{1/2}}$ are the rapidity of the first and second hardest jet, respectively.
The Zeppenfeld variable of the third jet, as well as the number of jets beyond two, are observables that are not defined at LO,
and are only non-zero thanks to the emissions of the parton shower.
It is thus expected that these feature a worse agreement than the previously discussed observables, as 
significantly different algorithms are employed by the parton shower generators considered.
Here the differences can reach up to $100\%$ in the central region.
The fact that {\sc Pythia} predicts more central jet activity has already been observed for the ${\rm W}^\pm{\rm W}^\pm{\rm j}{\rm j}$ 
signature in preliminary results of a comparative study \cite{Anders:2018gfr} but also for the 
${\rm Z}{\rm j}{\rm j}$ signature \cite{Aaboud:2017emo,Sirunyan:2017jej}, where the predictions are compared with data.
We note we have not tuned the parameters and algorithms
of the shower but rather consider the spread of predictions as reflective of the uncertainty 
of parton shower dominated observables. While variables such as the Zeppenfeld of the third jet
have known separation power between the EW and QCD induced production, tuning an experimental selection 
on this observable would introduce large theoretical uncertainties, especially if only LO predictions
are considered. A similar argument holds for a veto on extra jet activity.

\begin{figure}[p]
\begin{center}
   \includegraphics[width=0.48\textwidth]{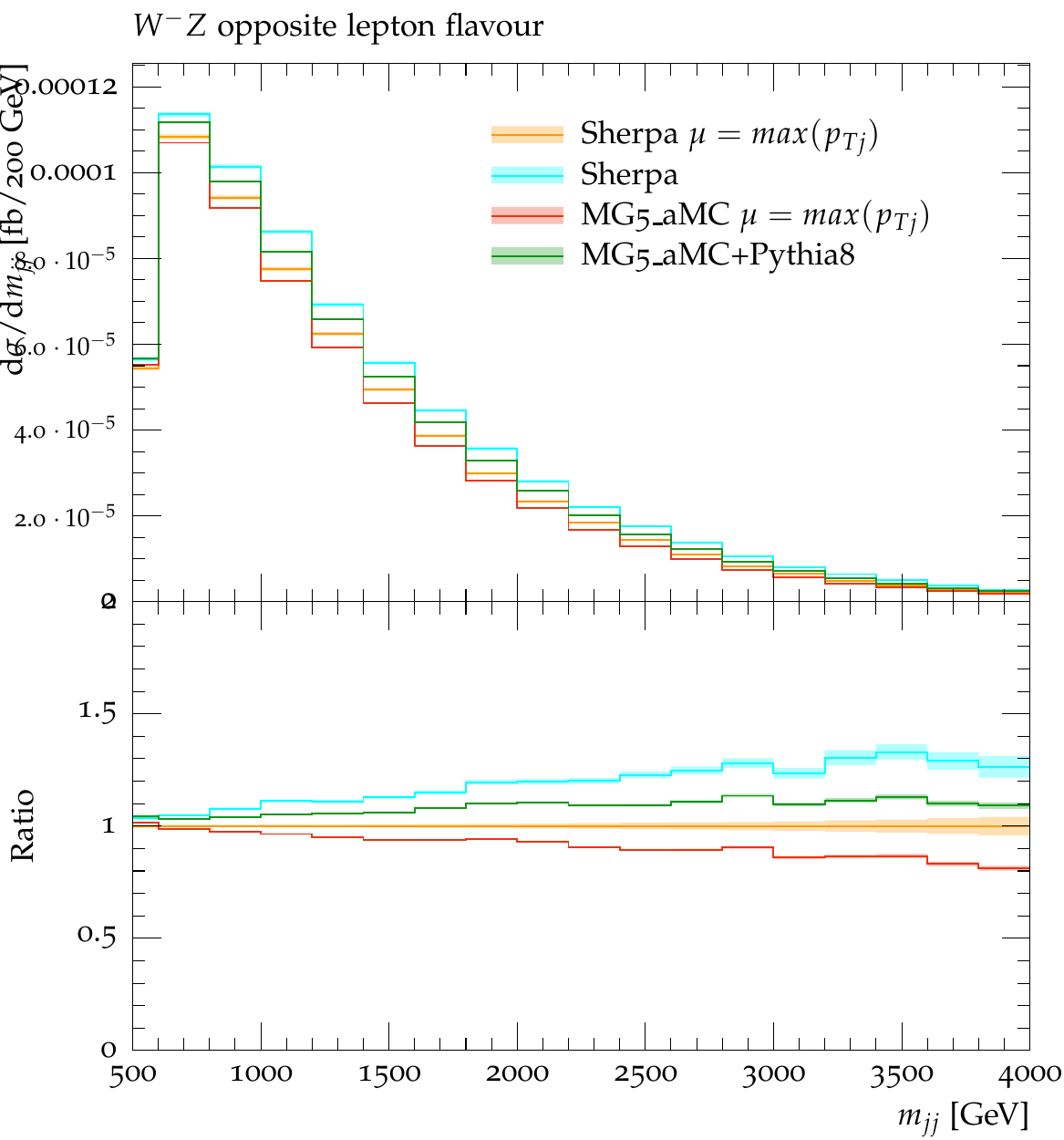}\hfill
   \includegraphics[width=0.48\textwidth]{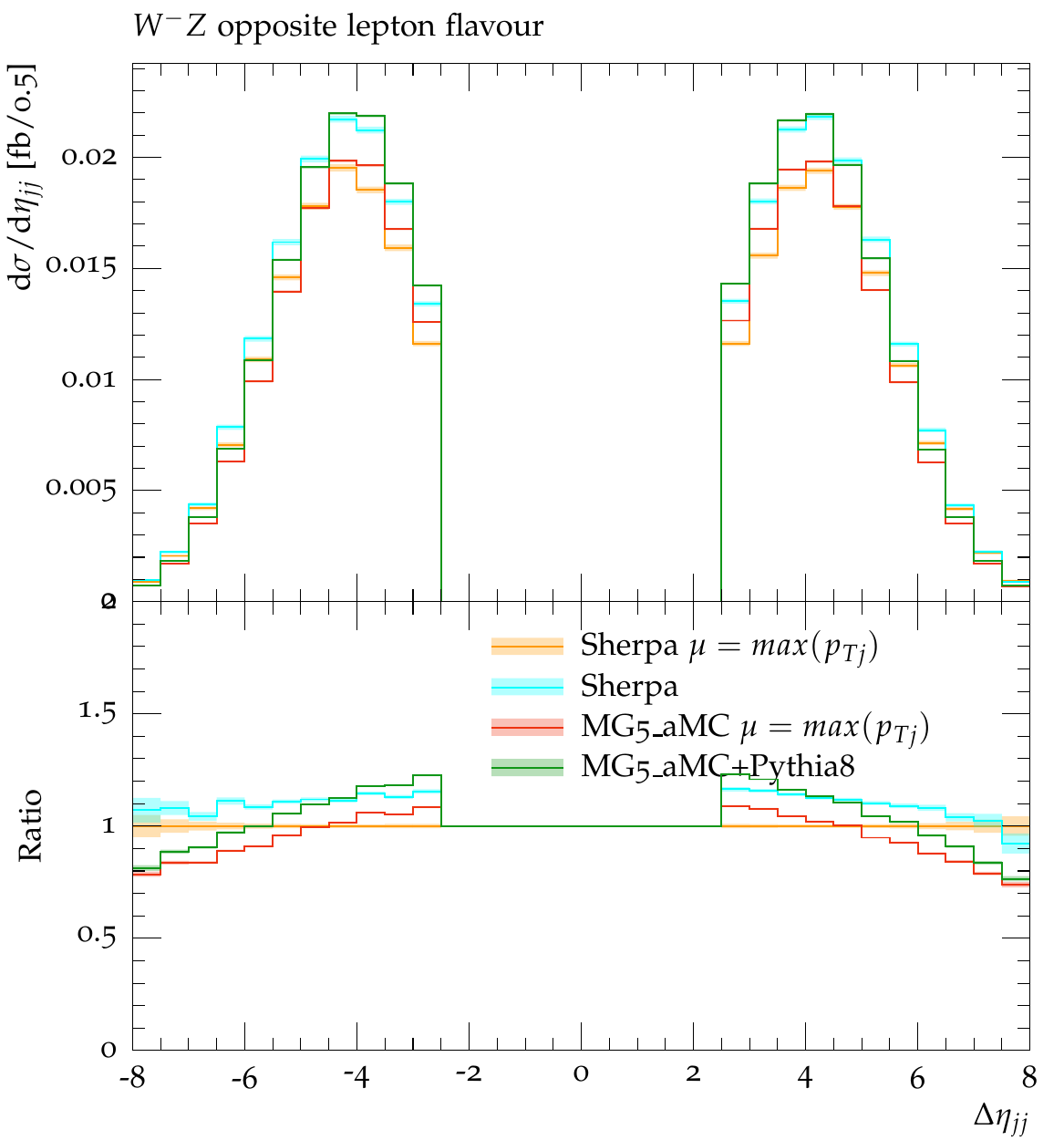}\\[2ex]
   \includegraphics[width=0.48\textwidth]{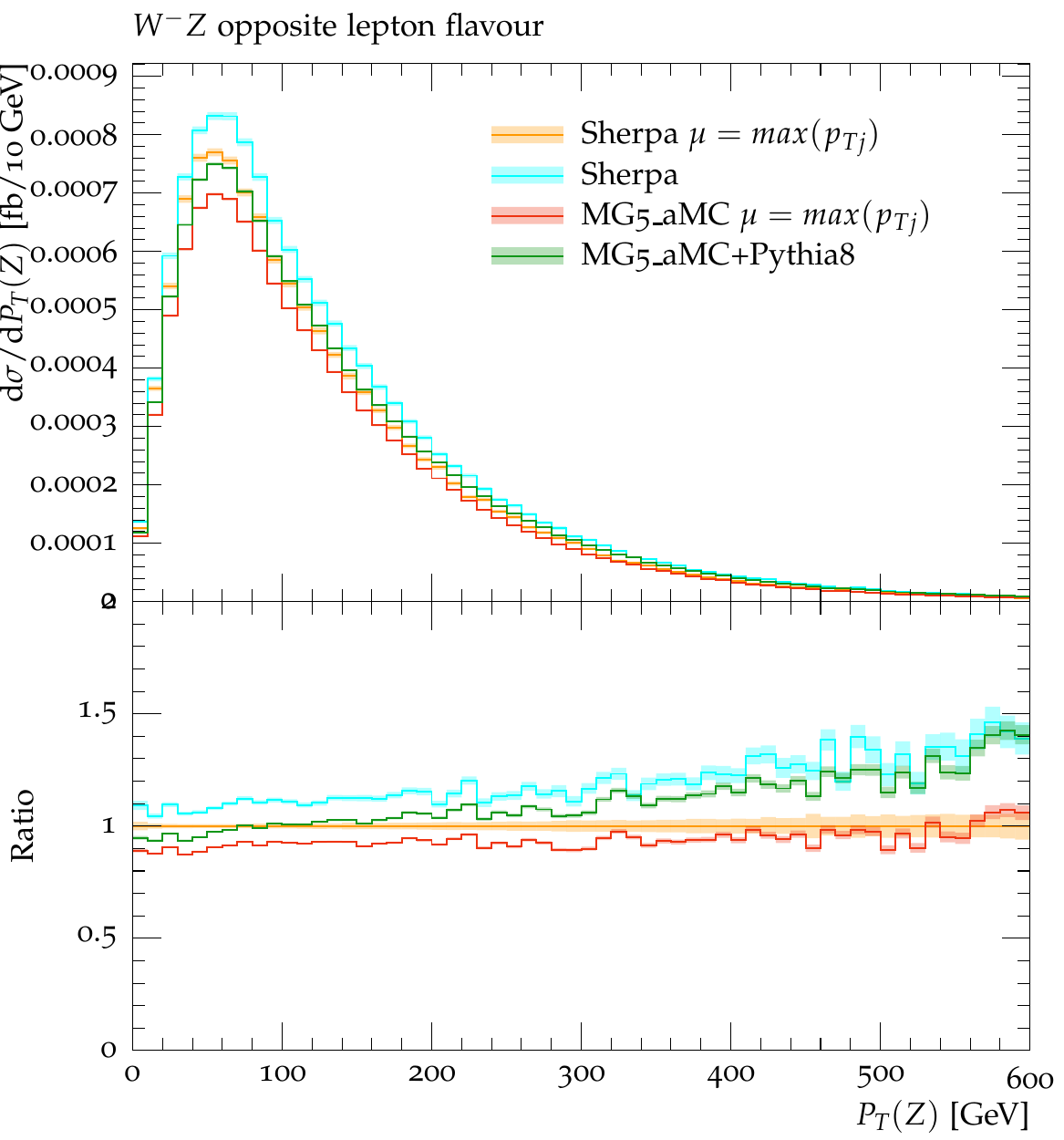}\hfill
   \includegraphics[width=0.48\textwidth]{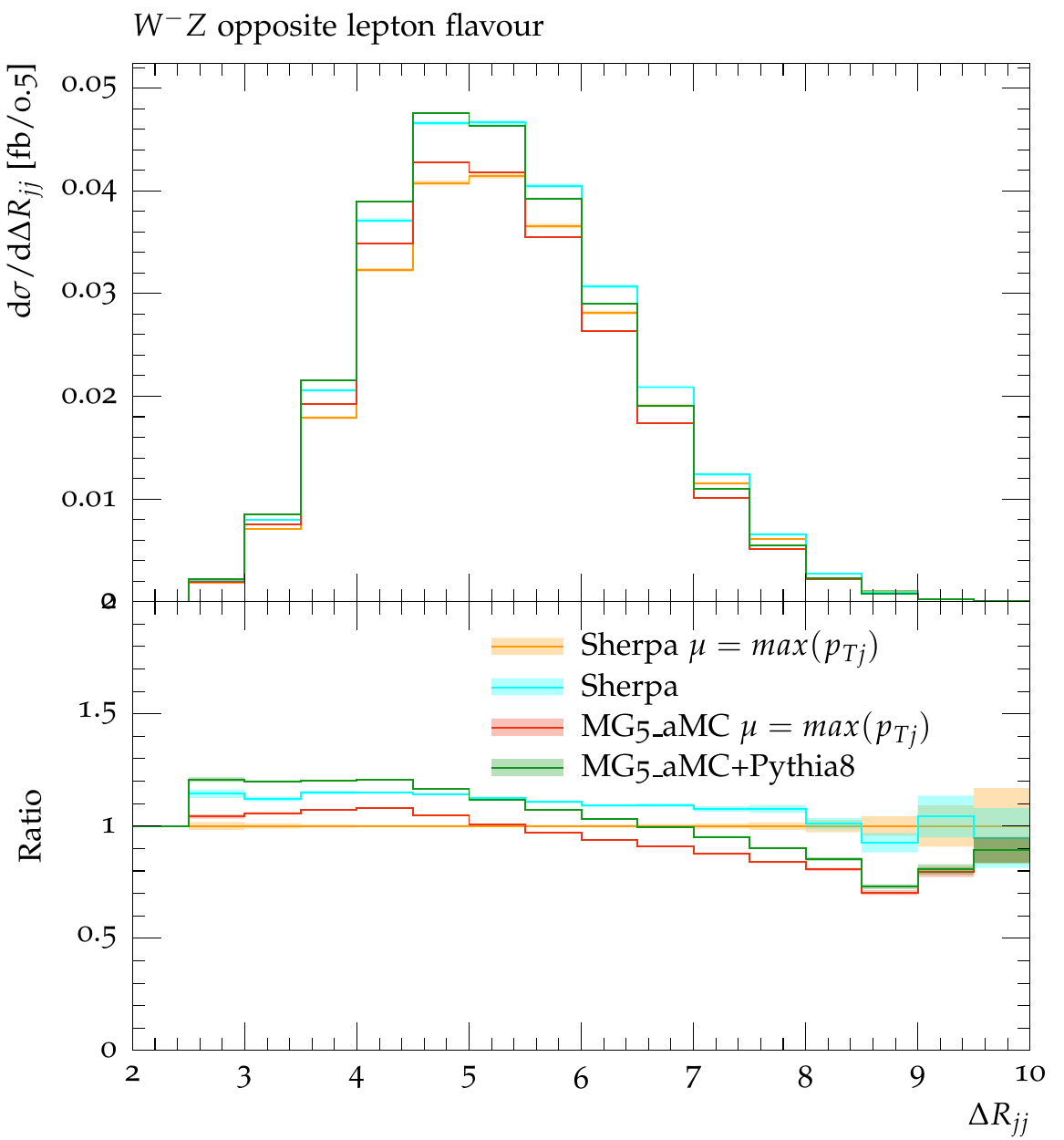}
   \vspace*{2ex}
\caption{Differential distributions at a centre-of-mass energy $\sqrt{s}=13{\rm TeV}$ at the LHC for ${\rm p} {\rm p}
  \to {\rm e}^-  \nu_{\rm e}  \mu^+ \mu^- {\rm j} {\rm j}$ at LO matched with parton shower with fixed and dynamic scale values:  
                invariant mass of the two jets~(top left),
                rapidity separation between the two jets~(top right),
                transverse momentum of the anti-muon--muon system~(bottom left), and
                distance between the two jets~(bottom right).
                In the lower plots, the predictions are normalised to the prediction of {\sc Sherpa} with
                dynamical scale. The error band represents
                the statistical error of the Monte Carlo integrations. For the two \MGaMC
                calculations this statistical error is largely correlated, as explained in the text.}
\label{fig:MC_vbs:shower_2a}
\end{center}
\end{figure}

\begin{figure}[p]
\begin{center}
   \includegraphics[width=0.48\textwidth]{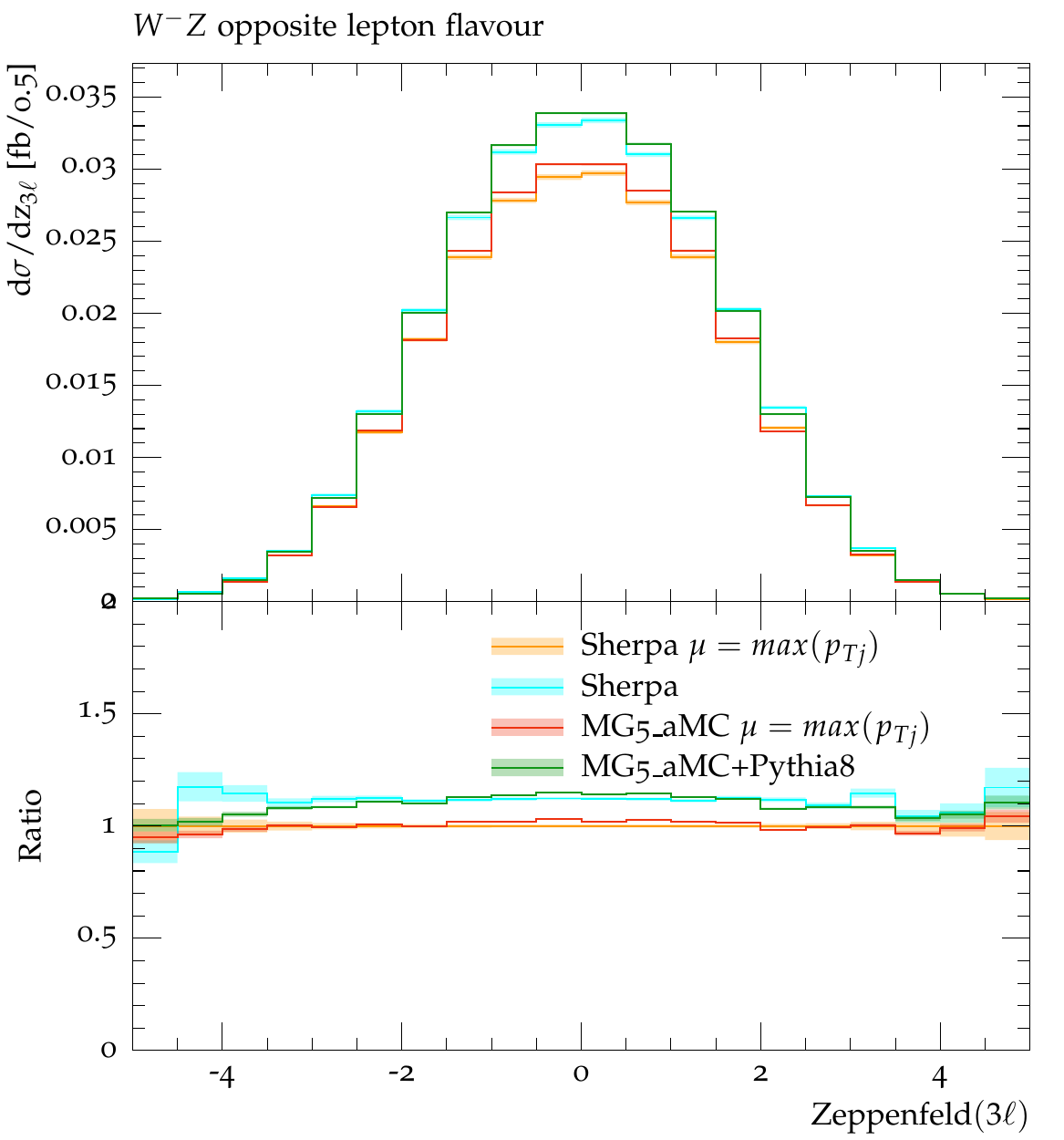}\hfill
   \includegraphics[width=0.48\textwidth]{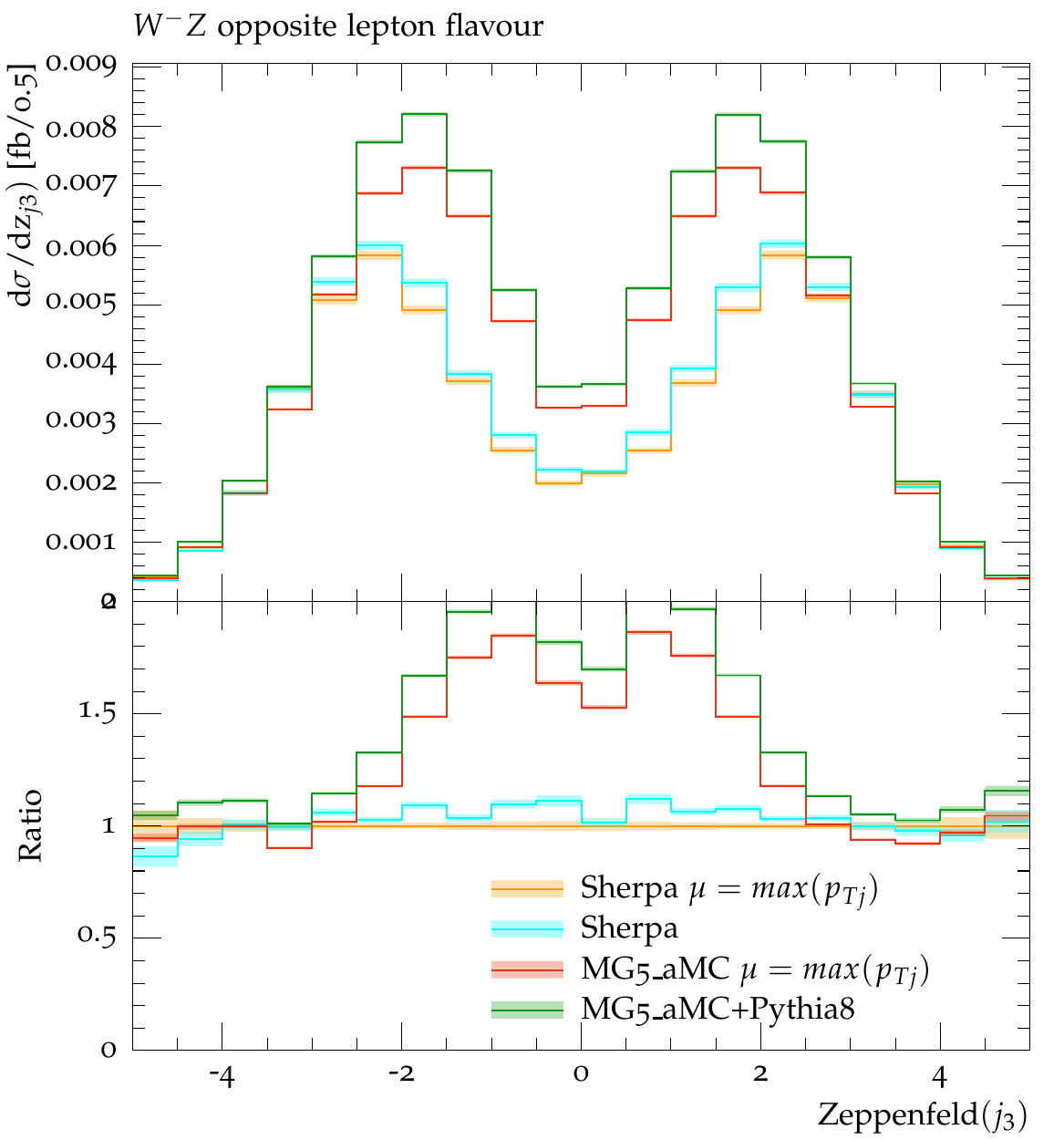}\\[2ex]
   \includegraphics[width=0.48\textwidth]{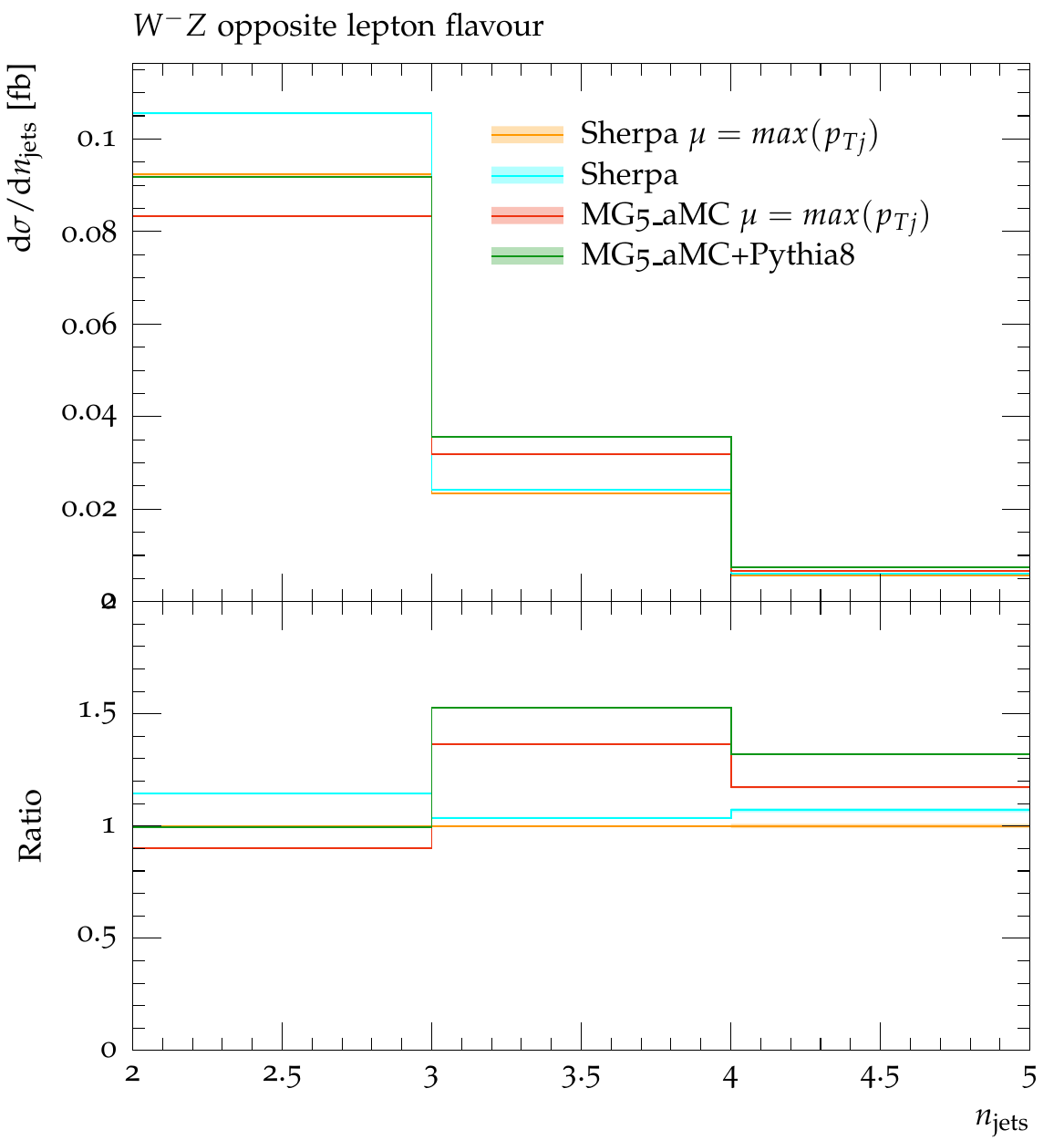}
   \vspace*{2ex}
\caption{Differential distributions at a centre-of-mass energy $\sqrt{s}=13{\rm TeV}$ at the LHC for ${\rm p} {\rm p} \to {\rm e}^-  \nu_{\rm e}  \mu^+ \mu^- {\rm j} {\rm j}$ at LO matched with parton shower with fixed and dynamic scale values: 
                Zeppenfeld variable for the three leptons~(top left),
                Zeppenfeld variable for the third jet~(top right), and
                number of jets~(bottom).
                In the lower plots, the predictions are normalised to the predictions of {\sc Sherpa} with dynamical scale. The error band represents
                the statistical error of the Monte Carlo integrations. For the two \MGaMC
                calculations this statistical error is largely correlated, as explained in the text.}
\label{fig:MC_vbs:shower_2b}
\end{center}
\end{figure}

In Figs.~\ref{fig:MC_vbs:shower_2a} and \ref{fig:MC_vbs:shower_2b} results obtained using the fixed scale $\mu = M_{\rm W}$ and dynamic scale $\mu = {\rm Max}\left[p_{\rm T, j}\right]$ are compared.
In particular, predictions for \MGaMC and {\sc Sherpa} for both scales are presented. For
\MGaMC this prediction is obtained by reweighting each event for the difference between the matrix-element calculation
with the fixed scale and the dynamic scale. The statistical uncertainty is therefore largely
correlated between the two predictions for this Monte Carlo.  
The observables displayed are the same as for the previous comparison.
For the invariant mass of the two jets as well as the transverse momentum of the anti-muon--muon system, for both generators, the use of fixed scale enhances the predictions toward high transverse momentum.
For the rapidity separation of the two jets as well as the distance between the two jets, the shape difference between fixed and dynamical scale is not present for \MGaMC.
On the other hand, in {\sc Sherpa}, the use of fixed scale enhances the predictions for small separations.

Concerning the last set of observables (Zeppenfeld variable of the three leptons and third jet as well as the number of jets) in Fig.~\ref{fig:MC_vbs:shower_2b}, the use of fixed or dynamical scale does not have a large impact.
The only clearly visible difference between fixed and dynamical scale is the normalisation.
This effect is already observed at the level of the fiducial cross section in Tables~\ref{tab:MC_vbs:xsectLOfix} and \ref{tab:MC_vbs:xsectLOdyn}.
The effect of the different scale amounts to a change in normalisation of about $12\%$ for both ${\rm p}{\rm p}\to{\rm e}^+\nu_{\rm e}\mu^+\mu^-{\rm j}{\rm j}$ and ${\rm p}{\rm p}\to{\rm e}^-\bar\nu_{\rm e}\mu^+\mu^-{\rm j}{\rm j}$ processes.

\subsection{Conclusions}
\label{sec:MC_vbs:concl}

The measurement of the EW component of the ${\rm p}{\rm p} \to {\rm W}^{\pm}{\rm Z}{\rm j}{\rm j}$ process constitutes a real challenge for experimental collaborations.
Such a measurement is complicated for several reasons including the low cross section and the large irreducible background.
Therefore, these measurements have to rely on theoretical predictions in order to extract a signal,
which makes theoretical predictions implemented in various Monte Carlo programs very important.
It is thus key to have a good control over these predictions.
To that end we have performed LO comparisons of different theoretical predictions with and without shower and hadronisation.
The comparisons have all been performed with pre-defined input values for the matrix element calculation and a generic fiducial region for VBS measurements.

The first finding of the present study is that the EW component is overwhelmingly dominated by its irreducible QCD background.
In the fiducial region that we have chosen, $80\%$ of the fiducial cross section in accounted by the QCD background.
This indicates the need for observables that enhance the EW contribution.
Beyond the di-jet invariant mass and the rapidity separation between the two jets, we found that the distance between the two jets could have a good discriminating power.
We also found that interferences between the QCD and EW amplitudes are also negligible in the fiducial region.

In the present proceedings, we have also performed a LO comparison.
At parton level, the agreement between the various theoretical predictions is generally very good.
This statement holds both for the inclusive cross section and differential distributions.
Nonetheless differences can appear especially with respect to \MGaMC predictions.
These can probably be attributed to the subtle miss-configuration of the runs presented here or underestimated statistical errors.
On the other hand, the differences between the full computations and the VBS-approximated one are relatively small.
This implies that, as for the ${\rm W}^\pm{\rm W}^\pm{\rm j}{\rm j}$ signature \cite{Anders:2018gfr}, given the present experimental precision, the VBS approximation is satisfactory in the typical fiducial regions used by experimental collaborations for their measurements.

The predictions at LO supplemented with parton shower have relied on the fixed-order configuration.
No tuning of the parton shower parameters have been performed and the default values have been taken.
For {\sc PYTHIA8}, the {\sc CUETP8M1} tune \cite{Khachatryan:2015pea} which is based on the Monash tune \cite{Skands:2014pea} has been used. 
The agreement is in general worse, in particular for observables that are defined only beyond LO, where the theoretical predictions diverge significantly.
Concerning the role of the fixed and dynamical scales, we have found that they can have a rather large influence on the predictions for the inclusive cross section and differential distributions.
The inclusion of higher order predictions may cure this behaviour.

Finally, we would like to stress that these results constitute only a preliminary study of theoretical predictions for the process ${\rm p}{\rm p} \to {\rm W}^{\pm}{\rm Z}{\rm j}{\rm j}$.
In particular, with a strictly fixed-order computation, some differences have been observed.
These propagate to LO predictions matched to parton shower.
In addition, for observables defined beyond LO, large differences due to the parton-shower algorithms appear.
This study demonstrates the difficulty to obtain consistent Monte Carlo predictions.
In particular, the choice of inputs and configuration can have a large impact on physics results.
The effects of NLO QCD and EW corrections have not been studied in this work.
They play a significant role in theoretical predictions for VBS and should be taken into account as much as possible in future studies.
Overall, this study presents a benchmark for the performance of leading-order Monte Carlo generators 
for the simulation of vector boson scattering, and motivates further efforts in the above-mentioned directions that have not 
been fully addressed here.

\subsection*{Acknowledgements}
The work of SB is supported by German Federal Ministry for Education and Research (BMBF) under contract 05H15MGCAA.
MP acknowledges financial support by the BMBF under contract no.~05H15WWCA1 and the German Science Foundation (DFG) under reference number DE 623/6-1. 
EY is supported by Chinese Academy of Sciences (CAS), MoST (Ministry of Science and Technology) through project no.~2013CB837801, and NSFC (National Natural Science Foundation of China) through project no.~11661141007.


\let\Sherpa\undefined
\let\Herwig\undefined
\let\Matchbox\undefined
\let\MGaMC\undefined
\let\MoCaNLO\undefined
\let\Recola\undefined
\let\VBFNLO\undefined



\section{Considerations for the underlying event in jet calibration at the LHC~\protect\footnote{
    J. Huston,
    P. Loch}{}}
\label{sec:SM_UE}


The jet reconstruction algorithms used by the two multi-purpose experiments ATLAS and CMS at the LHC provide jets calibrated at the level of stable particle jets.
These particles are produced by the fragmentation of the hard scattered partons and by the underlying event generated in the same proton--proton collision.
The corresponding calibration procedures apply corrections that not only remove the signal contribution from diffuse particle emissions from multiple proton collisions in the same bunch crossing, but also signal remnants from previous (or future) bunch crossings  affecting the detector signals.   
These corrections typically reduce or remove the underlying event contribution to the experimental jets to an indeterminate degree. It is desirable to add the underlying event contribution back into the truth jet definition, as the detector jet is supposed to be calibrated to the full particle level for physics analysis.
The recovery of this contribution in the applied jet calibration procedures is an inherent feature of the Monte Carlo based jet energy scale correction, which provides the absolute calibration for jets in simulation and experiment. 
The model-dependent nature of this approach is emphasized and the considerations of model variations in the determination of the systematic uncertainties are summarized.

\subsection{Introduction}
The jet calibration applied by both ATLAS \cite{PERF-2007-01} and CMS \cite{Chatrchyan:2008aa} includes a necessary correction to remove signal contributions from multiple proton--proton interactions happening in the same LHC bunch crossing (\emph{in-time pile-up}), together with contributions from preceding and possibly even following bunch crossings (\emph{out-of-time pile-up}).
In both Run 1 and Run 2, approaches based on the jet-area-based correction described in Ref.~\cite{Cacciari:2007fd} have been used.
Its stochastic nature does not distinguish between pile-up contributions and contributions from the underlying event (UE) when subtracting the estimated pile-up signal from a reconstructed jet. 
An additional consideration when discussing the UE signal inside and outside of jets is that
the significant pile-up at high LHC intensities, with a typical average number of pile-up collisions per bunch crossing of about $\langle\mu\rangle \approx 20$ towards the end of Run 1, and rising to around $\langle\mu\rangle \approx 60$ during the recently concluded 2017 data taking period of Run 2, may render the UE contribution insignificant. 
Nevertheless, after corrections are applied to the detector jet to mitigate the effect of pile-up at any level, the UE contribution approaches scales of relevance, in particular for jets with low transverse momentum.    
This is even more important particular true when considering the typical precision reached for jet energy measurements at LHC, which in large regions of phase space approaches~1\%.  

This note briefly summarises a typical jet calibration scheme and the resulting representation of the UE in the calibrated jet signal. 
The strategies are similar for CMS and ATLAS, even though the detector signals entering the jet calibration are different. 
The discussion here focusses on the UE signal and the jet calibration procedures applied in ATLAS.  

\subsection{Pile-up mitigation in jet reconstruction}

The jet reconstruction in the two multi-purpose experiments at the LHC uses particle-flow objects in CMS \cite{Sirunyan:2017ulk} and clusters of topologically connected calorimeter cell signals (topo-clusters) in ATLAS \cite{Aad:2016upy}.
The principal pile-up mitigation technique applied in both experiments is the (scalar) jet-area-based pile-up subtraction method introduced in Ref.~\cite{Cacciari:2007fd}. 
It is applied right after jet reconstruction, before any absolute calibration.

The set of topo-clusters in ATLAS reconstructed for a given event encompasses contributions from the hard scatter final state and the two softer contributions, UE and pile-up.
Both pile-up and UE generally manifest themselves in generating additional topo-clusters with low energies (\emph{direct} contribution).
The inherent possibility of merging of signals of particles from the three sources in the topo-cluster formation algorithm, together with the splitting of topo-clusters implicit to this algorithm, leads to complex hard scatter signal modifications by the two soft contributions (\emph{indirect} contribution) with significant variations from event to event, even with similar levels of soft activity.   
In addition, the overlap in phase space and the basically identical emission characteristics in terms of average transverse momentum flow between pile-up and UE makes it basically impossible to remove the pile-up signal contribution without affecting the UE contribution, in particular in high pile-up conditions like the ones observed at the end of Run 1 and during Run 2.

The formation of topo-clusters is guided by signal significance (signal-to-noise) patterns steered by thresholds adjusted to expected noise levels for a given run year at LHC.
The increasing $\langle\mu\rangle$ in LHC operations from early to late Run 1, from Run 1 to Run 2, and recently in Run 2 leads to a reduced sensitivity of the ATLAS calorimeter to UE in general.\footnote{The transverse momentum scale of UE contributions is set by $\hat{s}$ in each hard scatter proton--proton interaction, and thus changed only slightly in Run 1 when LHC moved from $\sqrt{s} = 7$ TeV to $\sqrt{s} = 8$ TeV. It changed a bit more significantly from Run 1 to Run 2, when $\sqrt{s} = 8$ TeV increased to $\sqrt{s} = 13$ TeV.}
The noise thresholds rise approximately proportional to $\sqrt{\langle\mu\rangle}$, which means that e.g. the energy needed to seed a topo-cluster at $\langle\mu\rangle = 40$ is two times higher than at $\langle\mu\rangle =10$. 
This renders a significant portion of the UE invisible at higher pile-up conditions, especially in signals outside of jets.     
 
All topo-clusters with a positive energy signal are used for finding jets. 
This means that the topo-clusters collected into one jet include direct and indirect contributions from all three sources, and the total jet kinematics, which is reconstructed from  the four-momentum recombination of all these clusters, is rather sensitive in particular to pile-up, which dominates at high proton intensities. 

Early in Run 1, when pile-up was at a relatively low level ($\langle\mu\rangle < 10$), ATLAS used a purely Monte Carlo (MC) simulation based mitigation technique, which actually formally preserved the UE contribution in a jet as much as it was visible in the detector at all.
In this approach the transverse momentum ($p_{\text T}$) contribution from pile-up was measured by the dependence of the ratio of the initial reconstructed jet -$p_{\text T}$ to the expected $p_{\text T}$ from a matched particle jet in MC simulations.
This dependence was evaluated with respect to the in-time pile-up represented by the number of reconstructed primary vertices $N_{\text PV}$ and to the out-of-time pile-up measured by the average number of pile-up collisions $\mu$ around the signal event (see Ref.~\cite{Aad:2016upy} for details).
The correction was then applied to each jet in MC simulations and experimental data using the average slopes $\partial p_{\text T}/\partial N_{\text PV}$ and $\partial p_{\text T}/\partial\mu$, together with the obvious assumption that the transverse momentum contribution from pile-up is expected to vanish for $N_{\text PV} = 1$ and $\mu = 0$.
As the UE contribution is expected to not depend on either $N_{\text PV}$ or $\mu$, this approach in principle does not remove it from the jet signal.    
It was purely MC based and required complex validation in data \cite{ATLAS-CONF-2014-018}.
In addition, this approach is only sensitive to the expected fluctuations in the in-time pile-up activity, where the number of additional proton--proton collisions fluctuates from bunch crossing to bunch crossing following a Poisson distribution around $\langle\mu\rangle$.
Similarly, the out-of-time pile-up induced calorimeter signal modifications are only corrected at an average level.  
Further fluctuations in the pile-up contribution to the reconstructed jet-$p_{\text T}$ even in events recorded in bunch crossings with the same $N_{\text PV}$ and $\mu$, which reflect varying $p_{\text T}$-flow patterns in the in-time pile-up collisions as well as in surrounding (previous) bunch crossings, are not addressed and limit the efficiency of this pile-up mitigation technique. 
 
 The recent approach implements the jet-area-based pile-up mitigation technique suggested in Ref.~\cite{Cacciari:2007fd}. 
 In this case the calorimeter signal itself is used to determine the level of pile-up in each event.
 This signal is sensitive to the in- and out-of-time $p_{\text T}$-flow patterns, and this technique thus reduces bunch crossing to bunch crossing fluctuations.  
The initially reconstructed small radius jet with transverse momentum $p_{\text{T}}^{\text{jet,raw}}$ is corrected by using its area $A^{\text{jet}}$ and the median transverse momentum density $\rho$ to a basic transverse momentum measure $p_{\text{T}}^{\text{jet,basic}}$ such that   
\begin{align}
	p_{\text{T}}^{\text{jet,basic}} = p_{\text{T}}^{\text{jet,raw}} - \rho \times A^{\text{jet}}\,.
	\label{eq:SM_UE:jab}
\end{align}
This procedure removes a signal of the scale of the $p_{\text T}$ contribution  from pile-up, together with the part of the UE which is represented by topo-clusters outside of the jet.\footnote{CMS uses the same technique but based on particle flow signals \cite{Chatrchyan:2011ds}.} 
The measurement of $\rho$ uses topo-clusters that are most likely outside of jets, due to its construction as a median transverse momentum density.   
Residual \emph{local} fluctuations of the signal contribution from pile-up to a given reconstructed jet are not corrected. 
Reconstructed calorimeter jets with $p_{\text{T}}^{\text{jet,basic}} < p_{\text{T}}^{\text{min}}$ are considered as originating from pile-up and thus are dropped.\footnote{The actual value of $p_{\text{T}}^{\text{min}}$ is adjusted to the pile-up conditions. It is typically of $\mathcal{O}(20\text{\ GeV})$.}

This UE contribution to $\rho$ in terms of topo-clusters is hard to determine independently in the experiment, even in low pile-up scenarios.\footnote{The $p_{\text{T}}$-flow from charged particles associated with the UE can be directly measured using reconstructed tracks from the hard scatter vertex, at least within the tracking detector acceptance of $|\eta|<2.5$, following the strategy lined out in Ref.~\cite{Field:2000dy}.}
In the presence of significant pile-up, with the already discussed higher clustering thresholds, the $\rho$ contribution from the UE is likely very small, in particular in terms of topo-clusters with a signal dominated by a particle from the UE. 
Here the indirect UE contribution merged into pile-up or hard scatter seeded clusters is more relevant, yet with a relatively small effect on the scale for $\rho$.
It is also expected to be different inside and outside of jets, due to the clustering algorithm, which leads to a higher probability of small signal survival in the presence of the larger signals especially in the core of the jet.
In this respect, UE and pile-up emissions close to the jet axis are likely to still contribute to the jet signal even after the jet-area-based correction, at a  harder scale.

In addition to the jet-area-based correction given in Eq.~\eqref{eq:SM_UE:jab}, MC simulation based corrections similar to the ones discussed for low pile-up conditions are applied to $p_{\text{T}}^{\text{jet,basic}}$ to mitigate residual dependencies on $N_{\text PV}$ and $\mu$.
These small corrections are again not expected to change the UE contribution.

\subsection{Calibration of narrow jets}

\begin{figure}[t!]\centering
\includegraphics[width=0.95\textwidth]{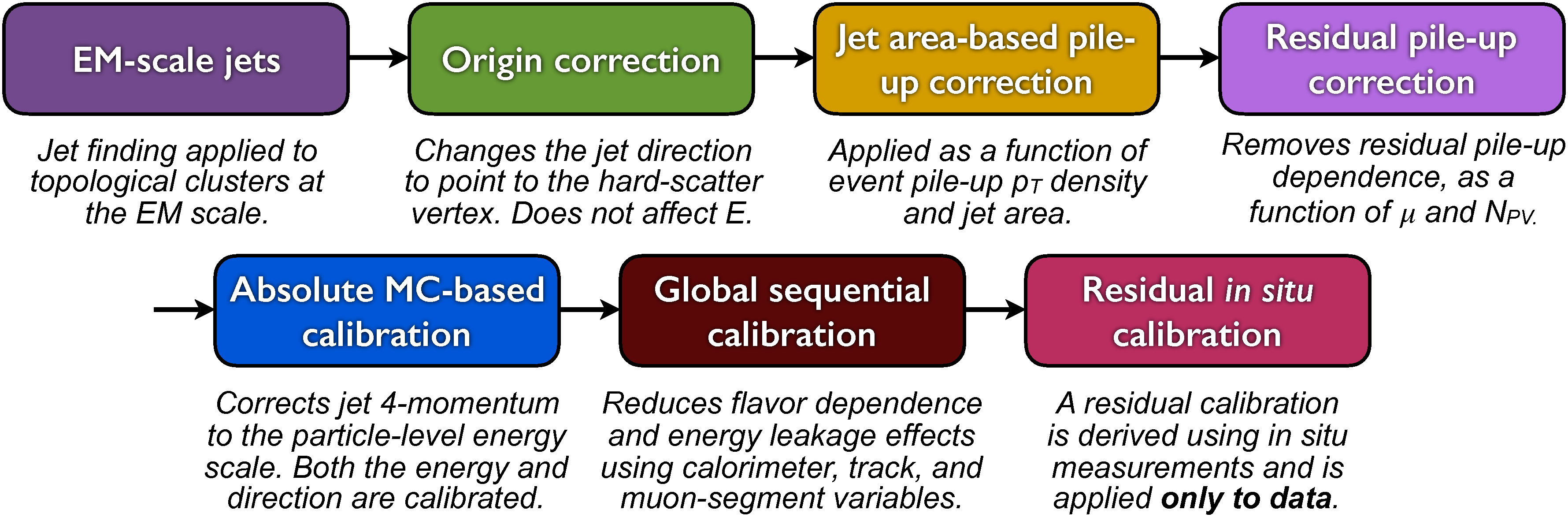}
\caption{The ATLAS jet calibration scheme for early (2015) LHC Run 2 (taken from Ref.~\cite{Aaboud:2017jcu}).}
\label{fig:SM_UE:cal}
\end{figure}
The full chain of jet calibration in ATLAS, as applied in 2015, is described in Ref.~\cite{Aaboud:2017jcu} and schematically shown in Fig.~\ref{fig:SM_UE:cal}.
When jets are clustered from topo-clusters on the electromagnetic energy scale \cite{Aad:2016upy} (\emph{EM-scale jets}), their direction is reconstructed with respect to the nominal detector centre. 
As a first step of the calibration chain, their direction is recalculated with respect to the reconstructed hard scatter vertex (\emph{origin correction}).
The \emph{jet-area-based pile-up correction}, as discussed in the previous section, is applied next, followed by the previously mentioned \emph{residual pile-up correction}.
After this, the jets are subjected to the \emph{absolute MC-based calibration}.
 
 This absolute jet energy scale (JES) calibration is determined by matching calorimeter jets with truth particle jets in full MC simulations including pile-up, in QCD dijet calibration samples generated with a default MC generator configuration\footnote{In 2015, the $2 \to 3$ matrix element calculation in Powheg-Box 2.0 \cite{Nason:2004rx,Frixione:2007vw} employing the CT10 PDF set \cite{Lai:2010vv} is used combined with parton showers and underlying event from Pythia8.1 \cite{Sjostrand:2007gs}, and the A14 tune of UE parameters \cite{ATL-PHYS-PUB-2014-021}.}   and a detector simulation configuration.
 The match is based on the angular distance of the two jets and thus purely geometrical.
 The calorimeter jet energies $E_{\text{calo}}^{\text{jet,raw}}$ include all contributions from in-time and out-of-time pile-up, and the UE. 
 After all corrections up to the ones for pile-up are applied ($E_{\text{calo}}^{\text{jet,raw}}\to E_{\text{calo}}^{\text{jet,basic}}$),
 the ratio of reconstructed calorimeter jet energy $E_{\text{calo}}^{\text{jet,basic}}$ to the truth jet energy $E_{\text{truth}}^{\text{jet}}$ is collected from the calibration sample in bins of $E_{\text{truth}}^{\text{jet}}$ and the truth jet pseudorapidity $\eta_{\text{truth}}^{\text{jet}}$,
 \begin{align*}
 \mathcal{R}(E_{\text{truth}}^{\text{jet}},\eta_{\text{truth}}^{\text{jet}}) = \dfrac{E_{\text{calo}}^{\text{jet,basic}}}{E_{\text{truth}}^{\text{jet}}}\,.
 \end{align*} 
The average $\langle\mathcal{R}\rangle(E_{\text{truth}}^{\text{jet}},\eta_{\text{truth}}^{\text{jet}})$ is determined in each bin of $E_{\text{truth}}^{\text{jet}}$ in a given slice of $\eta_{\text{truth}}^{\text{jet}}$.
Using numerical inversion to project
\begin{align*} 
\langle\mathcal{R}\rangle(E_{\text{truth}}^{\text{jet}},\eta_{\text{truth}}^{\text{jet}} \mapsto \langle\mathcal{R}\rangle(E_{\text{calo}}^{\text{jet,basic}},\eta_{\text{calo}}^{\text{jet,basic}})\,,
\end{align*} and inverting the averages 
\begin{align*}
\langle\mathcal{R}\rangle(E_{\text{calo}}^{\text{jet,basic}},\eta_{\text{calo}}^{\text{jet,basic}}) \mapsto \langle\mathcal{R}\rangle^{-1}(E_{\text{calo}}^{\text{jet,basic}},\eta_{\text{calo}}^{\text{jet,basic}})\,,
\end{align*}
yields a calibration function when fitting a smooth functional form over all energies in an $\eta_{\text{calo}}^{\text{jet,basic}}$ slice. 

The truth jet includes all stable particles generated in MC simulations, except for neutrinos, muons and any particle\footnote{Due to the large number of pile-up particles, those are not even stored in the calibration samples.} emerging from the overlaid in-time pile-up interactions. 
The remaining set comprises all particles with lifetimes $\tau$ given by $c\tau > 10$ mm in the laboratory frame.
Among those are particles generated by the fragmentation of the parton showers associated with the hard scatter partons as well as the particles generated by radiation, and by multiple parton interactions, all characteristic for the UE.
The total truth jet energy and, to a lesser extent, its direction are thus affected by the UE. 
With this respect, the UE energy contribution is reinstated in the calorimeter jet after this absolute calibration -- the calibrated calorimeter jet energy corresponds on average to the truth jet energy in the same phase space, and thus contains an average UE at particle level. 

This approach scales the calorimeter jet to a model dependent JES. 
Direct application of this calibration to data projects this model dependence into the jet measurement.
To understand the bias introduced by this calibration method, alternative generators with different orders of, or approaches to, the hard process calculation and/or different choices of the parton distribution functions (PDFs), fragmentation models and sets of tuned UE parameters are used in the simulation of  calorimeter jets, which are then subjected to the jet calibration derived with the default configuration.
Recent (2015) generators considered are Sherpa 2.1 \cite{Gleisberg:2008ta} employing multi-leg $2\to N$ matrix elements and their own default tuned set of soft event modelling parameter, and Herwig++ 2.7 \cite{Corcella:2000bw,Bahr:2008pv} with a $2\to 2$ matrix element and their own default tuned parameter set.
Any observed differences of the alternative calorimeter JES with respect to the default JES are absorbed into a systematic uncertainty.
Due to the small contribution that can be expected from the UE in recent pile-up conditions, no particular uncertainty component is attributed to this modelling feature.

It is notable that the actual survival of the UE in the calorimeter signal space does not really enter into the jet calibration, as long as the jets in the calibration samples are subjected to the same reconstruction chain, up to and including the full pile-up correction, as any other jet in MC simulations and in experimental data.
No matter how much UE signal might have been removed in this chain, the jets will consistently be calibrated including the UE.
Significant changes in the default truth jet generator or in the detector simulation require a redetermination of the calibration functions.

In addition, significant changes to any of the steps leading up to calibration typically require a new round of determining the absolute MC calibration functions, such that calibrated calorimeter jets on average maintain the same truth JES.  
In cases where the changes in the upstream  corrections are insignificant, the calibration functions may not be changed and additional systematic uncertainty components may be injected. 
Changes in the pile-up activity e.g. due to increasing beam intensities at the LHC, may not require a full recalibration if the pile-up mitigation maintains its hard scatter jet finding efficiency.  
Smaller changes in the simulated detector response also add an uncertainty component.
A rather complete evaluation is described in \cite{Aad:2014bia}.

The next step after the absolute MC calibration is the \emph{global sequential calibration}, which attempts to reduce fluctuations in the jet response by reducing its sensitivity to the jet flavour composition and possible longitudinal energy leakage in the calorimeters.
This step is designed to not change the average response in any given phase space bin anymore, but to improve the resolution of the response function.
  
The origin correction is independently applied in data and MC simulations, as it only needs the reconstructed hard scatter vertex.
The jet-area-based pile-up correction is also independently applied, as it only needs the reconstructed $\rho$ in any given  event. 
Assuming that this $\rho$ reflects the pile-up (and possibly UE) contribution to jets in the same way in data and MC simulations, the remaining jet signal should have the same quality (e.g. with respect to the level of pile-up suppression) in both environments. 
In this context it is not required that $\rho$ is actually the same in data and MC simulations (their distributions are typically found to be quite similar).
This introduces a certain level of insensitivity to the actual pile-up simulation, which is typically far from perfect, see e.g. case studies in Ref.~\cite{Aad:2016upy}.       

The residual pile-up correction, the absolute calibration and the global sequential calibration are all derived from MC simulations. 
A final set of  in-situ measurements provide calibrations that are applied only to data and lead to the same average jet response in data and MC simulations.  

\subsection{Large radius jets}

At LHC, jets with large radius ($R \geq 0.8$) are often considered in searches for new particles or boosted Standard Model particles.
The large catchment area in $(y,\phi)$-space associated with these jets, especially their large extension in rapidity $y$, prohibits the straight forward application of a jet-area-based pile-up correction at the typically required levels of precision (e.g., $\mathcal{O}(1\%)$ for the $p_{\text T}$ scale).  
This is largely due to the $y$ range covered by these jets and the fact that both the number density of particles emerging from minimum bias (soft) proton--proton collisions and the average $p_{\text T}$ are $y$-dependent.
This suggests a significant variation of $\rho$ inside the large radius jet.
Additional complications may arise from a changing calorimeter readout granularity across such a large detector object.

Often large radius jets are groomed with e.g. trimming \cite{} to extract hard radiation pattern inside the jets. 
Such techniques often remove pile-up and inadvertently UE contributions from the jet, with significant effects on the reconstructed jet kinematics, including the single jet mass.
While the UE contribution can indeed affect the jet mass in ungroomed jets, in particular those without two- or three-prong decay substructure, the grooming reduces the impact of signals from any soft emission into the large radius jet.

 The phase space of interest for large radius jet applications is usually determined by a search or decay tag goal. 
 Assuming an attempt to tag hadronic $W$ boson decays in a jet of $R =1.0$, a typical (lower) $p_{\text T}$ scale is given by $p_{\text T} \approx 2 \times M_{W}$, where $M_{W}$ is the mass of the  $W$ boson. 
 Higher masses introduce higher $p_{\text T}$ scales accordingly. 
 Even for the largest practical jet radii, the UE contribution to the overall $p_{\text T}$ is insignificant, in particular in high pile-up conditions.
 Any loss of UE signal after grooming seems to be acceptable when jet-mass-based searches are conducted.   
 
 Any grooming attempt introduces a non-trivial modification of the jet composition, and as such needs careful reflections using MC simulations and careful validation with experimental data. 
 In particular scenarios where UE may significantly affect a reconstructed jet feature, modelling dependencies have to be studied and incorporated into any relevant uncertainty or efficiency.
 
A cleaner inclusion of the UE into large radius jets by using jet re-clustering \cite{Nachman:2014kla} is emerging for ATLAS (see first evaluations in Ref.~\cite{ATLAS:2017csu}).
The clustering of narrow jets into large radius jets provides fully corrected and calibrated inputs to any grooming technique and reconstruction of substructure-related observables for these large jets.
Each narrow jet includes a contribution from the UE, as discussed in the previous section, and is corrected for pile-up as well. 
In analyses where softer radiation or energy flow between harder structural elements of the large radius jet are not explored, the re-summed large jet kinematics, including its mass, correctly include the UE within the physics model used for the narrow jet calibration.

\subsection{Conclusions}

Both ATLAS and CMS apply a MC simulation based absolute jet energy calibration that includes the particle-level underlying event contribution. 
Independent of how well this contribution is actually represented in the detector signals -- clusters of topologically connected calorimeter cell signals in ATLAS and particle flow objects in CMS -- the final jet energy scale reflects the underlying event as modelled by the physics models used in the determination of the absolute calibration.
A limited but well motivated set of models and modelling parameters is used to evaluate systematic uncertainties introduced to the calibration of the jets in the experiment by the choice of a certain default MC simulation.
The concern that any pile-up correction removes underlying event signals is valid but has no detectable influence on the jet energy scale, as long as all corrections and calibrations are derived and applied in a consist manner.
Final calibration functions derived from in-situ transverse momentum balance studies applied to data only shift the experimental jet energy scale to the one from the default MC simulation. 
In this the underlying event contribution is scaled in the same way as the overall jet energy scale. 

The complexity of the jet calibration procedure in both CMS and ATLAS, and the misunderstandings arising from it, encouraged this pedagogical contribution to the proceedings.







\newcommand{\Herwig}{H\protect\scalebox{0.8}{ERWIG}\xspace}
\newcommand{\Pythia}{P\protect\scalebox{0.8}{YTHIA}\xspace}
\newcommand{\Sherpa}{S\protect\scalebox{0.8}{HERPA}\xspace}
\newcommand{\pythia}{\Pythia}
\newcommand{\sherpa}{\Sherpa}

\newcommand{\nnlojet}{NNLO\protect\scalebox{0.8}{JET}\xspace}

\newcommand{\Vincia}{V\protect\scalebox{0.8}{INCIA}\xspace}
\newcommand{\Dire}{D\protect\scalebox{0.8}{IRE}\xspace}
\newcommand{\vincia}{\Vincia}
\newcommand{\dire}{\Dire}

\newcommand{\Rivet}{R\protect\scalebox{0.8}{IVET}\xspace}
\newcommand{\Professor}{P\protect\scalebox{0.8}{ROFESSOR}\xspace}
\newcommand{\eps}{\varepsilon}
\newcommand{\mc}[1]{\mathcal{#1}}
\newcommand{\mr}[1]{\mathrm{#1}}
\newcommand{\mb}[1]{\mathbb{#1}}
\newcommand{\tm}[1]{\scalebox{0.95}{$#1$}}


\section{Precision comparisons of predictions for Higgs boson + jet production at the LHC as a function of jet size~\protect\footnote{
           J.~Bellm,
           A.~G.~Buckley,
           X.~Chen,
           J.~R.~Currie, 
           A.~Gehrmann-De~Ridder,
           T.~Gehrmann, 
           E.~W.~N.~Glover,
           S.~H{\"o}che, 
           A.~Huss, 
           J.~Huston, 
           S.~Kuttimalai,
           J.~Pires,  
           S.~Pl{\"a}tzer, 
           M. Sch{\"o}nherr}{}}
\label{sec:SM_Higgs_jet_R}

  We perform a phenomenological study of Higgs-boson production
  in association with a jet at the Large Hadron Collider. We investigate
  in particular the dependence of the leading jet cross section on the
  jet radius as a function of the jet transverse momentum. Theoretical
  predictions are obtain using perturbative QCD calculations up to
  next-to-next-to-leading order. They are compared to
  results obtained from matching next-to-leading order calculations
  to parton showers and possibly including higher-order real radiative
  effects through multi-jet merging.

\subsection{Introduction}

During Les Houches 2015~\cite{Badger:2016bpw}, a detailed comparison of fixed
order (at NLO and NNLO) and matrix element plus parton shower (MEPS) predictions
for differential Higgs boson (+jet) production was carried out.  The goal was
multi-fold: using identical starting configurations, to check the consistency of
the MEPS predictions among themselves, and to demonstrate that the MEPS
predictions revert to their underlying fixed order predictions in non-Sudakov
kinematic regions. These comparisons largely showed good agreement among the
ME+PS predictions and that the ME+PS predictions agreed reasonably well with
their fixed-order counterparts. 

We have continued these studies in Les Houches 2017, but now looking in finer
detail at the level of agreement. In particular, the perturbative jet shape is
not as well modeled at fixed order as with the NLO matrix element plus parton
shower predictions that are available. This may have a quantitative impact on
the cross section predictions, especially for small jet sizes.   

The study is designed to use Higgs+jet, Z+jet and inclusive jet production,
taking advantage of the NNLO calculations available for all three processes.
The latter two processes are important for global PDF fits, where only fixed
order predictions (along with the relevant non-perturbative corrections) have
been used.  Due to time constraints, only the Higgs+jet process will be
considered in detail in these proceedings. The other two processes, along with
Higgs+jet,  will be included in a more comprehensive study in a future
publication. 

As a  further test of the impact of parton showers versus fixed order, the jet
size was varied over the values $R\in$ [0.3, 0.4, 0.5, 0.6, 0.7, 1.0], using the anti-$k_T$ jet
algorithm~\cite{Cacciari:2008gp}. Indirectly, this tests how well the one (two) extra gluon(s) at
NLO (NNLO) reproduce the perturbative aspects of the jet shapes, as embodied by
the parton shower. This is of particular interest as the Higgs boson + jets
measurements that have been performed at the LHC in Run 2 have used a jet size
of 0.4, which is slightly above the jet size region where small $R$ effects may be
important. Taking them into proper account would require resummation, as
discussed in~\cite{Dasgupta:2016bnd}. The MEPS predictions also basically
provide this resummation, through the parton showers. The MEPS predictions at
NLO can properly take these effects into account. However, the highest precision
for $H+\ge1$ jet is from the NNLO predictions, and there is no MEPS formalism
that works at this order.   As an additional motivation, there has been recent
speculation that the quark jet shape is not well-described at NLO. This study
provides samples of both quark and gluon jets to test this hypothesis.

  Predictions from MEPS programs were carried out at the parton shower level
(for easiest comparison to fixed order).%
\footnote{In the future publication,
comparisons will also be made at hadron level, as a way of comparing the
non-perturbative corrections as a function of jet radius.  As a reminder, the
non-perturbative corrections used for fixed order predictions are determined
from a comparison of the parton shower predictions with and without the
non-perturbative effects, as a function of jet radius. This implicitly requires
the integrated jet shape determined by fixed order predictions to agree
with those determined by parton showers.}

To the degree to which it was possible, the initial conditions have been
constrained to be the same for all predictions. Each calculation used the
PDF4LHCNNLO\_30 PDFs~\cite{Butterworth:2015oua}, with its central value of
$\alpha_s(m_Z)$ of 0.118. As its name implies, this PDF has 30 error PDF sets
that completely determine the PDF uncertainty. The scale choices for all
predictions have been designed to be as similar as possible, providing a greater
level of control than was available in the 2015 Les Houches study. More detail
will be provided in the sub-sections dealing with each prediction. 
A CMS Rivet~\cite{Buckley:2010ar} routine from the 13~TeV  inclusive jet
analysis~\cite{Khachatryan:2016wdh}, was modified to add the different $R$ values,
as well as additional variables. This Rivet routine was further modified for the
Higgs boson + jet (and Z boson + jet) studies.

\subsubsection{\texorpdfstring{\nnlojet}{NNLOJET} - NNLO Calculation of \texorpdfstring{$pp\to H + j$}{pp->H+j}}

The NNLO corrections to $pp \to H+j$ receive contributions from three types of
parton-level subprocesses: the $H$+5~parton tree-level
process~\cite{DelDuca:2004wt,Dixon:2004za,Badger:2004ty} (double-real
correction), the $H$+4~parton process at
one-loop~\cite{Badger:2009hw,Badger:2009vh,Dixon:2009uk} (real--virtual
correction), and the two-loop $H$+3~parton process~\cite{Gehrmann:2011aa}
(double-virtual correction).  Each of these contributions are separately
infrared (IR) divergent, while the divergences cancel upon integration over the
unresolved phase space by virtue of the Kinoshita--Lee--Nauenberg theorem.  In
order to arrive at a fully differential prediction, a procedure for the
subtraction of IR singularities is required to make this cancellation manifest.
To this end, we employ the antenna subtraction
formalism~\cite{GehrmannDeRidder:2005cm,GehrmannDeRidder:2005aw,GehrmannDeRidder:2005hi,Daleo:2006xa,Daleo:2009yj,Gehrmann:2011wi,
Boughezal:2010mc,GehrmannDeRidder:2012ja,Currie:2013vh} to extract the IR
singularities from the various contributions and to achieve their cancellation
prior to the phase-space integration.  As a result, the phase-space integration
over the contributions with different parton multiplicities are individually
rendered finite and can therefore be performed numerically using Monte Carlo
techniques.  The antenna subtraction formalism is implemented in the \nnlojet
framework. The application to NNLO corrections for $H+j$ production and its
validation against an independent calculation using the sector-improved residue
subtraction method~\cite{Boughezal:2015dra} was described in
Refs.~\cite{Chen:2014gva,Chen:2016vqn,deFlorian:2016spz}. 

For the current study, we employ the dynamical scale
\begin{equation}
  \mu_0 = \frac{1}{2} H_T = \frac{1}{2}\bigg(\sqrt{m_H^2+p_{T,H}^2}+\sum_{jet} p_{T,jet}\bigg) ,\label{eq:SM_Higgs_jet_R:htprimescale}
\end{equation}
as our central scale choice.  The renormalisation ($\mu_R$) and factorisation
($\mu_F$) scales are varied independently around $\mu_0$ by factors of
$\tfrac{1}{2}$ and $2$ to estimate the size of missing higher-order
contributions.  Here, the two extreme variations are excluded such that we
arrive at the custom 7-scale variation:
\begin{equation}
  (\mu_R,\mu_F) = \bigl\{ 
  (1,1), \;
  (2,2), \;
  (\tfrac{1}{2},\tfrac{1}{2}), \;
  (\tfrac{1}{2},1), \;
  (1,\tfrac{1}{2}), \;
  (2,1), \;
  (1,2)
  \bigr\} \times \mu_0 .
\end{equation}
The LO and NLO contributions using this dynamical scale choice were validated
against Sherpa and Herwig7 as described in other sections in this report. 

In order to obtain the results for the various cone sizes ($R=0.3$, $0.4$,
$0.5$, $0.6$, $0.7$, and $1.0$) required in this study, we have exploited the
fact that the Born-level kinematics is insensitive to $R$.  As a result, the
difference between two cone sizes can be obtained from a $H+2j$ calculation at
one order lower:
\begin{equation}
  \sigma^{NNLO}_{H+j}(R) - \sigma^{NNLO}_{H+j}(R') =
  \int \bigl[
  \mathrm{d}\sigma^{NLO}_{H+2j}(R) - \mathrm{d}\sigma^{NLO}_{H+2j}(R')
  \bigr]_{njets\geq1} .
  \label{eq:SM_Higgs_jet_R:multi-run}
\end{equation}
Note that the difference has to be taken at the level of the integrand, since
one term acts as a local counter-term of the other in all IR-divergent limits
where a jet becomes unresolved and $H+2j \to H+j$.  Using
Eq.~\eqref{eq:SM_Higgs_jet_R:multi-run}, all predictions for different $R$ values can be
obtained from a single NNLO computation by adding differences of $H+2j$
computations that are substantially cheaper to compute.

\subsubsection{Herwig}

We used \Herwig7~\cite{Bahr:2008pv,Bellm:2015jjp,Platzer:2011bc} based on
version 7.1.2  and ThePEG version 2.1.2 with minor changes to standard \Herwig7
scale settings to the agreed ones. The NLO matching was performed with matrix
elements form OpenLoops~\cite{Cascioli:2011va} and MadGraph~\cite{Alwall:2011uj}
interfaced with the BLHA2 standard~\cite{Alioli:2013nda}.  For parton
distributions the PDF interface from LHAPDF6~\cite{Buckley:2014ana} was used. In
the results we show matched $NLO \oplus PS$ predictions -- to emphasise the
similarities of matching and merging at high transverse momenta -- with the
$\tilde{Q}$-shower but tested with lower statistics that merging according 
to~\cite{Bellm:2017ktr} and matching to the \Herwig7 dipole
shower~\cite{Platzer:2009jq} show similar behaviour. Hadronisation and MPI models
are switched off and $\alpha_S$ of the hard process is synchronized with the PDF
set. We include effects of CMW scheme~\cite{Catani:1990rr} by an enhanced
shower $\alpha_S=0.124$.  The scale used for the core process in the Matching is
defined as in Eq.~\eqref{eq:SM_Higgs_jet_R:htprimescale}. As shower starting scale we use the
transverse momentum of the hardest jet.%
\footnote{The effect of changing this
scale to the one in Eq.~\eqref{eq:SM_Higgs_jet_R:htprimescale} is moderate and only noticeable
for transverse momenta smaller or in the range of the Higgs mass.}

\subsubsection{Sherpa}
We use a pre-release version of the \Sherpa Monte Carlo event 
generator~\cite{Gleisberg:2003xi,Gleisberg:2008ta}, based on version Sherpa-2.2.4.  The
NLO matching is performed in the S-MC@NLO
approach~\cite{Hoeche:2011fd,Hoeche:2012ft}, while the multi-jet merging
employed is based on the MEPS@NLO
formalism~\cite{Gehrmann:2012yg,Hoeche:2012yf}. We use a modified version of a
parton shower algorithm~\cite{Schumann:2007mg}, which is based on Catani-Seymour
dipole subtraction~\cite{Catani:1996vz,Catani:2002hc}. We use a running coupling
consistent with the PDF, and we employ the CMW scheme to include the two-loop
cusp anomalous dimension in the simulation~\cite{Catani:1990rr}.  In the merging
procedure a core scale needs to be defined.  To make the result comparable to
the FO result we use the same definition as in Eq.~\eqref{eq:SM_Higgs_jet_R:htprimescale}. If no
ordered clustering history can be defined,  the couplings in the effective
gluon-gluon-Higgs-vertex are evaluated at this core scale. 

\subsection{Results}
The analyses use the anti-$k_T$ jet algorithm, with varying jet size as described
above, with a jet transverse momentum threshold of 30~GeV, along with an
(absolute) rapidity cut on the jets of 4.5. To avoid generation cut effects, the
comparisons are performed above a jet transverse momentum value of 50~GeV.
\begin{figure}[t]
  \centerline{\includegraphics[width=\textwidth]{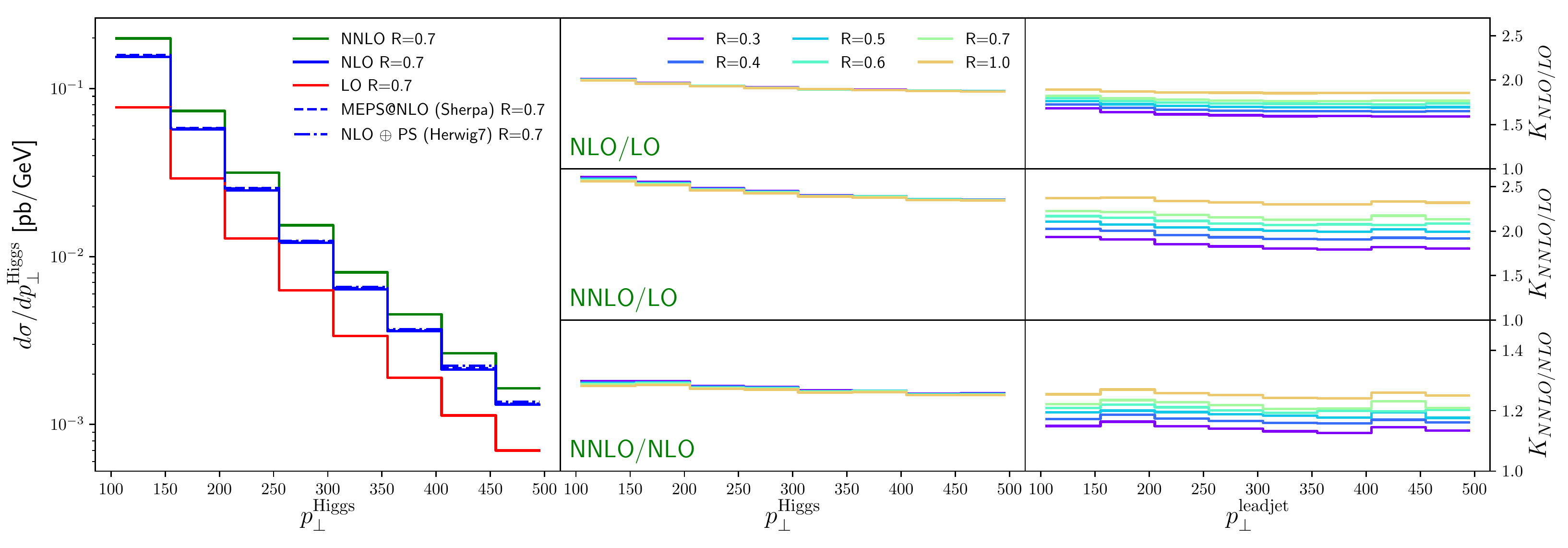}}
  \caption{
    Shown on the left panel is the transverse momentum spectra of the Higgs boson $(p^\mathrm{Higgs}_\perp)$ and the
    leading jet $(p^\mathrm{leadjet}_\perp)$ in Higgs plus jet production at $\sqrt{s}=13$~TeV. We compare
    LO, NLO, NNLO and multi-jet merged predictions for a jet radius of 0.7.
    \label{fig:SM_Higgs_jet_R:HiggspTSpectrum} 
    The $K$-factors (NLO/LO, NNLO/LO and NNLO/NLO) determined
    from the \nnlojet predictions in this study, as a function of $p^\mathrm{Higgs}_\perp$ $(p^\mathrm{leadjet}_\perp)$
    for various jet sizes is shown on the middle (right) panel.
    \label{fig:SM_Higgs_jet_R:KFactorsFO}}
\end{figure}

Figure~\ref{fig:SM_Higgs_jet_R:HiggspTSpectrum} (left) shows the transverse momentum spectrum
of the Higgs boson 
as predicted by the fixed-order LO, NLO and NNLO calculation, as well as the
result from a multi-jet merged computation using the \Sherpa event generator and
the NLO-matched \Herwig result.  We observe that above a $p_T$ value of the
order of the Higgs mass, the distributions agree in their shape, while the
normalization differs slightly due to the higher-order effects included in the
NNLO calculation and the differences between fixed-order and the MEPS results in
their choice of renormalization and factorization scales.  The Higgs boson $p_T$
spectrum is the most inclusive observable in this study, and the prediction
above a certain $p_T$ may therefore serve as a useful normalization when
comparing shapes of other distributions between fixed order and MEPS results.

Figure~\ref{fig:SM_Higgs_jet_R:KFactorsFO} (middle) shows the $K$-factors as a function of the
Higgs boson $p_T$; as expected there is no jet size dependence for this
variable.  Figure~\ref{fig:SM_Higgs_jet_R:KFactorsFO} (right) shows the local $K$-factors
(NLO/LO, NNLO/LO, NNLO/NLO) for $H+\ge1$ jet production, as a function of the
leading jet transverse momentum, for various jet sizes, obtained from \nnlojet. The
$K$-factors are relatively flat as a function of jet $p_T$.  They grow with
increasing jet size, due to inclusion of additional real radiation. 
\begin{figure}[t]
  \centering
  \begin{minipage}{0.357778\textwidth}
    \includegraphics[width=\textwidth]{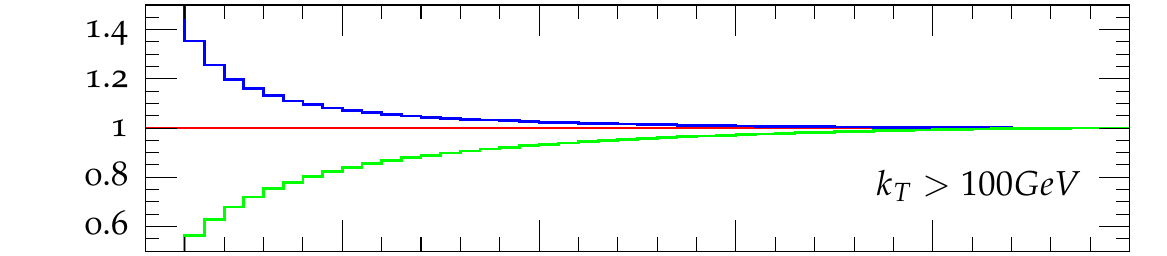}\\[-1mm]
    \includegraphics[width=\textwidth]{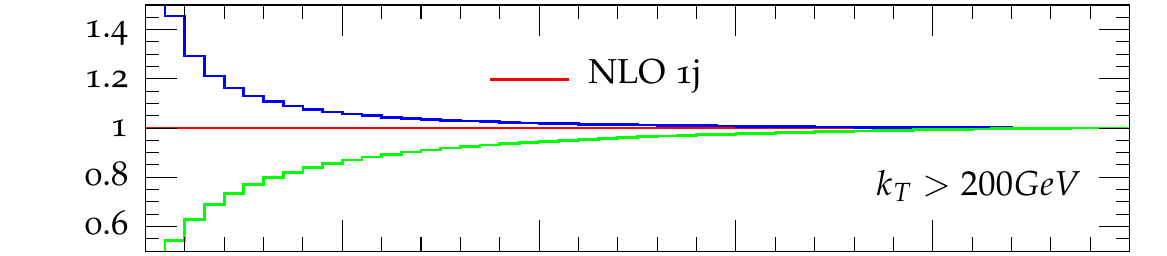}\\[-1mm]
    \includegraphics[width=\textwidth]{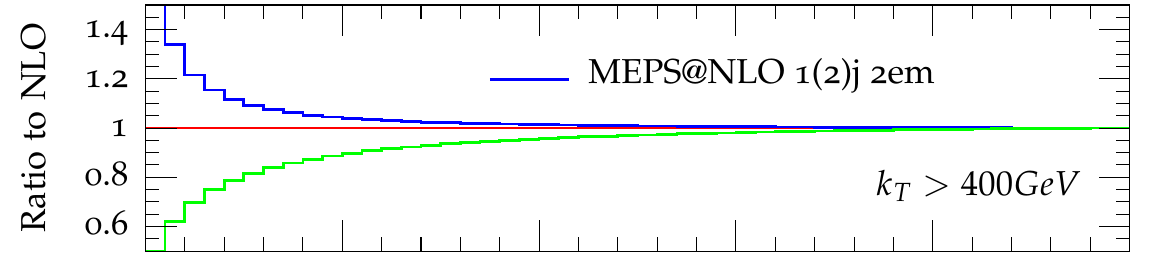}\\[-1mm]
    \includegraphics[width=\textwidth]{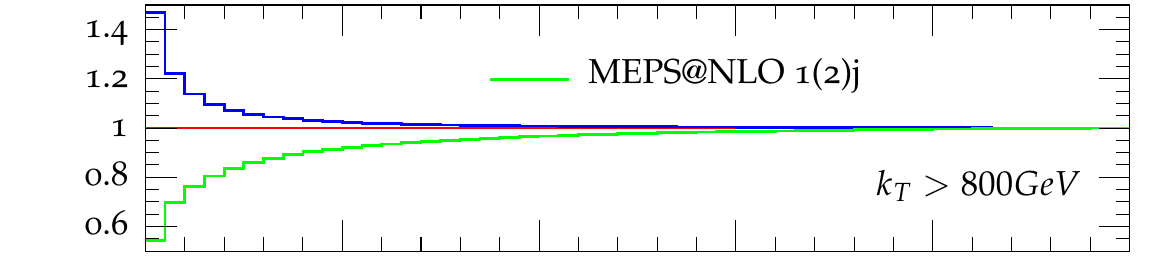}\\[-1mm]
    \includegraphics[width=\textwidth]{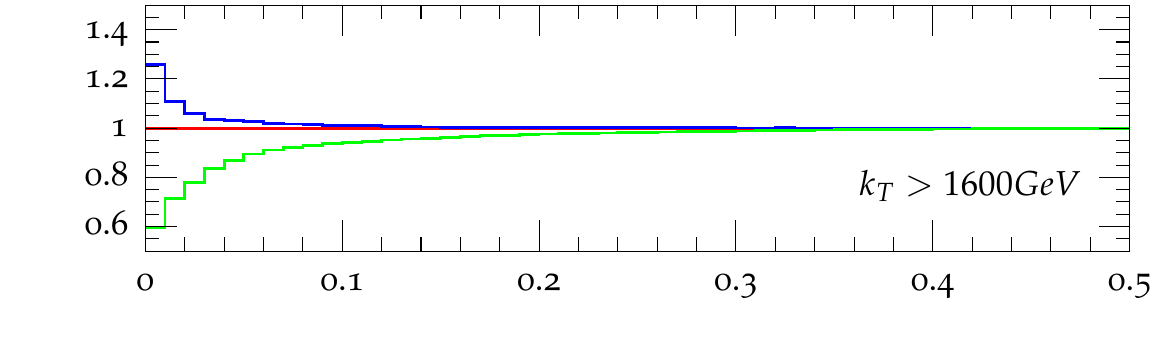}
  \end{minipage}
  \begin{minipage}{0.311111\textwidth}
    \includegraphics[width=\textwidth,clip,trim=1.4cm 0 0 0]{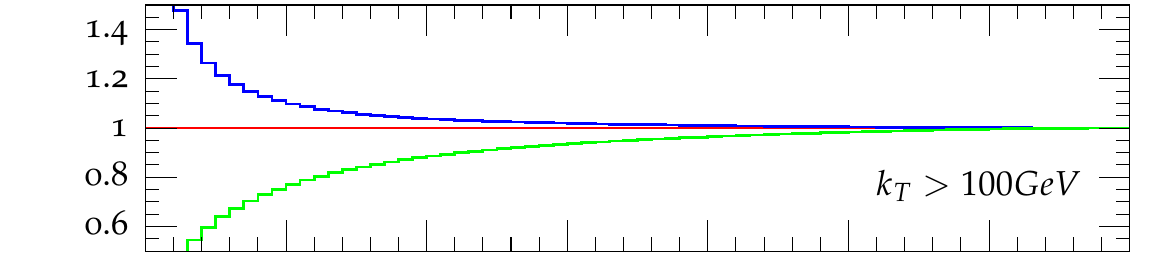}\\[-1mm]
    \includegraphics[width=\textwidth,clip,trim=1.4cm 0 0 0]{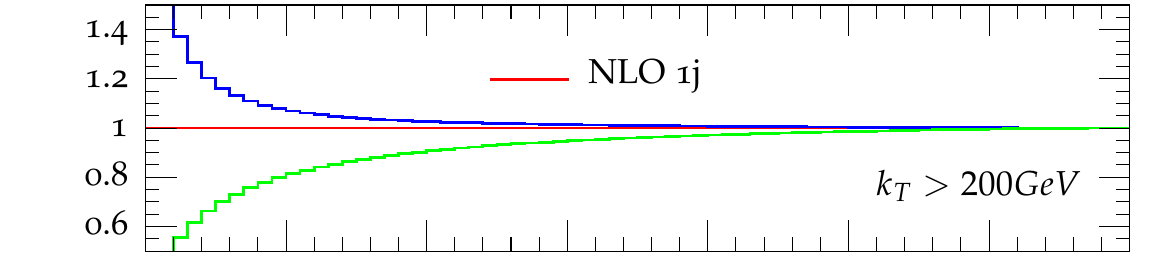}\\[-1mm]
    \includegraphics[width=\textwidth,clip,trim=1.4cm 0 0 0]{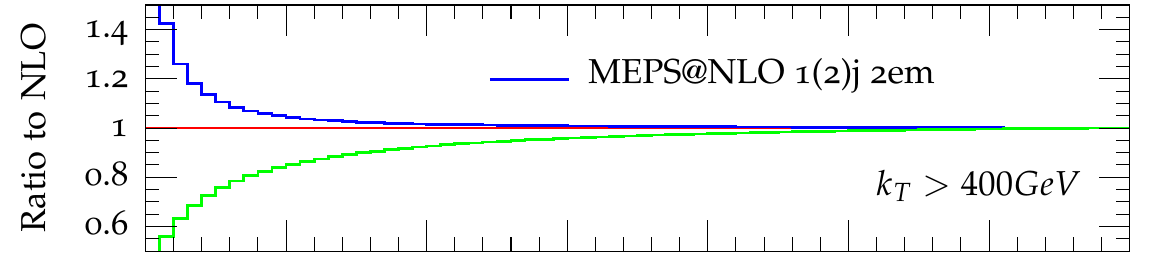}\\[-1mm]
    \includegraphics[width=\textwidth,clip,trim=1.4cm 0 0 0]{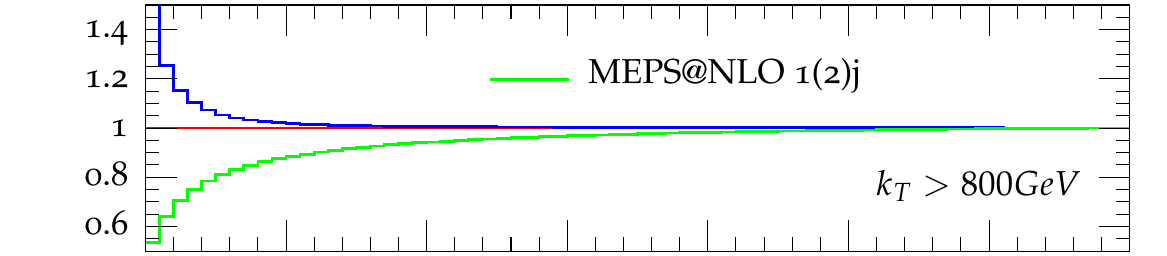}\\[-1mm]
    \includegraphics[width=\textwidth,clip,trim=1.4cm 0 0 0]{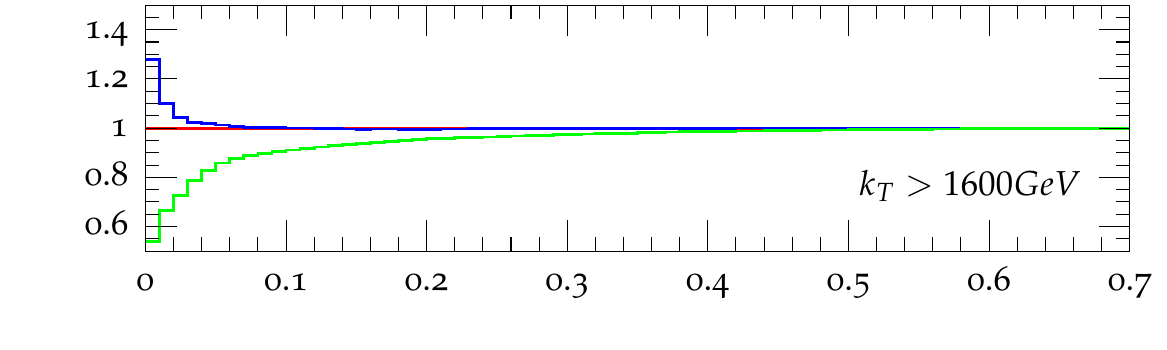}
  \end{minipage}
  \begin{minipage}{0.311111\textwidth}
    \includegraphics[width=\textwidth,clip,trim=1.3cm 0 0 0]{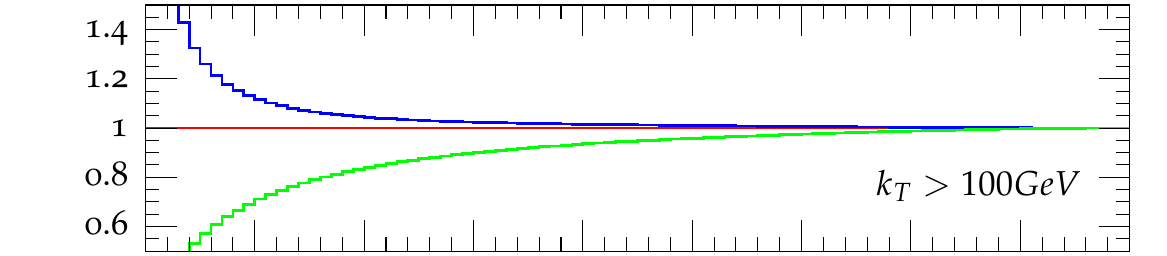}\\[-1mm]
    \includegraphics[width=\textwidth,clip,trim=1.3cm 0 0 0]{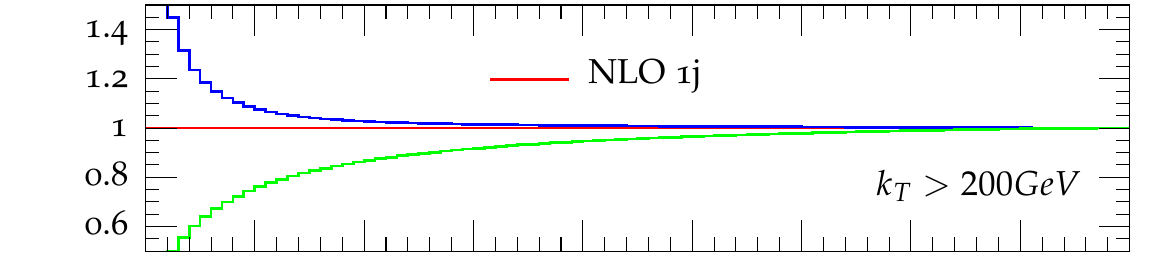}\\[-1mm]
    \includegraphics[width=\textwidth,clip,trim=1.3cm 0 0 0]{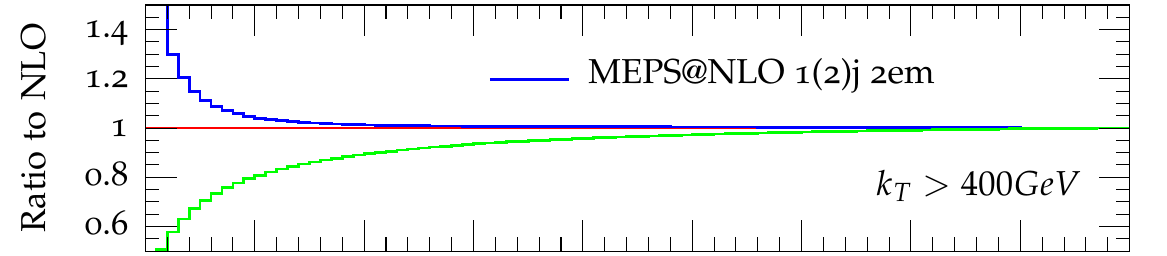}\\[-1mm]
    \includegraphics[width=\textwidth,clip,trim=1.3cm 0 0 0]{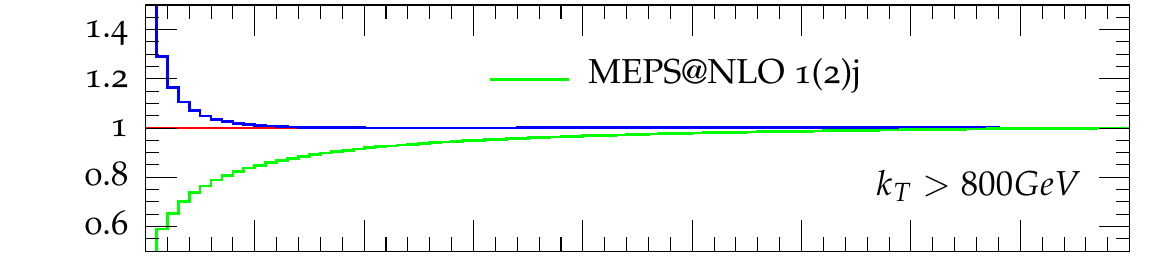}\\[-1mm]
    \includegraphics[width=\textwidth,clip,trim=1.3cm 0 0 0]{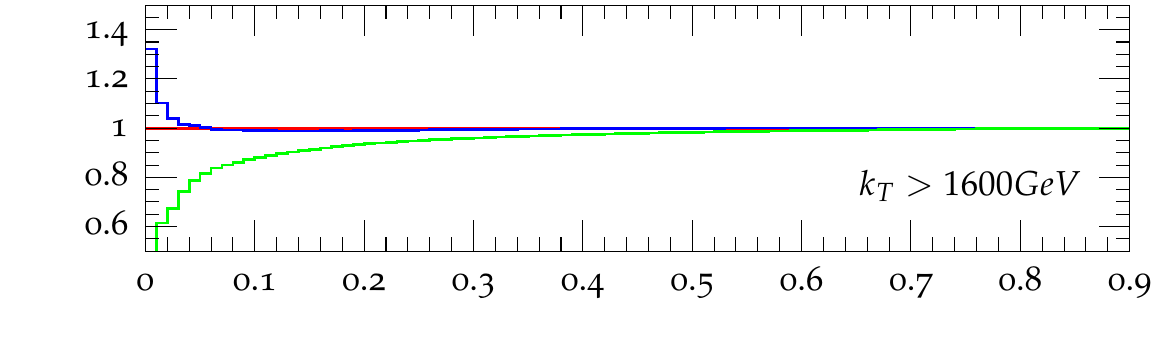}
  \end{minipage}
  \caption{Ratios of integrated jet shapes (see Eq.~\ref{eq:jetshape}) as function of $r$, for $R=0.5$, $R=0.7$ and $R=0.9$ (left to right).
    \label{fig:SM_Higgs_jet_R:shapes}}
\end{figure}

In Fig.~\ref{fig:SM_Higgs_jet_R:shapes} we investigate the difference between the fixed-order NLO
and multi-jet merged predictions for integrated jet shapes (Eq.~\ref{eq:jetshape}). 
The integrated jet shape is defined as
\begin{equation}
\Psi(r) = \frac{1}{N^\mathrm{jet}}\sum\limits_{\mathrm{jets}} \frac{p_\mathrm{T}(0,r)}{p_\mathrm{T}(0,R)}, 
\label{eq:jetshape}
\end{equation}
with $r$ being the radius of a cone which is concentric to the jet axis and $p_\mathrm{T}(r_1,r_2)$ being the magnitude of the vectorial momentum sum of all
particles in the annulus between radius $r_1$ and $r_2$. 
We also compare
to a truncated merged prediction, where the number of final-state partons generated
in the simulation is limited to at most two. This simulation presents the closest
possible approximation to the fixed-order result that we are able to generate using the
merging algorithms. It reflects the kinematical restrictions of the NLO calculation
(i.e.\ that only up to one additional final-state parton can be present), but it also
includes additional approximate higher-order virtual corrections by means of Sudakov
factors. Nevertheless we observe that the full NLO result and the truncated merged
result approach each other well within the jet cone, and the convergence is naturally
faster for larger jet transverse momenta. This strongly suggests that the differences
between the fixed-order predictions and merged results in
Fig.~\ref{fig:SM_Higgs_jet_R:ps_vs_fo_rpt_higgs} below are due to higher-multiplicity final
states. The discrepancies between fixed-order results and merged predictions for
small and large $R$ should therefore be reduced for higher-order perturbative
calculations. 

\begin{figure}[t]
  \centerline{\includegraphics[width=\textwidth]{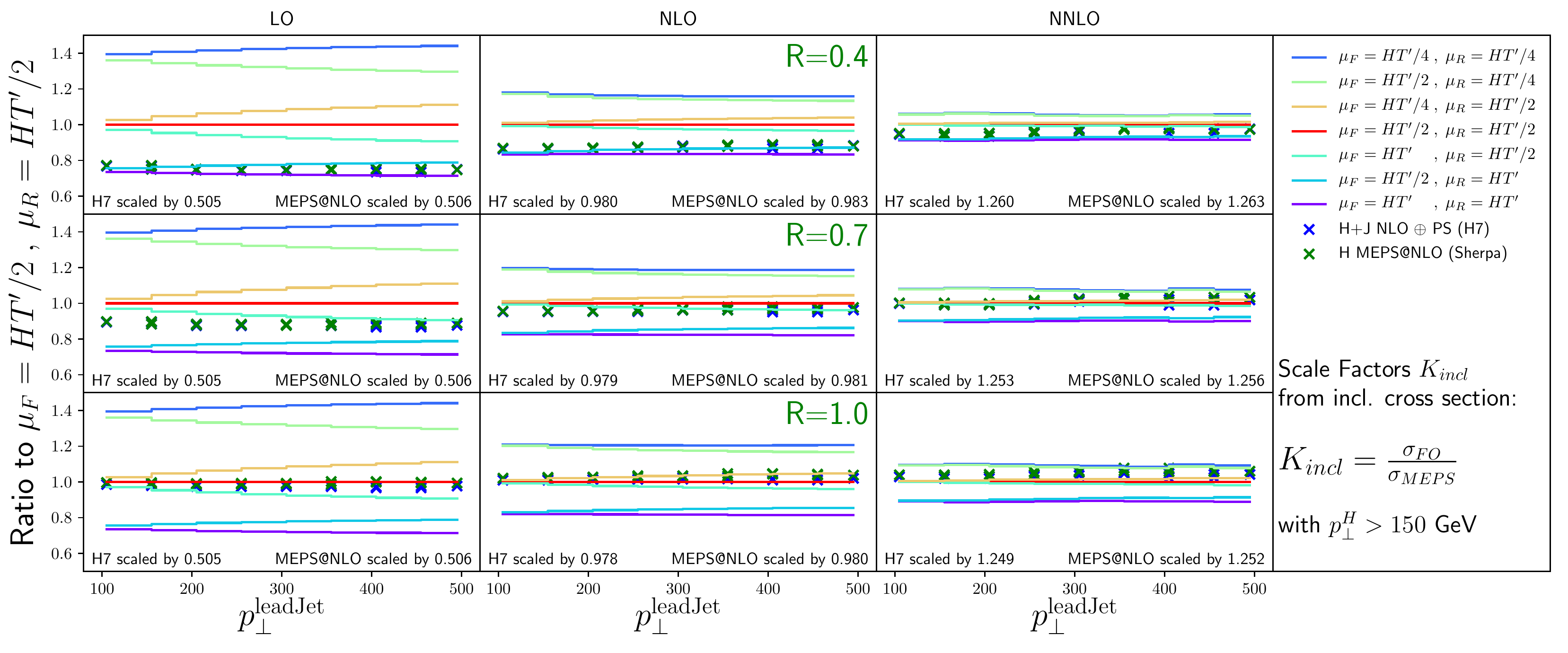}}
  \caption{The scale variations at LO, NLO, and NNLO from \nnlojet for 3 jets
size, as a function of leading jet transverse momentum, are shown.  For comparison,
the nominal NLO MEPS predictions are also shown. The generator predictions are
scaled with the inclusive $K_{incl}$ factor with Higgs $p^H_\perp > 150$~GeV,
see Fig.~\ref{fig:SM_Higgs_jet_R:KFactorsFO}.  \label{fig:SM_Higgs_jet_R:ScaleVariationsFO}}
\end{figure}

Figure~\ref{fig:SM_Higgs_jet_R:ScaleVariationsFO} shows the cross section scale variations at
LO, NLO and NNLO for $H+\ge1$ jet production, as a function of the leading jet
transverse momentum, for various jet sizes. The uncertainty band is given by the
highest and lowest cross section predictions at each order. As expected, the
uncertainty on the cross section decreases from LO to NLO to NNLO. It also
decreases as the jet size decreases, perhaps not unsurprising given that larger
jet radii lead to inclusion of more real radiative corrections.  Also shown for
comparison are the predictions from the two MEPS calculations (nominally of NLO
accuracy). We scale the MEPS predictions with the $K$-factors derived from the
Higgs pt distribution above 150~GeV, see discussion of Fig.~\ref{fig:SM_Higgs_jet_R:KFactorsFO}.

\begin{figure}[t]
  \centerline{\includegraphics[width=\textwidth]{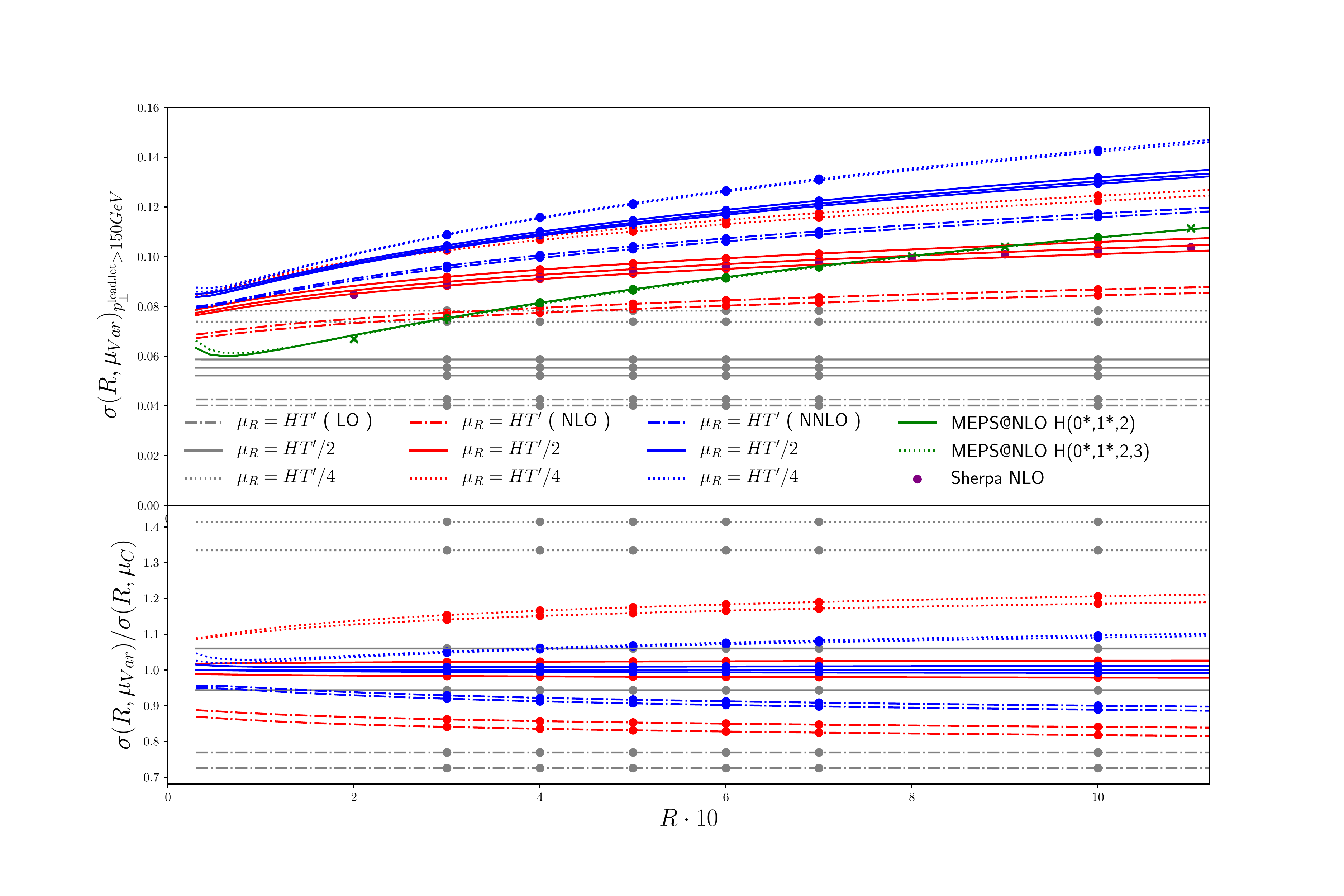}}
  \caption{The $R$-dependence of the cross sections at NLO, NNLO and MEPS are shown, for 
particular scale values, as a function of the jet radius, for leading jet transverse momenta above 150~GeV.   \label{fig:SM_Higgs_jet_R:AccidentalScaleComp}}
\end{figure}

Figure~\ref{fig:SM_Higgs_jet_R:AccidentalScaleComp} shows the leading jet $p_T$ cross sections for
the different scale choices, at NLO and NNLO, as a function of the jet size $R$.
In this case, a minimum transverse momentum requirement of 150~GeV has been
placed on the leading jet.  We assume this scale to be large enough such that $M_H$
is not the large scale in the process.  The dots for each scale choice have been
fit to a functional form motivated by the expected behavior for jet-vetoed cross
sections.
We assume the leading functional form:
\begin{equation}
f(R)=a+b\log(R)+c\log^2(R) \label{eq:SM_Higgs_jet_R:fit}
\end{equation}
as we expect a logarithmic behaviour for the scales induced by the effective
veto on the cross section by cutting with the jet cone $R$.  The lines in
Fig.~\ref{fig:SM_Higgs_jet_R:AccidentalScaleComp} are then interpolations with 
Eq.~\eqref{eq:SM_Higgs_jet_R:fit} and the fitted values. 
 
Again, the scale variation band is given by the upper and lower curves at each
order. It is notable that the scale uncertainty bands shrink as the jet size
decreases, as mentioned earlier. For very low values of $R$, this improvement in
the uncertainty can be regarded at least partially due to accidental
cancellations that stem from the restrictions in phase space.  Similar effects
were pointed out in the context of exclusive jet rate
measurements~\cite{Stewart:2011cf}. It can also be observed that for each
particular scale, the slope is greater at NNLO than at NLO. The MEPS predictions
are also plotted in the figure, and can be observed to have a greater slope than
even the NNLO predictions.  This can be seen as an effect of either including
(at large $R$) or not excluding (at small $R$) additional semi-hard real emissions,
which have a leading-order scale dependence and therefore induce a large change
in the cross section.

\begin{figure}[t]
  \centerline{\includegraphics[width=\textwidth]{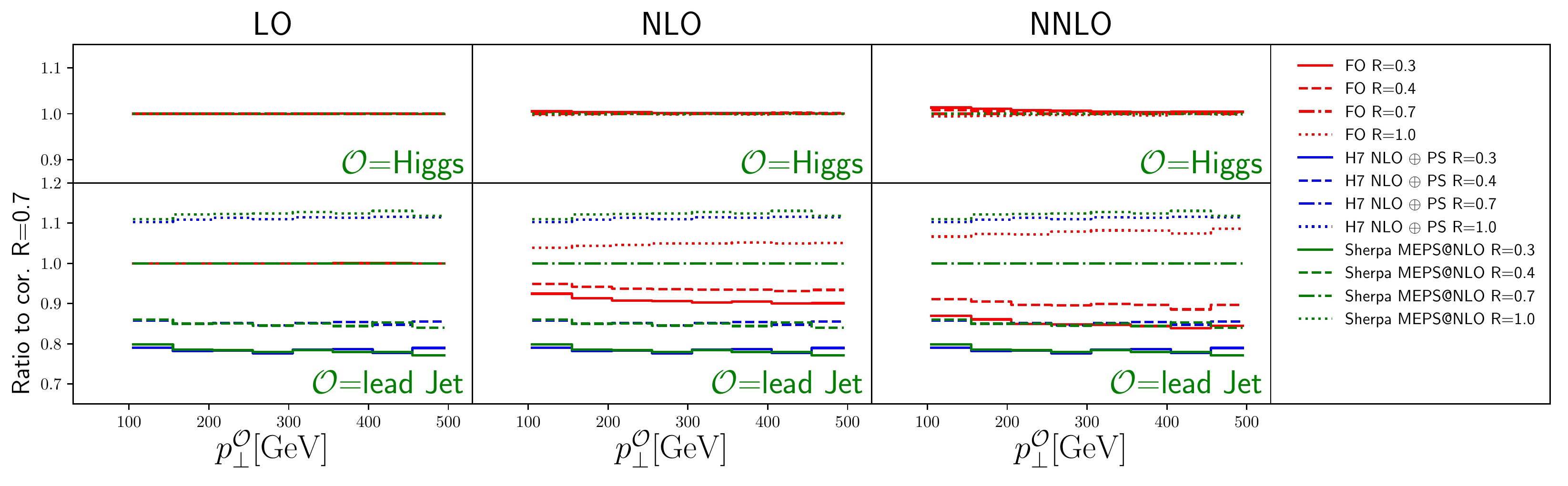}}
  \caption{The ratio of each cross section (either Higgs $p_T$ or lead
    jet $p_T$) for specific jet sizes, scaled to the cross section for
    each prediction for a jet size of
    $R=0.7$.  \label{fig:SM_Higgs_jet_R:multiratios_vs_NLO_07}}
\end{figure}

\begin{figure}[t]
  \centerline{\includegraphics[width=\textwidth]{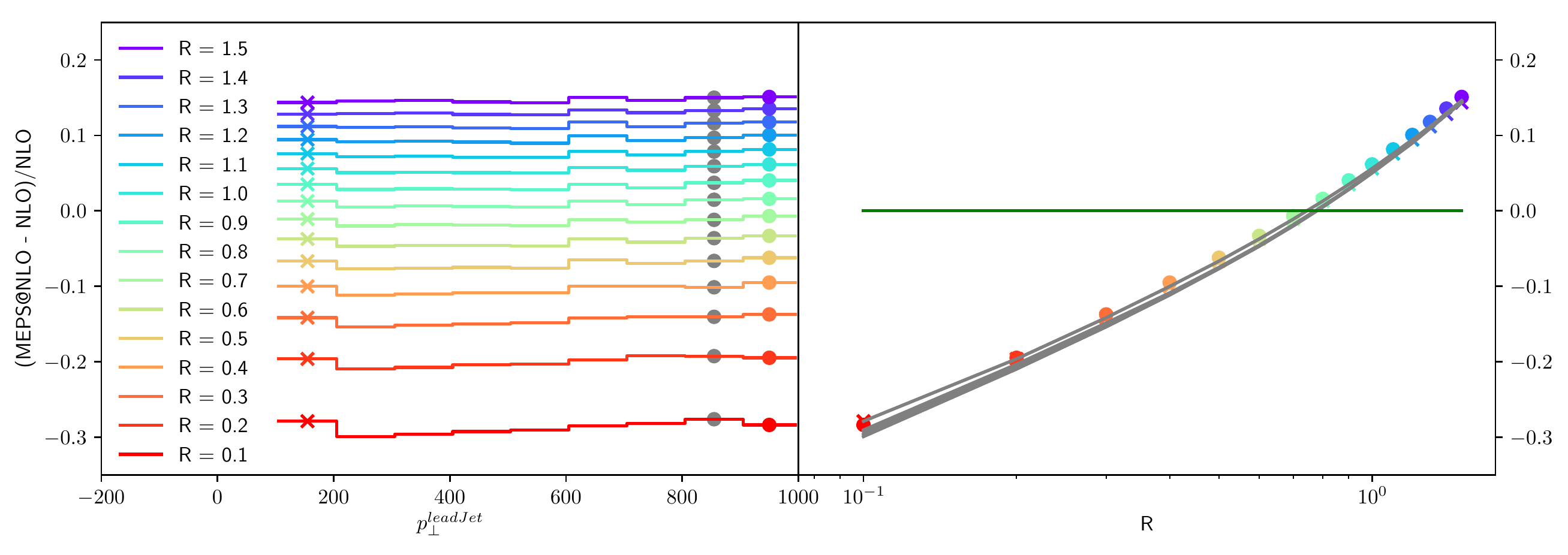}}
  \caption{Relative difference between the NLO-matched and multi-jet merged
    prediction to the fixed-order result as a function of the leading jet transverse
    momentum and the jet radius.\label{fig:SM_Higgs_jet_R:ps_vs_fo_rpt_higgs}}
\end{figure}

Figure~\ref{fig:SM_Higgs_jet_R:multiratios_vs_NLO_07} shows the cross sections for the Higgs
transverse momentum (top) and leading jet transverse momentum (bottom) for several
different jet sizes, at LO, NLO and NNLO (from \nnlojet) and from the two MEPS
predictions. All cross sections have been scaled to their respective value for the
reference jet size of $R=0.7$.  At this value we observe the best agreement
between fixed-order and multi-jet merged results, save for an overall
normalization which can be extracted from the Higgs transverse momentum
spectrum, cf.\ Fig.~\ref{fig:SM_Higgs_jet_R:HiggspTSpectrum}. The absolute value of the
difference between the fixed-order and the multi-jet merged predictions away
from $R=0.7$ increases roughly proportional to $\log (R/0.7)$
(cf.\ Fig.~\ref{fig:SM_Higgs_jet_R:ps_vs_fo_rpt_higgs}), which is expected due to the
higher-multiplicity real-emission corrections included in the multi-jet merged
calculation. Depending on kinematics they either enhance (at $R>0.7$) or reduce
(at $R<0.7$) the cross section.  The differences between the MEPS predictions
and those from \nnlojet decrease as the order is raised from NLO to NNLO.  The
difference is on the order of 5\% for $R=0.5$ at NLO and of the order of a few
percent at NNLO.

Given the better description of the jet shape provided by the MEPS predictions,
this is an indication of the theoretical uncertainty associated with the
truncation of the perturbative series. The uncertainty is reduced at NNLO as
expected.  It is noteworthy that the ratios in
Fig.~\ref{fig:SM_Higgs_jet_R:multiratios_vs_NLO_07} are relatively flat as a function of the
transverse momenta.%
\footnote{ The flatness may in fact be somewhat accidental as
we are using the EFT and needs to be confirmed upon including finite top mass
effects.}

Figure~\ref{fig:SM_Higgs_jet_R:ps_vs_fo_rpt_higgs} shows the dependence of the relative
difference between a NLO-matched, multi-jet merged prediction from Sherpa and
the NLO fixed-order result as a function of the leading jet transverse momentum
for varying jet radii.  The ratio is flat as a function of the leading jet
$p_T$.  In Fig.~\ref{fig:SM_Higgs_jet_R:AccidentalScaleComp} we compared integrated cross
sections, while here we observe interestingly a similar behaviour for the
differential cross sections.  In the right plot, the projection is with respect
to the radius,  and displays, in grey, the various transverse momentum intervals
and, in coloured, the lowest and highest energies.  Assuming the leading
behaviour is given by Eq.~\eqref{eq:SM_Higgs_jet_R:fit}, and with the flatness in the leading jet
transverse momentum, the linear, (but slightly quadratic) behaviour in the
logarithmic plot is expected.  We note the zero crossing of the curve on the
right-hand side, which corresponds to the best agreement between fixed-order and
matched/merged result, is located at $R \approx$0.7 (see the discussion of
Fig.~\ref{fig:SM_Higgs_jet_R:multiratios_vs_NLO_07}). In configurations where the jet rapidity
is zero, this corresponds to a roughly equal partitioning of the rapidity
phase-space into collinear sectors for color dipoles spanned between the
initial-state partons and the final-state jet. 

\begin{figure}[t]
  \centerline{\includegraphics[width=0.48\textwidth]{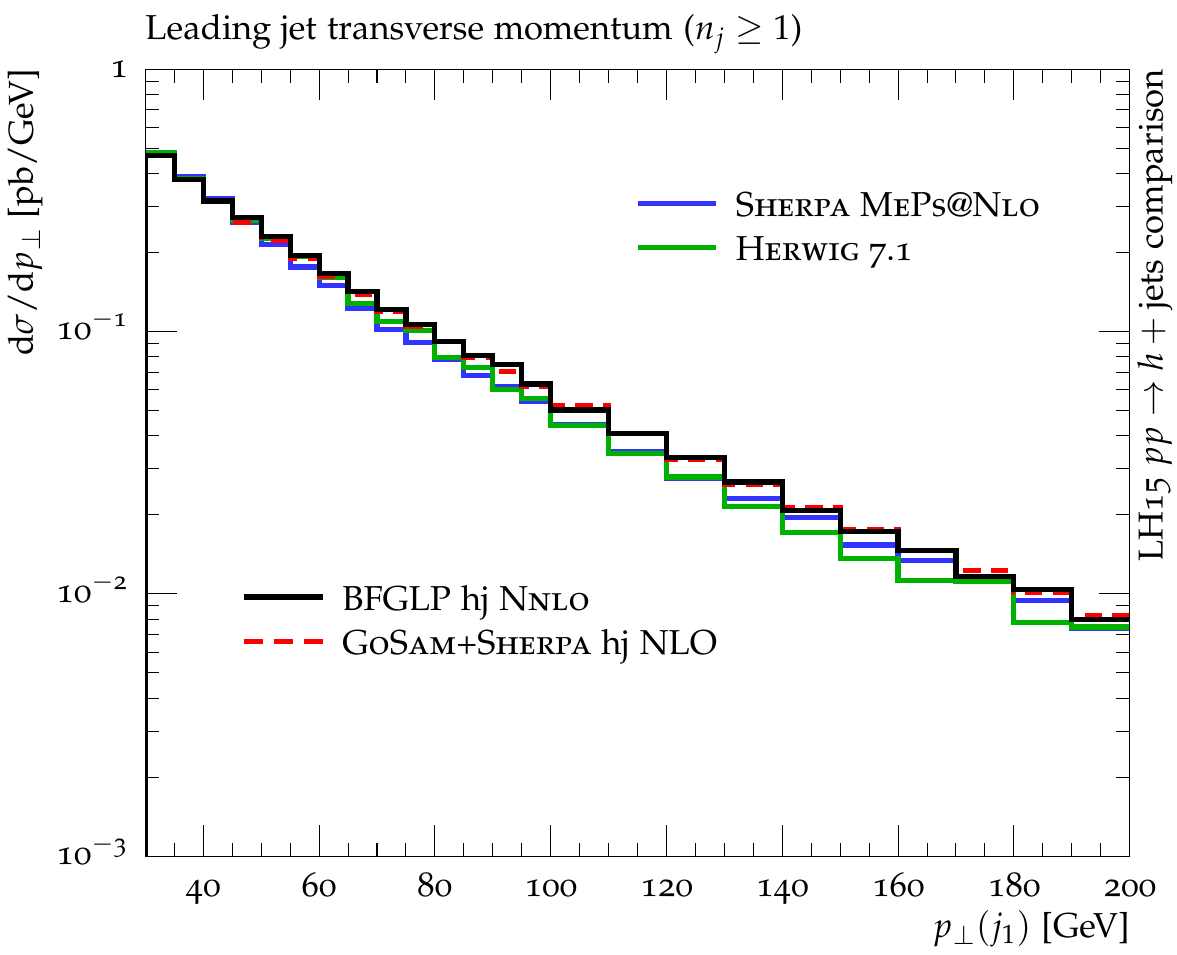}\hfill
    \includegraphics[width=0.48\textwidth]{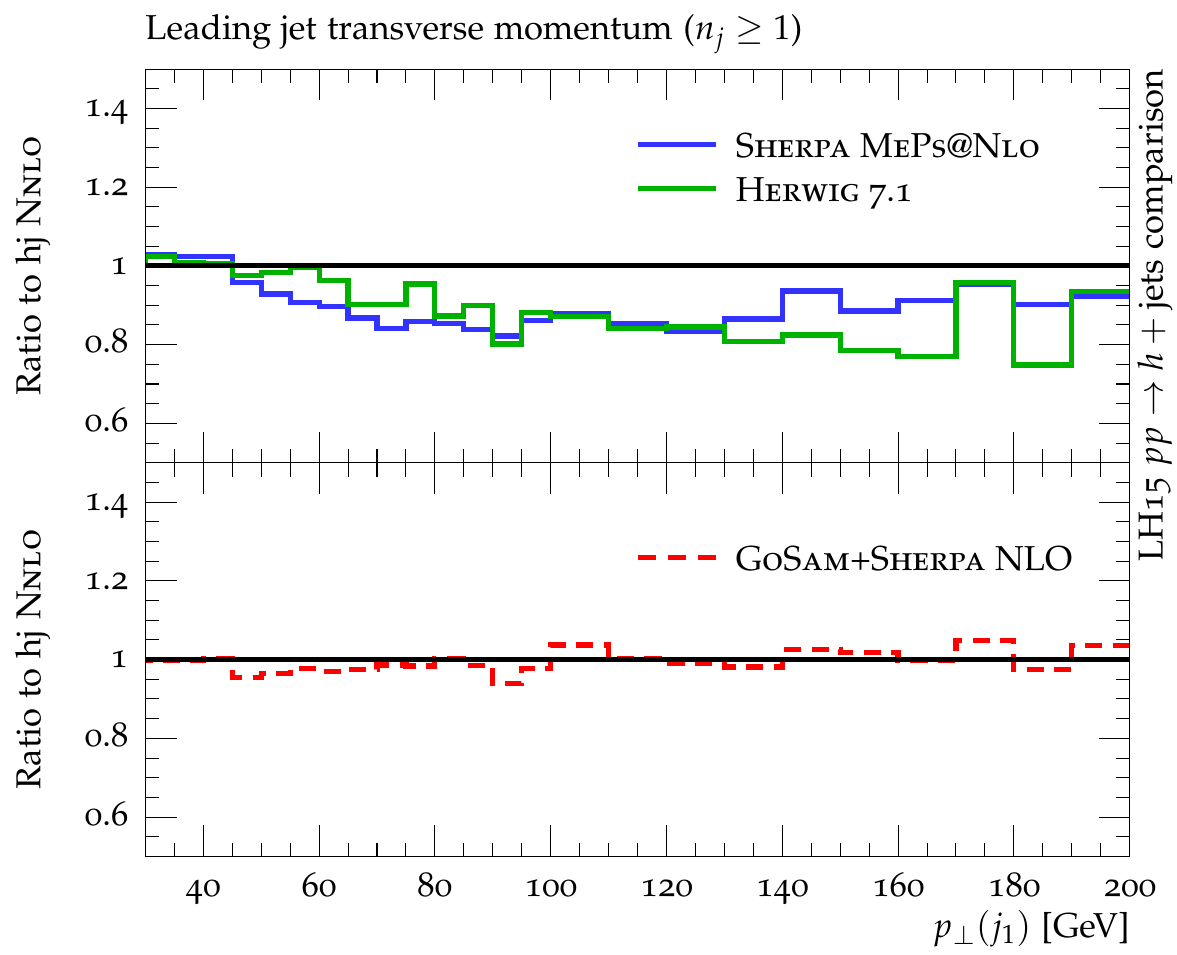}}
  \caption{Predictions for the leading jet transverse momentum cross sections, for NLO, 
    NNLO and MEPS calculations, from the 2015 Les Houches study~\cite{Badger:2016bpw}.\label{fig:SM_Higgs_jet_R:ratio}}
\end{figure}

Finally, for comparison, in Fig.~\ref{fig:SM_Higgs_jet_R:ratio}, we show a (simplified) plot
of the comparison of predictions carried out in Les Houches
2015~\cite{Badger:2016bpw} for the leading jet transverse momentum.  For these
particular scale choices, the NNLO~\cite{Boughezal:2015aha} and
NLO~\cite{vanDeurzen:2013rv,Cullen:2013saa,Greiner:2015jha} give essentially the
same result. The predictions from Sherpa and Herwig agree with NNLO at low
$p_T$, but fall 10-15\% below at higher $p_T$.  This can be due partially to
differences in the scale choices between each prediction in the figure,
differences between the scale choices used in~\cite{Badger:2016bpw} compared to
ours, and to the jet shape differences explored in detail in this note.

\subsection{Outlook}
Searches for new physics, as well as a better understanding of standard model
physics, require an increasing level of precision, both for measurement and for
theory. For differential distributions for $H+\ge1$ jet, the highest level of
precision is obtained with NNLO predictions. Matched NLO plus parton shower
predictions (MEPS) provide a more complete event description, but at a level of
precision one order of $\alpha_s$ lower (although with a more complete
logarithmic treatment).  Most physics measurements at the LHC make use of
relatively small jet sizes (anti-$k_T$ with $R=0.4$), and $H+\ge1$ jet production is no
exception.  There can be differences between fixed order and MEPS predictions
for the same observable just due to the different estimates of the amount of jet
energy contained in a jet of radius $R$. These differences can be comparable to
the size of the scale uncertainty for the cross section at that order. 

In this contribution, we have reported on an investigation of the impact of
different jet sizes on Higgs boson plus jet physics at the LHC, paying close
attention to the impact of the jet size on $K$-factors, on scale uncertainties,
and on differences between fixed order and MEPS predictions. A more detailed
study, intended for a future publication, will proceed from this contribution,
expanded to also include inclusive jet and Z boson + jet final states. Better
understanding of the issues described in this contribution may allow an
improvement in the accuracy, and precision,  of such predictions at the LHC.

\subsection*{Acknowledgements}
We thank the organizers for an inspiring workshop.  This work was supported by
the US Department of Energy under the contract DE--AC02--76SF00515.
This project has received funding from the European Research Council (ERC)
under the European Union's Horizon 2020 research and innovation programme, grant
agreement No 668679. We thank the University of Zurich S3IT (http://www.s3it.uzh.ch) for providing support and computational resources. This research was supported in part by the UK Science and Technology Facilities Council, by the Swiss National Science Foundation (SNF) under contracts 200020-175595, 200021- 172478, and CRSII2-160814, by a grant from the Swiss National Supercomputing Centre (CSCS) under project ID UZH10, by the Research Executive Agency (REA) of the European Union under the Grant Agreement PITN-GA-2012-316704 (“HiggsTools”) and the ERC Advanced Grant MC@NNLO (340983), and by the Fundacao para a Ciencia e a Tecnologia (FCT-Portugal), project UID/FIS/00777/2013. AB acknowledges support from Royal Society grants UF110191 and UF160548.

\let\Herwig\undefined
\let\Pythia\undefined
\let\Sherpa\undefined
\let\pythia\undefined
\let\sherpa\undefined
\let\Vincia\undefined
\let\Dire\undefined
\let\vincia\undefined
\let\dire\undefined
\let\Rivet\undefined
\let\Professor\undefined
\let\eps\undefined
\let\mc\undefined
\let\mr\undefined
\let\mb\undefined
\let\tm\undefined



\chapter{Monte Carlo studies}
\label{cha:mc}







\newcommand{\Herwig}{H\protect\scalebox{0.8}{ERWIG}\xspace}
\newcommand{\Pythia}{P\protect\scalebox{0.8}{YTHIA}\xspace}
\newcommand{\Sherpa}{S\protect\scalebox{0.8}{HERPA}\xspace}
\newcommand{\pythia}{\Pythia}
\newcommand{\sherpa}{\Sherpa}

\newcommand{\Vincia}{V\protect\scalebox{0.8}{INCIA}\xspace}
\newcommand{\Dire}{D\protect\scalebox{0.8}{IRE}\xspace}
\newcommand{\vincia}{\Vincia}
\newcommand{\dire}{\Dire}

\newcommand{\Rivet}{R\protect\scalebox{0.8}{IVET}\xspace}
\newcommand{\Professor}{P\protect\scalebox{0.8}{ROFESSOR}\xspace}
\newcommand{\eps}{\varepsilon}
\newcommand{\mc}[1]{\mathcal{#1}}
\newcommand{\mr}[1]{\mathrm{#1}}
\newcommand{\mb}[1]{\mathbb{#1}}
\newcommand{\tm}[1]{\scalebox{0.95}{$#1$}}



\section{On renormalization scale variations in parton showers~\protect\footnote{
      S.~H\"{o}che,
      S.~Mrenna,
      S.~Prestel,
      M.~Sch\"{o}nherr,
      P.~Z.~Skands}{}}
\label{sec:MC_PS_scalevariation}

We perform a phenomenological study of different schemes for varying
the renormalization scale in parton shower calculations. To this end,
we compare the conventional CMW scheme against a method which
explicitly includes higher-order corrections proportional to the QCD
beta function and the two-loop cusp anomalous dimension. The
techniques are implemented in both \sherpa and \pythia, and compared
to the default methods employed by these generators. In contrast to
earlier studies of the same topic, we find that the uncertainty
estimates agree well between the different parton-shower algorithms.

\subsection{Introduction}

General-Purpose Event Generators are tools that combine perturbative
calculations and phenomenological models to predict the final states
of particle-particle collisions.   The perturbative calculations
include a fixed-order, matrix-element based component and an all-order
(resummed) parton shower component.    Precise, fixed-order
predictions have become readily available due to improvements in
calculation methods and numerical techniques.   The limiting factor 
in testing predictions on data, particularly for observables that are
not fully inclusive, are uncertainties in the parton shower evolution.
Thus, realistic prescriptions to assess parton shower uncertainties are needed.
They should ensure that only the effect of \emph{uncalculated or ambiguous}
higher-order terms are estimated, while preserving the leading higher-order
corrections that have been included in the simulation due to their importance
for practical applications. The latter include the use of transverse momenta
as a scale in the evaluation of the strong coupling~\cite{Amati:1980ch},
and the use of the Catani-Marchesini-Webber (CMW) scheme~\cite{Catani:1990rr}
to recover known $\mathcal{O}(\alpha_s^2)$ corrections to soft gluon emission.

Changes in the renormalization scale are diagnostic of parts of the
parton shower uncertainty. Weighting techniques~\cite{Hoeche:2009xc,Platzer:2011dq,Lonnblad:2012hz}
have made it very convenient to perform the related variations~\cite{Bellm:2016voq,
  Mrenna:2016sih,Bothmann:2016nao}. A previous analysis has thus focused on
the comparison of uncertainties estimated by the various public parton-shower
generators~\cite{Badger:2016bpw}, with partially surprising results. It is timely
to perform such a comparison again, and to question how parton shower variations
should be implemented such as to obtain realistic, but not overly aggressive or
overly conservative uncertainty estimates. The technique investigated here is
motivated by comparison to analytic resummation, see for example~\cite{Banfi:2004yd}.
We perform a dedicated comparison of different implementations of the method in 
the two event generators \pythia and \sherpa, using three different parton
shower algorithms.

This contribution is organized as follows. First, we will motivate and define
our renormalization scale variation scheme in Sec.~\ref{sec:MC_PS_scalevariation:motivation}. Next, we specify the implementation
details for \pythia and \sherpa in Sec.~\ref{sec:MC_PS_scalevariation:psunc__tools}. We present first results and compare the
predicted uncertainties in jet production at LEP Run I and gluon-induced
Higgs-boson production at LHC Run II in
Sec.~\ref{sec:MC_PS_scalevariation:psunc__results}. The study concludes with a short
summary in Sec.~\ref{sec:MC_PS_scalevariation:results}.
Note that our aim is to better understand the impact of higher-order corrections
related to soft-gluon emission in a phenomenological study. While the underlying
computational technique is based on soft-gluon resummation at NLL, no claim of
formal theoretical improvements of the parton-shower method is implied.

\subsection{Motivation}
\label{sec:MC_PS_scalevariation:motivation}

The leading-order evolution kernels for a parton shower with evolution variable $t$
and splitting variable $z$ can be written in the universal form~\cite{Buckley:2011ms}
\begin{eqnarray}\label{eq:MC_PS_scalevariation:psunc__ps_general_kernel}
\mathcal{K}_{\widetilde{\imath\jmath}\rightarrow i,j}(t,z)
 = \frac{1}{t} \frac{\alpha_s(\mu)}{2\pi}
   P_{\widetilde{\imath\jmath}\rightarrow i,j}(t,z)
 + \mathcal{O}(\alpha_s^2)
\end{eqnarray}
In this context, $P_{\widetilde{\imath\jmath}\rightarrow i,j}^{(i)}(t,z)$ are the model specific
splitting probabilities for parton $\widetilde{\imath\jmath}$ branching into partons
$i$ and $j$, in the presence of a parton (or set of partons) $k$ that accounts for momentum
conservation. This parton is conventionally called the recoil partner.
Only the first-order pieces are usually considered in Eq.~\eqref{eq:MC_PS_scalevariation:psunc__ps_general_kernel},
and a judicious scale setting is applied, based on the analysis in~\cite{Amati:1980ch}.
For transverse-momentum ordered showers, for which the evolution variable reduces to the relative
transverse momentum $p_\perp^2$ between the partons $i$ and $j$ in the soft and collinear limits,
this leads to
\begin{eqnarray*}
\mathcal{K}_{\widetilde{\imath\jmath}\rightarrow i,j}(t,z)
 = \frac{1}{t} \frac{\alpha_s(bt)}{2\pi}
   P_{\widetilde{\imath\jmath}\rightarrow i,j}^{(0)}(t,z)
\end{eqnarray*}
The prefactor $b$ can be chosen according to the so-called CMW scheme~\cite{Catani:1990rr}
in order to reproduce the higher-logarithmic corrections arising from the two-loop cusp
anomalous dimension~\cite{Kodaira:1981nh,Davies:1984hs,Davies:1984sp,Catani:1988vd}.
This implies $b(t)=\exp\{K(t)/\beta_0(t)\}$, where
\begin{eqnarray*}\label{eq:MC_PS_scalevariation:def_cmw_fac}
  K(t)=\left(\frac{67}{18}-\frac{\pi^2}{6}\right)C_A-\frac{10}{9}T_R\,n_f(t)\;,
  \qquad
  \beta_0(t)=\frac{11}{3}C_A-\frac{4}{3}T_R\,n_f(t)\;.
\end{eqnarray*}
Note that $n_f(t)$ is the number of active flavours at scale $t$.
For five active flavors, $b=0.45$, thus roughly halving the scale.
For evolution kernels with soft gluon enhancement, this scale choice leads to
sub-leading logarithmic corrections. At NLO, they take the expected form
\begin{eqnarray}\label{eq:MC_PS_scalevariation:psunc__oa2_from_cmw}
  \left.\mathcal{K}_{\widetilde{\imath\jmath}\rightarrow i,j}(t,z)\right|_{\mathcal{O}(\alpha_s^2)}
  = \frac{1}{t} \left(\frac{\alpha_s(t)}{2\pi}\right)^2\,K(t)
  \left\{\frac{2}{1-z}+\ldots\right\}\;,
\end{eqnarray}
where the dots stand for terms that are non-singular in the soft-gluon limit.
Alternatively, using $b(t)$ can be interpreted as using a larger value
of the strong coupling at the $Z$ pole, i.e.\ $\alpha_s(M_Z)\approx 0.130$
instead of the world average $\alpha_s(M_Z)\approx0.118$.
In the resummation of the transverse momentum of Drell-Yan lepton pairs
and similar observables, Eq.~\eqref{eq:MC_PS_scalevariation:psunc__oa2_from_cmw} would lead to the
familiar resummation coefficient $A_2$. In addition, explicit logarithms of
the renormalization scale and the transverse momentum in a dipole picture, $k_T^2$,
appear in these calculations (see for example \cite{Banfi:2004yd}), such that
the double-logarithmic piece of the $\mathcal{O}(\alpha_s^2)$ correction to the parton-shower splitting
kernels should read
\begin{eqnarray}\label{eq:MC_PS_scalevariation:plsunc__kernels_soft_nlo_contribs}
  \left.\mathcal{K}_{\widetilde{\imath\jmath}\rightarrow i,j}(t,z)\right|_{\mathcal{O}(\alpha_s^2)}
  = \frac{1}{t} \left(\frac{\alpha_s(t)}{2\pi}\right)^2\,\left[\beta_0(t)\log\frac{k_T^2}{t}+K(t)\right]
    \frac{2}{1-z}\;.
\end{eqnarray}
The existing techniques for renormalization scale variations in parton showers
implement this equation in different ways that are discussed in Sec.~\ref{sec:MC_PS_scalevariation:psunc__tools}.
Depending on the precise treatment of higher-logarithmic contributions, the related
uncertainty estimates can exhibit a sizable variation. The phenomenological consequences
of this effect will be discussed in Sec.~\ref{sec:MC_PS_scalevariation:psunc__results}.
Our main demand is that variations around a central scale should not
ruin the NLO accuracy in the soft limit.    It will be necessary to
introduce a compensation factor to undo part of the naive scale variation.

In the current contribution, we will maintain a close 
correspondence with analytical results, but without any claim about the parton
shower accuracy\footnote{A formal comparison of other uncertainties and of
  the relation between parton showers and analytic NLL resummation for simple
  observables has been presented in~\cite{Hoeche:2017jsi}.}.
Thus, when varying the renormalization scale of parton shower splittings,
we will implement variants of Eq.~\eqref{eq:MC_PS_scalevariation:plsunc__kernels_soft_nlo_contribs},
which differ only in how $K(t)$ is included in the calculation.
Note, however, that the definition of the soft enhanced part of the
leading-order splitting function $P^{(0)}(z,t)$ differs between
parton showers with different definitions of $t$ and $z$, and that the correspondingly different single-pole
terms can also lead to visible differences. Nevertheless, we employ this 
minimalistic approach to define the least common denominator for three
rather different parton showers, assuming our method will still provide
insight that can be helpful for future studies.

\subsection{Monte-Carlo Event Generators for this study}
\label{sec:MC_PS_scalevariation:psunc__tools}
We have implemented the scheme outlined at the end of 
Sec.~\ref{sec:MC_PS_scalevariation:motivation} in \pythia and \sherpa. Before presenting our 
results, we will describe the current default variation schemes and the
precise technical realization of our new technique in the various generators.

\subsubsection{\sc Pythia}
\label{sec:MC_PS_scalevariation:pythiavars}

\pythia~8~\cite{Sjostrand:2014zea} is the most recent General-Purpose Event Generator
in the \pythia family~\cite{Sjostrand:1982fn,Sjostrand:2006za,Sjostrand:2007gs}. It includes a native $p_\perp$-ordered parton 
shower supplemented with matrix-element corrections and multiple choices
of recoil scheme, and further supports the \vincia~\cite{Giele:2011cb,Fischer:2016vfv} and \dire~\cite{Hoche:2015sya} 
parton shower plugins. For this study, we will use the native shower in
\pythia version 8.230 and the proper settings to enable the CMW scheme to
set the argument of $\alpha_s$, by scaling $\Lambda_{\mathrm{QCD}}(n_f)$ appropriately.

The \pythia implementation of scale variations~\cite{Mrenna:2016sih}
reflects the ambiguity in how compensation terms should be
applied~\cite{Badger:2016bpw}.
Similar to what is now assumed, it was decided that
compensation terms should only apply to gluon emission, and not to 
$g\to q\bar{q}$ splittings. Beyond that, the \pythia implementation of
compensation terms was conservative, i.e.\ the effect of 
compensation was minimized when confronted with choices beyond $\mathcal{O}(\alpha_s^2)$.
This led to modifying the parton-shower branching probability to
\begin{equation}
\mathcal{K}_{\widetilde{\imath\jmath}\rightarrow i,j}(t,z)
~=~ \frac{\alpha_s(k b t)}{2\pi}\left(1 + (1-\zeta)\frac{\alpha_s(b\mu_\mr{max})}{2\pi}\beta_0 \ln k\right)\frac{P(z)}{t}~,
\end{equation}
for gluon emission, where $P(z)$ is the full DGLAP splitting kernel, possibly 
including matrix-element corrections, $\mu_\mr{max}=\max(m^2_\mr{dip},kt)$ and
\begin{equation}
\zeta = \left\{ 
\begin{array}{ccl}
1-z&&\mbox{for splittings with a $1/(1-z)$ singularity}\\
\min(z,1-z)&&\mbox{for splittings with a $1/(z(1-z))$ singularity}
\end{array}
\right.~.
\end{equation}
The inclusion of the factor $(1-\zeta)$ was motivated by the fact that the
compensation is only theoretically motivated in the soft limit (i.e.\ for 
$\zeta\rightarrow 0$). Thus, to remain conservative, the compensation was 
explicitly linearly damped outside of the soft limit.

To implement the new scheme proposed in Sec.~\ref{sec:MC_PS_scalevariation:motivation}, we
change the branching probabilities to
\begin{equation}\label{eq:MC_PS_scalevariation:psunc__pythia_implementation}
\mathcal{K}_{\widetilde{\imath\jmath}\rightarrow i,j}(t,z)
~=~ \frac{\alpha_s(k b t)}{2\pi}\left(1 + \frac{\alpha_s( b t)}{2\pi}\beta_0 \ln k\right)\frac{1}{t}\frac{2C_F}{1-z}
~~+~~ \frac{\alpha_s(k b t)}{2\pi}\frac{1}{t} \left[ P_{qq}(z) - \frac{2C_F}{1-z}\right]
\end{equation}
for gluon emission off (anti)quarks, and to 
\begin{eqnarray}\label{eq:MC_PS_scalevariation:psunc__pythia_implementation2}
\mathcal{K}_{\widetilde{\imath\jmath}\rightarrow i,j}(t,z)
&=& \frac{\alpha_s(k b t)}{2\pi}\left(1 + \frac{\alpha_s(b t)}{2\pi}\beta_0 \ln k\right)\frac{1}{t}\left[\frac{2C_A}{1-z} + \frac{2C_A}{z}\right]\\
&+& \frac{\alpha_s(k b t)}{2\pi}\frac{1}{t} \left[ P_{gg}(z) - \frac{2C_A}{1-z} - \frac{2C_A}{z}\right]\nonumber
\end{eqnarray}
for gluon emission off gluons. For the first (and partially also subsequent)
parton-shower emissions, Eqs.~\eqref{eq:MC_PS_scalevariation:psunc__pythia_implementation} and \eqref{eq:MC_PS_scalevariation:psunc__pythia_implementation2}
are potentially rescaled with finite matrix-element correction factors.
 Note that the implementation does enforce that
the compensation terms are only applied to pieces yielding double-logarithmic
contributions. However, the two-loop cusp term is not only added to the
soft contribution, but also enters in the hard- and collinear phase sectors, as
is common for implementations of the CMW scheme. The difference w.r.t.\ 
Eq.~\eqref{eq:MC_PS_scalevariation:plsunc__kernels_soft_nlo_contribs} is of $\mathcal{O}(\alpha_s^3)$ in the soft region, and
of $\mathcal{O}(\alpha_s^2)$ in the hard and collinear regions.

\subsubsection{Sherpa}
\label{sec:MC_PS_scalevariation:sherpavars}

The \Sherpa Monte Carlo event generator \cite{Gleisberg:2008ta} in its latest
release, Sherpa-2.2.4, comprises two parton shower algorithms: CSS
\cite{Schumann:2007mg} and \Dire \cite{Hoche:2015sya}. Both are based on 
Catani-Seymour \cite{Catani:1996vz,Catani:2002hc} dipole splitting functions.
While CSS is constructed along the lines of a standard, transverse momentum ordered
parton shower, \Dire combines the standard treatment of collinear configurations
in parton showers with the resummation of soft logarithms in color dipole cascades.
The evolution kernels of CSS can be written in the form of Eq.~\eqref{eq:MC_PS_scalevariation:def_cmw_fac},
while in \Dire $b(t)\to 1$ and the two-loop cusp anomalous dimension multiplies
the soft enhanced terms of the splitting functions with $1+\alpha_s/(2\pi)K(t)$.

When varying the argument of the strong coupling constant, i.e.\ 
replacing $t\to kt$ with $k$ a constant, the higher-logarithmic structure
induced by the running of the coupling constant in the presence of the CMW
scale factor needs to be preserved in order not to change the resummation
implemented by the parton shower. Thus, in~\cite{Badger:2016bpw} the following
naive replacement was used
\begin{equation}
  \begin{split}\label{eq:MC_PS_scalevariation:psunc__tools_sherpa_asct}
    \alpha_s(b(t)\,t)
    \,\to\;& \alpha_s(b(t)\, k\,t)\,f(k,t)\;,
    \quad
    f(k,t)=1+\sum_{i=0}^{n_\text{th}+1}\frac{\alpha_s(b(t)\,t)}{2\pi}\,\beta_0(n_f(t))\,\log\frac{t_i}{t_{i-1}}
  \end{split}
\end{equation}
The sum runs over the number $n_\text{th}$ of parton mass thresholds in the
interval $[t,k\,t]$ with $t_0=t$, $t_{n_\text{th}+1}=k\,t$ and $t_i$ are the
encompassed parton mass thresholds. If $k<1$, the ordering is reversed,
recovering the correct sign. It is clear that this method will largely eliminate
the dependence of the overall prediction on $k$. In order to obtain a realistic
uncertainty estimate, we use Eq.~\eqref{eq:MC_PS_scalevariation:plsunc__kernels_soft_nlo_contribs}.

In the case of the CSS shower, we remove the explicit dependence on $K(t)$
and reweight instead with a factor $\alpha_s(b(t)\,t)/\alpha_s(t)$, which leads
to the definition of the final-state emitter, final-state spectator
branching probabilities
\begin{equation}
  \begin{split}
    \mathcal{K}_{\widetilde{\imath\jmath}\rightarrow i,j}(t,z)
    = \frac{\alpha_s(k b t)}{2\pi}\frac{1}{t}\frac{2C_F}{1-z(1-y)}\,f(k,t)
    + \frac{\alpha_s(k b t)}{2\pi}\frac{1}{t} \left[ V_{ij,k}(z,t) - \frac{2C_F}{1-z(1-y)}\right]
  \end{split}
\end{equation}
for gluon emission off (anti)quarks, and
\begin{eqnarray}
\mathcal{K}_{\widetilde{\imath\jmath}\rightarrow i,j}(t,z)
&=& \frac{\alpha_s(k b t)}{2\pi}\frac{1}{t}\left[\frac{2C_A}{1-z(1-y)} + \frac{2C_A}{y+z(1-y)}\right]\,f(k,t)\\
&+& \frac{\alpha_s(k b t)}{2\pi}\frac{1}{t} \left[ V_{ij,k}(z,t) - \frac{2C_A}{1-z(1-z)} - \frac{2C_A}{y+z(1-y)}\right]\nonumber
\end{eqnarray}
for gluon emission off gluons. The $V_{ij,k}$ are the splitting kernels
in~\cite{Catani:1996vz}. This will generate higher-logarithmic
corrections compared to Eq.~\eqref{eq:MC_PS_scalevariation:plsunc__kernels_soft_nlo_contribs},
which are, however, much smaller than the ones produced by
Eq.~\eqref{eq:MC_PS_scalevariation:psunc__tools_sherpa_asct}.

In the case of the \dire shower, we use the branching probabilities
\begin{equation}
  \begin{split}
    \mathcal{K}_{\widetilde{\imath\jmath}\rightarrow i,j}(t,z)
    = \frac{\alpha_s(k t)}{2\pi}\frac{C_F}{t}\left[\frac{2(1-z)}{(1-z)^2+\kappa^2}\,\left(f(k,t)+\frac{\alpha_s(t)}{2\pi}\,K(t)\right)-(1+z)\right]
  \end{split}
\end{equation}
for gluon emission off (anti)quarks, and
$\mathcal{K}_{\widetilde{\imath\jmath}\rightarrow i,j}(t,z)=
\mathcal{K}_{gg}(t,z)+(i\leftrightarrow j)$, for gluon emission off gluons
in the final state. The unsymmetrized gluon-to-gluon kernel is given by
\begin{eqnarray}
  \mathcal{K}_{gg}(t,z)
= \frac{\alpha_s(k b t)}{2\pi}\frac{2C_A}{t}\left[\frac{(1-z)}{(1-z)^2+\kappa^2}\,\left(f(k,t)+\frac{\alpha_s(t)}{2\pi}\,K(t)\right)-1+\frac{z(1-z)}{2}\right]\,.
\end{eqnarray}
Note that the definition of $\kappa^2$ depends on the
type of dipole and is given in~\cite{Hoche:2015sya}.
For gluon emission off gluons in the initial state we use
\begin{eqnarray}
  \mathcal{K}_{\widetilde{\imath\jmath}\rightarrow i,j}(t,z)
  = \frac{\alpha_s(k b t)}{2\pi}\frac{2C_A}{t}\left[\frac{(1-z)}{(1-z)^2+\kappa^2}\,\left(f(k,t)+\frac{\alpha_s(t)}{2\pi}\,K(t)\right)-2+\frac{1}{z}+z(1-z)\right]\;.
\end{eqnarray}

\subsection{Results}
\label{sec:MC_PS_scalevariation:psunc__results}

In the following, we will use the baseline parton-shower settings
$\alpha_{s}(M_Z)=0.118$, use two-loop running of $\alpha_s$, and employ the
CMW prescription for all parton showers except \dire. It is important to emphasize that
these settings are only chosen to aid the comparison of parton shower 
variations, and are not based on the resulting quality of data description.
With this in mind, we will also refrain from comparing our results to data, 
although we will employ the analyses~\cite{Heister:2003aj} and~\cite{Pfeifenschneider:1999rz}, as
implemented in the \Rivet framework~\cite{Buckley:2010ar}. We then
perform variations of $\alpha_{s}(k t)$ with $k=\{1/4,1,4\}$ (and $t$ of
dimension GeV$^2$) for both $e^+e^-$ collisions at LEP I and gluon-induced
Higgs-production at a 13 TeV LHC. For the latter, we do not include
multiple parton interactions, primordial $k_T$ modelling, QED effects or
hadronization in the predictions. We use the parton distribution functions given
in {\tt NNPDF30\_nlo\_as\_0118}~\cite{Ball:2014uwa} and implemented in {\tt LHAPDF}~\cite{Buckley:2014ana}.
Finally, for the LHC setup, we use $\mu_Q=125$ GeV as starting scale of the
parton shower. Please note again that these settings define a theory study,
and should \emph{not} be considered a recommendation when performing 
a detailed experimental analysis.

The results will be presented as a main plot and three adjoint ratios. The 
latter show the predictions of one shower, compared to the baseline prediction 
of this same shower. The gray band represents the current default variation
band of the tool, while the colored bands give the results of the schemes
proposed in Sec.~\ref{sec:MC_PS_scalevariation:motivation}. The hatched gray bands show naive 
variations without any compensation terms, as a baseline. From these, it is immediately obvious
that the inclusion of compensating terms has significant effects.

\subsubsection{LEP results}

In this section, we compare the MCEGs to each other, using
the analyses of~\cite{Pfeifenschneider:1999rz} and~\cite{Heister:2003aj}. We
include both parton showering and hadronization into the simulations.

\begin{figure}[t!]
  \centering
  \begin{minipage}{0.48\textwidth}
    \includegraphics[width=\textwidth]{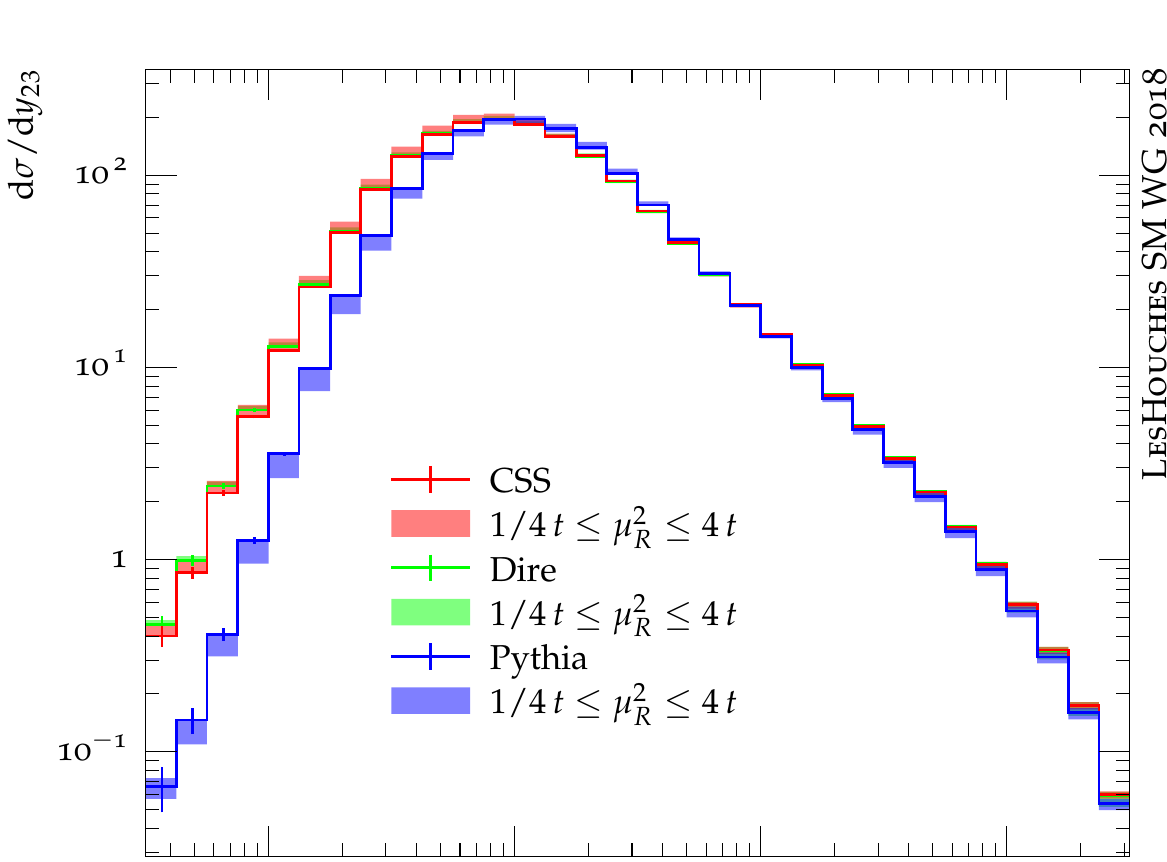}\\[-1mm]
    \includegraphics[width=\textwidth]{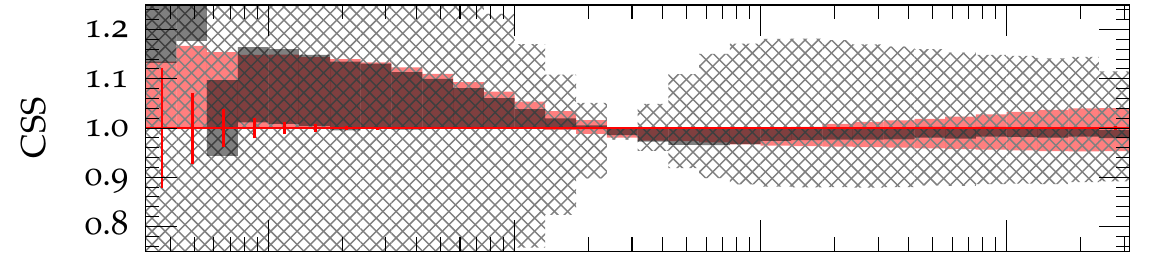}\\[-1mm]
    \includegraphics[width=\textwidth]{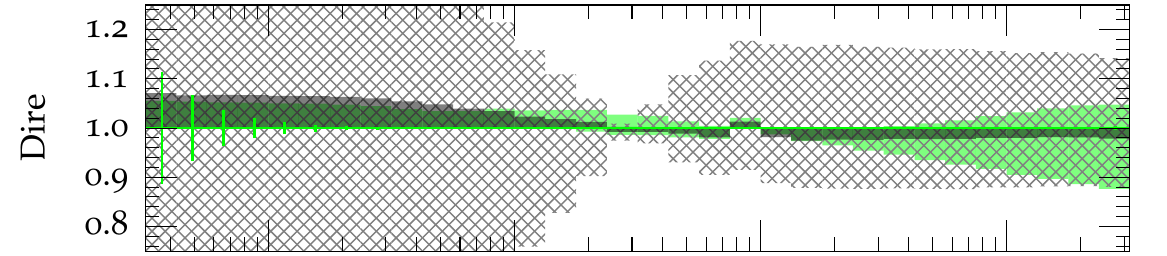}\\[-1mm]
    \includegraphics[width=\textwidth]{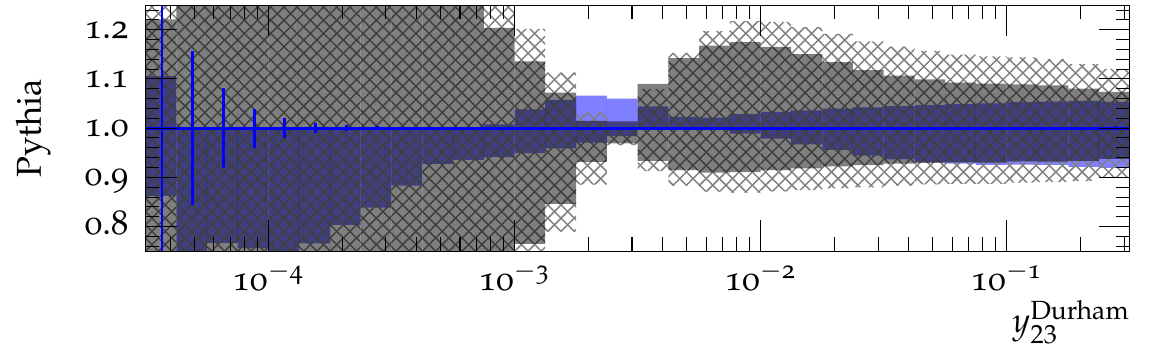}
  \end{minipage}\hskip 5mm
  \begin{minipage}{0.48\textwidth}
    \includegraphics[width=\textwidth]{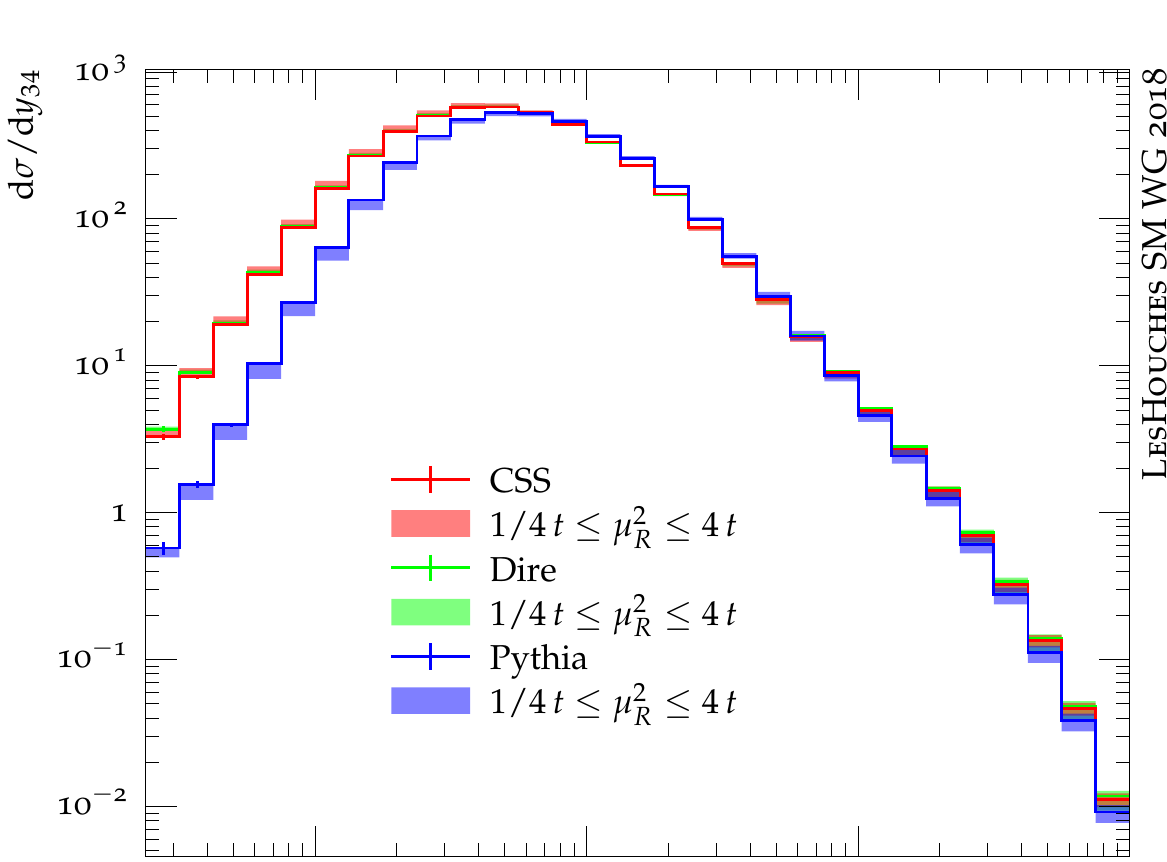}\\[-1mm]
    \includegraphics[width=\textwidth]{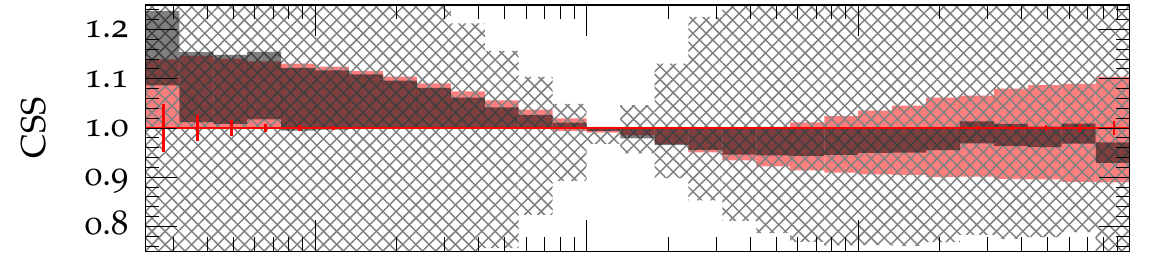}\\[-1mm]
    \includegraphics[width=\textwidth]{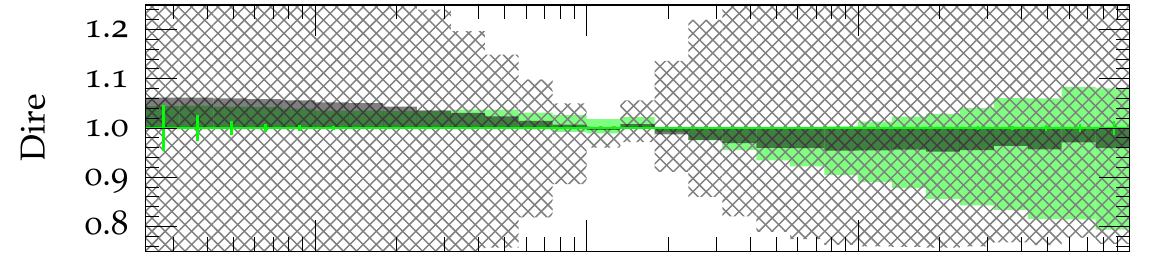}\\[-1mm]
    \includegraphics[width=\textwidth]{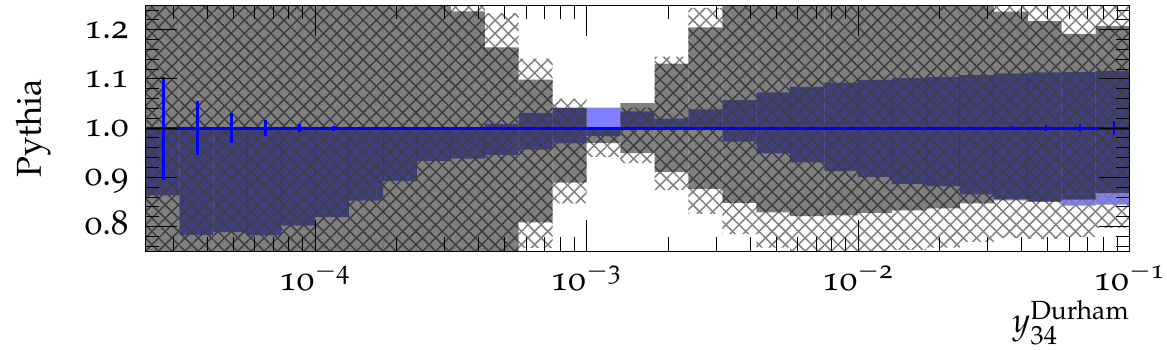}
  \end{minipage}
  \caption{Uncertainty estimates for the Durham $2\to3$ jet rate (left) and
    $3\to4$ jet rate (right) in the analysis setup of~\cite{Pfeifenschneider:1999rz}.
    Gray bands show the estimates using the method of \cite{Badger:2016bpw}
    for \sherpa and \dire, and of \cite{Mrenna:2016sih} for \pythia.
    Blue 
    bands give \pythia variations according to Sec.~\ref{sec:MC_PS_scalevariation:pythiavars}.
    Red and green bands give, respectively, \sherpa and \dire variations
    according to Sec.~\ref{sec:MC_PS_scalevariation:sherpavars}.
    Hatched gray bands show naive variations without any compensation terms.
    \label{fig:MC_PS_scalevariation:lep_jetrates}}
\end{figure}
\begin{figure}[t!]
  \centering
  \begin{minipage}{0.48\textwidth}
    \includegraphics[width=\textwidth]{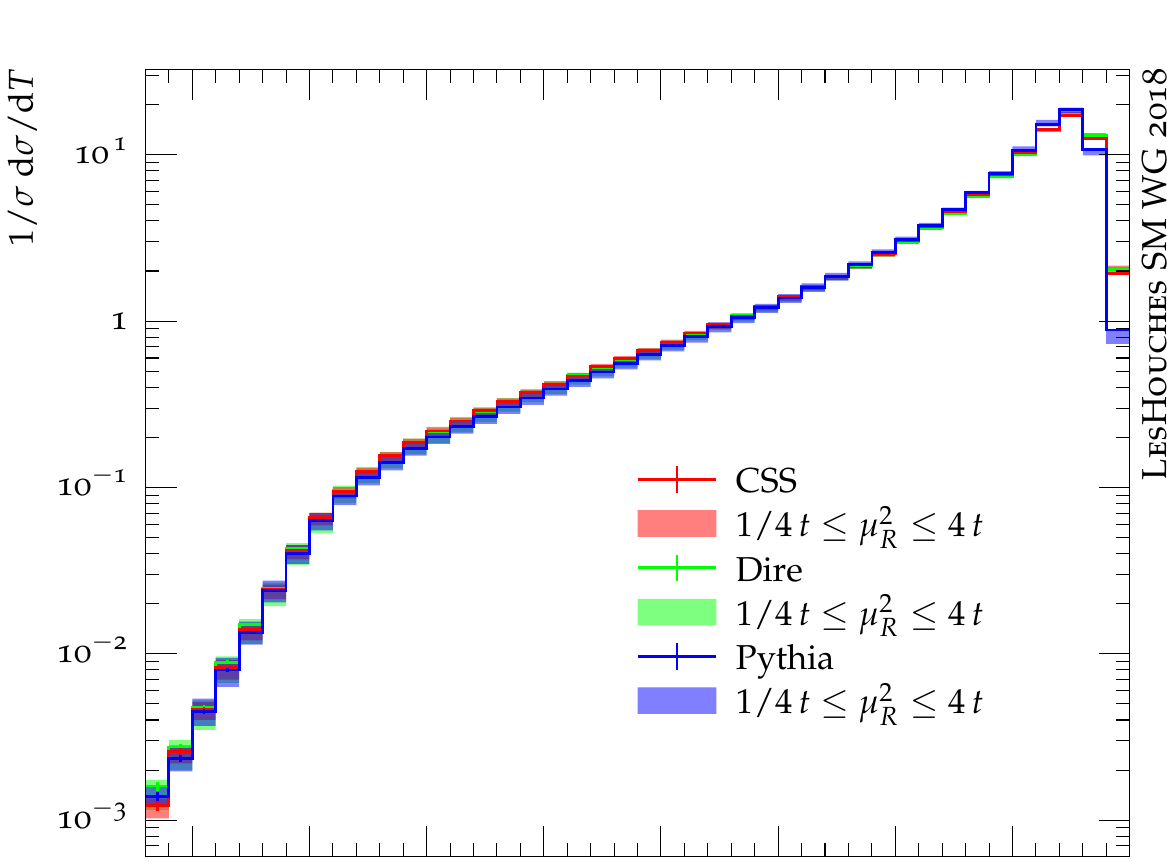}\\[-1mm]
    \includegraphics[width=\textwidth]{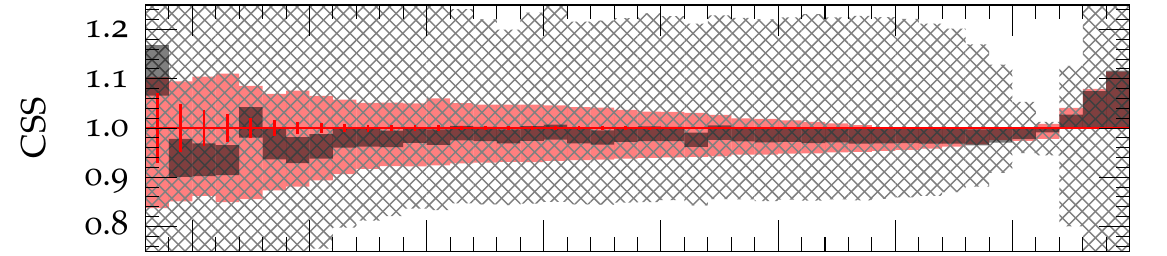}\\[-1mm]
    \includegraphics[width=\textwidth]{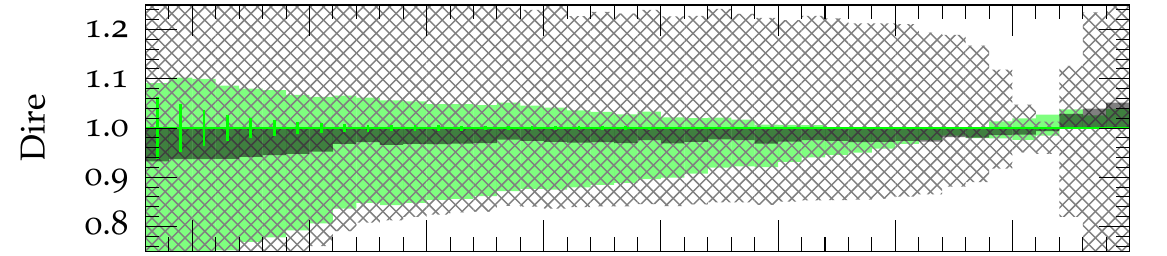}\\[-1mm]
    \includegraphics[width=\textwidth]{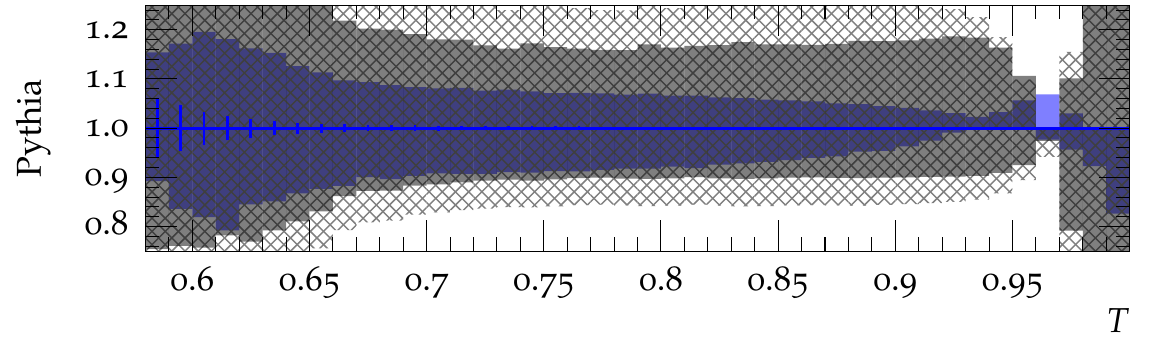}
  \end{minipage}\hskip 5mm
  \begin{minipage}{0.48\textwidth}
    \includegraphics[width=\textwidth]{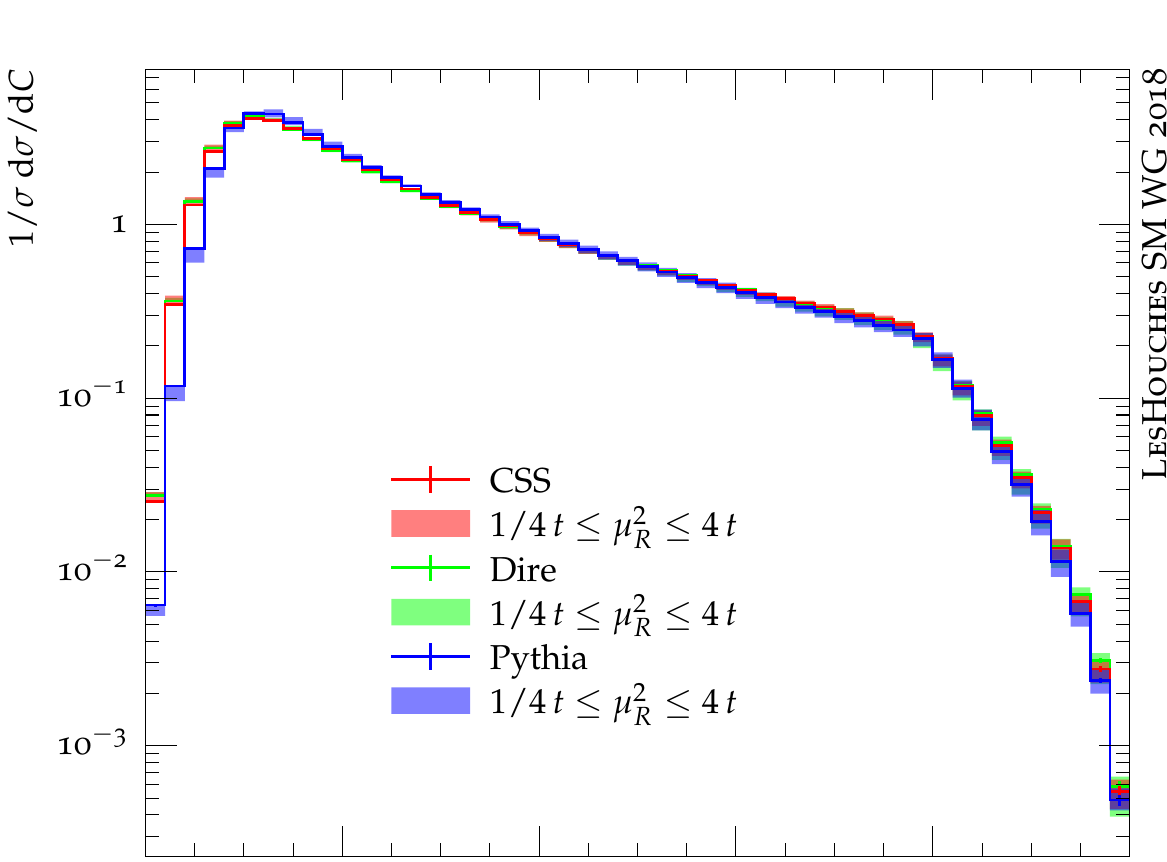}\\[-1mm]
    \includegraphics[width=\textwidth]{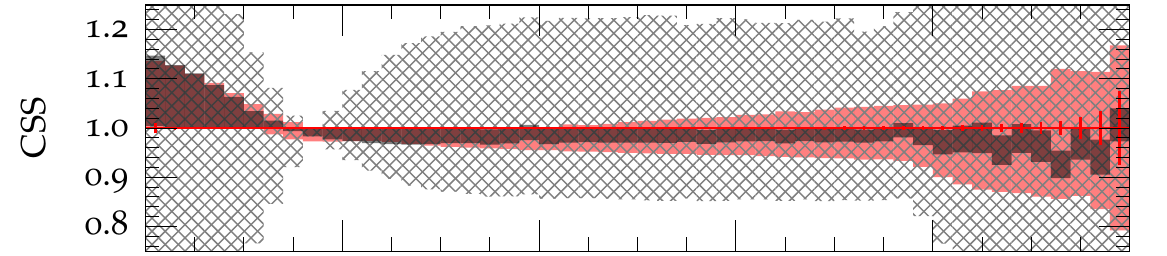}\\[-1mm]
    \includegraphics[width=\textwidth]{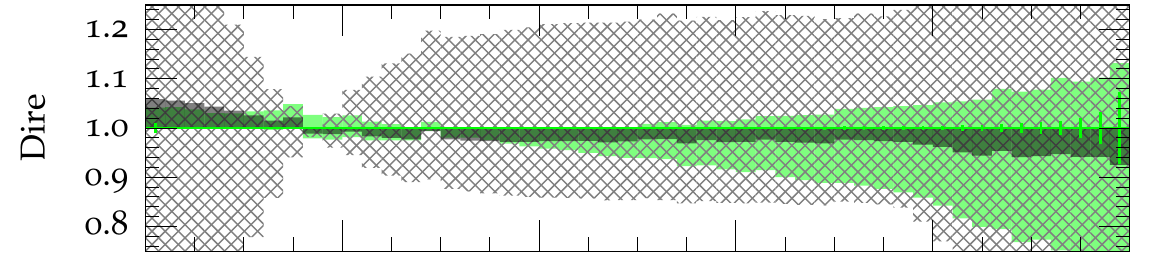}\\[-1mm]
    \includegraphics[width=\textwidth]{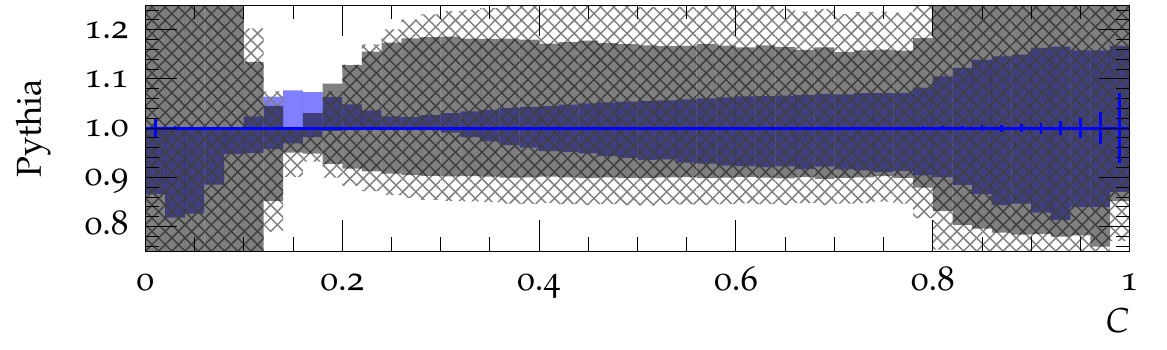}
  \end{minipage}
  \caption{Uncertainty estimates for the thrust (left) and C parameter (right) in
    the analysis setup of~\cite{Heister:2003aj}.
    The gray bands show the estimates using the method of \cite{Badger:2016bpw}
    for \sherpa and \dire, and of \cite{Mrenna:2016sih} for \pythia.
    Blue 
    bands give \pythia variations according to Sec.~\ref{sec:MC_PS_scalevariation:pythiavars}.
    Red and green bands give, respectively, \sherpa and \dire variations
    according to Sec.~\ref{sec:MC_PS_scalevariation:sherpavars}.
    Hatched gray bands show naive variations without any compensation terms.
    \label{fig:MC_PS_scalevariation:lep_shapes}}
\end{figure}

Figure~\ref{fig:MC_PS_scalevariation:lep_jetrates} shows the separation between jets in the Durham
algorithm. As expected, we observe that the default (gray) \sherpa variation is much
more aggressive, while default (gray) \pythia is much more conservative.
The variation bands of \pythia, CSS and \dire are much more similar in size, 
highlighting that although different single-logarithmic terms in different
showers, the minimal prescription leads to a convergence. 
The region below $\sim 10^{-3}$ is dominated by hadronization effects, and thus,
the differences between MCEGs here should not be considered problematic, since 
it heavily depends on how the event generator tune had been produced. It is 
however interesting to observe that the \pythia bands in the non-perturbative
region are reduced in the minimal prescription, which will clearly aid in
finding MCEG tunes including scale variations\footnote{Such a study was 
performed elsewhere in the proceedings of this workshop.}

Similar conclusions can be drawn from Fig.~\ref{fig:MC_PS_scalevariation:lep_shapes}. We again observe
that the minimal prescription leads to a variation band that is more conservative
than the default CSS and \dire scheme, and more aggressive than the default
\pythia variation. The width of the band becomes comparable in all three
showers. Note again that the regions of $\tau\sim1$ and $C<0.25$ show large
hadronization corrections, i.e.\ that the difference of the three tools in this
region is dominated by tuning (and by using $\alpha_s$ setups that are rather
different from those used in the tunes). Again, it is worth mentioning that
a smaller perturbative variation band in the non-perturbative region will
ease tuning in the presence of perturbative variations. 

These results -- and other distributions not shown here -- indicate that
at LEP, varying the renormalization scale in the parton shower by factors 
of $\frac{1}{4},1,4$ (i.e. $\frac{1}{2},1,2$ when applied to GeV-valued
scales) in the ``minimal" prescription leads to uncertainty bands of 
$\mathcal{O}(10-15\%)$ in all three showers considered here. This improves
over previous attempts that should be considered as too aggressive. 

\subsubsection{LHC results}

We will now discuss the impact of variations on distributions in gluon-induced
Higgs-boson production at a 13 TeV LHC. The results of this section are at
the parton level, i.e.\ we do not include the effect of multiple parton 
interactions, beam remnants, hadronization or other non-perturbative effects.

\begin{figure}[t!]
  \centering
  \begin{minipage}{0.48\textwidth}
    \includegraphics[width=\textwidth]{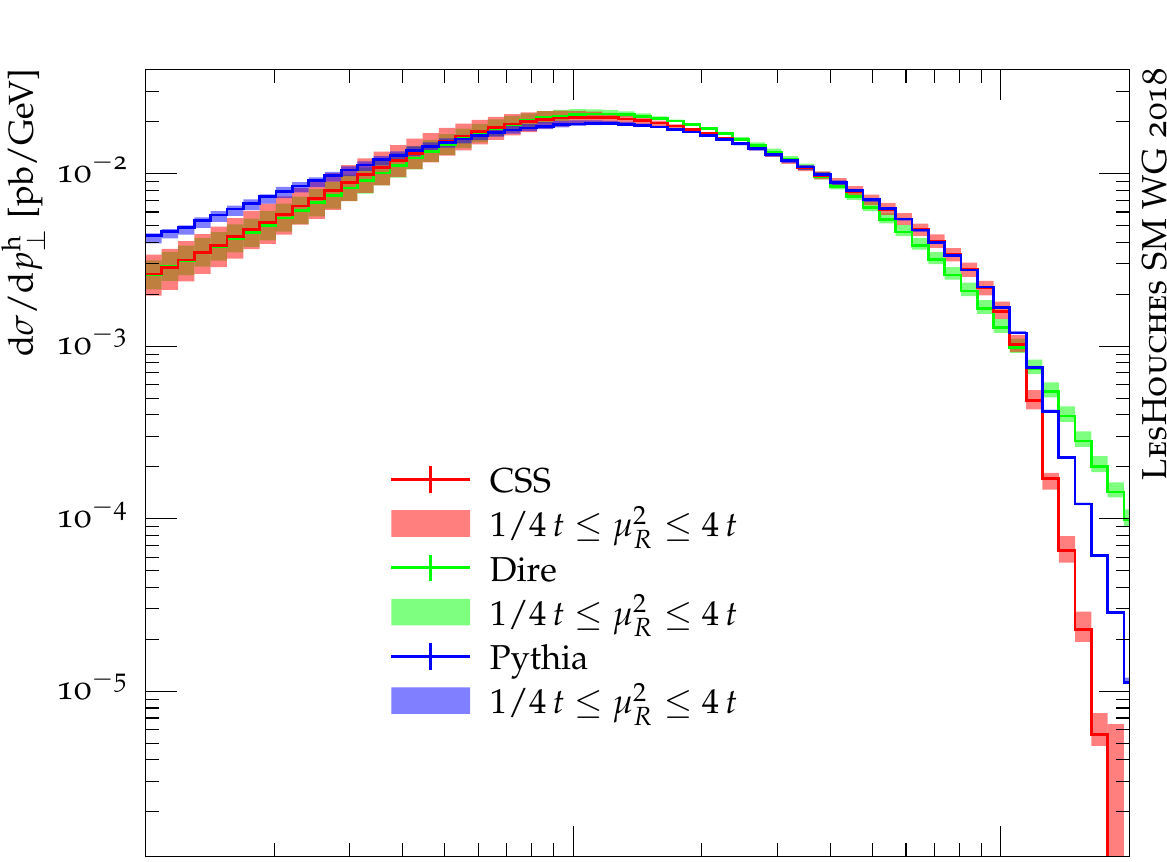}\\[-1mm]
    \includegraphics[width=\textwidth]{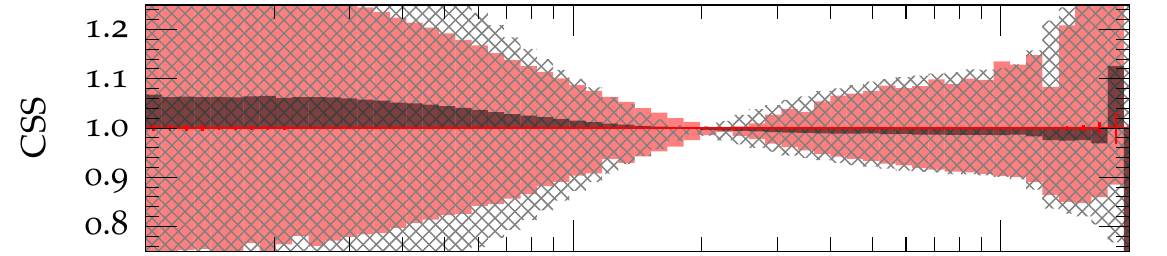}\\[-1mm]
    \includegraphics[width=\textwidth]{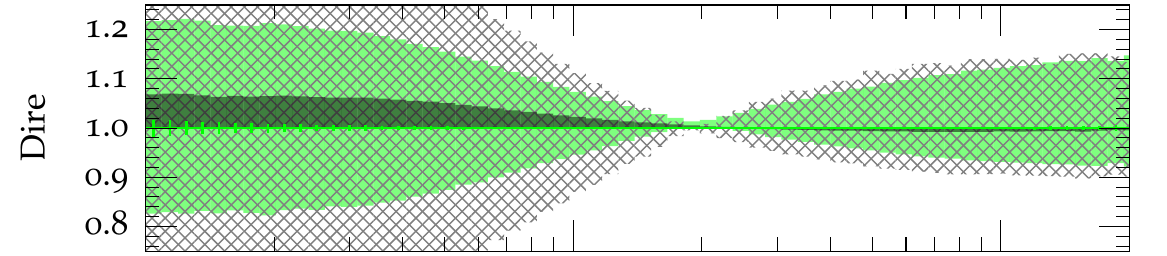}\\[-1mm]
    \includegraphics[width=\textwidth]{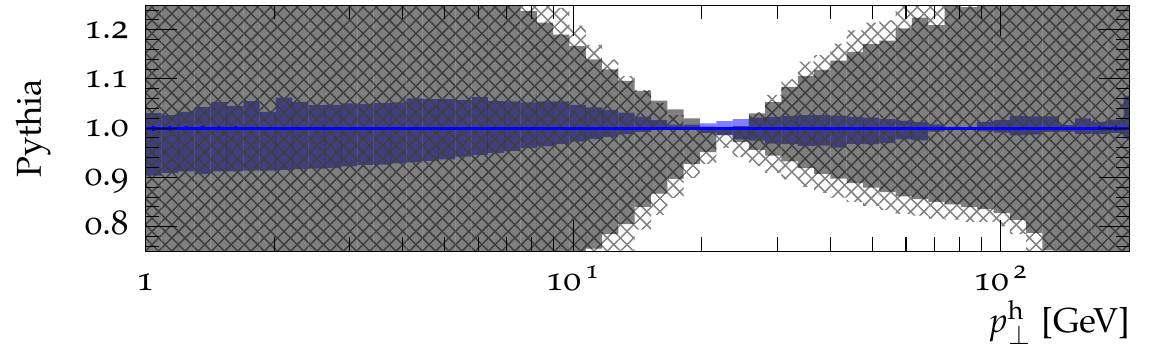}
  \end{minipage}\hskip 5mm
  \begin{minipage}{0.48\textwidth}
    \includegraphics[width=\textwidth]{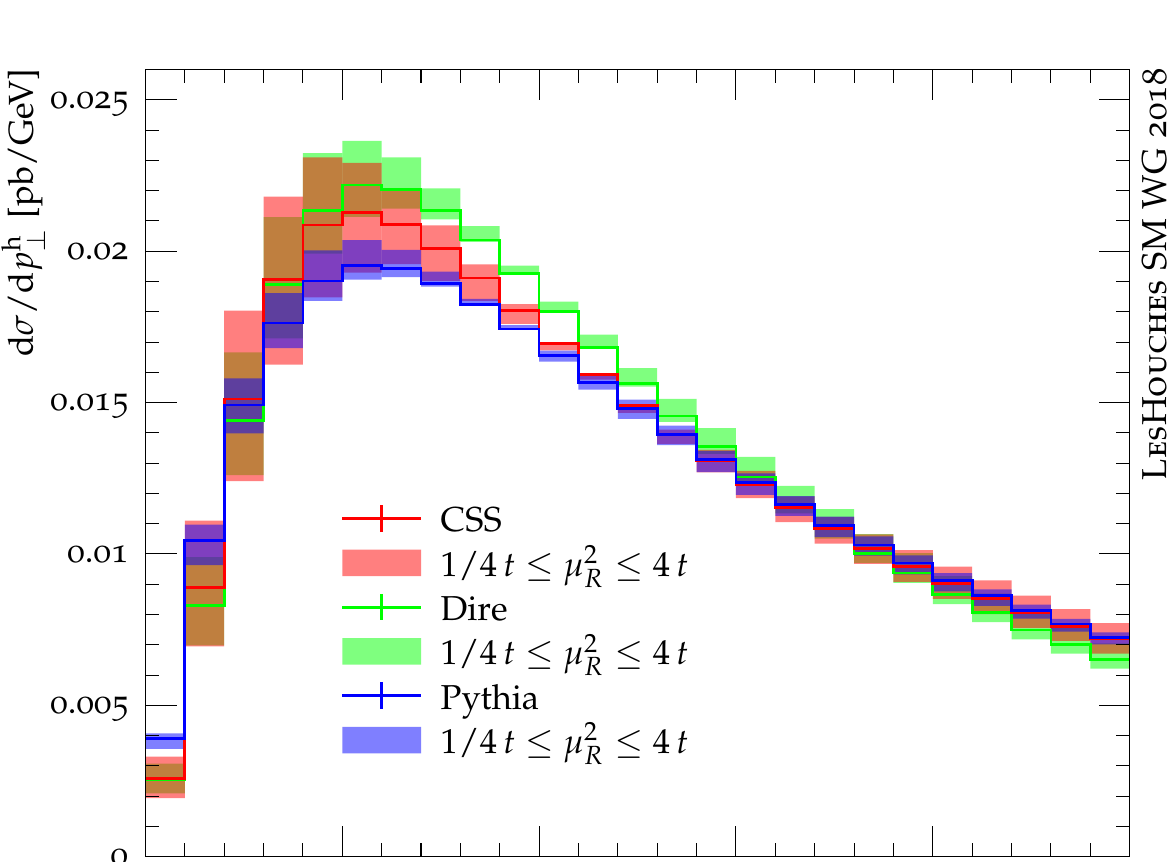}\\[-1mm]
    \includegraphics[width=\textwidth]{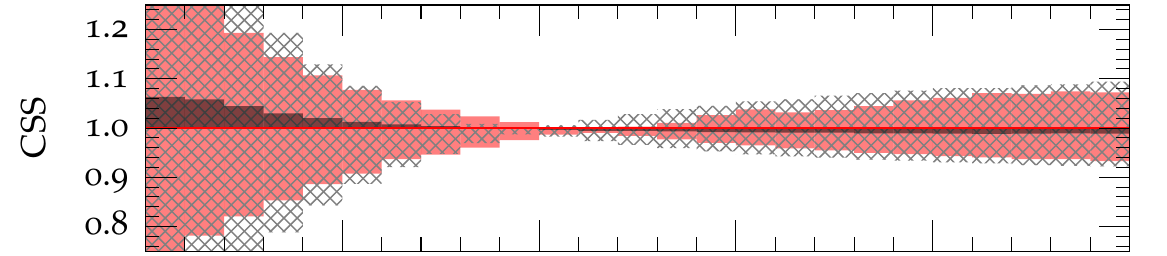}\\[-1mm]
    \includegraphics[width=\textwidth]{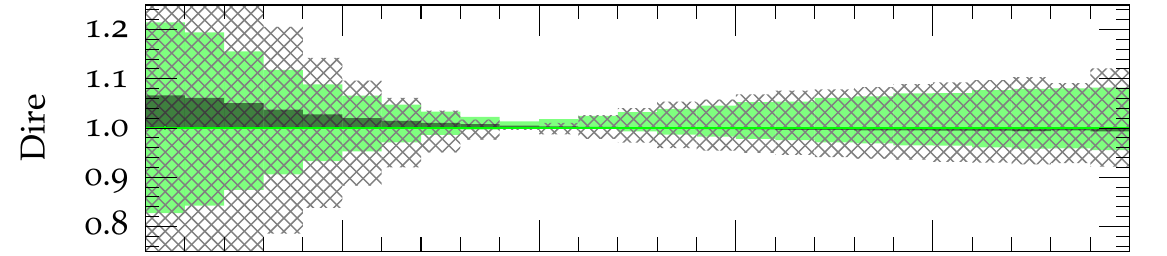}\\[-1mm]
    \includegraphics[width=\textwidth]{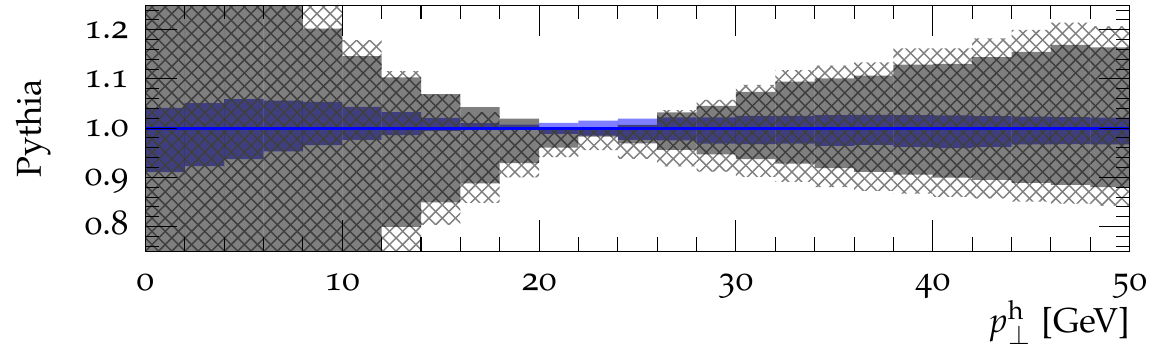}
  \end{minipage}
  \caption{Impact of shower variations on the Higgs $p_\perp$ spectrum.
    The gray bands show the estimates using the method of \cite{Badger:2016bpw}
    for \sherpa and \dire, and of \cite{Mrenna:2016sih} for \pythia.
    Blue 
    bands give \pythia variations according to Sec.~\ref{sec:MC_PS_scalevariation:pythiavars}.
    Red and green bands give, respectively, \sherpa and \dire variations
    according to Sec.~\ref{sec:MC_PS_scalevariation:sherpavars}.
    Hatched gray bands show naive variations without any compensation terms.
    \label{fig:MC_PS_scalevariation:higgs_pt}}
\end{figure}
\begin{figure}[t!]
  \centering
  \begin{minipage}{0.48\textwidth}
    \includegraphics[width=\textwidth]{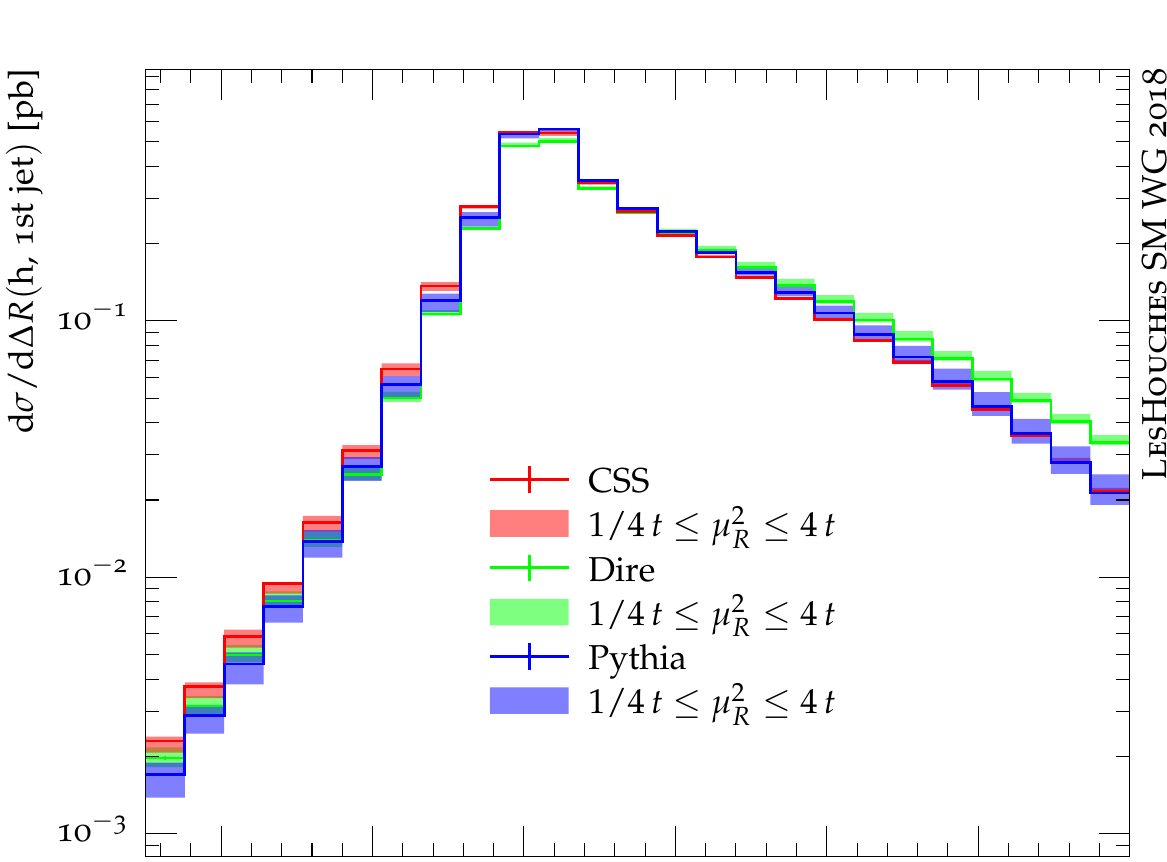}\\[-1mm]
    \includegraphics[width=\textwidth]{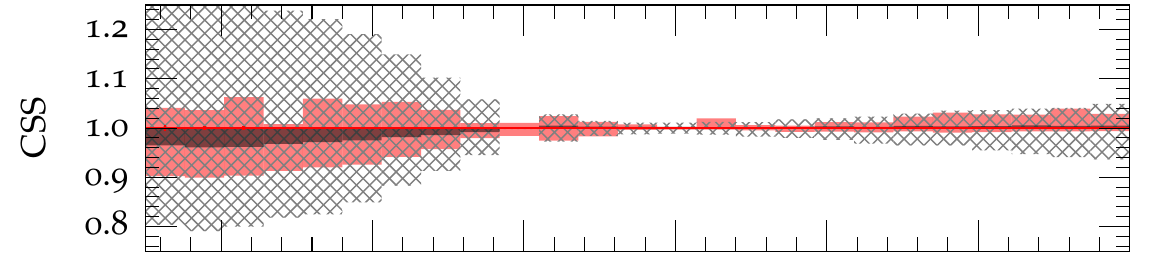}\\[-1mm]
    \includegraphics[width=\textwidth]{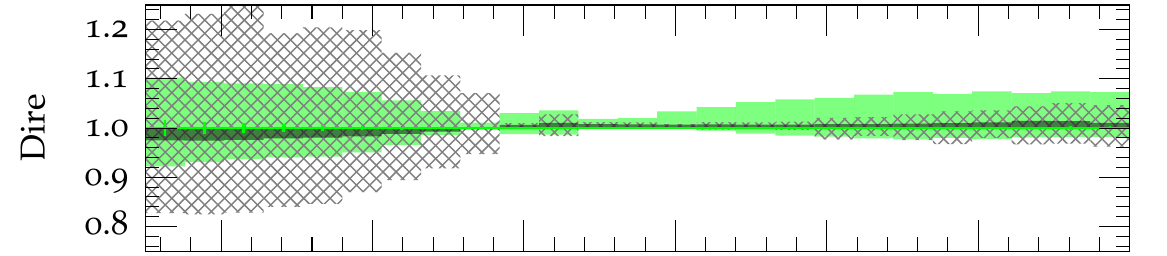}\\[-1mm]
    \includegraphics[width=\textwidth]{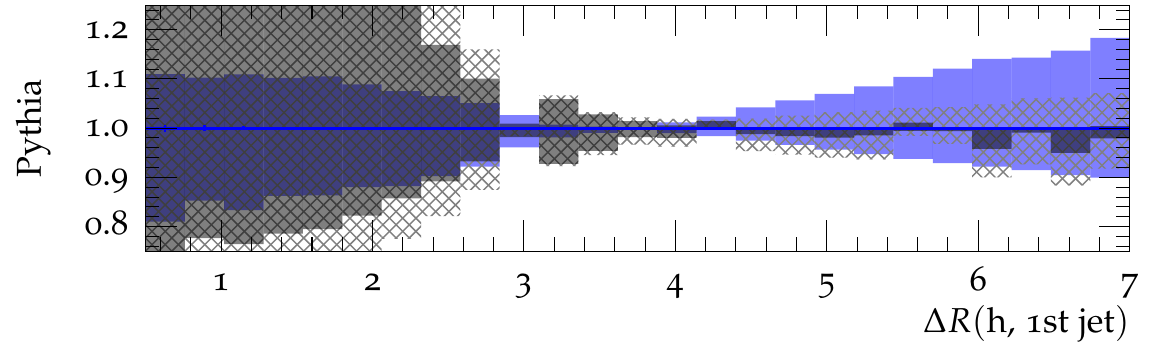}
  \end{minipage}\hskip 5mm
  \begin{minipage}{0.48\textwidth}
    \includegraphics[width=\textwidth]{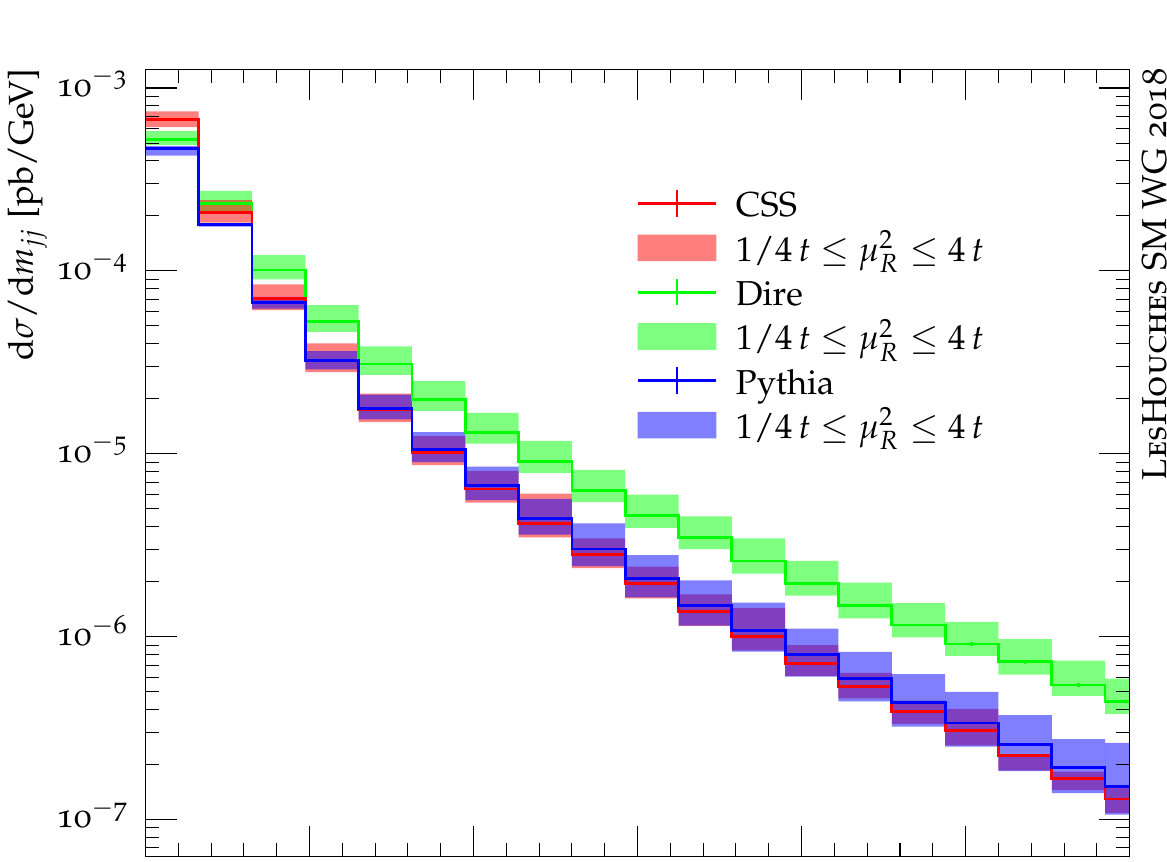}\\[-1mm]
    \includegraphics[width=\textwidth]{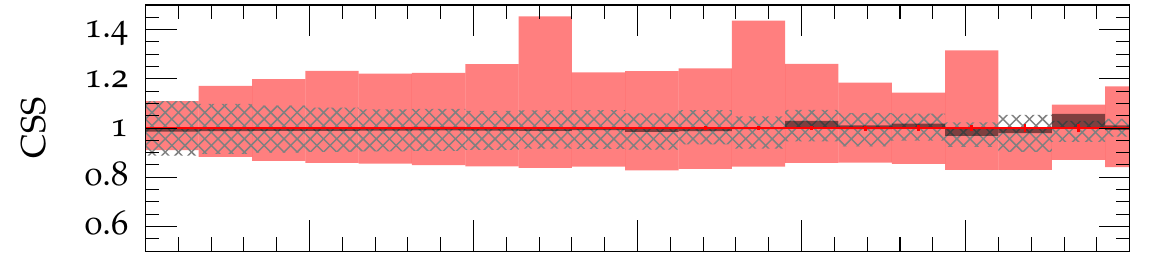}\\[-1mm]
    \includegraphics[width=\textwidth]{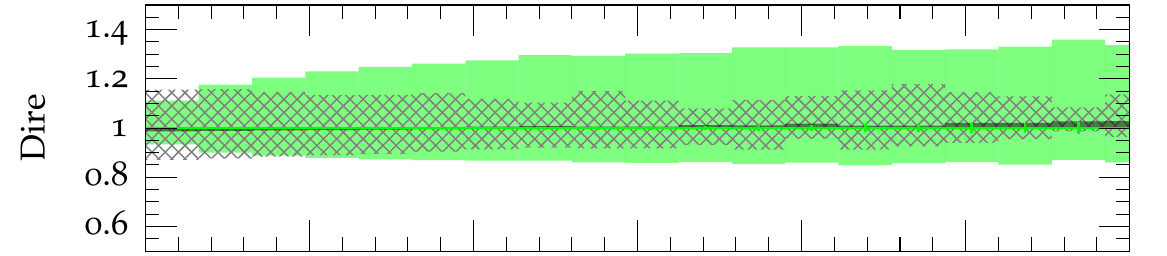}\\[-1mm]
    \includegraphics[width=\textwidth]{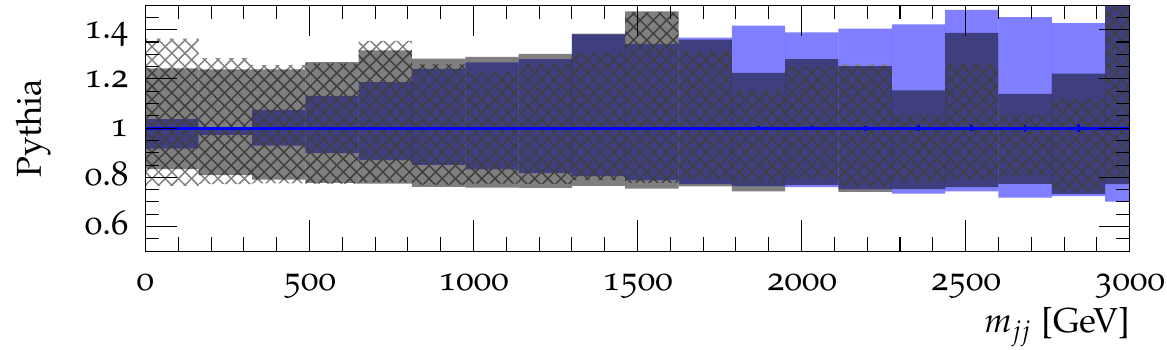}
  \end{minipage}
  \caption{Impact of shower variations on the Higgs-jet $\Delta R$ separation and
    dijet invariant mass in Higgs + jets events.
    The gray bands show the estimates using the method of \cite{Badger:2016bpw}
    for \sherpa and \dire, and of \cite{Mrenna:2016sih} for \pythia.
    Blue 
    bands give \pythia variations according to Sec.~\ref{sec:MC_PS_scalevariation:pythiavars}.
    Red and green bands give, respectively, \sherpa and \dire variations
    according to Sec.~\ref{sec:MC_PS_scalevariation:sherpavars}.
    Hatched gray bands show naive variations without any compensation terms.
    \label{fig:MC_PS_scalevariation:higgs_aux}}
\end{figure}

Figure~\ref{fig:MC_PS_scalevariation:higgs_pt} shows the Higgs-boson transverse momentum on 
logarithmic and linear scale, with the latter focussing on the low-$p_\perp$
region interesting for parton shower resummation. Before discussing the size of
the bands, it is interesting to see that \pythia and the \sherpa CSS shower
produce very similar results for $2~\textnormal{GeV} < p_{\perp^h} < 100
~\textnormal{GeV}$. The differences above $100~\textnormal{GeV}$ would be 
ameliorated upon NLO matching or multi-leg merging, while the differences below
$2~\textnormal{GeV}$ are covered by non-perturbative effects in a hadron-level
simulation. The \dire shower predicts a rather different shape of the $p_\perp$
spectrum above the Sudakov peak, which arises from the different choice of
ordering variable compared to the CSS and \pythia. This difference is not covered
by the respective uncertainty bands, because effects relating to momentum conservation
and ordering variables are an inherently different degree of freedom of the
parton shower implementation~\cite{Hoeche:2017jsi}.

The effect of employing the minimal compensation scheme on the variation
bands is very different in \pythia and \sherpa. As before, we see that the
minimal scheme squeezes the band w.r.t.\ the default scheme in \pythia, 
while the the minimal scheme widens the band in CSS and \dire. However, the
\pythia band within the minimal scheme reduces to 10\% or less, the 
CSS band increases beyond 50\% at low $ p_{\perp^h}$ and 20\% at high values,
and the \dire band spans up to 35\% variation at low $ p_{\perp^h}$ and 20\% 
at high values. These trends are reversed in the $\Delta R$ and $m_{jj}$
distributions shown in Fig.~\ref{fig:MC_PS_scalevariation:higgs_aux}. Here, the \pythia band
in the minimal scheme yields the largest variation, while the band of CSS
is almost vanishing.

These effects should make clear that, as anticipated in the motivation,
the effect of the compensation is observable-dependent. In particular, since 
the difference of parton shower evolution variables will lead to
different single-logarithmic terms, the effect of compensation can, for
one observable, be quantitatively different for different showers. 
Gluon-induced Higgs-boson production is, due to high parton-shower 
activity, an ideal laboratory to highlight these subtleties.


\subsection{Discussion and Summary}
\label{sec:MC_PS_scalevariation:results}

In this study, we have investigated renormalization scale variations in parton
showers. We have discussed compensation terms to ensure that such scale
variations do not deteriorate expressly introduced higher-order corrections, 
and attempt to define a common baseline for applying such compensation in
\pythia and \sherpa. We have then compared this new ``minimal" prescription
to traditional LEP observables and to the result of the previous 
default variation schemes in the generators. This comparison suggests that
in the minimal scale variations by factors $k=\frac{1}{4},4$ lead to reasonable agreement
between the three showers used in the study, while not impeding the ability
to tune the event generator due to large perturbative variations in 
hadronization-dominated phase space regions. The study of Higgs-boson production
at the LHC does not lead to conclusive results, since the shower- and 
observable-dependence of the variation bands is exposed by the high parton-shower
activity. In this case, we recommend to use the most conservative uncertainty
estimate available in the event generator.

\subsection*{Acknowledgements}
  We thank the organizers for an inspiring workshop.
  This work was supported by the US Department of Energy
  under the contract DE--AC02--76SF00515.
  This manuscript has been authored by Fermi Research Alliance, LLC under 
  Contract No. DE-AC02-07CH11359 with the U.S. Department of Energy, Office of 
  Science, Office of High Energy Physics.

\let\Herwig\undefined
\let\Pythia\undefined
\let\Sherpa\undefined
\let\pythia\undefined
\let\sherpa\undefined
\let\Vincia\undefined
\let\Dire\undefined
\let\vincia\undefined
\let\dire\undefined
\let\Rivet\undefined
\let\Professor\undefined
\let\eps\undefined
\let\mc\undefined
\let\mr\undefined
\let\mb\undefined
\let\tm\undefined




\newcommand{\Herwig}{H\protect\scalebox{0.8}{ERWIG}\xspace}
\newcommand{\Pythia}{P\protect\scalebox{0.8}{YTHIA}\xspace}
\newcommand{\Sherpa}{S\protect\scalebox{0.8}{HERPA}\xspace}
\newcommand{\Rivet}{R\protect\scalebox{0.8}{IVET}\xspace}
\newcommand{\Professor}{P\protect\scalebox{0.8}{ROFESSOR}\xspace}
\newcommand{\eps}{\varepsilon}
\newcommand{\mc}[1]{\mathcal{#1}}
\newcommand{\mr}[1]{\mathrm{#1}}
\newcommand{\mb}[1]{\mathbb{#1}}
\newcommand{\tm}[1]{\scalebox{0.95}{$#1$}}





\section{On the cross-talk of parameter optimization and perturbative variations in event generators~\protect\footnote{
    J.~Bellm,
    L.~L{\"o}nnblad,
    S.~Pl{\"a}tzer, 
    S.~Prestel,
    D.~Samitz,
    A.~Siodmok,
    A.~H.~Hoang}{}}
\label{sec:MC_tunes_variations}


  We discuss the interplay of parton-shower variations and optmimization of
  non-perturbative parameters in general-purpose event generators. To understand
  the correlation between infrared cut-offs and hadronization parameters with
  variations of $\alpha_s$, we investigate the interplay of tuning and 
  $\alpha_s$-variations for $e^+e^-$ colliders. For LHC collisions, we 
  investigate the cross-talk of $\alpha_s$- and PDF variations. 

\subsection{Introduction}

General-Purpose Monte Carlo Event Generators (MCEG)
\cite{Buckley:2011ms} are the backbone of collider-physics analysis
prototyping, while at the same time providing the accurate background
predictions to new-physics searches. With ever more detailed
measurements at the LHC, the focus of MCEG developments has been
providing and enabling precision calculations in perturbative QCD. At
this point, it is important to stress that the most accurate
simulation should predict unknown phenomena based on known
effects. The latter will include the results of previous measurements
as well as higher order calculations. Event generators combine both
into a simulation that allows predictions for new analyses and
collider setups by using higher order calculations and transferring,
or extrapolating, a realistic model of previous experimental data.  To
this end, MCEG parameters are commonly ``tuned'' to many different
data.  This ``tuning'' helps constrain parameters that cannot be fixed
by theory considerations alone, and encodes the global picture of
previous measurements into the MCEG.  On top of the (mostly
non-perturbative QCD) parameters that are constrained by this
exercise, MCEGs also heavily rely on pre-tabulated parton distribution
functions (PDFs)~\cite{Rojo:2015acz} to describe the longitudinal
structure of colliding hadrons. The non-perturbative component of
these PDFs is usually ``extracted" by fitting the parameters of the
PDF parametrizations to measurements of (mostly inclusive) scattering
cross-sections.  In either case, we expect strong correlations between
non-perturbative parametrizations and perturbative input.

This has made a comprehensive error budget for MCEGs a long-standing
problem.  Tuning will thus have the effect of mixing perturbative and
non-perturbative effects, and will potentially also account for
deficienscies in the perturbative prediction. Thus, a well-defined
event generator uncertainty should include both effects, and correctly
incorporate the cross-talk between them. This is a daunting task in
principle, but also in practice, since correlations and subtleties can
easily be masked by the overall statistical uncertainty inherent to
finite-time Monte Carlo simulation. With the advent of weighting
techniques \cite{Bellm:2016voq,Mrenna:2016sih,Bothmann:2016nao}, some
of the technical difficulties of correlated parton-shower variations
were recently ameliorated. Thus, we will investigate the correlation
of perturbative and non-perturbative MCEG pieces with these new
tools. In this study, we do not attempt to provide a theoretical
answer to the question of defining uncertainties. Instead, we provide
a phenomenological study of some cross-talk, in hopes of gaining
insight for the future.

This article is organized into two main sections followed by the summary. The main sections are in spirit intimately 
connected, in that we try to assess different perturbative-non-perturbative
correlations from different angles. The first part is devoted to the cross-talk
between the tuning of the parameters of hadronization models and final-state
parton shower variations. The second part investigates the correlation between
PDF choices and initial-state parton showering.


\subsection{Generator tuning vs. parton-shower variations}

In this section, we will investigate the cross-talk between non-perturbative
hadronization parameters and changes of $\alpha_s$ in the (perturbative)
parton shower evolution. Parameters in hadronization models are commonly
extracted at lepton colliders and then used unchanged at hadron 
colliders (with the primordial kT of partons inside of colliding hadrons being
one notable exception). To not over-complicate the study, we thus discuss the
cross-talk using LEP measurements.

\subsubsection{Methodology}

In order to investigate the interplay between generator tuning and parton-shower 
variations studied in~\cite{Mrenna:2016sih,Bothmann:2016nao,Bellm:2016rhh}
we use two different parton shower Monte Carlo event generators \Herwig 7~\cite{Bellm:2015jjp} 
and \Pythia 8~\cite{Sjostrand:2006za,Sjostrand:2014zea} to produce a fully exclusive simulation of the events 
at the Leading Order. Both generators provide an option for using
different shower modules, however at this stage of the project we only
use default parton showers algorithms\footnote{In the future, we plan
  to use also Dipole Shower of \Herwig~\cite{Platzer:2009jq} and
  Dire~\cite{Hoche:2015sya}}: the angular
ordered shower in \Herwig~\cite{Gieseke:2003rz} and $p_\perp$-ordered
shower in \Pythia~\cite{Sjostrand:2004ef}.  For generator tuning we
use the parametrization-based tune method provided by the \Professor
package~\cite{Buckley:2009bj}. The starting point for the tuning
procedure is the selection of a range
$[p_i^{\rm min},\, p_i^{\rm max}]$ for each of the $N$ tuning
parameters~$p_i$.

Since we want to explore the correlation of generator tuning and
parton-shower variations our selection of parameters consist of the
main parameters of non-perturbative hadronization models and the
parton shower cut-off scale. To be more precise in \Herwig we adjust
the shower cut-off (\texttt{pTMin}), and main hadronization parameters
of the cluster model~\cite{Webber:1983if}: gluon constituent mass
(\texttt{g:ConstituentMass}) and splitting parameter for clusters
\texttt{PSplit} (not flavour specific\footnote{Cluster model
  implemented in \Herwig has possibility to set splitting parameters
  separately for light (\texttt{PSplitLight}), bottom
  (\texttt{PSplitBottom}) and charm (\texttt{PSplitCharm}) clusters,
  see~\cite{Bahr:2008pv} for details.  However, in our exercise for
  simplicity we keep them all the same.}). Initially we also included
into the procedure flavor depended parameters: \texttt{ClMaxs} and
\texttt{ClPows}, however since the data selection was not really able
to constrain the parameters we decided to keep them close to their
default values and effectively remove from the tuning procedure.  In
\Pythia except for the shower cut-off (\texttt{TimeShower:pTmin}) we
also tune three main hadronization parameters of the Lund
model~\cite{Andersson:1983ia}: a- and b-parameter
(\texttt{StringZ:aLund} and \texttt{StringZ:bLund}) and
non-perturbative pt-width (\texttt{StringPT:Sigma}).

For both generators, event samples are generated for random points in
this multidimensional hypercube in their parameter space.  The number
of points sampled is chosen, depending on the number of input
parameters, to ensure good control of the final tune. Each generated
event is directly handed over to the Rivet
package~\cite{Buckley:2010ar}, which implements the experimental
analyses.  Thus the results for each observable are calculated at each
set of parameter values. \Professor parametrizes each bin of each
histogram as a function of the input parameters. It is then able to
find the set of parameters that fit the selected observables best. As
a user, one simply has to choose the set of observables that one
wishes to tune to and, optionally, their relative weights in the fit.

Usually, the hadronization and shower parameters are tuned to a wide
range of experimental data, however, most of them are from
LEP. Therefore, in our tuning effort, we also use LEP data. In
particular we use ALEPH data from \cite{Heister:2003aj}, however, we
have not tuned to all available distributions\footnote{The full list
  of selected observables and their weights are provided in
  Appendix~\ref{app:MC_tunes_variations:weights}.}. We have, \textit{e.g.}\ left out the
higher jet multiplicities while including the lower ones. We also note
that although the average multiplicity is included, it is only one bin
and will therefore not constrain the tuning very much. It should be
noted that the tunings done here are mainly for illustration and are
not optimized in any way nor do they reflect the optimal outcome with
respect to a global tuning effort.


To assess the effect of the correlation between infrared cut-offs and
hadronization parameters with variations of $\alpha_s$, for each MCEG
we perform two sets of tunes with the scale variation bands:
\begin{enumerate}\itemsep 0mm
\item The central (reference) tune, is obtained by using a "central"
  $\alpha_s(M_z)$ in the tuning procedure.  This is simply chosen to
  be the default value in the respective generator. Its scale
  variation band (the envelope of the tune) is estimated by shifting
  $\alpha_s(M_z)\!\to\! \alpha_s(\{\frac{1}{\xi},1,\xi\} M_z)$\linebreak
  $\to\! \alpha_s^\prime(M_z)$, with\footnote{We also
    performed studies for $\xi=\sqrt{2}$, however in order to show
    clearly the effect of the retuning we only present results for
    $\xi=2$.} $\xi=2$ without additional retuning for each
  $\alpha_s^\prime(M_z)$ variation.
\item Two more tunes, called ``retune'' and the corresponding scale
  variation band, are obtained by the same $\alpha_s(M_z)$ variation
  as for ''central`` tune but this time we also retune to ALEPH data
  for each $\alpha_s^\prime(M_z)$ variation.
\end{enumerate}
In the next section we compare in details the results from the both MCEG using their ''central`` and ''retune`` versions
of the tunes. 
\subsubsection{Results}
\begin{figure}[t]
\centering
\includegraphics[width=0.48\textwidth]{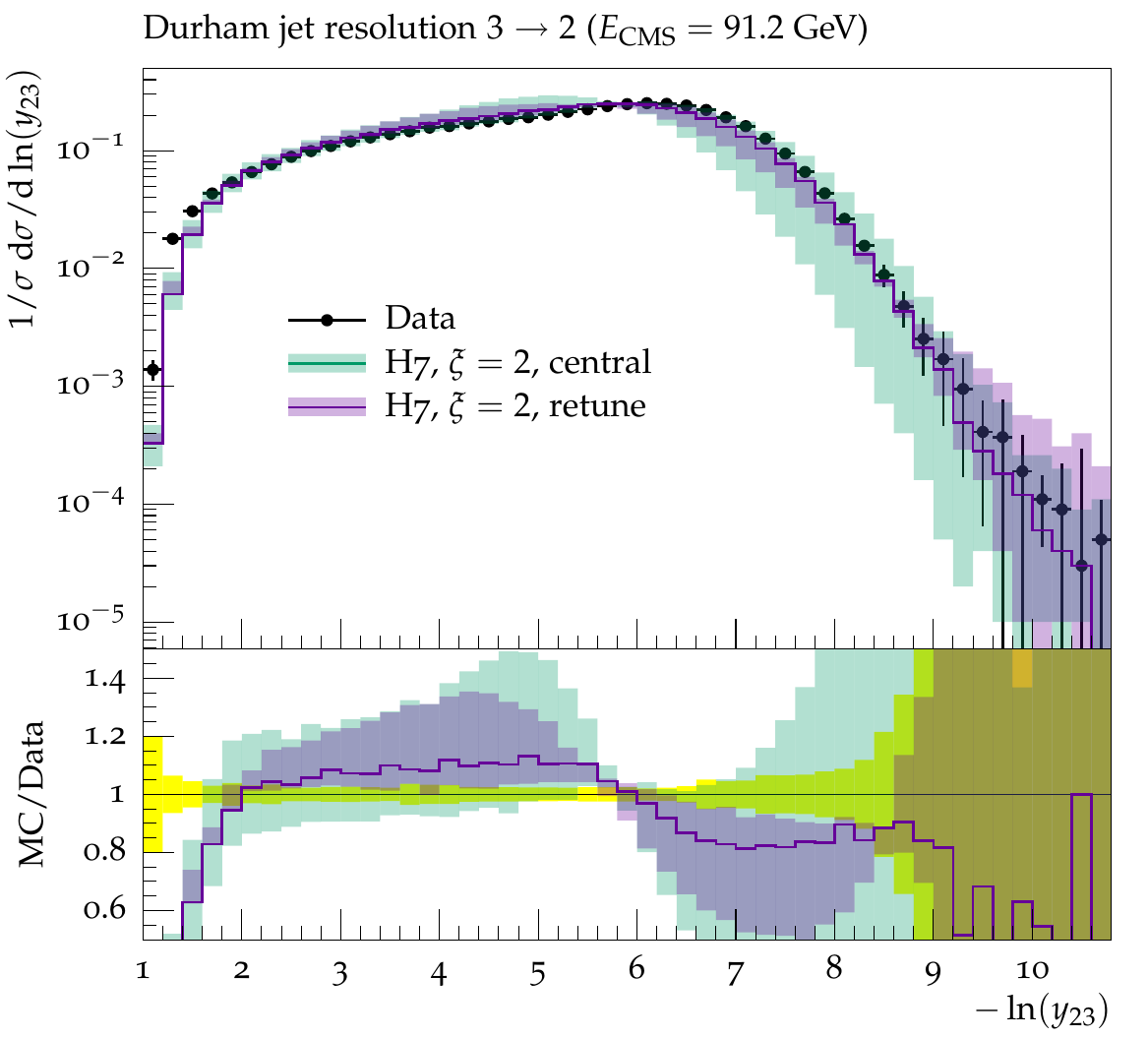}\hfill
\includegraphics[width=0.48\textwidth]{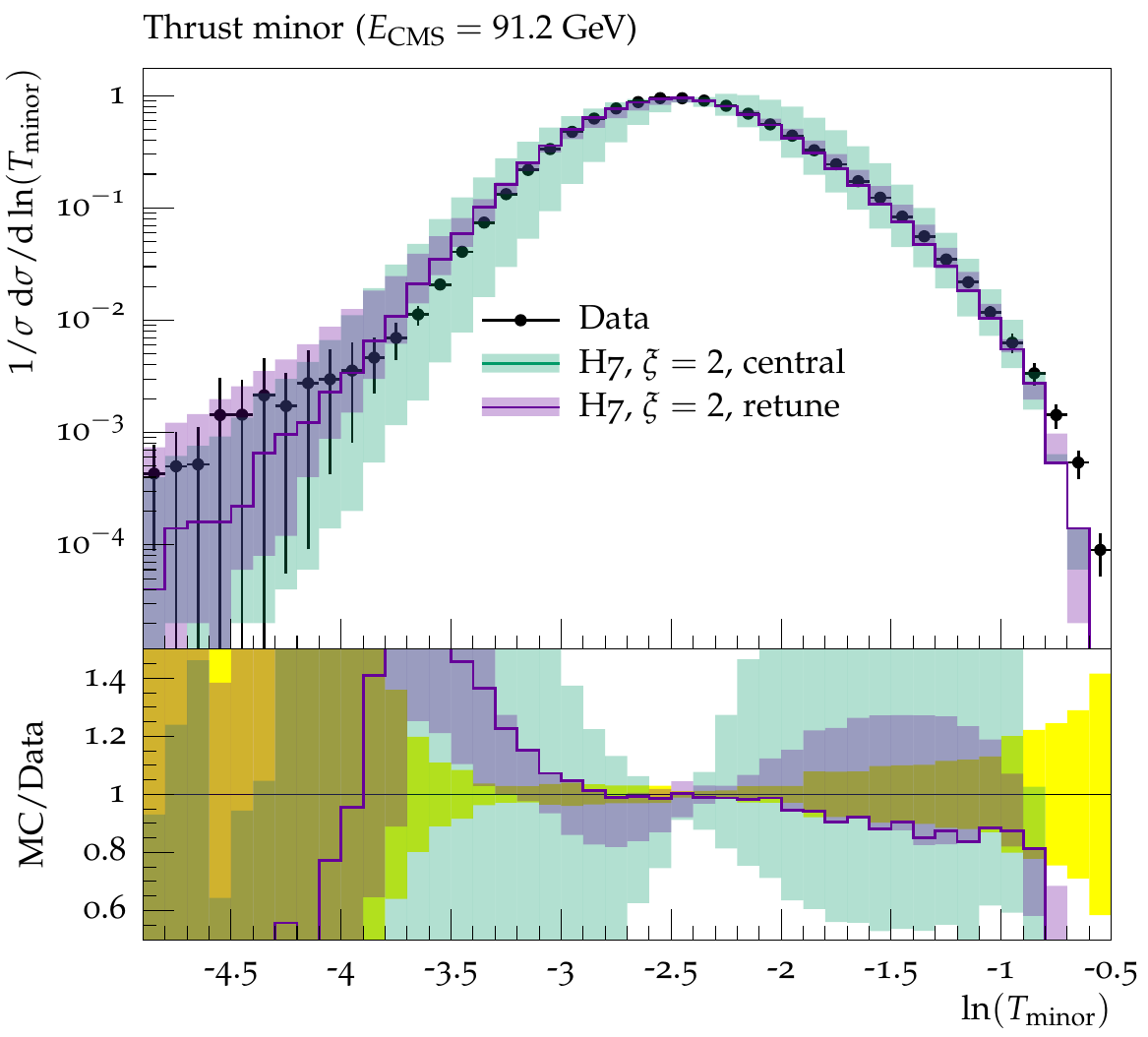}\\
\includegraphics[width=0.48\textwidth]{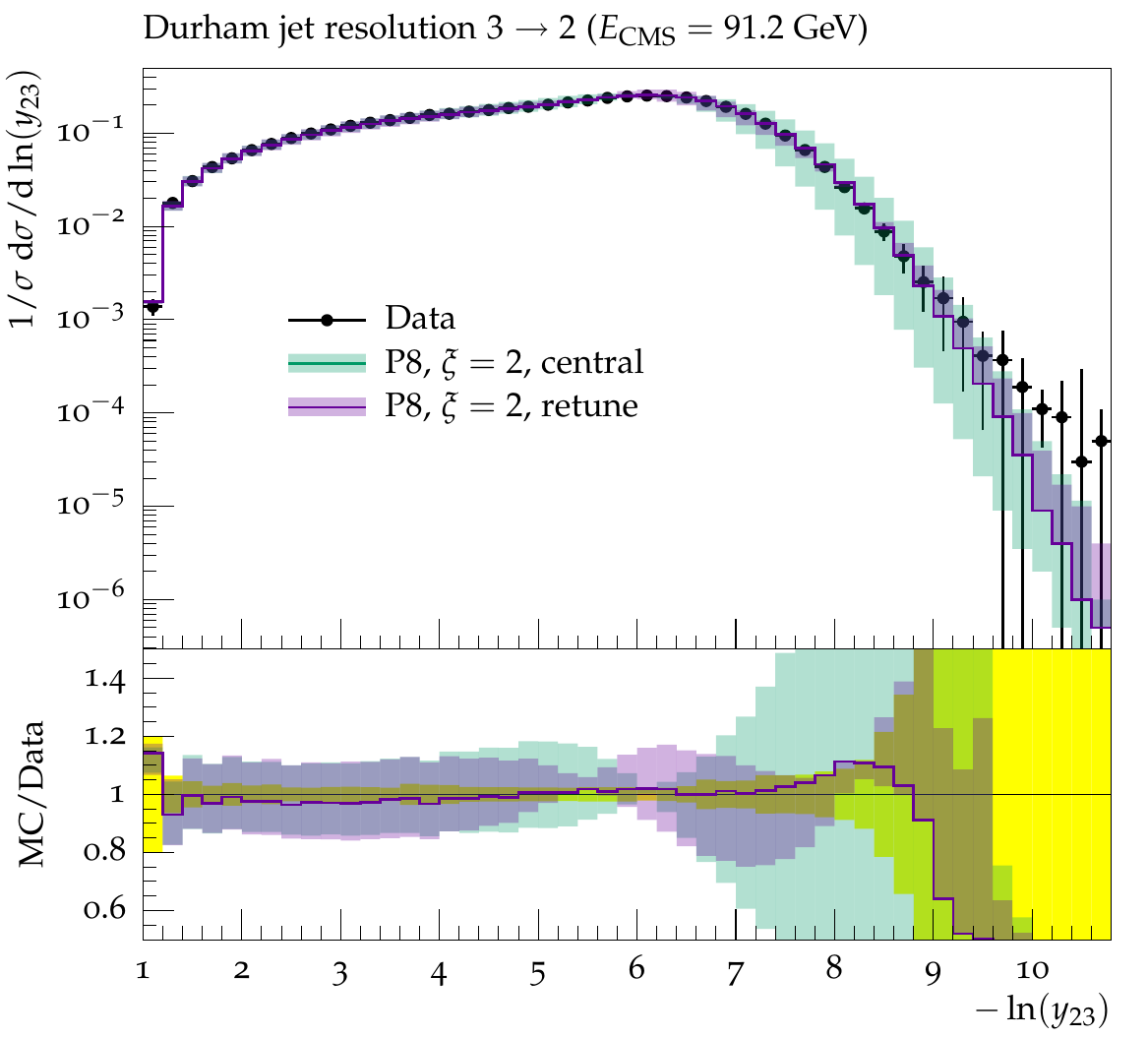}\hfill
\includegraphics[width=0.48\textwidth]{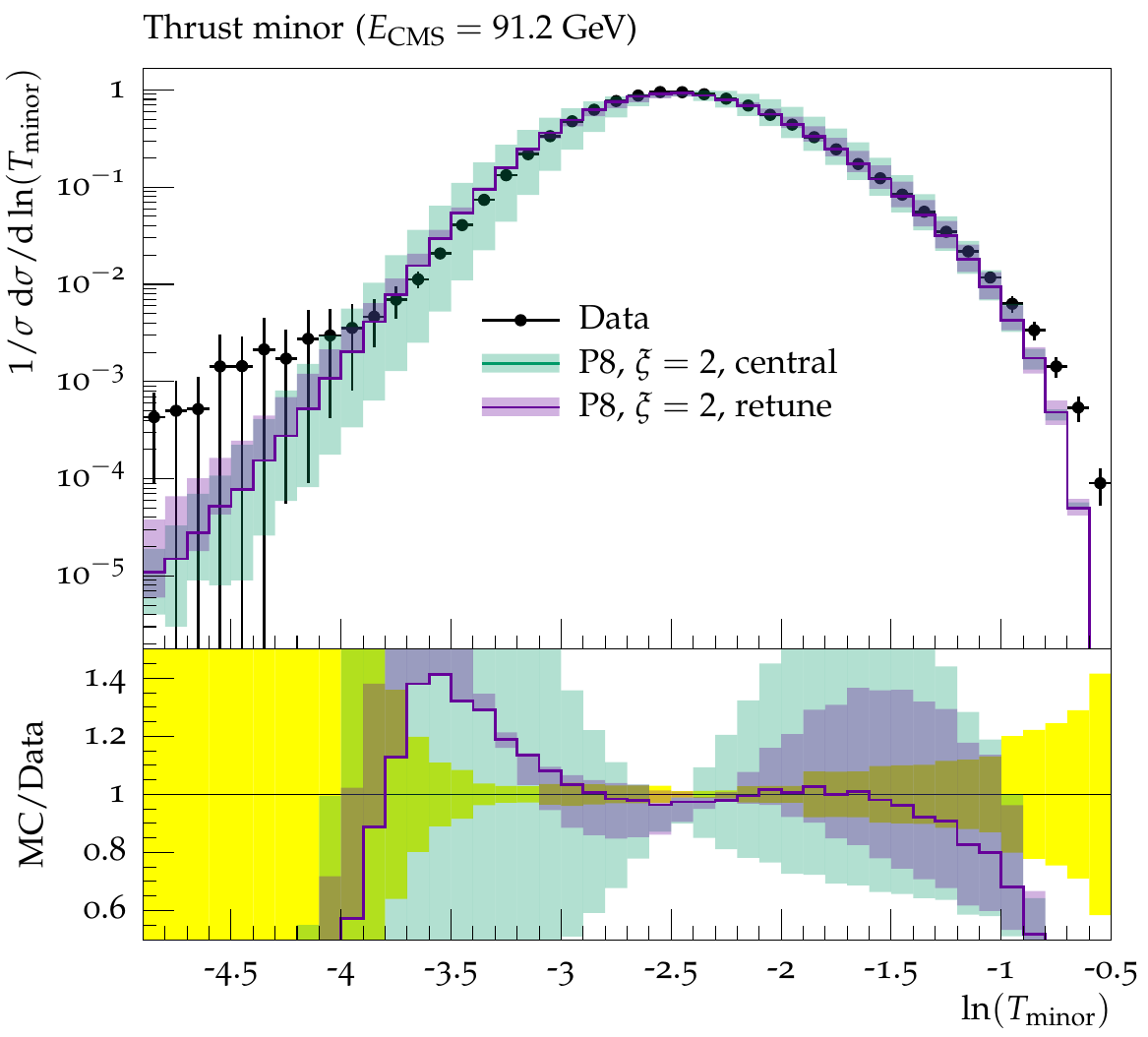}
\caption{Examples of the resulting tunes, ''central`` and ''retune``, with their uncertainty band compared to ALEPH data at 91.2~GeV. Top
  plots are \Herwig and bottom is \Pythia. Left plots are the
  Durham jet resolution $3\rightarrow2$ distribution, and the right
  plots are the Thrust minor distribution. The both observables were used in the tuning procedure.}
  \label{fig:MC_tunes_variations:DJR32}
\end{figure}

We start by looking at the distributions that were actually tuned
to. In Fig.~\ref{fig:MC_tunes_variations:DJR32} we show the distribution in jet
resolutions scales where an event with three jets would be clustered
into two, $y_{23}$, and the \textit{thrust minor} ($T_{\mr{minor}}$)
distribution (the thrust calculated wrt.\ the axis out of the event
plane spanned by the thrust and thrust major axes), respectively. We
see immediately that both generators are able to fit the distributions
fairly well and, as expected, the scale variations of the central tune
are much larger then the ones obtained when retuning for each scale
choice.

We note that since the distributions are normalised to unity, the
scale variations are artificially reduced around the center of the
distributions. Looking more closely, we see some imperfections in the
fits. For example, we see that both generators have problems
describing the region of large $T_{\mr{minor}}$ region, corresponding
to large activity out of the event plane driven by hard effects of
$\mc{O}(\alpha^2_s)$, where neither generator includes matrix element
corrections in our studies.  Similarly, \Herwig has problems
describing large $y_{23}$ values (small $-ln(y_{23}$)).  However, this
is as expected since the \Herwig simulations had switched-off matching
to the $\mc{O}(\alpha_s)$ tree-level matrix elements, while the
\Pythia simulation had it on.  We would also like to note that in the
region of large $y_{23}$ (small $-ln(y_{23}$)) the scale variations
for \Pythia are not reduced by the retuning. This is expected since
there is no mechanism by which the non-perturbative parameters in
\Pythia can compensate for the effect of scale changes on hard
jets. This is, however, not the case for \Herwig, where sometimes a
cluster with a large mass is allowed to decay isotropically into two
lighter clusters, then decaying into hadrons and ending up as jets in
the Durham algorithm. This is regulated by the \texttt{PSplit}
parameter which in this case can compensate somewhat for the scale
variations at large $y_{23}$.

\begin{figure}[t]
\centering
\includegraphics[width=0.48\textwidth]{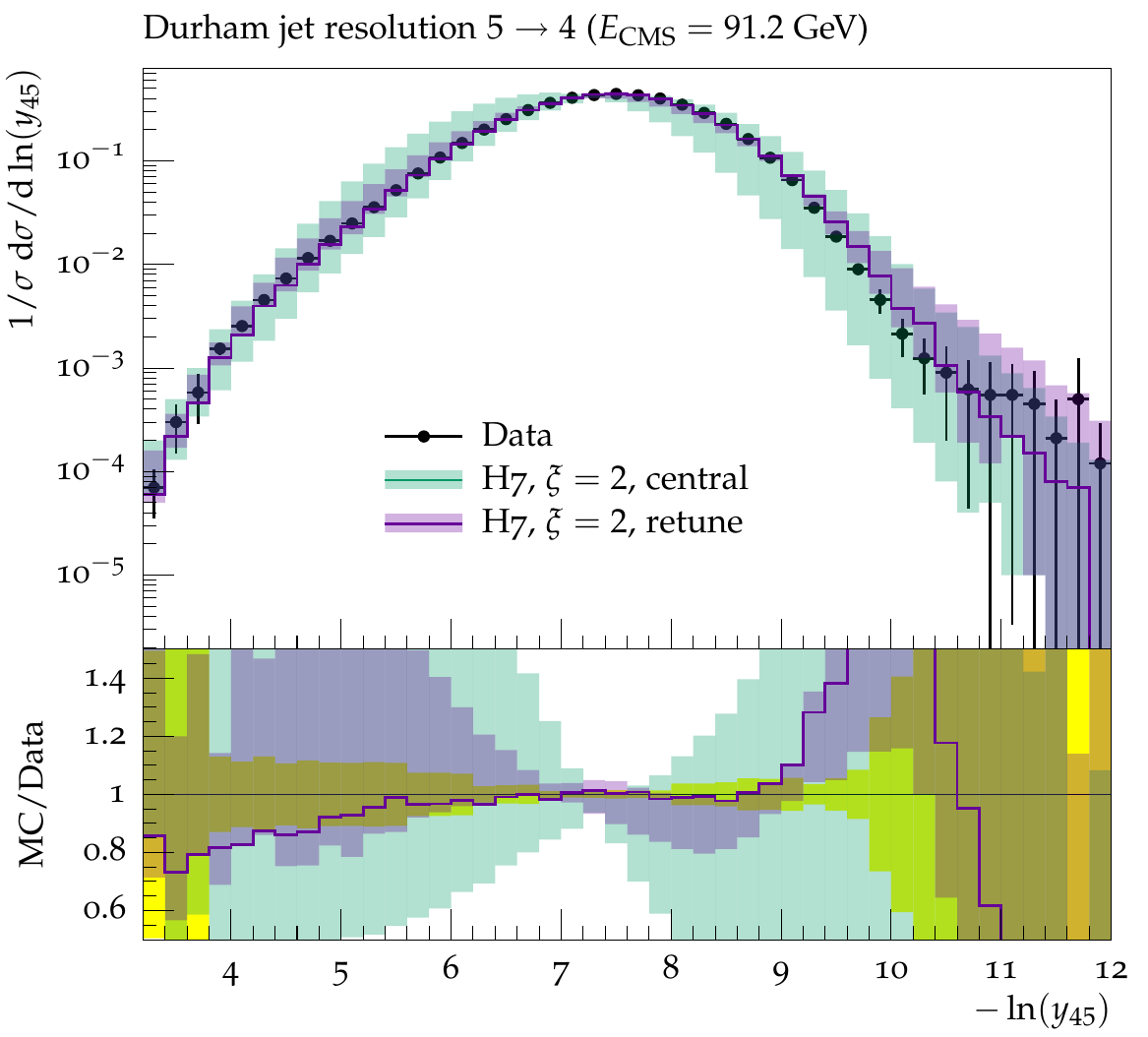}\hfill
\includegraphics[width=0.48\textwidth]{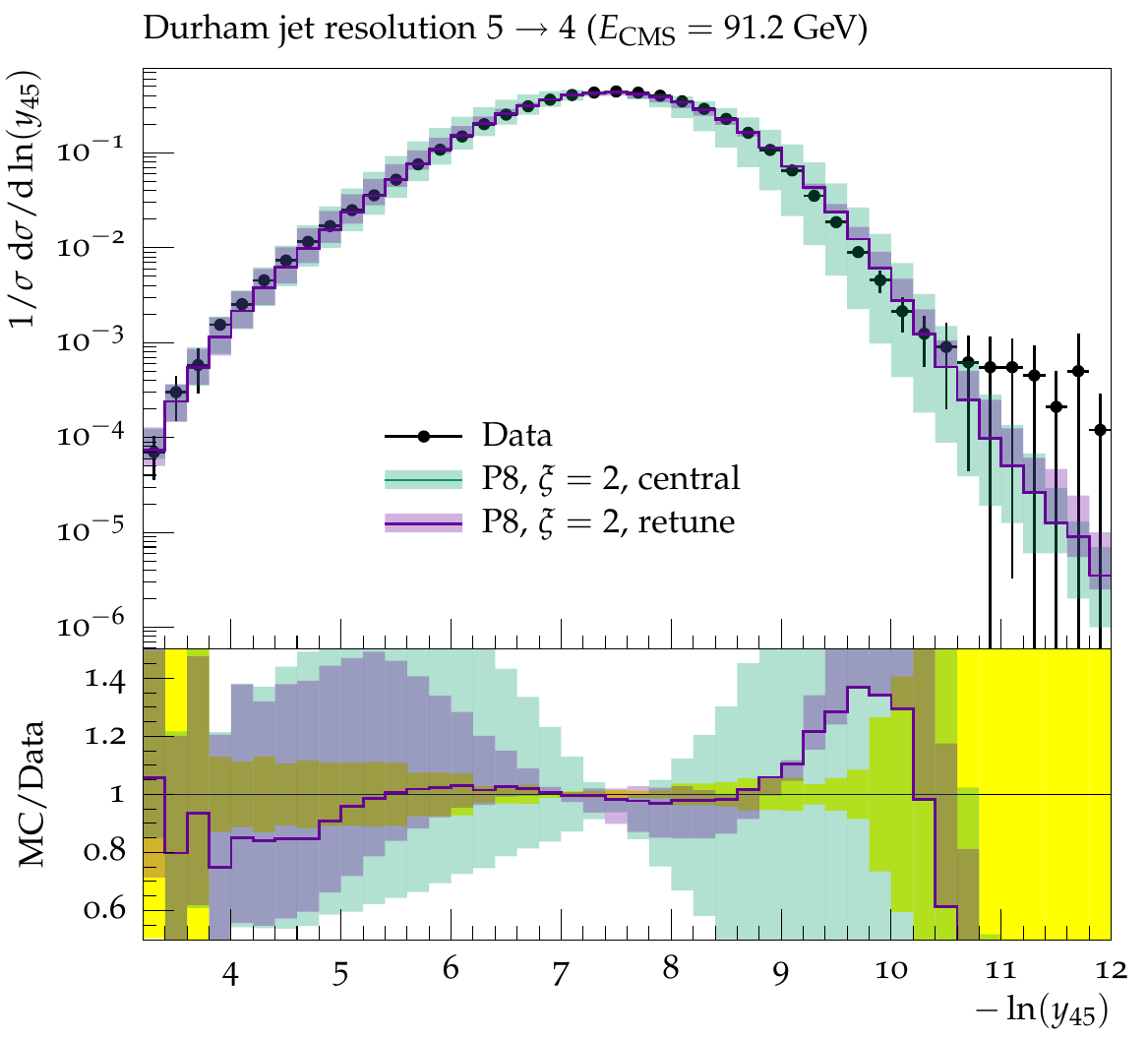}
\caption{The distribution of the Durham jet resolution scale
  $5\rightarrow4$ at LEP for \Herwig (left) and \Pythia (right).}
  \label{fig:MC_tunes_variations:DJR54}
\end{figure}

Turning now to the predictions for observables not tuned to, we first
stay at the $Z^0$ peak and look at the Durham $y_{54}$ distribution in
Fig.~\ref{fig:MC_tunes_variations:DJR54}. Here we see again that the retuned scale
variations are smaller than the central ones. We see the same pattern
as for the $y_{23}$ distribution, that for large $y_{45}$ the scale
variations are not much reduced for \Pythia, while for \Herwig the
reduction is significant. Comparing with the data, one could argue
that the central scale variations overestimate the uncertainties in
the predictions, while the retuned scale variations, where we use data to constrain the fit,
give a more
reasonable estimate of the uncertainties involved. Although in the
region of low $y_{45}$ where the retuned variations could be argued to
be too small.

\begin{figure}[t]
\centering
\includegraphics[width=0.31\textwidth]{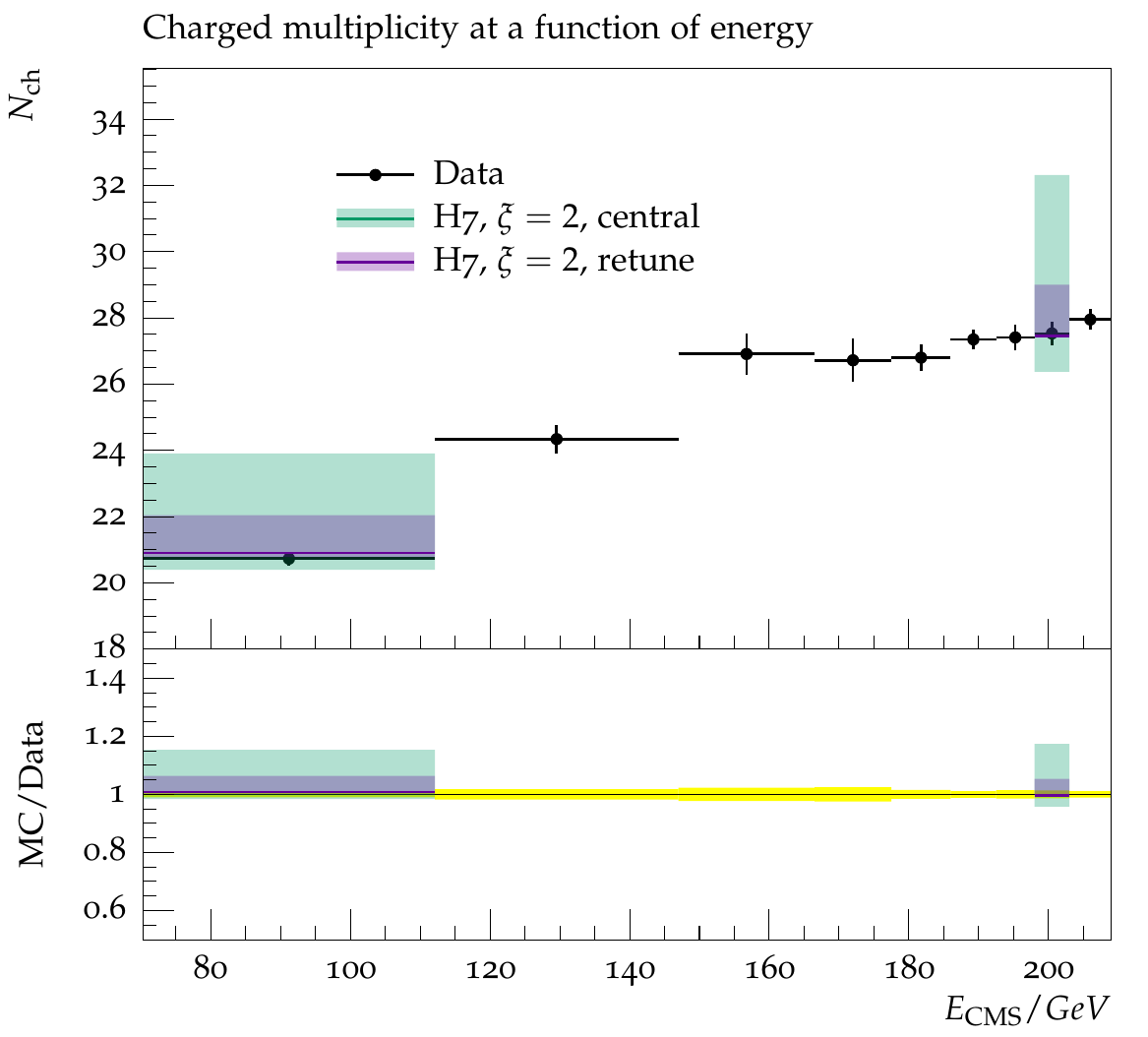}\hfill
\includegraphics[width=0.31\textwidth]{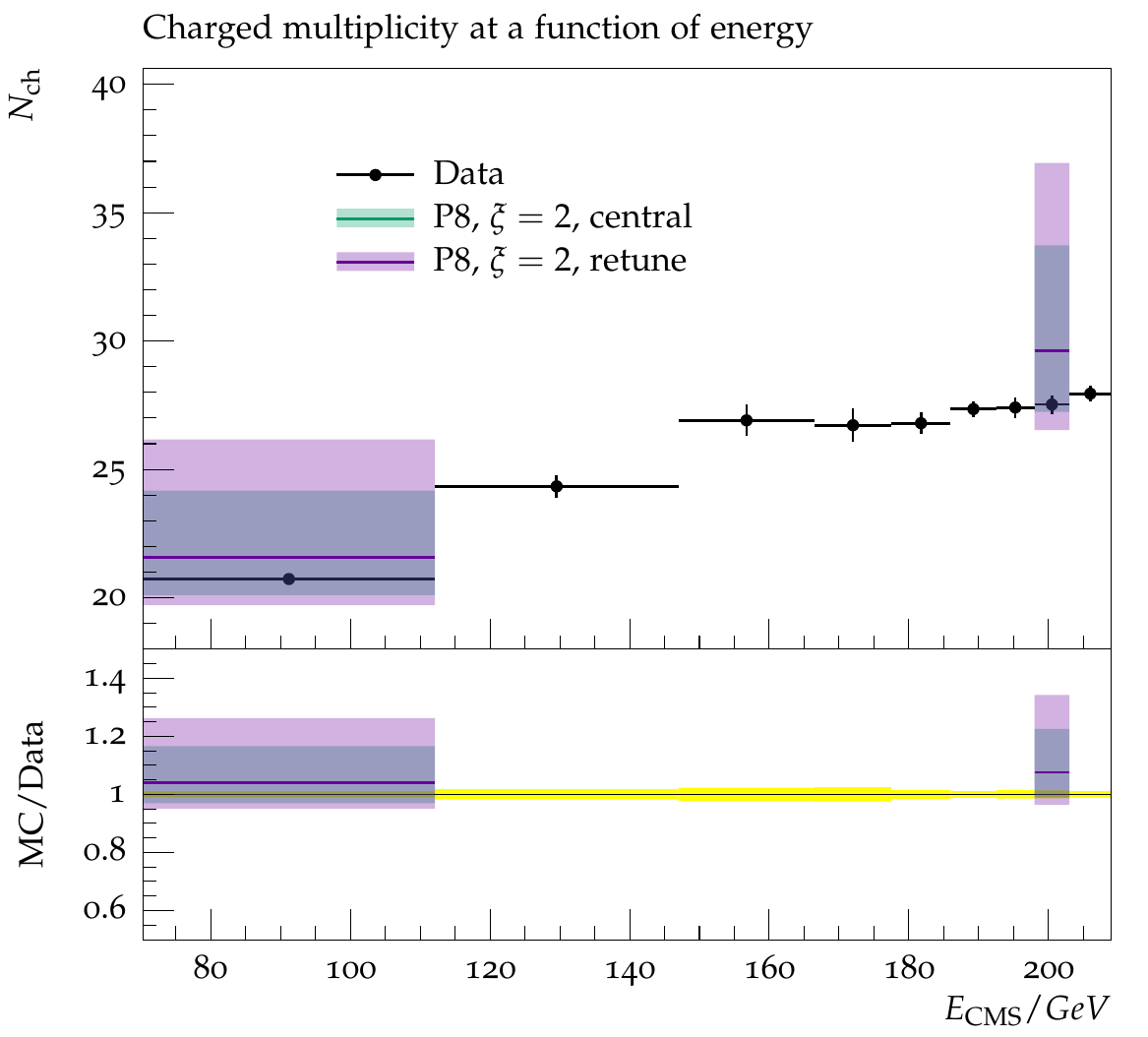}\hfill
\includegraphics[width=0.31\textwidth]{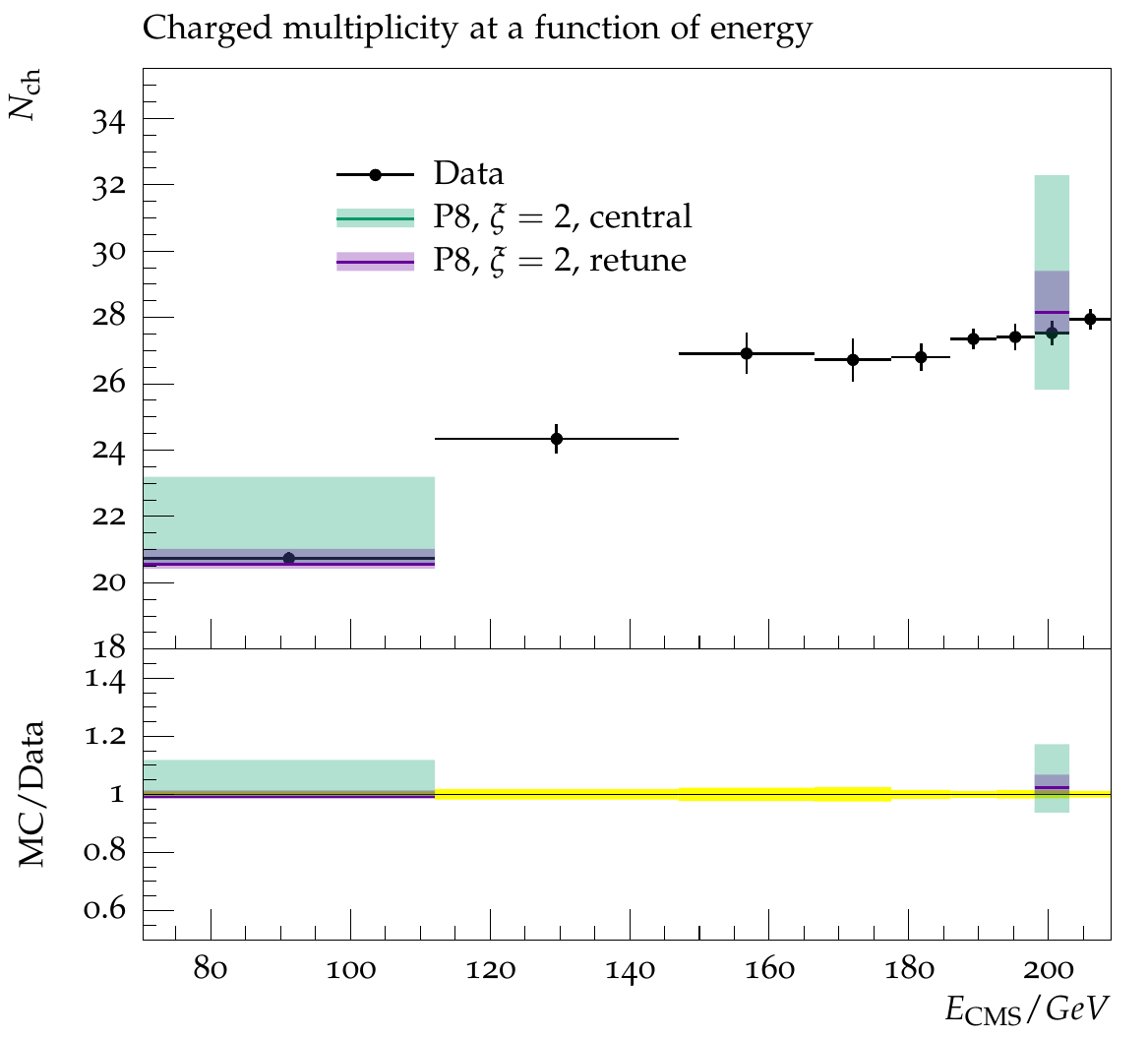}
\caption{Average charged multiplicity as a function of energy. Only
  two energies (91 and 200~GeV) are shown for \Herwig (left) and
  \Pythia (middle). The right plot shows an alternative tuning of
  \Pythia which includes the charged jet multiplicity at 91~GeV from
  L3 \cite{Achard:2004sv}.}
  \label{fig:MC_tunes_variations:Nch}
\end{figure}

Moving on to extrapolations in energy, we first look at the average
charged multiplicity in Fig.~\ref{fig:MC_tunes_variations:Nch}. Here we see an unexpected
behaviour for \Pythia in the middle plot. The scale variations are
actually increased with the tuning. This is because, even though the
91~GeV point is included in the tuning set, it is only one bin and the
tuning procedure has clearly decided to sacrify this bin in order to
fit many others in other distributions. There are indication that one
of the variations in \Herwig leave more space for other parameters to
still adopt to the multiplicity bin, though this issue deserves a more
detailed investigation. In the right plot, we have instead included
the full multiplicity distribution from L3 \cite{Achard:2004sv} in the
tunings and there we get the expected behaviour, similar to that of
\Herwig.

\begin{figure}[t]
\centering
\includegraphics[width=0.48\textwidth]{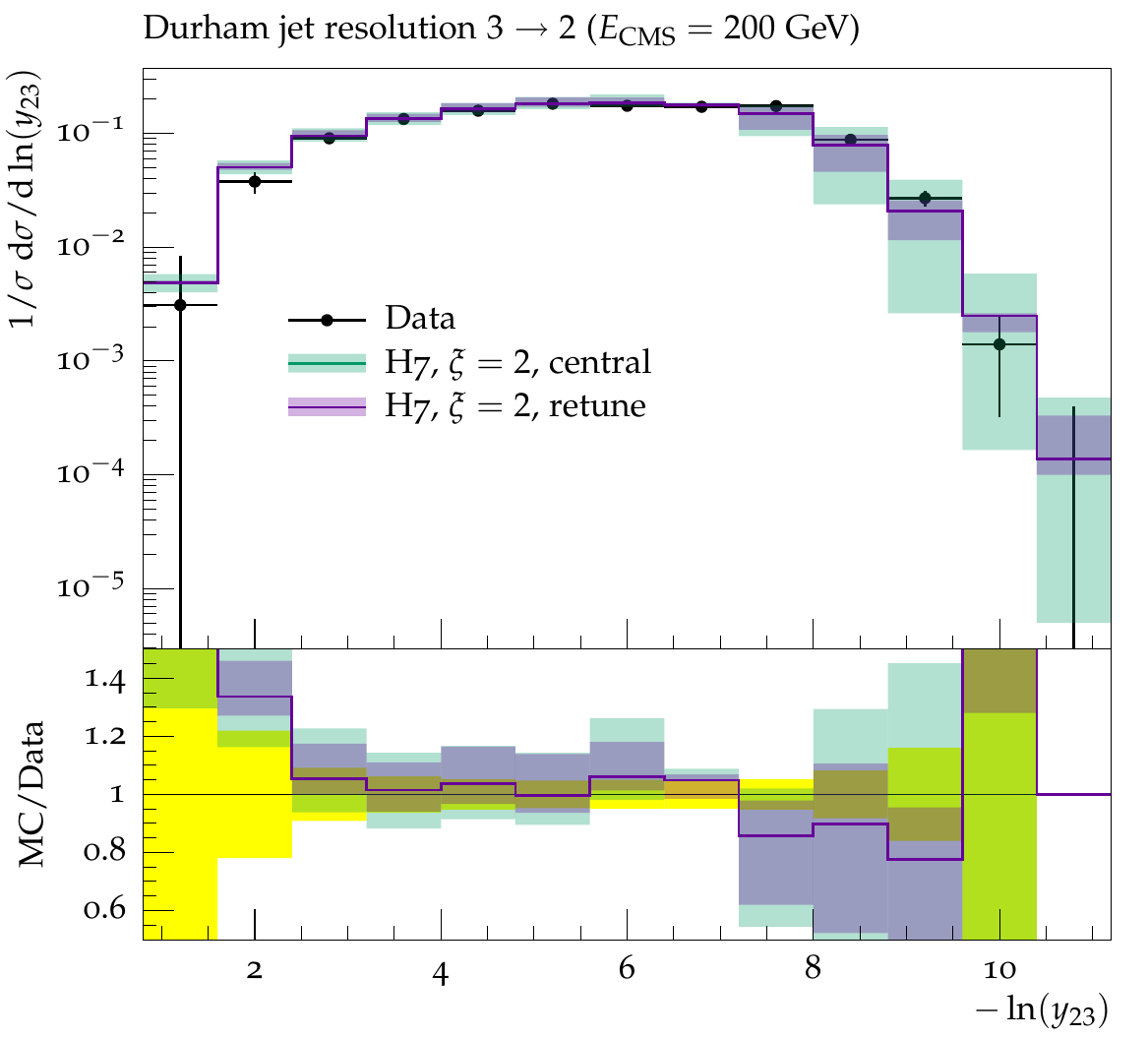}\hfill
\includegraphics[width=0.48\textwidth]{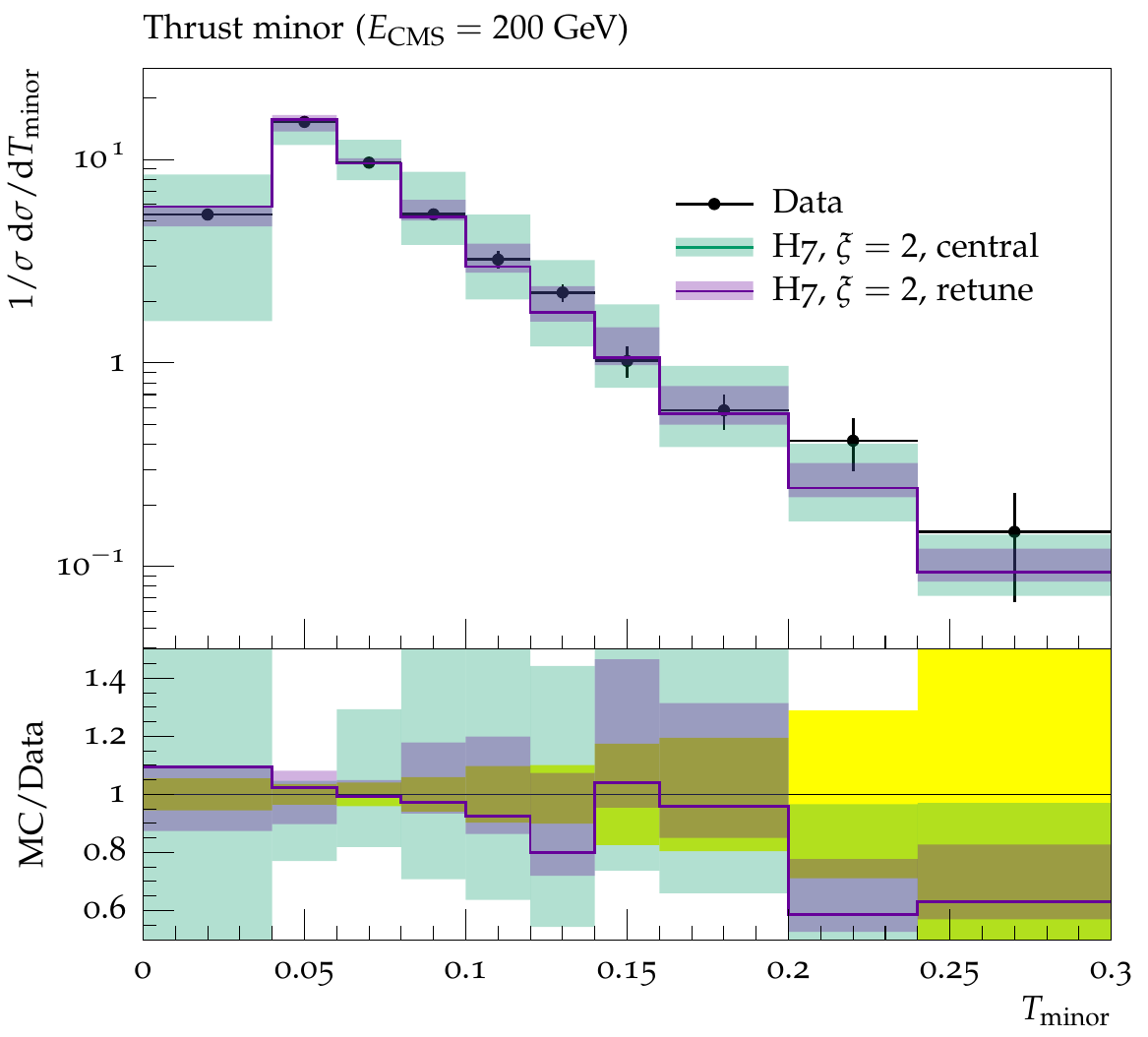}\\
\includegraphics[width=0.48\textwidth]{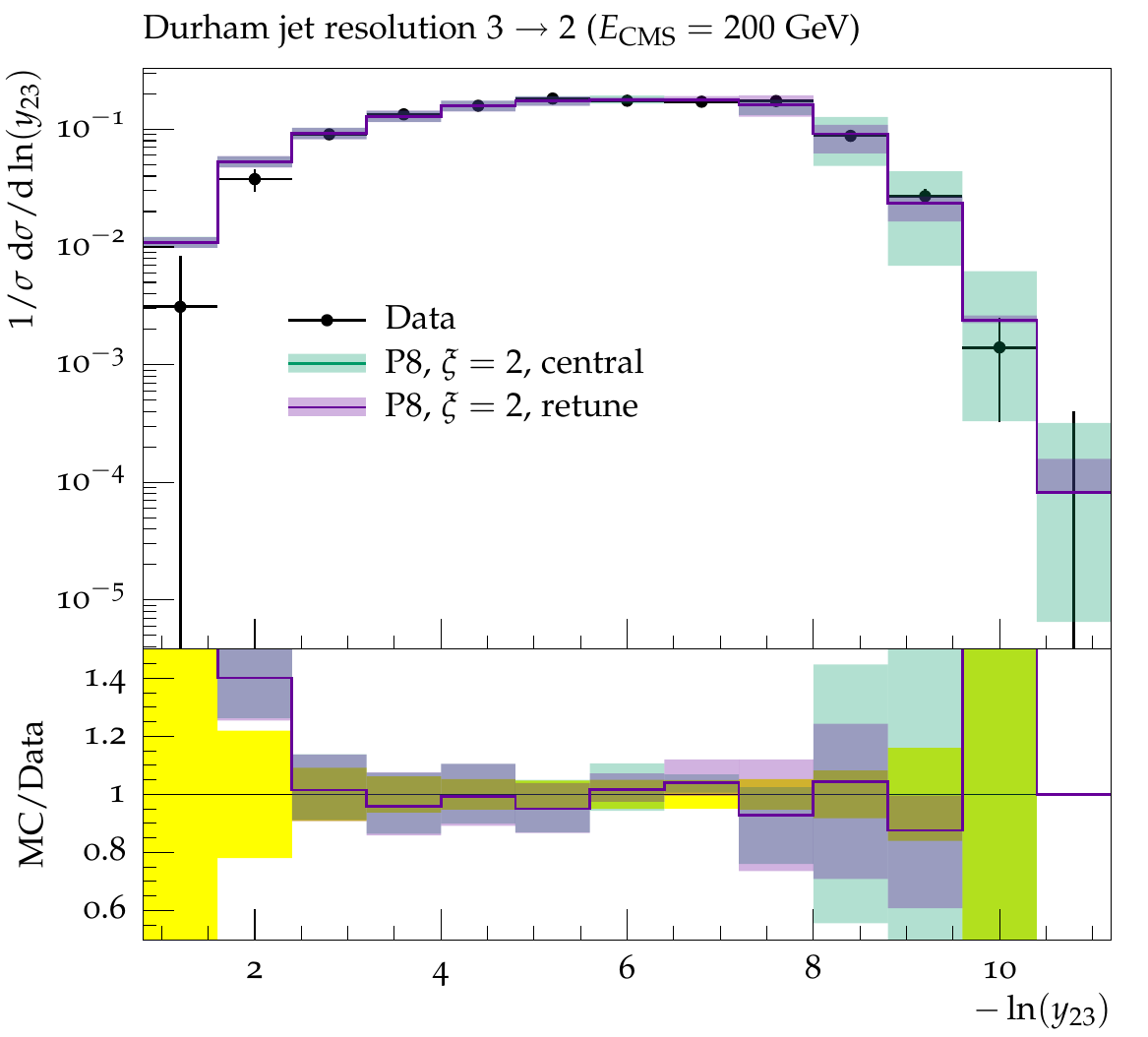}\hfill
\includegraphics[width=0.48\textwidth]{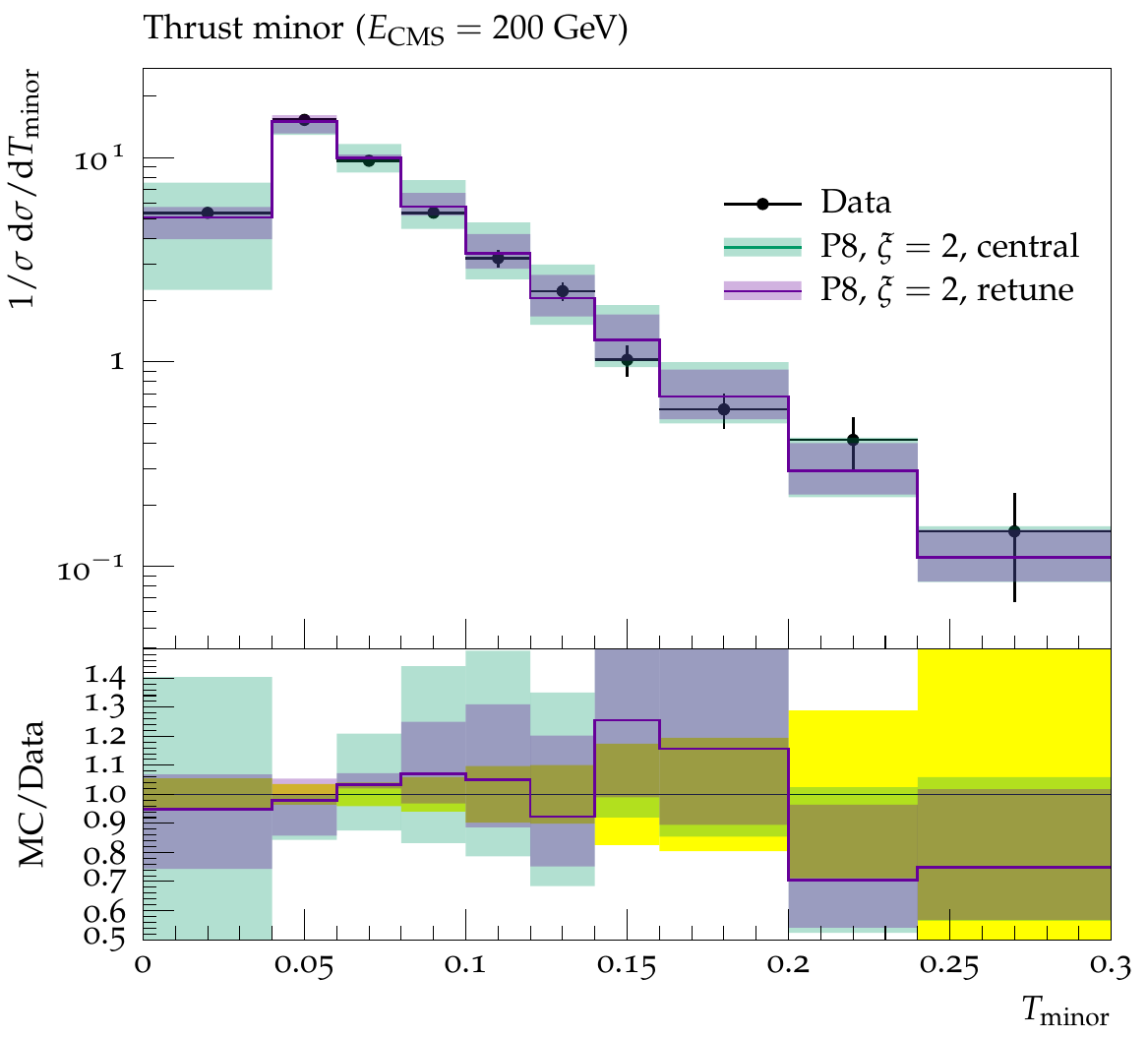}
\caption{Predictiona for 200~GeV from the tunes to ALEPH data at
  91~GeV. Top plots are \Herwig and bottom is \Pythia. Left plots are
  the tune Durham jet resolution $3\rightarrow2$ distribution, and the
  right plots are the Thrust minor distribution.}
  \label{fig:MC_tunes_variations:ThrustMinor}
\end{figure}

Finally we return to $y_{23}$ and $T_{\mr{minor}}$
in Fig.~\ref{fig:MC_tunes_variations:ThrustMinor},
but at the higher
collision energy, where we again see that the retuned scale variations
in the predictions gives a more reasonable estimate of the
uncertainties than the variations using the central tune. Again we see
that for \Pythia the variations are not reduced in the hard part of
the spectra (small $y_{23}$ and large $T_{\mr{minor}}$), while the
retuning in \Herwig is able to compensate for the scale variations.

\subsubsection{Discussion}

We think it is safe to say that using retuned scale variations gives a
better estimate of the uncertainties in the predictions of
shower/hadronisation models. This does however not mean that it gives
a better estimate of the uncertainties in general. It is, in any case,
an interesting exercise to make, as it can increase our understanding
of how the models behave and how the tunings work in more detail. From
our results it is \textit{e.g.}\ clear that not using leading order
matrix element matching in \Herwig can be compensated also in the hard
regions by non-perturbative parameters in the
hadronisation. Similarly, in \Pythia it is possible to compensate some
shortcomings in the description of several event shapes at the expense
of an accurate description of the total multiplicity.

\subsection{PDF fits and parton-shower variations}

In this section, to complement the previous section, we will investigate the 
cross-talk between (collinear) non-perturbative initial-state parton 
distributions and changes of $\alpha_s$ in the (perturbative, initial-state) 
parton shower evolution. We expect correlations due to the presence of
PDF ratios in the evolution kernels used in backward parton-shower 
evolution \cite{Sjostrand:1985xi}. We can illustrate this effect by dissecting the 
parton-shower result for the Drell-Yan lepton pair transverse momentum at a 
hadron collider.
In an initial-state parton shower that is ordered in kinematical $p_\perp$
with respect to the incoming proton beams, the first emission will, schematically,
yield the distribution\footnote{ Note that we only use this as a schematic illustration. For 
parton showers not ordered in (this particular definition of) $p_\perp$, the 
result will be more complicated, but the main message will still apply.
The symbols in Eq.~\eqref{eq:MC_tunes_variations:ptz} are defined by
$d\hat\sigma(pp\rightarrow Z; \Phi_0)$ $\rightarrow$ differential partonic cross section for 
Z-boson production at a ``Born" phase space point $\Phi_0$; 
$f_{p_i}(x_{i}, \mu)$ $\rightarrow$ PDF for parton flavour 
$\widetilde{p_i}$ with momentum fraction $x_{i}$, evaluated at scale $\mu$
with hadron moving in the $i=\pm p_z$ direction;
$\mu_Q$ $\rightarrow$ parton-shower starting or veto scale;
$z_i$ $\rightarrow$ energy fraction carried away from the initial line by 
emission off parton moving in $i$-direction;  
$P_{\widetilde{p_i}\rightarrow{p_i}}$ $\rightarrow$ probability of producing
parton $p_i$ from parton $\widetilde{p_i}$ by a parton-shower branching.
}
\begin{eqnarray}
\label{eq:MC_tunes_variations:ptz}
\frac{d^n\sigma}{d\Phi_0dp_\perp}
&=&
  d\hat\sigma(pp\rightarrow Z; \Phi_0)
  f_{\widetilde{p_+}}(x_{+}, \mu_F)
  f_{\widetilde{p_-}}(x_{-}, \mu_F)\\
&\otimes&
  \left\{
  \int dz_+
  \Pi(\mu_Q,p_\perp)
  \frac{\alpha_s(p_\perp)}{2\pi}
  \frac{f_{\widetilde{p_+}}(\frac{x_+}{z_+}, p_\perp)}{f_{\widetilde{p_+}}(x_{+}, p_\perp)}
  \frac{f_{\widetilde{p_-}}(x_{-}, p_\perp)}{f_{\widetilde{p_-}}(x_{-}, p_\perp)}
  P_{\widetilde{p_+}\rightarrow{p_+}}(z_+,p_\perp)
  + (-\leftrightarrow+)\nonumber
\right\}
\end{eqnarray}
with the parton-shower no-emission probability
\begin{eqnarray}
\Pi(\mu_Q,p_\perp)=&  \\
& \exp 
  \left\{
  - \int^{\mu_Q}_{p_\perp} d \bar{p}_\perp
  \int d\bar{z}_+
  \frac{\alpha_s(p_\perp)}{2\pi}
  \frac{f_{\widetilde{p_+}}(\frac{x_+}{\bar{z}_+}, \bar{p}_\perp)}{f_{\widetilde{p_+}}(x_{+}, \bar{p}_\perp)}
  \frac{f_{\widetilde{p_-}}(x_{-}, \bar{p}_\perp)}{f_{\widetilde{p_-}}(x_{-}, \bar{p}_\perp)}
  P_{\widetilde{p_+}\rightarrow{p_+}}(\bar{z}_+,\bar{p}_\perp)
  + (-\leftrightarrow+)
\right\}\nonumber
\end{eqnarray}
and where, to ease notation, the symbol $P(z,p_\perp)$ collects splitting 
kernels, propagator- and Jacobian factors. Assuming that the parton shower
is a faithful implementation of DGLAP evolution of $f_{\widetilde{p_+}}$
(and $f_{\widetilde{p_-}}$) from $\mu_Q=\mu_F$ to $p_\perp$, we can write
\begin{eqnarray}
\label{eq:MC_tunes_variations:PiVDelta}
\Pi(\mu_Q,p_\perp) 
&=&
\frac{f_{\widetilde{p_+}}({x_+}, {p}_\perp)}{f_{\widetilde{p_+}}(x_{+}, \mu_Q)}
\frac{f_{\widetilde{p_-}}({x_-}, {p}_\perp)}{f_{\widetilde{p_-}}(x_{-}, \mu_Q)}
\Delta(\mu_Q,p_\perp)\quad \textnormal{where}\\
\Delta(\mu_Q,p_\perp) 
&=& \exp 
  \left\{
  - \int^{\mu_Q}_{p_\perp} d \bar{p}_\perp
  \int d\bar{z}_+
  \frac{\alpha_s(p_\perp)}{2\pi}
  P_{\widetilde{p_+}\rightarrow{p_+}}(\bar{z}_+,\bar{p}_\perp)
  ~~+~~ (-\leftrightarrow+)
\right\}\nonumber
\end{eqnarray}
which leads to a result reminiscent of analytic resummation\footnote{See e.g.~\cite{Catani:2013tia,Becher:2010tm}
for overviews of analytic resummation of color-singlet boson $p_\perp$ spectra.}
\begin{eqnarray}
\frac{d^n\sigma}{d\Phi_0dp_\perp}
&=&
  d\hat\sigma(pp\rightarrow Z; \Phi_0)
  f_{\widetilde{p_+}}(x_{+}, {p}_\perp)
  f_{\widetilde{p_-}}(x_{-}, {p}_\perp)\\
 & &\left\{
  \int dz_+
  \Delta(\mu_Q,p_\perp)
  \frac{\alpha_s(p_\perp)}{2\pi}
  P_{\widetilde{p_+}\rightarrow{p_+}}(z_+,p_\perp)
  ~+~ (-\leftrightarrow+)
\right\}\qquad \nonumber
\end{eqnarray}
It is important to stress again that this result assumes that the parton shower
is a faithful implementation of DGLAP evolution of $f_{\widetilde{p}}$, which 
in particular suggests that the parton-shower evolution should
be adjusted for different PDF sets $f^\prime_{\widetilde{p}}$.
Conversely, it is sensible to assume that particular parton shower settings 
should correspond to a particular choice of PDF. 

In the following, we will investigate the cross-talk between the choice of
$\alpha_s$ in the parton shower and (input) PDF sets.
To prevent bias, we will investigate the effect of correlations
on Higgs-boson and Drell-Yan lepton pair $p_\perp$ distributions at two different 
energies at a $pp$ collider. 
The uncertainties on boson transverse momentum distributions
often contribute heavily to background modelling uncertainties in 
experimental new physics searches at the LHC. MCEGs are often used to 
study and model $p_\perp$ distributions. We expect that significant
changes in the uncertainty estimates due to the inclusion of correlations 
will provide useful information for the experimental communities.
We will further check our findings on the complementary observable
$\langle N_{jets}\rangle$, which should be very sensitive to the details
of the parton shower, but only mildly PDF dependent.


\subsubsection{Methodology}
\label{sec:MC_tunes_variations:pdf:methodology}

\begin{figure}[t]
\centering
\includegraphics[width=0.48\textwidth]{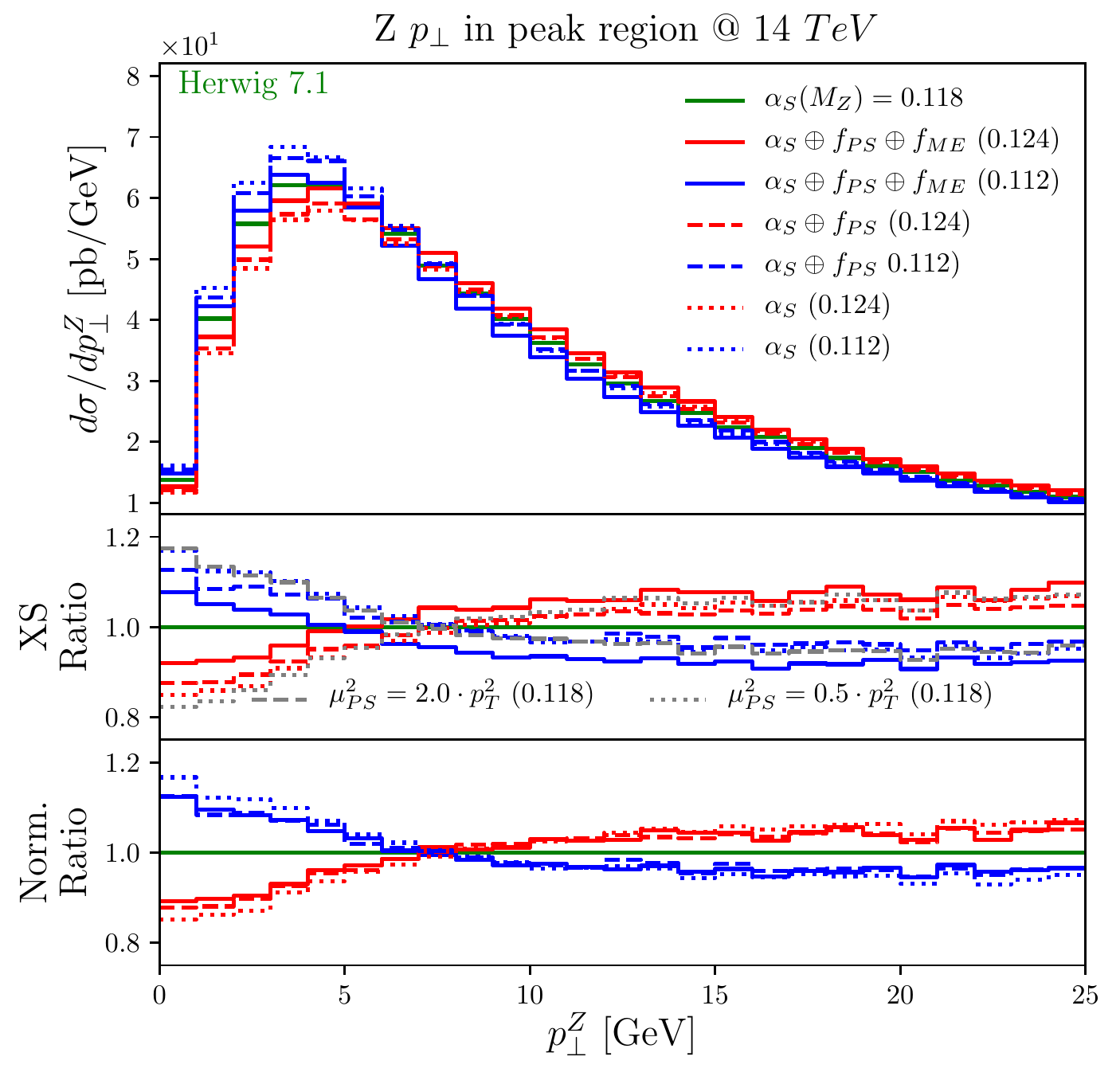}\hfill
\includegraphics[width=0.48\textwidth]{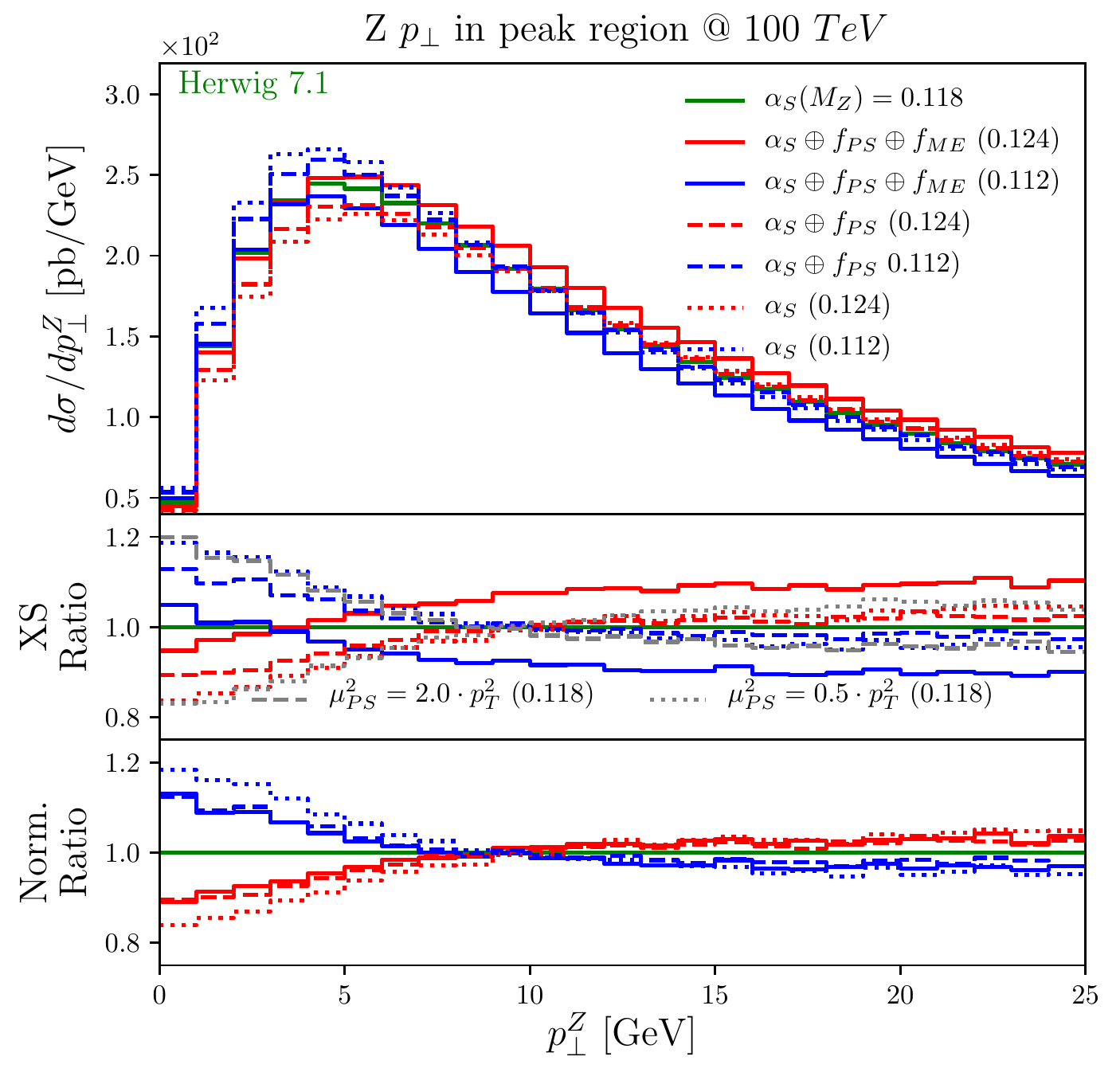}\\
\includegraphics[width=0.48\textwidth]{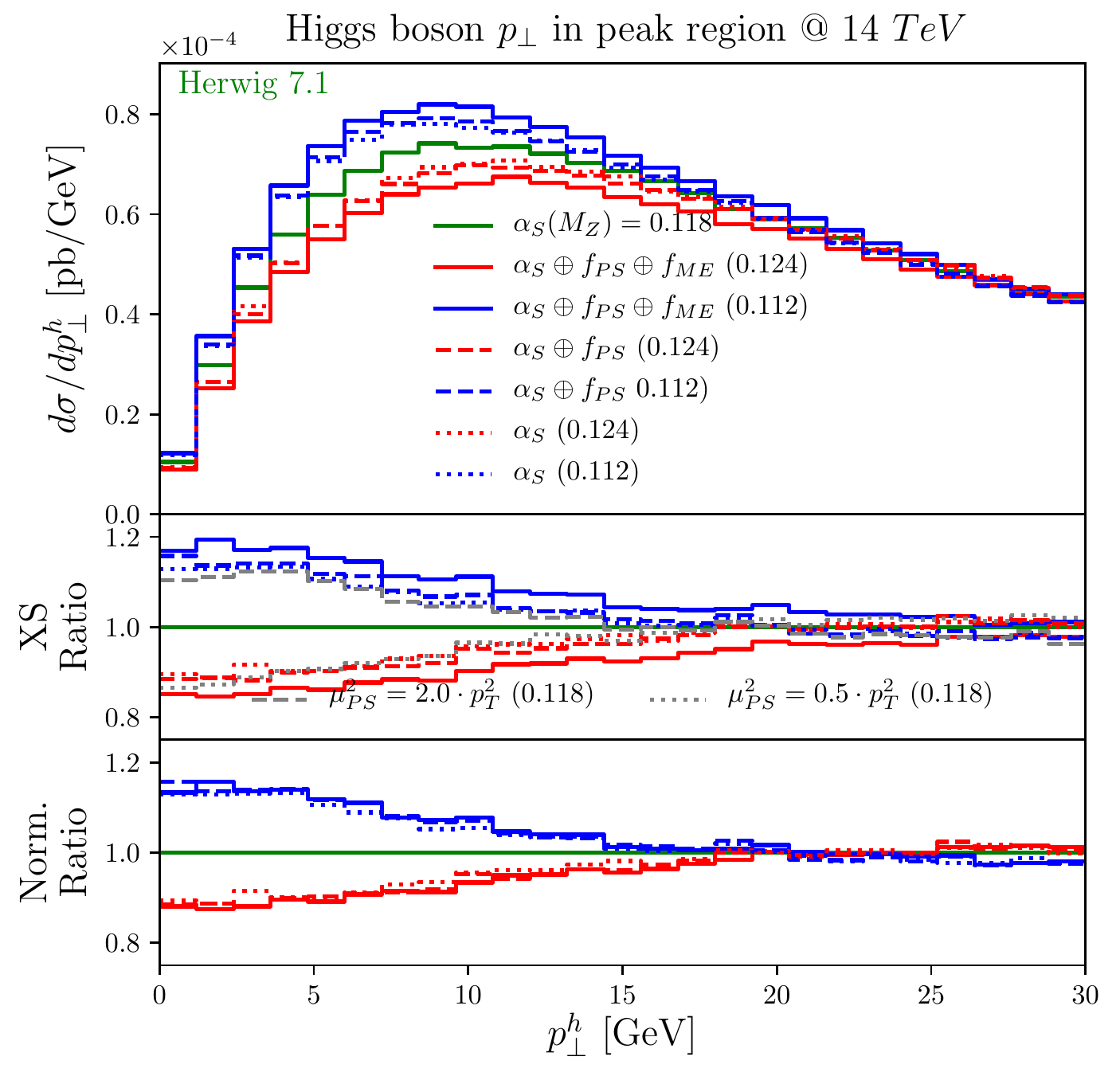}\hfill
\includegraphics[width=0.48\textwidth]{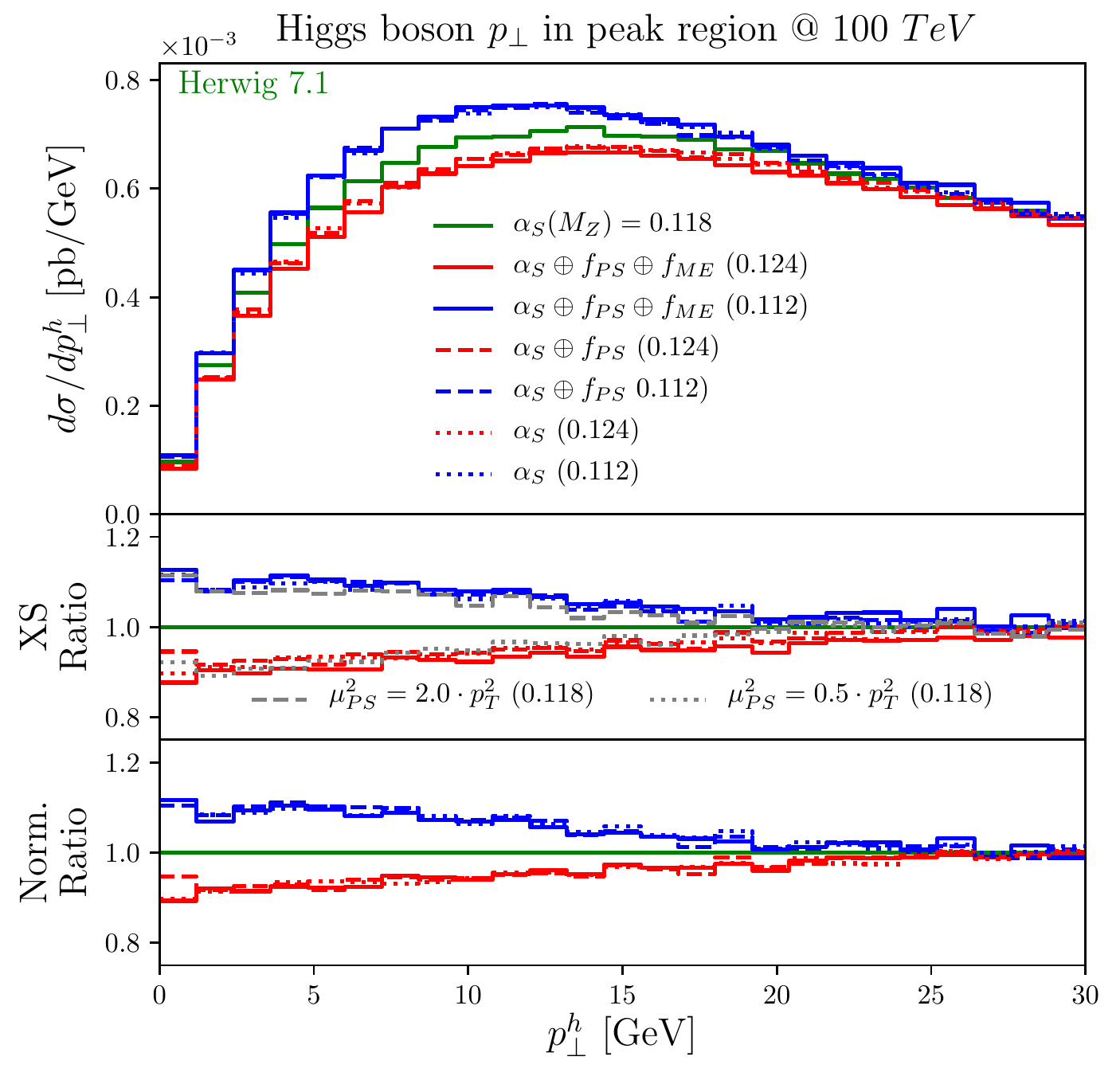}
\caption{Variation according to Table~\ref{tab:MC_tunes_variations:pdf-vars} for the bosons transverse 
momentum in  Z boson (top) and Higgs (bottom) production at 14 (left) and 
100 TeV(right). While the upper ratio plot shows the ratio to the baseline for differential cross section, the lower ratio plot illustrates the ratio w.r.t. normalized distributions. In the first ratio plot we add the lines for pure scale variations in the shower $\alpha_S$ to display the similarity to value variations. See Sec.~\ref{sec:MC_tunes_variations:pdf:discussion} for discussion.}
  \label{fig:MC_tunes_variations:pt:ZHvar}
\end{figure}

We will assess the correlations by investigating how $\alpha_s(M_Z)$ choices
in the parton shower and PDF choices interact. For this, we first
define the ``baseline" $\alpha^{(c)}_s(M_Z)=0.118$ and 
produce parton-shower results with $\alpha^{(c)}_s(k\cdot t)$ with 
$k\in\{\frac{1}{\sqrt{2}},1,\sqrt{2}\}$. We have verified that $a)$ the result of 
$\alpha^{(c)}_s(\frac{1}{\sqrt{2}} t)$ is numerically equivalent to using 
$\alpha^{\downarrow}_s (t)$ with 
$\alpha^{\downarrow}_s (M_Z)=\alpha^{(c)}_s(\frac{1}{\sqrt{2}} M_Z)\approx 0.124$, and
that $b)$ the result of $\alpha^{(c)}_s(\sqrt{2} t)$ is numerically equivalent to using 
$\alpha^{\uparrow}_s (t)$ with 
$\alpha^{\uparrow}_s (M_Z)=\alpha^{(c)}_s(\sqrt{2} M_Z)\approx 0.112$, see gray lines in second ratio plots of Fig.~\ref{fig:MC_tunes_variations:pt:ZHvar}.

This allows to reinterpret parton-shower renormalization scale variations
as variations of $\alpha_s(M_Z)$. For judiciously assigned values of
$\alpha_s(M_Z)$, we can find PDF sets that used this $\alpha_s(M_Z)$ when 
performing the fit. Such sets using a predetermined 
$\alpha_s$ have recently become supplied by several PDF fitting groups.
With these prerequisites, we compare the results with the settings
given in Table~\ref{tab:MC_tunes_variations:pdf-vars}.

\begin{table}
  \renewcommand{\arraystretch}{1.1}
  \centering
\setlength{\tabcolsep}{2.mm}
  \begin{tabular}{ | p{1.8cm} | p{4.8cm} | p{3.8cm} | p{3.8cm} |}
    \hline\hline
    {Name}               & Description & $\alpha_s$ and PDF set                           & Legend in plots \\ \hline
    {Baseline}:        &  Result of showering 
    using only default settings and $\alpha_{s}^{(c)}(M_Z)$.   & $\alpha_s(M_Z)=0.118$, baseline PDF set, e.g.~CT10nlo\_as\_0118  & $\alpha_s(M_Z)=0.118$ and green solid line\\ \hline
    {$\alpha_s^{\downarrow}$}  variation  & Result of 
    changing $\alpha_{s}^{(c)}(M_Z)\rightarrow\alpha_{s}^{(c)}(\frac{1}{\sqrt{2}}M_Z) $ when showering. & $\alpha_s(M_Z)=0.124$, baseline PDF set, e.g.~CT10nlo\_as\_0118  & $\alpha_s~(0.124)$ and red dotted line\\ \hline
    {$\alpha_s^{\uparrow}$} variation   & Result of 
    changing $\alpha_{s}^{(c)}(M_Z)\rightarrow\alpha_{s}^{(c)}(\sqrt{2}M_Z) $ when showering. & $\alpha_s(M_Z)=0.112$, baseline PDF set, e.g.~CT10nlo\_as\_0118  & $\alpha_s~(0.112)$ and blue dotted line\\ \hline

    Correlated PS variation &  Result of 
    simultaneously changing $\alpha_{s}^{(c)}(M_Z)\rightarrow\alpha_{s}^{(c)}(\frac{1}{\sqrt{2}}M_Z)$
    and PDF set used for showering.  & $\alpha_s(M_Z)=0.124$, PDF fitted with $\alpha_s(M_Z)=0.124$, e.g.~CT10nlo\_as\_0124  & $\alpha_s\oplus f_{PS}~(0.124)$ and red dashed line\\ \hline
    Correlated PS variation  & Result of 
    simultaneously changing $\alpha_{s}^{(c)}(M_Z)\rightarrow\alpha_{s}^{(c)}(\sqrt{2}M_Z)$
    and PDF set used for showering.  & $\alpha_s(M_Z)=0.112$, PDF fitted with $\alpha_s(M_Z)=0.112$, e.g.~CT10nlo\_as\_0112  & $\alpha_s\oplus f_{PS}~(0.112)$ and blue dashed line\\ \hline

    Correlated PS+ME variation & Result of 
    simultaneously changing $\alpha_{s}^{(c)}(M_Z)\rightarrow\alpha_{s}^{(c)}(\frac{1}{\sqrt{2}}M_Z)$,
    and the PDF set used for showering and in the calculation of the hard scattering cross section. & $\alpha_s(M_Z)=0.124$, PDF fitted with $\alpha_s(M_Z)=0.124$, e.g.~CT10nlo\_as\_0124  & $\alpha_s\oplus f_{PS}\oplus f_{ME}~(0.124)$ and red solid line\\ \hline
    Correlated PS+ME variation  & Result of 
    simultaneously changing $\alpha_{s}^{(c)}(M_Z)\rightarrow\alpha_{s}^{(c)}(\sqrt{2}M_Z)$
    and the PDF set used for showering and in the calculation of the hard scattering cross section. & $\alpha_s(M_Z)=0.112$, PDF fitted with $\alpha_s(M_Z)=0.112$, e.g.~CT10nlo\_as\_0112  & $\alpha_s\oplus f_{PS}\oplus f_{ME}~(0.112)$ and blue solid line\\\hline
  \end{tabular}
\caption{\label{tab:MC_tunes_variations:pdf-vars}List of settings to assess the correlations between PDFs and $\alpha_s$ choices. In the text, we will refer to the envelope of
{$\alpha_s^{\downarrow}$} and {$\alpha_s^{\uparrow}$} variation as ``(uncorrelated) $\alpha_s$ variation", refer to the combiation of the third and fourth items as ``correlated PS variation", and refer to the combination of the fifth and sixth curve as ``correlated PS+ME" variation.}
\end{table}


As discussed above, we will focus the discussion on transverse momentum 
spectra. To investigate the effects of correlating 
$\alpha_s$ and PDF further, we will check if changes in the $p_\perp$ 
distribution also translate to the average number of jets 
$\langle N_{jets}\rangle$. The number of jets $N_{jets}$ produced through 
parton showering is, by virtue of 
Eq.~\eqref{eq:MC_tunes_variations:PiVDelta}, only sensitive to the value of the PDF at the 
parton-shower cut-off. $\langle N_{jets}\rangle$ will thus only have a mild
PDF dependence, but showcase the cross-talk between PDFs and $\alpha_s$.

\subsubsection{Simulation Setup}

We use the default leading-order \Herwig setup for the simulations in this 
study. This includes matrix element corrections for the first jet in Higgs-
and Z-boson production. As the baseline we use the central value of 
$\alpha_s = 0.118$ by setting\\

\verb|set /Herwig/Shower/AlphaQCD:AlphaMZ 0.118;|.\\

\noindent
As PDF set, we use \texttt{CT10nlo\_as\_0118}, as implemented in  
LHAPDF\cite{Buckley:2014ana}. Since our goal is to study the 
compensation effect of fitting in the evolution of parton showers,
we limit most of the following to the ME+PS level.
To turn off hadronisation and multiple parton interaction we set\\

\verb|read Matchbox/PQCDLevel.in;|.\\

\noindent
If desired we will further, as described in Table~\ref{tab:MC_tunes_variations:pdf-vars}, correlate
PDF set and $\alpha_s$-values by using $\alpha_s^{\uparrow} (M_Z)=
\alpha^{(c)}_s(\sqrt{2} M_Z)\approx 0.112$ together with 
\texttt{CT10nlo\_as\_0112} PDF sets, and $\alpha_s^{\downarrow}(M_Z)=
\alpha^{(c)}_s(\frac{1}{\sqrt{2}} M_Z)\approx 0.124$ together with 
\texttt{CT10nlo\_as\_0124} PDF sets.

\subsubsection{Results}
\label{sec:MC_tunes_variations:pdf:results}

The effect of correlated $\alpha_s$ + PDF variation and uncorrelated 
$\alpha_s$ variation on color-singlet boson $p_\perp$ spectra is
illustrated in Fig.~\ref{fig:MC_tunes_variations:pt:ZHvar}. The top ratio plots give the ratio 
to the baseline configuration on cross section level, while the lower ratios 
are taken by dividing the corresponding normalised distributions.
For both for Higgs and Drell-Yan $p_\perp$ spectra at the cross-section
level (top ratio), the difference between correlated PS variation and 
uncorrelated $\alpha_s$ variation is smaller than the difference correlated PS+ME variation 
and uncorrelated $\alpha_s$ variation.
The normalized differential distributions, however, show that the difference 
in correlated PS and PS+ME variations can be traced back to variations in the 
overall cross-section, which is induced by the difference in the PDFs 
$f_{ME}$ used in the hard cross section calculation.
It is important to note that the difference between uncorrelated 
$\alpha_s$ variation and correlated PS variations remains visible at the level
of normalized spectra, i.e.~leads to $p_\perp$-dependent shape changes:
The correlated PS variation leads to a squeezed band w.r.t. uncorrelated 
$\alpha_s$ variation for the Drell-Yan $p_\perp$, and has little effect on the
$p_\perp$ of the Higgs boson.
 
\begin{figure}[t]
\centering
\includegraphics[width=0.48\textwidth]{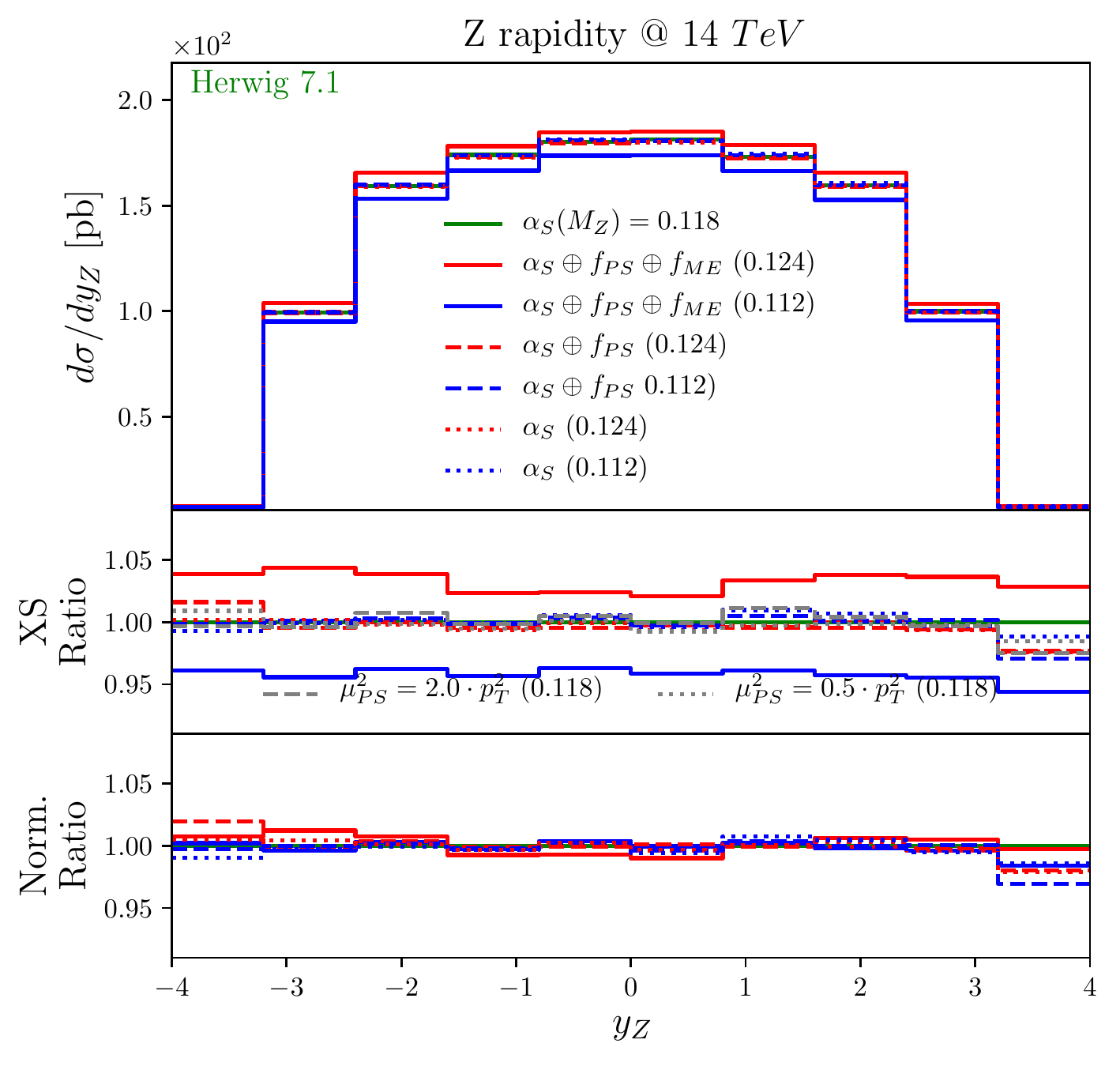}\hfill
\includegraphics[width=0.48\textwidth]{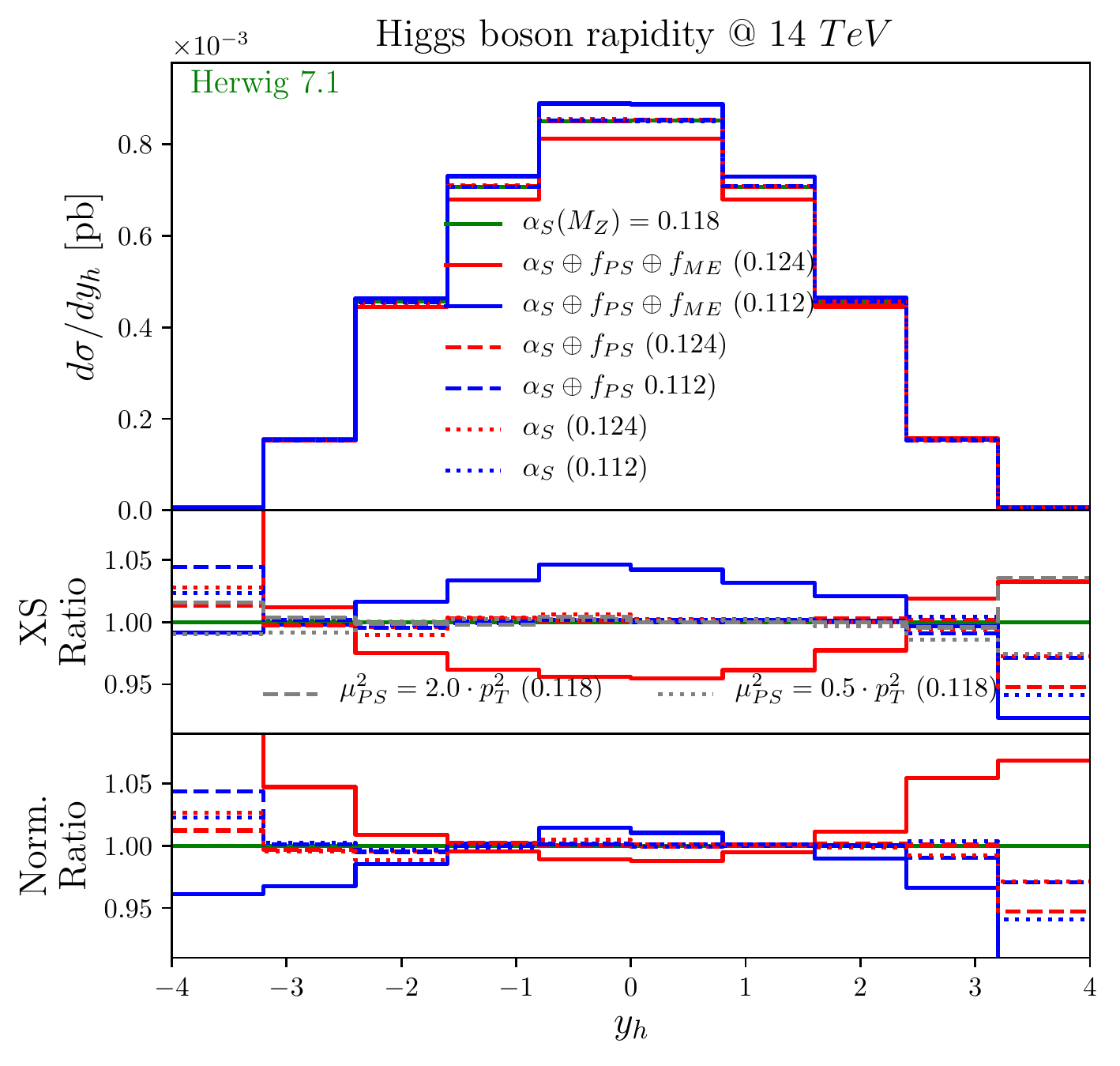}
\caption{Rapidity distributions for H-boson and Drell-Yan lepton pair production at
leading-order, for 14 TeV LHC.}
  \label{fig:MC_tunes_variations:y:ZHvar}
\end{figure}

\begin{figure}[t]
\centering
\includegraphics[width=0.7\textwidth]{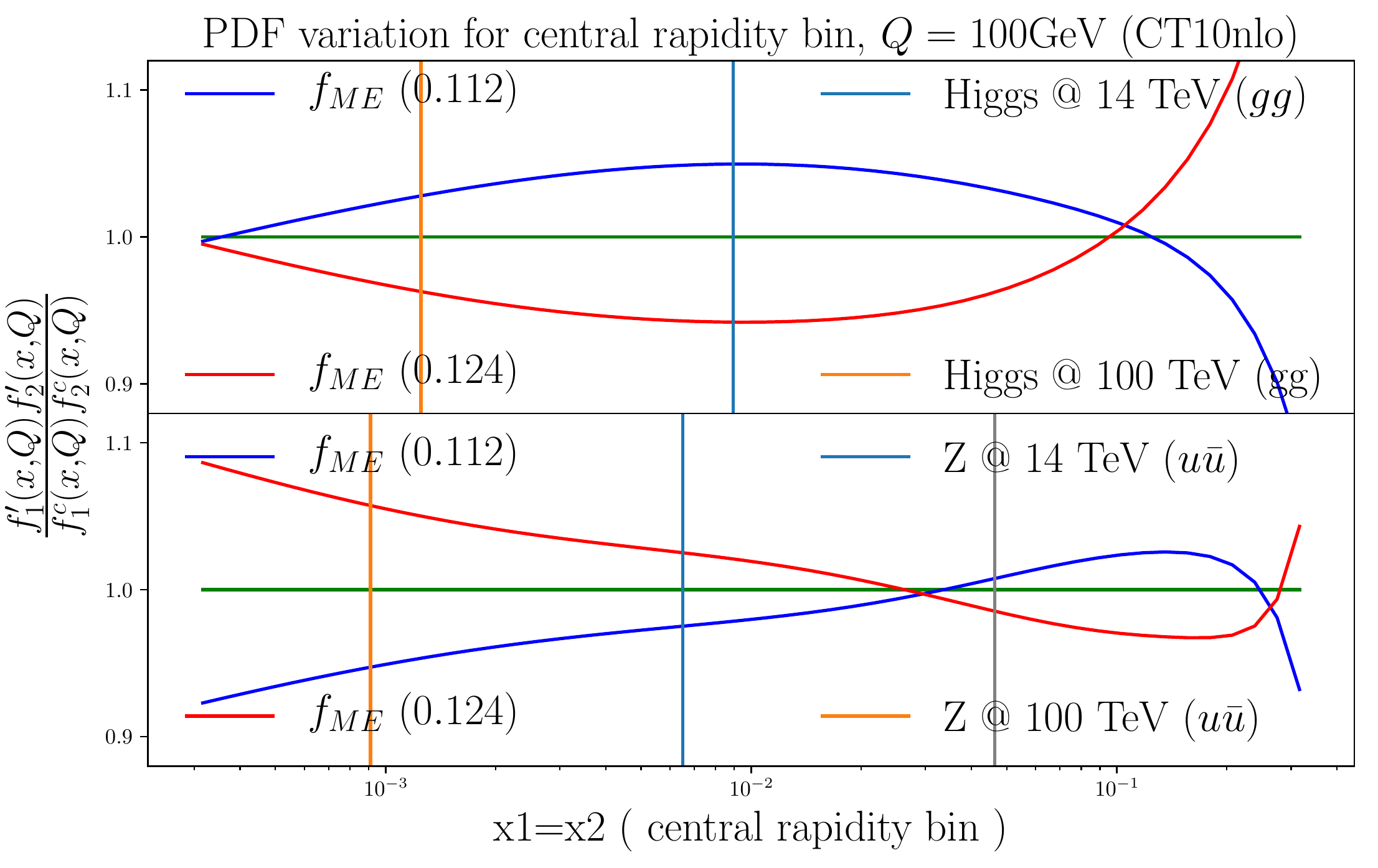}
\caption{Relative change \texttt{CT10nlo\_as\_0112} and 
\texttt{CT10nlo\_as\_0124} w.r.t.\ of \texttt{CT10nlo\_as\_0118}, as function of $x$.}
  \label{fig:MC_tunes_variations:x:pdfvar}
\end{figure}

The larger differences at the cross-section level can be understood by 
considering the rapidity distributions in Fig.~\ref{fig:MC_tunes_variations:y:ZHvar}, and the 
PDF changes in Fig.~\ref{fig:MC_tunes_variations:x:pdfvar}. 
As expected, the $\alpha_S$ variation and the correlated PS variation lead to a 
vanishing uncertainty band\footnote{We do not vary the $\alpha_S$ in the effective $ggH$-coupling.}, whereas the correlated PS+ME variations yield a 
band.
The normalization uncertainty of rapidity distributions in Fig.~\ref{fig:MC_tunes_variations:y:ZHvar} is more uniform for 
the Drell-Yan process. At the same time, the uncertainty for Higgs production
is very small at large rapidity, i.e.\ in the region that will most likely
lead to a significant contribution to the parton-shower radiation pattern and
thus the Higgs $p_\perp$ spectrum. This property then leads to smaller shape 
changes in the Higgs $p_\perp$ spectrum, compared to the Drell-Yan $p_\perp$
distribution - for Drell-Yan leptons, the correlated PS curves will uniformly
be rescaled by a O(5\%) cross section differences, thus changing the 
apparent ``shape" of the variation band, while for Higgs, the effect is 
smaller. Changes in the collider energy enhance this effect further, as can be
seen in the second column of Fig.~\ref{fig:MC_tunes_variations:pt:ZHvar}. 
In addition, for the typical $x$-values for different collider energies,
gluon and quark PDF uncertainty behave differently. 
This is indicated in Fig.~\ref{fig:MC_tunes_variations:x:pdfvar} for \texttt{CT10nlo} sets by 
plotting the PDF ratio
\begin{equation}
\frac{f'_1(x,Q)f'_2(x,Q)}{f^c_1(x,Q)f^c_2(x,Q)}\; \; \text{ and } Q=100 GeV\; ,
\end{equation}
for a configuration $x=x_1=x_2$ corresponding to a central rapidity bin in
Fig.~\ref{fig:MC_tunes_variations:y:ZHvar}. The vertical lines represent the values at 
$x=M_{Z/H}/\sqrt{S}$ for $\sqrt{S}=14$ and $100$ TeV
and in gray the central rapidity bin for Z-production at Tevatrons Run II 
energy $\sqrt{S}=1.96$ TeV. The spread in the PDF values for
quark-initiated Z-boson production grows with collider energy, while it 
diminishes for gluon-initiate Higgs-boson production. This explains why
the effect of PDF changes on the Z-boson $p_\perp$ grow with collider energy,
while they become smaller in the Higgs-boson $p_\perp$ distribution.
Note that, as indicated by the grey line, the uncertainty on the quark PDFs is
small at Tevatron energies, since the PDF fit includes Tevatron data. Also, 
note that had we investigated compensation effects in Drell-Yan $p_\perp$
spectra below Tevatron energies, the distributions would have painted a picture
similar to what we now observe in the Higgs $p_\perp$ case.

The normalized $p_\perp$ distributions expose that the difference between 
correlated PS variation and correlated PS+ME variation is in fact minimal.
The difference between correlated variation and uncorrelated $\alpha_{s, PS}$
is most visible in the DY process, where the band shrinks. As desired, the
inclusion of correlations reduces the overall shape uncertainty. No such
effect is visible for Higgs. This effect can again be traced to the different
PDF sets dominating the color-singlet boson production cross sections. 

We now move on to the question if the effect in normalized distributions is
robust to changing the observable, and when taking non-perturbative and
soft-physics phenomena into consideration. The average number of jets defined 
in Sec.~\ref{sec:MC_tunes_variations:pdf:methodology}, binned in rapidity provides a complementary
observable that is both very sensitive to the parton-shower evolution
and to multiple parton interactions (MPI), while at the same time insensitive
to the overall normalization.

\begin{figure}[t]
\centering
\includegraphics[width=0.47\textwidth]{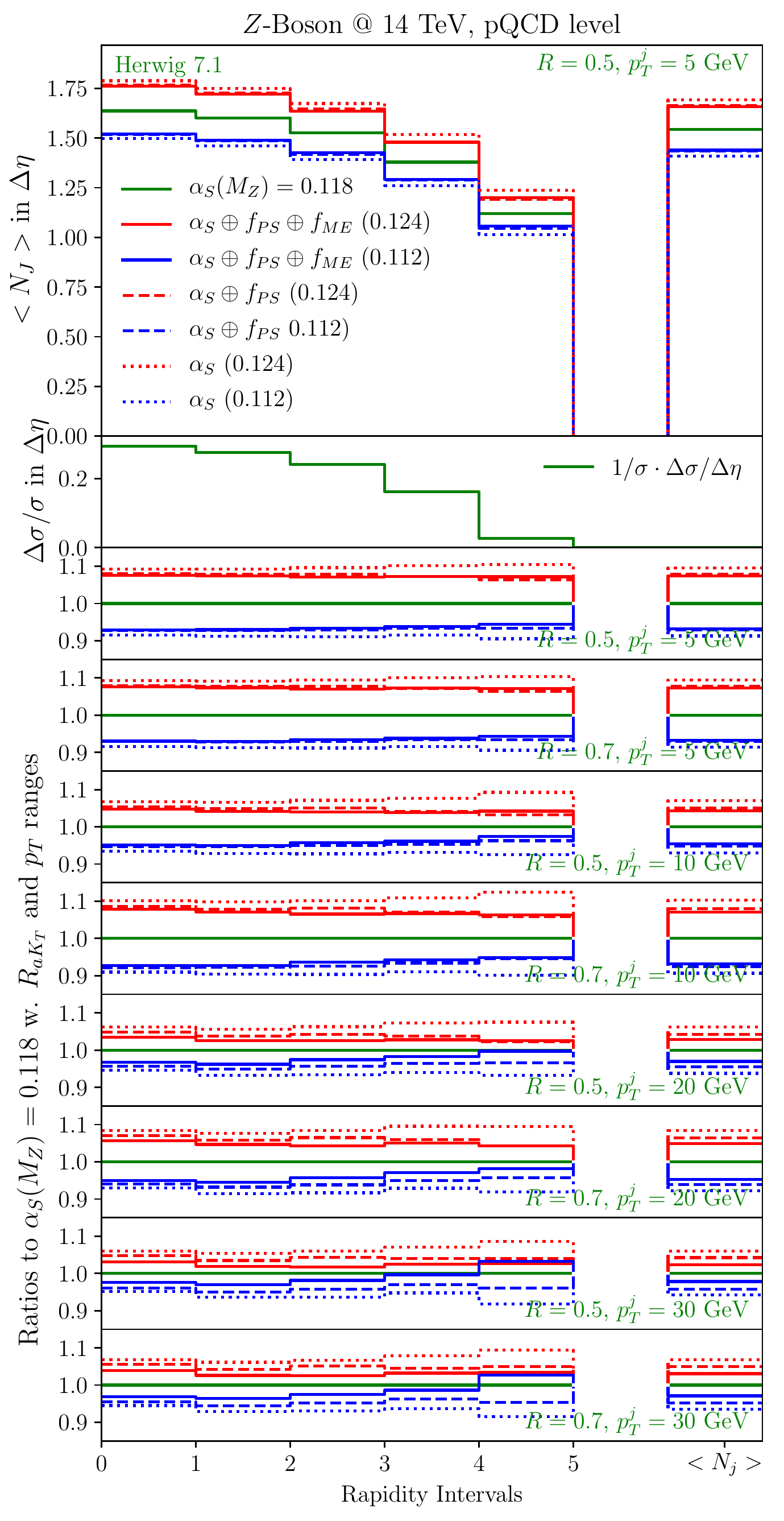}\hfill
\includegraphics[width=0.47\textwidth]{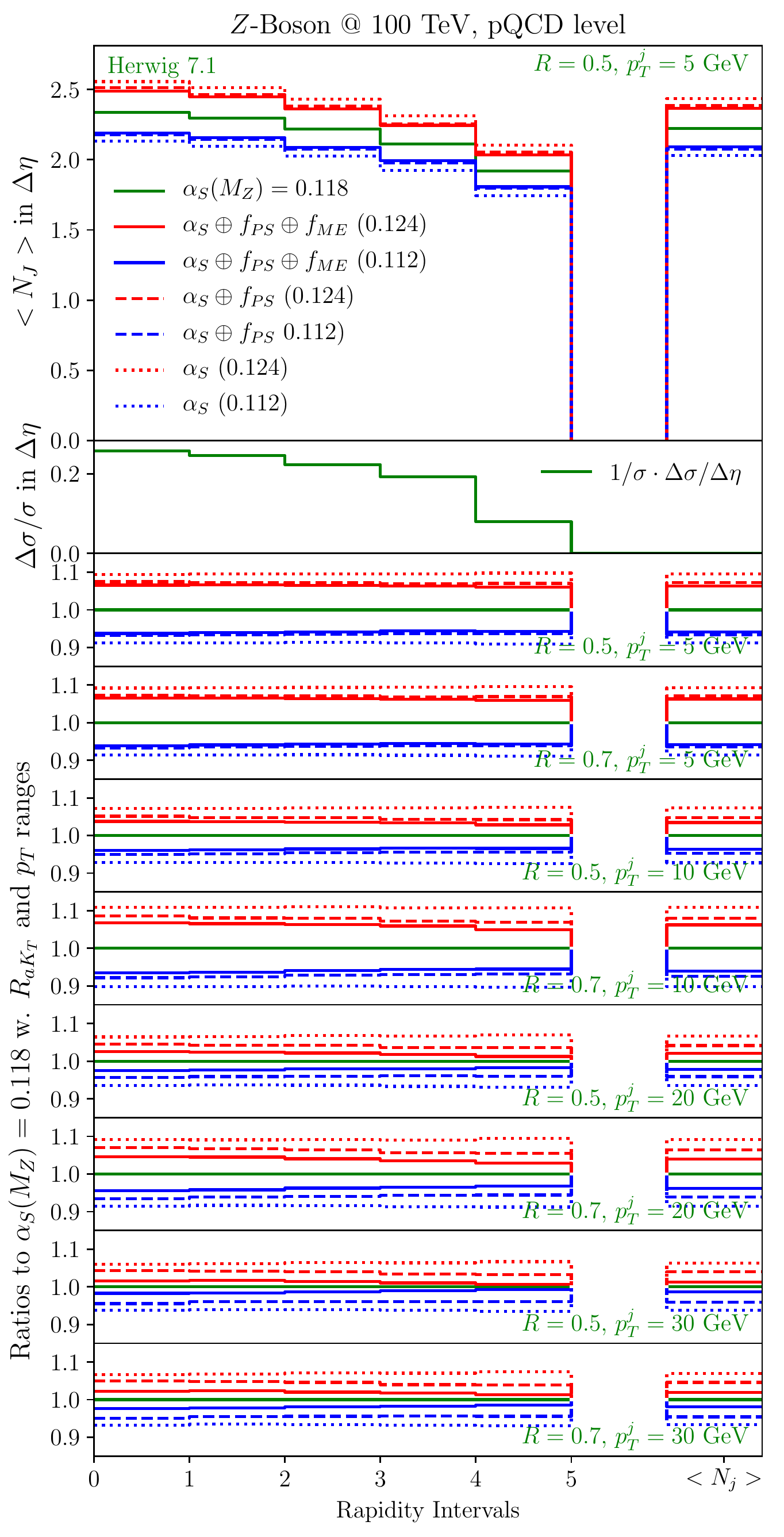}\\[1ex]
\caption{Average number of jets in intervals of Z boson rapidity $\Delta \eta$, 
left for $14$ TeV  collider energy, right for 100 TeV. 
The second plot shows the relative size of the cross section in the intervals.
The dependences of jet definition is illustrated as ratio plots w.r.t. 
the central prediction. 
The red and blue lines show the prediction for modified $\alpha_S$ values, 
$\alpha_S(M_Z)=0.124$ and $\alpha_S(M_Z)=0.112$ respectively.
For dotted lines we exclusively change the value of $\alpha_S(M_Z)$ used in the
showering process. To receive the dashed lines, also the pdf set used 
in the showering process was fitted with the corresponding $\alpha_S(M_Z)$ value.
The solid lines correspond to a full correlation of $\alpha_S(M_Z)$ values in shower and the pdf sets used in hard cross section and shower calculation. 
}
  \label{fig:MC_tunes_variations:pdf:njetsZ}
\end{figure}

\begin{figure}[t]
\centering
\includegraphics[width=0.47\textwidth]{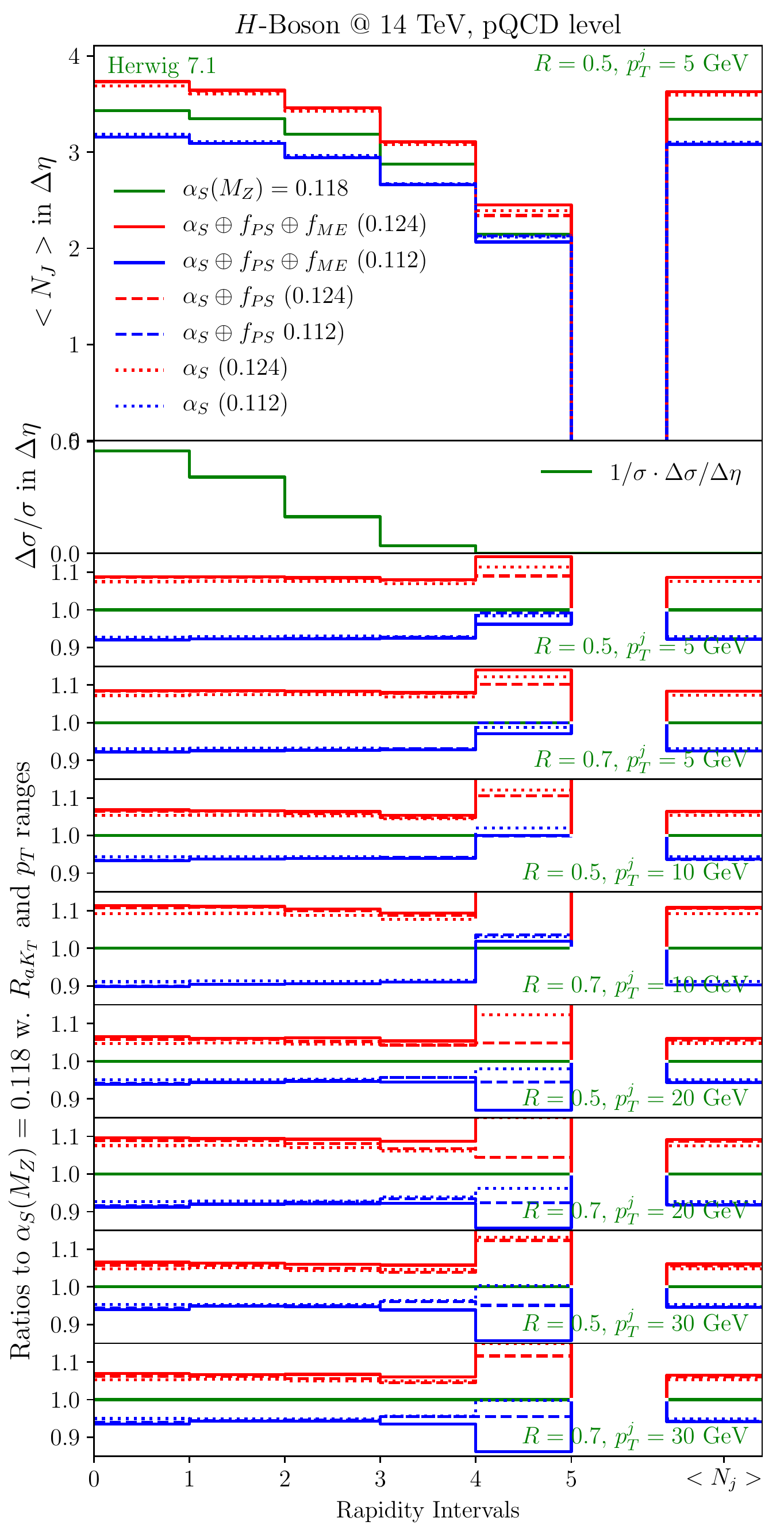}\hfill
\includegraphics[width=0.47\textwidth]{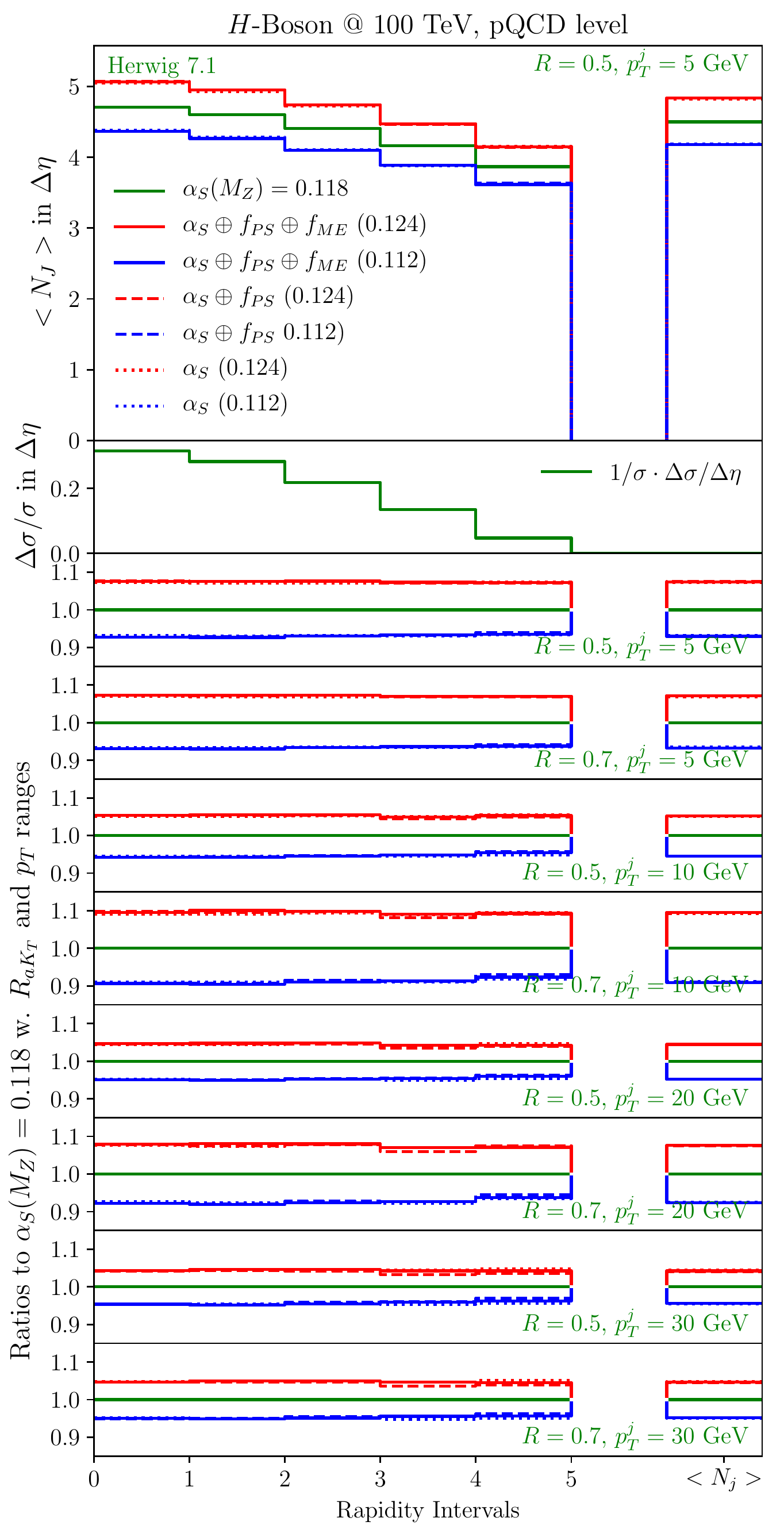}\\[1ex]
\caption{Same as Fig.~\ref{fig:MC_tunes_variations:pdf:njetsZ} here for Higgs boson production.} 
  \label{fig:MC_tunes_variations:pdf:njetsH}
\end{figure}

\begin{figure}[t]
\centering
\includegraphics[width=0.47\textwidth]{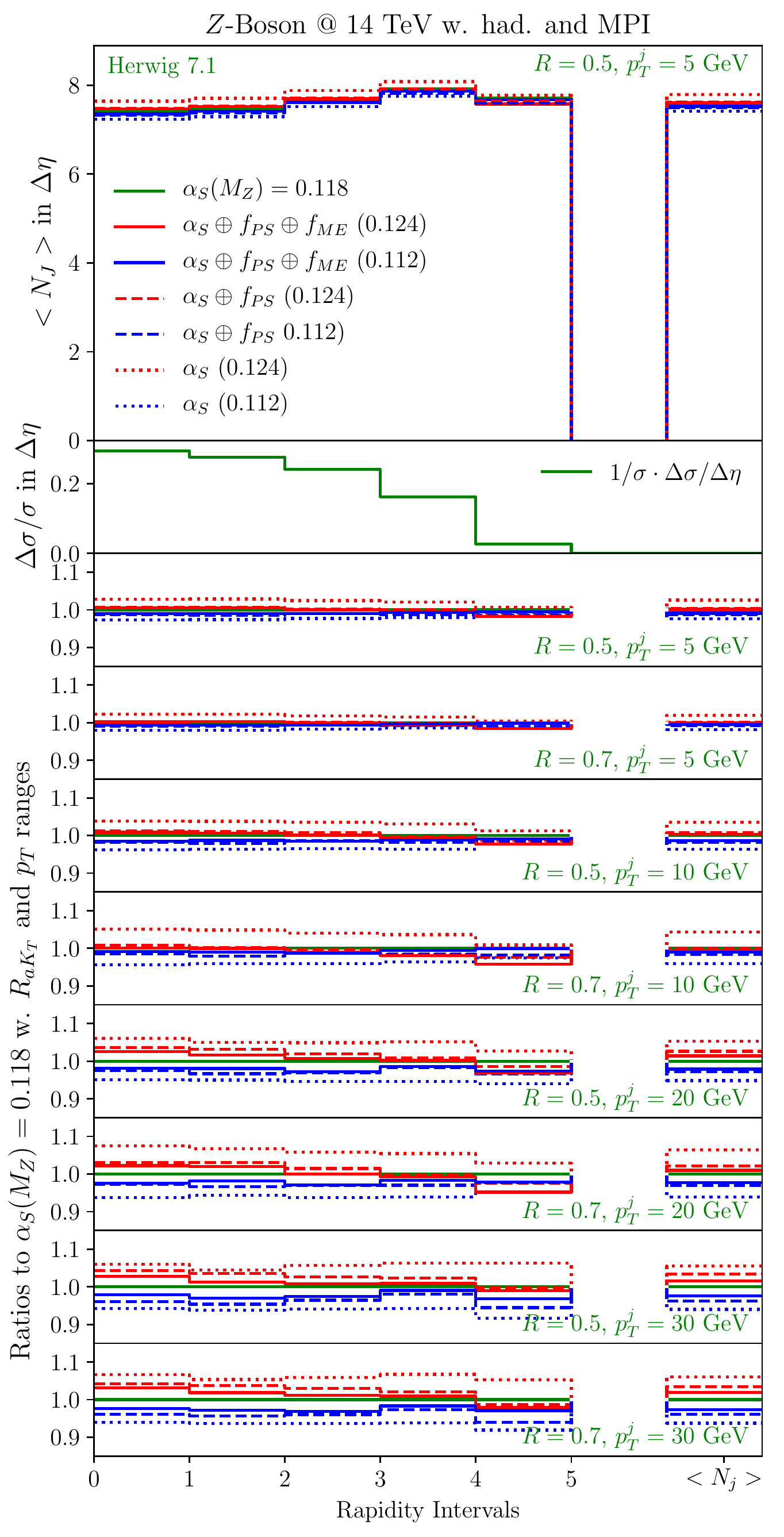}\\[1ex]
\caption{As Fig.~\ref{fig:MC_tunes_variations:pdf:njetsZ}. Here including hadronisation effects as well as
 multiple parton interactions. Compared to Fig.~\ref{fig:MC_tunes_variations:pdf:njetsZ} the average number
  of jets is less dependent on the value of the strong coupling constant. This is
   especially visible for low pt jet definitions. This reduction of the error band is
    explained by the available energy left after hard process calculation and parton
     shower evolution. 
While the enhanced  $\alpha_S(M_Z)$ radiates more and therefore leaves less energy for additional MPI interactions, reducing the value of $\alpha_S(M_Z)$ allows more additional MPI jets. This directly moderates the effect of value variation.   \label{fig:MC_tunes_variations:pdf:njetsZMPI}}
\end{figure}

In Figs.~\ref{fig:MC_tunes_variations:pdf:njetsZ} and~\ref{fig:MC_tunes_variations:pdf:njetsH} we show the average 
number of jets for Z- and Higgs-boson production as a function of the boson
rapidity. The first ratio plots (green line) is the relative cross section 
contribution in the rapidity intervals. 
The multiple ratios show the result when using different definitions
of anti-$K_T$ jets: $R_{aK_T} \in [0.5,0.7]$ and $p^j_T \in [5,10,20,30]$ GeV. 
The last bin in the distribution represents the inclusive average number of jets. 
Figure~\ref{fig:MC_tunes_variations:pdf:njetsZMPI} highlights the effect of including multiple 
parton interactions.

Compared to the transverse momenta this observable does not show the
usual effect of crossing error bands due to parton shower unitarity. 
We further observe, depending on the jet definition, a drastic reduction of 
variation when using fully correlated variations in Figs.~\ref{fig:MC_tunes_variations:pdf:njetsZ}
and~\ref{fig:MC_tunes_variations:pdf:njetsZMPI}.
In gluon dominated processes like Higgs and dijet production, this effect is not visible.
The comparison between Figs.~\ref{fig:MC_tunes_variations:pdf:njetsZ} and~\ref{fig:MC_tunes_variations:pdf:njetsZMPI} also 
shows an interesting reduction of the variation bands when including the effects 
of hadronisation and multi parton interaction (MPI). Although the difference 
between correlated and uncorrelated variation is hardly altered by including MPI it is 
worth noting that if the shower produces fewer emissions because of a reduced 
strong coupling, the MPI is more active and adds more partons, since more of the
hadron energy/momentum is left after evolution. Similar effects are expected
with the interleaved MPI employed by Pythia.


%
%


\subsubsection{Discussion}
\label{sec:MC_tunes_variations:pdf:discussion}

Before concluding, let us summarise and discuss the findings of 
Sec.~\ref{sec:MC_tunes_variations:pdf:results}. We find that unnormalized distributions suggest 
large differences between $\alpha_s$ variations and correlated PS+ME 
variations. These are directly linked to scaling with different PDFs used 
in the hard-scattering calculation. Since such uncertainties are beyond the 
scope of a (cross-section preserving) parton shower, it is prudent to base
conclusions on normalized distributions. In this case, the effect of correlations is
clearer. Including correlations by aligning the PDF with the $\alpha_s$ used in
the parton shower shrinks the envelope of variations. This effect is moderate
for the Drell-Yan pair $p_\perp$ distribution at a 14 TeV LHC, and small 
for the Higgs-boson $p_\perp$ distribution at the same setup. This might suggest
that it is not necessary to respect the correlation for a 14 TeV LHC. However,
aligning PDFs with $\alpha_s$ variations can lead to more pronounced effects
for observables that are sensitive to more details of the shower evolution.
For the average number of jets, included correlations lead to larger differences
 between variations and to a more pronounced uncertainty reduction. 
Since multiple parton interactions
are sensitive to the proton content \emph{after} showering, the effect remains
upon including MPI. In fact, the difference between correlated and uncorrelated
variations is enhanced and develops intricate phase-space dependences. This
is particularly visible for uncorrelated $\alpha_S$ variations, which,
after including MPI, exhibit a dependence on the $p_{T,min}$ cut used to
define jets. This is because larger shower variations lead to larger variations
in MPI activity. The situation is improved when performing correlated PS
variations. Thus, we conclude that for normalized distributions, it is 
reasonable to align the PDF and the $\alpha_s$ when performing shower 
variations.

\subsection{Summary}

In this contribution, we have tried to illustrate and assess the
cross-talk of perturbative parton-shower uncertainties and
non-perturbative modelling in the context of MCEGs. We have argued
that we should regard an MCEG as a tool that transfers our most
detailed and well-constrained knowledge of the theory and previous
data to new measurements. Adhering to this philosophy means that
perturbative and non-perturbative MCEG aspects should be considered
correlated. We have investigated this effect from two angles.

In the first part, we have argued that perturbative variations that clearly do 
not describe known data should be complemented with an adjustment of 
non-perturbative parameters to achieve a sensible description of known data.
This can be achieved by retuning the MCEG's description of hadronization-sensitive
data for each parton-shower variation. We have compared the results of the
envelope of such tunes with naive parton-shower scale variations around a
``central" tune, for both \Herwig and \Pythia. We find that, as expected, the
retuned results exhibit a smaller variation in phase-space regions dominated by
non-perturbative effects. The naive ''central`` variation bands of \Herwig and \Pythia are
rather different, with \Herwig being conservative-leaning and \Pythia being
more optimistic about the size of the variation band. The size of the retuned uncertainties is comparable between
the two simulations. This convergence of two very different models suggests 
that comparing retuned variations might provide a better assessment of MCEG
uncertainties in the future.

The second part of the study was motivated by cross-talk between PDF sets and
parton-shower variations due to initial-state parton-shower backward evolution.
In particular, we argued that the assumption of a faithful implementation of
PDF evolution suggested that $\alpha_s$ variations in the parton shower should
be accompanied changing to a corresponding PDF set. We find, depending on the
process, large changes in $p_\perp$ distributions when including the changes
in normalization due to changing PDF sets. These effects are much reduced
when discussing normalized distributions. In these, as expected, using  
correlated PS variations leads, in some cases, to a reduced variation band.
The impact is of order 5\% for the Drell-Yan $p_\perp$ spectrum at 14 TeV 
collider energy. Larger effects can be seen in the average number of
jets in boson rapidity intervals of $\Delta y=1$. In this observable, the
effect of correlated PS variations is of 5\% for the Drell-Yan process without
 the inclusion of muliple parton interactions, and more pronounced when
including the latter. Multiple parton interactions enhance the impact of 
correlations, since the probability for MPI is very sensitive to parton-shower
variations. Thus, we are led to conclude that it is prudent to
change the parton-shower PDF consistently when varying $\alpha_s$ in the parton
shower.

\subsection*{Acknowledgements}
We thank the organizers for an inspiring workshop.  This manuscript
has been authored by Fermi Research Alliance, LLC under Contract
No. DE-AC02-07CH11359 with the U.S. Department of Energy, Office of
Science, Office of High Energy Physics. This project has also received
funding from the European Research Council (ERC) under the European
Union's Horizon 2020 research and innovation programme, grant
agreement No 668679 and the MCnetITN3 H2020 Marie Curie Initial
Training Network, contract 722104. AHH and DS acknowledge partial
support by the FWF Austrian Science Fund under the Doctoral Program
No. W1252-N27 and the Project No. P28535-N27. AHH and SP also
acknowledge support by the COST action CA16201 PARTICLEFACE. AS
acknowledges support from the National Science Centre, Poland Grant
No. 2016/23/D/ST2/02605 and COST Action CA15213 THOR. LL acknowledges
support from the Swedish Research Council, contracts numbers
2016-03291 and 2016-05996.

%
%
%
\setcounter{subsection}{0}
\renewcommand\thesubsection{\thesection.\Alph{subsection}}
\subsection{Appendix}
\subsubsection{Generator tuning vs. parton-shower variations: Weights}
\label{app:MC_tunes_variations:weights}

For the tuning to LEP data we used the following weights in the
\Professor tune, for both \Herwig and \Pythia.

{\footnotesize
\begin{verbatim}
A_2004 := ALEPH_2004_S5765862
L_2004 := L3_2004_I652683

/A_2004/d01-x01-y01  1  # Charged multiplicity
/A_2004/d102-x01-y01 1  # Thrust minor 
/A_2004/d110-x01-y01 1  # Jet mass difference
/A_2004/d118-x01-y01 1  # Aplanarity
/A_2004/d133-x01-y01 1  # Oblateness
/A_2004/d141-x01-y01 1  # Sphericity
/A_2004/d149-x01-y01 1  # Durham jet 2->1
/A_2004/d157-x01-y01 1  # Durham jet 3->2
/A_2004/d165-x01-y01 1  # Durham jet 4->3
/A_2004/d187-x01-y01 1  # 1-jet fraction
/A_2004/d195-x01-y01 1  # 2-jet fraction
/A_2004/d203-x01-y01 1  # 3-jet fraction
/A_2004/d211-x01-y01 1  # 4-jet fraction
/A_2004/d54-x01-y01  1  # Thrust
/A_2004/d62-x01-y01  1  # Heavy jet mass
/A_2004/d70-x01-y01  1  # Total jet broadening
/A_2004/d78-x01-y01  1  # Wide jet broadening
/A_2004/d86-x01-y01  1  # C-parameter
/A_2004/d94-x01-y01  1  # Thrust major
\end{verbatim}
}
\noindent
For \Pythia we also did a separate tune where also the charged multiplicity distribution
 from L3 \cite{Achard:2004sv} was included.

{\footnotesize
\begin{verbatim}
/L_2004/d59-x01-y01  1  # Charged multiplicity
\end{verbatim}
}


\renewcommand\thesubsection{\thesection.\arabic{subsection}}


\let\Herwig\undefined
\let\Pythia\undefined
\let\Sherpa\undefined
\let\Rivet\undefined
\let\Professor\undefined
\let\eps\undefined
\let\mc\undefined
\let\mr\undefined
\let\mb\undefined
\let\tm\undefined


           






\section{Determination of nonperturbative correction factors and their dependence on Monte Carlo modeling~\protect\footnote{
    J.~Bellm,
    S.~Kuttimalai}{}}
\label{sec:MC_npfactors}


  In this contribution we discuss the influence of Monte Carlo
  modelling on the extraction and calculation of nonperturbative
  correction factors. In particular we address the question whether
  the choice of PDF sets and Monte Carlo models in the extraction of
  nonperturbative correction factors may bias measurements in which
  the correction factors are used, e.g. PDF measurements at the LHC.
  We determine the non perturbative factors for various settings and
  explain the outcome in detail.

\subsection{Introduction and Motivation}

With an increasing amount of data being recorded and getting analyzed
by the LHC experiments, it becomes possible to perform measurements
more and more differentially in the event kinematics. This allows for
the extraction of more information that can be used for a more precise
determinations of key quantities, like the parton distribution
functions (PDFs) of the proton.


At the same time, recent theoretical progress has lead to an increase
in precision of fixed-order
calculations~\cite{Currie:2017eqf,Currie:2016bfm} which can, in many
cases, be carried out at the next-to-next-to leading order in the
strong coupling constant. Using such high-precision calculations and
more differential observables for PDF fits requires the application of
corrections that capture nonperturbative (NP) effects. They are not
included in fixed-order calculations but their impact can be
comparable to the remaining fixed-order uncertainty. 

Modern Monte Carlo event generators (MCEGs) typically operate only at
the leading order (LO) or the next-to-leading order (NLO) but
supplement fixed-order calculations with an all-order treatment of
soft and collinear QCD radiation through parton shower (PS)
simulations and nonperturbative effects through models for
hadronization and multiple parton interactions (MPI). While
hadronization models account for the nonperturbative transition of QCD
partons to observable colorless hadrons, MPI models account for the
effect of interactions between the beam remnants which generally lead
to an increase of softer jet activity in the events.

In order to correct high-precision fixed-order calculations for NP
effects correction factors are often extracted MCEGs by performing
separate runs with both hadronization and MPI disabled and enabled.
The ratios of the corresponding predictions are then applied to
fixed-order calculation as differential correction factors of the form
\begin{align}
K^{\mathcal{O}}_\mathrm{NP} = \frac{\dif\sigma^{\mathrm{PS}+\mathrm{HAD}+\mathrm{MPI}} /\dif \mathcal{O}}{\dif\sigma^{\mathrm{PS}}/\dif \mathcal{O}}\,,\label{eq:MC_npfactors:k-np}
\end{align}
where
$\dif\sigma^{\mathrm{PS}+\mathrm{HAD}+\mathrm{MPI}}/\dif \mathcal{O}$
is the MCEG prediction differential in observable $\mathcal{O}$ with
parton showering, hadronization, and MPI
\cite{Bahr:2008dy,Alekhin:2005dx} enabled and
$\dif\sigma^{\mathrm{PS}}/\dif \mathcal{O}$ refers to the MCEG
prediction with hadronization and MPI disabled. Here $\mathcal{O}$ can
be multi differential.

It can be shown that the NP corrections, defined in Eq.~\eqref{eq:MC_npfactors:k-np}, can be
minimized by a smart choice of the jet radius in the observable
definition. This is due to the hadronization correction and the MPI
corrections exhibiting different scaling behaviors with $R$
\cite{Dasgupta:2007wa}. While decreasing the jet radius $R$ leads to
larger hadronization corrections, it decreases corrections due to MPI.
This can be intuitively understood since a large radius increases the
chance of the jet receiving contributions from uncorrelated MPI
interactions. A very small jet radius, on the other hand, leads to
energy getting emitted outside of the jet cone in the transition from
partons to hadrons, thereby increasing the impact of hadronization
corrections. An optimal cone size can thus be chosen in order to
minimize the sum MPI and hadronization effects.

When extracting PDFs from data it is crucial to ensure that the NP
corrections discussed above do not introduce any bias. In particular,
the NP correction factors should be independent of the PDFs and the
MCEGs that were used to calculate them.
\begin{figure}[t]
  \centering
  \includegraphics[width=\textwidth]{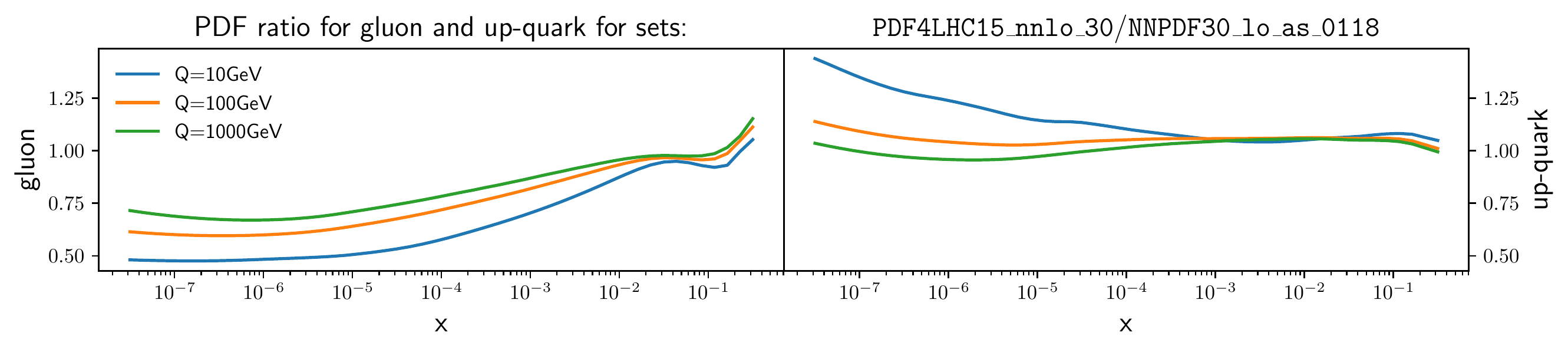}
  \caption{Gluon and up-quark pdf ratio for the NNLO PDF set $\texttt{PDF4LHC15\_nnlo\_30}$ 
\cite{Butterworth:2015oua} w.r.t. the leading order set $\texttt{NNPDF30\_lo\_as\_0118}$ \cite{Ball:2014uwa} this the same $\alpha_S(M_Z)=0.118$ value.} 
  \label{fig:MC_npfactors:pdfdif}
\end{figure}
In this note we study the
effect of different PDFs, MCEGs, and the perturbative order at which
the MCEGs operate (LO or NLO) on the the extracted NP factors and on
the jet radius that minimizes the NP corrections.
In order to do so we choose the NNLO PDF4LHC recommendation and the 
LO pdf set from $\texttt{NNPDF30\_lo\_as\_0118}$ as the ratio of gluon 
and quark pdfs can be large, see Fig.~\ref{fig:MC_npfactors:pdfdif}.
Despite the warning in \cite{TorbjornsWarning} and the discussion in \cite{Buckley:2016caq}
we explicitly choose very diverse gluon distributions to estimate an upper bound on the effect.

In \cite{Gieseke:2004tc} it was shown that parton shower evolution is
usually less effected by PDF variations and more by modifications of
the strong coupling, for example. More parton shower uncertainties
have been studied in \cite{Bellm:2016rhh,Bothmann:2016nao}. Here we
concentrate on the modification of the NP part of the simulation.




\subsection{Analysis and Observable}
\label{sec:MC_npfactors:analysis}

As a benchmark observable we use the inclusive jet transverse momentum
$p_T$ at the LHC. It was measured by the CMS collaboration for two
different jet radii $R=0.4$ and $R=0.7$ in several bins of jet
rapidity \cite{Khachatryan:2016wdh}. For this purpose of the presented
study we extended the corresponding validated Rivet
\cite{Buckley:2010ar} analysis to include more radii in the range
$R \in [0.1,0.2...1.1]$. In this analysis jets are defined using the
anti-$k_T$ \cite{Cacciari:2008gp} jet algorithm. We perform our analysis for the LHC
operating at a center-of-mass energy of
$\sqrt{s}=\SI{13}{\tera\electronvolt}$.

\subsection{Simulation Setup}

\label{sec:MC_npfactors:setup}
For the extraction of NP correction factors we use the two general
purpose MCEGs Herwig7 \cite{Bellm:2015jjp,Bellm:2017bvx} and Sherpa
\cite{Gleisberg:2008ta}. For parton showering we use the
$\tilde{Q}$-shower in Herwig7 \cite{Gieseke:2003rz} and the CS dipole
shower \cite{Schumann:2007mg}. We calculate the hard scattering
process of inclusive jet production at LO or at NLO and use adaptations of the MC@NLO
parton shower matching scheme \cite{Frixione:2002ik} for consistent
combination with the parton showers. In the calculation of the hard
scattering we set the factorization and renormalization scales to
$\mu_f=\mu_r=H_T/2$, where $H_T$ is the scalar sum of all parton
transverse momenta in the event. The hadronization and MPI models
implemented in Sherpa are detailed in \cite{Gleisberg:2008ta}.
Herwig7's models are documented in \cite{Bahr:2008pv}. For the study of
the impact of PDFs we use the $\texttt{PDF4LHC15\_nnlo\_30}$ 
\cite{Butterworth:2015oua} and the
$\texttt{NNPDF30\_lo\_as\_0118}$ \cite{Ball:2014uwa} sets as implemented in LHAPDF
\cite{Buckley:2014ana}. We replace all PDFs sets used in the modelling 
(hard cross section, parton shower, MPI and proton remnant handling) 
and all instances of strong coupling constants to the cossesponding 
$\alpha_S(M_Z)=0.118$ value. 

We would like to stress that our generator settings are not tuned to
data but rather aligned across both MCEGs for a comparison. The focus
of this study is not the nominal NP correction factor but rather it's
dependence on the parameters in the extraction.

\subsection{Results}

\begin{figure}[t]
\centering
\includegraphics[width=\textwidth]{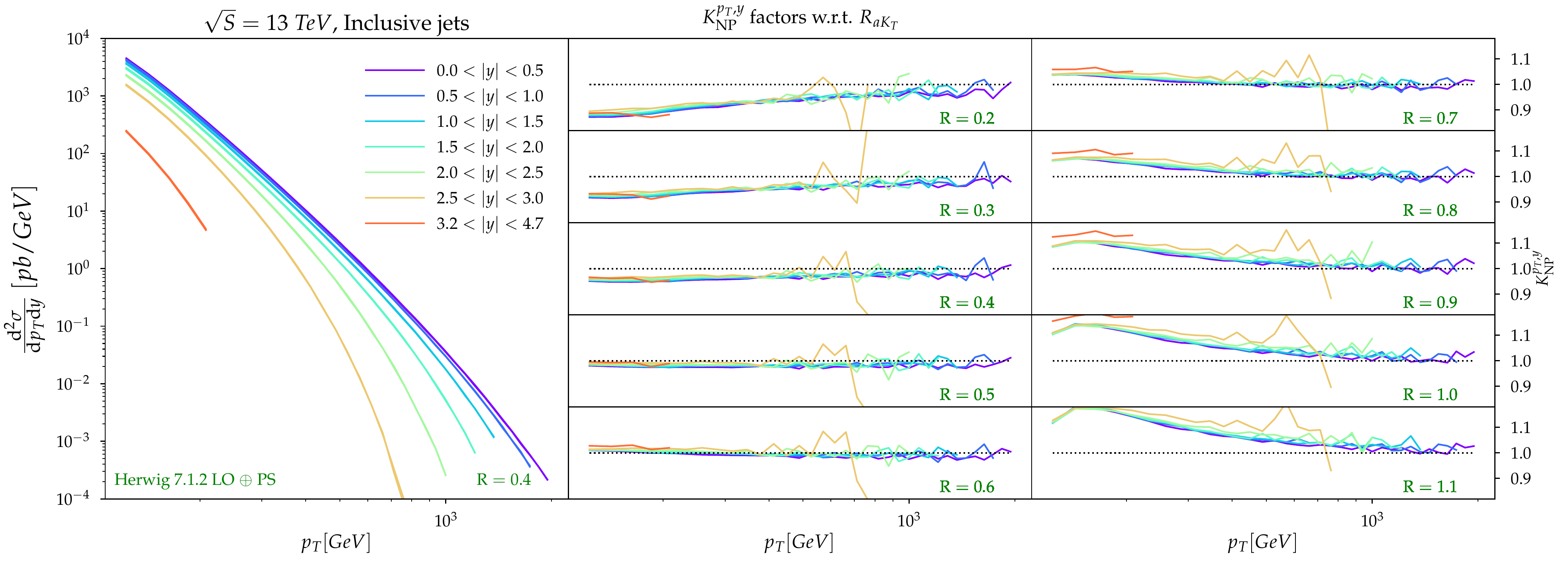}
\caption{Double differential cross section in transverse momentum and
  in intervals of rapidity. Left panel shows the differential cross
  section. The ratio plots in the center and to the right show the non
  perturbative factor for various jet opening angles $R_{antiK_T}$.
  While for small radii hadronisation effects tend to remove energy
  from thin jets, for large radii the jets tend to collect more MPI
  jets. Therefore the former get softer the latter harder. For broad
  jet definitions ($R>0.8$) even the generation cut is visible as the
  MPI addition allows to shift the cross section drop off into the
  analysis region. }
  \label{fig:MC_npfactors:mainHerwig}
\end{figure}

We start our discussion with results for the NP corrections for the
differential inclusive jet transverse momentum spectrum in bins of
pseudorapidity $y$ as obtained from MCEG runs using LO matrix elements
for the hard scattering process calculation. As shown in 
Fig.~\ref{fig:MC_npfactors:mainHerwig}
we observe large negative NP corrections of the
order of \SI{10}{\percent} in the region of small $p_T$ when choosing
a small jet radius of $R\approx 0.1$. The origin of these corrections
is the energy that gets radiated out of the jet in the process of
hadronization, as discussed above. At large values of $R$, we see
large positive corrections of up to \SI{20}{\percent} due to
additional MPI radiation entering the jets. At values of
$R\approx 0.5$ the NP corrections are minimized to the
\SI{1}{\percent} level. We note that the dependence of NP effects on
the pseudorapidity is very small and that NP effects generally vanish
at very large transverse momenta of the order of
\SI{1}{\tera\electronvolt}.

\begin{figure}[t]
  \centering
  \includegraphics[width=\textwidth]{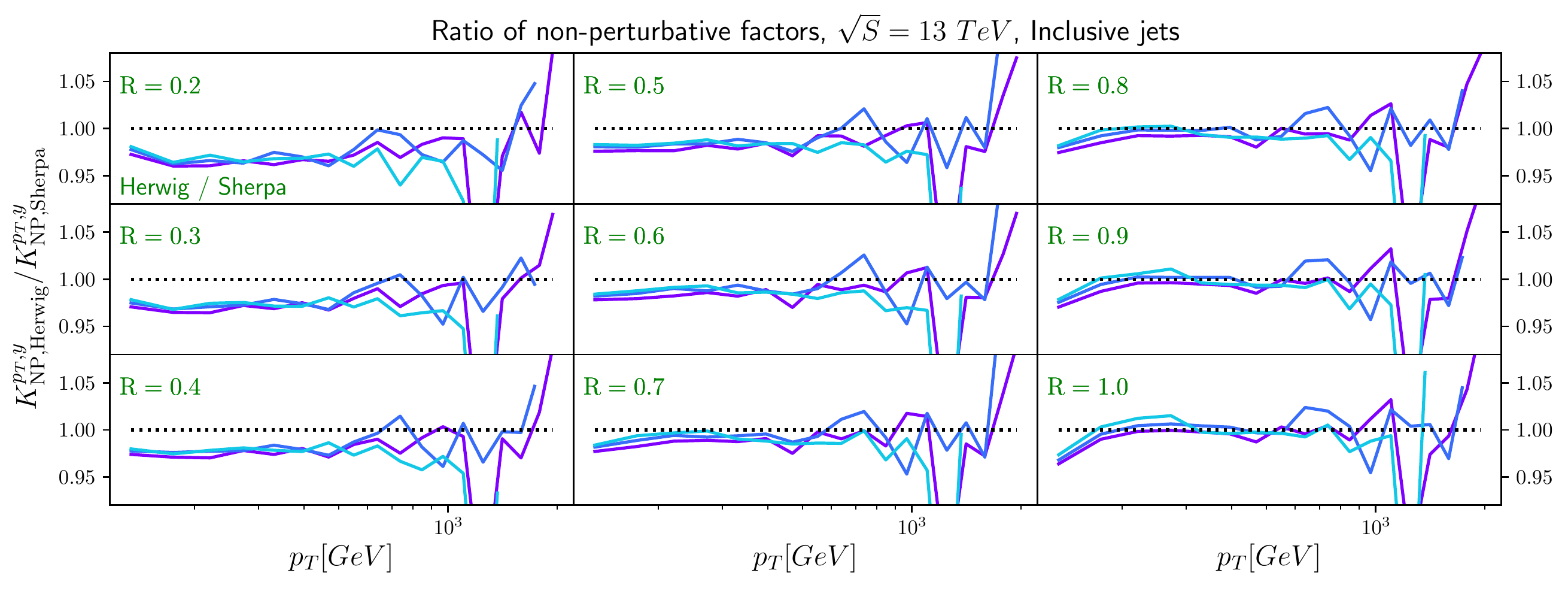}
  \caption{Ratio of NP factors calculated using Herwig7 and Sherpa. With
    the setup in use the MPI modification area of broad jets is less
    modified while the hadronisation region is altered by the
    different shower and hardonisation models. We stress again that
    the findings are setup dependent and effects might shrink with
    tuned values. }
  \label{fig:MC_npfactors:lo-herwig-sherpa}
\end{figure}

\begin{figure}[t]
\centering
\includegraphics[width=0.70\textwidth]{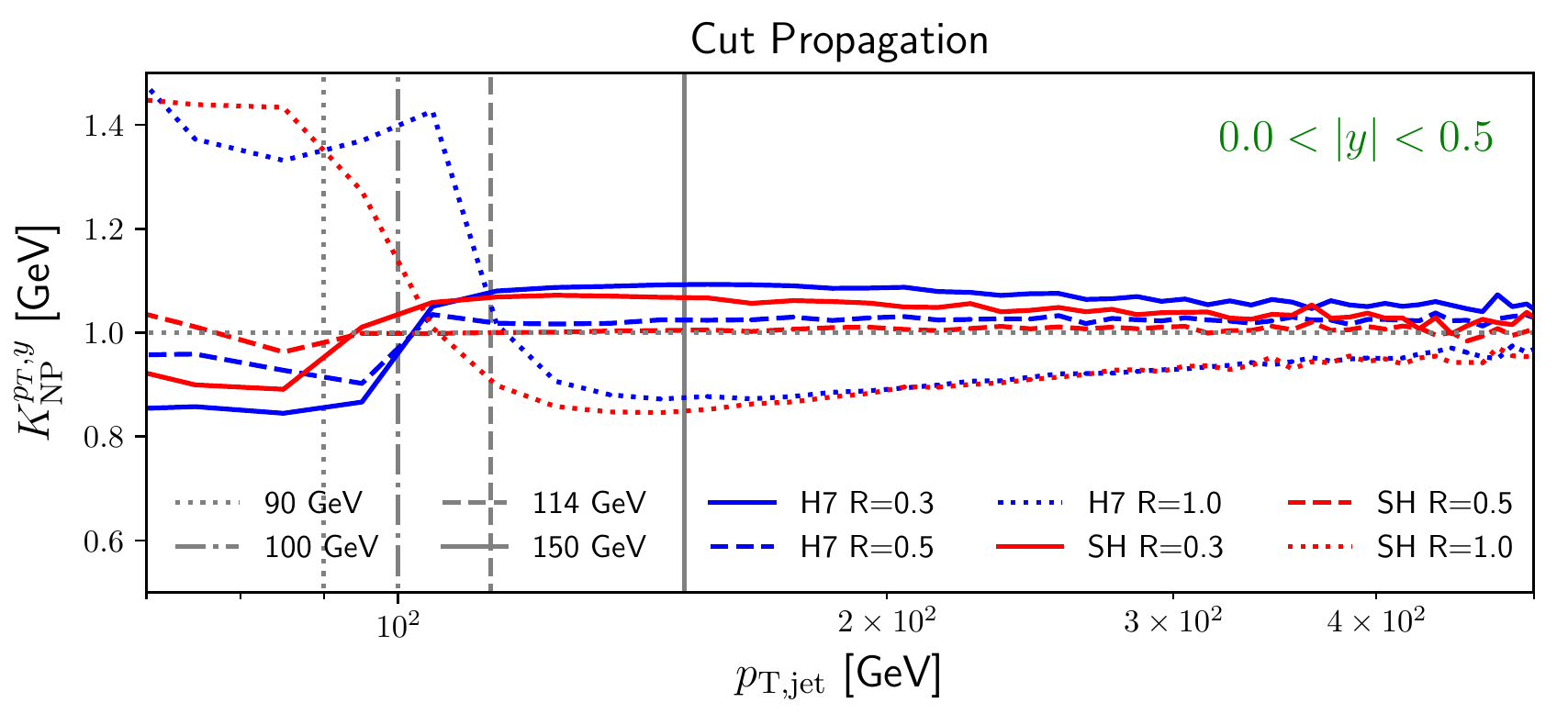}
\caption{Effect of event generation cuts on the NP factors. The cut on
  the leading jet transverse momentum was set to
  \SI{100}{\giga\electronvolt} and \SI{90}{\giga\electronvolt} for
  Herwig7 and Sherpa, respectively}
\label{fig:MC_npfactors:cuts}
\end{figure}

Having shown the qualitative features of the NP corrections
themselves, we now turn to the discussion of their dependence on the
MCEG parameters used for the extraction.
For this purpose, we show the
ratio of NP factors extracted with Herwig7 and Sherpa at leading order
in Fig.~\ref{fig:MC_npfactors:lo-herwig-sherpa}.
The observed differences are below the \SI{5}{\percent} level and most
pronounced at small jet radii. This suggests that the source of the
small discrepancies are mainly differences in the hadronization and
parton shower models. In addition we note that there are differences
at large $R>0.8$ and very small $p_T$ close to the analysis jet
cut. We were able to trace these discrepancies back to different event
generation cuts on the leading jet transverse momentum, which was set
to \SI{90}{\giga\electronvolt} and \SI{100}{\giga\electronvolt} for
Sherpa and Herwig7, respectively. This effect is illustrated in
Fig.~\ref{fig:MC_npfactors:cuts}, where the event generation cuts are
clearly visible as features in the NP factor. For larger jet radii
this effect gets washed out, thus extending it into the region shown
in Fig.~\ref{fig:MC_npfactors:lo-herwig-sherpa}.

\begin{figure}[t]
  \centering
  \includegraphics[width=\textwidth]{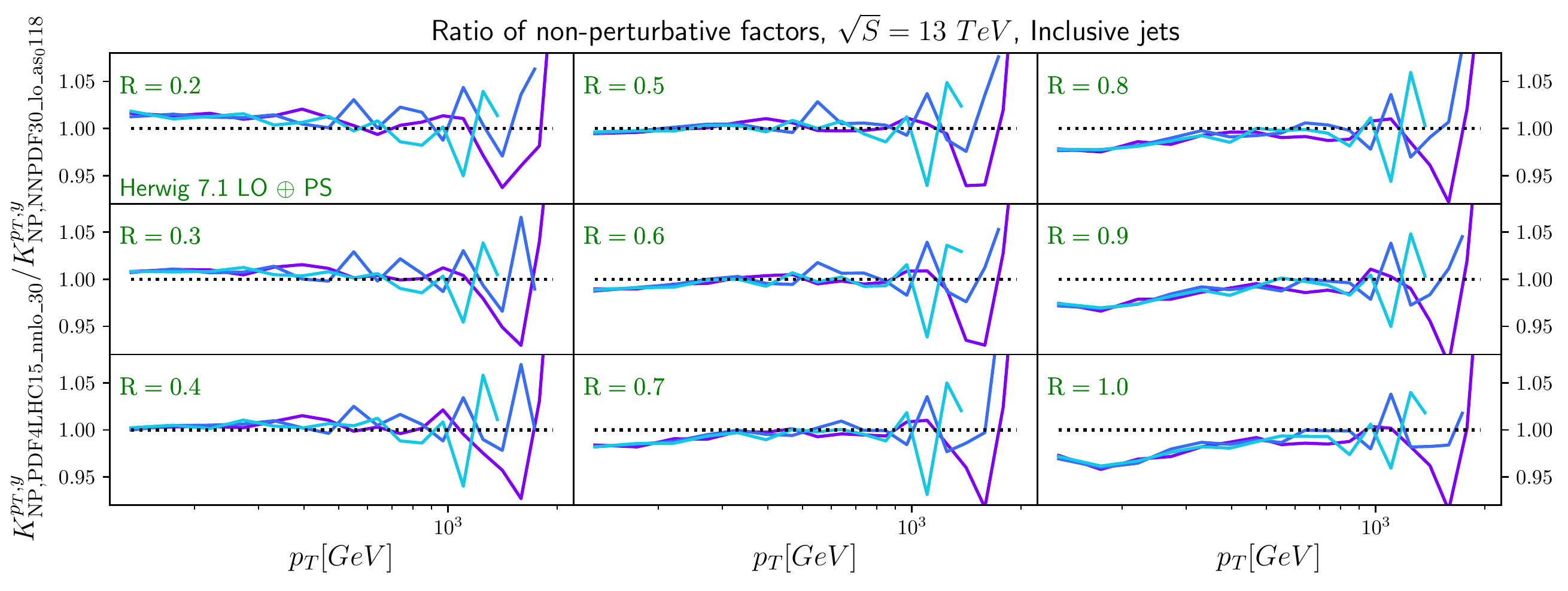}
  \caption{Ratio of NP factors calculated using two different PDF
    sets. The effect of removing energy from the jet due to emissions
    outside of the small cone in the hadronisation process is hardly
    effected by changing the PDF set. The differences for higher R
    values can be explained by modification of the MPI addition with
    modified PDFs. The strongly reduced NNLO gluon pdf at small x
    values allows less MPI activity and reduced the additional energy
    in the broad jets. Therefore $K^{p_T,y}_\mathrm{NP}(R)$ is reduced
    for large R values. }
  \label{fig:MC_npfactors:lo-pdf}
\end{figure}

In Fig.~\ref{fig:MC_npfactors:lo-pdf} we show the impact of using a
different PDF set. Hadronization corrections can be expected to be
independent of the PDF, as the hadronization models act only on final
state of partons after the full evolution through parton showers has
terminated.
Correspondingly, we see barely any dependence on the PDF at small $R$
where hadronization corrections dominate. At large $R$, we see larger
effects. This is to be expected since the MPI modelling, which mostly
affects large jet radii, crucially depends on the PDFs.  We note,
however, that MPI models are typically tuned to data based on a
specific PDF set. It is thus very likely that the observed differences
would have been much smaller, had we re-tuned the MPI models to the
respective PDFs. In that sense, the MPI models in MCEGs are to be
understood as a 'postdiction' or parametrization of MPI effects as
measured in data.

\begin{figure}[t]
\centering
\includegraphics[width=0.48\textwidth]{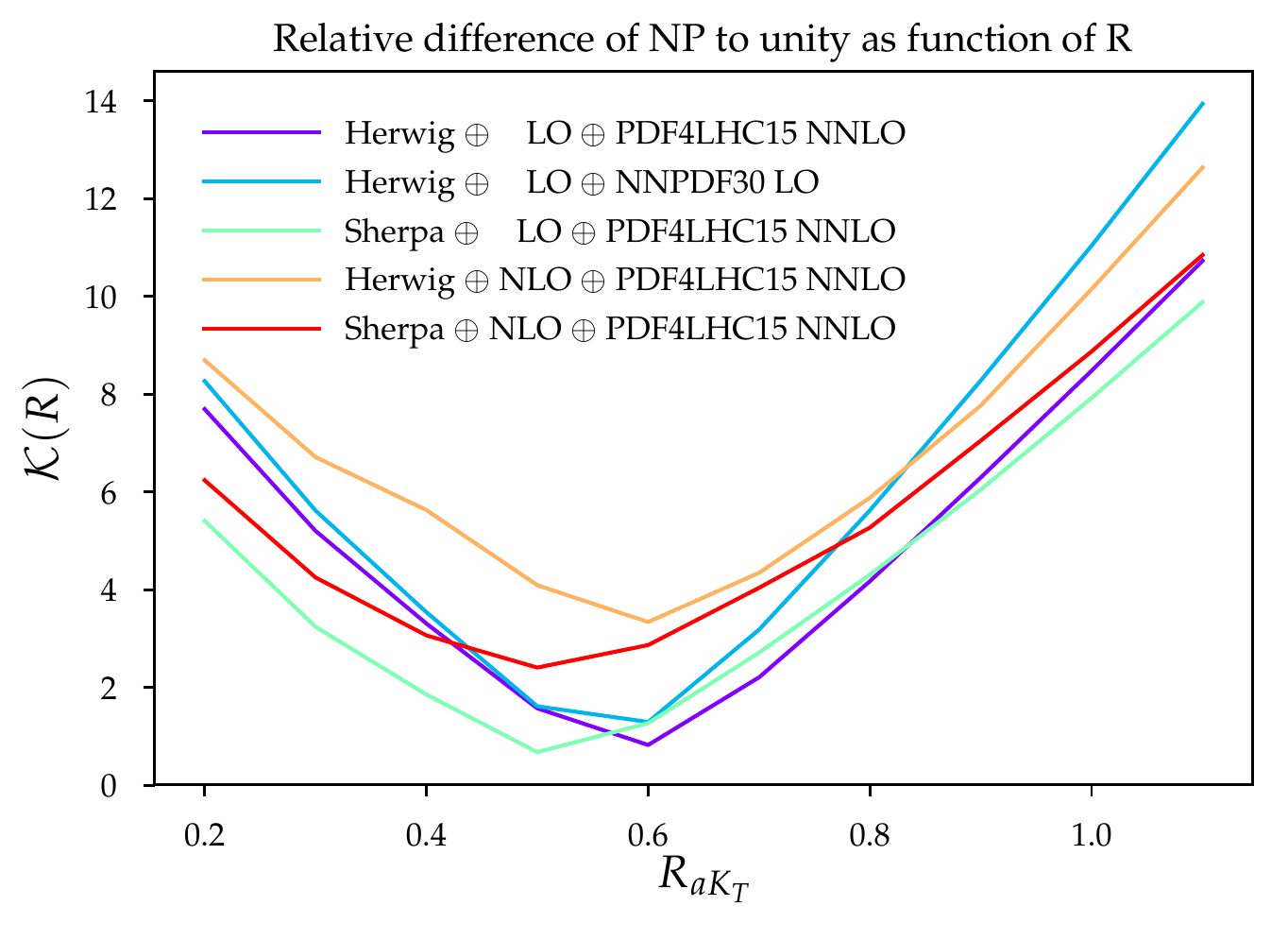}
\caption{For the simulation setup as described in Sec.~\ref{sec:MC_npfactors:setup} we
  calculate the measure according to Eq.~\eqref{eq:MC_npfactors:measures}. For Herwig
  we get an optimal value for $R \approx 0.6$ , for Sherpa slightly
  below. The pdf influence on the best R value is mild and matching to
  higher order matrix elements is even smaller. Note: The used setup
  is not representative for the tuned default setup. Therefore the
  absolute value of the optimal point might shift with standard
  parameters. Here we are interested in the shift due to PDF or higher
  perturbative accuracy. The hight of the distribution is driven by MC
  statistics.}
  \label{fig:MC_npfactors:optimal-r}
\end{figure}

In order to show how the optimal choice of $R$ is affected by the
variables considered in this study, we quantify the overall size of NP
corrections by the following measures
\begin{align}
\mathcal{K}(R)=\int \big{|} {\left( K^{p_T,y}_\mathrm{NP}(R) -1\right)} \big{|}\ \theta(p^{\max}_T - p_T) \dif p_T\dif y \;. \label{eq:MC_npfactors:measures}
\end{align}
Note that we do not apply a weight proportional to the double differential cross section\linebreak
$\dif\sigma/\dif p_T\dif y$ in the definition above since we would like to
minimize the NP corrections equally well in the entire phase space. As
shown in Fig.~\ref{fig:MC_npfactors:optimal-r}, the value of $R$ that minimizes
the NP corrections we find\footnote{Note that these can change with
  tuned values.} is $R\approx 0.6$ for Herwig and $R\approx 0.5$ for
Sherpa, and it is independent of PDF or perturbative order within
$\Delta R=0.05$.

\subsection{Conclusions}

In this contribution, we investigated the effect of PDF set choice and
simulation setup used to calculate nonperturbative correction factors.
In order to do so, we chose strongly diverse PDFs and calculated for
two different shower and hadronisation models as well as LO and NLO
plus parton shower simulations the NP factors. As expected we
confirmed in this study that these factors are only mildly affected by
the choice of model parameters or the perturbative order. The observed
differences in the extracted NP factors are of the order of only
\SI{5}{\percent} relative to the absolute value of the NP correction
themselves. We found that jet radii of $R\approx 0.5-0.6 $ minimize
the NP corrections to the level of only a few percent and the
phenomenlogical impact of the differences in the extracted NP factors
should therefore be very small. In addition we illustrated that event
generation matrix element cuts can have an impact on NP correction
factors if the opening angle of the jets allows the collection of much
MPI activity.

\subsection*{Acknowledgements}

We would like to thank Joey Huston and Stefan Prestel for useful
discussions and helpful comments. This project has received funding
from the European Research Council (ERC) under the European Union's
Horizon 2020 research and innovation programme, grant agreement No
668679. Furthermore, it was supported in part by the U.S. Department
of Energy under contract number DE-AC02-76SF00515.







\newcommand{\Sherpa}{S\protect\scalebox{0.8}{HERPA}\xspace}
\newcommand{\Comix}{C\protect\scalebox{0.8}{OMIX}\xspace}
\newcommand{\MEPSatNLO}{M\protect\scalebox{0.8}{E}P\protect\scalebox{0.8}{S}@N\protect\scalebox{0.8}{LO}\xspace}
\newcommand{\MGaMC}{M\protect\scalebox{0.8}{AD}G\protect\scalebox{0.8}{RAPH5}\_\protect\scalebox{0.8}{A}MC@NLO\xspace}
\newcommand{\Pyegt}{Pythia8\xspace}
\newcommand{\HW}{H\protect\scalebox{0.8}{ERWIG}7\xspace}



\section{Is there room for improvement in the description of the gluon splitting to heavy quarks?~\protect\footnote{
      J.~A.~McFayden,
      D.~Napoletano,
      E.~Re}{}}
\label{sec:MC_g2qqFSR}

  We perform a comparison of three different Monte Carlo generators
  for a process involving $B$-hadrons in the final state, with the aim
  of assessing to which extent NLO tools, based on a $4$-flavour
  scheme, are successful in simulating kinematic configurations
  sensitive to the gluon splitting to pair of (massive) $b$-quarks.

\subsection{Introduction}
The Monte Carlo modelling of processes featuring an heavy-quark pair
in the final state, possibly in association with an heavy
color-singlet, has been, for several years, one of the open issues for
Monte Carlo developers.

Due to their large mass, top quarks are always treated differently
with respect to the other QCD partons, i.e. they are always simulated
using exact matrix elements, and, nowadays, whenever possible, this is
done at NLO. Moreover, because of their large mass, it's nearly
impossible for a parton shower algorithm to produce a top-quark pair
from a gluon splitting, hence, even in (multijet-merged) NLO+PS
accurate predictions, the kinematics of top quarks is, to a very large
extent, dominated by the matrix-element description.

The situation is very different for bottom (and charm) quarks, and, in
fact, one of the open issues related to the treatment of heavy quarks
in event generators has to do with processes where $b$ quarks are
present. As detailed below, in this context one of the main problem is
the modelling of the $g\to Q\bar Q$ splitting, especially when
$b$-quarks are produced at small opening angles.  It turns out that
this region is relevant in several LHC analysis, not only when precise
SM measurements are performed and compared with
theory~\cite{Khachatryan:2016iob,Khachatryan:2016ipq,Aad:2014dvb,
  Aad:2013vka,Chatrchyan:2013zja}, but also when simulation tools
become necessary to simulate Higgs production (\emph{e.g.}
$b\bar{b}\to H$~\cite{Krauss:2016orf,Krauss:2017wmx})
or backgrounds with several $b$-jets. An important example belonging
to the latter category is the modelling of QCD production of $b\bar b$
pairs in association with vector bosons, which is one of the main
backgrounds for measuring associated Higgs-boson production with
$H\to b\bar b$ decay. In this case in fact, complications arise
from double $b$-taggs in a single jets, which  come from a
collinear $g\to b\bar b$ which is only seen as one jet.
Similarly, it is known~\cite{Cascioli:2013era,Jezo:2018yaf} that a proper
description of the $g\to b\bar b$ splitting (with mass effects) is very important
for the simulation of $t\bar t + b$ jets production, an irreducible
background for measurements of $t\bar t H$ production in the $H\to
b\bar b$ channel.

In this short contribution we want to have a critical look at how well
the production of final states involving $b$-quarks is described, by
means of a comparison among different generators, for the production
process $pp\to b\bar{b}$.

\subsection{Schematic outline of the problem}
Broadly speaking, the problem of simulating processes with $b$ quarks
can be formulated as follows: as the mass of the $b$ quark ($m_b$) is
relatively small, a $b$-quark is a quasi-massless particle. This means
that, a priori, in the limit $m_b\to 0$, a parton shower algorithm is
expected to simulate fairly well the kinematic regions dominated by
the splitting $g\to b\bar{b}$. One should also notice that, even in
the limit $m_b\to 0$, the splitting $g\to b\bar{b}$ does not contain
soft singularities, i.e. only a single collinear logarithm is present
(the splitting probability $dP_{g\to q\bar q}(q^2,z,\phi)$ is
proportional to $dq^2/q^2$, with $q^2$ being the invariant mass, or
the transverse momentum of the splitting).

When one needs to keep $m_b$ finite, it is natural to use a
four-flavour (4F) scheme, that is, a scheme where there are no PDF for
$b$-quarks (and hence no hard processes initiated by $b$-quarks). By
using these matrix elements, power suppressed effects are, by
definition, included, at the order of the matrix-element used.  For
phenomenological reasons, one also allows off-shell gluons, in a
time-like shower, to split into massive $b$-pairs. This last step is
intrinsically ambiguous, because the splitting is now, strictly
speaking, non-singular, whereas the basic principle underlying parton
shower algorithms is deeply connected to the logarithmic structure of
QCD splittings.  These ambiguities are related to the treatment of the
kinematics for a quasi-collinear splitting, as well as to the choice
of the renormalization scale $\mu$ at which the strong coupling is
evaluated. In modern parton showers, for massless splittings, the
latter is chosen to be the transverse momentum of the splitting
itself, as this choice guarantees that (known) subleading logarithmic
terms are dynamically taken into account by the algorithm. As soon as
the splitting contains massive quarks, several constraints based on
theoretical arguments are lifted, leaving room for several a-priori
legitimate choices~\cite{PY8gbb,Bellm:2015jjp,Schumann:2007mg}.

Despite the significant progress made in the last years to study and
improve the simulation of processes with $b$-quarks (or $b$-jets) in
the final
state~\cite{Krauss:2016orf,Krauss:2017wmx,GehrmannDeRidder:2011dm,
  Luisoni:2015mpa,Frederix:2011qg},
it's not yet clear what is the best strategy
to treat the $g\to b\bar b$ splitting in final-state showers. For
instance, the results presented in Ref.~\cite{Aaboud:2017vqt} or
refs.~\cite{Aad:2014dvb,Chatrchyan:2013zja} show that the data/theory
agreement for observables sensitive to the kinematics of $g\to b\bar
b$ splitting is not always satisfactory, especially when the splitting
is quasi-collinear. It's also worth noticing that this conclusion
often applies not only to simulations based on LO matrix elements, but
also on NLO ones. In short, it is fair to say that a better
understanding of the aforementioned splittings is still needed.

In this contribution, we decided to make a small step forward, by
performing a comparison, as fair as possible, among different
generators and by using different observables. In particular, for the
first time we will compare 4-flavour NLO predictions, matched to
different parton showers, for observables sensitive to the underlying
process $pp\to b\bar{b}$ (+ jets) at the LHC. Data from a recent ATLAS
measurements will also be included. In the following sections, we
describe the tools used and the analysis considered, and comment on
our findings. As we will observe, also for NLO-accurate tools,
differences are observed mostly when the dominant kinematics is
collinear $g\to b\bar b$ splitting, calling therefore for further
investigations. 

\subsection{Settings}
\subsubsection{Measurements and analysis}

Monte Carlo predictions are compared to a recent ATLAS measurement of
the production of $B$-hadron pairs in proton-proton collisions at
$\sqrt{s} = 8$ TeV~\cite{Aaboud:2017vqt}. The measurement considers
the $B(\rightarrow J/\psi[\rightarrow\mu\mu]+X)B(\rightarrow\mu+X)$
decay mode where normalised differential cross sections are measured
for observables designed to be sensitive to the different $B$-hadron
production modes. For example, $B$-hadron production is measured as a
function of $\Delta\phi$ and $\Delta R = \sqrt{\Delta\eta^2 -
  \Delta\phi^2}$. This includes measurement down to $\Delta R=0$, a
region sensitive to gluon-splitting to $b\bar{b}$. As described above,
these measurements can help developing more effective phenomenological
models for the $g\rightarrow Q\bar{Q}$ splittings.

These are the parameters that we will focus on for the remainder of this
study. The predictions described below are analysed at
$B$-hadron-level and the results are translated to the 3-muon-level
via transfer functions provided by in Ref.~\cite{Aaboud:2017vqt}.

A similar study was carried out in the context of~\cite{Badger:2016bpw},
where many set-ups and generator where compared against one another,
and data, in the case of the associated production of a vector boson
$Z,W$ with $b$-quarks. However in that case the most relevant effects
where coming from space-like (or initial state) splittings.

\subsubsection{Configuration of event generators used}
In the following we describe the generator setups used throughout. As
we are mainly interested in the region where the two $b$-flavoured
objects are not widely separated, we expect kinematical mass effects
to play an important role.  For this reason we chose to show only 4F
scheme results.  In the 4F scheme, $b$-quarks are decoupled from the
evolution of QCD, and can only appear as, massive, final state
particles as the result of a ``hard'' $g\to b\bar{b}$ splitting.

\Sherpa--\MEPSatNLO~\cite{Gehrmann:2012yg,Hoeche:2012yf}:
$pp\rightarrow b\bar{b} +$ up to $3j$, merging scale $Q_{Cut}=20$~GeV.
Renormalization, factorization and shower starting scales are
set according to \Sherpa's default choice.
Each emission is clustered until a suitable core, $2\rightarrow 2$,
process is found. Scales appearing in the
fixed order matrix elements are calculated as
\begin{equation}
  \mu_R =\mu_F =\mu_Q =-\frac{1}{\sqrt{2}}\left(\frac{s+t+u}{s\,t\,u}\right)\,,
\end{equation}
while for each emission from the shower the scale appearing in
$\alpha_s$, or in the Jacobian PDF ratio, is set to be the $k_T$ of
the splitting. Instead for the $g\rightarrow Q\bar{Q}$ splitting the
scales are set equal to the invariant mass of the $Q\bar{Q}$ pair.
Jets are clustered using the anti-$k_T$~\cite{Cacciari:2008gp}
algorithm with a $p_T$ cut of 5~GeV. Finally, results are obtained
using the NNPDF30 NNLO PDF~\cite{Ball:2014uwa} set with
$\alpha_s(m_Z) = 0.118$, the default choice in \Sherpa.

\MGaMC{}~\cite{Alwall:2014hca}+\Pyegt{}~\cite{Sjostrand:2007gs}:
\MGaMC{}v2.5.5 is used to provide production of $pp\rightarrow
b\bar{b}$ events at NLO in the 4-flavour scheme with massive
$b$-quarks that are subsequently showered by \Pyegt{}.230. In the
matrix element calculation a dynamical renormalization and factorization scale, defined as the
sum of the transverse masses divided by two of all final state
particles and partons, and the NNPDF30 NNLO PDF set with
$\alpha_s(m_Z) = 0.118$ are used. In \Pyegt the $g\to b\bar b$ splitting
scale is ``Option 1'' as described in Ref.~\cite{PY8gbb} which uses
$k_T$-based scale. The A14 tune~\cite{ATL-PHYS-PUB-2014-021} is used.

\HW{}~\cite{Bahr:2008pv,Bellm:2015jjp,Bellm:2017bvx}: \HW{}.1 is used to
provide $pp\rightarrow b\bar{b}$ events at NLO in the 4-flavour scheme
with massive $b$-quarks. The MC@NLO matching scheme is used along with
the angular-ordered parton shower. In the matrix element calculation
the maximum jet $p_T$ is used for the renormalization and factorization scales,
and the MMHT14 PDF set~\cite{Harland-Lang:2014zoa} with $\alpha_s(m_Z) = 0.118$ is
used. The default \HW tune is used and the $g\to b\bar b$ splitting
scale is defined as the virtuality of the splitting.

\subsection{Comparisons and discussion}
\begin{figure}[t]
  \begin{center}
    \includegraphics[width=0.48\textwidth]{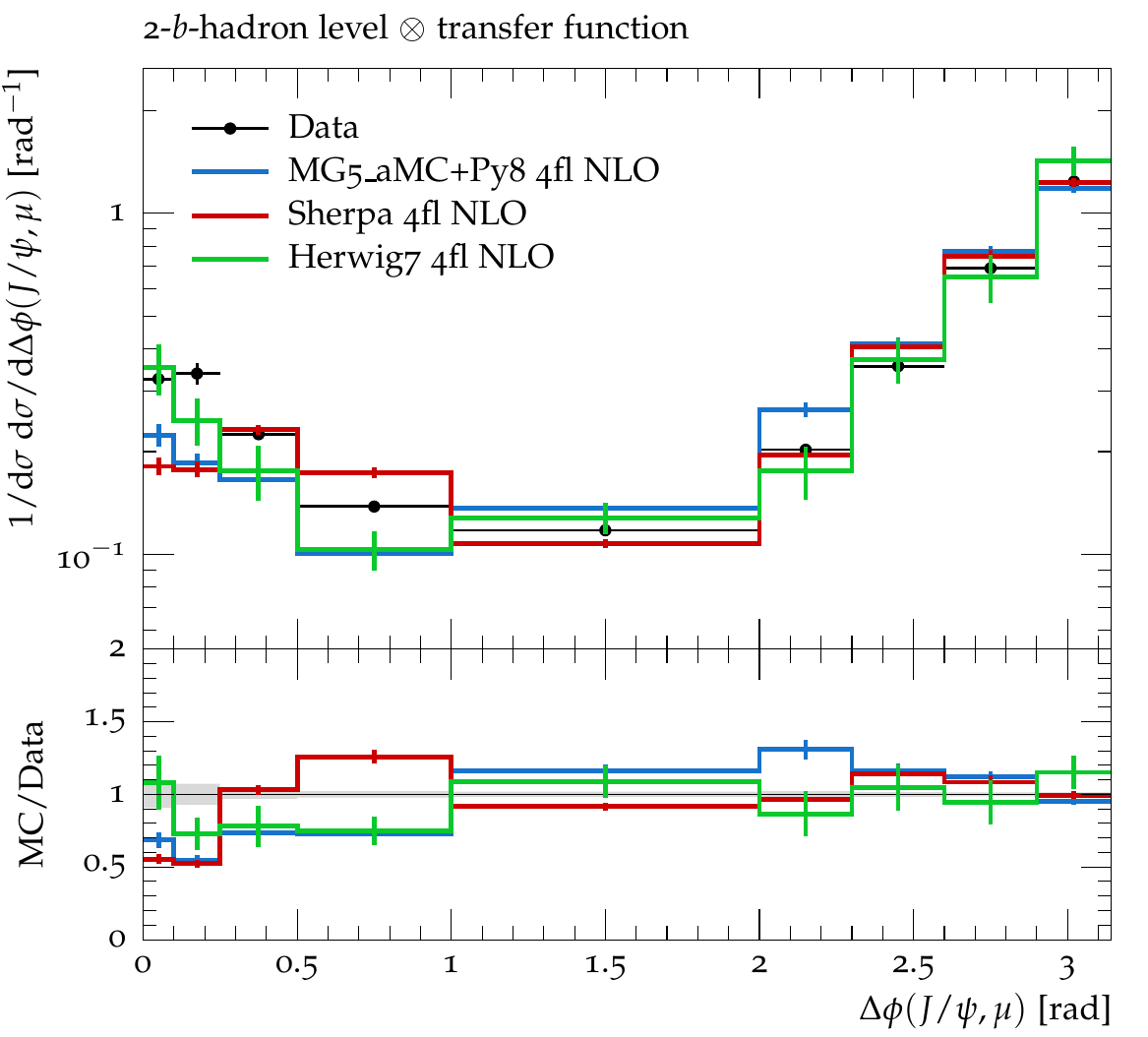}\hfill
    \includegraphics[width=0.48\textwidth]{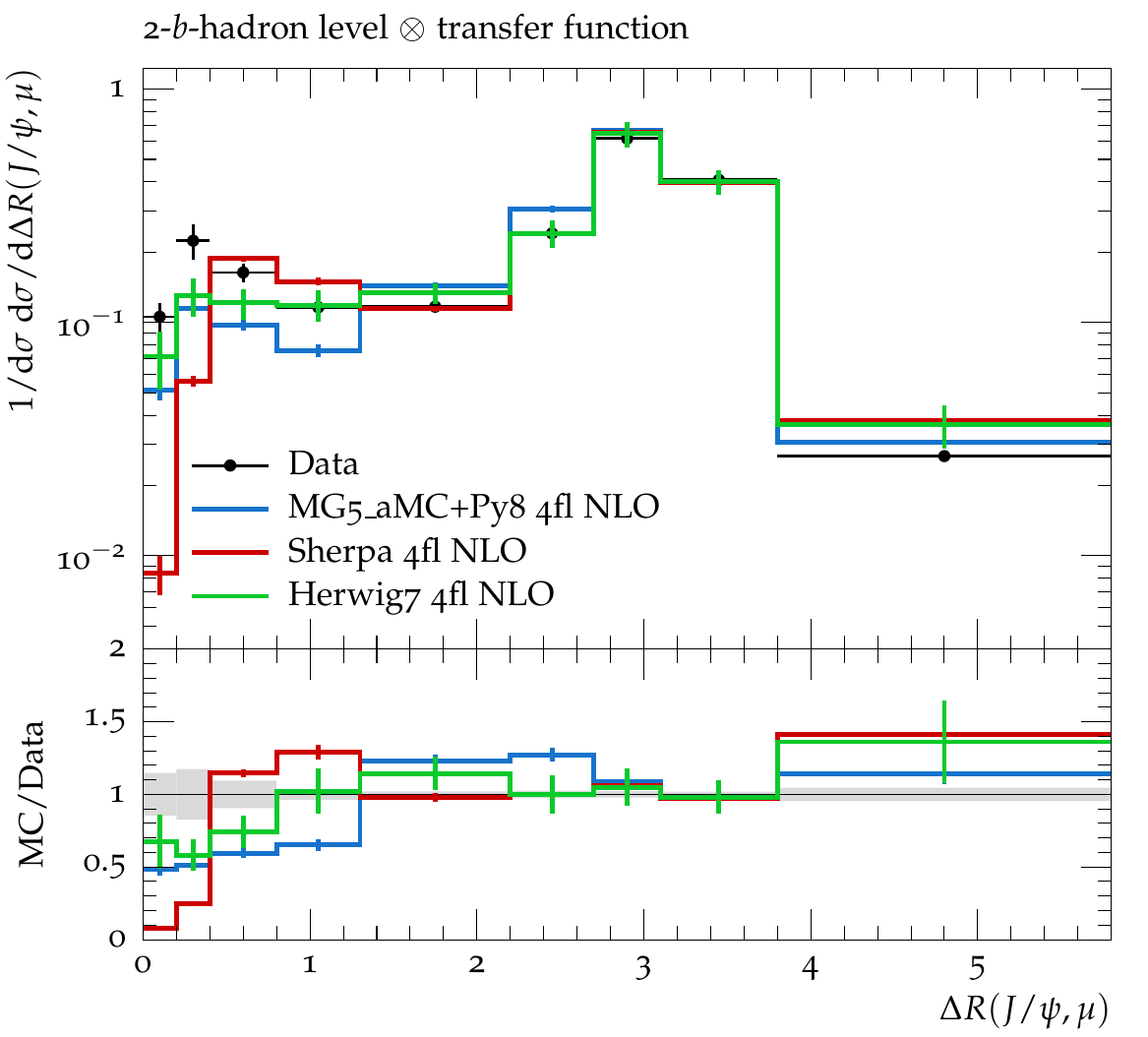}\\
    \includegraphics[width=0.48\textwidth]{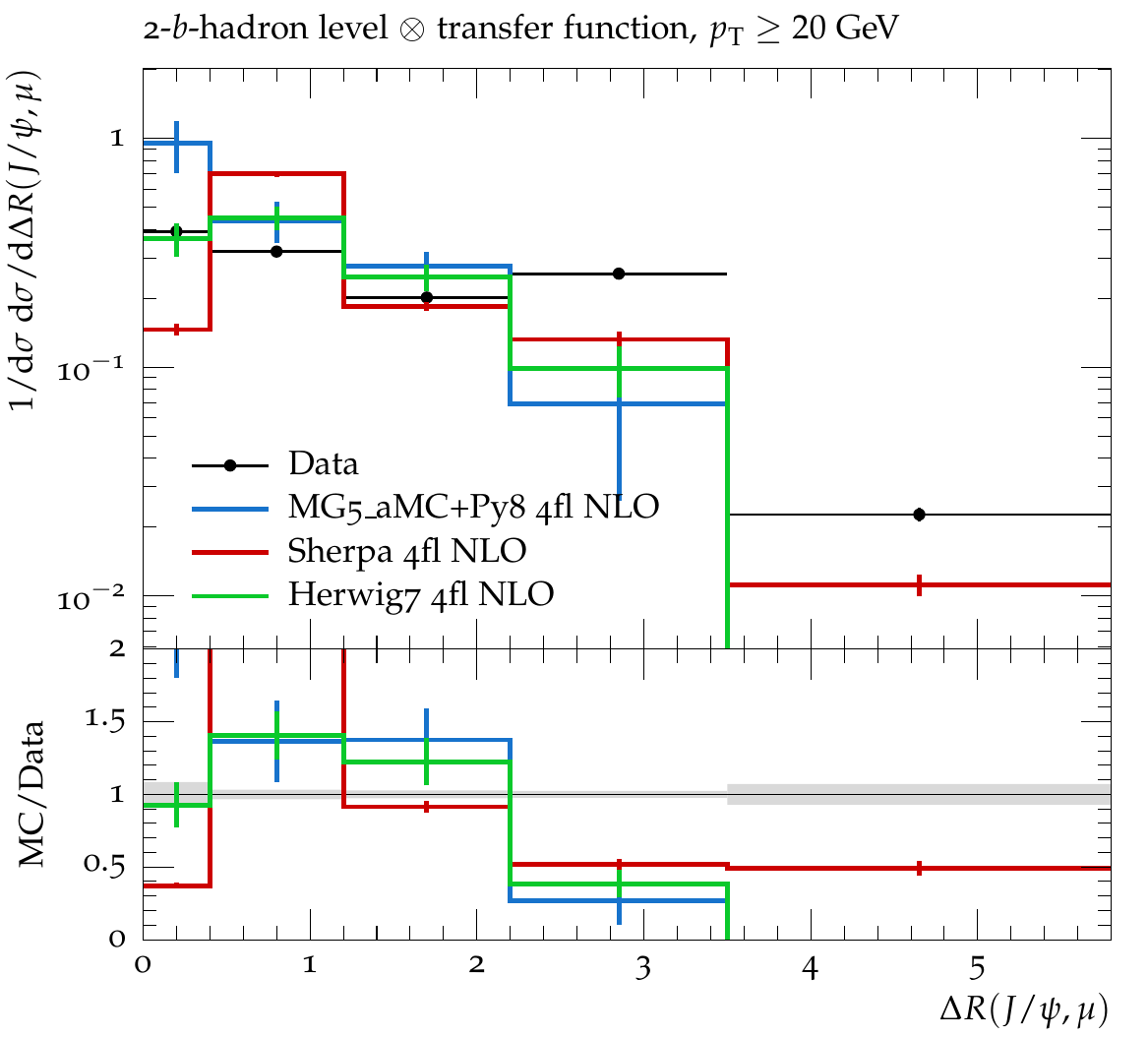}
    \caption{Top-left plot: $\Delta\phi$ distance between the $J/\psi$ and
      the third muon; Top-right plot: $\Delta R$ distance between the
      $J/\psi$ and the third muon, in the fully inclusive case;
      Bottom plot: $\Delta R$ distance for $p_T(J/\psi,\mu)\geq20$~GeV}
  \label{fig:MC_g2qqFSR:dphi_dr}
  \end{center}
\end{figure}

Results described in the previous section, compared to experimental
results are collected in Fig.~\ref{fig:MC_g2qqFSR:dphi_dr}, where we show
normalized distributions for the azimuthal angular separation
$\Delta\phi = \phi_2 -\phi_1$ between the $J/\psi$ and the third muon in the
azimuthal plane (top left), and the $\Delta R$ distance among the
same two final states, both in the inclusive case (top right), and for
$p_T(J/\psi,\mu)\geq20$~GeV (bottom).
We start our discussions by noting that in all three plots two regions
can be identified: a matrix element region and a parton shower one.
Leaving, for a moment, the transition between the two regimes aside,
it appears that while in the former a good level of agreement is shown
among the three Monte Carlo predictions\footnote{modulo large
  statistical uncertainties}, and between Monte Carlo and data, the
latter present visible discrepancies.  The former statement is somehow
obvious: apart from scale and PDF choices there are not many places in
which the three generators can be much different. Note that scale/PDF
variation bands are not shown, however it is to be expected that they
would cover the difference in the matrix element region,~\emph{i.e.}
the region $\Delta \phi\gtrsim 0.3$, $\Delta R\gtrsim 0.5$.

On the other hand, in the shower dominated regime,
one could argue that the spread in Monte Carlos
is a testament of the size of theory uncertainties.
To see how such large difference can come up,
let us remind the reader that the splitting
kernel for the $g\rightarrow Q\bar{Q}$, that enters the Sudakov form factor
splitting can, symbolically, be written as~\cite{PY8gbb}
\begin{equation}
  P_{g\rightarrow Q\bar{Q}} \propto \frac{\alpha_s(X)}{2\,\pi}\left[
    P_{g\to q\bar{q}}^{(m_q =0)}\,
    +\,\frac{m_Q^2}{Y_{Q\bar{Q}}}\right]\,.
\end{equation}
While the ambiguity in the exact choice of $Y_{Q\bar{Q}}$ is a genuine
subleading contribution in the shower, the actual value of $X$, while
being formally higher order, can be quite large in specific phase-space
corners~\cite{Aaboud:2017vqt}, like the in the small
$\Delta \phi$ and $\Delta R$ region.

Crucially, the three shower implementations we compare here,
as discussed in the previous section, differ in the choice
of both $X$ and $Y$, and this is most likely the reason for
the difference displayed in the plots we show here.

\subsection{Conclusions}

In this study we have observed that even 4-flavour NLO computations
for $pp\to b\bar b$, matched to parton showers, don't describe very
accurately the $\Delta R$ distribution among $B$-hadrons, in the
region that receives an important contribution from the kinematical
conigurations associated to a (quasi)-collinear $g\to b\bar b$
splittings. Given that all the simulations considered contain the same
(or very similar) perturbative accuracy at the level of hard matrix
elements, it is very likely that a large part of the observed
disagreement can be attributed to the parton showering part. This is
also reflected by the spread of predictions obtained with different
tools.

It would be highly desirable to perform further, dedicated studies to
ameliorate the theoretical understanding of the $g\to b\bar b$
splitting, and/or to find processes and observables that enhance the
contribution of gluon splitting. In fact, having a better
discrimination power among different choices for the modelling of
gluon splitting into heavy quarks seems necessary in order to rule out
some a-priori legitimate choices by means of a comparison with
measured distributions. For instance the methods proposed in
Ref.~\cite{Ilten:2017rbd} look very promising. We also note that to
better understand the mechanics of $b$-quark pair production in
regimes important for $H\to b\bar b$ production, the $p_T$ dependence
of $g \to b \bar b$ production would be important to measure. We hope
that our study will help as a motivation and starting point for future
studies on this topic.

\subsection*{Acknowledgements}
The work of DN has received funding from the European Union's Horizon
2020 research and innovation programme as part of the Marie
Sk\l{}odowska-Curie Innovative Training Networks ``MCnetITN3'' (grant
agreement no. 722104) and ``HiggsTools'' (grant agreements
PITN-GA2012- 316704) and by the ERC Advanced Grant MC@NNLO (340983).
The work of ER is supported by a Maria Sk\l{}odowska-Curie Individual
Fellowship of the European Commission's Horizon 2020 Programme under
contract number 659147 PrecisionTools4LHC.


\let\Sherpa\undefined
\let\Comix\undefined
\let\MEPSatNLO\undefined
\let\MGaMC\undefined
\let\Pyegt\undefined
\let\HW\undefined


\clearpage

\bibliographystyle{atlasnote}
\bibliography{LH17}



\end{document}